\documentclass[1p,authoryear]{elsarticle}

\usepackage{amssymb}
\usepackage{epsfig}
\usepackage{amsmath}
\usepackage{enumerate}
\usepackage{subfigure}
\usepackage{natbib}

\newcommand{\be}{\begin{equation}}
\newcommand{\ee}{\end{equation}}
\newcommand{\ba}{\begin{eqnarray}}
\newcommand{\ea}{\end{eqnarray}}
\newcommand{\grad}{\ensuremath{\vec{\nabla}}}
\newcommand{\lap}{\ensuremath{\Delta}}
\newcommand{\adotoa}{\ensuremath{{\cal H}}}
\newcommand{\tadotoa}{\ensuremath{{\cal \tilde{H}}}}
\newcommand{\Lie}[1]{\ensuremath{{\cal L}_{#1}}}
\newcommand{\twoform}[1]{\ensuremath{\underline{\underline{#1}}} }
\newcommand{\tr}{\ensuremath{\mbox{tr} } }
\newcommand{\oneform}[1]{\ensuremath{\underline{#1}}} 
\newcommand{\GammaGI}{\ensuremath{\hat{\Gamma}}}
\newcommand{\curv}[1]{\ensuremath{#1\kappa}}

\newcommand{\PsiGI}{\ensuremath{\hat{\Psi}}}
\newcommand{\PhiGI}{\ensuremath{\hat{\Phi}}}

\newcommand{\etal}{{\it et al.}}
\newcommand{\phib}{\ensuremath{\bar{\phi}}}
\newcommand{\metE}{\ensuremath{\tilde{g}}}
\newcommand{\metM}{\ensuremath{g}}
\newcommand{\RiemE}{\ensuremath{\tilde{R}}}
\newcommand{\volE}{\ensuremath{\sqrt{-\metE}}}
\newcommand{\volM}{\ensuremath{\sqrt{-\metM}}}
\newcommand{\metS}{\ensuremath{\hat{g}}}
\newcommand{\connE}{\ensuremath{\tilde{\nabla}}}
\newcommand{\EinE}{\ensuremath{\tilde{G}}}

\newcommand{\Vp}{\ensuremath{\frac{dV}{d\mu}}}
\newcommand{\PhiE}{\ensuremath{\tilde{\Phi}}}
\newcommand{\PsiE}{\ensuremath{\tilde{\Psi}}}
\newcommand{\zetaE}{\ensuremath{\tilde{\zeta}}}
\newcommand{\nablac}{\ensuremath{\nabla}}
\newcommand{\nablas}{\ensuremath{ \overset{*}{\nabla}}}
\newcommand{\nablag}{\ensuremath{{\;}^{\Gamma}\nabla}}
\newcommand{\RR}{\ensuremath{\;^{\Gamma}R}}
\newcommand{\GG}{\ensuremath{\;^{\Gamma}G}}

\bibpunct{[}{]}{,}{n}{}{}

\def\R{{\cal R}}
\def\T{{\cal T}}

\def\G{{\cal G}}
\def\g{\gamma}
\def\mn{{\mu\nu}}

\def\ab{{ab}}
\def\AB{{AB}}
\def\TT{\tilde{T}}
\def\half{\frac{1}{2}}
\def\quarter{\frac{1}{4}}
\def\del{\partial}

\def\apj{ApJ}%
\def\apjl{ApJ}%
\def\apjs{ApJS}%
\def\apss{Ap\&SS}%
\def\aap{A\&A}%
\def\jcap{J. Cosmology Astropart. Phys.}%
\def\mnras{MNRAS}%
\def\prd{Phys.~Rev.~D}%
\def\nat{Nature}%
\def\grl{Geophys.~Res.~Lett.}%
\journal{Physics Reports}

\begin{document}


\begin{frontmatter}

\title{Modified Gravity and Cosmology}
\author[Ox]{Timothy~Clifton}
\author[Ox]{Pedro~G.~Ferreira}
\author[Nott]{Antonio~Padilla}
\author[Nott]{Constantinos~Skordis}
\address[Ox]{Department of Astrophysics, University of Oxford, UK.}
\address[Nott]{School of Physics and Astronomy, University of Nottingham, UK.}
\begin{abstract}
In this review we present a thoroughly comprehensive survey of recent
work on modified theories of gravity and their cosmological consequences.  
Amongst other things, we cover General Relativity, Scalar-Tensor,
Einstein-Aether, and Bimetric theories, as well as TeVeS, $f(R)$,
general higher-order theories, Ho\v rava-Lifschitz gravity, Galileons,
Ghost Condensates, and models of extra dimensions including
Kaluza-Klein, Randall-Sundrum, DGP, and higher co-dimension
braneworlds. We also review attempts to construct a Parameterised
Post-Friedmannian formalism, that can be used to constrain deviations
from General Relativity in cosmology, and that is suitable for
comparison with data on the largest scales. 
These subjects have been intensively studied over the past decade,
largely motivated by rapid progress in the field of observational
cosmology that now allows, for the first time, precision tests
of fundamental physics on the scale of the observable Universe. 
The purpose of this review is to provide a reference tool for researchers and
students in cosmology and gravitational physics, as well as a
self-contained, comprehensive and up-to-date introduction to the
subject as a whole.
\end{abstract}

\end{frontmatter}


\newpage

\tableofcontents


\newpage

\section{Introduction}
\label{intro}

The General Theory of Relativity is an astounding accomplishment:
Together with quantum field theory, it is now widely considered to be one
of the two pillars of modern physics.  The theory itself is couched in the language of
differential geometry, and was a pioneer for the use of modern mathematics in
physical theories, leading the way for the gauge theories
and string theories that have followed.  It is no exaggeration to say
that General Relativity set a new tone for what a physical theory can
be, and has truly revolutionised our understanding of the Universe.

One of the most striking facts about General Relativity is that,
after almost an entire century, it remains completely unchanged: The
field equations that Einstein communication to the Prussian Academy of
Sciences in November 1915 are still our best
description of how space-time behaves on macroscopic scales.  These are
\begin{eqnarray}
G_{\mu\nu}=\frac{8\pi G}{c^4} T_{\mu\nu}
\end{eqnarray}
where $G_{\mu\nu}$ is the Einstein tensor, $T_{\mu\nu}$ is the energy momentum tensor, $G$ is
Newton's constant,  and $c$ is the speed of light.  It is these
equations that are thought to govern the expansion of the Universe,
the behaviour of black holes, the propagation of gravitational waves,
and the formation of all structures in the Universe from planets and
stars all the way up to the clusters and super-clusters of galaxies
that we are discovering today.  It is only in the microscopic world of
particles and high energies that General Relativity is thought to be
inadequate.  On all other scales it remains the gold standard.

The great success of General Relativity, however, has not stopped alternatives
being proposed.  Even during the very early days after Einstein's
publication of his theory there were proposals being made
on how to extend it, and incorporate it in a larger, more unified theory. Notable examples
of this are Eddington's theory of
connections, Weyl's scale independent theory, and the higher dimensional theories of Kaluza and Klein. To some
extent, these early papers were known to have been influential on
Einstein himself.  They certainly influenced the physicists who came
after him.

The ideas developed by Eddington during this period were later picked
up by Dirac, who pointed out the apparent coincidence between the
magnitude of Newton's constant and the ratio of the mass and scale of
the Universe.  This relationship between a fundamental constant and
the dynamical state of a particular solution led Dirac to conjecture
that Newton's constant may, in fact, be varying with time.  The
possibility of a varying Newton's constant was picked up again in the 1960s by
Brans and Dicke who developed the prototypical version of what are now
known as scalar-tensor theories of gravity.  These theories are still
the subject of research today, and make up Section
\ref{scalartensorsection} of our report.

Building on the work of Hermann Weyl, the Soviet physicist Andrei
Sakharov proposed in 1967 what would prove to be one of the most
enduring theories of modified gravity. In Sakharov's approach, the
Einstein-Hilbert action, from which the Einstein field equations can
be derived, is simply a first approximation to a much more complicated
action: Fluctuations in space-time itself lead to higher powers
corrections to Einstein's theory. In 1977 Kellogg Stelle showed
formally that these theories are renormalizable in the presence of
matter fields at the one loop level. This discovery was
followed by a surge of interest, that was boosted again later on by the
discovery of the potential cosmological consequences of these theories, as found by
Starobinsky and others.  In Section \ref{HD} we review this work.

The idea of constructing a quantum field theory of gravity
started to take a front seat in physics research during the 1970s and 80s, with the rise of
super-gravity and super-string theories.  Both of these proposals rely
on the introduction of super-symmetry, and signalled a resurgence in
the ideas of Kaluza and Klein involving higher dimensional spaces.
Boosted further by the discovery of D-branes as fundamental objects in
string theories, this avenue of research led to a vastly richer set of
structures that one could consider, and a plethora of proposals were
made for how to modify the effective field equations in four dimensions.  In
Section \ref{HD2} we review the literature on this subject.

By the early 1970s, and following the `golden age' of general
relativity that took place in the 1960s, there was a wide array of candidate theories
of gravity in existence that could rival Einstein's.  A formalism was
needed to deal with this great abundance of possibilities, and this
was provided in the form of the Parameterised Post-Newtonian (PPN) formalism by
Kenneth Nordtvedt, Kip Thorne and Clifford Will.  The PPN formalism was built on the
earlier work of Eddington and Dicke, and allowed for the numerous theories
available at the time to be compared to cutting edge astrophysical
observations such as lunar laser ranging, radio echo, and, in 1974,
the Hulse-Taylor binary pulsar.  The PPN formalism provided a clear structure within which one could compare
and assess various theories, and has been the benchmark for how theories
of gravity should be evaluated ever since.  We will give an outline of
the PPN formalism, and the constraints available within it today, in
Section \ref{GR}.

The limits of General Relativity have again come into focus with the
emergence of the `dark universe' scenario.  For almost thirty years there has existed evidence
that, if gravity is governed by Einstein's field equations, there
should be a substantial amount of `dark matter' in galaxies and
clusters.  More recently, `dark energy' has also been found to be required in
order to explain the apparent accelerating expansion of the
Universe.  Indeed, if General Relativity is correct, it now seems that
around  $96\%$ of the Universe should be in the form of energy
densities that do not interact electromagnetically.  Such an odd
composition, favoured at such high confidence, has led some to
speculate on the possibility that General Relativity may not, in fact,
be the correct theory of gravity to describe the Universe on the
largest scales.  The dark universe may be just another signal that
we need to go beyond Einstein's theory.

The idea of modifying gravity on cosmological scales has really taken
off over the past decade.  This has been triggered, in part, by
theoretical developments involving higher dimensional theories,
as well as new developments in constructing renormalizable theories of
gravity.  More phenomenologically, Bekenstein's relativistic
formulation of Milgrom's Modified Newtonian Dynamics (MoND) has
provided a fresh impetus for new study: What was previously a rule
of thumb for how weak gravitational fields might behave in regions of low
acceleration, was suddenly elevated to a theory that could be used to study
cosmology.  Insights such as Bertschinger's realisation that
large-scale perturbations in the Universe can be directly related to
the overall expansion rate have also made it possible to
characterise large classes of theories simply in terms of how they
make the Universe evolve.  Finally, and just as importantly, there has been tremendous progress
observationally.  A key step here has been the measurement of the growth of
structure at redshifts of $z\simeq 0.8$, by Guzzo and his
collaborators. With these measurements one can test, and reject, a
large number of proposals for modified gravity. This work is
complemented by many others that carefully consider the impact of
modifications to gravity on the cosmic microwave background, weak
lensing and a variety of other cosmological probes. As a result,
testing gravity has become one of the core tasks of many current, and
future, cosmological missions and surveys.

In this report we aim to provide a comprehensive exposition of the
many developments that have occurred in the field of modified gravity
over the past few decades.  We will focus on how these theories differ
from General Relativity, and how they can be distinguished from it, as
well as from each other.  A vast range of modified theories now exist
in the literature. Some of these have extra scalar, vector or tensor
fields in their gravitational sector; some take Sakharov's idea
in an altogether new direction, modifying gravity in regions of
low, rather than high, curvature; others expand on the ideas first put
forward by Kaluza and Klein, and take them into new realms by invoking
new structures.  Indeed, as the reader will see from our table of
contents, there are now a great many possible ways of modifying gravity
that can, in principle, be tested against the real Universe.  We will
attempt to be as comprehensive in this report as we consider it
reasonably possible to be.  That is, we will attempt to cover as many aspects
of as many different theories as we can.

To be able to efficiently assess the different candidate theories of gravity we have opted
to first lay down the foundations of modern gravitational physics and
General Relativity in Section \ref{GR}.  We have aimed to make this a
self-contained section that focuses, to some extent, on why general
relativity should be considered `special' among the larger class of
possibilities that we might consider.  In this section we also survey the current
evidence for the `dark universe', and explain why it has become the
standard paradigm.  From here we move on to discuss and compare
alternative theories of gravity and their observational consequences.
While the primary focus of this report is to elucidate 
particular theories, we will also briefly delve into the recent
attempts that have been made to construct a formalism, analogous to
the PPN formalism, for the cosmological arena.  We
dub these approaches `Parameterised Post Friedmannian'.

Let us now spell out the conventions and definitions that we will use
throughout this review.
We will employ the `space-like convention' for the metric, such that
when it is diagonalised it has the signature $(-+++)$. We will choose
to write space-time indices using the Greek alphabet, and space
indices using the Latin alphabet.  Where convenient, we will also
choose to use units such that speed of light is equal to $1$.  Under
these conventions the line-element for Minkowski space, for example,
can then be written 
\be
ds^2=\eta_{\mu \nu} dx^{\mu} dx^{\nu} = -dt^2 + dx^2+dy^2+dz^2.
\ee
For the Riemann and Einstein curvature tensors we will adopt the
conventions of Misner, Thorne and Wheeler \cite{MTW}:
\begin{eqnarray}
R^{\mu}_{\phantom{\mu} \nu \alpha \beta} &=& \partial_{\alpha}
\Gamma^{\mu}_{\phantom{\mu} \nu \beta}- \partial_{\beta}
\Gamma^{\mu}_{\phantom{\mu} \nu \alpha} + \Gamma^{\mu}_{\phantom{\mu}
  \sigma \alpha} \Gamma^{\sigma}_{\phantom{\sigma} \nu \beta} -
\Gamma^{\mu}_{\phantom{\mu} 
  \sigma \beta} \Gamma^{\sigma}_{\phantom{\sigma} \nu \alpha}\\
G_{\mu \nu} &=& R_{\mu \nu} - \frac{1}{2} g_{\mu \nu} R,
\end{eqnarray}
where $R_{\mu \nu} = R^{\alpha}_{\phantom{\alpha} \mu \alpha \nu}$ and
$R=R^{\alpha}_{\phantom{\alpha} \alpha}$.  The energy-momentum tensor
will be defined with respect to the Lagrangian density for the matter
fields as
\be
T^{\mu \nu} = \frac{2}{\sqrt{-g}} \frac{\delta \mathcal{L}_m}{\delta
  g_{\mu \nu}},
\ee
where the derivative here is a functional one.  Throughout this review
we will refer to the energy density of a fluid as $\rho$, and its
isotropic pressure as $P$.  The equation of state, $w$, is then
defined by
\be
P = w \rho.
\ee
When writing the Friedmann-Lema\^{i}tre-Robertson-Walker (FLRW)
line-element we will use $t$ to denote the `physical time' (proper 
time of observers comoving with the fluid), and $\tau = \int dt/a(t)$
to denote the `conformal time' coordinate.  Unless otherwise stated,
when working with linear perturbations about an FLRW background we
will work in the conformal Newtonian gauge in which
\be
ds^2 = a^2(\tau) \left[ -(1+2 \Psi) d\tau^2+(1-2 \Phi) q_{ij}
  dx^i dx^j \right],
\ee
where $q_{ij}$ is the metric of a maximally symmetric 3-space
with Gaussian curvature $\kappa$:
\be
ds_{(3)}^2=q_{ij} dx^i dx^j = \frac{dr^2}{1-\kappa r^2} +r^2
d\theta^2+r^2 \sin^2 \theta d \phi^2.
\ee
When dealing with time derivatives in cosmology we will use the dot
and prime operators to refer to derivatives with respect to physical
and conformal time, respectively, such that
\begin{eqnarray}
\dot{\;} &\equiv& \frac{d}{dt}\\
\;^{\prime} &\equiv& \frac{d}{d\tau}.
\end{eqnarray}
In four dimensional space-time we will denote covariant derivatives
with either a semi-colon or a $\nabla_{\mu}$.  The four dimensional
d'Alembertian will then be defined as
\be
\Box \equiv g^{\mu \nu} \nabla_{\mu} \nabla_{\nu}.
\ee
On the conformally static three-dimensional space-like hyper-surfaces
the grad operator will be denoted with an arrow, as $\grad_i$, while
the Laplacian will be given by
\be
\Delta \equiv q^{ij} \grad_i \grad_j.
\ee
As is usual, we will often make use of the definition of the Hubble
parameter defined with respect to both physical and conformal time as
\begin{eqnarray}
H &\equiv& \frac{\dot{a}}{a}\\
\mathcal{H} &\equiv& \frac{a^{\prime}}{a}.
\end{eqnarray}
The definitions we have made here will be restated at various points
in the review, so that each section remains self-contained to
a reasonable degree.  The exception to this will be Section \ref{HD2},
on higher dimensional theories, which will require the introduction of
new notation in order to describe quantities in the bulk.

Let us now move onto the definitions of particular terms.  We choose
to define the equivalence principles in the following way:
\begin{itemize}
\item
Weak Equivalence Principle (WEP):  All uncharged, freely falling test 
particles follow the same trajectories, once an initial position and 
velocity have been prescribed.
\item
Einstein Equivalence Principle (EEP):  The WEP is valid, and 
furthermore in all freely falling frames one recovers (locally, and up 
to tidal gravitational forces) the same laws of special relativistic physics, 
independent of position or velocity.
\item
Strong Equivalence Principle (SEP):  The WEP is valid for massive 
gravitating objects as well as test particles, and in all freely 
falling frames one recovers (locally, and up to tidal gravitational 
forces) the same special relativistic physics, independent of position 
or velocity.
\end{itemize}
Of these, the EEP in particular is known to have been very influential
in the conception of General Relativity.  One may note that some
authors refer to what we have called the EEP as the `strong
equivalence principle'.

Let us now define what we mean by `General Relativity'.  This term is
often used by cosmologists to refer simply to Einstein's
equations.  Particle physicists, on the other hand, refer to
any dynamical theory of spin-2 fields that incorporates general
covariance as `general 
relativity', even if it has field equations that are different from
Einstein's\footnote{Note that under this definition the Einstein-Hilbert and
  Brans-Dicke Lagrangians, for example, represent different models of
  the same theory, which is called General Relativity.}. 
In this report when we write about `General Relativity' we
refer to a theory that simultaneously exhibits general covariance,
and universal couplings to all matter fields, as well as satisfying Einstein's field equations.  When
we then discuss `modified gravity' this will refer to any modification
of any of these properties.  However, it will
be clear from reading through this report that almost all the
proposals we report on preserve general covariance, and the universality of
free fall.  Let us now clarify further what exactly we mean by `modified'
theories of gravity.

As we will discuss in the next section, the effect of gravity on matter is
tightly constrained to be mediated by interactions of the matter
fields with a single rank-2 tensor field.  This does not mean that
this field is the only degree of freedom in the theory, but that
whatever other interactions may occur, the effect of gravity on the
matter fields can only be through interactions with the rank-2 tensor
(up to additional weak interactions that are consistent with the
available constraints).  The term `gravitational theory' can then be
functionally defined by the set of field equations obeyed by the rank-2
tensor, and any other non-matter fields it interacts with.  If these
equations are anything other than Einstein's equations, then we
consider it to be a `modified theory of gravity'.  We will not appeal
to the action or Lagrangian of the theory itself here; our definition is an
entirely functional one, in terms of the field equations alone.

While we have constructed the definition above to be as simple as
possible, there are of course a number of ambiguities involved.
Firstly, exactly what one should consider as a `matter field' can
be somewhat subjective.  This is especially true in terms of the exotic
fields that are sometimes introduced into cosmology in order to try
and understand the apparent late-time accelerating expansion of the
Universe.  Secondly, we have not defined exactly what we mean by
`Einstein's equations'.  In four dimensions it is usually clear what
this term refers to, but if we allow for the possibility of
extra dimensions then we may choose for it to refer either to the
equations derived from an Einstein-Hilbert action in the higher
dimensional space-time, or to the effective set of equations in
four dimensional space-time.  Clearly these two possible definitions are not
necessarily consistent with each other.  Even in four dimensions it is
not always clear if `Einstein's equations' include the existence of a
non-zero cosmological constant, or not.

To a large extent, the ambiguities just mentioned are a matter of
taste, and have no baring on the physics of the situation.  For
example, whether one chooses to refer to the cosmological constant as a
modification of gravity, as an additional matter field, or as part of
Einstein's equations themselves makes no
difference to its effect on the expansion of the Universe.  In this
case it is only convention that states that the Einstein equations
with $\Lambda$ is not a modified theory of gravity.  Although less
established than the case of the cosmological constant, similar
conventions have started to develop around other modifications to the
standard theory. For example, quintessence fields that are minimally
coupled to
the metric are usually thought of as additional matter fields, whereas
scalar fields that non-minimally couple to the Einstein-Hilbert term
in the action are usually thought of as being `gravitational' fields
(this distinction existing despite what numerous studies call
non-minimally coupled quintessence fields).  Although not always
clear, we try to follow what we perceive to be the conventions that exist
in the literature in this regard.  We therefore include in this review
a section on non-minimally coupled scalar-tensor theories, but not a
section on minimally coupled quintessence fields. 

\newpage 

\section{General Relativity, and its Foundations}
\label{GR}

General Relativity is the standard theory of gravity.  Here we will
briefly recap some of its essential features, and foundations.
We will outline the observational tests of gravity that have been
performed on Earth, in the solar system, and in other astrophysical
systems, and we will then explain how and why it is that General Relativity
satisfies them. We will outline why General Relativity should be
considered a special theory in the more general class of theories that
one could consider, and will present some of the theorems it obeys as well as the
apparatus that is most frequently used to parameterise deviations away
from it.  This will be followed by a discussion of the cosmological
solutions and predictions of the concordance general relativistic
$\Lambda$CDM model of the Universe.

\subsection{Requirements for Validity}

In order to construct a relativistic theory of gravity it is of primary
importance to establish the properties it must satisfy in order
for it to be considered viable.  These include foundational
requirements, such as the universality of free fall and the isotropy of space,
as well as compatibility with a variety of different observations involving the
propagation of light and the orbits of massive bodies.
Today, radio and laser signals can be sent back and forth from the Earth to
spacecraft, planets and the moon, and detailed observations of the orbits
of a variety of different astrophysical bodies allow us to
look for ever smaller deviations from Newtonian gravity, as well as entirely new
gravitational effects. It is in this section that we will discuss the gravitational
experiments and observations that have so far been performed in
these environments.  We will discuss what they can tell us about
relativity theory, and the principles that a theory must obey in order
for it to stand a chance of being considered observationally viable.  

\subsubsection{The foundations of relativistic theories}

First of all let us consider the equivalence principles.  We will not
insist immediately that any or all of these principles are valid, but
will rather reflect on what can be 
said about them experimentally. This will allow us to separate out
observations that test equivalence principles, from observations that
test the different gravitational theories that obey these principles --
an approach pioneered by Dicke \cite{dickebook}.

The least stringent of the equivalence principles is the WEP.  The
best evidence in support of the WEP
still comes from E\"{o}tv\"{o}s type experiments that use a torsion
balance to determine the relative acceleration of two different
materials towards distant astrophysical bodies.  In reality these
materials are self-gravitating, but their mass is usually
small enough that they can effectively be considered to be
non-gravitating test particles in the gravitational field of
the astrophysical body.  Using beryllium and titanium the
tightest constraint on the relative difference in accelerations of the
two bodies, $a_1$ and $a_2$, is currently \cite{eotwash}
\be
\label{eotwash}
\eta = 2 \frac{\vert a_1 - a_2 \vert}{\vert a_1 + a_2 \vert} = (0.3
\pm 1.8) \times 10^{-13}.
\ee
This is an improvement of around $4$ orders of magnitude on the original
results of E\"{o}tv\"{o}s from 1922 \cite{eotvos}.  It is expected that this can
be improved upon by up to a further $5$ orders of magnitude when space based
tests of the equivalence principle are performed \cite{STEP}.  These
null results are generally considered to be a very tight constraint on the
foundations of any relativistic gravitational theory if it is to be thought
of as viable:  The WEP must be satisfied, at least up to the accuracy
specified in Eq. (\ref{eotwash}).

Let us now briefly consider the gravitational redshifting of light.  This is
one of the three ``classic tests'' of General Relativity, suggested by
Einstein himself in 1916 \cite{classic}.  It is not, however, a
particularly stringent test of relativity theory.  If we accept energy-momentum
conservation in a closed system then it is only really a test of the WEP, and
is superseded in its accuracy by the E\"{o}tv\"{o}s experiment we
have just discussed.  The argument for this is the following
\cite{dickebook,Haugan1979}: Consider an atom that initially has an
inertial mass $M_i$, and a gravitational mass $M_g$.  The atom starts
near the ceiling of a lab of height $h$, in a static gravitational
field of strength $g$, and with an energy reservoir on the lab floor
beneath it.  The atom emits a photon of energy $E$ that then travels down
to the lab floor, such that its energy has been blue-shifted by
the gravitational field to $E'$ when it is collected in the reservoir.
This process changes the inertial and gravitational masses of the atom
to $M_i'$ and $M_g'$, respectively.
The atom is then lowered to the floor, a process which lowers its
total energy by $M_g' g h$.  At this point, the atom re-absorbs a photon
from the reservoir with energy $E'=(M_i''- M_i')c^2$ and is then raised to
its initial position at the ceiling. This last process raises its energy by
$M_g'' g h$, where here $M_i''$ and $M_g''$ are the inertial and
gravitational masses of the atom after re-absorbing 
the photon.  The work done in lowering and raising the atom in this
way is then $w=(M_g'' - M_g') g h$.
Recalling that the energy gained by the photon in travelling from the
lab ceiling to the lab floor is $E'-E$, the
principle of energy conservation then tells us that $(E'-E)= w = (M_g''
- M_g') g h $.  Now, if the 
WEP is obeyed then $M_i=M_g$, and this equation simply becomes $(E'-E) =
E' g h$.  This, however, is just the usual expression for
gravitational redshift. Crucial here is the assumption that local
position invariance is valid, so that both $M_i$ and $M_g$ are
independent of where they are in the lab.
If the laws of physics are position independent, and energy is
conserved, gravitational redshift then simply tests the equivalence of
gravitational and inertial masses, which is what the E\"{o}tv\"{o}s
experiment does to higher accuracy.  Alternatively, 
if we take the WEP to be tightly constrained by the E\"{o}tv\"{o}s
experiment, then gravitational redshift experiments can be used to 
gain high precision constraints on the position dependence of the laws
of physics \cite{redshift}.  The 
gravitational redshift effect by itself, however, does {\it not}
appear to be able to 
distinguish between different theories that obey the WEP and local
position invariance.  In Dicke's approach it should therefore be
considered as a test of the foundations of relativistic gravitational
theories, rather than a test of the theories themselves.

The next most stringent equivalence principle is the EEP.  Testing
this is a considerably more demanding task than was the case for the
WEP, as one now not only has to show that different test particles follow
the same trajectories, but also that a whole set of special relativistic
laws are valid in the rest frames of these particles.  Despite the
difficulties involved with this, there is still very compelling
evidence that the EEP should also be considered valid to high
accuracy.  The most accurate and direct of this evidence is due to the Hughes-Drever
experiments \cite{hughes,drever}, which test for local spatial anisotropies by carefully
observing the shape and spacing of atomic spectral lines.  The basic
idea here is to determine if any gravitational fields beyond a single
rank-2 tensor are allowed to couple directly to matter fields.  To
see why this is of importance, let us first consider a number
of point-like particles coupled to a single rank-2 tensor,
$g_{\mu\nu}$.  The Lagrangian density for such a set of particles is
given by
\be
\label{Lpart}
\mathcal{L} = \sum_I \int m_I \sqrt{-g_{\mu\nu} u^\mu u^\nu} d \lambda,
\ee
where $m_I$ are the masses of the particles, and $u^\mu=d
x^\mu/d\lambda$ is their 4-velocity measured with respect to some
parameter $\lambda$.  The Euler-Lagrange equations derived from
$\delta \mathcal{L}=0$ then tell us that the particles in
Eq. (\ref{Lpart}) follow geodesics of the metric $g_{\mu\nu}$, and
Riemannian geometry tells us that at any point we can choose
coordinates such that $g_{\mu\nu}=\eta_{\mu\nu}$ locally.  We
therefore recover special relativity at every point, and the EEP is
valid.  Now, if the matter fields couple to {\it two} rank-2 tensors
then the argument used above falls apart.  In this case the Lagrangian
density of our particles reads
\be
\label{Lpart2}
\mathcal{L} = \sum_I \int \left[ m_I \sqrt{-g_{\mu\nu} u^\mu u^\nu}+
  n_I \sqrt{-h_{\mu\nu} u^\mu u^\nu} \right] d \lambda,
\ee
where $h_{\mu\nu}$ is the new tensor, and $n_I$ is the coupling of each
particle to that field.  The particles above can now no longer be
thought of as following the geodesics of any one metric, as the
Euler-Lagrange Equations (\ref{Lpart2}) are not in the form of
geodesic equations.  We therefore have no Riemannian geometry with
which we can locally transform to Minkowski space, and the EEP is
violated.  The relevance of this discussion for the Hughes-Drever
experiments is that EEP violating couplings, such as those in
Eq. (\ref{Lpart2}), cause just the type of spatial anisotropies
that these experiments constrain.  In this case the 4-momentum
of the test particle in these experiments becomes
\be
p_\mu = \frac{m g_{\mu\nu} u^\nu}{\sqrt{-g_{\alpha\beta} u^\alpha
  u^\beta}} + \frac{n h_{\mu\nu} u^\nu}{\sqrt{-h_{\alpha\beta}
  u^\alpha u^\beta }},
\ee
and as $g_{\mu\nu}$ and $h_{\mu\nu}$ cannot in general be made to be
simultaneously spatially isotropic, we then have that $p_\mu$ is
spatially anisotropic, and should cause the type of shifts and
broadening of spectral lines that Hughes-Drever-type experiments are
designed to detect.  The current tightest constraints are around $5$
orders of magnitude tighter than the original experiments of Hughes and
Drever \cite{HD1, HD2}, and yield constraints of the order
\be
n \lesssim 10^{-27} m,
\ee
so that couplings to the second metric must be very weak in order to
be observationally viable.  This
result strongly supports the conclusion that matter fields must be
coupled to a single rank-2 tensor only.  It then follows that
particles follow geodesics of this metric, that we can recover
special relativity at any point, and hence that the EEP is valid.  It
should be noted that these constraints do not apply to gravitational
theories with multiple rank-2 tensor fields that couple to matter in a
linear combination, so that they can be written as in Eq. (\ref{Lpart})
with $g_{\mu\nu} = \sum_I c_I h^{(I)}_{\mu\nu}$, where $c_I$ are a set of $I$
constants.  Local spatially isotropy, and the EEP, is always recovered
in this case.

Beyond direct experimental tests, such as Hughes-Drever-type
experiments, there are also theoretical reasons to think that the EEP
is valid to high accuracy.  This is a conjecture attributed to Schiff,
that states {\it `any complete and self-consistent gravitational theory that
obeys the WEP must also satisfy the EEP'}.  It has been shown using
conservation of energy that preferred frame and preferred location
effects can cause violations of the WEP \cite{Haugan1979}.  This goes some
way towards demonstrating Schiff's conjecture, but there is as yet
still no incontrovertible proof of its veracity.  We will not consider
the issue further here.

The experiments we have just described provide very tight constraints
on the WEP, the EEP, and local position invariance.  It is, of course,
possible to test various other aspects of relativistic
gravitational theory that one may consider as `foundational' (for
example, the constancy of a constant of nature \cite{constants}).
For our present purposes, however, we are mostly interested in the
EEP.  Theories that obey the EEP are often described as being `metric'
theories of gravity, as any theory of gravity based on a
differentiable manifold and a metric tensor that couples to matter, as
in Eq. (\ref{Lpart}), can be shown to have test particles that follow
geodesics of the resulting metric space.  The basics of
Riemannian geometry then tells us that at every point in the manifold
there exists a tangent plane, which in cases with Lorentzian
signature is taken to be Minkowski space.  This allows us to recover special
relativity at every point, up to the effects of second
derivatives in the metric (i.e. tidal forces), so that the EEP is
satisfied.  Validity of the EEP can then be thought of as implying
that the underlying gravitational theory should be a metric one \cite{willmetric}.

\subsubsection{Observational tests of metric theories of gravity}

In what follows we will consider gravitational experiments
and observations that can potentially be used to distinguish between
different metric theories of gravity.
\newline
\newline
\noindent
{\it Solar system tests}
\newline

As well as the gravitational redshifting of light that we have already mentioned,
the other two `classic tests' of General Relativity are the bending of
light rays by the Sun, and the anomalous perihelion precession of Mercury.
These can both be considered tests of gravitational theories {\it
  beyond} the foundational issues discussed in the previous section.
That is, each of these tests is (potentially) able to distinguish between different
metric theories of gravity.  As well as these two tests, there are also a
variety of other gravitational observations that can be performed in
the solar system in order to investigate relativistic gravitational
phenomena. A viable theory of gravity must be compatible with all of them.
For convenience we will split these into tests involving null
trajectories (such as light bending) and tests involving time-like
trajectories (such as the perihelion precession of planets).

First of all let us consider tests involving null geodesics.  As
already mentioned, the most famous of these is the spatial deflection
of star light by the Sun. 
In General Relativity the deflection angle, $\theta$, of a photon's
trajectory due to a mass, $M$, with impact parameter $d$, is given by
\be
\theta = \frac{2 M}{d} (1+ \cos \varphi ) \simeq 1.75^{\prime \prime},
\ee
where $\varphi$ is the angle made at the observer between the direction of the incoming
photon and the direction of the mass.  The $1.75^{\prime \prime}$
is for a null trajectory that grazes the limb of the Sun.  This result
is famously twice the size of the effect that one might naively
estimate using the equivalence principle alone \cite{einstein1911}.
The tightest observational constraint to date on $\theta$ is due to
Shapiro, David, Lebach and Gregory who use around $2500$ days worth of
observations taken over a period of $20$ years.  The data in this
study was taken using $87$ VLBI sites and $541$ radio sources, yielding
more than $1.7 \times 10^6$ measurements that use standard
correction and delay rate estimation procedures.  The result of
this is \cite{sdlg}
\be 
\theta = (0.99992 \pm 0.00023) \times 1.75^{\prime \prime},
\ee
which is around $3$ orders of magnitude better than the observations
of Eddington in 1919.

A further, and currently more constraining, test of metric theories of gravity
using null trajectories involves the Shapiro time-delay effect
\cite{shapiro}.  Here the deflection in time is taken into account
when a photon passes through the gravitational field of a massive
object, as well as the deflection in space that is familiar from the
lensing effects discussed above.  The effect of this in General Relativity is
to cause a time delay, $\Delta t$,  for a light-like signal reflected off a
distant test object given by
\be
\Delta t = 4 M \ln \left[ \frac{4 r_1 r_2}{d^2} \right] \simeq 20 \left(
12 - \ln \left[ \left( \frac{d}{R_{\odot}}\right)^2 \left(
\frac{au}{r_2} \right) \right] \right) \; \mu s, 
\ee
where $r_1$ and $r_2$ (both assumed $\gg d$) are the distances of the
observer and test object from an object of mass $M$, respectively.
The second equality here is the approximate magnitude of this effect
when the photons pass close by the Sun, and the observer is on
Earth. Here we have written $R_{\odot}$ as the radius of the Sun, and $au$ as
the astronomical unit.  The best constraint on gravity using this
effect is currently due to Bertotti, Iess and Tortora using
radio links with the Cassini spacecraft between the 6th of June and
the 7th of July 2002 \cite{bit}.  These observations result in the
constraint
\be
\Delta t = (1.00001 \pm 0.00001) \Delta t_{GR},
\ee
where $\Delta t_{GR}$ is the expected time-delay due to general
relativity.  The Shapiro time-delay effect in fact constrains the same
aspect of relativistic gravity as the spatial deflection of light
(this will become clear when we introduce the parameterised
post-Newtonian formalism later on).  This aspect is sometimes called
the `unit curvature' of space.

Let us now consider tests involving time-like trajectories.  The
`classical' test of General Relativity that falls into this category
is the anomalous perihelion precession of Mercury (this is called a
test, despite the fact that it was discovered long before General
Relativity \cite{mercuryold}).  In Newtonian physics the perihelion of a test
particle orbiting an isolated point-like mass stays in a fixed
position, relative 
to the fixed stars.  Adding other massive objects into the system perturbs
this orbit, as does allowing the central mass to have a non-zero
quadrupole moment, so that the perihelion of the test particle's orbit
slowly starts to precess.  In the solar
system the precession of the equinoxes of the coordinate system
contribute about $5025^{\prime \prime}$ per century to Mercury's
perihelion precession, while the other planets contribute about
$531^{\prime \prime}$ per century.  The Sun also has a non-zero
quadrupole moment, which contributes a further $0.025^{\prime
  \prime}$ per century.  Taking all of these effects into account,
it still appears that the orbit of Mercury in the solar system has an
anomalous perihelion precession that cannot be explained by the
available visible matter, and Newtonian gravity alone.  Calculating
this anomalous shift exactly is a complicated matter, and depends on the
exact values of the quantities described above.  In Table
\ref{merctab} we display the observed anomalous perihelion precession of
Mercury, $\Delta \omega$, as calculated by various different groups.
For a more detailed overview of the issues involved, and a number of other
results, the reader is referred to \cite{merc}.
\begin{table}[tbh]
\begin{center}
\begin{tabular}{|l|c|}
\hline
\qquad Source & \quad $\Delta \omega$ /($\;^{\prime \prime}$ per century)
\\ \hline
Anderson \textit{et al.} \cite{91} & $42.94 \pm 0.20$ \\ 
Anderson \textit{et al.} \cite{92} & $43.13 \pm 0.14$  \\ 
Krasinsky \textit{et al.} \cite{93a}: EPM$1988$ & $42.984 \pm 0.061$ \\ 
\hspace{106pt} DE$200$ & $42.977 \pm 0.061$ \\ 
Pitjeva \cite{93b}: EPM$1988$ & $42.963 \pm 0.052$ \\ 
\hspace{60pt} DE$200$ & $42.969 \pm 0.052$ \\ \hline
\end{tabular}%
\end{center}
\caption{The value of the perihelion precession of Mercury obtained from
observations by various authors. The acronyms
EPM$1988$ and DE$200$ refer to different numerical ephemerides, which are reviewed in 
\protect\cite{pit01}.}
\label{merctab}
\end{table}
In relativistic theories of gravity the additional post-Newtonian
gravitational potentials mean that the perihelion of a test particle
orbiting an isolated mass is no longer fixed, as these potentials do
not drop off as $\sim 1/r^2$.  There is therefore an additional
contribution to the perihelion precession, which is sensitive to the
relative magnitude and form of the gravitational potentials, and hence
the underlying relativistic theory.  For General Relativity, the
predicted anomalous precession of a two body system is given by
\be
\Delta \omega = \frac{6 \pi M}{p} \simeq 42.98^{\prime\prime},
\ee
where $m$ is the total mass of the two bodies, and $p$ is the
semi-latus rectum of the orbit.  The last equality is for the
Sun-Mercury system, and is compatible with the observations shown in
Table \ref{merctab}.  Each relativistic theory predicts its own
value of $\Delta \omega$, and by comparing to observations such as
those in Table \ref{merctab} we can therefore constrain them.  This
test is an additional one beyond those based on null geodesics alone
as it tests not only the `unit curvature' of space, but also the
non-linear terms in the space-time geometry, as well as preferred
frame effects.

Another very useful test involving time-like geodesics involves
looking for the `Nordtvedt effect' \cite{nordtvedt}.  This effect is
the name given to violations of the SEP.  In the previous section we
only considered tests of the WEP and EEP, which provide strong evidence that
viable gravitational theories should be `metric' ones.  Now, it is
entirely possible to satisfy the WEP and EEP, with a metric theory of
gravity, while violating the SEP.  Such violations do not occur in
General Relativity, but do in most other theories.  Every test of the
Nordtvedt effect is  therefore a potential killing test of general
relativity, if it delivers a non-null result.  To date, the most
successful approach in searching for SEP violations is to use the
Earth-Moon system in the gravitational field of the Sun as a giant
E\"{o}tv\"{o}s experiment.  The difference between this and the
laboratory experiments described in the previous section is that while
the gravitational fields of the masses in WEP E\"{o}tv\"{o}s
experiments are entirely negligible, this is no longer the case with
the Earth and Moon.  By tracking the separation of the Earth and Moon
to high precision, using lasers reflected off reflectors left on the
Moon by the Apollo 11 mission in 1969, it is then possible to gain
the constraint \cite{nordtvedtobs}
\be
\eta= (-1.0 \pm 1.4 ) \times 10^{-13},
\ee
where $\eta$ is defined as in Eq.~(\ref{eotwash}).  This is indeed a
null result, consistent with General Relativity, and is tighter even
than the current best laboratory constraint on the WEP.  It can
therefore be used to constrain possible deviations from
General Relativity, and in fact constrains a similar (but not identical) set of
gravitational potentials to the perihelion precession described previously.

A third solar system test involving time-like geodesics is the
observation of spinning objects in orbit.  While currently less
constraining than the other tests discussed so far, these observations
allow insight into an entirely relativistic type of gravitational
behaviour:  gravitomagnetism.  This is the generation of gravitational
fields by the rotation of massive objects, and was discovered in the
very early days of General Relativity by Lense and Thirring
\cite{lt1,lt2}.  The basic idea here is that massive objects should `drag'
space around with them as they rotate, a concept that is in good keeping
with Mach's principle.  Although one can convincingly argue that the same aspects of
the gravitational field that cause frame-dragging are also being
tested by perihelion precession and the Nordtvedt effect, it is {\it
  not} true that in these cases the gravitational fields in question are being communicated
through the rotation of matter. Now, in the case of General Relativity it can be shown that
the precession of a spin vector ${\bf S}$ along the trajectory of a freely-falling gyroscope in orbit
around an isolated rotating massive body at rest is given by
\be
\frac{d {\bf S}}{d \tau} = {\bf \Omega} \times {\bf S},
\ee
where
\be
\label{omega}
{\bf \Omega} = \frac{3}{2} {\bf v} \times \nabla U -\frac{1}{2} \nabla \times {\bf g} .
\ee
Here we have written the vector ${\bf g} = g_{0i}$, and have taken ${\bf v}$
and $U$ to be the velocity of the gyroscope and the Newtonian
potential at the gyroscope, respectively.  The first term in
(\ref{omega}) is called `geodetic precession', and is caused by the
`unit curvature' of the space.  This effect exists independent of the
massive bodies rotation.  The second term in (\ref{omega}) is the
Lense-Thirring term, and causes the frame-dragging discussed above.
The most accurate measurement of this effect claimed so far is at the
level of $5-10\%$ accuracy, and has been made using the LAser
GEOdynamics Satellites (LAGEOS) \cite{lageos} (there has, however, been
some dispute of this result \cite{lageos2,lageos3}).  The Gravity Probe B
mission is a more tailor made experiment which was put in orbit around
the Earth between April 2004 and September 2005.  The current accuracy
of results from this mission are at the level of $\sim 15\%$
\cite{gpb}, although this could improve further after additional
analysis is performed.

All of the tests discussed so far in this section have been for
long-ranged modifications to Newtonian gravity.  As well as these,
however, there are a host of alternatives to General Relativity
that also predict short-ranged deviations from $1/r^2$ gravity.  These
range from extra-dimensional theories \cite{Kaluza, Klein2}, to fourth-order
theories \cite{cliftonppn} and bimetric theories \cite{CliftonBanadosSkordis2010}, all of which
predict `Yukawa' potentials of the form
\be
U =\alpha \int \frac{\rho(x^{\prime}) e^{-\vert x-x^{\prime}
    \vert/\lambda}}{\vert x-x^{\prime} \vert} d^3 x^{\prime 3},
\ee
where $\alpha$ parameterises the `strength' of the interaction, and
$\lambda$ parameterises its range.  The genericity of these
potentials, often referred to as `fifth-forces', provides strong
motivation for experimental attempts to detect them.  Unfortunately,
due to the their scale dependence, one can no longer simply look for
the extra force on one particular scale, and then extrapolate the
result to all scales.  Instead, observations must be made on a whole range
of different scales, so that we end up with constraints on $\alpha$ at
various different values of $\lambda$.  These observations are taken
from a variety of different sources, with the scale of the phenomenon
being observed typically constraining $\lambda$ of similar size.  So,
for example, on the larger end of the observationally probed scale we
have planetary orbits \cite{planet} and lunar laser ranging \cite{nordtvedtobs}
constraining $\alpha \lesssim 10^{-8}$ between $10^8 m \lesssim
\lambda \lesssim 10^{12} m$.  On intermediate scales the LAGEOS
satellite, and observations of gravitational accelerations at the top
of towers and under the oceans provide constraints of $\alpha$ that
range from $\alpha \lesssim 10^{-8}$ at $\lambda \sim 10^7 m$
\cite{lag} to $\alpha \lesssim 10^{-3}$ at $10^{-1} m \lesssim \lambda
\lesssim 10^4 m$ \cite{geo1, geo2}. At smaller scales laboratory
searches must be performed, and current constraints in this regime
range from $\alpha \lesssim 10^{-2}$ at $\lambda \sim 10^{-2} m$, to
$\alpha \lesssim 10^{6}$ at $\lambda \sim 10^{-5} m$ \cite{eot, col,
  sta}.  Weaker constraints at still smaller scales are available
using the Casimir effect.  For a fuller discussion of these searches,
and the experiments and observations involved, the reader is referred
to the reviews by Fischbach and Talmadge \cite{fifth1}, and
Adelberger, Heckel and Nelson \cite{fifth2}.
\newline
\newline
\noindent
{\it Gravitational waves, and binary pulsars}
\newline

A generic prediction of all known relativistic theories of gravity is the
existence of gravitational waves: Propagating gravitational
disturbances in the metric itself.  However, while all known
relativistic gravitational theories predict gravitational radiation,
they do not all predict the same type of radiation as the quadrupolar, null
radiation that we are familiar with from General Relativity.  It is
therefore the case that while the mere existence of gravitational
radiation is not itself enough to effectively discriminate between
different gravitational theories, the {\it type} of gravitational radiation that is
observed is.  The potential differences between different types of
gravitational radiation can take a number of different forms, which we will now discuss.

Firstly, one could attempt to determine the propagation speed of
gravitational waves.  In General Relativity it is the case that
gravitational waves have a velocity 
that is strictly equal to that of the speed of light in vacuum.
Generically, however, this is not true:  Some theories predict null
gravitational radiation, and others do not.  So, for example, if one
were able to detect gravitational waves from nearby supernovae, then
comparing the arrival time of this radiation with the arrival
time of the electromagnetic radiation would provide a potentially
killing test of General Relativity.  There are, however, a number of
different theories that predict null gravitational radiation.  Tests
of the velocity of gravitational waves therefore have the potential to
rule out a number of theories, but by themselves are not sufficient to
distinguish any one in particular.

A second, more discriminating test, is of the polarity of gravitational
radiation.  General Relativity predicts radiation with helicity modes
$\pm 2$ only, and so far is the only proposed theory of gravity that
does so.  In general, there are six different polarisation states --
one for each of the six `electric' components of the Riemann tensor,
$R_{0i0j}$.  These correspond to the two modes familiar from General
Relativity, as well as two modes with helicity $\pm 1$, and two
further modes with helicity $0$.  One of these helicity-$0$ modes
corresponds to an additional oscillation in the plane orthogonal to
the wave vector $k^\mu$, while the remaining 3 modes all correspond to
oscillations in a plane containing $k^\mu$.  The extent to which
observations of these modes can constrain gravitational theory depends
on whether or not the source of the radiation can be reliably identified.  If
the source can be identified, then the vector $k^\mu$ is known,
and one should then be able to uniquely identify the individual
polarisation modes discussed
above.  We then have 6 different tests of relativistic
gravitational theory -- one for each of the modes.  In the
absence of any knowledge of $k^\mu$, however, one cannot necessarily uniquely
identify all of the modes that are present in a gravitational wave, although it may still
be possible to constrain the modes being observed to a limited number
of possibilities.

Direct observations of gravitational waves, of the kind discussed
above, provide an excellent opportunity to further constrain
gravity.  Indeed, some theories can be shown to be indistinguishable
from General Relativity using post-Newtonian gravitational
phenomena in the solar system alone, while being easily distinguishable when
one also considers gravitational radiation.  This is the case with
Rosen's bimetric theory of gravity \cite{Rosen1973,Will1976,WillEardley1977}.  To
date, however, the direct detection of gravitational radiation has
yet to be performed.  At present the highest accuracy
null-observations of gravitational radiation are those of 
the Laser Interferometer Gravitational-wave Observatory (LIGO).  This
experiment consists of two sites in the USA (one in Livingston,
Louisiana and one in Richland, Washington).  Each site is an
independent interferometer constructed from two 4 km arms, along which
laser beams are shone.  The experiment has an accuracy capable of
detecting oscillations in space at the level of $\sim 1$ part in
$10^{21}$, but has yet to make a positive detection.  Further
experiments are planned for the future, including Advanced LIGO, which
is scheduled to start in 2014, and the Laser Interferometer Space
Antenna (LISA).  Both Advanced LIGO and LISA are expected to make
positive detections of gravitational waves.

Another way to search for gravitational waves is to look for 
their influence on the systems that emitted them.  In
this regard binary pulsar systems are of particular interest.  Pulsars
are rapidly rotating neutron stars that emit a beam of electromagnetic
radiation, and were first observed in 1967 \cite{pulsar}.  When these
beams pass over the Earth, as the star rotates, we observe regular
pulses of radiation.  The first pulsar observed in a binary system was
PSR B1913+16 in 1974, by Russell Hulse and Joseph Taylor \cite{HT}.
This is a particularly `clean' binary system of a pulsar with rotational
period $\sim 59$ms in orbit around another neutron star.
Binary pulsars are of particular significance for gravitational
physics for a number of reasons.  Firstly, they can be highly
relativistic.  The Hulse-Taylor binary system, for example, exhibits a
relativistic periastron advance that is more than $30 \; 000$ times that
of the Mercury-Sun system.  In this regard they provide an important
compliment to the observations of post-Newtonian gravity that we
observe in the solar system.  Secondly, they are a source of
gravitational waves.  Given the high degree of accuracy to which
the orbits of these systems are known, the change in angular
momentum due to gravitational radiation can be determined and
observed.  In the Hulse-Taylor system the observed decrease in orbital period
over the past 30 years is $0.997\pm0.002$ of the rate predicted by
General Relativity \cite{newjoel}.
Finally, neutron stars are composed of a type of compact
matter that is of particular interest for the study of
self-gravitational effects.  For a review of pulsars in this context
the reader is referred to \cite{lrpulsar}.

There are large number of relativistic parameters that can be probed
by observations of binary pulsar systems \cite{binaryparameters}.  To
date, however, the most constrained are the 5 `post-Keplerian' effects,
which are the rate of periastron advance, the rate of change
of orbital period, the gravitational redshift, and two
Shapiro time-delay effects.  These effects are familiar from the solar system
tests discussed above, apart from the change in orbital period that is negligible in
the solar system.  One further effect that has been measured only
relatively recently is the `geodetic' precession of the pulsar spin
vector about its angular momentum vector \cite{KramerPrec}.  This is a
purely relativistic effect that is observed via changes in the
observed pulse profile over a period of time that can be attributed to
our line of sight to the pulsar crossing the emitting region at
varying positions due to the precession.  The determination of the
precession rate using these observations is, however, complicated
somewhat by a degeneracy between the {\it a priori} unknown shape of the
emitting region and the geometry of the system as a whole \cite{TW,CW}.

Not all of the post-Keplerian effects are always apparent in any given
binary system, and not all provide independent tests of gravity.  For
example, in the Hulse-Taylor binary only three of these effects can be
observed (the inclination angle of the system on the sky is too large
to observe any significant Shapiro delay), and there are two unknown quantities in the system (the
masses of the pulsar, and that of its companion).  The Hulse-Taylor binary
therefore provides only $3-2=1$ test of relativistic gravity.  The
recently discovered `Double Pulsar' PSR J0737-3039A/B \cite{dp1}, however, does
significantly better \cite{dp2}.  All five post-Keplerian effects are visible in
this system, and because both neutron stars are observable as pulsars
the ratio of their masses can be directly inferred from their orbits.
This leaves only one unknown quantity, and hence gives $5-1=4$
independent tests of relativistic gravity.  So far, all binary pulsar tests of gravity,
including those of the double pulsar, are consistent with General Relativity.

Finally, let us return to constraining gravitational theory through
the emission of gravitational waves.  The effect of emitting
gravitational radiation from a binary system is to change its orbital
period.  In General Relativity we know that only quadrupole radiation
with positive energy should be emitted from a system.  For most
relativistic theories, however, dipole gravitational radiation is also
expected, and sometimes this can carry away {\it negative} energy.
The existence of dipole radiation is sometimes attributed to
violations of the SEP, whereby the centre of the mass responsible for
gravitational radiation is no longer the same as the centre of
inertial mass.  If the centre of inertial mass is what stays fixed,
then the centre of mass responsible for the gravitational radiation can move
and generate dipole radiation.  Dipole radiation is expected to be
most dominant in binary systems with high eccentricity, and where the
companion mass is a white dwarf.  No evidence for dipolar radiation
yet exists \cite{nodp1, nodp2}.  Null observations that attest to this result therefore
allow for experimental limits to be set on theories that predict
positive energy dipolar radiation.  The lack of any observation of
dipolar radiation can also be used to rule out with high confidence
theories that allow negative energy dipolar radiation, such as Rosen's
theory \cite{WillEardley1977}.

\subsubsection{Theoretical considerations}
\label{sec:intro-theory}

As he developed the Special Theory of Relativity, it is often assumed
that Einstein's inspiration came from  experiments pointing towards the
constancy of the speed of light. It is true that he was certainly
aware of these experiments, but he was also inspired by theory,
specifically his faith in the principle of relativity and the validity
of Maxwell's equations in any inertial frame.  So too, in developing
models of modified gravity, we should not only take our lead from
observation but also from theory. Indeed, theoretical considerations
are a very powerful tool in testing new models. Typically these
involve the study of classical and quantum fluctuations about
classical solutions. Do the classical fluctuations  propagate
super-luminally? Can we excite a ghost? Do the quantum fluctuations
become strongly coupled at some unacceptably low energy scale?
\newline
\newline
\noindent
{\it Ghosts} \label{sec:intro-ghosts}
\newline

Ghosts are a common feature of many modified gravity models that hope
to explain dark energy.  Intuitively it is easy to see why this might
be the case. To get cosmic acceleration we need an additional
repulsive force to act  between massive objects at large distances. If
this force is to be mediated by a particle of even spin, such as a
scalar (spin 0) or a tensor (spin 2), then  the kinetic term
describing this must have the ``wrong" sign\footnote{In our
conventions, the Lagrangian for a canonical scalar is  ${\cal
L}=-\half (\del \psi)^2$, whereas a ghost has ${\cal L}=+\half(\del
\psi)^2$.}, that is, it must be a ghost.

We should be clear about the distinction between the kind of ghost
that arises in certain modified gravity models and the Faddeev-Popov
ghost used in the quantisation of non-abelian gauge theories. The
latter is introduced in the path integral to absorb unphysical gauge
degrees of freedom. It does not describe a physical particle and can
only appear as an  {\it internal} line in Feynman diagrams.   In
contrast, the ghosts that haunt modified gravity describe  {\it
physical} excitations and can appear as {\it external} lines in
Feynman diagrams.  

When  a physical ghost is present one has a choice: Accept the
existence of negative norm states and abandon unitarity, or else
accept that the energy eigenvalues of the ghost are {\it negative}
\cite{Cline-phantom}. Since the former renders the entire quantum
description completely non-sensical, one usually accepts the latter.
However, it now follows that the ghost will generate instabilities if
it  couples to other, more conventional, fields. When these fields are
already excited,  the ghost can and will continually dump its energy
into the ``conventional"  sector through classical processes, since
its energy is unbounded from below. Even in vacuum, one will get the
spontaneous (quantum) production of ghost-non-ghost pairs, and in a
Lorentz invariant theory, the production  rate is divergent
\cite{Cline-phantom}. 

There are a few ways to try to exorcise the ghost. One is to isolate
it somehow, such that it completely decouples from other
fields. Another option is to make it heavy, so much so that its mass
exceeds the cut-off for the effective theory describing the relevant
fluctuations, and one can happily integrate it out. A third option is
to break Lorentz invariance, perhaps spontaneously, so that one can
introduce an explicit Lorentz non-invariant cut-off to regulate  the
production rate of ghost-non-ghost pairs (see, for example,
\cite{Izumi-no-de-sitter}).  However, perhaps the safest way to deal
with a ghost is to dismiss as unphysical those solutions of a theory
upon which the ghost can fluctuate.  This school of thought is
exploited to good effect in the ghost condensate model
\cite{Hamedetal2004}. 
\newline
\newline
\noindent
{\it Strong coupling}
\newline

Some modified gravity models are said to suffer from ``strong
coupling" problems.  Given a classical solution to the field
equations, this refers to quantum fluctuations on that solution
becoming strongly coupled at an unacceptably low scale. For example,
in DGP gravity, quantum fluctuations on the Minkowski vacuum  becomes
strongly coupled at around $\Lambda \sim 10^{-13}$ eV $\sim 1/(1000
\textrm{km})$. In other words, for scattering processes above $\Lambda$,
perturbative quantum field theory on the vacuum is no longer well
defined, and one must sum up the contribution from all the multi-loop
diagrams. One then has complete loss of predictivity.  Furthermore, the
classical solution itself is meaningless at distances below
$\Lambda^{-1}$ since it would require a scattering process involving
energies above  the cut-off to probe its structure.  

The strong coupling scale is, of course, dependent on the background
classical solution, and may even depend on position in space-time.
Whether the inferred strong coupling scale is acceptable, or not, again
depends on the background. For example,  strong coupling at 1000 km on
the Minkowski vacuum of DGP gravity is not really an issue as
Minkowski space does not  represent a good  approximation to the
classical solution in the vicinity of the Earth.  Indeed, for the
classical solutions sourced by the Earth to leading order,  quantum
fluctuations will become strongly coupled at some scale that depends
on the radial distance from the Earth's centre. Computed at the
Earth's surface one should require that this lies below an meV  since
quantum gravity effects have yet to show up in any lab based
experiments up to this scale.  

It has actually been argued that strong coupling on the vacuum can be
a virtue in modified gravity models \cite{Dvali-predictive}. This is
because it can be linked to a breakdown of classical perturbation
theory, which is necessary for the successful implementation of the
Vainshtein mechanism \cite{vainshtein, Deffayet-nonperturbative}.  We
discuss the Vainshtein mechanism and strong coupling in some detail in
the context of DGP gravity in Section \ref{sec:dgp-sc}. Here we will
make some generic statements. Consider a model of gravity that
deviates  from GR at large distances. To be significant in terms of
understanding dark energy, this deviation must be at least ${\cal
O}(1)$ on cosmological scales, but be suppressed down to $\lesssim
{\cal O}(10^{-5})$ on Solar System scales. Therefore, the field or
fields that are responsible for the modification must be screened
within the Solar System. How can this screening occur? One way is for
the fields to interact so strongly that they are frozen together, so
much so that they are unable to propagate freely. This is the idea
behind the Vainshtein mechanism -- higher order derivative
interactions help to suppress the extra modes near the source (the
Sun).  

Alternative ways to screen  the extra fields have been suggested in
the form of the chameleon \cite{cham1,cham2}, and the symmetron
\cite{symmetron} mechanisms. Both methods exploit the dependence of
the effective potential on  the environment. For the chameleon, the
mass of the field is environmentally dependent, getting heavy in the
Solar System. For the symmetron, the strength of the matter coupling
is  (indirectly) environmentally dependent, tending to zero near a
heavy source. 

\subsection{Einstein's Theory}

Having considered the requirements that must be satisfied by a viable
relativistic theory of gravity, let us now consider Einstein's theory
of General Relativity in particular.  General Relativity satisfies all
of the requirements described in the previous section, either by
construction (for the foundational requirements) or by trial (in the
case of tests of metric theories of gravity).  

General Relativity is a gravitational theory that
treats space-time as a 4-dimensional manifold.  The 
connection associated with covariant differentiation,
$\Gamma^{\mu}_{\phantom{\mu}\alpha\beta}$, should be viewed as an additional structure
on this manifold, which, in general, can be decomposed into parts that
are symmetric or antisymmetric in its last two indices:
\be
\Gamma^{\mu}_{\phantom{\mu} \alpha\beta} = \Gamma^{\mu}_{\phantom{\mu} (\alpha\beta)} +
\Gamma^{\mu}_{\phantom{\mu} [\alpha\beta]}.
\ee
In General Relativity we take
$\Gamma^{\mu}_{\phantom{\mu} [\alpha\beta]}=0$, or, in the language of
differential geometry, we assume that torsion vanishes.  We are then
left with only the symmetric part of the connection, which describes
the curvature of the manifold.

Now, to define distances on the manifold one also requires a
metric tensor, $g_{\mu\nu}$.  Along the curve $\gamma$ this gives the
measure of distance
\be
s = \int_{\gamma} d \lambda \sqrt{g_{\mu\nu} \dot{x}^\mu \dot{x}^\nu},
\ee
where $\lambda$ is a parameter along the curve, $x^\mu=x^\mu(\lambda)$,
and over-dots here mean differentiation with respect to $\lambda$.  The metric
should also be considered as an additional structure on the manifold, which
is in general independent from the connection.  The
relationship between the connection and the metric is defined via the
non-metricity tensor, $Q_{\mu \alpha\beta} \equiv \nabla_\mu
g_{\alpha\beta}$.  In General Relativity it is assumed that the
non-metricity tensor vanishes.
We can now use the metric to define the Levi-Civita connection, which has
components given by the Christoffel symbols:
\be
\Gamma^{\mu}_{\phantom{\mu}\alpha\beta} = {\genfrac{\{}{\}}{0pt}{}{\mu}{\alpha\beta}}
\equiv \frac{1}{2} g^{\mu\nu} \left( g_{\alpha \nu, \beta} + g_{\beta
  \nu, \alpha} - g_{\alpha\beta, \nu} \right). 
\label{chris}
\ee

To summarise, as a consequence of the two assumptions $Q_{\mu\alpha\beta}=0$ and $\Gamma^{\mu}_{\phantom{\mu}[\alpha\beta]}=0$ , the components of the connection are uniquely given by the
Christoffel symbols via (\ref{chris}), and so the connection, and all geometric
quantities derived from it, are defined entirely in terms of the
metric.  In General Relativity, therefore, the metric tells us
everything there is to know about both distances and parallel
transport in the space-time manifold.

The resulting set of structures is known as a Riemannian manifold (or, more
accurately, pseudo-Riemannian in the case where the metric is not
positive definite, as is required to recover special relativity in the
tangent space to a point in space-time).  Riemannian manifolds have a
number of useful properties including tangent vectors
being parallel to themselves along geodesics, the geodesic
completeness of space-time implying the metric completeness of
space-time, and a particularly simple form for the contracted Bianchi
identities:
\be
\nabla_\mu \left( R^{\mu\nu}- \frac{1}{2} g^{\mu\nu} R \right) = 0,
\ee
where $R_{\mu\nu}$ and $R=g^{\mu\nu} R_{\mu\nu}$ are the Ricci tensor
and scalar curvature, 
respectively .  This last equation is of great significance for
Einstein's equations.

\subsubsection{The field equations}

Having briefly discussed the geometric assumptions implicit in General
Relativity, let us now display the field equations of this theory:
\be
\label{fieldequations}
R_{\mu\nu}- \frac{1}{2} g_{\mu\nu} R = 8 \pi G T_{\mu\nu}- g_{\mu\nu} \Lambda.
\ee
Here $T_{\mu\nu}$ is the energy-momentum tensor of matter fields in the
space-time, and $\Lambda$ is the cosmological constant.  These
equations are formulated such that energy-momentum is a conserved
quantity (due to the contracted Bianchi identity and metric-compatibility of
the connection), so that special relativity can be recovered in the
neighbourhood of every point in space-time (up to tidal forces), and
so that the usual Newtonian Poisson equation for weak gravitational
fields is recovered in non-inertial frames kept at a fixed space-like
distances from massive objects (up to small corrections).

The Field Equations (\ref{fieldequations}) are a set of 10
generally covariant, quasi-linear second-order PDEs in 4 variables, for
the 10 independent 
components of the metric tensor.  They constitute 4
constraint equations and 6 evolution equations, with the contracted
Bianchi identities ensuring that the constraint equations are always
satisfied.  Furthermore, the conserved nature of $T_{\mu \nu}$ and the
Riemannian nature of the manifold ensure that the WEP and EEP are
always satisfied: Massless test particles follow
geodesics, and in any freely falling frame one can always choose `normal
coordinates' so that local space-time is well described as Minkowski
space.

\subsubsection{The action}

As with most field theories, the Field Equations
(\ref{fieldequations}) can be derived from the variation of an
action.  In the case of General Relativity this is the
Einstein-Hilbert action:
\be
\label{EH}
S = \frac{1}{16 \pi G} \int \sqrt{-g} (R-2 \Lambda) d^4x + \int
\mathcal{L}_m(g_{\mu\nu}, \psi) d^4x, 
\ee
where $\mathcal{L}_m$ is the Lagrangian density of the matter fields,
$\psi$, and the gravitational Lagrangian density has been taken to be
$\mathcal{L}_g = \sqrt{-g} (R-2 \Lambda)/16 \pi G$.  Let us now assume
the Ricci scalar to be a function of the metric only, so that
$R=R(g)$. Variation of Eq. (\ref{EH}) with respect to the metric tensor then gives
the Field Equations (\ref{fieldequations}), where
\be
T^{\mu\nu} \equiv  \frac{2}{\sqrt{-g}} \frac{\delta \mathcal{L}_m}{\delta g_{\mu\nu}}.
\ee
The factors of $\sqrt{-g}$ are included in Eq. (\ref{EH}) to ensure that
the $\mathcal{L}$s transform as scalar densities under coordinate
transformations, i.e. as
\be
\bar{\mathcal{L}} = det \left( \frac{\partial x^\mu}{\partial
  \bar{x}^\nu} \right) \mathcal{L}, 
\ee
under coordinate transformations $\bar{x}^\mu=\bar{x}^\mu (x^\nu)$.  This
property ensures $S$ is invariant under general coordinate
transformation, and that the resulting tensor field equations are
divergence free (i.e. the contracted Bianchi identities and
energy-momentum conservation equations are automatically satisfied).

We have outlined here how Einstein's equations can be obtained from
the variation of an invariant action with respect to the metric, once
it has been assumed that the space-time manifold is
Riemannian.  The vanishing of torsion and non-metricity then
tell us that the metric is the only independent structure on the
manifold, and the invariant action principle ensures that we end up
with a set of tensor field equations in which
energy-momentum is conserved.  Because of this formulation the WEP and EEP are
satisfied identically.  Now, when considering alternative theories of
gravity one often wants to modify the field equations while conserving
these basic properties. Modified theories of gravity are therefore often
formulated in a similar way; from the metric variation of an invariant
action principle under the assumption of Riemannian geometry, with a
universal coupling of all matter fields to the same metric.

\subsection{Alternative Formulations}

The discussion in the previous section involved deriving Einstein's
equations under the {\it a priori} assumption of Riemannian geometry
(i.e. assuming to begin with that the torsion vanishes
and that the connection is metric compatible).  In this case the metric
is the only remaining geometric structure, and a simple metric
variation of the action is the only option.  We can, however, be less
restrictive in specifying the type of geometry we wish to consider.
For the case of the Einstein-Hilbert action, Eq. (\ref{EH}), this usually
still leads to the Einstein Equations (\ref{fieldequations}).  For
alternative theories of gravity, however, this is often not the
case:  Different variational procedures, and different assumptions
about the geometric structures on the manifold, can lead to different
field equations.  It is for this reason that we now outline some
alternative formulations of General Relativity. A large collection of
many such formulations can be found in~\cite{Peldan1993}.

\subsubsection{The Palatini procedure}

The most well known deviation from the metric variation approach is
the `Palatini procedure' \cite{pal}.  Here the connection is no longer
immediately assumed to be metric compatible, but is still assumed to be symmetric and thus torsionless.
In addition, all matter fields are still taken to couple universally
to the metric only\footnote{This assumption has limited validity,
  however, as it cannot be applied to tensor fields without using a covariant derivative.}. 
 The action to be varied is then
\be
\label{palact}
S = \frac{1}{16 \pi G} \int \sqrt{-g} \left[ g^{\mu\nu}\;
  \RR_{\mu\nu}-2 \Lambda\right] d^4x + \int \mathcal{L}_m(g_{\mu\nu},
\psi) d^4x,
\ee
where $\RR_{\mu\nu}$ indicates that the Ricci tensor
here is defined with respect to the connection and not the metric (at
this stage the metric and connection are still independent
variables), and is given by
\begin{equation}
\RR_{\mu\nu} = 
\partial_\alpha \Gamma^\alpha_{\phantom{\alpha}\mu\nu} - \partial_\mu \Gamma^\alpha_{\phantom{\alpha}\alpha\nu} 
+ \Gamma^\beta_{\phantom{\beta}\beta\alpha} \Gamma^\alpha_{\phantom{\alpha}\mu\nu} 
-\Gamma^\alpha_{\phantom{\alpha}\mu\beta} \Gamma^\beta_{\phantom{\beta}\alpha\nu}.
\label{Ricci_asym}
\end{equation} 
The Ricci tensor defined above, as well as the Einstein tensor derived
from it, are in general asymmetric. However, they become symmetric as
soon as we assume the connection is symmetric. Variation of Eq. (\ref{palact}) with respect
to the connection gives the condition that the connection is in fact
the Levi-Civita connection. Variation with respect to the metric then
recovers the Einstein equations.

If the torsionless condition on the connection is dropped then
complications arise, as has been shown by Hehl and Kerlick~\cite{hehl}.
The general form of the connection can be shown to be given by
\be
\Gamma^{\mu}_{\phantom{\mu} \alpha\beta} = 
{\genfrac{\{}{\}}{0pt}{}{\mu}{\alpha\beta}} + J^\mu_{\phantom{\mu}\alpha\beta}
= {\genfrac{\{}{\}}{0pt}{}{\mu}{\alpha\beta}} +
K^{\mu}_{\phantom{\mu} \alpha\beta} +L^{\mu}_{\phantom{\mu} \alpha\beta}.
\ee
The tensor field $K^{\mu}_{\phantom{\mu} \alpha\beta}$ is the
contorsion tensor, that can be defined in terms of the antisymmetric
components of the connection, known as the torsion, as
\be
K^{\mu}_{\phantom{\mu} \alpha\beta} \equiv S^{\mu}_{\phantom{\mu}\alpha\beta}
-S_{\alpha\beta}^{\phantom{\alpha\beta}\mu}
-S_{\beta\alpha}^{\phantom{\alpha\beta}\mu},
\label{contorsion}
\ee
where $S^\mu_{\phantom{\mu}\alpha\beta} = \Gamma^{\mu}_{\phantom{\mu} [\alpha\beta]}$ is the torsion tensor.
The tensor field $L^{\mu}_{\phantom{\mu} \alpha\beta}$ is defined in terms of the
non-metricity tensor as
\be
L^{\mu}_{\phantom{\mu} \alpha\beta} \equiv \frac{1}{2} \left(
Q^{\mu}_{\phantom{\mu} \alpha\beta} - Q^{\phantom{\alpha\beta}
  \mu}_{\alpha\beta}  
- Q^{\phantom{\alpha\beta} \mu}_{\beta\alpha} \right).
\ee
To avoid confusion, we continue to denote the covariant derivative
associated with the Levi-Civita connection as $\nablac_\mu$, while we use $\nablag_\mu$ to denote
 the covariant derivative associated with $\Gamma^\mu_{\;\;\alpha\beta}$.

Varying the action with respect to the metric $g_{\mu\nu}$ we find the
analogue of the Einstein equations:
\be
G_{(\mu\nu)} + \Lambda g_{\mu\nu} = 8\pi G  \tilde{T}_{\mu\nu}.
\label{Ein_palatini}
\ee
One should note that only the symmetric part of the Einstein tensor
appears here, and that we have used $\tilde{T}_{\mu\nu}$ rather than $T_{\mu\nu}$ to emphasise the fact that
$\tilde{T}_{\mu\nu}$ is defined at constant $\Gamma^\mu_{\phantom{\mu}\alpha\beta}$ 
   in the variation, i.e. $\tilde{T}_{\mu\nu} =
   -\frac{2}{\sqrt{-g}}\frac{\delta {\cal L}_m}{\delta
     g^{\mu\nu}}\big|_{\Gamma}$. On the other hand  
 $T_{\mu\nu} = -\frac{2}{\sqrt{-g}}\frac{\delta {\cal L}_m}{\delta
     g^{\mu\nu}}\big|_{J}$.  This is not an important distinction at
   this stage, as we have assumed that matter field do not couple to
   the connection, and hence $\tilde{T}_{\mu\nu} = T_{\mu\nu}$. It
   will, however, be important in the following subsection and in
   Section \ref{newEC}.

Varying with respect to the connection defines the Palatini tensor as
 $  P_{\mu}^{\phantom{\mu}\alpha\beta} = \frac{8\pi G}{\sqrt{-g}}
\frac{\delta (\sqrt{-g} R)}{\delta
  \Gamma^\mu_{\phantom{\mu}\alpha\beta}}$, that can be written as
\be
P_{\alpha\mu\beta} = 
S_{\mu\alpha\beta}
+ 2 g_{\mu[\alpha} S_{\beta]}  
+ g_{\mu[\alpha} Q_{\beta]}
- g_{\mu[\alpha} \bar{Q}_{\nu]\beta}^{\;\;\;\;\nu} ,
\label{palatini_tensor}
\ee
where $S_\mu = S^\alpha_{\phantom{\alpha}\mu\alpha}$, and where we
have split the non-metricity tensor into trace and traceless parts as
$Q_{\mu\alpha\beta} = Q_\mu g_{\alpha\beta} + \bar{Q}_{\mu\alpha\beta}$, with $g^{\alpha\beta}\bar{Q}_{\mu\alpha\beta}=0$.
The Palatini tensor has only $60$ independent components because it is
identically traceless:
$P_{\alpha}^{\phantom{\alpha}\mu\alpha}=0$. Now, the second field
equation is the vanishing of the Palatini tensor,
\be
 P_{\mu}^{\phantom{\mu}\alpha\beta}   =  0,
\ee
but this provides only $60$ constraints among the $64$ independent components of the connection.
In fact it may be shown that the equation $ P_{\mu}^{\phantom{\mu}\alpha\beta}   =  0$ is equivalent to
the connection taking the following form~\cite{hehl}:
\begin{equation}
\Gamma^{\mu}_{\phantom{\mu} \alpha\beta} =
      {\genfrac{\{}{\}}{0pt}{}{\mu}{\alpha\beta}} - \frac{1}{2}
      Q_\alpha \delta^\mu_{\phantom{\mu}\beta} 
=  {\genfrac{\{}{\}}{0pt}{}{\mu}{\alpha\beta}} + \frac{2}{3} S_{\alpha} \delta^\mu_{\phantom{\mu}\beta}.
\end{equation}
Clearly then, there are $4$ degrees of freedom left undetermined by
the field equations. Thus the Palatini approach in its most 
general form does not lead to a unique set of field
equations\footnote{It is often said that the Palatini procedure
  uniquely recovers GR. As we have seen, however, this is a myth. 
It does so only after further assumptions, for instance that the
torsion vanishes, or that the connection is metric compatible, or that
$Q_\alpha=0$. To make the Palatini variation well 
defined one has to impose such conditions in the action by means of Lagrange multipliers.}.

The constraint $Q_\mu=0$ is sufficient to produce a consistent
theory. This, however, has to be imposed as a Lagrange multiplier in the action via a
term $\int d^4x \sqrt{-g} \lambda^\alpha Q_\alpha$. Once this is done, one recovers General Relativity uniquely.
For theories of gravity other than General Relativity the difference between the metric variation and the Palatini
procedure is even more significant:  The resulting field equations are, in
general, different.  This will be spelt out explicitly for some specific
theories in the sections that follow.

\subsubsection{Metric-affine gravity and matter}
A further generalisation of the metric variation approach is to 
keep the metric and connection completely independent, as discussed above, and further allow
matter to couple not only to the metric, but also the connection \cite{hehl}. 
  In this case the action takes the form
\be
S = \frac{1}{16 \pi G} \int \sqrt{-g} (g^{\mu\nu} \RR_{\mu\nu}-2
\Lambda) d^4x + \int
\mathcal{L}_m(g_{\mu\nu},\Gamma^{\mu}_{\phantom{\mu} \alpha\beta},
\psi) d^4x, 
\label{met_aff}
\ee
where $\Gamma^{\mu}_{\phantom{\mu} \alpha\beta}$ and $g_{\mu\nu}$ are once again independent. 
Performing the variations we recover Eq. (\ref{Ein_palatini}) as before,
and
\be
 P_{\mu}^{\phantom{\mu}\alpha\beta}   = 8\pi G  \Delta_{\mu}^{\phantom{\mu}\alpha\beta} ,
\label{conn_aff}
\ee
where $\Delta_{\mu}^{\phantom{\mu}\alpha\beta} = -\frac{1}{\sqrt{-g}}
\frac{\delta{\cal L}_m}{\delta \Gamma^\mu_{\phantom{\mu}\alpha\beta}}$
is called the hypermomentum tensor~\cite{hehl,ECSK}.

In this case $T_{\mu\nu} \ne \tilde{T}_{\mu\nu}$, but it is straightforward to find that
\begin{equation}
T_{\mu\nu} = \tilde{T}_{\mu\nu} +  \nablac_\rho\left[
  \Delta^\rho_{\phantom{\rho}(\mu\nu)} -
  \Delta^{\phantom{(\mu}\rho}_{(\mu\phantom{\rho}\nu)}   -
  \Delta_{(\mu\nu)}^{\phantom{(\mu\nu)}\rho}\right], 
\label{T_relation}
\end{equation}
where $\nablac_\mu$ is the covariant derivative associated with the
Levi-Civita connection.

Equation (\ref{conn_aff}) can be shown to be self-inconsistent for reasonable forms of matter,
 as the Palatini tensor is invariant under projective transformations of the form
$\Gamma^{\mu}_{\phantom{\mu} \alpha\beta} \rightarrow
 \Gamma^{\mu}_{\phantom{\mu} \alpha\beta} + \lambda_\alpha
 \delta^\mu_{\phantom{\mu} \beta}$, while there is no reason to
 suspect this invariance is exhibited by the matter fields and hence
 the hypermomentum. 
Equivalently, the Palatini tensor obeys the identity
$P_{\alpha\mu}^{\phantom{\alpha\mu}\alpha}=0$, while there is no
reason that this should identically hold 
for the hypermomentum\footnote{Consider for example a simple Einstein-{\AE}ther model for which the matter action is 
$S_{M} = \int d^4x\sqrt{-g}\left[\alpha \nabla_\mu A^\nu \nabla_\nu A^\mu + \lambda (A_\mu A^\mu+1)\right]$. The hypermomentum 
is $\Delta_\mu^{\phantom{\mu}\alpha\beta}=-2A^\beta\nabla_\mu
A^\alpha$ which clearly does not obey
$\Delta_{\alpha\mu}^{\phantom{\alpha\mu}\alpha}=0$. The variation done
this way is inconsistent. On the other hand using the Lagrange
constraint $\int d^4x\sqrt{-g}
\beta_\mu^{\phantom{\beta}\alpha\beta}J^\mu_{\phantom{\mu}\alpha\beta}$
in the action imposes
$J^\mu_{\phantom{\mu}\alpha\beta}=0$, and hence the vanishing of the Palatini tensor.
This leads to a modified Eq. (\ref{conn_aff}), as $
P_{\mu}^{\phantom{\mu}\alpha\beta}  =0 = 8\pi G (
\Delta_{\mu}^{\phantom{\mu}\alpha\beta}-
\beta_\mu^{\phantom{\beta}\alpha\beta})$,
and to a modified Eq. (\ref{Ein_palatini}), which now includes
derivatives of $\beta_\mu^{\phantom{\beta}\alpha\beta}$. After using
Eq. (\ref{T_relation}), however,
the resulting equations are completely equivalent to the metric
variation.}.  One way to impose self-consistency
is to demand that both torsion and non-metricity must vanish (by using Lagrange multipliers in the action),
 leading again to General Relativity. This type of self-consistency
 is, however, very strong, and weaker constraints have been found in~\cite{hehl}.
One such weaker constraint leads to the
Einstein-Cartan-Sciama-Kibble theory~\cite{ECSK}, that we shall
briefly describe in Section \ref{newEC}.

\subsubsection{Other approaches}
\label{sec:otherGR}

There are a variety of other formalisms that one can use to derive
Einstein's equations.  We will not go into the full details of all of these
here, but merely mention some of the approaches that exist in the
literature.  For brevity we will only consider vacuum general
relativity here.

In the `{\it vierbein}' formalism the Einstein-Hilbert action can be
written
\be
\label{vb}
S = \int d^4x  \; e\;  e^\mu_{\hat{\alpha}}  e^\nu_{\hat{\beta}}  R_{\mu\nu}^{\phantom{\mu\nu}\hat{\alpha} \hat{\beta}}
,
\ee
where indices with hats correspond to a basis in the tangent space
defined by the set of contravariant vectors, $e^{\phantom{\hat{\mu}} \mu}_{\hat{\mu}}$, with determinant $e = \det[ e^{\phantom{\hat{\mu}} \mu}_{\hat{\mu}} ]$.
The inverse of $e^{\phantom{\hat{\mu}} \mu}_{\hat{\mu}}$ 
is $e^{\phantom{\mu} \hat{\mu}}_{\mu}$, such that 
$e^{\phantom{\mu} \hat{\mu}}_{\rho} e^{\phantom{\hat{\mu}} \rho}_{\hat{\nu}} = \delta^{\hat{\mu}}_{\phantom{\hat{\mu}}\hat{\nu}}$,
and $e^{\phantom{\hat{\mu}} \mu}_{\hat{\rho}} e^{\phantom{\mu} \hat{\rho}}_{\nu} = \delta^{\mu}_{\phantom{\mu}\nu}$. 
The metric tensor is constructed as  $g_{\mu\nu} = \eta_{\hat{\mu}\hat{\nu}} e^{\phantom{\mu} \hat{\mu}}_{\mu} e^{\phantom{\mu} \hat{\nu}}_{\nu}$.
The  spin connection $\omega_{\mu}^{\phantom{\mu} \hat{\alpha} \hat{\beta}}$ then defines a space-time and Lorentz covariant derivative, ${\cal D}_\mu$, as 
${\cal D}_\mu v_\nu^{\hat{\rho}} = \nabla_\mu v_\nu^{\hat{\rho}} + \omega_{\mu\phantom{\hat{\rho}}\hat{\lambda}}^{\phantom{\mu}\hat{\rho}}  v_\nu^{\hat{\lambda}} $,
where $\nabla_\mu$ is the Levi-Civita connection\footnote{Given a metric, $g_{\mu\nu}$, the Levi-Civita connection can always be defined. The question is whether
that is the connection that is used to define parallel transport.}.
The curvature tensor $ R_{\mu\nu}^{\phantom{\mu\nu}\hat{\alpha} \hat{\beta}}$ is defined in terms of the spin connection as
\be
R_{\mu\nu}^{\phantom{\mu\nu}\hat{\mu}\hat{\nu}} \equiv 
 \partial_{\mu} \omega_{\nu}^{\phantom{\nu} \hat{\mu} \hat{\nu}}
- \partial_{\nu} \omega_{\mu}^{\phantom{\mu} \hat{\mu} \hat{\nu}}
 + \omega_{\mu}^{\phantom{\mu} \hat{\mu}\hat{\rho}} \; \omega_{ \nu \hat{\rho}}^{\phantom{\mu \vert \hat{\rho}} \hat{\nu}}
 - \omega_{\nu}^{\phantom{\mu} \hat{\mu}\hat{\rho}} \; \omega_{ \mu \hat{\rho}}^{\phantom{\mu \vert \hat{\rho}} \hat{\nu}}.
\ee
Variation now proceeds as in the Palatini formalism  by assuming that the spin connection and vierbein are independent fields, from which one obtains
the two field equations
\be
\mathcal{D}_{[\mu} e^{\hat{\alpha}}_{\phantom{\hat{\mu}} \nu]}=0 ,
\label{vbfe}
\ee
and
\be
 G^\alpha_{\hat{\rho}} \equiv e^\alpha_{\hat{\alpha}} 
 e^\mu_{\hat{\rho}} e^\nu_{\hat{\beta}} R_{\mu\nu}^{\phantom{\mu\nu} \hat{\alpha}\hat{\beta} } 
- \frac{1}{2}  ( e^\mu_{\hat{\alpha}} e^\nu_{\hat{\beta}} R_{\mu\nu}^{\phantom{\mu\nu} \hat{\alpha}\hat{\beta} } )
 e^\alpha_{\hat{\rho}} 
= 0,
\ee
where $G_{\hat{\nu}}^{\mu}$ is the Einstein tensor. Equation (\ref{vbfe}) can be used to obtain the spin connection in terms of the partial derivatives
of the vierbein, and the resulting relation  implies that  $\omega_{\mu}^{\phantom{\mu} \hat{\alpha} \hat{\beta}}$ is torsion-less, i.e. 
one recovers Cartan's first structure equation, $d e^{\hat{\mu}} + \omega^{\hat{\mu}}_{\phantom{\hat{\mu}}\hat{\nu}} \wedge e^{\hat{\nu}}=0$.
The second equation says that the vacuum Einstein equations are recovered.

Another interesting alternative formulation of General Relativity is
given by the Plebanski formalism \cite{pleb}.  It is derived from the action
\be
S= \int \Sigma^{AB} \wedge R_{AB} - \frac{1}{2} \Psi_{ABCD}
\Sigma^{AB} \wedge \Sigma^{CD},
\ee
where upper case indices denote two component spinor indices to be
raised and lowered with $\epsilon^{AB}$ and its inverse, and where
the exterior product $\wedge$ acts on space-time indices, which have
been suppressed. The curvature 2-form $R_{AB} \equiv d \omega_{AB}
+ \omega_A^{\phantom{A} C} \wedge \omega_{CB}$ is
defined with respect to a spin connection 1-form
$\omega_A^{\phantom{A} B}$.  Variation of this action with respect to
$\Psi_{ABCD}$ and $\omega_{AB}$ then tells us that the
2-form $\Sigma^{AB}$ is the exterior product of some set of 1-forms
that we can identify with the tetrad $\theta^{A A^{\prime}}$, and that
the connection $\omega_{AB}$ is torsion-free with respect to
$\Sigma^{AB}$.  Using this together with the variation of the action
with respect to $\Sigma^{AB}$ then gives the vacuum Einstein
equations, where the metric is given by $g=\theta^{A A^{\prime}}
\otimes \theta_{A A^{\prime}}$.

One further alternative formulation of General Relativity is the purely
affine `Eddington formalism' \cite{eddington}.  In previous
subsections we outlined how one can either treat the metric as the
only independent structure on the manifold, or treat the metric and
connection as being two independent structures.  Another approach is
to take the connection as the only structure on the manifold.  In this
case, the simplest way of constructing a Lagrangian density with the correct
weight (and without a metric) is to simply take the square root of the
determinant of the Ricci tensor itself:
\be
S = \int \sqrt{-det[ R_{\mu\nu} (\Gamma) ]} d^4 x.
\ee
Varying this action with respect to the connection then gives the
field equations
\be
\nabla_\rho \left( \sqrt{-det[ R_{\alpha\beta} (\Gamma) ]} R^{\mu \nu} \right)
=0,
\ee
which can be shown to be equivalent to Einstein's equations in vacuum
with a cosmological constant, if we take the connection to be the
Levi-Civita connection.  Due to the lack of a metric in the action for
this theory, however, it is not a trivial matter to introduce matter
fields into the theory \cite{edmat}.

Finally, let us mention that approaches exist that treat gravity
as simply a spin-2 field on flat space \cite{bary1,bary2}.  It has
been conjectured that one could reconstruct the Einstein-Hilbert
action in such an approach by considering consistency conditions
order by order in perturbation theory.  This will, of course, be an
invalid treatment when gravity is strong, and in cosmology.

\subsection{Theorems}

There a number of theorems in General Relativity that are of great
importance for the structure of the theory itself, as well as for the
solutions to the field
equations.  These theorems underpin a lot of the acquired intuition on
how gravity should function in different environments, and what
the resulting phenomenology should be.  In
alternative theories of gravity, however, the theorems of General
Relativity often fail, allowing new behaviours that would otherwise be
impossible.

Here we briefly recap what we consider to be some of the most important
theorems of General Relativity.  In later sections we will show how
these theorems are violated in alternative theories, and discuss the
consequences of this.

\subsubsection{Lovelock's theorem}
\label{sec:lovelock-thm}

Lovelock's theorem \cite{love1, love2} limits the theories that one
can construct from the metric tensor alone.  To enunciate this
theorem, let us begin by assuming that the metric tensor is the only
field involved in the gravitational action.  If the action can be
written in terms of the metric tensor $g_{\mu\nu}$ alone, then we can write
\be
S=\int d^4 x \mathcal{L}(g_{\mu\nu}).
\ee
If this action contains up to second derivatives of $g_{\mu\nu}$, then
extremising it with respect to the metric gives the
Euler-Lagrange expression
\be
E^{\mu\nu}[\mathcal{L}]= \frac{d}{d x^\rho} \left[ \frac{\partial
    \mathcal{L}}{ \partial g_{\mu\nu, \rho}} - \frac{d}{d x^\lambda} \left(\frac{\partial \mathcal{L}}{ \partial g_{\mu\nu, \rho\lambda}} \right) \right] -
  \frac{\partial \mathcal{L}}{\partial g_{\mu\nu}},
\label{E_L}
\ee
and the Euler-Lagrange equation is $E^{\mu\nu}(\mathcal{L})=0$.
Lovelock's theorem can then be stated as the following:
\newtheorem{theorem}{Theorem}[section]
\begin{theorem}
\emph{(Lovelock's Theorem)}
\newline
The only possible second-order Euler-Lagrange expression obtainable
in a four dimensional space from a scalar density of the form
$\mathcal{L}=\mathcal{L}(g_{\mu\nu})$ is
\be
E^{\mu\nu} = \alpha \sqrt{-g} \left[ R^{\mu\nu}-\frac{1}{2} g^{\mu\nu} R \right]+ \lambda \sqrt{-g} g^{\mu\nu},
\ee
where $\alpha$ and $\lambda$ are constants, and $R_{\mu\nu}$ and $R$ are the Ricci tensor and scalar curvature, respectively.
\end{theorem}
This powerful theorem means that if we try to create any gravitational
theory in a four-dimensional Riemannian space from an action principle involving
the metric tensor and its derivatives only, then the only field equations
that are second order or less are Einstein's equations and/or a cosmological constant.  This does not,
however, imply that the Einstein-Hilbert action is the only
action constructed from $g_{\mu\nu}$ that results in the Einstein
equations.  In fact, in four dimensions or less one finds that the most
general such action is
\be
\nonumber
\mathcal{L} = \alpha \sqrt{-g} R - 2 \lambda \sqrt{-g} + \beta
\epsilon^{\mu\nu\rho\lambda} R^{\alpha\beta}_{\phantom{\alpha\beta} \mu\nu} R_{\alpha\beta \rho\lambda}+ \gamma \sqrt{-g}
\left(R^2-4 R^{\mu}_{\phantom{\mu} \nu} R^{\nu}_{\phantom{\nu} \mu} +R^{\mu \nu}_{\phantom{\mu\nu} \rho\lambda} R^{\rho\lambda}_{\phantom{\rho\lambda} \mu\nu} \right),
\ee
where $\beta$ and $\gamma$ are also constants.  The third and fourth
terms in this expression do not, however, contribute to the
Euler-Lagrange equations as 
\ba
E^{\mu\nu}\left[\epsilon^{\alpha\beta \rho\lambda} R^{\gamma\delta}_{\phantom{\gamma\delta} \alpha\beta} R_{\gamma\delta\rho\lambda}\right] &=& 0\\
E^{\mu\nu} \left[ \sqrt{-g} \left(R^2-4 R^{\alpha}_{\phantom{\alpha} \beta} R^{\beta}_{\phantom{\beta} \alpha} 
+R^{\alpha\beta}_{\phantom{\alpha\beta} \rho\lambda} R^{\rho\lambda}_{\phantom{\rho\lambda} \alpha\beta} \right) \right] &=& 0,
\ea
where the action of $E^{\mu\nu}$ on any function $X$ is defined as in
Eq. (\ref{E_L}).
The first of these equations is valid in any number of dimensions, and
the second is valid in four dimensions only.

Lovelock's theorem means that to construct metric theories of gravity with
field equations that differ from those of General Relativity we must
do one (or more) of the following:
\begin{itemize}
\item Consider other fields, beyond (or rather than) the metric tensor.
\item Accept higher than second derivatives of the metric in the field
  equations.
\item Work in a space with dimensionality different from four.
\item Give up on either rank (2,0) tensor field equations, symmetry of
  the field equations under exchange of indices, or divergence-free
  field equations.
\item  Give up locality.
\end{itemize}
The first three of these will be the subject of the next three
sections of this report.  The fourth  option requires giving up on
deriving field equations from the metric variation of an action
principle, and will not be considered further here.

\subsubsection{Birkhoff's theorem}

Birkhoff's theorem\footnote{This theorem is commonly attributed to
  Birkhoff, although it was already published two years earlier by Jebsen
  \cite{jeb}. It is not to be confused with {\it Birkhoff's pointwise
  ergodic theorem}.} is of great significance for the weak-field limit
  of General Relativity.  The theorem states \cite{birk}
\begin{theorem}
\emph{(Birkhoff's Theorem)}
\newline
All spherically symmetric solutions of Einstein's equations in vacuum
must be static and asymptotically flat (in the absence of $\Lambda$).
\end{theorem}
Strictly speaking, there are very few situations in the real Universe
in which Birkhoff's theorem is of direct applicability:  Exact
spherical symmetry and true vacuums are rarely, if ever, observed.
Nevertheless, Birkhoff's theorem is very influential
in how we understand the gravitational field around (approximately)
isolated masses.  It provides strong support for the relativistic
extension of our Newtonian
intuition that far from such objects their gravitational influence
should become negligible, or, equivalently, space-time should be
asymptotically flat\footnote{Of course, in a cosmological setting
  asymptotic regions are never realised as we will eventually come
  across the other masses in the Universe.}.  We can therefore proceed
with some confidence in treating the weak-field limit of General
Relativity as a perturbation about Minkowski space.  Birkhoff's theorem also tells us
that certain types of gravitational radiation (from a star that
pulsates in a spherically symmetric fashion, for example) are not possible.

As we will show below, Birkhoff's theorem does not hold in many
alternative theories of gravity.  We therefore have less
justification, aside from our own intuition, in treating the weak
field limit of these theories as perturbations about Minkowski space.
We must instead be more careful, as the space-time we perform our expansion
around can have asymptotic curvature, leading to either time or
space-dependence of the background (or some combination of the two).
What is more, the perturbations themselves may be time-dependent, and
their form can be sensitive to the type of asymptotic curvature that
the background exhibits.  Behaviours such as these are not expected in
General Relativity \cite{LueStarkman2003}.

\subsubsection{The no-hair theorems}

These theorems are named after the phrase coined by Wheeler that
``{\it black holes have no hair}''.  The first of these theorems was
given by Israel and showed that the only static uncharged
asymptotically flat black hole solution to Einstein's equations is the
Schwarzschild solution \cite{isbh1}.  He later extended this theorem
to include charged objects \cite{isbh2}, and Carter extended it to
black holes with angular momentum \cite{cabh}.  The theorem is
therefore often stated today as ``{\it the generic final state of
  gravitational collapse is a Kerr-Newman black hole, fully specified
  by its mass, angular momentum, and charge }'' \cite{wabh}.

Complementary to the black hole no-hair theorems is the no-hair
`theorem' of de Sitter space.  The claim here is that
in the context of General Relativity with a cosmological constant all
expanding universe solutions should evolve towards de Sitter space.
This has been shown explicitly by Wald for all Bianchi type
models\footnote{Except type-IX universes with large amounts of spatial
  curvature.} \cite{wald}.

These theorems play an important role in General Relativity and
cosmology.  Some progress has been made in extending them to
alternative theories of gravity, but there have also been explicit
examples of them being violated in particular theories.  This
will be discussed further in subsequent sections.

\subsection{The Parameterised Post-Newtonian Approach}
\label{sec:PPN}

This section is a recap of the Parameterised Post-Newtonian
(PPN) formalism that is widely used by both theoretical and
observational gravitational physicists.  The idea here is to create a
construction that encompasses a wide array of different gravitational
theories, and that contains parameters that can be constrained by
observations in a reasonably straightforward fashion.  In this way labour
can be saved on both the theoretical and observational ends of the
spectrum:  Observers can apply their results to constrain a wide array
of theories without having to trawl through the details of the
individual theories themselves, and theorists can straightforwardly
constrain their new theories by comparing to the already established
bounds on the PPN parameters without having to re-calculate individual
gravitational phenomena.  To date, this approach has been highly
successful, and in the following sections of this report we will often
refer to it.  We will therefore outline here how the PPN formalism proceeds.
For a more detailed explanation of the principles and
consequences of this formalism the reader is referred to \cite{tegp}. 

\subsubsection{Parameterised post-Newtonian formalism}

The PPN formalism is a perturbative treatment of weak-field gravity,
and therefore requires a small parameter to expand in.  For this
purpose an ``order of smallness'' is defined by
\begin{equation*}
U \sim v^2\sim\frac{P}{\rho} \sim \Pi \sim O(2),
\end{equation*}
where $U$ is the Newtonian potential, $v$ is the 3-velocity of a fluid
element, $P$ is the pressure of the fluid, $\rho$ is its rest-mass
density and $\Pi$ is the ratio of energy density to rest-mass
density.  Time derivatives are also taken to have an order of
smallness associated with them, relative to spatial derivatives:
\begin{equation*}
\frac{\vert \partial/\partial t \vert}{\vert \partial/\partial x
  \vert} \sim O(1).
\end{equation*}
Here we have chosen to set $c=1$.  The PPN formalism now
proceeds as an expansion in this order of smallness.

For time-like particles coupled to the metric only
the equations of motion show that the level of approximation
required to recover the Newtonian limit is $g_{00}$ to $O(2)$, with no
other knowledge of other metric components beyond the background level
being necessary.  The 
post-Newtonian limit for time-like particles, however, requires a knowledge of
\begin{align*}
&g_{00} \qquad \text{to} \qquad O(4)\\
&g_{0i} \qquad \text{to} \qquad O(3)\\
&g_{ij} \qquad \text{to} \qquad O(2).
\end{align*}
Latin letters here are used to denote spatial indices.  To obtain the
Newtonian limit of null particles we only need to know the metric to
background order:  Light follows straight lines, to Newtonian accuracy.
The post-Newtonian limit of null particles requires a knowledge of
$g_{00}$ and $g_{ij}$ both to $O(2)$.

The way in which the PPN formalism then proceeds is as follows.  First
one identifies the different fields in the theory.  All dynamical
fields should then be perturbed from their expected background values,
and the perturbations assigned an appropriate order of smallness each.
For theories containing a metric the appropriate expansion is usually
\ba
g_{00} &=& -1 + h^{(2)}_{00} + h^{(4)}_{00} +O(6)\\
g_{0i} &=& h^{(3)}_{0i} +O(5)\\
g_{ij} &=& \delta_{ij} + h^{(2)}_{ij} +O(4),
\ea
where superscripts in brackets denote the order of smallness of the
term.  If, for example, the theory contains an additional scalar
field, then the usual expansion for this quantity is
\be
\phi = \phi_0 + \varphi^{(2)} + \varphi^{(4)} +O(6),
\ee
where $\phi_0$ is the constant background value of $\phi$.
Additional vector and tensor gravitational fields can be specified in a
corresponding way.  

The energy-momentum tensor in the PPN formalism is then taken to be
that of a perfect fluid.  To the relevant order, the
components of this tensor are given by
\begin{align}
\label{T00}
&T_{00} = \rho (1+\Pi +v^2 - h_{00}) +O(6)\\
\label{T0i}
&T_{0i} = -\rho v_{i} +O(5)\\
\label{Tij}
&T_{ij} = \rho v_{i} v_{j} +P \delta_{ij} +O(6).
\end{align}
Taking these expressions, the field equations for the theory in
question, and substituting in the perturbed expressions for the
dynamical fields in the theory, as prescribed above, the field
equations can then be solved for order by order in the smallness
parameter.

The first step in such calculations is usually to solve for
$h_{00}^{(2)}$.  With this solution in hand, one then proceeds to
solve for $h_{ij}^{(2)}$ and $h_{0i}^{(3)}$ simultaneously, and
finally $h_{00}^{(4)}$ can be solved for.  If additional fields exist,
beyond the metric, then these quantities must also be solved for to
increasing order of smallness as the calculation proceeds.  In finding
$h_{ij}^{(2)}$, $h_{0i}^{(3)}$ and $h_{00}^{(4)}$ one needs to
specify a gauge.  
%
%
After such a specification one still, of course, has the freedom to make
additional gauge transformations of the form $x^{\mu} \rightarrow x^{\mu} + \xi^{\mu}$, where $\xi^\mu$ is $O(2)$ or smaller.  This freedom
should be used at the end of the
process to transform the metric that has been obtained into the
``standard post-Newtonian gauge''.   This is a gauge in which the spatial part of
the metric is diagonal, and terms containing time derivatives are
removed.   Once this has been done then one is in
possession of the PPN limit of the theory in question.

We have so far outlined the procedure that one needs to follow in
order to gain the appropriate form of the metric that couples to matter fields in
the weak-field limit.  Once done, the result can then be compared to
the `PPN test metric' below:
\ba
\hspace{-30pt} g_{00} &=& -1 +2 G U-2 \beta G^2 U^2-2 \xi G^2 \Phi_W +(2 \gamma +2 +
\alpha_3 +\beta_1 -2 \xi) G \Phi_1 \nonumber \\ &&
 +2 (1+3 \gamma -2 \beta +\beta_2 +\xi) G^2 \Phi_2 +2 (1+\beta_3)
G \Phi_3 -(\beta_1-2 \xi) G \mathcal{A} \nonumber \\ &&
 +2 (3 \gamma+3 \beta_4-2 \xi) G \Phi_4 \nonumber\\
\hspace{-30pt} g_{0i} &=& -\frac{1}{2} (3+4 \gamma +\alpha_1-\alpha_2
+\beta_1-2 \xi) G V_i -\frac{1}{2}(1+\alpha_2-\beta_1+2 \xi) G W_i \nonumber\\
\hspace{-30pt} g _{ij} &=& (1+2 \gamma G U) \delta_{ij} \nonumber.
\ea
Here $\beta$, $\gamma$, $\xi$, $\beta_1$, $\beta_2$, $\beta_3$,
$\beta_4$, $\alpha_1$, $\alpha_2$ and $\alpha_3$ are the
`post-Newtonian parameters', $U$ is the Newtonian gravitational
potential that solves the Newtonian Poisson equation, and $\Phi_W$, $\Phi_1$,
$\Phi_2$, $\Phi_3$, $\Phi_4$, $\mathcal{A}$, $V_i$ and $W_i$ are
the `post-Newtonian gravitational potentials' (the precise form of
these potentials is given in \cite{tegp}).  The particular combination
of parameters before each of these potentials is chosen here
so that they have  particular physical significance, once
gravitational phenomena have been computed.

\subsubsection{Parameterised post-Newtonian constraints}

Comparison of the weak field metric of a particular theory
with the PPN test metric above allows one to read off
values for the PPN parameters $\beta$, $\gamma$, $\xi$, $\beta_1$,
$\beta_2$, $\beta_3$, $\beta_4$, $\alpha_1$, $\alpha_2$ and $\alpha_3$
for the theory in question.  The test metric has been constructed to
include the type of potentials that often appear when one modifies
gravity\footnote{It is not, however, an exhaustive collection of all possible
potentials, and in some theories it is occasionally necessary to
include additional terms.}.
The great utility of the PPN formalism is that observers can take the
PPN test metric above and constrain the parameters {\it without having
  a particular theory in mind}.  These constraints can then be applied
directly to a large number of gravitational theories, without having
to work out how complicated gravitational phenomena work in each theory
individually.

In General Relativity we have that $\beta=\gamma=1$ and $\xi =\beta_1
=\beta_2 =\beta_3 =\beta_4 =\alpha_1 =\alpha_2 =\alpha_3 =0$.  Other
theories will predict other values for these parameters, and we will
discuss these on a case by case basis in the sections that follow.
Observationally, one can use the gravitational phenomena discussed in
Section \ref{intro} to impose the constraints that follow.

As already discussed, observations that involve only null geodesics
are sensitive to the Newtonian part of the metric, $g_{00}^{(2)}$, and
the term $g_{ij}^{(2)}$ only.  These two terms involve the
PPN parameter $\gamma$ only.  We can now use constraints on the
bending of light by the Sun to get a constraint on $\gamma$.  Using
the PPN test metric the predicted 
bending of light that one should observe is \cite{tegp}
\be
\theta = 2 (1+\gamma) \frac{m}{r} = \frac{(1+\gamma )}{2} \theta_{GR},
\ee
where $m$ is the mass of the Sun, $r$ is its radius, and $\theta_{GR}$
is the general relativistic prediction.  Using the observed value of
$\theta$ given in Section \ref{intro} then gives \cite{sdlg}
\be
\gamma -1 = (-1.7 \pm 4.5) \times 10^{-4},
\ee
which is consistent with the general relativistic value of $\gamma
=1$.  Similarly, we can use the PPN test metric to find that the Shapiro time delay
effect is given by \cite{tegp}
\be
\Delta t = \frac{(1+\gamma)}{2} \Delta t_{GR},
\ee
where subscript $GR$ again means the value of this quantity as
predicted by General Relativity.  Taking the observed value of $\Delta t$
given in Section \ref{intro} then gives the even tighter constraint \cite{bit}
\be
\label{gamcas}
\gamma -1 = (2.1 \pm 2.3) \times 10^{-5},
\ee
again consistent with $\gamma = 1$.
It can now be clearly seen that the bending of light by the Sun, and
the Shapiro time delay effect do, in fact, constrain the same aspect
of space-time geometry.  They can therefore be considered as
complimentary to each other.

If we now consider observations of gravitational phenomena that
involve time-like geodesics then we are able to observe, potentially,
all of the post-Newtonian potentials in the PPN test metric.  This
becomes clear from the expression for perihelion precession, which now
becomes
\be
\nonumber
\Delta \omega =\frac{6 \pi M}{p} \left[ \frac{1}{3} (2+2 \gamma -
  \beta) + \frac{1}{6} (2 \alpha_1-\alpha_2+\alpha_3 +2 \beta_2)
  \frac{\mu}{M} + J_2 \left( \frac{r^2}{2 M p} \right) \right],
\ee
where $M$ is the total mass of the two bodies involved, $\mu$ is their
reduced mass, and $p$ is the semi-latus rectum of the orbit.  The
affect of modifying the geometry can be seen here to be degenerate
with the effect due to the solar quadrupole moment, $J_2$.  Once the
value of this quantity is known, however, then one is able to gain
constraints on the above combination of $\beta$, $\gamma$, $\alpha_1$,
$\alpha_2$, $\alpha_3$ and $\beta_2$.  This can be done for any or all
of the observations of the perihelion precession of Mercury given in
Section \ref{intro}, and if we take the value of $\gamma$ to be that
given by Eq. (\ref{gamcas}), as well as\footnote{These values will be
  given some justification shortly.} $\alpha_1 \sim \alpha_2 \sim \alpha_3 \sim
\beta_2 \sim 0$ and a reasonable value of $J_2 \sim 10^{-7}$, then
this gives constraints on $\beta$ of the order $\beta-1 \sim O(10^{-3})$
or $O(10^{-4})$.  However, as already noted, these constraints are
somewhat sensitive to a number of assumptions about the orbits of the
other planets, as well as the solar quadrupole moment.

The Nordtvedt effect is similarly an observation of time-like
geodesics.  In this case it is convenient to define the `Nordtvedt
parameter'
\be
\eta_N \equiv 4 \beta - \gamma-3 -\frac{10}{3} \xi
-\alpha_1+\frac{2}{3} \alpha_2 -\frac{2}{3} \beta_1 - \frac{1}{3}
\beta_2,
\ee
which is not to be confused with the equivalence principle violation
parameter $\eta$ defined in Equation (\ref{eotwash}).  The
observations of Williams, Turyshev and Boggs \cite{nordtvedtobs} then
give the constraint $\eta_N = (4.4 \pm 4.5) \times 10^{-4}$, which, if
we again take $\gamma$ to be given by observations of the Shapiro time
delay effect with all other PPN parameters being zero, gives us
\be
\beta-1 = (1.2 \pm 1.1) \times 10^{-4},
\ee
which is a much cleaner constraint on $\beta$ than those which can be
derived from observations of the perihelion precession of Mercury.

In `conservative' theories of gravity it is usually only the PPN
parameters $\beta$ and $\gamma$ (and sometimes $\xi$) that vary from
their general relativistic values.  These quantities are often
interpreted as the degree of non-linearity in the gravitational
theory, and the amount of spatial curvature per unit mass that is
produced, respectively.  The other parameters $\xi$, $\alpha_i$ and $\beta_i$ are
usually interpreted as corresponding to preferred location effects, preferred
frame effect and the violation of conserved quantities.
When considering theories in which such effects are expected to be
absent it is therefore usual to assume that these parameters are all
zero, and to search instead for constraints on $\beta$ and $\gamma$.

Of course, one can subject the  $\xi$, $\alpha_i$ and $\beta_i$
parameters to observational scrutiny in a number of ways.  The table
below gives a selection of the tightest constraints currently
available:
\begin{table}[tbh]
\begin{center}
\begin{tabular}{|c|c|l|}
\hline
\qquad Parameter & Limit & Source
\\ \hline
$\xi$ & $10^{-3}$ & Ocean tides \cite{tegp} \\ 
$\alpha_1$ & $10^{-3}$ & Lunar laser ranging \cite{a1}\\ 
$\alpha_2$ & $4 \times 10^{-7}$ & Alignment of Sun's spin axis with ecliptic \cite{a2}\\ 
$\alpha_3$ & $4 \times 10^{-20}$ & Pulsar acceleration \cite{a3} \\ 
\hline
\end{tabular}
\end{center}
\end{table}

\noindent
Further constraints and discussion on the $\beta_i$ parameters can be
found in \cite{tegp}.  For more details of the observations leading to
these constraints on $\xi$ and $\alpha_i$ the reader is
referred to the source material cited above and \cite{tegp}.

The constraints on the PPN parameters that we have discussed above are
all, to date, in reasonably good agreement with General Relativity,
and it is likely that future observations of, for example, the `double pulsar'
\cite{dp1, dp2} will tighten these constraints even further in
coming years.  This excellent concordance of numerous different
physical phenomena means that one must reconcile any alterations to
General Relativity with observations in weak field systems that appear
to be narrowing down on a general relativistic description.  As we
will describe in the sections that follow, this places tight
constraints on a variety of different modified theories of gravity:  It must be
the case that any alternative theories that we consider should
reproduce General Relativity in the appropriate weak field limit, or at least
something very close to it.

There are a number of mechanisms that have been considered in the
literature that allow for a general relativistic weak field limit even
in theories that are, in general, very different from General
Relativity.  These include the
Vainshtein mechanism~\cite{vainshtein} which occurs when large derivative interactions are present, the
Chameleon mechanism of theories with non-minimal coupling to scalar
fields \cite{cham1}, as well as the attractor mechanism of Damour and
Nordtvedt \cite{attractor}.  These different approaches allow, potentially,
for theories that deviate considerably from General Relativity to
exist without disturbing gravitational physics in the solar system to
a large extent.  They are thought to be successful in a
number of different environments, and have sometimes been applied to
situations that are quite different to the ones in which they were
originally conceived.

As well as successful reproductions of general relativistic behaviour, however,
there have also been a number cases found in the literature of
theories that produce weak field gravity that is surprisingly inconsistent with the
predictions of General Relativity.  Perhaps the most famous of these is the van
Dam-Veltman-Zakharov (vDVZ) discontinuity that was
originally found in the context of Pauli-Fierz gravity
\cite{vdvz1,vdvz2} (a theory with
one dynamical metric, and one non-dynamical {\it a priori} specified
metric).  Here the graviton acquires a mass through the introduction
of terms into the gravitational Lagrangian that, in the weak field
limit, look like mass terms for the perturbations $h_{\mu\nu}$ around
Minkowski space,  i.e. like $m^2 h^{\mu\nu}h_{\mu\nu}$.  Naively one might then
expect in the limit $m \rightarrow 0$, when the graviton becomes
massless, that the zero mass theory of General Relativity should
be recovered.  This is, however, not the case.  Instead one finds from
the study of linear perturbations around Minkowski space that
$\gamma \rightarrow 1/2$, which can be seen from the constraints above
to be in strong disagreement with a number of different observations,
including light bending and time delay effects.  The general
relativistic limit in this case is therefore a singular one, and any
finite but non-zero graviton mass, no matter how small, appears to give results
that are incompatible with observations.  Similar results have also been
found in some theories of gravity constructed from more general functions
of the Ricci curvature than the Einstein-Hilbert action \cite{chiba},
and are expected in other theories as well.
In these cases one must either abandon the theory as being
incompatible with observations, or show that the treatment being
applied is unsatisfactory because, for example, one of the mechanisms
discussed previously should be applied.

Issues such as those just discussed can make the study of weak field
gravity in modified theories a more complicated subject than it is
in General Relativity.  One must be careful to make sure that the
treatments being applied are justifiable, that the limits of the
theory take the expected form (rather than being singular), and that
non-linear mechanisms and non-perturbative effects are being fully
taken into account.  How this should be done for specific modified
theories of gravity will be the subject of subsequent sections.  In
some cases it is still an active area of research.

\subsection{Cosmology}
\label{sec:cosmology}

We now turn to cosmology, which forms  a major part of this review. In
this section we first describe
cosmology from the point of view of General Relativity, including
Friedmann-Lema\^{i}tre-Robertson-Walker (FLRW) solutions,
cosmic distance measures and cosmological perturbation theory. We then
consider the observational evidence that has led to
the rise of the ``Dark Sector'', thus arriving at the so-called
$\Lambda$CDM `concordance model'.  We end this section with a short
discussion of the successes of $\Lambda$CDM, its predictions and
potential shortcomings.

\subsubsection{The Friedmann-Lema\^{i}tre-Robertson-Walker solutions}
The Robertson-Walker metric in the synchronous coordinate system is
\begin{equation}
 ds^2 = -dt^2 + a^2(t) q_{ij}dx^i dx^j ,
\end{equation}
where $q_{ij}$ is a maximally symmetric $3$-metric of Gaussian
curvature $\kappa$.  In a spherically symmetric coordinate system this
can be written
$$
ds^2 =  -dt^2 + a^2 \left[ \frac{dr^2}{1 - \kappa r^2}+ r^2
d\Omega\right] 
$$ 
where $\kappa$ is a real constant. If $\kappa=0$, the hyper-surfaces
of constant $t$ are flat, if $\kappa>0$ they are positively curved,
and if $\kappa<0$ they are negatively curved.  The function $a(t)$ is
called the scale factor, and we assume it to be normalised to unity today.  

The Hubble parameter is defined as
\begin{equation}
 H = \frac{1}{a} \frac{da}{dt},
\end{equation}
with $H_0 = H(a=1)$ being the Hubble constant, i.e. the value of the
Hubble parameter at the present time. 

The dynamics of the scale factor is given by the Friedmann equation
\begin{equation}
3H^2 = 8\pi G \sum_i\rho_i
\end{equation}
where $\rho_i$ are the energy densities of all possible fluids,
including photons $\rho_\gamma$,  neutrinos $\rho_\nu$ (possibly with
mass $m_\nu$), pressureless matter $\rho_m$, and spatial curvature
$\rho_{\kappa}$. We may also define the relative densities 
\begin{equation}
\Omega_i = \frac{\rho_i}{\rho_T},
\end{equation}
where $\rho_T$ is the total energy density $\rho_T = \frac{3H^2}{8\pi
G} =   \sum_i\rho_i$. The Friedmann equation then becomes the
constraint $\sum_i \Omega_i = 1$.

If each fluid is uncoupled then energy-momentum conservation gives 
\begin{equation}
  \dot{\rho}_i + 3 H (1 + w_i) \rho_i = 0,
\label{eq_rho_uncoupled}
\end{equation}
where $w$ is the equation of state, defined by $P = w \rho$.  For the
known forms of matter $w_\gamma = \frac{1}{3}$, $w_m = 0$, $w_{\kappa}
= -\frac{1}{3}$, and $w_\nu$ is in the range  $[0,\frac{1}{3}]$.
We may solve Eq. (\ref{eq_rho_uncoupled}) for a few cases of interest,
and then determine the dynamics of the scale factor. For radiation we
obtain $\rho_r = \rho_{0r} a^{-4}$, for pressureless matter $\rho_m =
\rho_{0m} a^{-3}$, for curvature $\rho_{\kappa} = -\frac{3\kappa}{8\pi
  G}  a^{-2}$, and for a cosmological constant $\rho_\Lambda =
\frac{1}{8\pi G}\Lambda$. 

A general analytic solution in the case where all the above fluids are
present is impossible.  However, analytic solutions can be found in
certain special cases.  If a single fluid is present and $w$ is constant then
$a^{3(1+w)/2}  = \frac{3(1+w)}{2} H_0t$, provided $w\ne -1$. For the
case of radiation we get $a =  \sqrt{2 H_0t}$, and for pressureless
matter $a  = \left(\frac{3H_0t}{2}\right)^{2/3}$. The case of a
cosmological constant is special: One obtains $a = e^{H_0 t}$, the de
Sitter solution, in which space (in this coordinate system) is
exponentially expanding.

In many cases of interest it is convenient to use a different time
coordinate, the conformal time, $\tau$, defined by $dt = a d\tau$.  In
a radiation dominated universe we then have $a(\tau) = H_0 \tau$, in a
matter dominated universe $a = \frac{1}{4} (H_0\tau)^2$, and for the
de Sitter universe $ a= \frac{1}{H_0(\tau_{\infty} - \tau)}$, where
$\tau_\infty$ is the value of the conformal time at $a\rightarrow
\infty$.  In a universe filled with both radiation and matter we get
$a =  \sqrt{\Omega_{0r}}H_0\tau  + \frac{\Omega_{0m}}{4} (H_0
\tau)^2$.  A summary of these solutions is shown in Table \ref{Tab_FRW_solutions}.

\begin{table}[tbh]
\begin{center}
\begin{tabular}{|l|l|l|}
\hline
 Matter type  &   $a(t)$ & $a(\tau)$ 
\\
\hline
 radiation   &  $a =  (2H_0 t)^{1/2}  $ & $a = H_0 \tau$
\\
 dust   &     $ a = (\frac{3}{2} H_0 t)^{2/3}$ & $a = \frac{1}{4} (H_0 \tau)^2$
\\
 radiation \& dust   &      complicated & $a = \sqrt{\Omega_{0r}} H_0\tau  + \frac{\Omega_{0m}}{4} (H_0 \tau)^2$
\\
 $\Lambda$  &  $a = e^{H_0t}$ & $a =
  \frac{1}{H_0(\tau_{\infty} - \tau)}$\\
\hline
\end{tabular}
\end{center}
\caption{A summary of particular solutions to the Friedmann equation.}
\label{Tab_FRW_solutions}
\end{table}

\subsubsection{Cosmological distances}

Given a Friedmann universe obeying Einstein's field equations, it is
useful to define observables that characterise the background
evolution. Distances play an important role if we are to map out its
behaviour (see \cite{Hogg99} for a more detailed explanation). Hubble's law 
$
v=H_0 d $
allows us to define a 
{\it Hubble time},
$
t_H=\frac{1}{H_0}=9.78\times 10^9~h^{-1}~{\rm yr} 
$
and the {\it Hubble distance},
$
D_H=\frac{c}{H_0}= 3000~h^{-1}~{\rm Mpc}
$. 
We can also integrate along a light ray to get the {\it comoving distance}:
\begin{eqnarray}
D_C=c\int^{t_0}_t \frac{cdt'}{a(t')}. \nonumber
\end{eqnarray}
From $-\kappa=\Omega_{\kappa}/D^2_H$, and performing
the radial integral (assuming the observer is at $r=0$), we have
\begin{eqnarray}
D_C=\int_0^{D_M}\frac{dr}{\sqrt{1-\kappa r^2}}=\left\{\begin{array}{ll} 
\frac{D_H}{\sqrt{\Omega_{\kappa}}}\sinh^{-1}[\sqrt{\Omega_{\kappa}} D_M/D_H] & \mbox{for} \ \ \Omega_{\kappa}>0 \\
D_M & \mbox{for} \ \ \Omega_{\kappa}=0 \\
\frac{D_H}{\sqrt{|\Omega_{\kappa}|}}\sin^{-1}[\sqrt{|\Omega_{\kappa}|} D_M/D_H] & \mbox{for} \ \ \Omega_{\kappa}<0, \end{array} \right. \nonumber
\end{eqnarray}
where the {\it proper motion distance} (also known as the {\it
  transverse comoving distance}) is $D_M$.  This can be rewritten as
\begin{eqnarray}
D_M=\left\{\begin{array}{ll} 
\frac{D_H}{\sqrt{\Omega_{\kappa}}}\sinh[\sqrt{\Omega_{\kappa}} D_C/D_H] & \mbox{for} \ \ \Omega_{\kappa}>0 \\
D_C& \mbox{for} \ \ \Omega_{\kappa}=0 \\
\frac{D_H}{\sqrt{|\Omega_{\kappa}|}}\sin[\sqrt{|\Omega_{\kappa}|} D_C/D_H] & \mbox{for} \ \ \Omega_{\kappa}<0. \end{array} \right. \nonumber
\end{eqnarray}
It is then possible to find an expression for  the {\it angular
  diameter distance}: 
$$ 
D_A=\frac{D_M}{1+z}.
$$
Hence, if we know the proper size of an object and its redshift we can
work out, for a given universe, the angular diameter distance,
$D_A$. If we measure the brightness or luminosity of an object, we
know that the flux of that object at a distance $D_L$ is given by $
F=\frac{L}{4\pi D^2_L} $, where $D_L$ is aptly known as the {\it
  luminosity distance} and is related to other distances through: 
$$ 
D_L=(1+z)D_M=(1+z)^2 D_A.
$$
This relation is a consequence of Etherington's
theorem~\cite{Etherington1933}, and holds in any metric theory of
gravity, irrespective of the field equations. It is however violated
if the photon number is not conserved (e.g. due to photon-axion
mixing), or if photons are extinguished due to the presence of dust.
It turns out that in astronomy one often works with a logarithmic
scale, i.e. with {\it magnitudes}. One can then define the {\it
  distance modulus}:
$$
DM\equiv 5\log\left(\frac{D_L}{10~{\rm pc}}\right), 
$$
which can be measured from the {\it apparent magnitude}, $m$, (related
to the flux at the observer), and the {\it absolute magnitude}, $M$,
(what the magnitude would be if the observer was at $10$ pc from the source)
through $m=M+DM$.

Finally, let us consider Hubble's law. Take two objects that are a
distance $d$ apart, and Taylor expand the scale factor today to find
\begin{eqnarray}
a(t)=a(t_0)+{\dot a}(t_0)[t-t_0]+\frac{1}{2}{\ddot a}(t_0)[t-t_0]^2 
+ \cdots . \nonumber
\end{eqnarray}
On small scales the distance to the emitter is roughly related to the
time of emission, $t$, by $d\simeq c(t_0-t)$. We can then rewrite the
above expression as
\begin{eqnarray}
(1+z)^{-1}=1-H_0\frac{d}{c}-\frac{q_0H^2_0}{2}\left(\frac{d}{c}\right)^2  
+ \cdots , \nonumber
\end{eqnarray}
where $q_0=-{\ddot a}a/{\dot a}^2|_{t=t_0}$ is the {\it deceleration
parameter}.  On small scales and at small redshifts we then have
Hubble's law, $cz=H_0d$.

To constrain the background evolution it is necessary to have good
distance measurements. So, for example, with measurement of supernovae
light curves at different redshifts it is, in principle, possible to 
measure $D_L(z)$. Alternatively one might try to measure $D_A(z)$
by observing known length scales in the universe. This has been done
spectacularly well with the sound horizon of the cosmic microwave
background at redshift $z\simeq 1100$. More tentatively, there is a
constraint on a combination of $D_A(z)$ and $H(z)$ using the imprint
of acoustic oscillations of baryons on galaxy clustering at moderate 
to low redshifts, $z\simeq 0.1-0.3$.

\subsubsection{Perturbation theory}

We now turn to perturbation theory, which is an indispensable tool for
making predictions for a variety of cosmological observations.  For
extensive treatments of cosmological perturbation theory the reader is
referred
to~\cite{Bardeen1980,KodamaSasaki1984,ellisbruni,MukhanovFeldmanBrandenberger1992}.
 Here we shall only consider scalar fluctuations, for which the
 perturbed FLRW metric can be written
\begin{equation}
ds^2 = a^2 
\left\{ 
-(1 -  2 \Xi) d \tau^2 - 2 (\grad_i \beta ) d \tau dx^i 
  + \left[  \left( 1 + \frac{1}{3}\chi \right) q_{ij} + D_{ij} \nu
    \right] dx^i dx^j
\right\} ,
\end{equation}
where $D_{ij} \equiv \grad_i \grad_j - \frac{1}{3} q_{ij} \Delta$ is a
trace-less spatial derivative operator.  We note that $\grad_i$ is the
covariant derivative compatible with the $3$-metric $q_{ij}$.  Perfect
fluids with shear have energy-momentum tensors that can be written as
\begin{equation}
T_{\mu\nu} = (\rho+P) u_\mu u_\nu + P g_{\mu\nu} + \Sigma_{\mu\nu},
\end{equation}
where $\rho$ is the energy density, $P$ is the pressure, $u_\mu$ the
4-velocity of the fluid (normalised to $u^\mu u_\mu=-1$), and
$\Sigma_{\mu\nu}$ is the anisotropic stress tensor which obeys
$u^\mu \Sigma_{\mu\nu}= \Sigma^\mu_{\phantom{\mu}\mu}=0$.  In a
homogeneous and isotropic space $\Sigma_{\mu\nu}=0$, and $u_\mu$ is
aligned with the time direction such that in the coordinate system
used above it has components $u_\mu = (a, \vec{0})$. For first order
scalar perturbations we can parameterise $T^\mu_{\phantom{\mu}\nu}$ as
\begin{eqnarray}
T^0_{\phantom{0}0} &=& -\rho \delta
\\
T^0_{\phantom{0}i} &=& -(\rho+P) \grad_i \theta
\\
T^i_{\phantom{i}0} &=& (\rho+P) \grad^i(\theta-\beta)
\\
T^i_{\phantom{i}j} &=&  \delta P \delta^i_{\phantom{i}j} +  (\rho + P)
D^i_{\phantom{i}j} \Sigma ,
\end{eqnarray}
while the fluid velocity is $u_\mu = a ( 1 - \Xi, \grad_i \theta)$.
Here $\delta P$ is the  pressure perturbation, and $\Sigma$ the scalar
anisotropic stress.

For any variable ${\bf X}$, its perturbation $\delta {\bf X}$ is not
necessarily an observable quantity, and may depend on a gauge. In
particular, one can always define a new perturbation $\delta {\bf X}'
= \delta {\bf X} + \Lie{\xi}{\bf \bar{X}}$ through the Lie derivative
acting on the background tensor ${\bf \bar{X}}$ through a vector field $\xi^\mu$.
The perturbations $\delta {\bf X}$ are thus in general
gauge-dependent\footnote{The Stewart-Walker lemma~\cite{StewartWalker1974}
states that the only gauge-invariant perturbed tensors are those that
have background values that are either zero or a constant multiple of
the identity matrix.}.  For scalar perturbations we can write $\xi_\mu =
a ( -\xi, \grad_i \psi )$, and then find how our variables transform
under gauge transformations using the Lie derivative.  All of them,
apart from $\Sigma$, are gauge-dependent, with transformations given by

\begin{table}[tbh]
\begin{center}
\begin{tabular}{l l}
$\Xi \rightarrow\Xi - \frac{{\xi}'}{a}$
& \qquad
$\beta \rightarrow \beta + \frac{1}{a}\left[ \xi + {\cal H}\psi - {\psi}'\right]$
\\
$\chi \rightarrow \chi  + \frac{1}{a}\left[ 6 {\cal H} \xi + 2\lap \psi\right]$
& \qquad
$\nu \rightarrow \nu + \frac{2}{a}\psi$
\\
$\delta \rightarrow \delta -\frac{3}{a}(1+w) {\cal H}\xi$
 & \qquad $\theta \rightarrow \theta + \frac{1}{a}\xi$ 
\\
$\frac{\delta P}{\rho} \rightarrow \frac{\delta P}{\rho} + \frac{1}{a} \left[{w}' - 3w(1+w) {\cal H} \right]\xi$
& \qquad
$\Sigma \rightarrow \Sigma$.
\end{tabular}
\label{gauge_trans}
\end{center}
\end{table}
where $\adotoa= \frac{a'}{a}$.

Given our set of perturbation variables, two linear combinations of them can be removed\footnote{One has to be careful and not over constrain the
gauge by removing two combinations that transform with the same gauge variable, e.g. $\delta$ and $\theta$ both transform with $\xi$ and therefore cannot be
set to zero simultaneously.}  (set to zero). Popular gauges are 
\begin{itemize}
 \item Newtonian gauge: $\nu = \beta =0$. The remaining metric perturbations give rise to the Newtonian potentials $\Phi = -\frac{1}{6}\chi$ and $\Psi = -\Xi$.
 \item Synchronous gauge: $\Xi = \beta =0$ (this does not completely fix the gauge). The remaining metric perturbations are related to the Ma-Bertschinger~\cite{MaBertschinger1995} 
variables as $\chi = h$ and $-k^2 \nu = h + 6\eta$.
 \item Comoving gauge: $\theta = \nu =0$. Strictly speaking there is a multitude of comoving gauges depending on which velocity $\theta$ is set to zero. Thus we may speak of
a "baryon comoving gauge" if $\theta_b=0$, a "photon comoving gauge" if $\theta_\gamma=0$,
   the total matter comoving gauge if $\theta_T = \frac{\sum_X (\rho_X + P_X) \theta_X}{\sum_X (\rho_X + P_X)} =0$, etc.
 \item Uniform density gauge: $\delta = \nu = 0$. Once again there is a multitude of uniform density gauges depending on which density fluctuation is set to zero, as in the
comoving gauges above.
 \item Spatially flat gauge: $\chi = \nu = 0$.
\end{itemize}
It is possible to find combinations of perturbation variables that are
gauge invariant, but note that there are an infinite number of them
as any linear combination of gauge-invariant variables is also
gauge-invariant. Two popular gauge-invariant metric variables are the
Bardeen potentials $\PhiGI$ and $\PsiGI$:
\begin{equation}
\label{BarGR1}
\PhiGI \equiv -\frac{1}{6}( \chi - \lap \nu  ) + \frac{1}{2}{\cal H} (
       {\nu}' +2 \beta), 
\end{equation}
and 
\begin{equation}
\label{BarGR2}
\PsiGI \equiv - \Xi  - \frac{1}{2}({\nu}'' +2 {\beta}')  -
\frac{1}{2}{\cal H} (  {\nu}' +2 \beta ). 
\end{equation}
The Newtonian gauge is special in this case as $\PhiGI = \Phi$ and
$\PsiGI = \Psi$.  From now on we will refer to $\Phi$ and $\Psi$
without a ``hat'' as the Newtonian gauge potentials.  The Einstein equations in the Newtonian gauge give
\begin{eqnarray}
 2 (\Delta + 3 \kappa) \Phi - 6 {\cal H}({\Phi}' +  {\cal
 H}\Psi) &=& 8\pi G a^2  \sum_i \rho_i  \delta_i 
\label{E_Delta}
\\
 2({\Phi}' +  {\cal H}  \Psi) &=&  8\pi G a^2 \sum_i (\rho_i + P_i)\theta_i 
\label{E_Theta}
\end{eqnarray}
\begin{eqnarray}
&&
  {\Phi}'' +  \adotoa {\Psi}' + 2\adotoa {\Phi}' + \left(2\adotoa'  + \adotoa^2 + \frac{1}{3}\Delta \right) \Psi - (\frac{1}{3}\Delta  + \kappa )\Phi = 4\pi G a^2 \sum_i \delta P_i
\label{E_P}
\end{eqnarray}
and
\begin{eqnarray}
 \Phi - \Psi =  8 \pi G a^2 \sum_i (\rho_i + P_i)\Sigma_i 
\label{E_Sigma}
\end{eqnarray}
Combining Eqs. (\ref{E_Delta}) and (\ref{E_Theta}) we can find $\Phi$
in terms of the matter variables as
\begin{equation}
2 (\Delta + 3 \kappa ) \Phi =  3 {\cal H}^2  \sum_i \Omega_i
\left[\delta_i + 3{\cal H} (1 + w_i)\theta_i\right] 
\end{equation}
while $\Psi$ is then obtained using Eq. (\ref{E_Sigma}).

Finally, all scalar modes can be decomposed in terms of a complete set
of eigen-modes of the Laplace-Beltrami operator. For example, a
variable $A$ can be decomposed as
$$
A(x^i,t) = \int d^3k \; Y(x^j,k_k)\; \tilde{A}(k_i,t),
$$
where the eigen-modes, $ Y(x^j,k_k)$, obey $\left(\Delta + k^2
\right)Y =0$ \label{def_k}. In the special case of topologically
trivial and spatially flat hyper-surfaces of constant $t$, we simply
have $Y  = e^{i k_j x^j}$.  The integral transform above is then a
Fourier transform.  The value of $k$ depends on the geometry and
topology of the spatial hyper-surfaces: In the case of trivial
topology $k$ takes values $k = \sqrt{k_{*}^2 - \curv{}}$, where
$k_{*}$ is continuous and obeys $k_{*}\ge 0$ for zero or negative
spatial curvature, while $k_{*} = N\sqrt{\kappa}$ for positive spatial
curvature, where $N \ge 3$ is an integer.

\subsubsection{Gravitational potentials and observations}

One of the main sources of information in cosmology is through the
observation of perturbations about a Friedmann background.  Such
perturbations can be probed through their effects on the dynamics of
particles and light, which we will now describe (see \cite{Peacock99}
for further details):
%
%

\begin{itemize}
\item {\it Density fluctuations}: Fluctuations in the matter density
  field, $\delta({\bf x})$, will reflect various properties of the
  cosmological model. The simplest approach is to assume that
  $\delta({\bf x})$ is a multivariate Gaussian random field that is
  entirely described by the {\it power spectrum}, $P(k)$, defined by
\begin{eqnarray}
\langle|\delta_{\bf k}|^2\rangle\equiv P(k),
\end{eqnarray} 
where $\delta_{\bf k}$ is the Fourier transform of $\delta({\bf
  x})$. The shape of the power spectrum contains a wealth of
information: The amplitude of clustering as a function of scale, its
redshift dependence, how its shape on small scales is distorted by
small scale velocities (known as redshift space distortions), and acoustic
features imprinted by the baryons from pre-recombination (known as
baryon acoustic oscillations) can all be used as distance indicators.  
The power spectrum can be estimated from surveys of galaxies or
clusters of galaxies, the clustering properties of which can
be directly related to the amplitude of fluctuations in the density
field (under certain assumptions of how galaxies (or clusters) trace
the density field (known as bias)). 
\item 
{\it Peculiar velocities}: The motion of galaxies relative to the
Hubble flow, $v^{i}$, is described by the non-relativistic geodesic
equation given above. In the linear regime, the peculiar velocity can
be related directly to the density field via the gravitational
potential:
\begin{eqnarray}
v^i_{\bf k}=-iafH_0\frac{k^i\delta_{\bf k}}{k^2}, \nonumber
\end{eqnarray}
where $f\equiv d\ln\delta/d\ln a$ and we have assumed the general
relativistic result $\Phi=\Psi$. Peculiar velocities will be
observable through their effects on the redshift of objects, either in
redshift galaxy surveys (through their distortion of $P(k)$), or when
supplemented  with independent distance measurements of each object
(using the Tully-Fisher relation or supernova light curves) in
peculiar velocity surveys.
\item
{\it Anisotropies in the Cosmic Microwave Background (CMB)}: The CMB
will be sensitive to density fluctuations, peculiar velocities, and
the gravitational potentials. It is usual to characterise anisotropies
in the CMB in terms of $\frac{\delta T({\hat n})}{T}$, the dimensionless
deviations of the black-body temperature of the Universe in a
direction given by the unit vector, ${\hat n}$. We can expand 
$$
\frac{\delta T({\hat n})}{T} =\sum_{\ell m}a_{\ell m}Y_{\ell m}({\hat
  n}),
$$ 
where we have spherical harmonics, $Y_{\ell m}({\hat n})$, and define
the angular power spectrum $C_\ell = \frac{1}{2\ell+1}\sum_m
\langle|a_{\ell m}|^2\rangle$.  Like $P(k)$, the $C_{\ell}$s contain a
wealth of information about the cosmological model. It is now
instructive to delve slightly further into the form of $\frac{\delta
  T({\hat n})}{T}$. We can schematically split 
CMB anisotropies into three cosmological contributions, 
$$
\frac{\delta T({\hat n})}{T}= 
\frac{\delta T({\hat n})}{T}\big|_{\rm LS}+ \frac{\delta T({\hat
    n})}{T}\big|_{\rm ISW}+ \frac{\delta T({\hat n})}{T}\big|_{\rm
  SEC} ,
$$
where the first term encompasses all effects from the surface of last
scattering, the second term (the {\it
Integrated Sachs Wolfe effect}) is due to integrated effects along the
line of sight, and the last term encompasses secondary effects such as
weak lensing of the CMB, the Sunyaev-Zel'dovich effect and other such
contributions. Let us focus on the effect of the gravitational
potentials, the consequence of which we can see through the geodesic equation for light
rays given above \cite{Bertschinger11}. The accumulated redshift of a
beam of light along the line of sight is given by
\begin{eqnarray}
  (1+z)=\frac{E_{\rm obs}}{E_{\rm em}}=1+\Psi_{\rm  em}-\Psi_{\rm obs}
    -\int_0^z(\dot\Phi+\dot\Psi)(d\tau/dz)\,dz\ + \textrm{higher order
  terms}. \nonumber
\end{eqnarray}
where factors such as the integrated visibility function have been ignored for simplicity.
The first term is the Sachs-Wolfe effect and, in the case of the CMB,
will give a redshift to the photons as they climb out of potential
wells at the surface of last scattering. The second term is the ISW
effect, and depends on the time dependence of the gravitational potentials
along the line of sight, as advertised above. 
\item 
{\it Weak lensing}: Lensing arise when photon light rays are deflected
due to the gravitational potentials along the line of sight. The
deflection angle is given by $\delta {\vec \theta}=-{\vec
\nabla}_\perp (\Phi+\Psi)d\tau $, and allows us to relate the true
position, ${\vec \theta}_{\rm true}$, to the deflected position, ${\vec
\theta}_{\rm def}$, via ${\vec \theta}_{\rm true}={\vec \theta}_{\rm
def}-\frac{r_{LS}}{r_L}\delta {\vec \theta}$, where $r_L$ ($r_{LS}$) is
the distance to the lens (between the lens and the source). In
practise we probe the gradient of the deflection through the inverse
magnification matrix:
\begin{eqnarray}
{\bf M}^{-1}=\frac{\partial {\vec \theta}_{\rm true}}{\partial {\vec
    \theta}_{\rm def}}={\bf I}+
    \int_0^{z_S}\frac{r_Lr_{LS}}{r_S}{\vec\nabla}_\perp{\vec\nabla}_\perp(\Phi+\Psi)
    d\tau . \nonumber
\end{eqnarray}
This two by two matrix is parameterised by the convergence, $K$, and
shear parameters $\gamma_1$ and $\gamma_2$. In the case of small
deflections this gives
\begin{eqnarray}
\end{eqnarray}
\begin{equation}
  {\bf M}=\left(
    \begin{array}{cc}
      1+K+\gamma_1 & \gamma_2 \\
      \gamma_2 & 1+K-\gamma_1
    \end{array}
    \right)\ .
\end{equation}
This information can be extracted from imaging surveys of distant
galaxies. The galaxy shapes (or ellipticities) will be distorted by
the intervening gravitational potentials. These distortions will induce
correlations between the galaxy shapes that will reflect the
underlying cosmology. Lensing will, of course, also affect the CMB
photons as they pass through potential wells. 
\end{itemize}

\subsubsection{The evidence for the $\Lambda$CDM model}

There is currently a consensus that in an FLRW Universe that is
governed by Einstein's field equations, roughly $95\%$ of the overall
energy density must be `dark' in order to be compatible with observations.
The current best fit model claims that about $25\%$ of this dark
material is in the form of a non-relativistic, non-interacting form
of matter called {\it dark matter}, and that the remaining $70\%$ is
in the form of a non-clustering form of energy density with a negative
equation of state known as {\it dark energy}.

The broad case for Cold Dark Matter (CDM) is as follows~\cite{Peacock99}:
\begin{itemize}
\item The rotation curves of galaxies tend to flatten out at large
  radii. This flattening can be explained if the baryonic part of the
  galaxy resides in a halo of dark matter with a density profile that
  falls of as $1/r^2$.
\item Clusters of galaxies appear to have deeper potential wells than would be
inferred from baryonic matter. This is manifest in the motions of
galaxies, as well as the X-ray temperature of gas, and weak lensing measurements
of the integrated gravitational potentials. Dark matter halos
surrounding clusters explain all these observations.
\item Diffusion damping during recombination is expected to wipe out
  all small-scale structure in baryons, preventing
the formation of galaxies at late times. Dark matter, however, can sustain
structure during the damping regime, and will seed the formation of
galaxies. 
\end{itemize}

The case for dark energy has been around since the early 1980s. After the proposal of
the original models of inflation, the idea that the Universe should
have Euclidean spatial geometry became ever more entrenched in the
standard lore. Given that baryons made up a small fraction of the
total energy budget, and that dark matter makes up about $25\%$, there
was clearly a shortfall of pressureless matter at late
times. Furthermore, estimates of the ages of globular clusters of
around 12-14 billion years were incompatible with a flat, matter
dominated universe \cite{TurnerSteigmanKrauss84}.

There was also tentative evidence from large-scale clustering that a
flat, cold dark matter dominated Universe could not explain some of
the observations. Most notably, an analysis of the APM galaxy
catalogue in \cite{EfstathiouSutherlandMaddox90} seemed to show that a
Universe with a cosmological constant might explain the amount of
galaxy clustering on a wide range of scales.  Now, with the advent of
what has been dubbed ``precision cosmology'' in the late 1990s, the evidence for dark
energy has become even more compelling. In particular, the following
results make a strong case for presence of an energy density with
negative equation of state:
\begin{itemize}
\item
Measurements of the luminosity distance of type Ia supernovae are
consistent with a universe with a cosmological constant, and
inconsistent with a flat, matter dominated universe or an open
universe \cite{PerlmutterEtAl1998,Riess98}. The latest results seem to
constrain the equation of state, $w=P/\rho\simeq
-1.068^{+0.080}_{-0.082}$ \cite{Conley11,Sullivan11}.  
\item 
Measurements of the CMB anisotropies from large to small scales
\cite{Komatsu11,Dunkley10}, combined with measurements of galaxy
clustering from the Sloan Digital Sky Survey (SDSS) \cite{Reid10},
greatly  favour a model with $\Omega_{\Lambda}=0.725\pm0.016$ and $w =
-1.10\pm0.14$.
\item
The cross correlation between the ISW effect from the CMB and a
variety of surveys of large-scale structure favour
$w=-1.01^{+0.30}_{-0.40}$, at around $4\sigma$ \cite{Giannantonio08, Ho08}.
\item 
The number density of clusters of galaxies as a function of redshift
disfavour a flat, matter dominated universe. The presence of massive
clusters at high redshift point accelerating expansion out to redshift
$z\simeq 2$ \cite{Allen11}.
\end{itemize}
Although  each individual observation may be subject to a variety of
interpretations, and different systematic effects, the overall
concordance is remarkable. Indeed, the model that best fits these
observations is now known as {\it the concordance model}, or
$\Lambda$CDM. 

\subsubsection{Shortcomings of the $\Lambda$CDM model}
\label{sec:pitfalls}

Perhaps the most serious problem with $\Lambda$CDM is the cosmological constant 
problem:  That the observed value of $\Lambda$ is around 120 orders of
magnitude smaller than the naive expectation that it should be of the
Planck Mass, $M_{\rm Pl}^4$.  Super-Symmetric (SUSY) theories can
lower this expectation to that of the SUSY breaking scale, but this
still required a bare $\Lambda_0$ to cancel the vacuum energy coming
from the SUSY symmetry breaking scale to about 60 decimal places. One
could consider arguing that some unknown physics at high energies may provide a mechanism for
achieving this level of fine-tuning, but this seems unlikely as the
problem already manifests itself at low energies.

Now, suppose that we want to describe all physics up to scales just above the electron mass. Then the contribution 
to the vacuum energy $\Lambda$ will include a bare term $\Lambda_1$, a term coming from 
the electron and a term coming from the neutrino. This is schematically 
given by 
$$
\Lambda = \Lambda_1 + c_e m_e^4 + c_\nu m_\nu^4  \ldots,
$$
 where $c_e$ and $c_\nu$ are coefficients.  If we now lower the energy below the electron mass, and integrate out the electron, we instead have 
$$
\Lambda = \Lambda_0+ c_\nu m_\nu^4  \ldots,
$$
 for a new bare term $\Lambda_0$. To get the same observable  vacuum
 energy, $\Lambda$, we must now have that $\Lambda_1$ and $\Lambda_0$
 cancel to 32 decimal places.

It may be thought that there could exist some mechanism that relaxes
the effective cosmological constant\footnote{By effective cosmological
  constant we mean the Ricci curvature of the vacuum.} to zero dynamically, but
Weinberg~\cite{Weinberg1989} has shown that this is impossible.  Suppose that there is a set of $N$ scalars, $\phi^A$, that are
responsible for driving the effective $\Lambda$ to zero. These scalars will contribute an effective potential, $V(\phi^A)$, to the
cosmological constant.  
 If we are to approach a global Minkowski metric at these energy levels, then $V(\phi^A)$ must cancel
the other contributions to $\Lambda$ to high accuracy as the fields settle to the minimum. However, this is hardly a readjustment mechanism:
 If the cosmological constant changes slightly, then the mechanism fails.  This proof assumes Poincar{\'e} invariance in the scalar sector, which 
could, however, be considered an unnecessary assumption (see Horndeski's theory
in Section \ref{genSTsec}).

The present value of $\Lambda$, as implied by cosmological observations, 
has another potential problem associated with it:  It has an energy density of the same
order of magnitude as the 
average matter density in the Universe today, 
$$
\rho_\Lambda\vert_{a=1} \sim \rho_m\vert_{a=1}.
$$
These two quantities scale with the size of the
Universe in very different ways, and so their similarity at the present
time appears naively to be somewhat of a coincidence.  Hence, this problem
is sometimes referred to as {\it the coincidence problem}.

Aside from the problems of the cosmological constant, there are some 
problems that plague dark matter as well. The first is another
coincidence problem:
Why is the dark matter energy density so close to the baryon energy
density? This is actually worse than it might seem. Baryons 
are produced non-thermally, out of equilibrium. CDM is usually thought to 
be produced thermally, as weak interaction cross-sections naturally
give rise to the right dark matter abundance via thermal production. 
But how can two components that have very different production
mechanisms have very similar energy densities\footnote{There are also
  non-thermal candidates for dark matter, e.g. axions, but this does
  not change the argument.}? Solutions 
to this puzzle have been proposed~\cite{BarrChivukulaFarhi1990,Kaplan1992,Thomas1995,KitanoMurayamaRatz2008,KaplanLutyZurek2009,DavoudiaslEtAl2010,
BuckleyRandall2010,HallMarch-RussellWestEtAl2010,HabaMatsumoto2010,McDonald2010,AllahverdiDuttaSinha2011}
but they typically require additional particles
 to those that form the dark matter, and there is as yet no well
 accepted mechanism. 

Other problems with dark matter are observational, and we will discuss
them only briefly.  The density profile of CDM, as determined from
N-body simulations, is inferred to be cuspy.  For example the Navarro-Frenk-White
(NFW) profile~\cite{NavarroFrenkWhite1995} gives $\rho_{CDM} \propto
\frac{1}{r}$ close to the centre of a halo.
Other simulations give similar results:  $\rho_{CDM} \propto
r^{-\alpha}$ with $\alpha \sim 0.7 - 1.5$.  Galaxies, however, are observed to
have cores such that $\rho$  flattens out at the centre. This is the
cusp problem~\cite{deBlok2010} and proposed solutions within the CDM
paradigm include self-interacting dark matter~\cite{SpergelSteinhardt2000},
fuzzy dark matter~\cite{HuBarkanaGruzinov2000}, or various feedback
processes that expel dark matter.
Note that simulations do not have enough resolution
to probe the small scales where the problem manifests itself, but rely
instead on extrapolations.  However, simulations with increasingly smaller
resolutions (although still above the probed scales) have not
indicated any kind of alleviation to the cusp problem.

Another problem is that of missing
satellites~\cite{Mateo1998,Bullock2010}.  The CDM paradigm predicts a
rich sub-structure within the main galactic halo that should lead to
numerous dwarf galaxies orbiting the main galaxy.  Indeed, simulations
indicate that about $500$ satellite galaxies should be orbiting the
Milky way~\cite{MooreEtAl1999}.  On the contrary, however, only about
$30$ such dwarfs have been observed. A possible resolution within the
CDM paradigm is that most of these 
galaxies are dark galaxies, i.e. have very little or no stars in them,
and are instead completely dominated by dark matter~\cite{SimonGeha2007}. 

A third problem is the tight correlation between dark matter and
baryons in galaxies that manifests itself in a universal acceleration scale,
$a_0\sim 1.2\times 10^{-10}m \;s^{-2}$~\cite{Milgrom1983,SandersMcGaugh2002}, 
 the Tully-Fisher relation~\cite{McGaughEtAl2000,McGaugh2011}, and the
 Faber-Jackson relation~\cite{Sanders2010}. 
Within the CDM paradigm, such correlations are not expected to be
present, as baryons should not know how the dark matter behaves. 
For further apparent discrepancies between $\Lambda$CDM and small
scale observations the reader is referred to~\cite{KroupaEtAl2010}. 

On cluster scales and larger, the $\Lambda$CDM model can boast of
success coming from a host of observations: Strong and weak lensing of
clusters, X-ray observations of clusters, the CMB angular power spectrum,  the
matter power spectrum, $P(k)$, and supernova data. Yet there are a few
cases of interesting discrepancies.  The collision velocity of the
bullet cluster~\cite{CloweEtAl2006} may be so large  that the
probability of it occurring in a $\Lambda$CDM scenario is at best $\sim
10^{-9}$~\cite{LeeKomatsu2010}.  In~\cite{ForeroRomeroEtAl2010},
however, the opposite conclusion is reached, so this appears far from
settled.  Cosmological voids seem to be more empty of galaxies than
expected, as has been championed  by Peebles~\cite{Peebles2001}.  The
CMB angular power spectrum has a lack of large-scale power above
$60^{o}$~\cite{SpergelEtAl2003}  (although the statistical
significance of this is debatable, due to cosmic variance). Certain
violations of statistical isotropy or other anomalies on large scales
in the CMB  have also  been
reported~\cite{HansenEtAl2008,CopiEtAl2010,YohoEtAl2010}. It remains
to be seen whether these are really problems with $\Lambda$CDM, if
they are due to systematic effects, or if they are statistical flukes.
These difficulties do, however, provide some motivation for looking at
alternatives to $\Lambda$CDM.

\newpage 

\section{Alternative Theories of Gravity with Extra Fields}
\label{extrafields}

In General Relativity the gravitational force is mediated by a single
rank-2 tensor field, or a massless spin-2 particle in the
quantum field theory picture.  While there is good reason to couple
matter fields to gravity in this way, there is less reason to think
that the field equations of gravity should not contain other fields,
and one is in general free to speculate on the existence of such
additional fields in the gravitational sector.  The simplest scenario
that one could consider in
this context is the addition of an extra scalar field, but one might also choose to
consider extra vectors, tensors, or even higher rank fields
\cite{Fradkin-higherspins, Vasiliev-higherspins}. Of course, the
effect of such  additional fields needs to be suppressed at scales
where General Relativity has been well tested, such as in the lab or
solar system. This is usually achieved making couplings very weak,
although novel screening mechanisms such
as the chameleon mechanism \cite{cham1, cham2} and Vainshtein
mechanism \cite{vainshtein} have also been explored. 

This section represents an overview of four-dimensional gravity
theories with extra fields, focusing on additional scalars, vectors
and tensors.  We note that some theories in other sections of this
review can also be considered as theories with extra fields
(e.g. $f(R)$ gravity, galileons, and ghost condensates).  The reader
is referred to later sections for details of this.
 
\subsection{Scalar-Tensor Theories}
\label{scalartensorsection}

The scalar-tensor theories of gravity are some of the most established and
well studied alternative theories of gravity that exist in the
literature.  They are often used as the prototypical way in which
deviations from General Relativity are modelled, and are of particular
interest as the relatively simple structure of their field equations
allow exact analytic solutions to be found in a number of physically
interesting situations.  Scalar-tensor theories arise naturally as
the dimensionally reduced effective theories of higher dimensional
theories, such as Kaluza-Klein and string models.  They are also often
used as simple ways to self-consistently model possible variations in
Newton's constant, $G$.

\subsubsection{Action, field equations, and conformal transformations}

A general form of the scalar-tensor theory can be derived from the
Lagrangian density \cite{Ber68, Nor70, Wag70}
\begin{equation}
\label{ST}
\mathcal{L} = \frac{1}{16 \pi} \sqrt{-g} \left[ f(\phi) R-g(\phi) \nabla_\mu \phi \nabla^{\mu}\phi -2 \Lambda(\phi) \right] +
\mathcal{L}_m(\Psi, h(\phi) g_{\mu\nu}),
\end{equation}
where $f, g, h$ and $\Lambda$ are arbitrary functions of the scalar
field $\phi$ and $\mathcal{L}_m$ is the Lagrangian density of the
matter fields $\Psi$.  The function $h(\phi)$ can be absorbed into the
metric by a conformal transformation of the form \cite{dicke62}
\begin{equation}
\label{htrans}
h(\phi) g_{\mu\nu} \rightarrow g_{\mu\nu}.
\end{equation}
The conformal frame picked out by this choice is one in
which there is no direct interaction between the scalar field and
matter fields, and is usually referred to as the Jordan
frame. As discussed in previous sections, test-particles in this
conformal frame follow geodesics of the metric to which they are coupled,
and the weak equivalence principle is satisfied for massless test particles. 
The effect of this transformation on the remainder of the Lagrangian can
then be absorbed into redefinitions of the as yet unspecified
functions $f$, $g$ and $\Lambda$.

By a redefinition of the scalar field $\phi$ we can now set $f(\phi)
\rightarrow \phi$, without loss of generality.  The Lagrangian density
(\ref{ST}) can then be written as
\begin{equation}
\label{ST2}
\mathcal{L} = \frac{1}{16 \pi} \sqrt{-g} \left[\phi R-\frac{\omega(\phi)}{\phi} \nabla_\mu \phi \nabla^{\mu}\phi -2 \Lambda(\phi)\right] +
\mathcal{L}_m(\Psi, g_{\mu\nu}),
\end{equation}
where $\omega(\phi)$ is an arbitrary function, often referred to as
the `coupling parameter',  and $\Lambda$ is a $\phi$-dependent generalisation
of the cosmological constant.  This theory reduces to the well known
Brans-Dicke theory \cite{BD} in the limit $\omega \rightarrow$ constant
and $\Lambda \rightarrow 0$, and approaches General Relativity with a
cosmological constant in the
limit $\omega \rightarrow \infty$, ${\omega}'/\omega^2 \rightarrow 0$
and $\Lambda \rightarrow$ constant.

The variation of the action  derived from integrating (\ref{ST2}) over all space,  with respect to $g^{\mu\nu}$, gives the field equations
\begin{equation}
\label{STfields}
\phi G_{\mu\nu}
+\left[ \square \phi +  \frac{1}{2}\frac{\omega}{\phi} (\nabla \phi)^2+ \Lambda\right]   g_{\mu\nu} 
-    \nabla_\mu \nabla_\nu \phi
-\frac{\omega}{\phi} \nabla_\mu \phi \nabla_\nu \phi
 = 8\pi T_{\mu\nu}.
\end{equation}
Now, as well as the metric tensor $g_{\mu\nu}$, these theories also
contain the dynamical scalar field $\phi$, and so we must vary
the action derived from Eq. (\ref{ST2}) with respect to this
additional degree of freedom.  After eliminating $R$  with the trace of  (\ref{STfields}), this yields
\begin{equation}
\label{STfields2}
(2 \omega +3) \square \phi 
+ \omega' (\nabla \phi)^2 
+4 \Lambda
-2 \phi \Lambda'
= 8 \pi T.
\end{equation}
where primes here denote differentiation with
respect to $\phi$.  These are the field equations of the scalar-tensor theories of gravity.

It is well known that these theories admit the very useful property
of being `conformally equivalent' to General Relativity.  By this we
mean that under a transformation of the metric that alters scales, but
not angles, one can find a new metric that obeys the Einstein
equation, with the scalar contributing as an ordinary matter field.  This does not, however, mean that scalar-tensor theories
are the same as General Relativity, as the metric
that couples to matter fields must also transform.  The theory that is
recovered after conformally transforming is one in which the metric
obeys a set of fields equations similar to Einstein's, 
but with an unusual matter content that does not follow
geodesics of the new metric (with the exception of radiation fields,
or null geodesics, which are themselves conformally invariant).  This
property of scalar-tensor theories can sometimes allow their field equations
to be manipulated into more familiar forms, that allow solutions to
be found more readily.

To be explicit, a conformal transformation of the metric $g_{\mu\nu}$
into $\bar{g}_{\mu\nu}$ can be written
\begin{equation}
\label{ctran}
g_{\mu\nu} = e^{2 \Gamma (x)} \bar{g}_{\mu\nu},
\end{equation}
where $\Gamma (x)$ is an arbitrary function of the space-time
coordinates $x^{\mu}$.  The line-element is then correspondingly
transformed as $ds^2 = e^{2 \Gamma (x)} d \bar{s}^2$,
and the square root of the determinant of the metric as
$\sqrt{-g}= e^{4 \Gamma} \sqrt{-\bar{g}}$ (in four dimensions). After
performing such a transformation we can use the term `conformal frame' to
distinguish the new, rescaled metric from the 
original.  

Among the infinite possible conformal frames we can then
identify two which have particular significance:
The Jordan frame and the Einstein frame.  The Jordan frame is the one
in which the energy-momentum tensor is covariantly conserved and in
which test-particles follow geodesics of the metric. This
is the frame picked out by the transformation (\ref{htrans}), and is
the one in which scalar-tensor theories are most usually formulated.
The Einstein frame is the conformal frame in which the field
equations of the theory take the form of the Einstein equations with the scalar contributing as an ordinary scalar field, as
discussed above.

Under the transformation (\ref{ctran}) it can be shown that the
Ricci tensor and Ricci scalar transform as
\begin{align}
\label{cRab}
R_{\mu\nu} &= \bar{R}_{\mu\nu} - 2 \bar{\nabla}_\mu \bar{\nabla}_\nu \Gamma +2 \bar{\nabla}_\mu \Gamma \bar{\nabla}_\nu \Gamma
- \left(2   \bar{\nabla}_\alpha \Gamma \bar{\nabla}^\alpha \Gamma+ \bar{\square} \Gamma \right) \bar{g}_{\mu\nu}  \\
\label{cR}
e^{2 \Gamma} R &= \bar{R} -6  \bar{\nabla}_\mu \Gamma \bar{\nabla}^\mu\Gamma  - 6 \bar{\square} \Gamma ,
\end{align}
while the d'Alembertian transforms as $e^{2 \Gamma} \square \phi =
\bar{\square} \phi +2 \bar{\nabla}_\mu \Gamma  \bar{\nabla}^\mu \phi$.
Here, over-bars on operators or indices denote that they are defined
using the metric $\bar{g}_{\mu\nu}$.  Under these transformations we will
now show how the scalar-tensor theories defined by the Lagrangian
(\ref{ST2}), in the Jordan frame, can all be transformed into the
Einstein frame. 

First, consider the term in Eq. (\ref{ST2}) containing the Ricci scalar,
which under the conformal transformation (\ref{ctran}) becomes
\begin{equation}
\label{L1}
\mathcal{L}_1 = \frac{1}{16 \pi} \sqrt{-\bar{g}} \phi e^{2 \Gamma}
(\bar{R}-6 \bar{g}^{\mu\nu} \Gamma_{, \mu} \Gamma_{, \nu}-6 \bar{\square}
\Gamma).
\end{equation}
The non-minimal coupling to the Ricci scalar can now be removed by
making the choice of conformal factor $e^{2 \Gamma} = \phi^{-1}$ such
that $g_{\mu\nu}= \bar{g}_{\mu\nu}/\phi$.  This defines the conformal
transformation between Jordan and Einstein frames in the
scalar-tensor theories. Applying the transformation (\ref{ctran}) to
the rest of Eq. (\ref{ST2}) then gives
\begin{equation}
\label{ST3}
\mathcal{L} = \frac{1}{16 \pi} \sqrt{-\bar{g}} \left( \bar{R}- 2 (3+2
\omega) \bar{\nabla}_\mu \Gamma \bar{\nabla}^\mu \Gamma - 2 e^{4 \Gamma} \Lambda \right) +\mathcal{L}_m(\Psi, e^{2 \Gamma} \bar{g}_{\mu\nu}).
\end{equation}
Now, by making the definitions $\sqrt{4 \pi/(3+2 \omega)}\equiv
\partial \Gamma/\partial \psi$ and $8 \pi V (\psi)\equiv e^{4 \Gamma}
\Lambda$, for the scalar $\psi$ and the function $V(\psi)$, we can
write the transformed Lagrangian (\ref{ST3}) as
\begin{equation}
\label{ST4}
\mathcal{L} = \frac{1}{16 \pi} \sqrt{-\bar{g}} \bar{R} -\sqrt{-\bar{g}} \left(
  \frac{1}{2} \bar{\nabla}_\mu \psi  \bar{\nabla}^\mu \psi + V(\psi) \right)+
  \mathcal{L}_m(\Psi, e^{2 \Gamma} \bar{g}_{\mu\nu} ).
\end{equation}
In the absence of any matter fields the scalar-tensor theories can now
be clearly seen to be conformally related to Einstein's theory in the
presence of a scalar field in a potential.  This potential
disappears when $\Lambda=0$  

In the Brans-Dicke theory \cite{BD} the coupling parameter $\omega$ is constant, and the scalar fields $\phi$ and $\psi$ are therefore related by
\begin{equation*}
\ln \phi = \sqrt{\frac{16 \pi}{(3+2 \omega)}} \psi.
\end{equation*}
For more general theories with $\omega=\omega(\phi)$ the definition of
$\psi$ must be integrated to obtain a relation between
$\phi$ and $\psi$.
By extremising the action (\ref{ST4}) with respect to
$\bar{g}_{\mu\nu}$ and $\psi$ we get the Einstein frame field equations
\begin{equation}
  \bar{G}_{\mu\nu} 
= 8\pi\left[   \bar{T}_{\mu\nu} 
+    \bar{\nabla}_\mu \psi  \bar{\nabla}_\nu\psi   
-   \left( \frac{1}{2}  \bar{\nabla}_\alpha \psi  \bar{\nabla}^\alpha\psi + V\right) \bar{g}_{\mu\nu} 
\right]
\end{equation}
and
\begin{equation}
     \tilde{\square} \psi   - \frac{dV}{d\psi} = -     \sqrt{4\pi}\alpha \tilde{T} 
\end{equation}
where $\alpha^{-2} = 3+2\omega$ and where we have defined the energy-momentum tensor $\bar{T}_{\mu\nu}$ with respect to $\bar{g}_{\mu\nu}$ so that
$\bar{T}^{\mu\nu}=e^{6 \Gamma} T^{\mu\nu}$.  It can now be
explicitly seen that while the Jordan frame energy-momentum tensor is
covariantly conserved, $\nabla_\mu T^{\mu\nu} =0$, its counterpart in
the Einstein frame is not, $\bar{\nabla}_\mu \bar{T}^{\mu\nu} =\sqrt{4\pi} \alpha  \bar{T} \bar{\nabla}^\nu\psi$.  

\subsubsection{Brans-Dicke theory}
\label{bdgravitysection}

The Brans-Dicke theory is given by the
Lagrangian density (\ref{ST2}) with $\omega=$constant, and $\Lambda=0$
\cite{BD}.  The behaviour of this theory in the vicinity of isolated
masses is well understood, and in the case of static and spherical symmetry
can be solved exactly by the line-element \cite{Bra62}
\begin{equation*}
ds^2 = -e^{2 \alpha} dt^2+e^{2 \beta} (dr^2+r^2 d\Omega^2)
\end{equation*}
where $\alpha=\alpha(r)$ and $\beta=\beta(r)$ are given by one of
the following four solutions:
\begin{flalign*}
\textbf{I} & & e^{\alpha} &= e^{\alpha_0}
\left[\frac{1-\frac{B}{r}}{1+\frac{B}{r}}\right]^{\frac{1}{\lambda}}  & \\
& & e^{\beta} &= e^{\beta_0} \left(1+\frac{B}{r} \right)^2
\left[\frac{1-\frac{B}{r}}{1+\frac{B}{r}}\right]^{\frac{(\lambda-C-1)}{\lambda}} & \\
& & \phi &= \phi_0
\left[\frac{1-\frac{B}{r}}{1+\frac{B}{r}}\right]^{\frac{C}{\lambda}} &
\\
\textbf{II} & & \alpha &= \alpha_0 + \frac{2}{\Lambda} \tan^{-1} \left(
\frac{r}{B} \right) & \\
& & \beta &= \beta_0 - \frac{2 (C+1)}{\Lambda}  \tan^{-1} \left(
\frac{r}{B} \right) - \ln \left[\frac{r^2}{(r^2+B^2)} \right] & \\
& & \phi &= \phi_0 e^{\frac{2 C}{\Lambda} \tan^{-1} \left( \frac{r}{B}
  \right)} & \\
\textbf{III} & & \alpha &= \alpha_0 -\frac{r}{B} & \\
& & \beta &= \beta_0 -2 \ln \left( \frac{r}{B} \right) + \frac{(C+1)
  r}{B} & \\
& & \phi &= \phi_0 e^{-\frac{C r}{B}} & \\
\textbf{IV} & & \alpha &= \alpha_0 -\frac{1}{B r} & \\
& & \beta &= \beta_0 + \frac{(C+1)}{B r} & \\
& & \phi &= \phi_0 e^{-\frac{C}{ Br}}. & 
\end{flalign*}
Here we have defined $\lambda^2 \equiv (C+1)^2 - C \left(1-\omega C/2 \right)
> 0$ in solution I, and $\Lambda^2 \equiv C \left(1-\omega C/2 \right) -
(C+1)^2 >0$ in solution II.  The constant $C$ is arbitrary in I and
II, and given by $C = (-1 \pm \sqrt{-2 \omega -3})/(\omega+2)$ in III
and IV.  The constants $B$, $\alpha_0$, $\beta_0$ and $\phi_0$ are
arbitrary throughout.  

Now, while solution $I$ is valid for all values of $\omega$, solutions $II$,
$III$ and $IV$ are only valid for $\omega< -3/2$.
Solution $I$ is also known to be conformally related to the minimally
coupled massless scalar field solution of Buchdahl \cite{buch59}.
It can be seen that these solutions are not all independent of each
other.  By a transformation of the form $r \rightarrow 1/r$
and some redefinition of constants, solution $II$ can be
transformed into the $\omega <-3/2$ range of solution $I$ \cite{Bhad} and
solution $III$ can be transformed into solution $IV$
\cite{Bhad2}.  It was also shown in \cite{Bhad} that the independent
solutions $I$ and $IV$ are both conformally related to the general
solution of the static, spherically symmetric case in the Einstein
frame, as found by Wyman \cite{Wy}.

These solutions are very useful for understanding the gravitational fields around
an isolated body in Brans-Dicke theory, but are {\it not} the only
spherically symmetric vacuum solutions of the Brans-Dicke field
equations.  A non-static spherically symmetric vacuum exact solution is also
known \cite{CBM}:
\begin{multline}
d\bar{s}^{2}=-A(r)^{\alpha (1-\frac{1}{\sqrt{3}\beta })}dt%
^{2}\\+A(r)^{-\alpha (1+\frac{1}{\sqrt{3}\beta })}t^{\frac{2(\beta -%
\sqrt{3})}{3\beta -\sqrt{3}}} \left[ dr^{2}+A(r)r^{2}(d\theta
^{2}+\sin^{2}\theta 
d\phi ^{2})\right],  \label{JBDsolution2}
\end{multline}%
with
\begin{equation}
\phi(r,t)=\left( 1-\frac{2C}{r} \right)^{\pm\frac{1}{2\beta}}t^{2/(%
\sqrt{3}\beta -1)},  \label{phi2}
\end{equation}
where we have $A(r) = 1-\frac{2C}{r}$, $\alpha = \pm
\frac{\sqrt{3}}{2}$, $\beta = \sqrt{2\omega +3}$, and $C=$constant.
This solution reduces to a flat vacuum FLRW metric in 
the limit $C\rightarrow 0$ (an inhomogeneous solution requires $C\neq 0$).
The metric (\ref{JBDsolution2}) is spatially homogeneous at large $r$
and has singularities
at $t=0$ and $r=2 C$; the coordinates $r$ and $t$ therefore cover
the ranges $0\leq t<\infty $ and $2 C\leq r<\infty $.  This
solution is known to be conformally related to \cite{Husain94}, and
shows explicitly the lack of validity of Birkhoff's theorem in
Brans-Dicke theory.  It also reduces to the Schwarzschild solution
when $\omega \rightarrow \infty$.  Black hole solutions in Brans-Dicke
theory with a power-law potential have been investigated in \cite{fextra1}.

Let us now consider the weak field limit of this theory.  Following
the PPN prescription outlined in previous sections one can
straightforwardly find that the relevant values for the PPN parameters
are:
\be
\beta_{PPN} = 1
\qquad \qquad \text{and} \qquad \qquad
\gamma_{PPN} = \frac{1+\omega }{2+\omega } ,
\label{gambd}
\ee
with all other parameters equalling zero.  The value of Newton's
constant can also be shown to be given by
\be
G= \left( \frac{4+2 \omega}{3+2 \omega} \right) \frac{1}{\phi_0},
\ee
where $\phi_0$ is the background (unperturbed) value of the scalar
field.  It can be seen that in the general relativistic limit $\omega
\rightarrow \infty$ we then recover the usual values of the PPN
parameters, and that for finite $\omega$ the only parameter that
deviates from its general relativistic value is $\gamma$.  

This value of $\gamma$ is valid for both the static and non-static exact
solutions shown above.  It is interesting to note, however, that it is
{\it not} the value of $\gamma$ that one should expect to measure
outside of a black hole that has formed from gravitational collapse in
this theory.  Such an object can be shown to have an external
gravitational field with $\gamma=1$ \cite{BDhole}, as predicted by
Hawking \cite{BDhole2}.  This does not, however, mean that
gravitational collapse to a black hole proceeds in the same way in
Brans-Dicke theory as it does in General Relativity. In the
Brans-Dicke case apparent horizons
are allowed to pass outside of the event horizon, scalar gravitational
waves are emitted during the collapse, and the surface area of the
event horizon can decrease with time.  Such behaviour does not occur
in General Relativity, and is allowed here because Brans-Dicke theory
can violate the condition $R_{\mu\nu}k^ak^{\nu}\geq 0$, where $k^ak_a=0$.
The problem of understanding black hole thermodynamics in Brans-Dicke
theory has been addressed in \cite{BDentropy1}.  Here it was found
that the expression for the entropy of a black hole with an horizon
$\Sigma$ of area $A$ is given by
\be
S_{BH} = \frac{1}{4} \int_{\Sigma} d^2x \sqrt{g^{(2)}} \phi =
\frac{\phi A}{4},
\ee
such that $S_{BH}$ is always non-decreasing, even if the area
decreases.  This shows that the second law of black hole
thermodynamics can indeed be extended to Brans-Dicke theories,
with the effective gravitational constant being replaced by $1/\phi$.
For an intuitive interpretation of this result in the Einstein frame,
and for further discussion on this topic, the reader is referred to
\cite{BDentropy2}. 

Having discussed the gravitational fields of point-like objects in Brans-Dicke
theory, let us now proceed to use observations of weak field phenomena
to constrain the theory.  This can be done most effectively using the constraint on
$\gamma$ given in Equation (\ref{gamcas}), derived from observations
of the time delay of radio signals from the Cassini spacecraft as it
passed behind the Sun.  Together with the expression (\ref{gambd}),
shown above, this gives the $2 \sigma$ constraint on the coupling parameter
\be
\omega \gtrsim 40\; 000.
\ee
This is a very restricting constraint on the theory, and
shows that deviations of this kind from General Relativity must be
very small indeed (see the following subsection, however, for a
discussion of scalar-tensor theories that can evade this bound while
still exhibiting significantly different behaviour to General
Relativity in the early universe).

Let us now proceed to discuss the cosmology of Brans-Dicke theory.
Using the usual FLRW line-element, and assuming a perfect fluid matter
content, the field equations reduce to:
\ba
\label{BDFRW1}
H^2 &=& \frac{8 \pi \rho}{3 \phi} - \frac{\kappa}{a^2} - H
\frac{\dot{\phi}}{\phi} + \frac{\omega}{6} \frac{\dot{\phi}^2}{\phi^2}\\
\frac{\ddot{\phi}}{\phi} &=& \frac{8 \pi}{\phi} \frac{(\rho-3P)}{(2
\omega +3)} -3 H \frac{\dot{\phi}}{\phi},
\label{BDFRW2}
\ea
where over-dots denote differentiation with respect to the proper time
of a comoving observer, $H=\dot{a}/a$, and $\dot{\rho}+3 H (\rho+P)=0$.
The general solutions to Eq. (\ref{BDFRW1}) and (\ref{BDFRW2}) are now
fully understood \cite{Gur73,STexact2}.  At
early times the vacuum solutions of O'Hanlon
and Tupper \cite{Tup} are recovered, while at late-times one
approaches the power-law solutions of Nariai \cite{Nar68} (when $\kappa=0$):
\begin{equation}
a(t)=t^{2 [1+\omega (1-W)]/[4+3\omega (1-W^2)]}  \label{bds1}
\end{equation}
\begin{equation}
\phi (t)=\phi _{0}t^{[2(1-3W)]/[4+3\omega (1-W^2)]},
\label{bds2}
\end{equation}
where\footnote{Note that in this section we use an upper case $W$ to denote the
equation of state of the fluid, rather than the usual lower case $w$
used in the rest of the review.  This is to avoid confusion with the
coupling parameter $\omega$.} $p=W \rho$.  These solutions can be considered
``Machian'' in the sense that the matter fields are driving the
expansion of the Universe, rather than $\phi$.

Let us now consider the general FLRW solutions in terms of a
transformed time coordinate $\eta=\eta (t)$.  Such solutions can be found
any equations of state $W$ \cite{Gur73}, but here let us consider only
the radiation dominated solutions with $W=1/3$.  In this case the new
time coordinate $\eta$ is simply
the conformal time $\tau$ given by $a d\tau \equiv dt$, and the general
solution for $\omega > -3/2$ is
\begin{align}
a(\tau)&=a_1 (\tau +\tau_{+})^{\frac{1}{2}+
  \frac{1}{2\sqrt{1+\frac{2}{3} \omega}}} (\tau
  +\tau_{-})^{\frac{1}{2} -
  \frac{1}{2\sqrt{1+\frac{2}{3} \omega}}} \label{aw>}\\
\phi(\tau)&=\phi_1 (\tau +\tau_{+})^{-
  \frac{1}{2\sqrt{1+\frac{2}{3} \omega}}} (\tau +\tau_{-})^{+
  \frac{1}{2\sqrt{1+\frac{2}{3} \omega}}} \label{phiw>}
\end{align}
where $\tau_{\pm}$, $a_1$ and $\phi_1$ are integration constants, and where
$8 \pi \rho_{r0}/3 \phi_1 a_1^2=1$. For $\omega < -3/2$, however, we
instead find 
\begin{align}
a(\tau)&=a_1
\sqrt{(\tau+\tau_{-})^2+\tau_{+}^2} \exp \left(
  \frac{-1}{\sqrt{\frac{2}{3}\vert\omega\vert -1}} \tan^{-1}
  \frac{\tau+\tau_{-}}{\tau_{+}} \right), \label{aw<}\\
\phi(\tau)&=\phi_1
\left(\frac{2}{\sqrt{\frac{2}{3}\vert\omega\vert -1}} \tan^{-1}
  \frac{\tau+\tau_{-}}{\tau_{+}} \right) \label{phiw<}.
\end{align}

For $\omega >-3/2$ ($\omega <-3/2$) we see that the scale factor here
undergoes an initial period of rapid (slow)
expansion and at late times is attracted towards the solution $a(\tau)
\propto \tau$, or, equivalently, $a(t) \propto t^{\frac{1}{2}}$.  Similarly, $\phi$
can be seen to be 
changing rapidly at early times and slowly at late times.  These two
different behaviours, at early and late times, can be attributed to
periods of free scalar--field domination and radiation domination,
respectively.
If $\rho_{r0}=0$ is chosen then these solutions become vacuum ones that
are driven by the $\phi$ field alone, and for $\omega<-3/2$ the
initial singularity can be seen to be avoided.  
Corresponding behaviour can also be shown to exist for
other equations of state, $W$.  For a more
detailed discussion of this phenomenon we refer the reader to
\cite{Bar04b}.

Unlike in General Relativity, in the Brans-Dicke theory it is also
possible to have spatially flat and positively curved exact vacuum
solutions. Spatially flat solutions can be found by 
assuming $\phi \propto t^{x}$ and $a\propto t^{y}$, and by
setting $a(0)=0$.  When $\kappa=0$ the vacuum Brans-Dicke equations are
then solved by \cite{Tup}
\ba
a(t)&\propto&t^{\frac{1}{3}(1+2(1-\sqrt{3(3+2\omega )})^{-1})},\\
\phi (t)&\propto& \left( \frac{t}{t_{0}}\right)
^{-2(1-\sqrt{3(3+2\omega )})^{-1}}. 
\ea
For spatially closed solutions one can follow the method prescribed in
\cite{STexact2}. Here one defines a new quantity $y\equiv \phi a^{2}$,
and uses the conformal time coordinate $\tau$, to write the field
equations as
\ba
\nonumber
(\ln \phi ),_\tau  = \sqrt{3} A y^{-1} (2 \omega +3)^{-1/2}
\qquad \text{and} \qquad
y,_\tau^2 &=& -4\kappa y^2 + A^2,
\ea
where $A$ is a constant.  For $\kappa>0$ these equations can be integrated
to find $y=(A/ 2\sqrt{\kappa}) \sin(2\sqrt{\kappa}(\tau-\tau_0))$, which then
gives the solutions 
\ba
\phi (\tau ) &\propto& \tan ^{\sqrt{\frac{3}{(2\omega
+3)}}}(\sqrt{\kappa}(\tau -\tau_0)),\\
a(\tau) &\propto&
\frac{\sin^{1/2}(2\sqrt{\kappa}(\tau-\tau_0))}{\tan^{\sqrt{\frac{3}{4(2\omega
+3)}}}(\sqrt{\kappa}(\tau-\tau_0))}.
\ea
Spatially flat and closed vacuum FLRW solutions such as those shown
here do not exist in General Relativity, and show the potential for
interesting new behaviour at early times in scalar-tensor theories of
gravity.  Phase plane analyses of perfect fluid FLRW solutions to the
Brans-Dicke field equations have been performed in \cite{stnew4,stnew6,BDphase}.

A number of anisotropic cosmological solutions of the Brans-Dicke
field equations are also known.  Bianchi type-$I$ solutions have
been found in \cite{BI,BIb}, type-$II$ in \cite{BII,BIIb,BIIBVIIIBIX},
type-$III$ solutions in \cite{BKS}, type-$V$ solutions in
\cite{BV,BVa,BVb,BVc,BVd},
type-$VI_0$ and $VI_h$ solutions in \cite{BVIh,BVI0BVIh,belinchon1},
type-$VII_h$
solutions in \cite{BVIIh,BVIIhb}, type-$VIII$ solutions in
\cite{BIIBVIIIBIX,BVIIIBIX}, type-$IX$ solutions in
\cite{BIIBVIIIBIX,BVIIIBIX}, and Kantowski-Sachs solutions in
\cite{BKS}.  Inhomogeneous cosmological solutions have also been found
\cite{CBM}, and braneworld cosmologies have been considered in
\cite{farh6,farh9}.  We will not reproduce any of these
solutions here, but rather refer the reader to the citations above,
and references therein.  For a discussion of the cosmic no-hair
theorems in Brans-Dicke theory the reader is referred to
\cite{BDhair}, where it is shown that these theorems are valid without
imposing any strong constraints on the coupling constant, $\omega$, so
that initially anisotropic universes can evolve towards an isotropic
final state.

Now let us consider perturbed FLRW space-times, within which
cosmological observations are usually interpreted.  For the
Brans-Dicke theory these equations have been studied many times
before, starting with \cite{narBD}.  Here we will present these
equations in the synchronous gauge and with $\kappa=0$, as found in
\cite{CKbd}.  In this case the equations take on a simpler form.  For
the more general case the reader is referred to \cite{BDpert}, or to the
$\omega=$constant limit of Eqs. (\ref{genSTpert1})-(\ref{genSTpert2})
in Section \ref{genSTsec} for the corresponding equations in the
conformal Newtonian gauge.  Now, the perturbed metric can be written as
\be
g_{\mu \nu} = \bar{g}_{\mu \nu} + a^2 (\tau) h_{\mu \nu},
\ee
where $a(\tau)$ is the FLRW scale factor, $\bar{g}_{\mu \nu}$ is the
unperturbed FLRW metric with $\kappa=0$, and
$h_{\mu \nu}$ is the perturbation that satisfies $h_{00}=h_{0\nu}=0$
in the synchronous gauge.  We can then proceed as normal, and decompose
the remaining non-zero $h_{ij}$ perturbations into harmonic
modes, and decouple the scalar, vector and tensor components.  
The scalar part of the perturbations can be written as in Section \ref{GR}
\be
h_{ij} = \frac{1}{3} h q_{ij} + D_{ij} \nu,
\ee
where the synchronous gauge has been adopted, while  $\delta \phi$ is the perturbation to the Brans-Dicke scalar.  
The $D_{ij}$ operator, as in Section \ref{GR}, is defined by $D_{ij}=\grad_i \grad_j- \frac{1}{3} q_{ij} \lap$.
As usual we define $\eta = -(h+k^2\nu)/6$ (see section \ref{GR}). The
perturbed equations are\footnote{Various typos in the corresponding
  equations in~\cite{CKbd} have been corrected. }:
\ba
-2k^2 \eta + \left(\adotoa + \frac{1}{2}  \frac{\phi'}{\phi}  \right)h'
 &=&\frac{8\pi a^2}{\phi} \sum_f \rho_f\delta_f
+ \left(\omega\frac{\phi'}{\phi} -  3 \adotoa   \right) \frac{\delta \phi'}{\phi}
\nonumber 
\\
&& 
\ \ \ \
-  \left[ k^2 + 3\adotoa^2 +\frac{\omega}{2} \frac{ {\phi'}^2 }{\phi^2} \right] \frac{\delta \phi}{\phi}
\\
 2\eta' &=& \frac{8\pi a^2}{\phi} \sum_f (\rho_f+P_f)  \theta_f + \frac{1}{\phi} \delta \phi' - \frac{1}{\phi}\left( \adotoa - \omega \frac{\phi'}{\phi}\right) \delta \phi
\quad
\\
\frac{1}{2} \nu''  +  \left(\adotoa  +\frac{\phi'}{2\phi}\right) \nu' + \eta &=& \frac{8\pi a^2}{\phi} (\rho+P)  \Sigma_f +   \frac{\delta \phi}{\phi}
\ea
and
\be
{\delta \phi}^{\prime \prime}+2 \adotoa {\delta \phi}^{\prime} +k^2 \delta \phi
 + \frac{1}{2} \phi' h^{\prime}  = \frac{8 \pi a^2}{2 \omega +3} \sum_f(\delta \rho_f-3 \delta P_f).
\ee
Here the perturbations to the energy density and pressure of the
non-interacting fluids $f$ are written as $\delta \rho_f$ and $\delta P_f$,
with the peculiar velocity potentials and anisotropic stress written as
$\theta_f$ and $\Sigma_f$, respectively.  Primes denote differentiation with
respect to conformal time, $\tau$.

For the tensor modes we can write the metric perturbations as
$h_{ij}=\tilde{h}_T Q_{ij}$, where $Q_{ij}$ is a harmonic function,
and with no tensor component involved in $\delta \phi$.
The evolution equation for $\tilde{h}_T$ is then given by
\be
\tilde{h}_T^{\prime \prime} +2\adotoa \tilde{h}_T^{\prime} + k^2 \tilde{h}_T =
\frac{8 \pi a^2}{\phi} \sum_f (\rho_f + P_f) \tilde{\Sigma}_f,
\ee
where $\tilde{\Sigma}_f$ is the tensor contribution to the anisotropic
stress of the fluid $f$.  We will not write the vector perturbation
equations here, which are not expected to be significant for most
cosmological applications.  For perturbation equations written in
terms of gauge invariant variables the reader is referred to \cite{Wu1} for
the covariant approach, or \cite{BDpert} for the Bardeen variable
approach (for the Brans-Dicke theory one should take $\omega=$constant in this
last reference).

The background cosmological evolution and perturbations can be used to
place constraints on Brans-Dicke theory from a number of different
sources.  The CMB is one such source, and can be used to place
constraints on the coupling parameter, $\omega$.  This has been
done a number of times in the literature \cite{Nag,Acq,Wu}, with the
latest results based on constraints given by the WMAP 5 year data, the
ACBAR 2007 data, the CBI polarisation data, and the BOOMERanG 2003
flight, together with large-scale structure data from the SDSS data
release 4, giving $\omega > 97.8$ or $\omega <-120.0$ to $2 \sigma$
\cite{Wu}.  This is in keeping with the results of \cite{Acq}, but
significantly weaker than those claimed by \cite{Nag} of $\omega
>1000$ to $2 \sigma$ based on the WMAP first year data.  Among the
detailed processes that lead to these constraints one can see that the
change in the horizon size at matter-radiation equality is altered in
Brans-Dicke theory due to the different expansion rates \cite{Mazum}.
This length scale is imprinted on the spectrum of perturbations as
during the radiation era perturbations inside the horizon are
effectively frozen, while during matter domination perturbations grow
on all length scales.  Different expansion rates also affect the
horizon size at recombination, which affects the level of `Silk
damping' that occurs on small scales due to viscosity and heat
conduction.  What is more, the thickness of the last scattering surface
is also changed, which affects anisotropy on small scales through the
exponential damping which has its cutoff determined by this quantity.
The upcoming data from the Planck satellite is, of
course, expected to tighten the constraints given above still further.

Another cosmological probe that has been extensively applied to
Brans-Dicke theory is that of the primordial nucleosynthesis of light
elements \cite{BDnuc1,BDnuc2,BDnuc3,BDnuc4,BDnuc5,BDnuc6,BDnuc7}.  In
the Brans-Dicke theory the scalar field $\phi$ is approximately
constant during the epoch of radiation domination.  Nucleosynthesis
therefore proceeds largely as in a general 
relativistic cosmology (up to the effect of `kicks' on the scalar
field due to the annihilation of electron-positron pairs
\cite{kicks}), but with a different value of $G$ during this process,
and hence a different expansion rate. Of course, the time at which
weak interactions freeze out in the early universe is determined by
equality between the rate of the relevant weak interactions and the
Hubble rate.  When the weak interaction rate is the greater then the ratio of neutrons
to protons it tracks its equilibrium value, while if the
Hubble rate is greater than the weak--interaction rate
then the ratio of neutrons to
protons is effectively `frozen--in', and $\beta$--decay is the only
weak process that still operates with any efficiency.  This is the
case until the onset of deuterium formation, at which time the
neutrons become bound and $\beta$--decay ceases.  Now, the onset of
deuterium formation is primarily determined by the photon to
baryon ratio, $\eta_{\gamma}$, which inhibits the formation of
deuterium nuclei until the critical temperature for photodissociation
is past.  As the vast 
majority of neutrons finally end up in $^4$He the primordial abundance of this
element is influenced most significantly by the number of neutrons at
the onset of deuterium formation, which is sensitive to the
temperature of weak--interaction freeze--out, and hence the Hubble
rate, and so $G$, at this time.  Conversely, the primordial abundances of
the other light elements are mostly sensitive to the temperature at
deuterium formation, and hence $\eta_{\gamma}$, when
nuclear reactions occur and the light elements form.  The reader is
referred to \cite{BDnuc8} for further discussion of these points.  The
typical bounds that can be achieved on the coupling parameter from
observations of element abundances are then given by $\omega \gtrsim
300$ or $\omega \lesssim -30$, assuming the power-law solutions
(\ref{bds1}) and (\ref{bds2}).  By using the general solutions
(\ref{aw>})-(\ref{phiw<}), however, these bounds can be somewhat
relaxed or tightened, depending on the behaviour of $\phi$ in the
early universe \cite{BDnuc7}.

While the cosmological bounds discussed above are weaker than those
derived in the solar system, and in binary pulsars, they probe a very
different physical environment and scale.  They are
therefore usually considered complimentary to the constraints imposed
from observations of weak field gravity, and a useful consistency
check.  After all, one may wish to consider theories in which the
coupling parameter $\omega$ varies throughout cosmic history.
Theories in which such behaviour can occur explicitly are the subject
of Section \ref{genSTsec}.

\subsubsection{General scalar-tensor theories}
\label{genSTsec}

The Brans-Dicke theory that has so far been considered is a very special
scalar-tensor theory, with only a single constant parameter.  The
more general class of scalar-tensor theories contains two free
functions, given by $\omega (\phi )$ and $\Lambda (\phi )$ in
Eq. (\ref{ST2}).  Let us now consider these more general theories.

First of all let us consider the case in which $\Lambda (\phi ) =0$.
Such theories have been well studied in the literature, and are often
used to model the possibility of having a coupling parameter
$\omega$ in the early universe that is small enough to have
interesting effects, while being large enough in the late universe to
be compatible with the stringent bounds imposed upon such couplings by
observations of gravitational phenomena in the solar system, and other
nearby astrophysical systems.  This interest is bolstered by the
presence of an attractor mechanism that ensures General Relativity is
recovered as a stable asymptote at late times in FLRW cosmology
\cite{attractor}.  We will explain this attractor in more detail
below.

Of course, in generalising the Brans-Dicke theory we want to know
what the consequences are for constraints imposed in the weak field
limit.  The extra complication caused by
allowing $\omega$ to be a function of $\phi$ means that
exact solutions are hard to find.  Perturbative analyses can still be readily
performed, however, leading to the PPN parameters
\be
\beta_{PPN} =  1+ \frac{d \omega/d \phi}{(4+2 \omega )(3+2 \omega )^2}
\qquad \qquad \text{and} \qquad \qquad 
\gamma_{PPN} =  \frac{1+\omega}{2+\omega} ,
\ee
with all other parameters equalling zero. The value of $\gamma$ here
can be seen to be the same as in the Brans-Dicke theory, while the
value of $\beta$ reduces to the Brans-Dicke (and General Relativity)
value of unity when $\omega=$constant.  Observations from the Cassini
satellite therefore place upon $\omega$ the same tight constraint
as in Brans-Dicke theory ($\omega \gtrsim 40\; 000$ to $2 \sigma$).  This
constraint, however, now only applies to
the {\it local} value of $\omega$ (i.e. with the present day value of
$\phi$ in the solar system).  The variation of $\omega$ with $\phi$ can then
be constrained by observations of post-Newtonian
phenomena that constrain $\beta$, such as the lunar laser ranging experiments
described in previous sections.  To constrain $\omega$ for other
values of $\phi$, however, requires making observations in other physical
environments, such as in the early universe, or near black holes.

Let us now consider the cosmological solutions of these theories.  It
has been shown by Clarkson, Coley and O'Neill in \cite{stEGS} that the
Ehlers-Geren-Sachs theorem can be extended to cover scalar-tensor theories
of gravity.  Taking the FLRW line-element, and assuming a perfect fluid
matter content, the field equations in this case reduce to
\ba
\label{STFRW1}
H^2 &=& \frac{8 \pi \rho}{3 \phi} - \frac{\kappa}{a^2} - H
\frac{\dot{\phi}}{\phi} + \frac{\omega}{6} \frac{\dot{\phi}^2}{\phi^2}\\
\frac{\ddot{\phi}}{\phi} &=& \frac{8 \pi}{\phi} \frac{(\rho-3P)}{(2
\omega +3)} -3 H \frac{\dot{\phi}}{\phi} - \frac{(d\omega/d\phi )
\dot{\phi}^2}{(2 \omega + 3)\phi},
\label{STFRW2}
\ea
where over-dots again denote differentiation with respect to the proper time
of comoving observers.  These equations are similar to those of the
Brans-Dicke theory, Eqs. (\ref{BDFRW1}) and (\ref{BDFRW2}), except for
the extra term on the RHS of Eq. (\ref{STFRW2}).
Exact solutions with $\kappa =0$ have been found to Eqs. (\ref{STFRW1}) and
(\ref{STFRW2}) in \cite{stnew1,stnew5,STexact1,STexact6}, and vacuum and 
radiation dominated solutions for arbitrary spatial curvature have
been found in \cite{STexact2,STexact3,STexact7}.  Some of
the methods used in these papers
are extended to anisotropic cosmologies in \cite{STexact4}, and the
asymptotics of FLRW cosmologies in scalar-tensor theories have been studied
in \cite{STexact5,STexact8}.  Exact homogeneous and
anisotropic solutions are found in \cite{stC1,stC3} that
act as past and 
future attractors for the general solution.  Exact homogeneous
self-similar solutions are found in \cite{belinchon2},
and inhomogeneous self-similar solutions are found in \cite{stC2}. We will
not reproduce these
solutions here, some of which can be quite complicated, but will
instead return to the attractor mechanism expounded in
\cite{attractor}.  

This mechanism is most easily seen in the Einstein
conformal frame, given by the Lagrangian (\ref{ST4}), such that for a
spatially flat FLRW geometry the evolution equation for the scalar
field can be written as
\be
\frac{8 \pi}{(3-4 \pi \psi^{\prime 2})} \psi^{\prime \prime} +
4 \pi (1- w) \psi^{\prime} +\sqrt{4\pi}(1-3 w) \alpha =0,
\label{STatt}
\ee
where here primes denote differentiation with respect to the natural log of
the Einstein frame scale factor, $\bar{a}$, and $w$ is the
equation of state $P=w \rho$.  The reader will recall that
$\psi=\sqrt{(3+2 \omega)/16 \pi} \ln \phi$ is the scalar field in the
Einstein frame, and $\alpha^{-2}=3+2 \omega$ denotes the strength of
coupling between the scalar and tensor degrees of freedom.  Equation
(\ref{STatt}) is clearly the equation for a simple harmonic oscillator
with a dynamical mass, a damping force given by $-4 \pi (1-w)$,
and a driving force given by the gradient of a potential $(1-3 w)
\Gamma$, where the reader will recall $\Gamma = \sqrt{4 \pi} \int
\alpha d \psi$.  This interpretation of $\Gamma$ as an effective
potential is often used to justify an expansion of the form
\be
\label{STexpand}
\Gamma = \alpha_0 (\psi-\psi_0 ) + \frac{\beta_0}{2} (\psi-\psi_0)^2 +
O((\psi-\psi_0)^3),
\ee
where $\psi_0$ is an assumed local minimum of $\Gamma (\psi )$, and
$\alpha_0$ and $\beta_0$ are constants.  In terms of this
parameterisation the PPN parameters $\beta_{PPN}$ and $\gamma_{PPN}$
then become
\ba
1-\beta_{PPN} &=& - \frac{\alpha_0^2 \beta_0}{2 (1+\alpha_0^2)^2}\\
1-\gamma_{PPN} &=& \frac{2 \alpha_0^2}{1+\alpha_0^2}.
\ea
The requirement of positive mass in (\ref{STatt})
can also be seen to be equivalent to the requirement of positive energy
density, $\bar{\rho}$, in the Einstein frame.

The cosmological dynamics that result from
Eq. (\ref{STatt}) are that $\psi$, and hence $\phi$, approach a
constant value during the radiation dominated epoch.  This is due to
the vanishing of the `potential' in (\ref{STatt}) when $w=1/3$,
and the negativity of the effective `damping force'.  Once radiation
domination ends, however, and matter domination begins, then the
scalar field rolls down to the minimum, $\psi_0$, of the now non-zero
potential $\Gamma (\psi )$
(assuming such a minimum exists).  Once this minimum is reached, after
some possible oscillations in the case of an under-damped system, then
we are left with $\alpha =0$, which is the general relativistic limit
of these theories.  This is a very useful general property of any scalar-tensor
theory which has a local minimum in its parameter $\Gamma (\psi )$,
and means that interesting new behaviour is possible at early times,
while still being (potentially) compatible with observations that
appear to point towards General Relativity at late-time.

Using order-of-magnitude approximations, the authors of
\cite{attractor} claim that this attractor mechanism is powerful
enough to drive the value of the PPN parameter $1-\gamma$ down to
values of as low as $\sim 10^{-7}$.  This is a couple of order of
magnitudes below the level that is probed by even the observations of
the Cassini spacecraft, but is not inconceivably small.  In
particular, it may be that upcoming observations of binary pulsar
systems could achieve such levels.  Further predictions of this
scenario are a possible oscillation in the effective value of Newton's
constant near the beginning of the matter dominated epoch of the
Universe's history, as well as a prediction for the locally measured
value of $\beta_{PPN}$ given by
\be
\beta_{PPN}-1 =\frac{\beta_0}{32 \pi} (1-\gamma^2),
\ee
where $\beta_0$ is defined in Eq. (\ref{STexpand}).  The validity and
limitations of these results are extended, and
are further studied in \cite{attractor2, attractor3}.

Let us now consider perturbations around a general FLRW background, in
these generalised theories.  We will work in the conformal
Newtonian gauge, which has the usual correspondence with Bardeen's
gauge invariant variables.  Tensor perturbations on cosmological
backgrounds have been studied in \cite{stnew2}, while the scalar part of
the perturbed line-element takes the form  
\be
ds^2 = a^2 \left[ -(1+2 \Psi) d\tau^2 +(1-2 \Phi) q_{ij} dx^i
dx^j \right],
\ee
where we have used conformal time, $\tau$, and $q_{ij}$ is now
the metric of a static 3-space with constant curvature.  Perturbations
to the  scalar field and energy momentum tensor are given by $\delta
\phi$ and
\ba
\delta T^0_{\phantom{0}0} &=& -\delta \rho  \\
\delta T^0_{\phantom{0}i} &=&  -(\rho+P) \grad_i \theta \\
\delta T^i_{\phantom{i}j} &=&  \delta P \delta^i_{\phantom{i}j}
+(\rho+P) D^i_{\phantom{i}j} \Sigma,
\ea
where $\rho$, $P$ and $\theta$ are the total energy density, pressure and
peculiar velocity of the matter fields.  The first-order perturbation
equations are then given by \cite{BDpert}
\ba
&&\hspace{-50pt} \frac{2}{a^2} \left[ 3 \left( \frac{a^{\prime}}{a} \right)^2 \Psi+3
\frac{a^{\prime}}{a} \Phi^{\prime}+(k^2-3\kappa) \Phi \right] 
+\frac{3 \delta \phi}{a^2 \phi} \left[
\left( \frac{a^{\prime}}{a} \right)^2 +\kappa \right] \label{genSTpert1}
\\ \nonumber
&=& -\frac{8 \pi}{\phi} \delta \rho  - \frac{1}{a^2 \phi}
\left[ \left[6 \left( \frac{a^{\prime}}{a} \right) \Psi +3
\Phi^{\prime} \right] \phi^{\prime}-3 \left( \frac{a^{\prime}}{a}
\right) \delta \phi^{\prime}-k^2 \delta \phi \right] \\ \nonumber
&& -\frac{\delta \phi}{2a^2} \left( \frac{\phi^{\prime}}{\phi} \right)^2
 \frac{d\omega}{d\phi} + \frac{\omega}{a^2 \phi} \left[ \frac{\delta \phi}{2}
 \left( \frac{\phi^{\prime}}{\phi} \right)^2 -\left(
 \frac{\phi^{\prime}}{\phi} \right) \delta \phi^{\prime} +\frac{\phi^{\prime
 2}}{\phi} \Psi \right],
\ea
\ba
&& \hspace{-50pt} \frac{2}{a^2} \left[ \frac{a^{\prime}}{a} \Psi^{\prime}+ \left[2
 \left( \frac{a^{\prime}}{a} \right)^{\prime}+ \left(
 \frac{a^{\prime}}{a} \right)^2-\frac{k^2}{3} \right]
 \Psi+\Phi^{\prime \prime}+2 \frac{a^{\prime}}{a} \Phi^{\prime} +
 \frac{k^2}{3} \Phi - \kappa \Phi \right]\\ \nonumber
&=& \frac{8 \pi \delta P}{\phi} + \frac{\delta \phi}{a^2 \phi} \left[2 \left(
 \frac{a^{\prime}}{a} \right)^{\prime} +\left( \frac{a^{\prime}}{a}
 \right)^2+\kappa \right] \\ \nonumber
&&- \frac{1}{a^2 \phi} \left[2 \phi^{\prime \prime}
 \Psi + \phi^{\prime} \left[\Psi^{\prime}+2 \frac{a^{\prime}}{a}
 \Psi+2 \Phi \right]-\delta \phi^{\prime \prime}-\frac{a^{\prime}}{a}
 \delta \phi^{\prime}- \frac{2k^2 \delta \phi}{3} \right]
\\ \nonumber
&&+\frac{\delta \phi}{2a^2} \left( \frac{\phi^{\prime}}{\phi} \right)^2
 \frac{d\omega}{d\phi} - \frac{\omega}{a^2 \phi} \left[ \frac{\delta \phi}{2}
 \left( \frac{\phi^{\prime}}{\phi} \right)^2 - \left(
 \frac{\phi^{\prime}}{\phi} \right) \delta \phi^{\prime} + \frac{\phi^{\prime
 2}}{\phi} \Psi \right],
\ea
and
\be
\frac{2}{a^2} \left[ \frac{a^{\prime}}{a} \Psi + \Phi^{\prime}
 \right] = \frac{8 \pi}{\phi} (\rho+P) \theta - \frac{1}{a^2 \phi} \left[
 \left( \frac{a^{\prime}}{a} \right) \delta \phi +\phi^{\prime}
 \Psi-\delta \phi^{\prime} \right] + \frac{\omega \phi^{\prime} \delta
 \phi}{a^2 \phi^2}.
\ee
We also have the perturbed scalar field equation
\ba
&&\hspace{-30pt} \delta \phi^{\prime \prime} + 2 \frac{a^{\prime}}{a}
 \delta \phi^{\prime}+k^2
 \delta \phi-2 \phi^{\prime \prime} \Psi- \phi^{\prime} \left( \Psi^{\prime}
 + 4 \frac{a^{\prime}}{a} \Psi +3 \Phi^{\prime} \right)
-\frac{8 \pi a^2 \left(\delta \rho-3 \delta P \right)}{(2 \omega+3)}
\\ \nonumber
&=& -\frac{a^2}{(2 \omega+3)} \Bigg[ 
 \frac{d^2 \omega}{d\phi^2} \frac{\phi^{\prime
 2}\delta \phi}{a^2}+ \frac{2}{a^2} \frac{d \omega}{d \phi} \left( \phi^{\prime}
 \delta \phi^{\prime} - \phi^{\prime 2} \Psi \right)
+ \frac{2}{a^2}
 \frac{d \omega}{d\phi} \left( \phi^{\prime \prime}+2
 \frac{a^{\prime}}{a} \phi^{\prime} \right)\delta \phi \Bigg],
\ea
as well as the condition
\be
\Phi-\Psi = \frac{8 \pi a^2}{\phi} (\rho+P)\Sigma + \frac{\delta \phi}{ \phi}.
\label{genSTpert2}
\ee
This last equation shows that, unlike in General Relativity, $\Phi
\neq \Psi$ when anisotropic stresses vanishes (unless the perturbations
to the scalar field also vanish).  Primes here denote differentiation
with respect to the conformal time, $\tau$, and $k$ is the wave-number
of the perturbation.

Using the equations given above with $\kappa=0$, an analysis of the first year WMAP
data has been performed and used to constrain the parameters
$\alpha_0$ and $\beta_0$ of the attractor model in \cite{Nag}.  The
authors of this study find that the following constraint can be imposed on these
parameters at the $2 \sigma$ level of significance:
\be
\alpha_0 < 5 \times 10^{-4-7 \beta_0}.
\ee
One should bear in mind here that as $\beta_0 \rightarrow 0$
Brans-Dicke theory is recovered, and as $\alpha_0 \rightarrow 0$
General Relativity is recovered.  As a corollary of this result these
authors also constrain the value of Newton's constant 
at recombination to be no more than $5\%$ different from the value
measured in the solar system today, at the $2 \sigma$ confidence level.  The effect of
allowing a non-zero spatial curvature should be expected to weaken
these bounds.

Big bang nucleosynthesis has also been explored in the context of
general scalar-tensor theories \cite{stnew3,kicks,STnuc1,STnuc2}.  In
\cite{kicks} it is found that the inferred upper bound on the baryon
density in the Universe is relatively insensitive to the presence of a
gravitational scalar field, and that the parameters of the attractor
model must satisfy the constraint
\be
\alpha_0^2 \lesssim 10^{-6.5} \beta_0^{-1} \left( \frac{\Omega_m
h^2}{0.15} \right)^{-1.5},
\ee
when $\beta_0 \gtrsim 0.5$.  For $\beta_0 \lesssim 0.5$ these bounds
are weakened by a few orders of magnitude.  These results are extended
and refined in \cite{STnuc1}, who also allow for a non-zero
self-interaction potential for the scalar field.  The apparent tension
between observed and theoretically predicted abundances of Lithium-7
is addressed in the context of scalar-tensor theories in
\cite{STnuc2}.  Here the authors point out that a period of expansion
slower than in General Relativity before primordial nucleosynthesis,
together with a period of more rapid expansion during nucleosynthesis,
can resolve this conflict.  They find such behaviour in numerous
scalar-tensor gravity theories, both with and without self-interaction
potentials.

Inflation in scalar-tensor theories of gravity has been extensively
studied, often under the name `extended inflation', as coined by La and
Steinhardt for the case of Brans-Dicke theory \cite{lastein}.  The
motivation behind this is the possibility of producing a
successful inflationary phase transition from a false vacuum state,
thus avoiding the fine tuning problems associated with `new
inflation'.  Unfortunately, it was soon found that bubble collisions at
the end of inflation produce unacceptable fluctuations in the CMB
\cite{ext1,ext4,ext33}.  Suggestions to improve this
situation were to include a self-interaction potential for the scalar field
\cite{ext2}, generalise the couplings of the Brans-Dicke scalar to other
fields \cite{ext6} (see also \cite{ext10, ext18}), include
quantum effects \cite{ext16}, add additional couplings between the
inflaton and the space-time curvature \cite{ext40}, or to
consider more general scalar-tensor theories of gravity
\cite{ext3,ext5}.  The latter of these approaches was dubbed
`hyper-extended inflation'. The inflationary solutions of
general scalar-tensor theories have been studied in detail in
\cite{ext7,ext19,ext35}, and specific models that
could be compatible with observations were proposed in \cite{ext48}.
Density perturbations in inflationary scalar-tensor scenarios have
been investigated extensively in \cite{ext17,ext22,ext23,ext24,ext25,ext26,ext29,ext31,ext32,ext34,ext42,ext44,ext49,ext53,ext59}.  Studies of topological defects
\cite{ext9,ext56}, black holes \cite{ext11}, gravitational
waves \cite{ext12,ext58}, baryogenesis \cite{ext13,ext14}, baryon asymmetry \cite{ext15}, dark matter \cite{ext20,ext37}, the formation of voids \cite{ext21,ext28}, bubble
nucleation rates and dynamics \cite{ext27,ext39}, reheating
\cite{ext36}, stochastic inflation \cite{ext41,ext43,ext46,ext54,ext55}, slow roll inflation \cite{ext45,ext47,ext50}, non-Gaussianity \cite{newSTinf},
isotropisation of the Universe \cite{ext51,ext52}, and quantum cosmology \cite{ext57} have all also been performed
in the context of inflation in scalar-tensor theories.  The initial
conditions for inflation in scalar-tensor theories have been considered in
\cite{qext,ext8,ext30}.  For further details the
reader is referred to 1993 review of extended inflation by Steinhardt
\cite{ext38}. 

Theories of gravity with non-minimally coupled scalar fields and
non-zero self-interaction potentials have been studied by a number of
authors under the name `extended quintessence'
\cite{quin1,quin2,quin3,quin4,STlense}.  Such theories can act as dark
energy as well as model possible deviations from General Relativity at
early times. These papers include studies of small angle CMB temperature and
polarisation power spectra, the integrated Sachs-Wolfe effect, the
matter power spectrum, supernovae observations and the affects that
should be expected on weak lensing observations.  The FLRW solutions of
theories with power-law self-interaction potentials have been studied in further
detail in \cite{quinphase}, where the attractor mechanism to general
relativity is investigated, as well the presence of periods of
accelerating expansion at late and early times.  Late-time
acceleration in models without a potential for the scalar field is
studied in \cite{farh8}.

Another interesting possibility in scalar-tensor theories of
gravity is the idea of `gravitational memory', proposed by Barrow in
\cite{mem1}.  The idea here is that when a black hole forms one of two
things can happen (or some combination of them).  Firstly, the
Schwarzschild radius of a black hole, which is given by $r_S=2 G(t) m$,
could vary as the value of the scalar field controlling the value of
$G$ varies in the background universe.  In this case there is no such
thing as a static black hole solution to the gravitational field
equations, unless the black hole exists in a static universe.
Secondly, the Schwarzschild radius of a black hole could be frozen in
at its value when the black hole formed, so that $r_S=2G(t_f) m$, where
$t_f$ is the time when the black hole formed.  In this case black
holes that formed early on in the Universe's history would remember,
in some sense, the conditions of the early universe, this being
reflected in the value of $G(t_f)$.  As Barrow points out, these two
possibilities have consequences for the evaporation, and explosion, of
black holes in the late universe.  This idea has motivated a number of
studies on the gravitational field of collapsed objects in
scalar-tensor theories of gravity
\cite{mem2,mem3,mem4,mem5,mem6,mem7,mem8,mem9}.  One particularly
interesting approach is that of matched asymptotic expansions, which suggests
that the first option is followed, and black holes do not have any
gravitational memory \cite{doug1,doug2,doug3}.
\newline
\newline
\noindent
{\it Horndeski's theory}
\newline

The most general  four dimensional scalar-tensor theory with
second-order field equations was worked out by Horndeski in
\cite{horndeski}. It has the following Lagrangian 
\ba
\label{eq:hornyLagrangian}
{\cal L}_{H} &=& \delta^{\alpha\beta\gamma}_{\mu\nu\sigma}\left[\kappa_1\nabla^\mu\nabla_\alpha\phi R_{\beta \gamma}^{\;\;\;\;\nu\sigma}
           -\frac{4}{3}\kappa_{1,X}\nabla^\mu\nabla_\alpha\phi\nabla^\nu\nabla_\beta \phi\nabla^\sigma\nabla_\gamma\phi  \right.   \nonumber\\
           && \qquad\qquad  \left.  +\kappa_3\nabla_\alpha\phi\nabla^\mu\phi R_{\beta \gamma}^{\;\;\;\;\nu\sigma} \nonumber
           -4\kappa_{3,X}\nabla_\alpha\phi\nabla^\mu\phi\nabla^\nu\nabla_\beta \phi\nabla^\sigma\nabla_\gamma\phi \right]\\\nonumber
        &~&+\delta_{\mu\nu}^{\alpha \beta }\left[(F+2W)R_{\alpha \beta }^{\;\;\;\;\mu\nu}
           -4F_{,X}\nabla^\mu\nabla_\alpha\phi\nabla^\nu\nabla_\beta \phi+2\kappa_8\nabla_\alpha\phi\nabla^\mu\phi\nabla^\nu\nabla_\beta \phi\right] \nonumber\\
           &&
        -3[2(F+2W)_{,\phi}+X\kappa_8]\nabla_\mu\nabla^\mu\phi
          +\kappa_9(\phi,X),
\ea
where $X=\nabla_\mu\phi\nabla^\mu\phi$, and
$\delta^{\nu_1\nu_2...\nu_n}_{\mu_1\mu_2...\mu_n}=n!\delta^{[\nu_1}_{\mu_1}\delta^{\nu_2}_{\mu_2}...\delta^{\nu_n]}_{\mu_n}$. 
The theory depends on   four arbitrary functions of $\phi$ and $X$,
$\kappa_i=\kappa_i(\phi,X)$  as well as $F=F(\phi, X)$, which  is
constrained so that 
$
F_{,X}=\kappa_{1,\phi}-\kappa_3-2X\kappa_{3,X}.
$
Note that $W=W(\phi)$, which means that it can be absorbed into a
redefinition of $F(\phi, X)$. This paper is not very well known, and
as a  result  Horndeski's theory has not been well explored.  It has,
however, been recently resurrected in \cite{fabfour}, where aspects of
the theory on FLRW backgrounds were studied. The effective Lagrangian
describing the cosmology in the minisuperspace approximation is given
by
\be
\label{eq:effectiveLag}
L_H^\textrm{eff}(a, \dot a, \phi, \dot
\phi)=a^3\sum_{n=0}^3\left(A_n-B_n\frac{\kappa}{a^2}\right)H^n ,
\ee
where $H=\dot a/a$ is the Hubble parameter, and where we have
\ba
A_0&=&-\tilde{Q}_{7,\phi}\dot\phi+\kappa_9\\
B_0&=&\tilde Q_{1,\phi}\dot\phi+ 12\kappa_3\dot\phi^2-12F\\
A_1&=&-12F_{,\phi}\dot\phi+3(Q_7\dot\phi-\tilde{Q}_7)+6\kappa_8\dot\phi^3\\
B_1&=&-Q_1\dot\phi+\tilde{Q}_1\\
A_2&=&-12F-12F_{,A}\dot\phi^2\\
A_3&=&8\kappa_{1,A}\dot\phi^3 ,
\ea
where 
\be
Q_1 = \frac{\del \tilde{Q}_1}{\del\dot\phi}=-12\kappa_1 , \qquad {\rm
  and} \qquad
Q_7 =\frac{\del
  \tilde{Q}_7}{\del\dot\phi}=6F_{,\phi}-3\dot\phi^2\kappa_8 .
\ee
It is assumed that matter is minimally coupled to the metric
$g_{\mu\nu}$, and not to the scalar. Indeed, it is argued that if  the
equivalence principle is to hold then this can be assumed without
further loss of generality. The cosmological field equations are
presented implicitly as a generalised Friedmann equation:
\be \label{eq:H}
\frac{1}{a^3}\left[\dot a \frac{\del L_H^\textrm{eff}}{\del \dot
    a}+\dot \phi \frac{\del L_H^\textrm{eff}}{\del \dot
    \phi}-L_H^\textrm{eff}\right]=-\rho ,
\ee
and the scalar equation of motion
\be
\frac{\del L_H^\textrm{eff}}{\del \phi} -\frac{d}{dt} \left[\frac{\del
    L_H^\textrm{eff}}{\del \dot  \phi}\right]=0 , \label{eq:E}
\ee
where $\rho$ is the energy density of the cosmological fluid.

In \cite{fabfour} the authors look for those corners of Horndeski's
theory that admit a self-tuning mechanism. They demand that  the
vacuum space-time is Minkowski, irrespective of the value of the
cosmological constant, and that this should remain true even after a
phase transition in which the cosmological constant changes by some
amount.  This is  not in violation of Weinberg's theorem since
Poincar\'e invariance is explicitly broken by the scalar. These
considerations reduce Horndeski's theory to four base Lagrangians
known as {\it the Fab Four}:
 \begin{eqnarray}
\label{eq:john}
{\cal L}_{{\rm john}} &=& \sqrt{-g} V_{{\rm john}}(\phi)G^{\mu\nu} \nabla_\mu\phi \nabla_\nu \phi \\
\label{eq:paul}
{\cal L}_{{\rm paul}} &=&\sqrt{-g}V_{{\rm paul}}(\phi)   P^{\mu\nu\alpha \beta} \nabla_\mu \phi \nabla_\alpha \phi \nabla_\nu \nabla_\beta \phi \\
\label{eq:george}
{\cal L}_{{\rm george}} &=&\sqrt{-g}V_{{\rm george}}(\phi) R\\
\label{eq:ringo}
{\cal L}_{{\rm ringo}} &=& \sqrt{-g}V_{{\rm ringo}}(\phi) \hat G ,
\end{eqnarray}
where $\hat G=R_{\mu\nu \alpha \beta}R^{\mu\nu \alpha
  \beta}-4R_{\mu\nu}R^{\mu\nu}+R^2$ is the Gauss-Bonnet combination, and
$$
P^{\mu\nu}_{\phantom{\mu\nu} \alpha \beta} = -
R^{\mu\nu}_{\phantom{\mu\nu}\alpha \beta} 
+2 R^{\mu}_{\phantom{\mu} [\alpha} \delta^{\nu}_{\phantom{\nu} \beta]}
-2 R^{\nu}_{\phantom{\nu} [\alpha} \delta^{\mu}_{\phantom{\mu} \beta]}
-R \delta^{\mu}_{\phantom{\mu} [ \alpha} \delta^{\nu}_{\phantom{\nu} \beta]}
$$
is the double dual of the Riemann tensor. These terms
give rise to self-tuning cosmologies for $\kappa<0$. The relevant
cosmological field equations are given by 
\be
{\cal H}_{{\rm john}}+{\cal H}_{{\rm paul}}+{\cal H}_{{\rm george}}+{\cal
  H}_{{\rm ringo}}=-\left[\rho_\Lambda+\rho_\textrm{matter}\right] ,
\ee
where we have separated  the net cosmological constant contribution,
$\rho_\Lambda$, and the matter contribution, $\rho_\textrm{matter}$,
and where
\ba
&&{\cal H}_{{\rm john}}=3V_{{\rm john}}(\phi)\dot\phi^2\left(3H^2+\frac{\kappa}{a^2}\right) \nonumber\\
&&{\cal H}_{{\rm paul}}=-3V_{{\rm paul}}(\phi)\dot\phi^3H\left(5H^2+3\frac{\kappa}{a^2}\right) \nonumber\\
&&{\cal H}_{{\rm george}}=-6V_{{\rm
    george}}(\phi)\left[\left(H^2+\frac{\kappa}{a^2}\right)+H\dot\phi
  \frac{V'_{{\rm george}}}{V_{{\rm george}}}\right]\qquad \nonumber\\
&&{\cal H}_{{\rm ringo}}=-24V'_{{\rm ringo}}(\phi)\dot\phi
H\left(H^2+\frac{\kappa}{a^2}\right) . \nonumber
\ea
The scalar equations of motion are ${\cal E}_{{\rm john}}+{\cal
  E}_{{\rm paul}}+{\cal E}_{{\rm george}}+{\cal E}_{{\rm ringo}}=0$ where 
\ba
&&{\cal E}_{{\rm john}}= 6{d \over dt}\left[a^3V_{{\rm
      john}}(\phi)\dot{\phi}\Delta_2\right]  - 3a^3V_{{\rm john}}'(\phi)\dot\phi^2\Delta_2
 \nonumber\\
&&{\cal E}_{{\rm paul}}= -9{d \over dt}\left[a^3V_{{\rm
       paul}}(\phi)\dot\phi^2H\Delta_2\right]  +3a^3V_{{\rm paul}}'(\phi)\dot\phi^3H\Delta_2
 \nonumber\\
&&{\cal E}_{{\rm george}}= -6{d \over dt}\left[a^3V_{{\rm
       george}}'(\phi)\Delta_1\right]  +6a^3V_{{\rm george}}''(\phi)\dot\phi \Delta_1  \nonumber \\
&&\qquad\qquad\qquad\qquad+6a^3V_{{\rm george}}'(\phi)\Delta_1^2  \nonumber\\
&&{\cal E}_{{\rm ringo}} = -24 V'_{{\rm ringo}}(\phi) {d \over dt}\left[a^3\left(\frac{\kappa}{a^2}\Delta_1 +\frac{2}{3}  \Delta_3 \right) \right] , \nonumber
\ea
and we define
$\Delta_n=H^n-\left(\frac{\sqrt{-\kappa}}{a}\right)^n$. We see that
the self-tuning is achieved at the level of the scalar equation of
motion, since on a Minkowski solution one has $H^2=-\frac{\kappa}{a^2}
\implies \Delta_n=0$ for $n \geq 1$. In vacuum, the cosmological
constant controls the value of the scalar via the  generalised
Friedmann equation. A detailed study of the phenomenology of the fab four
has yet to be carried out, but the authors of \cite{fabfour} argue
that the `john' and `paul'
terms are expected to play a crucial role, as their derivative
interactions could give rise to Vainshtein effects that could help
pass solar system constraints. The Vainshtein mechanism is discussed
in detail in Section \ref{sec:dgp-sc}.

Note that it has been shown that Horndeski's general theory is equivalent to \cite{cedric} in four dimensions \cite{Kobayashi1}. Aspects of cosmological perturbations are studied in \cite{Kobayashi1} that may be applied to {\it the Fab Four} in the appropriate special case.

\subsubsection{The chameleon mechanism}
\label{scalartensorsection2}

The `chameleon mechanism' was introduced as a concept in gravitational
physics by Khoury and Weltman in \cite{cham1,cham2}.  The
basic concept here is that if we consider theories with a
non-minimally coupled scalar field, then in the presence of other
matter fields these scalars can acquire an effective mass parameter
that is environmentally dependent.  One can then potentially satisfy
the tight constraints on non-minimally coupled scalar degrees of
freedom that are imposed in relatively dense environments, such as
exist in the solar system, while still having interesting new behaviour in
less dense environments, such as those that can exist in cosmology.  

This mechanism is usually formulated in the Einstein conformal frame,
where the coupling between the scalar curvature and scalar field is
minimal, but where the scalar field couples non-minimally to matter
fields.  The relevant action is then
\be
\label{chamL}
\mathcal{L} = \sqrt{-\bar{g}} \left\{ \frac{1}{16 \pi} \bar{R} -
\frac{1}{2} \bar{g}^{\mu \nu} \psi_{,\mu} \psi_{,\nu} -V(\psi)
\right\} + \mathcal{L}_m (\Psi_i, e^{2 \sqrt{8 \pi} \beta_i \psi}
\bar{g}_{\mu \nu}), 
\ee
where, in the notation used in Eq. (\ref{ST4}), we have taken $\Gamma=
\sqrt{8 \pi} \beta_i \psi$, and where the $\beta_i$ are a set of constants
denoting the coupling of $\psi$ to each of the $i$ matter fields
$\Psi_i$.  In the scalar-tensor theories so far discussed the scalar
field should be considered coupled to each of the matter fields with
the same universal coupling, which in the Brans-Dicke theory is given
by $\beta^{-2}=2(3+2 \omega)$.  Assuming such a coupling, the
non-relativistic limit of the scalar field equation can then be
written as
\be
\nabla^2 \psi = \frac{d V_{\textrm{eff}}}{d \psi},
\ee
where $V_{\textrm{eff}}(\psi)\equiv V(\psi) + \rho e^{\sqrt{8 \pi} \beta \psi}$.  This new
`effective potential' can be seen to be dependent on the ambient energy
density, and if $\psi$ and $\beta$ are both positive then any runaway
potential with $dV/d\psi<0$ will result in an effective potential with
a local minimum whose position depends on $\rho$.  What is more, for
couplings of the type specified in Eq. (\ref{chamL}) the local
effective mass of  the scalar field $\psi$, given by
$m_{\psi}=d^2V_{\textrm{eff}}/d\psi^2 $, can be seen to generically increase with
increasing $\rho$.  Hence, the name `chameleon'.

The behaviour of scalar fields outside of massive objects, when the
chameleon mechanism is present, can be shown to be crucially dependent
on the ratio of $\Delta \psi$ to $\Phi_c$. Here $\Delta \psi$ denotes
the difference in the value of the scalar field inside the object,
$\psi_c$, and asymptotically, $\psi_{\infty}$, while $\Phi_c$ is the
value of the Newtonian potential at the surface of the object, where
$r=R_c$.  More precisely, when one satisfies the condition
\be
\label{ts}
\frac{\sqrt{8 \pi} (\psi_{\infty}-\phi_c)}{6 \beta \Phi_c} \ll 1
\ee
then the resulting configuration of gravitational fields is found to
be one in which $\psi$
occupies the minimal of the effective potential inside the bulk of the
massive object, except for a thin region of depth $\Delta R_c$ just
below its surface where the value of $\psi$ rises.  Outside of the
object $\psi$ increases further, and approaches its asymptotic value
$\psi_{\infty}$ as 
\be
\label{ex1}
\psi \simeq \psi_{\infty}-\frac{2 \beta}{\sqrt{8\pi}} \left(
\frac{3\Delta R_c}{R_c}\right) \frac{M_c e^{-m_{\infty}(r-R_c)}}{r},
\ee
where $m_{\infty}$ is the effective mass of the field at
asymptotically large distances from the object of mass $M_c$.
Now, the ratio of the thickness of the shell just below the object's
surface to the object's overall radius, $\Delta R_c/R_c$, can be shown
to be well approximated by the LHS of Eq. (\ref{ts}).  The condition
given in (\ref{ts}) is then equivalent to the condition that a `thin
shell' should be present, with $\Delta R_c/R_c \ll 1$.

If the `thin shell' condition is not met then one instead has
$\psi \sim \psi_{\infty}$  everywhere, and the exterior solution is
given by 
\be
\label{ex2}
\psi \simeq \psi_{\infty}-\frac{2 \beta}{\sqrt{8\pi}} \frac{M_c
e^{-m_{\infty}(r-R_c)}}{r}.
\ee
A comparison of Eq. (\ref{ex2}) with Eq. (\ref{ex1}) immediately shows that
without a thin shell variations in $\psi$ are no longer suppressed
by the small factor of $3 \Delta R_c/R_c$, and that we should
therefore expect in this case more obvious consequences to the
existence of $\psi$ within the vicinity of massive objects.  
Khoury and Weltman proceed to argue that in order to avoid violations of the
weak equivalence principle, and unacceptable deviations from the
predictions of General Relativity in the solar system, we should
require that the Earth, and other astrophysical bodies, should satisfy
the thin shell condition \cite{cham2}.


This idea of a scalar field with an environmentally dependent mass has
sparked widespread interest since it was proposed.  In particular, it
allows for the possibility of measuring fifth forces, or violations of
the weak equivalence principle, that are different in space than they are on Earth,
\cite{cham1,cham2,cham3,cham4,cham5,cham6,cham9,cham10,cham12,cham14,cham26}.
 It can act as dark energy \cite{chamDE1,chamDE2}, and has been
 studied in the context of structure formation \cite{chamSF1,cham25},
 as well as a number of other cosmological scenarios
 \cite{cham4,cham6,cham9,chamDE1,cham22,cham24,cham28}. The effect of `chameleon
 particles' on searches for axion-like particles and experiments
 involving magnetic fields have been studied in
 \cite{champart1,champart2,champart3}, and their effect on the propagation
 of light in astrophysics in \cite{champart4,cham11}.  Experimental
 searches for chameleons have now been performed by GammeV
 \cite{champart5,cham13,cham18,cham30}, and ADMX \cite{cham23}, which
 have started to constrain the viable parameter space of these theories.  Other
 tests of this scenario are also proposed in
 \cite{cham16,cham17,cham27,cham19,cham21,cham29}.

\subsection{Einstein-{\AE}ther Theories}
\label{aesubsection}

Vector-tensor theories, in the  form of Einstein-{\ae}ther theories, have
had a revival over the past decade, and are now often used as a
counterfoil to test General Relativity. They have the particular
property that they single out a preferred reference frame and have
become somewhat of a theoretical workhorse for studying violations of
Lorentz symmetry in gravitation.  In the Einstein-{\ae}ther theory
\cite{Jacobson2008a} violations of Lorentz invariance arise within
the framework of a diffeomorphism-invariant theory, and their modern
incarnations are a refinement of the gravitationally coupled vector field theories 
first proposed by Will and Nordtvedt in the 1970s
\cite{WillNordvedt1972a,WillNordvedt1972b}.
The presence of a Lorentz-violating vector field, henceforth called
the {\it {\ae}ther}, can dramatically affect cosmology: 
It can lead to a renormalisation of the Newton constant \cite{CarrollLim2004}, leave
an imprint on perturbations in the early universe
\cite{Lim2005,KannoSoda2006}, and in more elaborate actions it can
even affect the growth rate of structure in the Universe
\cite{ZlosnikFerreiraStarkman2007,ZuntzFerreiraZlosnik2010,HalleZhaoLi2007}.

\subsubsection{Modified Newtonian dynamics}

Some of the theories that we will discuss in this subsection and the
next have been constructed to give modifications to Newtonian gravity
on galactic scales. To be more specific, they should lead to Milgrom's
Modified Newtonian Dynamics \cite{Milgrom1983}, also known as MOND, in
regimes of low acceleration. Given its relevance for Einstein-{\ae}ther
theories, we will now briefly describe the motivation for MOND, and
how it works. We also briefly mention some of its successes and failures.

MOND was first proposed as a possible explanation of the need for dark
matter in galaxies, based on observations of their rotational
velocities.  With Newtonian gravity and the visible baryonic matter in
galaxies only one expects that the rotational velocity, $v_r$, should
depend on the distance from the centre of the galaxy, $r$, as
$v_r\propto r^{-1/2}$. What is in fact found in observations of spiral
galaxies is that $v_r$ is approximately constant at large radii. The
conventional answer to this problem is to posit that galaxies sit in
halos of dark matter, with energy density profiles that vary as $\rho
\sim r^{-2}$ for large $r$.  Milgrom's proposal was that,
alternatively, Newton's inverse square law of gravity could be
modified in the low-acceleration regime of galactic dynamics.  Such a
modification, it was ventured, may be able to account for the
anomalously high rotational velocities in spiral galaxies without
invoking any new matter fields.

In MOND the spherically symmetric gravitational potential has two
regimes: High acceleration and low acceleration. In regions of high
acceleration (where $|{\vec a}|\gg a_0$, for constant $a_0$), it
simply satisfies Newton's second law: ${\vec a}=-\nabla \Phi$ where
$\Phi$ is the gravitational potential. On the other hand, in the low
acceleration regime (where $|{\vec a}|\ll a_0$), Newton's second law
is modified to $(|{\vec a}/a_0){\vec a}=-\nabla \Phi$. Albeit a simple
rule of thumb, Milgrom's proposal is remarkably successful at fitting
a large range of spiral galaxy observations. Furthermore, it can be
used to explain the Tully-Fisher relation that relates the velocity of
rotation of a spiral galaxy with its intrinsic luminosity.
Unfortunately MOND is unable to explain the dynamics of clusters of
galaxies without recourse to additional dark matter (possibly in the
form of neutrinos), and the behaviour of dwarf spheroidals in
different environments is also problematic.  Nevertheless, it is an
interesting proposal that has had a renewed surge of interest in the
past decade.

The non-relativistic Poisson equation in MOND can be written as
\begin{eqnarray}
\grad \cdot \left[\mu\left(\frac{|\grad \Phi|}{a_0}\right)\grad\Phi\right]=4\pi G\rho,
\end{eqnarray}
where $\rho$ is the energy density in baryons, and the function
$\mu(x)\rightarrow 1$ as $x\rightarrow \infty$ and $\mu(x)\rightarrow
x$ as $x\rightarrow 0$.  There are a variety of proposals for the
precise form of $\mu(x)$ that fit observations of galaxies to a
greater or lesser degree. As a theory of modified gravity, however,
MOND's greatest limitation is that it is restricted to
non-relativistic regimes. It therefore cannot be used to make
prediction on cosmological scales, nor can it be used to calculate
fundamentally relativistic observables, such as lensing.  Many of the
theories that follow in this section have been constructed to address
this deficiency:  They are relativistic gravitational theories that
have MOND as a non-relativistic limit.

\subsubsection{Action and field equations}

As the name suggests, vector-tensor theories involve the introduction
of a space-time 4-vector field, $A^\mu$. A general action for such
theories is given by
\be
\label{aef1}
S=\int d^4x \sqrt{-g}\left[\frac{1}{16\pi G}R+{\cal L}(g^{\mu\nu},A^{\nu})\right]
+S_{M} (g^{\mu \nu},\Psi ),
\ee
where $S_m$ is the matter action.  Note that the matter fields $\Psi$
in $S_M$ couple only to the metric $g_{\mu\nu}$, and not to $A^{\nu}$.

Let us now focus on Einstein-{\ae}ther theories, and hence forth consider
$A^\mu$ to have a time-like direction. The simplest
(and most thoroughly studied) version of the Einstein-{\ae}ther theory is quadratic in
derivatives of $A^{\nu}$, and has the form
\be
\label{aef2}
{\cal L}_{EA}(g^{\mu\nu},A^{\nu})\equiv\frac{1}{16\pi
G}[K^{\mu\nu}_{\phantom{\mu\nu}\alpha\beta}
\nabla_\mu A^{\alpha}\nabla_{\nu}A^{\beta}
+\lambda(A^{\nu}A_\nu+1)]  ,
\ee
where $K^{\mu\nu}_{\phantom{\mu\nu}\alpha\beta}\equiv c_1g^{\mu\nu}g_{\alpha \beta}
+c_2\delta^{\mu}_{\phantom{\mu}\alpha}\delta^{\nu}_{\phantom{\nu}\beta}+
c_3\delta^{\mu}_{\phantom{\mu}\beta}\delta^{\nu}_{\phantom{\nu}\alpha}
-c_{4}A^{\mu}A^{\nu}g_{\alpha \beta}$ and $\lambda$ is a Lagrange
multiplier.  In what follows we will use the notation $c_{12 \dots
}\equiv c_{1}+c_{2}+ \dots$. We call the theory derived from
Eqs. (\ref{aef1}) and (\ref{aef2}) the {\it linear} Einstein-{\ae}ther theory.

A more general, non-linear Lagrangian for the {\ae}ther field can be written in the form
\be
\label{eq:Lagrangian}
{\cal L}_{GEA}(g^{\mu\nu},A^{\mu})=\frac{M^2}{16\pi G}
	 F(K) +\frac{1}{16\pi G}\lambda(A^{\mu}A_\mu+1),
\ee
where $K=K^{\mu\nu}_{\phantom{\mu\nu}\alpha\beta}\nabla_\mu
A^{\alpha}\nabla_{\nu}A^{\beta}$, and $M$ has the dimension of
mass. We shall call this a {\it generalised} Einstein-{\ae}ther theory. 

Such actions arise from Lorentz violating physics in quantum
gravity. Indeed, the linear Einstein-{\ae}ther theory can be constructed
using the rules of effective field theory, and has been shown to be
stable with regard to quantum effects \cite{Withers2009}. Such
theories, however, can suffer from instabilities at the classical
level, with the onset of caustics in a finite time
\cite{ContaldiWisemanWithers2008}. This raises the question of whether
the vector field in such theories are merely an effective (possibly
composite) degrees of freedom, or whether they are genuine fundamental
fields.

The gravitational field equations for this theory, obtained by varying
the action for the Generalised Einstein-{\ae}ther theory with respect to
$g^{\mu\nu}$  are given by
\begin{eqnarray}
G_{\mu\nu}&=&\tilde{T}_{\mu\nu}+8\pi GT^{\mathrm{matter}}_{\mu\nu}, \\
\tilde{T}_{\mu\nu} &=& \frac{1}{2}\nabla_{\alpha}
(F_{K}(J_{(\mu}^{\phantom{\mu}\alpha}A_{\nu)}-
J^{\alpha}_{\phantom{\alpha}(\mu}A_{\nu)}-J_{(\mu\nu)}A^{\alpha}))
\nonumber \label{lagrangian equation} \\ && -F_{K}Y_{(\mu\nu)}
+\frac{1}{2}g_{\mu\nu}M^{2}F+\lambda A_{\mu}A_{\nu},
 \end{eqnarray}
where $F_{K} \equiv  \frac{dF}{dK}$ and $J^{\mu}_{\phantom{\mu}\alpha} \equiv
(K^{\mu\nu}_{\phantom{\mu\nu}\alpha\beta}+
K^{\nu\mu}_{\phantom{\nu\mu}\beta\alpha})\nabla_{\nu}A^{\beta}$.
Brackets around indices denote symmetrisation, and $Y_{\mu\nu}$ is
 defined by the functional derivative
$
Y_{\mu\nu} =\nabla_{\alpha}A^{\rho}\nabla_{\beta}A^{\sigma}
\frac{\partial(K^{\alpha \beta}_{\phantom{\alpha \beta}\rho\sigma})}{\partial 
g^{\mu\nu}}
$.
The equations of motion for the vector field, obtained by varying with
respect to $A^{\nu}$, are
\begin{eqnarray}
\label{eq:veceq1}
\nabla_{\mu}(F_{K}J^{\mu}_{\phantom{\mu}\nu})
+F_{K} y_{\nu}&=&2\lambda A_{\nu},
\label{vectoreom}
\end{eqnarray}
where we have defined $y_{\nu}=\nabla_{\alpha}A^{\rho}\nabla_{\beta}A^{\sigma}
\frac{\partial(K^{\alpha \beta}_{\phantom{\alpha \beta}\rho\sigma})}
{\partial A^\nu}$. Finally, variations of the action with respect to
$\lambda$ fix $A^{\nu}A_\nu=-1$.

\subsubsection{FLRW solutions}

In a  homogeneous and isotropic universe with perfect fluid matter
content, the vector field  will only have a non-vanishing `$t$'
component, so that $A^{\mu}= (1,0,0,0)$.  The equations of motion then
simplify dramatically, so that $\nabla_{\mu}A^{\mu}= 3H$ and ${K} =
3\frac{\alpha H^{2}}{M^{2}}$, where $\alpha \equiv
c_{1}+3c_{2}+c_{3}$.  Note that the $\alpha$ we have defined here has
the same sign as $K$.  The field equations then reduce to
\begin{eqnarray}
\left[1-\alpha{ K}^{1/2}\frac{d}{d{K}}\left(\frac{ F}{{
K}^{1/2}}\right)\right]H^2 &=& \frac{8\pi G}{3}\rho, \label{00m} \\
\frac{d}{dt}(-2H+F_{K}\alpha H)&=&8 \pi G (\rho+P) .\label{summ}
\end{eqnarray}
\begin{figure}
\begin{center}
\epsfig{file=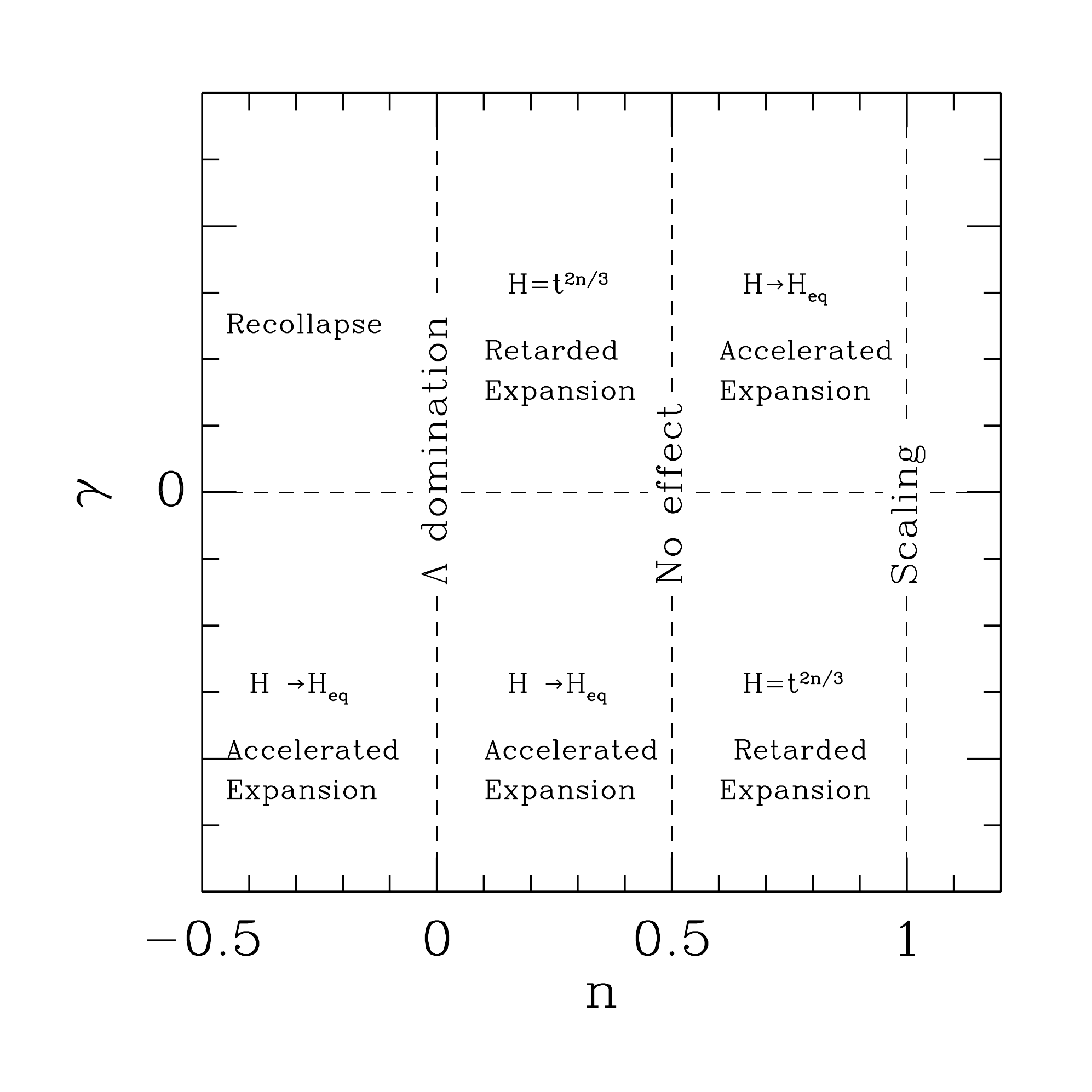,height=8cm}
\caption{A schematic representation of the late-time evolution of FLRW
solutions as a function of $n$ and $\gamma$, for $n<1$.}
\end{center}
\label{AEcosm}
\end{figure}
If we now take ${F}(x)=\gamma x^n$, the modified Friedmann equations become
\begin{eqnarray}
\label{eq:modfri}
\left[1+\epsilon\left(\frac{H}{M}\right)^{2(n-1)}\right]H^2=\frac{8\pi G}{3}\rho  ,
\end{eqnarray}
where $\epsilon \equiv (1-2n)\gamma(-3\alpha)^{n}/6$.  We also get the
relationship
\be
\gamma=\frac{6(\Omega_{m}-1)}{(1-2n)(-3\alpha)^n}\left(\frac{M}{H_0}\right)^{2(n-1)},
\ee
where $\Omega_{m}\equiv 8\pi G\rho_{0}/3 H_0^2$, and $H_0$ is the
Hubble constant today.  Let us now consider a few special cases: If
$n=1/2$, the Friedmann equations are unchanged ($\epsilon=0$) and there is no
effect on the background cosmology;  with  $n=1$ we have that 
$\epsilon=\gamma\alpha/2$ and  Newton's constant is rescaled by a
factor of $1/(1+\epsilon)$ \cite{CarrollLim2004}; if $n=0$ we recover
a cosmological constant, $\Lambda\simeq sign(-\gamma)  M^2$.
More generally, we will obtain different regimes depending on the relative size
of each term in the modified Friedmann equation. We can summarise
these behaviours in Figure \ref{AEcosm}. 

\subsubsection{Cosmological perturbations}

The four-vector $A^{\mu}$ can be perturbed as $A^{\mu}= (1- \Psi,\frac{1}{a}\grad^i V)$, where $V$ is a small quantity. 
  Perturbing ${K}$ to linear order then gives ${ K} = {K}_0+ {K}_1$, where 
${K}_0 = 3\frac{\alpha H^{2}}{M^{2}}$ and 
${K}_1 = -2\frac{\alpha
H}{M^{2}}(k^{2}\frac{V}{a}+3H\Psi+3\dot{\Phi})$.  The gravitational
potentials $\Psi$ and $\Phi$ come from the perturbed metric:
\be
ds^2 = a^2(\tau) \left[ -(1+2 \Psi) d\tau^2 + (1-2 \Phi) q_{ij} dx^i
dx^j \right],
\ee
were $q_{ij}$ is the unperturbed conformal metric of the
hyper-surfaces of constant $\tau$.

The evolution equation for the perturbations in the vector field are
\begin{eqnarray}
\label{vecp}
0 &=& c_{1}[V''+k^{2}V+2{\cal H}V'+2{\cal H}^{2}V+\Psi'+\Phi '+2{\cal H}\Psi] \\
\nonumber && +c_2[k^{2}V+6{\cal H}^{2}V-3\frac{a''}{a}V+3\Phi'+3{\cal H}\Psi]\\
\nonumber && +c_3[k^{2}V+2{\cal H}^{2}V-\frac{a''}{a}V+\Phi'+{\cal H}\Psi]\\
\nonumber && +\frac{F_{KK}}{F_{K}}[-K_1\alpha{\cal
H}-K_0'(-c_{1}(V'+\Psi)+3c_{2}{\cal H} V+c_{3}{\cal H} V)] .
\end{eqnarray}
The perturbation in the vector field is sourced by the two gravitational
potentials $\Phi$ and $\Psi$.  The first-order perturbations to the
vector field's stress-energy tensor are 
\begin{eqnarray}
a^{2}\delta \tilde{T}^{0}_{\phantom{0}0}&=& F_{K}c_{1}[-{\cal
H}k^{2}V-k^{2}V'-k^{2}\Psi] \\ \nonumber &&+ F_{K}\alpha [ {\cal H} k^2 V+3{\cal H}
\Phi' +3 {\cal H}^2 \Psi ] -3 F_{KK} \alpha {\cal H}^2 K_1 \\
\nonumber &=& F_{K}c_{1}[-{\cal H}k^{2}V-k^{2}V'-k^{2}\Psi]
+F_{K}\alpha(2n-1)[{\cal H}k^{2}V+3{\cal H}\Phi '+3{\cal H}^{2}\Psi],\\
a^{2}\delta \tilde{T}^{0}_{\phantom{0}i} &=& ik_{i}F_{K}c_{1} \left[
V''+2{\cal H}V'+\frac{a''}{a}V   +\Psi '+{\cal H}\Psi \right]
\\\nonumber &&+ik_{i}F_{K}\alpha \left[ 2{\cal H}^{2}V-\frac{a''}{a}V
\right]  +ik_{i}F_{KK}{ K}_{0}'[c_{1}({\cal H}V+V'+\Psi)  - \alpha {\cal
H}V], \nonumber\\ \label{eq::tij}
a^{2}\delta \tilde{T}^{i}_{\phantom{i}j} &=& F_{K}c_{2}k^{2}[2{\cal
H}V+V']\delta^{i}_{\phantom{i}j} +F_{K}(c_{1}+c_{3})[2{\cal H}V+V']k^{i}k_{j}
\\ \nonumber &&+ F_{K}\alpha
\left[ 2 {\cal H} \Phi' +\Phi''+2 \frac{a''}{a} \Psi - {\cal H}^2 \Psi
+ {\cal H} \Psi' \right] \delta^i_{\phantom{i}j}+F_{KK}(c_{1}+c_{3})K_0'Vk^{i}k_{j}\\
&&-F_{KK}[\alpha K_1\frac{a''}{a}+(c_{1}+c_{2}+c_{3}) K_1{\cal H}^{2} 
+\alpha {\cal H}K_1'
\nonumber 
\\&&\qquad \qquad \qquad-\alpha K_0'\Phi '-2\alpha K_0'{\cal H}\Psi
+\alpha \ln(F_{KK})'K_1{\cal
H}-c_{2}K_0'k^{2}V]\delta^{i}_{\phantom{i}j},  \nonumber
\end{eqnarray}
where the second expression for $a^{2} \delta
\tilde{T}^{0}_{\phantom{0}0}$ assumes the monomial form for $F({ K})$. 
In the absence of anisotropic stresses in the matter fields, we may
obtain an algebraic relation between the metric potentials $\Phi$ and
$\Psi$ by computing the transverse, traceless part of the perturbed
Einstein equations.  This gives
\begin{eqnarray}
k^{2}(\Psi-\Phi) &=&
\frac{3}{2}a^{2}(\hat{k}_{i}\hat{k}_{j}-\frac{1}{3}\delta_{ij})(\delta\tilde{T}^{i}_{j})
\label{GRTT} \\ 
\nonumber &=& (c_{1}+c_{3})k^{2}[F_{K}(2{\cal H}V+V')+F_{KK}K_0'V] .
\end{eqnarray}
We then find the following expression for the perturbed field equations:
\begin{eqnarray}
\label{eqn:Poisson}
k^{2}\Phi&=&-\frac{1}{2}F_{K}c_{1}k^{2}[V'+\Psi+(3+2\tilde{c}_{3}){\cal H}V]\\
\nonumber && -4\pi
Ga^{2}\sum_{a} \left( \bar{\rho}_{a}\delta_{a}+3(\bar{\rho}_{a}+\bar{P}_{a}){\cal
H}\theta_{a} \right) .
\end{eqnarray}

Before we look at the cosmological consequences of these theories, and
constraints that can be imposed on them, it is instructive to study
the effect of the vector field during matter domination.  This
should allow us some insight into how the growth of structure proceeds
in the generalised Einstein-{\ae}ther case.  First let us consider the
simplest case in which the dominant contribution to the energy density
is baryonic, so that we can treat it as a pressureless perfect fluid.
Let us also introduce the new variable $V' \equiv E $.  For
illustrative ease we will initially consider only the case where $V$
is described by a growing monomial, such that
$V=V_{0}\left(\frac{\tau}{\tau_{0}}\right)^{p}$,
During the matter dominated era we then have $ a^{2}\delta
T^{0}_{\phantom{0}0} \simeq -l_{E}\xi(k)k^{2}\tau^{5+p-6n} \label{too}
$ and $k^{2}(\Psi-\Phi) \simeq -l_{S}\xi(k)k^{2}\tau^{5+p-6n} $, where
$l_{E} \equiv -(c_{1}(2+p)n+2\alpha(1-2n)n)$, $l_{S} \equiv
-(c_{1}+c_{3})n(6n-p-10)$, and 
\be
\xi(k) \sim \gamma
V_{0}(k)\left(\frac{1}{\tau_{0}}\right)^{p}k_{hub}^{6-6n}\left(3
\alpha\Omega_{m}\left(\frac{H_{0}}{M}\right)^{2}\right)^{n-1},
\label{xixi}
\ee
where $k_{hub}\equiv 1/\tau_{today}$. Hence, the vector field affects
the evolution equations for the matter and metric perturbations only
through its contribution to the energy density and anisotropic stress. 
On large scales, $k \tau \ll1$, and assuming adiabatic initial
conditions for the fields $\delta,\Phi,\theta$, this leads to 
$\delta= C_{1}(k)+\frac{6l_{S}\xi(k)}{(10+p-6n)}\tau^{5+p-6n}$,
where $C_{1}$ is a constant of integration and we have omitted the decaying mode.
Therefore, even before horizon crossing, the anisotropic stress term
due to the vector field can influence time evolution of the baryon
density contrast.  On small scales, $k\tau \gg1$, we find
$\delta(k,\tau) = C_{2}(k)\tau^{2}
+\frac{(\frac{1}{2}l_{E}+l_{S})}{(5+p-6n)(10+p-6n)}\xi(k)(k\tau)^{2}\tau^{5+p-6n}$,  
where $C_{2}(k)$ is another constant of integration. Hence, for
sub-horizon modes, the influence of the vector field on the evolution
of $\delta$ is a combination of its affect on the energy density and
anisotropic stress contributions, though both, in this limit, result
in the same contributions to the scale dependence and time evolution
of the density contrast.

\subsubsection{Observations and constraints}

Let us now consider the constraints that can be imposed on these
theories.  First of all we will consider the linear Einstein-{\ae}ther
theory, and then we will consider the generalised Einstein-{\ae}ther models.

In the case of the linear Einstein-{\ae}ther theory, a number of non-cosmological
constraints on the $c_i$ have been derived:
Most notably, a Parameterised Post-Newtonian (PPN) analysis
of the theory leads to a reduction in the dimensionality
of parameter space.  This is occurs due to the requirement that $c_2$
and $c_4$ must be expressed in terms of the other two parameters in
the theory as $c_2=(-2c^2_1-c_1c_3+ c^2_3)/3c_1$ and $c_4=-c^2_3/c_1$.
Additionally, the squared
speeds of the gravitational and {\ae}ther waves with respect to
the preferred frame must be greater than unity, so as to
prevent the generation of vacuum \v{C}erenkov radiation by
cosmic rays. 
A final constraint arises from considering the effects of the
{\ae}ther on the damping rate of binary pulsars. The rate of
energy loss in such systems by gravitational radiation
agrees with the prediction of General Relativity to one
part in $10^3$. In the case of the Einstein-{\ae}ther
theory it has been shown that to agree with General Relativity in these
systems we must require that $c_+\equiv c_1+c_3$ and $c_-\equiv c_1-c_3$
are related by an algebraic constraint. A more exotic, but viable, subset of the
parameter space can be considered if we set $c_1=c_3=0$.
The PPN and pulsar constraints are then no longer applicable, and a
cosmological analysis is potentially the only way of constraining
the values of the coupling constants. 
\begin{figure}[htbp]
\begin{center}
\epsfig{file=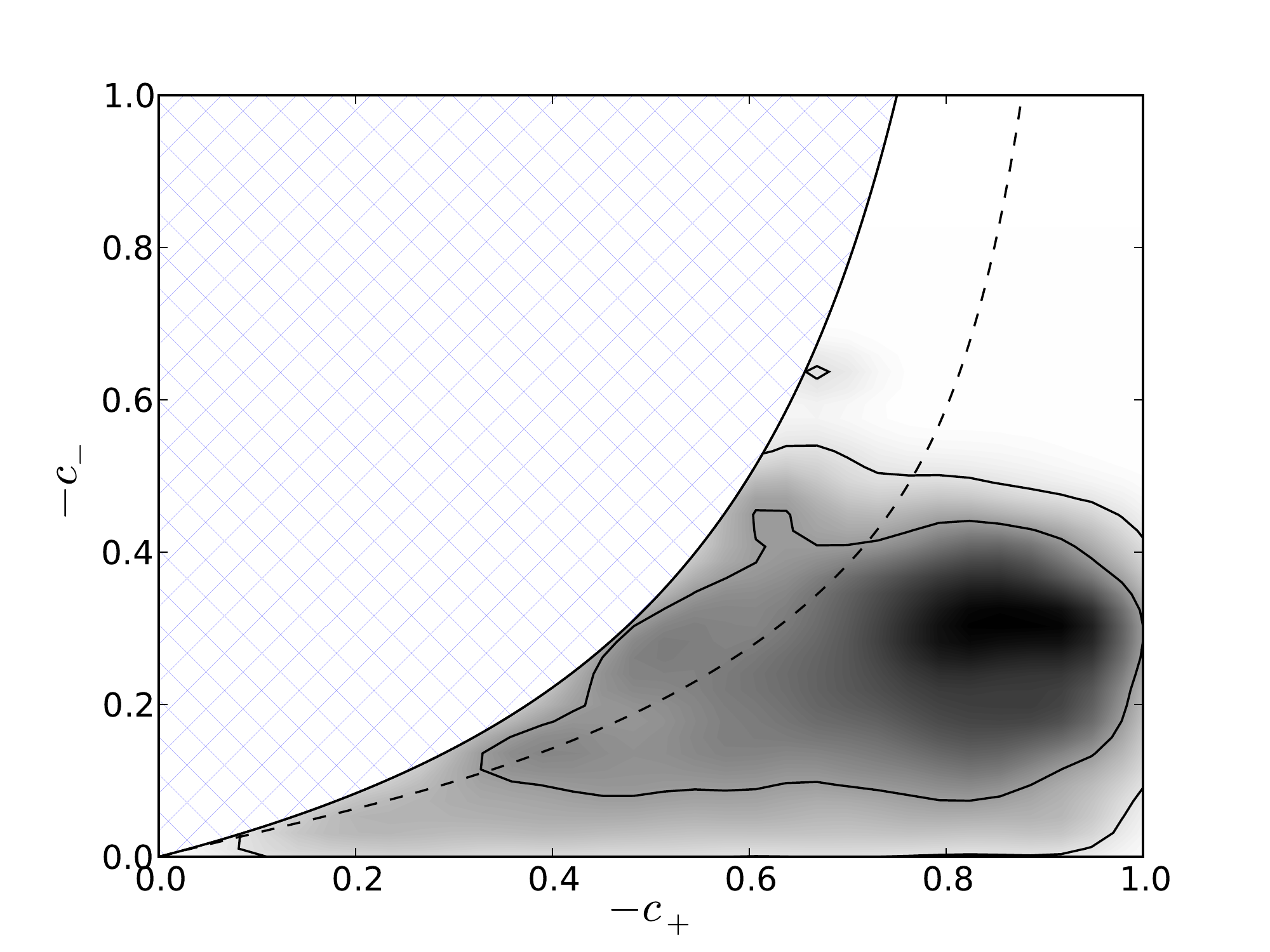,scale=0.5}
\caption{Likelihood plot in the parameter space of $-c_+$ and $-c_-$
from observations of the CMB and large-scale structure.
The black lines are the $1$ and $2\sigma$ contours, for which we have
marginalised over the values of the other parameters.  The hatched
region is excluded by \v{C}erenkov constraints.  The dashed line
indicates the constraints available from binary pulsars.}
\label{linEA2}
\end{center}
\end{figure}

Using a combination of CMB and large-scale structure data
 \cite{ZuntzFerreiraZlosnik2008} it is possible to impose constraints
 on the coefficients of the theory, $c_i$, as well as the overall
 energy density in the {\ae}ther 
field.  The main effect on the evolution of perturbations is 
through the change in the background evolution, and not
necessarily through the presence of perturbations in the
vector field. Indeed, artificially switching off the perturbations
in the {\ae}ther field has essentially no effect on the
power spectrum of large-scale structure, and a small effect (of approximately
10$\%$) on the angular power spectrum of the CMB.  In Figure \ref{linEA2}
we plot the join constraints on $c_+$ and $c_-$ that can be imposed
 from these observables.

If we consider the generalised Einstein-{\ae}ther theory, we find that the effect on
the CMB is much more pronounced. Let us first consider a Universe with
no dark matter, and in which the perturbations in the {\ae}ther field
simultaneously mimic a perturbed pressureless fluid in the formation
of large scale structure, whilst behaving entirely differently in the
cosmological background  \cite{ZuntzFerreiraZlosnik2010}.
The first requirement for successful perturbation evolution is that
structure can form at all.  One necessary condition for this is that
the sound speed of the structure seed is not too large, since this
would wash out structure.  It is therefore required that the sound
horizon in the models  we are considering should be less than the
smallest scales where linear structure can form: $C_S k_{\mathrm{max}}
\tau  \lesssim 1$, where $k_\mathrm{max} \sim 0.2 h/M\mathrm{pc}$.
For matter power observations at $\tau \sim 3\times 10^{4}$, which is
the present epoch, this yields $C_S \lesssim 10^{-4}$.

There are two underlying physical processes that can constrain these models.
The first is a change in the rate of growth of the amplitude of
perturbations.  This can cause discrepancies between the amplitudes we
expect in the matter power spectrum and the CMB, since the evolution
between the two is different.  It can also lead to an integrated
Sachs-Wolfe effect during the matter era, as $\Phi$ becomes time dependent.
The second process is due to the increased magnitude $\Phi-\Psi$.
This also leads directly to a non-negligible integrated Sachs-Wolfe
effect in the matter dominated era.  The details of each of these
processes depends on the functional form of $F$, the time-dependence
of the $\xi$ growing mode, and the choice of the parameters
$c_{i}$. It is extremely challenging to find combinations of the
parameters that allows for a realistic growth of structure, while
simultaneously ensuring the integrated Sachs-Wolfe effect is
acceptably small.

A consequence of these two effects is that it is impossible to find
models where the {\ae}ther field replaces the dark matter that fit the
available cosmological data.  This is not due to the matter power
spectra, which can be reasonably fitted to the SDSS data, but from the
CMB.  In the low-$\ell$ regime a large ISW effect is clearly present,
destroying any chance of fitting the CMB data at large scales.  The
positions of the peaks are also poorly fit by the model. Finally, to
fit the matter power spectrum to the data requires rescaled by a factor $0.02$,
which corresponds to a galaxy bias of $0.14$.  Such a scaling is
considered to be improbably small, on physical grounds.    All these
effects cause severe problems when attempting to simultaneously fit
the CMB and large-scale structure.

Finally, let us consider the possibility of late-time accelerating
expansion.  A detailed comparison with the data seems to allow a range
of values for the index $n$, and the three coupling terms of the
theory, which can produce this behaviour.  In the limit
$n_{\mathrm{ae}}\rightarrow 0$, however, the {\ae}ther field behaves
exactly as a cosmological constant term.

\subsection{Bimetric Theories}
\label{bigravitysection}

In this section we will consider theories that involve two rank-2
tensors.  These are often referred to as ``bimetric'', or ``tensor-tensor'',
theories of gravity. The first formulation of a bimetric theory
appears to be due to Rosen~\cite{Rosen1940,Rosen1973}, and involves
the addition of an extra non-dynamical rank-2 tensor into the theory.
Rosen's theory, however, is now known to lead to the existence of
states that are unbounded from below in their energy. As a result
Rosen's theory predicts the spin {\it up} of pulsars, as gravitational
waves with negative energy are emitted.  This severely violates
the constraints on these systems that have been imposed by observations of
millisecond pulsars \cite{Will1976}.

Following in Rosen's footsteps, there were a number of proposals over
the years of how one could formulate a viable bimetric theory of
gravity.  Here we highlight what we consider to be some of the most
interesting cases. These include Drummond's bimetric (or
``bi-vierbein") theory, which is claimed to mimic the dark
matter in spiral galaxies  \cite{Drummond2001}, as well
as arguments by Magueijo that bimetric theories could exhibit a
variable speed of light, thus providing a way to  model time-varying
fundamental constants. More recently, Ba\~{n}ados and collaborators have
shown that a general form of bigravity, which includes specific forms
previously proposed in\cite{Isham1973,DamourKogan2002} might allow one
to account for some aspects of the dark sector \cite{BanadosEtAl2008,Banados2009}.
Finally, Milgrom has recently proposed a bimetric theory that reduces
to MOND in the appropriate limits. In what we follows, we will briefly
outline each of these theories.


The basic idea behind bimetric, or tensor-tensor, theories is the
introduction of a second `metric' tensor into the theory\footnote{A second
rank-2 tensor would probably be a more accurate description of what is
actually being added here, as the term `metric' implies a
particular geometric function.  Nevertheless, the term `metric' for
this additional field is commonly used, and so we follow this
convention here.}: a dynamical metric, $g_{\mu\nu}$, and a second
metric, ${\tilde g}_{\alpha\beta}$.  The first of these is usually
universally coupled to the matter fields, and is used to construct the
energy-momentum tensor of the non-gravitational fields.  It is this
field is used to define the geodesic equations of test particles.  The
equations that govern $g_{\mu \nu}$, however, are {\it not} the
Einstein field equations:  They invariably involve ${\tilde g}_{\alpha\beta}$ as well.

If the ${\tilde g}_{\alpha\beta}$ is not dynamical, then it is usually
taken to be highly symmetric (i.e. exhibiting the maximal 10 Killing
vectors $X$, such that $\mathcal{L}_X {\tilde g}_{\alpha\beta} =0$).
An obvious choice for ${\tilde g}_{\alpha\beta}$ is the Minkowski
metric, $\eta_{\mu \nu}$, so that all components of the Riemann tensor
constructed from ${\tilde g}_{\alpha\beta}$ vanish.  Rosen's bimetric
theory is a particular example of such a construction, as are some
attempts to construct a massive theory of gravity.  If ${\tilde
g}_{\alpha\beta}$ is to be dynamical, then a kinetic term of the
Einstein-Hilbert form is required in the gravitational action.
Coupling terms are then also required between ${\tilde g}_{\alpha\beta}$
and $g_{\mu\nu}$, with the matter fields usually coupling to either
one, or a combination of both, metrics.

\subsubsection{Rosen's theory, and non-dynamical metrics}

As advertised, Rosen's bimetric theory is constructed with an extra
flat metric, ${\tilde g}_{\alpha\beta}=\eta_{\alpha \beta}$, such
that\footnote{Note that this equation can be derived from an action
principle by including a space-time dependent rank-2 tensor as a
Lagrange multiplier.  We will not go into the details of how to do
this here.}
\begin{eqnarray}
{\tilde R}^\mu_{\phantom{\mu}\nu\rho\sigma} ({\tilde g}_{\alpha\beta})=0.
\end{eqnarray}
We can now define a covariant derivative in terms of ${\tilde
g}_{\alpha\beta}$, which we will call ${\tilde \nabla}_\mu$, such that the
field equations for the dynamical metric can be written
\begin{eqnarray}
\frac{1}{2}{\tilde g}^{\alpha\beta}{\tilde \nabla}_\alpha {\tilde
\nabla}_\beta g_{\mu \nu}
-\frac{1}{2}{\tilde g}^{\alpha\beta}g^{\delta\epsilon}
{\tilde \nabla}_\alpha g_{\delta\mu} {\tilde \nabla}_\beta g_{\epsilon\nu}
=-8\pi G\frac{\sqrt{-g}}{\sqrt{-{\tilde
g}}}(T_{\mu\nu}-\frac{1}{2}g_{\mu\nu}g^{\alpha\beta}T_{\alpha\beta})
\end{eqnarray}
The energy-momentum tensor satisfies the conventional conservation
equation $\nabla^{\mu}T_{\mu\nu}=0$. 

Rosen's theory has been the subject of a number studies over the
years. It has been found to be extremely successful when subjected to a PPN analysis, and
compared to Solar System observations \cite{Will1976}. In fact, almost
all the PPN parameters in Rosen's theory are indistinguishable from
those of General Relativity.  The only exception is
$\alpha_2=v_g^2/c^2-1$, where $v_g$ is the speed of gravitational
waves in Rosen's theorem, and $c$ is the speed of light.  One should
note here, however, that $v_g$ is not uniquely determined by the
theory, but rather by the cosmological solutions to the theory.  One
can then adjust the initial conditions of the Universe in
order to tune $\alpha_2$.  If this is done then the theory is
observationally indistinguishable from General Relativity in the weak
field, low velocity regime of post-Newtonian gravitational physics.

Rosen's theory fails, however, when its predictions for the emission
of gravitational waves are compared to observations of binary
pulsars. Will and Eardley  found that unless the binary system under consideration obeys
very specific properties, in terms of masses and mass differences, then
Rosen's theory leads to the emission of a large amount of dipole
gravitational radiation \cite{WillEardley1977}.  This in turn results
in a sizeable {\it increase} in the orbital period of the system,
which is not observed.  Binary pulsar observations are therefore
incompatible with this theory.  Rosen later proposed replacing flat
space metric by an {\it a priori} specified, but time-varying,
cosmological background \cite{Rosen1978}.
Unfortunately this does not circumvent the pulsar problem.

Other bimetric theories that also have been proposed with an
additional {\it a priori} specified, non-dynamical metric field.
These include Rastall's theory \cite{Rastall76,Rastall77} and
Lightman and Lee's theory \cite{LightLee73}, for which the PPN limits
of both theories are known.  It has been conjectured by Will, however,
that all such theories that incorporate prior specified geometry could
suffer the same deficiency as Rosen's, when it comes to calculating
the emission of gravitational radiation from binary systems \cite{tegp}.

\subsubsection{Drummond's theory}

Let us now turn to a more recent formulation of the bimetric
theory. In \cite{Drummond2001} it was proposed to work in the vierbein
formulation with $g_{\alpha\beta}=\eta_{AB}e^A_\alpha e^B_\beta$ and
${\tilde g}_{\alpha \beta}=\eta_{{\tilde A}{\tilde B}}{\tilde
e}^{\tilde A}_{\alpha} {\tilde e}^{\tilde B}_{\beta}$. In this case
{\it both} sets of vierbein are dynamical. While ${\tilde e}^{\tilde
A}_{\alpha}$ is used to construct the Einstein-Hilbert action,
$e^B_\beta$ is used to construct the action from which the
energy-momentum tensor is derived. The missing pieces of the theory
are then a transformation tensor, $M^{\tilde A}_{B}$, and a scalar,
$\phi$, which relate ${\tilde e}^{\tilde A}_\mu$ and $e^B_\mu$ by
\begin{eqnarray}
{\tilde e}^{\tilde A}_\mu= e^\phi M^{\tilde A}_{\phantom{\tilde A}B} e^B_\mu.
\end{eqnarray}
Finally, we then need to define a ``linking action''. This is given by
Drummond as
\begin{eqnarray}
S_{L}&=&\frac{1}{16 \pi G_1}\int d^4 x\sqrt{-{\tilde g}}{\tilde
g}^{\mu\nu}{\rm Tr}(j_\mu j_\nu) \nonumber \\ 
& &+\frac{1}{16 \pi G_2}\int d^4 x\sqrt{-{\tilde g}}{\tilde
g}^{\mu\nu}(\partial_\mu\phi\partial_\nu\phi) \nonumber \\ 
& &+\frac{1}{16\pi G_1}\int d^4x \sqrt{-{\tilde g}}\frac{m^2}{4}\left(
M^{\tilde A}_AM_{\tilde A}^A+M_{\tilde A}^AM^{\tilde
A}_A-\gamma\right) \nonumber \\ & & -\frac{1}{16 \pi G_2}\int d^4x
\sqrt{-{\tilde g}}m^2\phi^2, 
\end{eqnarray}
where the current $j_\mu$ is defined as
\begin{eqnarray}
j^{{\tilde A}{\tilde B}}_\mu \equiv {\tilde g}^{{\tilde B}{\tilde
C}}[{\tilde \nabla}_\mu  M^{\tilde A}_{\phantom{\tilde A}B}]
M^B_{\phantom{A}\tilde C}, 
\end{eqnarray}
and where $G_1$ and $G_2$ are new gravitational constants, $m$ is a
mass parameter, and $\gamma$ is a free parameter which corresponds to
the cosmological constant. Note that the action for $M$ is similar to
that of the non-linear sigma model found in meson physics.

Drummond has shown that his bimetric theory has a well defined
Newtonian limit and so, in principle, can satisfy the time delay
measurements from radio signals. He also claims that the higher order
correction is exactly what is needed to satisfy Solar System
constraints from the precession of the orbit of Mercury, and that in
the weak field limit the dynamical metric $g_{\alpha\beta}\simeq
\eta_{\alpha\beta}+h_{\alpha\beta}$ gives rise to a potential of the
form 
\begin{eqnarray}
h_{00}=-\frac{GM}{r}\left(1+\frac{G_1}{G}e^{-mr}\right).
\end{eqnarray}
Hence, for $mr\ll 1$ the effective Newton's constant is $G_N=G+G_1$, while for
large scales $G_N\simeq G$. Such a correction can alleviate the problem of
flat galactic rotation curves that arises in standard Newtonian
gravity with no dark matter, but does not completely resolve
it. Albeit an intriguing proposal for a theory of modified gravity,
there has been little progress in studying the various astrophysical
and cosmological consequences of Drummond's theory. 

\subsubsection{Massive gravity}
\label{sec:massive} 

The theory of a single massive spin-2 field can also be considered as
a bimetric theory, with a non-dynamical background metric $\tilde
g_{\alpha\beta}$ and a dynamical fluctuation given by $
g_{\alpha\beta}=\tilde g_{\alpha\beta}+h_{\alpha\beta}$. Taking the
background to be Minkowski space, for simplicity, we can then generate
a mass for the spin-2 field $h_{\alpha\beta}$ by  adding the
Pauli-Fierz (PF) term to the Einstein-Hilbert action \cite{Fierz},
resulting in 
\be
S_{PF}=\frac{1}{16\pi G} \int d^4x \sqrt{-g}R+\frac{m^2}{4} \sqrt{-g}
\left[g^{\mu\nu}g^{\alpha\beta}-g^{\mu\alpha} g^{\mu\beta}\right]
h_{\mu\nu} h_{\alpha\beta}, \label{PF}
\ee
where $m$ is a constant mass parameter.  It is well known that, in
four dimensions, a massive spin-2 field  ought to have five
propagating degrees of freedom: Two of  helicity 2, two of helicity 1,
and one of helicity 0.  However, a {\it generic} mass term with
arbitrary coefficients will result in higher derivative terms for the
helicity-0 mode,  giving rise to an additional ghost-like degree of
freedom. The form of the PF mass term is specifically chosen so that
this is not the case to  linear order.  Massive gravity exhibits some
interesting phenomenology, not least the so-called vDVZ discontinuity,
and its possible resolution via the Vainshtein mechanism. These will
be discussed in more detail in the context of DGP gravity in Section
\ref{sec:dgp}.

Unfortunately, the PF Lagrangian by itself cannot describe a
consistent theory because the ghost-like mode reappears at non-linear
order \cite{BDghost}. This mode is often referred to as the
Boulware-Deser ghost, and it was believed that one could not  find
generalisations of the  theory that succeeded in eliminating it to all
orders \cite{Creminelli-ghosts}.  There has, however, been some recent
progress on this issue by de Rham and Gabadadze and collaborators
\cite{Gabadadze-aux, deRham-massive, deRham-self, deRham-gen,
deRham-resum} who have  proposed the following action
\cite{deRham-gen, deRham-resum}:
\be
S_{GPF}=\frac{1}{16\pi G} \int d^4x \sqrt{-g}R+m^2\sqrt{-g} U(g, h),
\ee
where $U(g, h)=\sum_{n=2}^4 a_m \delta^{\mu_1}_{[\nu_1} \ldots
\delta^{\mu_m}_{\nu_m]} K^{\mu_1}_{\nu_1} \cdots K^{\mu_m}_{\nu_m}$
and where the $a_m$ are constants, and
\be
K^\mu_\nu=\sqrt{\delta^\mu_\nu+g^{\mu \alpha}h_{\alpha
\nu}}-\delta^\mu_\nu=\frac{1}{2}g^{\mu \alpha}h_{\alpha
\nu}-\frac{1}{8}g^{\mu \alpha}h_{\alpha \beta } g^{\beta
\gamma}h_{\gamma \nu }+\ldots.
\ee
It is now clear that the leading order part of the potential gives the
PF mass term upon choosing $a_2=2$. To study the behaviour of the
theory beyond linear order it is convenient to restore general
coordinate invariance by means of the Stuckelberg trick.  To this end
one can perform the following field redefinition \cite{deRham-gen,
ArkaniHamed-effective}, 
\ba
h_{\mu\nu} &=& \frac{\hat
h_{\mu\nu}}{M_{pl}}+\eta_{\mu\nu}-\eta_{\alpha \beta}\frac{\partial
\phi^\alpha}{\partial x^\mu}\frac{\partial \phi^\beta}{\partial x^\nu}
\label{st1}\\ 
&=& \frac{\hat h_{\mu\nu}}{M_{pl}}+2\frac{\del_{\mu}\del_{\nu} \pi
}{\Lambda_3^3}-\frac{\eta^{\alpha\beta}
\partial_{\alpha}\del_{\mu}\pi\partial_{\beta}\del_{\nu}
\pi}{\Lambda_3^6},  \label{st2}
\ea
where in going from Eq. (\ref{st1}) to Eq. (\ref{st2}) we have set
$\phi^\alpha=x^\alpha-\eta^{\alpha \beta} \del_\beta\pi/\Lambda_3$, in
order to focus on the dynamics of the helicity zero mode. Note that
$\Lambda_3=(M_{pl}m^2)^{1/3}$ and $M_{pl}=1/\sqrt{8\pi G}$. For the
original PF action, Eq. (\ref{PF}), the Boulware-Deser ghost reveals
itself by expanding the action in terms of $\hat h_{\mu\nu}$. At
zeroth order  one finds higher derivative terms for $\pi$ that
contribute to the equations of motion, indicating the presence of the
ghostly extra mode. In contrast, the generalised PF action is chosen
so that the resulting higher derivative terms contribute a total
derivative at zeroth order in $\hat h_{\mu\nu}$. This is a crucial
first step in avoiding the extra mode. One can go further and study
the theory in the so-called decoupling limit, $m\to 0, M_{pl} \to
\infty$, and $\Lambda_3=$constant. After some suitable field
redefinitions one finds that the theory contains the quintic galileon
Lagrangian \cite{deRham-gen}.  Note that it does not reproduce the
galileon theory discussed in Section \ref{galileons} exactly, since
generically there is  mixing with a graviton of the form $\hat  h^{\mu\nu}
X^{3}_{\mu\nu}$, where   $X^{3}_{\mu\nu}$ is cubic in $\pi$. This
mixing cannot be eliminated by a local field redefinition and may have
important phenomenological consequences. In particular, when this coupling comes in with a particular sign it can prevent the recovery of GR inside the Vainshtein radius around a heavy source \cite{Koyama-anal, Koyama-strong, Chka}. In any event, one can confidently say that the
Boulware-Deser ghost does not appear in the decoupling limit. Of
course, it is possible that this limit corresponds to taking its mass
to infinity, and that it will reemerge in the full theory. Whether or
not this is the case has yet to be established. 

Self-accelerating and self-tuning cosmologies were studied de Rham and
Gabadadze's theory in \cite{deRham-cosmic}, whilst spherically
symmetric solutions have also been considered recently  \cite{Koyama-anal, Koyama-strong, Chka}.

\subsubsection{Bigravity}

A class of theories that were first proposed in the 1970s by Isham,
Salam, Stradthee \cite{Isham1973}, and revisited a few years ago by
Kogan \cite{DamourKogan2002} and collaborators, have recently been
resurrected by Ba\~{n}ados and collaborators
\cite{Banados2008,BanadosEtAl2008,Banados2009} (see
for~\cite{SkordisProcEBI} a short overview).  
Further studies of bi-gravity include weak-field solutions and gravitational waves \cite{Ber2008a}, exact
spherically symmetric solutions \cite{Ber2008b} and the energy of
black holes \cite{Com2011}.
The starting point~\cite{Banados2008} is an extension of Eddington's affine theory (see section \ref{GR})  so that
the dynamical fields are a metric $g_{\mu\nu}$ (with curvature scalar $R$) and a connection $C^\alpha_{\mu\nu}$ with Ricci tensor $K_{\mu\nu}[C]$.
The action is
\be
S[g,C] = \frac{1}{16\pi G} \int d^4x \sqrt{-g} (R - 2 \Lambda) + \frac{2}{\alpha \ell^2}\sqrt{- det[g_{\mu\nu} - \ell^2 K_{\mu\nu}]} 
\ee
where $\Lambda$ is a cosmological constant, $\alpha$ is a dimensionless parameter and $\ell$ is a length scale.
It may be shown~\cite{BanadosEtAl2008} by introducing a 2nd metric $\tilde{g}_{\mu\nu}$ corresponding to the connection $C^\alpha_{\mu\nu}$ that
the above theory is a special case of bigravity with action given by
\begin{eqnarray}
S=\frac{1}{16 \pi G}\int d^4 x\left [\sqrt{-g}(R-2\Lambda)+\sqrt{-{\tilde g}}
({\tilde R}-2{\tilde \Lambda})-\sqrt{-{\tilde
g}}\frac{1}{\ell^2}({\tilde
g}^{-1})^{\alpha\beta}g_{\alpha\beta}\right ], 
\end{eqnarray}
where ${\tilde \Lambda} = \frac{\alpha}{\ell^2}$ is a cosmological constant term.
In these theories, {\it both} metrics are used to build Einstein-Hilbert
actions even though only one of them couples to the matter content.

Such bigravity theories lead to interesting behaviour on cosmological
scales~\cite{BanadosEtAl2008,Banados2009}. The homogeneous and isotropic FLRW metrics can be written as
$g_{\alpha\beta}=diag(-1,a^2,a^2,a^2)$ and 
${\tilde g}_{\alpha\beta}=diag(-X^2,Y^2,Y^2,Y^2)$, where $X$ and $Y$
are functions of $t$ alone. The corresponding Friedmann equations are
then of the form
\begin{eqnarray}
H^2=\frac{8\pi G}{3}({\tilde \rho}+\rho),
\end{eqnarray}
where ${\tilde \rho}=Y^3/(8\pi G \ell^2 X a^3)$. This fluid satisfies
a conventional conservation equation of the form 
\be
\frac{d{\tilde \rho}}{dt}=-3(1+{\tilde w}){\tilde \rho},
\ee
where ${\tilde w}$, satisfies a somewhat intricate evolution equation, given by
\begin{eqnarray}
\frac{d{\tilde w}}{dt}=2{\tilde w}\left[1+3{\tilde w}+\sqrt{4(-{\tilde
w})^{3/2}{\tilde \Omega}\alpha-2\frac{(1+3{\tilde
w})\rho_{\ell}}{\rho_c}}\right],
\end{eqnarray}
where $\rho_c=\rho+{\tilde \rho}$, $\rho_\ell = (8\pi G \ell^2)^{-1}$ and ${\tilde \Omega}={\tilde \rho}/\rho_c$.
The extra metric here can lead to a range of interesting behaviours
and, in particular, can drive the expansion to a de Sitter phase, or
mimic the effects of dark matter. Anisotropic universes in these models were studied in~\cite{Rodrigues2008}.

The cosmological evolution of perturbations in these theories has been worked out in
some detail, and it turns out that the perturbations in the auxiliary
field can be rewritten in the form of a generalised dark matter fluid~\cite{Hu1998}
with density, momentum, pressure and shear that obey evolution equations. As
a result, it is possible to work out cosmological observables such as
CMB anisotropies and large-scale structure. In \cite{Banados2009} it
was found that distinctive 
signatures emerge during periods of accelerated expansion.
If the ${\tilde \rho}$ field dominates, and is responsible for cosmic
acceleration, there is a clear instability in the gravitational
potentials; they not only grow but diverge  leading
very rapidly to an overwhelming integrated Sachs-Wolfe
effect on large scales. It is difficult to reconcile the angular
power spectrum of fluctuations and the power spectrum of
the galaxy distribution predicted by a bimetric theory that
unifies the dark sector with current data. If we restrict
ourselves to a regime in which ${\tilde \rho}$ simply behaves
as dark matter, however, then the best-fit bimetric
model is entirely indistinguishable from the standard CDM scenario.

Bigravity theory can also be extended to consider more complicated actions,
such as
\begin{eqnarray}
S_L=-\frac{1}{16\pi G}\int d^4 x \sqrt{-{\tilde
g}}\left[\kappa_0({\tilde
g}^{-1})^{\alpha\beta}g_{\alpha\beta}+\kappa_1(({\tilde
g}^{-1})^{\alpha\beta}g_{\alpha\beta})^2
+ \kappa_2({\tilde g}^{-1})^{\alpha\beta}({\tilde g}^{-1})_{\alpha\beta}\right],
\end{eqnarray}
and, although a full analysis of its PPN parameters has been
undertaken \cite{CliftonBanadosSkordis2010}, its cosmology remains to
be explored.  Black holes, and their thermodynamics, have been studied
in bimetric gravity in \cite{fextra2}.

\subsubsection{Bimetric MOND}

A bimetric theory of MOND, somewhat akin to bigravity, has recently
been proposed by Milgrom \cite{Milgrom2008}. 
The action for bimetric MOND, or BIMOND, is of the form
\begin{eqnarray}
S&=&\frac{1}{16\pi G}\int d^4x \left[ \beta \sqrt{-g}R+\alpha \sqrt{-\tilde g}{\tilde R} 
-2({\tilde g}g)^{1/4}a_0^2{\cal M}\right] 
\nonumber \\
& & - {\tilde S}_M({\tilde g}_{\mu\nu},{\tilde \psi})- {S}_M({g}_{\mu\nu},{\psi})
\end{eqnarray}
where ${\cal M}$ is the interaction term that connects the two metrics, and $\psi$ and
${\tilde \psi}$ are the matter fields that couple to $g_{\mu\nu}$ and
${\tilde g}_{\mu\nu}$, respectively.  The factor ${\cal M}$ is a 
non-linear function of the tensor
$\Upsilon_{\mu \nu}$, given by
\begin{eqnarray}
\Upsilon_{\mu\nu}=C^{\alpha}_{\phantom{\alpha}\mu\beta}
C^{\beta}_{\phantom{\beta}\nu\alpha}-C^{\alpha}_{\phantom{\alpha}\mu\nu}
C^{\alpha}_{\phantom{\alpha}\alpha\alpha},   
\end{eqnarray}
where
\begin{eqnarray}
C^{\alpha}_{\phantom{\alpha}\mu\nu}={\tilde
\Gamma}^{\alpha}_{\phantom{\alpha}\mu\nu}-\Gamma^{\alpha}_{\phantom{\alpha}\mu\nu},
\end{eqnarray}
and $\Gamma^{\alpha}_{\phantom{\alpha}\mu\nu}$ and
${\tilde \Gamma}^{\alpha}_{\phantom{\alpha}\mu\nu}$ are the Christoffel symbols
constructed from $g_{\mu\nu}$ and ${\tilde g}_{\mu\nu}$, respectively.
Note that the even though the $\Gamma$s are not tensors, $C$,
constructed in this way, is a tensor. The constants $\alpha$ and
$\beta$ can be kept unrelated, leading to different gravitational
couplings in the two sectors. If we set $\alpha=\beta$, however, but
leave $\beta$ arbitrary, then we get the field equations
\be
\beta G_{\mu\nu}+S_{\mu\nu}=-8\pi G T_{\mu\nu}
\ee
and
\be
\beta {\tilde G}_{\mu\nu}+{\tilde S}_{\mu\nu}=-8\pi G {\tilde T}_{\mu\nu},
\ee
which look like the conventional Einstein equations, except for the
contributions from $S_{\mu\nu}$ and ${\tilde S}_{\mu\nu}$ which contain
the interaction terms between the two metrics. These tensors are
quadratic in $C^{\alpha}_{\phantom{\alpha}\mu\nu}$, and are non-linear
functions of $g$, ${\tilde g}$ and $g^{\mu\nu}{\tilde g}_{\mu\nu}$.

This theory has been constructed to reproduce MOND phenomenology on
small scales, in the weak field and low acceleration regime. Its
cosmological implications have been studied in 
\cite{CliftonZlosnik2010,MilBiPert}, where it was shown that
in the high acceleration regime BIMOND reproduces conventional FLRW
behaviour. In low acceleration regime, however, we have that the scale
factor $a(\tau)$ (where $\tau$ is conformal time) can take the form
the form $a\simeq \tau^p$,  where \cite{CliftonZlosnik2010}
\begin{eqnarray}
p=\frac{1-3{\tilde w}}{(1+2w+8nw-3{\tilde w}-8n{\tilde w}-6w{\tilde w})}.
\end{eqnarray}
and where $w$ and ${\tilde w}$ are the equations of state for the
matter coupled to $g_{\mu\nu}$ and ${\tilde g}_{\mu\nu}$, and $n$ is
one of the parameters in ${\cal M}$. This leads to an interesting
range of behaviours. For example, it is possible to have a dust filled
universe that is static, if the matter coupled to the second metric is
radiation.  It is shown in \cite{MilBiPert} that to calculate
fluctuations about an FLRW background in either metric requires a
knowledge of the matter coupled to both metrics.  It is also shown
that the growth of fluctuations does not proceed in a purely Newtonian
way, but has a MOND contribution as well.

\subsection{Tensor-Vector-Scalar Theories}

In General Relativity, the space-time metric $g_{\mu\nu}$ is the sole
dynamical agent of gravity.  We have seen above that scalar-tensor
theories extend this by adding a scalar field that mediates a spin-$0$
gravitational interaction, while in Einstein-{\ae}ther theories one
makes use of a vector field.  TeVeS has both of these types of fields
as extra degrees of freedom: A scalar field, $\phi$, and  a (dual)
vector field, $A_\mu$, both of which participate in the gravitational sector.
Like GR, it obeys the Einstein equivalence principle, but unlike GR it
violates the strong equivalence principle.

TeVeS is a product of past antecedent theories, namely the Aquadratic
Lagrangian theory of gravity (AQUAL) and its relativistic
version~\cite{BekensteinMilgrom1984}, the phase-coupling
gravitation~\cite{Bekenstein1988}, the disformal relativistic scalar
field theory~\cite{BekensteinSanders1994}, and the Sanders' stratified
vector field theory~\cite{Sanders1997}.  Since its
inception~\cite{Bekenstein2004a} TeVeS has been intensively researched,
including studies of
cosmology~\cite{Bekenstein2004a,HaoAkhoury2005,SkordisEtAl2006,Diaz-RiveraSamushiaRatra2006,DodelsonLiguori2006,Skordis2006,BourliotEtAl2006,Zhao2006a,Skordis2008a,FerreiraSkordisZunkel2008},
spherically symmetric solutions~\cite{Bekenstein2004a,Giannios2005,JinLi2006,SagiBekenstein2008,LaskySotaniGiannios2008,SkordisZlosnik2011},
gravitational collapse and stability~\cite{Seifert2007,ContaldiWisemanWithers2008}, 
solar system tests~\cite{Bekenstein2004a,Giannios2005,BekensteinMagueijo2006,Tamaki2008,Sagi2009a},
gravitational
lensing~\cite{Bekenstein2004a,ChiuKoTian2005,ChenZhao2006,ZhaoEtAl2006,Zhao2006b,Chen2007,FeixFedeliBartelmann2007,XuEtAl2007,ShanEtAl2008,ChiuTianKo2008,MavromatosSakellariadouYusaf2009},
issues of superluminality~\cite{Bruneton2006},
and the travel time of gravitational
waves~\cite{KahyaWoodard2007,Kahya2008,DesaiKahyaWoodard2008}. A
thorough and up-to-date review of TeVeS can be found
in~\cite{Skordis2009a}. Here we will concentrate mostly on
cosmological features of the theory.

\subsubsection{Actions and field equations}

The original and most common way to specify TeVeS is to write the
action in a mixed frame. That is, we write the action in the
``Bekenstein frame'' for the gravitational fields, and in the
`physical frame', for the matter fields. In this way we ensure that the
Einstein equivalence principle is obeyed. The three gravitational
fields are the metric, $\metE_{\mu\nu}$ (with connection
$\tilde{\nabla}_a$), that we refer to as the Bekenstein metric, 
the Sanders vector field, $A_\mu$, and the scalar field, $\phi$.
To ensure that the  Einstein equivalence principle is obeyed, we write
the action for all matter fields using a single `physical metric', 
$g_{\mu\nu}$ (with connection $\nabla_\mu$), that we call the
`universally coupled metric'~\footnote{Some work on TeVeS, including
the original articles by Sanders~\cite{Sanders1997} and
Bekenstein~\cite{Bekenstein2004a}, refer to the Bekenstein frame
metric as the ``geometric metric'', and denote it as $g_{\mu\nu}$,
while the universally coupled metric is referred to as the ``physical
metric'', and is denoted by $\tilde{g}_{\mu\nu}$. Since it is more
common to denote the metric which universally couples to matter as
$g_{\mu\nu}$, in this review we interchange the tilde.}. The
universally coupled metric is algebraically defined via a disformal
relation~\cite{Bekenstein1993} as
\begin{equation}
   \metM_{\mu\nu} = e^{-2\phi}\metE_{\mu\nu} - 2\sinh(2\phi)A_\mu A_\nu.
   \label{eq:metric_relation}
\end{equation}
The  vector field is further enforced to be unit-time-like with
respect to the Bekenstein metric, i.e.
\begin{equation}
 \metE^{\mu\nu} A_\mu A_{\nu}= -1.
\label{eq:A_unit}
\end{equation}
The unit-time-like constraint is a phenomenological requirement for
the theory to give an acceptable amount of light bending.  Using the
unit-time-like constraint, Eq. (\ref{eq:A_unit}), it can be shown that
the disformal transformation for the inverse metric is 
\begin{equation}
   \metM^{\mu\nu} = e^{2\phi}\metE^{\mu\nu} + 2\sinh(2\phi)A^{\mu}A^\nu,
   \label{eq:inv_metric_relation}
\end{equation}
where  $A^{\mu}= \metE^{\mu\nu}A_\nu$.  The existence of a scalar and
a vector field may seem odd at first, but they are both the product of
a series of extensions from older theories, based on theoretical and
phenomenological constraints.
\newline
\newline
\noindent
{\it Actions}
\newline

TeVeS is based on an action, $S$, which is split as 
\be
S = S_{\metE} + S_A + S_{\phi}+S_m, 
\ee
where $S_{\metE}$, $S_A$, $S_{\phi}$ and $S_m$ are the actions for
$\metE_{\mu\nu}$, the vector field, $A_\mu$, the scalar field, $\phi$,
and matter fields, respectively. 

As already discussed, the action for  $\metE_{\mu\nu}$, $A_\mu$, and
$\phi$ is written using only the Bekenstein metric, $\metE_{\mu\nu}$,
and not $\metM_{\mu\nu}$, and is such that $S_{\metE}$ is of
Einstein-Hilbert form
\begin{equation}
   S_{\metE} = \frac{1}{16\pi G}\int d^4x \; \volE \; \RiemE,
\label{eq:S_EH}
\end{equation}
where $\metE$ and $\RiemE$ are 
the determinant and scalar curvature of $\metE_{\mu\nu}$,
respectively, while $G$ 
is the bare gravitational constant. The relation between $G$ and the
measured value of Newton's constant, $G_N$, will be elaborated on
below, in Section \ref{sec_quasistatic}. 

The action for the vector field, $A_\mu$, is given by
\begin{equation}
	S_A = -\frac{1}{32\pi G}  \int d^4x \; \volE \; \left[ K
	F^{\mu\nu}F_{\mu\nu}   - 2\lambda (A_\mu A^{\mu}+ 1)\right],
\end{equation}
where  $F_{\mu\nu} = \nabla_\mu A_{\nu}- \nabla_{\nu}A_\mu$ leads to a
Maxwellian kinetic term, $\lambda$ \label{def_lambda} is a
Lagrange multiplier that ensures the unit-timelike constraint on $A_\mu$, and $K$
is a dimensionless constant.  Indices on $F_{\mu\nu}$ are moved using
the Bekenstein metric, i.e. $F^\mu_{\;\;\nu} = \metE^{\mu\alpha}
F_{\alpha\nu}$. This form of a vector field action has been considered
by Dirac as a way of incorporating electrons into the  electromagnetic
potential~\cite{Dirac1951a,Dirac1951b,Dirac1954}. More recently it has
been considered as a natural generalisation of GR, in the
Einstein-{\ae}ther theories discussed in Section \ref{aesubsection}
\cite{JacobsonMattingly2000,Jacobson2008a}.

The action for the scalar field, $\phi$, is given by
\begin{equation}
    S_{\phi} = -\frac{1}{16\pi G} \int d^4x  \volE \left[ 
    \mu \; \metS^{\mu\nu}\connE_\mu\phi \connE_\nu\phi +   V(\mu) \right],
\end{equation}
where  $\mu$ \label{def_mu} is a non-dynamical dimensionless scalar
field, $\metS^{\mu\nu}$ is a new metric defined by
\begin{equation}
\metS^{\mu\nu} = \metE^{\mu\nu} - A^{\mu}A^\nu,
\end{equation}
and $V(\mu)$ is an arbitrary  function which typically depends on a
scale, $\ell_B$. Not all choices of $V(\mu)$ give the correct
Newtonian or MONDian limits in a quasi-static situation. The allowed
choices are presented in Section \ref{sec_quasistatic}.  The metric
$\metS^{\mu\nu}$ is used in the scalar field action, rather than
$\metE^{\mu\nu}$, to avoid the superluminal propagation of
perturbations. Note that it is possible to write the TeVeS action
using $\metS^{\mu\nu}$, with the consequence of having more general
vector field kinetic terms (see the appendix of~\cite{Skordis2009a}).

It is also easier in some cases to work with an alternative form for
the scalar field action that does not have the non-dynamical field,
$\mu$, but rather has the action written directly in terms of a
non-canonical kinetic term for $\phi$ given by a free function $f(X)$,
with $X$ defined by
\begin{equation}
 X = \ell_B^2 \hat{g}^{\mu\nu}  \nabla_\mu \phi \nabla_{\nu}\phi.
\label{eq_X}
\end{equation}
The field $\mu$ is then given in terms of $f(X)$ by $\mu =
\frac{df}{dX}$, while $f(X)$ can be related to $V$ by $f = \mu X +
\ell_B^2 V$.
 
The matter fields in the action are coupled only to the `universally
coupled metric', $\metM_{\mu\nu}$, and thus their action is of the form 
\begin{equation}
 S_m[\metM,\chi^A, \nabla \chi^A] = \int d^4x \; \volM \; L[ \metM
 ,\chi^A, \nabla \chi^A],
\end{equation}
for some generic collection of matter fields, $\chi^A$. The matter
stress-energy tensor is then defined with respect to $\delta S_m$ in
the usual way.

It should be stressed that the action for the scalar field has been
constructed such that the theory has a MONDian non-relativistic limit,
under the right conditions, for specific choices of functions $V(\mu)$
(or equivalently $F(X)$). The action for the vector field has no
particular significance other than the fact that it is simple.  More
general actions can be considered without destroying the MOND limit,
but that in addition provide new features or improved phenomenology.
\newline
\newline
\noindent
{\it The field equations}
\newline

The field equations of TeVeS are found using a variational principle.
This gives two constraint equations, namely the unit-timelike
constraint, given in Eq. (\ref{eq:A_unit}), and the $\mu$-constraint:
\begin{equation}
      \metS^{\mu\nu}\nabla_\mu\phi \nabla_\nu\phi  = -\Vp,
\label{eq:mu_con} 
\end{equation}
that is used to find $\mu$ as a function of $\nabla_\mu\phi$.
The field equations for $\metE_{\mu\nu}$ are given by
\begin{eqnarray}
   \EinE_{\mu\nu} &=& 8\pi G\left[ T_{\mu\nu} + 2(1 -
  e^{-4\phi})A^{\alpha}T_{\alpha(\mu} A_{\nu)}\right] 
  \nonumber \\ 
&&        
+ \mu \left[ \connE_\mu \phi \connE_\nu\phi  - 2
  A^\alpha\connE_\alpha\phi \; A_{(\mu}\connE_{\nu)}\phi  \right]
        + \frac{1}{2}\left(\mu V' -  V\right) \metE_{\mu\nu}   
  \nonumber \\ 
&&        
+   K\left[F^\alpha_{\;\;\mu} F_{\alpha\nu} - \frac{1}{4} F^{\alpha
  \beta} F_{\alpha \beta} \metE_{\mu\nu}\right] - \lambda A_\mu A_\nu,
\end{eqnarray}
where $\EinE_{\mu\nu}$ is the Einstein tensor \label{def_G}
constructed from $\metE_{\mu\nu}$.
The field equations for the vector field, $A_\mu$, are
\begin{equation}
K \connE_\alpha F^\alpha_{\;\;\mu}
=  -\lambda A_\mu - \mu A^\nu\connE_\nu\phi \connE_\mu\phi  + 8\pi G
(1 - e^{-4\phi})A^{\nu}T_{\nu\mu} \label{eq:A_eq},
\end{equation}
and the field equation for the scalar field, $\phi$, is 
\begin{eqnarray}
   \connE_\mu \left[   \mu \metS^{\mu\nu} \connE_\nu\phi \right] &=&
 8\pi G e^{-2\phi}\left[\metM^{\mu\nu} + 2e^{-2\phi}
 A^{\mu}A^\nu\right] T_{\mu\nu}.
 \label{eq:Phi_eq}
\end{eqnarray}
The Lagrange multiplier can be solved for by contracting
Eq. (\ref{eq:A_eq}) with $A^{\mu}$.

\subsubsection{Newtonian and MOND limits}
\label{sec_quasistatic}

We now describe the quasi-static limit, as relevant for establishing
the existence of the Newtonian and MONDian limits.  The details of the
derivation we consider here can be found in~\cite{Bekenstein2004a},
while an alternative shorter derivation in the spirit of the PPN
formalism is given in~\cite{Skordis2009a}.

It can be shown that the PPN parameter $\gamma$ is unity in TeVeS,
hence the universally coupled metric can be written as
\be
ds^2 = - (1 + 2\Phi)dt^2 + (1 - 2\Phi) \delta_{ij} dx^i dx^j,
\ee
where $\Phi$ is related to the acceleration of particles, $\vec{a}$,
by $\vec{a} =  - \grad \Phi$.  The scalar field is perturbed as $\phi
= \phi_c + \varphi$, where $\phi_c$ is the cosmological value of $\phi$.
The Bekenstein metric takes a similar form to $g_{\mu \nu}$:
\be
d\tilde{s}^2 = - e^{-2\phi_c}(1 + 2\tilde{\Phi})dt^2 + e^{2\phi_c} (1
- 2\tilde{\Phi}) \delta_{ij} dx^i dx^j.
\ee
The vector field does not play a role at this order of perturbations,
and is simply given by 
\be
A_\mu = e^{-\phi_c}(-1-\tilde{\Phi},0,0,0).
\ee
The field equations to $O(v^2)$ are then
\begin{eqnarray}
 \Delta \tilde{\Phi} = \frac{8\pi G}{2-K} \rho,
\label{eq_Phi_tilde}
\end{eqnarray}
and
\begin{eqnarray}
  \grad \cdot \left[  \mu \grad\varphi \right] &=& 8\pi G  \rho,
\label{eq_varphi_aqual}
\end{eqnarray}
while the potential $\Phi$ is given via the disformal transformation
$\Phi = \tilde{\Phi} +\varphi$.  Eqs. (\ref{eq_Phi_tilde}) and
(\ref{eq_varphi_aqual}) can be solved for any quasi-static situation,
regardless of the boundary conditions or the symmetry of the system in
question, provided a function $\mu (|\grad\varphi|)$ is supplied.

To find the Newtonian and MONDian limits we can consider, for
simplicity, spherically symmetric situations. In this case we can
combine Eqs. (\ref{eq_Phi_tilde}) and (\ref{eq_varphi_aqual}) into
a single equation for $\Phi$, called the AQUAL equation:
\begin{equation}
\grad \cdot \left[ \mu_m \grad \Phi\right] = 4 \pi G_N \rho,
\end{equation}
where 
\begin{equation}
\mu_m = \frac{G_N}{2G} \frac{\mu}{\left( 1 + \frac{\mu}{2-K} \right) }.
\end{equation}
The ratio $G_N/G$ is not free, but is found by taking the limit
$\mu_m\rightarrow 1$, i.e. the Newtonian limit. Consistency requires
that $\mu\rightarrow\mu_0$ which is then a constant\footnote{The
constant $\mu_0$ is related to the constant $k$ introduced by
Bekenstein as $\mu_0 = \frac{8\pi}{k}$.} contained in the function $f$
(or $V$).  This gives the relation
\begin{equation}
 \frac{G_N}{G} = \frac{2}{\mu_0} + \frac{2}{(2-K)}.
\end{equation}
The MOND limit is now clearly recovered as $\mu_m \rightarrow
\frac{|\grad\Phi_N|}{a_0}$, and we get
\begin{eqnarray}
\mu \rightarrow   \frac{2G}{G_N}  \frac{|\grad \varphi|}{a_0} = 
\frac{2G}{G_N} \frac{1}{\ell_B a_0} e^{\phi_c}  \sqrt{X},
\end{eqnarray}
where $X$ 
is given in Eq. (\ref{eq_X}). Since $\mu = \frac{df}{dX}$, we may
integrate the above equation to find the function $f(X)$, which in the
MOND limit should be given by
\begin{eqnarray}
 f \rightarrow \frac{2}{3\ell_B a_0} \frac{1}{ \left( \frac{1}{\mu_0}
 + \frac{1}{2-K} \right) } e^{\phi_c}   X^{3/2} ,
\end{eqnarray}
where the integration constant has been absorbed into the cosmological
constant associated with the metric $\metE_{\mu\nu}$.  Since both $X$
and $f$ are dimensionless we may define a new constant $\beta_0$, such
that $a_0$ is a derived quantity given by
\begin{eqnarray}
a_0  = \frac{2}{3\beta_0\ell_B} \frac{1}{\left( \frac{1}{\mu_0}  +
\frac{1}{2-K} \right) } e^{\phi_c},
\label{eq_def_a_0}
\end{eqnarray}
and the function $f$ has the MONDian limit $f \rightarrow \beta_0
X^{3/2}$.  Since in the Newtonian limit we have $f\rightarrow \mu_0
X$,  there are at least three constants that can appear in $f(X)$,
namely $\mu_0$, $\beta_0$ and $\ell_B$.  

In terms of the function $\frac{dV}{d\mu}$ the  MONDian limiting case
implies that $\frac{dV}{d\mu}\rightarrow-\frac{4}{9\beta_0^2\ell_B^2}
\mu^2$ as $\mu\rightarrow 0$, while it diverging as $\mu\rightarrow
\mu_0$ in the Newtonian limit.  This second limit is imposed if
$\frac{dV}{d\mu}\rightarrow (\mu_0 - \mu)^{-m}$, for some constant
$m$. Bekenstein chooses this to be $m=1$, although other choices are
equally valid, even functions that have essential singularities.

It is clear from Eq. (\ref{eq_def_a_0}) that  $a_0$  depends on the
cosmological boundary condition, $\phi_c$, which can differ for each
system, depending on when it was formed. It could thus be considered
as a slowly varying function of time.   This possibility has been
investigated by Bekenstein and Sagi~\cite{BekensteinSagi2008}, and by
Limbach et al.~\cite{LimbachPsaltisFeryal2008}.

The two limiting cases for $f(X)$ are somewhat strange. In particular
we require that $f(X)\rightarrow X$ for $X\gg 1$ to recover the
Newtonian limit, and that  $f(X)\rightarrow X^{3/2}$ for $X\ll 1$
(i.e. a higher power) to  recover the MONDian limit.  This signifies
that in this kind of formulation of relativistic MOND (i.e. in terms
of a scalar field) the function $f(X)$ should be non-analytic.   It
further signifies that $f(X)$ can be expanded in positive powers of
$\sqrt{X}$ for small $X$, and in positive powers of $\frac{1}{X}$ for
large $X$, but that these two expansions cannot be connected. In other
words, it is impossible to perturbatively connect the Newtonian regime
with the MONDian regime via a perturbation series in $|\grad \varphi|$.

The Bekenstein free function in~\cite{Bekenstein2004a} is given in the
notation used in this review by
\begin{equation}
  \frac{dV}{d\mu} = -\frac{3}{32\pi \ell_B^2 \mu_0^2}\frac{\mu^2
  (\mu-2\mu_0)^2}{(\mu_0-\mu)},
\label{eq_beke_fn}
\end{equation}
which means that $\beta_0 =  \frac{4}{3} \sqrt{\frac{2\pi\mu_0}{3}}$,
and thus 
\begin{eqnarray}
a_0  = \frac{\sqrt{3}}{ 2 \sqrt{2\pi\mu_0}\ell_B} \frac{1}{\left(
\frac{1}{\mu_0}  + \frac{1}{2-K} \right) } e^{\phi_c}. 
\label{eq_a_0_bek}
\end{eqnarray}
This is in agreement with~\cite{BekensteinSagi2008} (the authors
of~\cite{BourliotEtAl2006} erroneously inverted a fraction in their
definition of $a_0$).

\subsubsection{Homogeneous and isotropic cosmology}
\label{sec_FLRW}

Homogeneous and isotropic Friedmann-Lemaitre-Robertson-Walker (FLRW)
solutions to the field equations of TeVeS have been extensively
studied~\cite{Bekenstein2004a,HaoAkhoury2005,SkordisEtAl2006,Diaz-RiveraSamushiaRatra2006,DodelsonLiguori2006,Skordis2006,BourliotEtAl2006,Zhao2006a,Skordis2008a,FerreiraSkordisZunkel2008}.
In this case the universally coupled metric can be written
in the conventional synchronous form as
\be
ds^2= -dt^2 + a^2(t) q_{ij} dx^i dx^j,
\ee
where $a(t)$ is the `physical scale factor'.  Here we assume for
simplicity that the hyper-surfaces of constant $t$ are spatially flat
(see~\cite{Skordis2006} for the curved case).  The Bekenstein metric
then has a similar form, and can be written as
\begin{equation}
 d\tilde{s}^2 = -d\tilde{t}^2 + b^2(\tilde{t}) q_{ij}dx^i dx^j,
\end{equation}
for a second scale factor $b(\tilde{t})$.  The disformal
transformation relates the two scale factors by $a = b e^{-\phi}$,
while the two time coordinates $t$ and $\tilde{t}$ are related by $dt = e^{\phi}d\tilde{t}$. The physical Hubble parameter is defined as
usual as by $H=\frac{\dot{a}}{a}$, while the Bekenstein frame Hubble
parameter is $\tilde{H} = \frac{d\ln b}{d\tilde{t}} e^\phi H + \frac{d\phi}{d\tilde{t}}$. 
 Cosmological evolution is governed by the analogue of the Friedmann equation:
\begin{equation}
3\tilde{H}^2 = 8\pi G e^{-2\phi} \left( \rho_\phi + \rho\right),
\label{eq_beke_friedmann}
\end{equation}
where $\rho$ is the energy density of physical matter, which obeys the
energy conservation equation with respect to the universally coupled
metric, and where the scalar field energy density, $\rho_{\phi}$, is
given by
\begin{equation}
\rho_\phi = \frac{e^{2\phi}}{16\pi G}\left( \mu \Vp + V \right).
\end{equation}
Similarly, one can define a scalar field pressure by
\begin{equation}
P_\phi = \frac{e^{2\phi}}{16\pi G}\left( \mu \Vp - V \right) 
\end{equation}
The scalar field evolves according to the two differential equations:
\begin{equation}
 \frac{d\phi}{d\tilde{t}} = -\frac{1}{2\mu}\Gamma,
\end{equation}
and
\begin{equation}
 \frac{d\Gamma}{d\tilde{t}} + 3 \tilde{H} \Gamma = 8\pi G e^{-2\phib} (\rho + 3 P),
\end{equation}
where $\mu$ is found by inverting ${\phi'}^2 = \frac{1}{2}\frac{dV}{d\mu}$.

It is important to note that the vector field must point to the time
direction, so that it can be written as $A_{\mu} =
(\sqrt{\metE_{00}},\vec{0}) $.  In this case it does not contain any
independent dynamical information, and it does not explicitly
contribute to the energy density. Its only effect is on the disformal
transformation which relates the Bekenstein-frame Friedmann equation,
Eq. (\ref{eq_beke_friedmann}), with the physical Friedmann equation.
This is also true in cases where the vector field action is
generalised, and where the only effect is a constant rescaling of the left-hand-side
of the  Bekenstein-frame Friedmann equation, as discussed  in~\cite{Skordis2008a}.
\newline
\newline
\noindent
{\it FLRW solutions with the Bekenstein function}
\newline
\label{sec_bekenstein_fn}

In the original TeVeS paper~\cite{Bekenstein2004a} Bekenstein studied
the cosmological evolution of an FLRW universe by assuming that the
free function is given by Eq. (\ref{eq_beke_fn}).  He showed that the
scalar field contribution to the Friedmann equation is very small, and
that $\phi$ evolves very little from the early universe until
today. He noted that with this choice of function, a cosmological
constant term has to be added in order to have an accelerating
expansion today, as appears to be required by cosmological observations.

Many other studies on cosmology in TeVeS have also used the Bekenstein
function, see for
example~\cite{HaoAkhoury2005,SkordisEtAl2006,DodelsonLiguori2006,Skordis2008a}.
In particular, Hao and Akhoury noted that the integration constant
obtained by integrating Eq. (\ref{eq_beke_fn}) can be
used to get a period of accelerating expansion, and that TeVeS
therefore has the potential to act as dark
energy~\cite{HaoAkhoury2005}. However, such an integration constant
cannot be distinguished from a bare cosmological constant term in the
Bekenstein frame, and so it is somewhat dubious as to whether this can
really be interpreted as dark energy arising from TeVeS. Nevertheless,
it would not be a surprising result if some other TeVeS functions
could, in fact account for dark energy, as the scalar field action in
TeVeS close resemblances that of
k-essence~\cite{Armendariz-PiconDamourMukhanov1999,Armendariz-PiconMukhanovSteinhardt2000}.
Zhao has investigated this issue further~\cite{Zhao2006a}(see below).

Exact analytical and numerical solutions with the Bekenstein free
function, Eq. (\ref{eq_beke_fn}), have been found by Skordis \etal
in~\cite{SkordisEtAl2006}, and by Dodelson and Liguori
in~\cite{DodelsonLiguori2006}. It turns out that not only, is the
scalar field is subdominant, as Bekenstein noted, but its energy
density also tracks the matter fluid energy density.  The ratio of the
energy density in the scalar field to that of ordinary matter then
remains approximately constant, so that the scalar field tracks the
matter dynamics.  One then gets that 
\begin{equation}
\Omega_\phi = \frac{(1+3w)^2}{6(1-w)^2\mu_0},
\end{equation}
where $w$ is the equation of state of the matter fields\footnote{Note
that this excludes the case of a stiff fluid with $w=1$.}.  Since
$\mu_0$ is required to be very large, the energy density in the scalar
field is always small, with values typically less than $\Omega_\phi \sim 10^{-3}$
in a realistic situation. Tracker solutions are also present for this
choice of function in versions of TeVeS with more general vector field
actions~\cite{Skordis2008a}.

In realistic situations, tracking in the radiation era is almost never
realised, as has been noted by Dodelson and
Liguori~\cite{DodelsonLiguori2006}. Rather, during the radiation era,
the  scalar field energy density is subdominant but slowly growing,
such that $\phi \propto a^{4/5}$.  However, upon entering the matter era $\phi$ settles into the
tracker solution. This transient solution in the radiation era has
been generalised by Skordis to arbitrary initial conditions for
$\phi$, more general free functions (see below), and a general vector
field action~\cite{Skordis2008a}. It should be stressed that the
solution in the radiation era is important for setting up initial
conditions for the perturbations about FLRW solutions that are
relevant for studying the CMB radiation and Large-Scale Structure
(LSS).

From Eq. (\ref{eq_def_a_0}) we see that $a_0$ for a quasi-static system
depends on the cosmological value of the scalar field at the time the
system broke off from the expansion, and collapsed to form a bound
structure. It is then possible that different systems could exhibit
different values of $a_0$ depending on when they formed. The impact of evolving $a_0$ on
 observations has been investigated in~\cite{BekensteinSagi2008,LimbachPsaltisFeryal2008}.

Finally, note that the sign of $\dot{\phi}$ changes between the matter
and cosmological constant eras. In doing so, the energy density of the
scalar field goes momentarily through zero, since it is purely kinetic
and vanishes for zero $\dot{\phi}$~\cite{SkordisEtAl2006}.
\newline
\newline
\noindent
{\it FLRW solutions by generalising the Bekenstein function}
\newline

Bourliot \etal~\cite{BourliotEtAl2006} studied more general free
functions, that have the Bekenstein function as a special case.  In
particular they introduced two new parameters, a constant, $\mu_a$,
and a power index, $n$, such that the free function is generalised to
\begin{equation}
\frac{dV^{(n)}}{d\mu} = 
 -\frac{3}{32\pi \ell_B^2 \mu_0^2}\frac{\mu^2 (\mu-\mu_a\mu_0)^n}{\mu_0-\mu}. 
\end{equation}
This function\footnote{Note that~\cite{BourliotEtAl2006} uses a
different normalisation for $V$, and their results can be recovered by
rescaling the $\ell_B$ used in this report by $\ell_B\rightarrow\ell_B
\sqrt{\frac{3}{2}\mu_0^{n-3}}$.} reduces to the Bekenstein function
when $n=2$ and $\mu_a=2$.  It retains the property of having a
Newtonian limit as $\mu\rightarrow \mu_0$ and a MOND limit as
$\mu\rightarrow 0$.  The cosmological evolution depends on the power
index, $n$. More general functions can also be constructed by
considering the sum of the above prototypical function with arbitrary
coefficients, i.e. by taking $\frac{dV}{d\mu} = \sum_n c_n
\frac{dV^{(n)}}{d\mu}$~\cite{BourliotEtAl2006}.

\label{sec_bourliot}
Clearly $\frac{dV}{d\mu}(\mu_a\mu_0) = 0$, and at this point $\dot{\phi}\rightarrow 0$.
Now suppose that the integration constant is chosen such that
$V(\mu_a\mu_0)=0$ as well. Then, just like the case of the Bekenstein
function, one finds tracker solutions: The function $\mu$ is driven to
$\mu=\mu_a\mu_0$, at which point $\dot{\phi}=0$.  There are no
oscillations around that point, but it is approached slowly so that it
is exactly reached only in the infinite future. The scalar field
relative density is now given by
 \begin{equation}
\Omega_\phi = \frac{(1+3w)^2}{3\mu_a(1-w)^2\mu_0},
\end{equation}
independent of the value of $n$. It should also be pointed out that
the evolution of the physical Hubble parameter, $H$, can be different
than the case of $GR$ even in the tracking
phase~\cite{BourliotEtAl2006}. For example in the case $w=0$ we have
$H \propto a^{-n_h}$, where $n_h =\frac{1+3\mu_a\mu_0}{2(\mu_a\mu_0-1)}$.

Furthermore, just like the Bekenstein case, the radiation era tracker
is untenable for realistic cosmological evolutions, for which $\mu_0$
must be large so that $\Omega_\phi$ is small ($\lesssim 10^{-3}$). In
this case we once again get a transient solution where the scalar
field evolves as $\phi \propto a^{4/(3+n)}$~\cite{Skordis2008a}.
In the case that the integration constant is chosen such that
$V(\mu_a\mu_0)\ne0$ one has an effective cosmological constant
present. Thus, once again, we get tracker solutions until the energy
density of the Universe drops to values comparable with this
cosmological constant, at which time tracking comes to an end, and the
Universe enters a de Sitter phase.

The cases $-2<n <0$ turn out to be pathological as they lead to
singularities in the cosmological
evolution~\cite{BourliotEtAl2006}. The case $n=-3$ is well behaved
when the matter fluid is a cosmological constant, but is also
pathological when $w=-1$~\cite{BourliotEtAl2006}.  The cases for which
$n\le-4$ are  well behaved in the sense that no singularities occur in
the cosmological evolution. Contrary to the $n\ge 1$ cases, the
cosmological evolution drives the function $\mu$ to
infinity. Moreover, these cases do not display the tracker solutions
of $n\ge 1$, but rather the evolution of $\rho_\phi$ is such that it
evolves more rapidly than the matter density, $\rho$, and so quickly
becomes subdominant. The general relativistic Friedmann equation is
thus recovered, such that $3 H^2 = 8\pi G \rho$.  This also results in
$\tilde{H} = H$, which means that the scalar field is slowly rolling.

The evolution of the scalar field variables $\Gamma$, $\phi$ and $\mu$
then depends on the equation of state of the matter fields.  If the
background fluid is a cosmological constant, then we get de Sitter
solutions for both metrics, and it can be shown that $\Gamma= 2H(e^{-3Ht} -1)$.
For the case of a stiff fluid, with $w=1$, we get that $\Gamma$ has
power-law solutions that are inverse powers of $t$, so that $\Gamma =
\frac{6}{t} + \frac{\Gamma_0}{t^3}$. A similar situation arises when
$-1 < w < 1$, for which we get $\Gamma = \frac{2(1+3w)}{1-w} H$,  and
the Hubble parameter evolves as $H  =
\frac{2}{3(1+w)}\frac{1}{t}$. Notice that the limit $w\rightarrow 1$,
for the $-1<w<1$ case, does not smoothly approach the $w=1$ case.

Mixing different powers of $n\ge 1$ leads once again to tracker
solutions. One may have to add an integration constant in order to
keep $V(\mu_a \mu_0) = 0$, although for certain combinations of powers
$n$ and coefficients $c_i$ this is not necessary.
Mixing $n=0$ with some other $n\ge 1$ cannot remove the pathological
situation associated with the $n=0$ case. Mixing $n=0$ with both
positive and negative powers could however lead to acceptable
cosmological evolution since the effect of the negative power is to
drive $\mu$ away  from the $\mu=\mu_a \mu_0$ point. In general, if we
mix an arbitrary number of positive and negative powers we get tracker
solutions provided we can expand the new function in positive
definite powers of $(\mu - \mu_a'\mu_0)$, where $\mu_a'$ is some
number different from the old $\mu_a$.  

The observational consequences for the CMB and LSS have not been
investigated for this class of function, unlike the case of the
Bekenstein function.
\newline
\newline
\noindent
{\it Inflationary/accelerating expansion for general functions}
\newline

Diaz-Rivera, Samushia and Ratra~\cite{Diaz-RiveraSamushiaRatra2006} have 
studied cases where TeVeS leads to inflationary, or self-accelerating,
solutions.  They first consider the vacuum case, in which they find
that de Sitter solutions exist with $b \sim e^{\tilde{H}_0
\tilde{t}}$, where the Bekenstein frame Hubble constant $\tilde{H}_0$
is given by the free function as $\tilde{H}_0 = \sqrt{\frac{\mu_0^2
V}{6}}$, and where $\frac{dV}{d\mu}=0$ (i.e. the scalar field is
constant, $\phi = \phi_i$).  Such a solution will always exist in
vacuum provided that the free function satisfies
$\frac{dV}{d\mu}(\mu_v)=0$ and $V(\mu_v)\ne0$, for some constant
$\mu_v$. In that case, the general solution is not de Sitter since
both $\phi$ and $\mu$ will be time-varying, but will tend to de Sitter
as $\mu \rightarrow \mu_v$.  Indeed, the $n\ge 1$ case of Bourliot
\etal~\cite{BourliotEtAl2006} with an integration constant is
precisely this kind of situation.

In the non-vacuum case, for a fluid with equation of state $P =
w\rho$, they make the ansatz $b^{3(1+w)} = e^{(1+3w)\phi}$.  This
brings the Friedmann equation into the form $3\tilde{H}^2 = 8\pi G
\rho_0 + \frac{1}{2}(\mu \frac{dV}{d\mu} + V)$, where $\rho_0$ is the matter
density today.  Once again, they assume that the
free-function-dependent general solution drives $\mu$ to a constant
$\mu_v$, but $\phi$ is evolving. Thus, we must have that $\phi =
\phi_1 \tilde{t} + \phi_2$, such that $\dot{\phi} = \phi_1$ is a
constant. In order for $\phi_1$ to be non-zero we must have
$\frac{dV}{d\mu}(\mu_v)\ne 0$.  However,  there is a drawback to this
approach. As they point out, consistency with the scalar field
equation requires that $w<-1$.  Furthermore, although this solution is
a de Sitter solution in the Bekenstein-frame, it corresponds to a
power-law solution for the universally coupled metric. In order for
this power-law solution to lead to acceleration, they find that  $
-5/3 < w < -1$.  This range of $w$ corresponds to a phantom fluid.
\newline
\newline
\noindent
{\it Accelerated expansion in TeVeS}
\newline

The simplest case of accelerated expansion in TeVeS is provided by a
cosmological constant term. This is equivalent to adding an
integration constant to
$V(\mu)$~\cite{HaoAkhoury2005,BourliotEtAl2006}, and it corresponds to
the accelerated expansion considered by Diaz-Rivera, Samushia and
Ratra~\cite{Diaz-RiveraSamushiaRatra2006} in both the vacuum and
non-vacuum cases (see above).  These solutions therefore suffer from
the usual fine-tuning and coincidence problems, and so it is of
interest to look for accelerating solutions without such
a constant, simply by employing the scalar field (these need not be de
Sitter solutions).

Zhao used a function $\frac{dV}{d\mu} \propto \mu^2$ to obtain
solutions which provide acceleration, and compared his solution with
the SN1a data~\cite{Zhao2006a}, finding good agreement. However, it is
not clear whether other observables, such as the CMB angular power
spectrum, or observations of LSS are compatible with this function.
Furthermore, this function is not realistic, as it does not have a
Newtonian limit (it is always MONDian).  Although no further studies
of accelerated expansion in TeVeS have been performed, it is plausible
to think that certain choices of function could lead to
acceleration. This is because the scalar field action has the same form 
as a k-essence/k-inflation
action~\cite{Armendariz-PiconDamourMukhanov1999,Armendariz-PiconMukhanovSteinhardt2000},
which has been considered as a candidate theory for acceleration.
More precisely, the system of cosmological equations corresponds to k-essence
coupled to matter.  It is  not known in general whether this type of
model has similar features as the uncoupled k-essence models, although Zhao's study
indicates that this a possibility. 
\newline
\newline
\noindent
{\it Realistic FLRW cosmology}
\newline

In TeVeS, cold dark matter is absent.  Therefore in order to get
acceptable values for the physical Hubble constant today (i.e. around
$H_0 \sim 70 \; \textrm{km} \; \textrm{s}^{-1} \; \textrm{Mpc}^{-1}$) , 
we have to supplement the absence of CDM with something else. The
reason for this is simply that if all the energy density in the
Universe today was in the form of baryons, then the Hubble constant
would be lower than what is observed by a factor of $\sim$5.
Possibilities of what this supplementary material could be include the
scalar field itself, massive
neutrinos~\cite{SkordisEtAl2006,FerreiraSkordisZunkel2008}, and a
cosmological constant. At the same time, one has to get the right
angular diameter distance to
recombination~\cite{FerreiraSkordisZunkel2008}. These two requirements
can place severe constraints on the allowed form of the free functions. 

\subsubsection{Cosmological perturbation theory}
\label{sec_linear}

Cosmological perturbation theory in TeVeS has been formulated to
linear order in \cite{Skordis2006}, and in variants of TeVeS with more
general vector field actions in~\cite{Skordis2008a}.  The
scalar modes of the linearly perturbed universally coupled metric are
given in the conformal Newtonian gauge, as usual, by
\begin{equation}
ds^2 = -a^2(1+2\Psi) d\tau^2 + a^2 (1-2\Phi) q_{ij} dx^i dx^j,
\end{equation}
where $\tau$ is conformal time, defined by $dt = a d\tau$.  Here, we
 will assume, for simplicity, that the spatial curvature is zero.
 The reader is referred to~\cite{Skordis2006,Skordis2008a} for 
 the curved cases, as well as for an enunciation of vector and tensor perturbations.
The scalar field is perturbed as $\phi = \bar{\phi} + \varphi$, where
 $\bar{\phi}$ is the FLRW background scalar field, and $\varphi$ is
 the perturbation. The vector field  is perturbed as $A_\mu = a
 e^{-\bar{\phi}}(1 + \Psi -\varphi , \grad_i \alpha)$, such that the
 unit-timelike constraint is satisfied.  This removes
the time component of $A_{\mu}$ as an independent dynamical degree of
 freedom.  Thus, there are two additional dynamical degrees of
 freedom, when comparing to cosmological perturbation theory in GR:  The scalar field
perturbation, $\varphi$, and the vector field scalar mode, $\alpha$.

The perturbed field equations for the scalar modes can be found in the
conformal Newtonian gauge in~\cite{SkordisEtAl2006}, and in more form
(including in the synchronous gauge) in~\cite{Skordis2006}.
Perturbation equations for more general TeVeS actions are given
in~\cite{Skordis2008a}. Here we only present the Newtonian gauge equations of the original TeVeS formulation. 
We  define the following variables: $\PhiE = \Phi - \varphi  $, $\PsiE = \Psi - \varphi$, $\zetaE = -(1-e^{-4\bar{\phi}})\alpha$ 
and $\tadotoa = \frac{b'}{b}$.
The scalar field obeys the two first order equations
\begin{eqnarray}
 \gamma' &=& -3\tadotoa \gamma +  \frac{\bar{\mu}}{a}e^{-3\bar{\phi}}k^2\left(\varphi + \bar{\phi}' \alpha\right)
  - \frac{e^{\bar{\phi}}}{a}\bar{\mu}\bar{\phi}'\left[6\PhiE' +2k^2\zetaE \right] 
\nonumber 
\\
&&
 + 8\pi G a e^{-3\bar{\phi}}\left[\sum_f \delta\rho_f + 3\delta P_f  + \sum_f (\bar{\rho}_f + 3\bar{P}_f)\left( \PsiE - 2\varphi\right)\right]  
\end{eqnarray}
and
\begin{equation}
   \varphi' = -\frac{1}{2U}ae^{-\bar{\phi}}\gamma +  \bar{\phi}' \PsiE.
 \label{eq:varphi_dot}
\end{equation}
The vector field equations are given by
\begin{equation}
K\left(E' + \tadotoa E \right) =   - \bar{\mu} \bar{\phi}'     (\varphi -\bar{\phi}' \alpha)  + 8\pi G  a^2 (1-e^{-4\bar{\phi}}) 
\sum_f(\bar{\rho}_f+\bar{P}_f) (\theta_f -\alpha) 
\end{equation}
and
\begin{equation}
 \alpha'  =  E + \PsiE +  \left(\bar{\phi}' - \adotoa\right)  \alpha 
\end{equation}
and finally the Einstein equations are given by the Hamiltonian constraint
\begin{eqnarray}
&&
-2\left(k^2 - 3\kappa\right)\PhiE
   - 2e^{4\bar\phi}\tadotoa\left[3\PhiE' + k^2\zetaE + 3\tadotoa\PsiE\right] + a e^{3\bar{\phi}} \bar{\phi}' \gamma 
\nonumber 
\\
&&
\ \ \ \
 - K k^2E 
 = 8\pi G a^2 \sum_f\bar{\rho}_f \left[ \delta_f  - 2 \varphi\right]  
\label{eq:TE_den}
\end{eqnarray}
the momentum constraint equation
\begin{equation}
\PhiE' + \kappa \zetaE + \tadotoa \PsiE -\bar{\mu}\bar{\phi}'\varphi =
   4\pi G a^2  e^{-4\bar{\phi}}\sum_f(\bar{\rho}_f + \bar{P}_f) \theta_f
\label{eq:TE_vel}
\end{equation}
and the two propagation equations
\begin{eqnarray}
&&  6\PhiE'' +  2k^2\left(\zetaE' - e^{-4\bar{\phi}}\PsiE\right)
 + 2e^{-4\bar{\phi}}\left(k^2 - 3\kappa\right)\PhiE  
+ 2 \tadotoa \left[6\PhiE' + 3 \PsiE' + 2k^2\zetaE \right] 
\nonumber \\
 &&
  + 4\bar{\phi}' \left[ 3\PhiE' + k^2\zetaE \right] 
+ 3\frac{\mu}{U}a e^{-\bar{\phi}} \bar{\phi}' \gamma
+6\left[2\tadotoa' + \tadotoa^2 + 4\bar{\phi}' \tadotoa \right]\PsiE
\nonumber \\
 &&
  = 24\pi G a^2e^{-4\bar{\phi}} (\delta P - 2 P \varphi) 
\end{eqnarray}
and
\begin{equation}
\PhiE - \PsiE +  e^{4\bar{\phi}}\left[ \zetaE' + 2\left(\tadotoa +\bar{\phi}' \right)\zetaE \right] =    8\pi G a^2 \sum_f (\bar{\rho}_f + \bar{P}_f)\Sigma_f
\end{equation}


Let us now turn to the problem of specifying initial conditions for
the scalar modes, which in general should depend on the chosen form of
the free function.  The exact adiabatic growing mode in TeVeS, and
generalised variants, have been found by Skordis
in~\cite{Skordis2008a}, but only for the case of the generalised
Bekenstein function.  If the free function is such so that the scalar
field contribution to the background expansion during the radiation
era is very small, however, then the adiabatic modes for other free
functions should be only marginally different from the ones found 
in~\cite{Skordis2008a}. In particular, the only effect should be a
difference in the initial conditions of $\varphi$, which is not
expected to make any difference to observations.

The only study that has been performed of the observational signatures of TeVeS in the CMB
radiation and in LSS is due to Skordis, Mota, Ferreira and
B{\oe}hm~\cite{SkordisEtAl2006}.  Here the initial conditions were
chosen such that both $\varphi$ and $\alpha$, as well as their
derivatives, are initially zero. While this is not a purely adiabatic
initial condition, it turns out that it is close enough to ensure that no
observable difference can be seen from isocurvature contamination.
Detailed studies of isocurvature modes in TeVeS have not yet been conducted.
In the light of the problems that TeVeS has with observations of the
CMB radiation~\cite{SkordisEtAl2006}, however, it may be important to
investigate what effects isocurvature modes are likely to have.
Preliminary studies by Mota, Ferreira and Skordis have shown that
setting the vector field perturbations to be large initially can have
a significant impact~\cite{MotaFerreiraSkordis_edinburg}.

In addition to the four regular isocurvature modes that exist in GR,
there could in principle exist four further isocurvature modes in
TeVeS: Two associated with the scalar field, and two
associated with the vector field. Preliminary studies by Skordis have
shown that none of the scalar field isocurvature modes are regular in
either the synchronous or conformal Newtonian gauges.  Conversely, under certain
conditions of the vector field parameters one of the vector field
isocurvature modes can be regular, while the other one is never
regular. Thus, it may be possible to have one regular isocurvature
mode in the TeVeS sector. The observational consequences of this mode
are still unknown, as is its generation method from early universe inflation.
Studies of the observable spectra based on vector or tensor modes
are also yet to be conducted although the necessary equations can be found in~\cite{Skordis2006,Skordis2008a}.

\subsubsection{Cosmological observations and constraints}

Let us now consider the observational signatures of the perturbation
theory discussed above, and how they can be used to constrain TeVeS.
\newline
\newline
\noindent
{\it Large-scale structure observations}
\newline

A traditional criticism of MOND-type theories was their lack of a dark
matter component, and therefore their presumed inability to form
large-scale structure compatible with current observational data.
This criticism was based on intuition formed from a general relativistic
universe filled with baryons only. In that case it is well known that,
since baryons are coupled to photons before recombination, they do not
have enough time to grow into structures on their own.  Furthermore,
their oscillatory behaviour at recombination is preserved, and is
visible as large oscillation in the observed galaxy power spectrum
$P_{gg}(k)$.    Finally, on scales smaller than the diffusion damping
scale they are exponentially suppressed due to Silk damping. Cold dark
matter solves all of these problems because it does not couple to
photons, and therefore can start creating potential wells early on in
the Universe's history, into which the baryons can fall.  This is
enough to generate the right amount of structure, erase most of the
oscillations, and overcome the Silk damping. 

TeVeS contains two additional fields, not present in GR, that change the structure of the equations significantly.  The
first study of large-scale structure observations in TeVeS was
conducted by Skordis, Mota, Ferreira and B{\oe}hm
in~\cite{SkordisEtAl2006}.  Here the perturbed TeVeS equations were
solved numerically for the case of the Bekenstein function, and the
effects on the matter power spectrum, $P(k)$, were determined. It was
found that TeVeS can indeed form large-scale structure compatible with
observations, depending on the choice of TeVeS parameters in the free
function. In fact, the form of the matter power spectrum, $P(k)$, in
TeVeS looks quite similar to the corresponding spectrum in $\Lambda$CDM.
Thus, one has to turn to other observables to distinguish the TeVeS
from General Relativity.

Dodelson and Liguori~\cite{DodelsonLiguori2006} provided an analytical
 explanation of the growth of structure that was found numerically in~\cite{SkordisEtAl2006}.
 They concluded the growth in TeVeS cannot be  due to the scalar
 field, as the scalar field perturbations are Bessel functions that
 decaying in an oscillatory fashion.  Instead, they reasoned, the
 growth of large-scale structure in TeVeS is due to the vector field perturbation.

Let us see how the vector field leads to growth.  Using the tracker
solutions in the matter era, from Bourliot
\etal~\cite{BourliotEtAl2006}, we can find the behaviour of the
background functions $a$, $b$ and $\bar{\phi}$.  Using these in the perturbed field equations, 
after setting the scalar field perturbations to zero, it can be shown
that in the matter era the vector field perturbation $\alpha$ obeys the equation
\begin{equation}
\alpha'' +  \frac{b_1}{\tau} \alpha' + \frac{b_2}{\tau^2} \alpha =  S(\Psi,\Psi',\theta)
\label{eq_vec_growth}
\end{equation}
in the conformal Newtonian gauge, where 
\begin{eqnarray}
b_1&=&\frac{4(\mu_0\mu_a-1)}{\mu_0\mu_a+ 3},
\\
b_2 &=& \frac{2}{(\mu_0\mu_a + 3)^2}    \left[ \mu_0^2\mu_a^2 -
\left(5+\frac{4}{K}\right)\mu_0\mu_a +  6  \right],
\end{eqnarray}
and where $S$ is a source term which does not explicitly depend on $\alpha$.
If we simultaneously take the limits $\mu_0\rightarrow\infty$ and
$K\rightarrow 0$, for which $\Omega_\phi \rightarrow 0$, meaning that
the TeVeS contribution is absent, then we get $b_1 \rightarrow 4$ and
$b_2 \rightarrow 2$.  In this case the two homogeneous solutions to
Eq. (\ref{eq_vec_growth}) are $\tau^{-2}$ and $\tau^{-1}$, which are
decaying. 

\begin{figure}
\epsfig{file=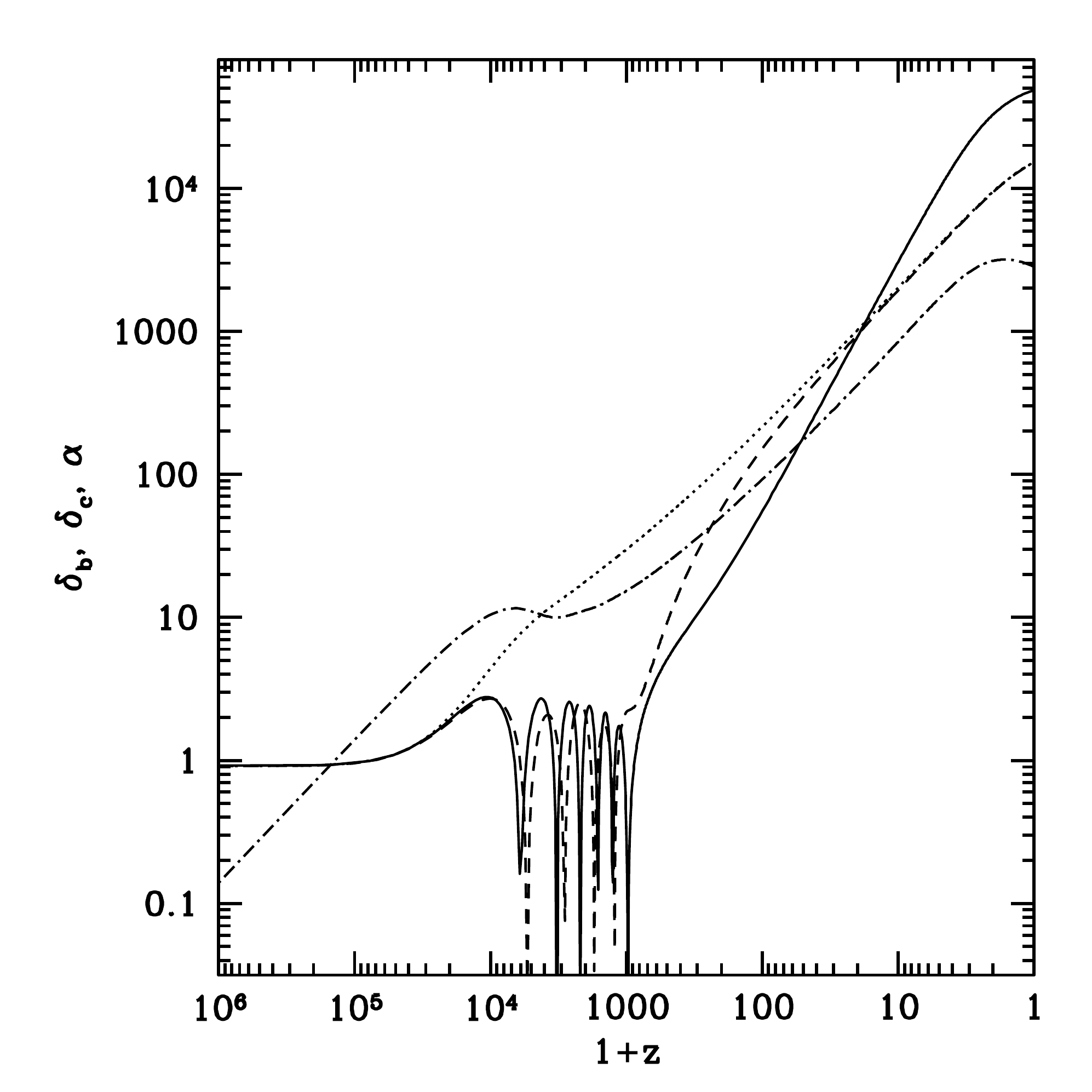,width=6.5cm}
\epsfig{file=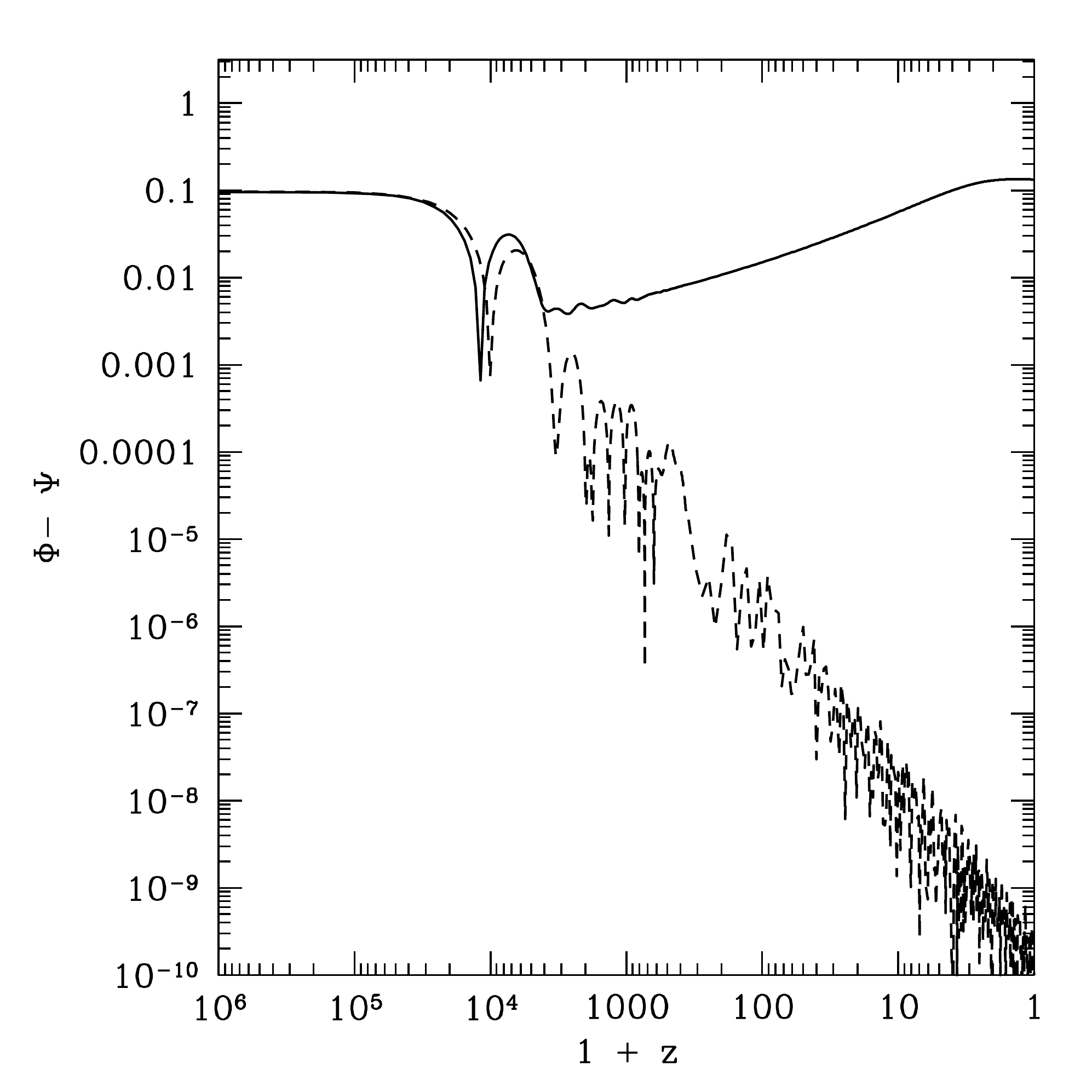,width=6.5cm}
\caption{
LEFT: The evolution in redshift of baryon density fluctuations in TeVeS (solid
line), and in $\Lambda$CDM (dashed line) for a wavenumber
$k=10^{-3}\textrm{Mpc}^{-1}$.  In both cases, the baryon density fluctuates
before recombination, and grows afterwards.  In the case of
$\Lambda$CDM, the baryon density eventually follows the CDM density
fluctuation (dotted line), which starts growing before
recombination. In the case of TeVeS, the baryons grow due to the
potential wells formed by the growing mode in the vector field,
$\alpha$ (dot-dashed line).  RIGHT: The difference of the two
gravitational potentials, $\Phi - \Psi$, for a wavenumber
$k=10^{-3}\textrm{Mpc}^{-1}$ as a function of redshift for both TeVeS
(solid line), and $\Lambda$CDM (dotted line).
}
\label{fig_phi_psi}
\end{figure}

Dodelson and Liguori show that the source term
$S(\Psi,\dot{\Psi},\theta)$ is not sufficient to create a growing mode
in the general solution to Eq. (\ref{eq_vec_growth}), and that in
the general relativistic limit TeVeS does not, therefore, provide
enough growth for structure formation.  Now let us consider the
general case.  Assuming that the homogeneous solutions to
(\ref{eq_vec_growth}) can be written as $\tau^n$, it can be shown that
for the generalised Bekenstein function of Bourliot
\etal~\cite{BourliotEtAl2006} we can get
\begin{equation}
n \approx -\frac{3}{2} + \frac{1}{2}  \sqrt{ 1 +\frac{32}{K\mu_0\mu_a  } }.
\end{equation}
Thus, we can have $n>0$, provided that for fixed $\mu_0\mu_a$ we also
have\footnote{Smaller values of $\mu_0\mu_a$ can also raise this
threshold.}
\begin{equation}
K \lesssim  0.01 .
\end{equation}
If this condition is met then there can exist a growing mode in
$\alpha$, which in turn feeds back into the perturbed Einstein
equation and sources a non-decaying mode in $\Phi$ that can drive structure formation.
This is displayed graphically in the left panel of Figure~\ref{fig_phi_psi}.
It is a striking result that even if the contribution of the TeVeS
fields to the background FLRW equations is negligible ($\sim 10^{-3}$
or less), one can still get a growing mode that drives structure
formation. 
\newline
\newline
\noindent
{\it CMB observations}
\newline

A general relativistic universe dominated by baryons cannot fit the
most up to date observations of anisotropies in the
CMB~\cite{NoltaEtAl2008}.  This is true even if a cosmological
constant and/or three massive neutrinos are incorporated into the
matter budget, so that the first peak of the CMB angular power
spectrum is at the right position\footnote{In this case the third peak is
unacceptably lower than the second.}.
This, however, is not proof that only a theory with CDM can fit CMB
observations (as claimed in~\cite{SlosarMelchiorriSilk2005,SpergelEtAl2006}).
A prime example to the contrary is the Eddington-Born-Infeld
theory~\cite{Banados2009}.  However, the linear
Boltzmann equation, and the resulting CMB angular power spectrum, have
been calculated in TeVeS, using initial conditions that are close to
adiabatic~\cite{SkordisEtAl2006}.  The resulting fits to the data were
poor, at least for the Bekenstein free function, showing that CMB
observations are, nevertheless, problematic for TeVeS.
It may be that the new isocurvature modes discussed above 
can provide a richer phenomenology but it remains to be seen whether this can save this theory.
\newline
\newline
\noindent
{\it The difference in the gravitational potentials: $\Phi - \Psi$}
\newline

The result of Dodelson and Liguori~\cite{DodelsonLiguori2006} have a
further direct consequence: The perturbation equations in TeVeS that
relate the difference of the two gravitational potentials, $\Phi
-\Psi$, to the shear of matter, have additional contributions coming
from the perturbed vector field, $\alpha$. This is not due to the
existence of the vector field {\it per se}, but comes from the disformal transformation in which the
vector field plays an important part. Indeed, in a single metric
theory where the vector field action is Maxwellian, as in TeVeS, there
is no contribution from the vector field to $\Phi - \Psi$.  Now, as
the vector field is required to grow in order to drive structure
formation, it will also inevitably lead to growth in $\Phi - \Psi$. This is 
precisely what we see numerically in the right panel of Figure~\ref{fig_phi_psi}.
If $\Phi - \Psi$ can be measured observationally, then it will provide
an excellent test of TeVeS.  This possibility is discussed in more detail in section~\ref{PPF}.

\subsection{Other Theories}

\subsubsection{The Einstein-Cartan-Sciama-Kibble Theory}
\label{newEC}

In this subsection we will briefly describe the
Einstein-Cartan-Sciama-Kibble (ECSK)
theory~\cite{Cartan1,Cartan2,Cartan3,Sciama1,Sciama2,Kibble}. 
The ECSK theory is in many cases equivalent to General Relativity and
departs from GR only when at least one matter field has intrinsic
spin. The ECSK theory has been reviewed by
Hehl \etal~\cite{ECSK}, and more recently by Trautman~\cite{Trautman2006}.
\newline
\newline
\noindent
{\it The ECSK theory as a theory with torsion}
\newline

The ECSK theory is basically General Relativity with the addition of torsion.
The connection is assumed to be metric compatible, but has non-zero torsion, and is thus given by
\be
\Gamma^{\mu}_{\phantom{\mu} \alpha\beta} = {\genfrac{\{}{\}}{0pt}{}{\mu}{\alpha\beta}} + K^{\mu}_{\phantom{\mu} \alpha\beta} ,
\label{ECSK_conn}
\ee
where $K^{\mu}_{\phantom{\mu} \alpha\beta}$ is the contorsion tensor given in terms of the torsion $S^{\mu}_{\phantom{\mu} \alpha\beta}$ 
by Eq. (\ref{contorsion}).
 The Riemann tensor is antisymmetric in both the first and the last
two indices, and hence the Ricci tensor is its only unique non-vanishing
contraction. It is, however, asymmetric, and is given by Eq. (\ref{Ricci_asym}).

The addition of torsion to the connection has a direct consequence on
the geometry of curves. In this case, autoparallels (straightest
lines) are not necessarily extremals (shortest or longest lines) as they are in GR. The former are given by
\begin{equation}
\frac{d^2x^\mu}{ds^2} + \Gamma^\mu_{\phantom{\mu}\alpha\beta} \frac{dx^\alpha}{ds} \frac{dx^\beta}{ds}  = 0,
\end{equation} 
while the latter are found by minimising the proper length $\int_\gamma \sqrt{-g_{\mu\nu} dx^\mu dx^\nu}$, and are given by
\begin{equation}
\frac{d^2x^\mu}{ds^2} +  {\genfrac{\{}{\}}{0pt}{}{\mu}{\alpha\beta}} \frac{dx^\alpha}{ds} \frac{dx^\beta}{ds}  = 0.
\end{equation} 
Spin-less test particles and gauge-fields (e.g. photons) do not feel the torsion and follow the extremals. However,
 spinning test particles do feel the torsion and obey analogues of the Mathisson-Papapetrou equations~\cite{Mathisson1937,Papapetrou1951}.
 The notions of autoparallel and extremal curves coincide if and only
 if the torsion is totally antisymmetric. The reader is referred
 to~\cite{Hehl1971,Trautman1972c,AdamowiczTrautman1975} for further
 discussion.
\newline
\newline
\noindent
{\it The action and field equations}
\newline

The action for the ECSK theory is the same as the one for the metric-affine gravity described in Eq. (\ref{met_aff}),
with the additional assumption that the connection is metric
compatible. To make sure that no inconsistency arises one has to
impose $Q_{\mu\alpha\beta}=0$ with a Lagrange multiplier. Variation
with respect to  $\Gamma^\mu_{\;\;\alpha\beta}$ then proceeds in the usual way.

We can follow a different approach, however, and assume from the start
that $Q_{\mu\alpha\beta}=0$, and that the connection is given by Eq. (\ref{ECSK_conn}).
We can then vary the action, Eq. (\ref{met_aff}), without the Lagrange
constraint, but instead take as independent variables the metric and
the contorsion. Variation with respect to the inverse metric at
constant contorsion then gives
\begin{equation}
 \GG_{(\mu\nu)} 
+    \nablac_\alpha \left[  P_{(\mu\nu)}^{\phantom{(\mu\nu)}\alpha}  - P^\alpha_{\;\;(\mu\nu)} \right] = 8\pi G T_{\mu\nu} ,
\label{ECSK_T}
\end{equation}
where the matter stress-energy tensor is given by $ T_{\mu\nu} =
-\frac{1}{2\sqrt{-g}}  \frac{\delta {\cal L}_m}{\delta
  g^{\mu\nu}}\big|_{K}$.  Variation with respect to the contorsion gives
\begin{equation}
 P_{\mu}^{\phantom{\mu}\alpha\beta}   = 8\pi G  \tau_{\mu}^{\phantom{\mu}\alpha\beta} ,
\label{ECSK_Ptau}
\end{equation}
where $\tau_\mu^{\;\;\alpha\beta} = -\frac{1}{\sqrt{-g}} \frac{\delta {\cal L}_m}{\delta K^\mu_{\;\;\alpha\beta}}$ is the spin angular momentum tensor of matter.
Due to the anti-symmetry of the contorsion in the first and third
indices we also have that $\tau_{\alpha\mu\beta} =
-\tau_{\beta\mu\alpha}$. Note, however, that the metricity assumption also means that the Palatini tensor simplifies to
\be
P_{\alpha\mu\beta} = S_{\mu\alpha\beta} + 2 g_{\mu[\alpha} S_{\beta]}  ,
\label{ECSK_palatini}
\ee
which is also antisymmetric in the first and third indices. Hence no
inconsistency arises from Eq. (\ref{ECSK_Ptau}), as it does in the general metric-affine case.
Now, Eq. (\ref{ECSK_T}) only determines the symmetric part of the
Einstein tensor, but we also need the anti-symmetric part. Using
Eqs. (\ref{ECSK_palatini}) and (\ref{Ricci_asym}) gives the anti-symmetric part of the Einstein tensor as 
\begin{equation}
G_{[\mu\nu]} = R_{[\mu\nu]} = \nablas_\alpha P_{\mu\;\;\nu}^{\;\;\alpha},
\label{GG_anti}
\end{equation}
where $\nablas_\mu = \nablag_\mu + 2 S_\mu$.
Equations (\ref{ECSK_T}), (\ref{ECSK_Ptau}) and (\ref{GG_anti}) form a consistent set of field equations for the ECSK theory. 

We can proceed one further step, however, and consider the torsion rather than the contorsion as the 2nd independent variable. This leads to the definition of the spin-energy potential, 
$\mu_{\nu}^{\;\;\alpha\beta} \equiv  -\frac{1}{\sqrt{-g}} \frac{\delta {\cal L}_m}{\delta S^\mu_{\;\;\alpha\beta}}$.
 Due to the antisymmetry of the torsion in the last two indices,
the spin-energy potential also obeys $\mu_{\nu\alpha\beta} = -\mu_{\nu\beta\alpha}$. It is straightforward to show that the spin-energy potential  and the spin angular momentum tensors
are related to each other by
\begin{equation}
 \tau_{\alpha\mu\beta} = -\mu_{[\alpha\beta]\mu} 
\end{equation}
and
\begin{equation}
\mu_{\mu\alpha\beta} = \tau_{\mu\alpha\beta}  - \tau_{\beta\mu\alpha} + \tau_{\alpha\beta\mu}.
\end{equation}
Carrying out the variation this way defines a new stress-energy
tensor, $\sigma_{\mu\nu}$, at constant $S^\mu_{\;\;\alpha\beta}$ by
$\sigma_{\mu\nu} \equiv -\frac{1}{2\sqrt{-g}}  \frac{\delta {\cal
    L}_m}{\delta g^{\mu\nu}}\big|_{S}$. This is related to $T_{\mu\nu}$ by
\begin{eqnarray}
\sigma_{\mu\nu} = T_{\mu\nu} -2 \left[ S_{\beta\alpha(\nu}  \tau_{\mu)}^{\phantom{\mu}\alpha\beta} 
+ S_{\alpha\beta(\nu}   \tau_{\mu)}^{\phantom{\mu}\alpha\beta}
  -  \tau^{\beta\alpha}_{\;\;\;\;(\mu} S_{\nu)\alpha\beta} 
 -  \tau^{\alpha\;\;\beta}_{\;\;(\mu} S_{\nu)\beta\alpha} 
\right].
\end{eqnarray}
The field equations obtained from varying the action with respect to the metric are
\begin{eqnarray}
&&
 \GG_{(\mu\nu)} 
+    \nablac_\alpha \left[  P_{\mu\nu}^{\phantom{\mu\nu}\alpha}  +   P_{\nu\mu}^{\phantom{\mu\nu}\alpha} \right]
\nonumber
\\
&&
+ 2\left[ P^{\beta\alpha}_{\;\;\;\;(\mu}  S_{\nu)\alpha\beta} 
+  P^{\alpha\;\;\beta}_{\;\;(\mu}  S_{\nu)\beta\alpha} 
- S_{\beta\alpha(\nu}   P_{\mu)}^{\phantom{\mu}\alpha\beta}
- S_{\alpha\beta(\nu}   P_{\mu)}^{\phantom{\mu}\alpha\beta}
\right]
 = 8\pi G \sigma_{\mu\nu} .
\end{eqnarray}
After some algebra, and using Eq. (\ref{GG_anti}) to form the full
asymmetric Einstein tensor, we find that the Einstein equations
simplify to 
\begin{equation}
\GG_{\mu\nu} = 8\pi G \Sigma_{\mu\nu},
\label{ECSK_field}
\end{equation}
where 
\begin{equation}
\Sigma_{\mu\nu} = \sigma_{\mu\nu} +\nablas_\alpha  \mu_{\mu\;\;\nu}^{\;\;\alpha}
\end{equation}
is a new stress-energy tensor. This new stress-energy tensor has a
very important interpretation~\cite{ECSK}: It is none other than the
canonical stress-energy tensor. Thus, in the ECSK theory the usual
symmetrisation procedure of the canonical stress-energy tensor to
obtain the stress-energy tensor that enters the Einstein equations is not necessary.

Equations (\ref{ECSK_field}) and (\ref{ECSK_Ptau}) form a consistent
set of field equations that determine the geometry of the space-time
from the matter stress-energy 
and spin distribution. They are supplemented by the conservation laws
\begin{equation}
\nablas_\nu\Sigma_\mu^{\;\;\nu} = 2 \Sigma_\alpha^{\;\;\beta}
S^\alpha_{\;\;\mu\beta} - \tau_\alpha^{\phantom{\alpha}\beta\lambda}
R^\alpha_{\;\;\mu\beta\lambda} 
\end{equation}
and
\begin{equation}
\nablas_\alpha \tau^{\;\;\alpha}_{\mu\;\;\nu} = \Sigma_{[\mu\nu]}.
\end{equation}

Let us now discuss a further property of the ECSK theory. By
inspecting of the field equations (\ref{ECSK_field}) and (\ref{ECSK_Ptau}) we notice that there are no derivatives
of the torsion appearing  anywhere.  Thus, the torsion in the ECSK theory is non-dynamical. Its presence is directly given in terms of the spin-angular momentum of
matter by Eq. (\ref{ECSK_Ptau}). 
This means that it completely vanishes in vacuum, or in cases where matter does not couple to contorsion (e.g. scalar and gauge fields).
 Since $S^\alpha_{\phantom\alpha\mu\nu}$ is algebraically determined one can eliminate it from all of the field equations. 
The final form of the field equations after eliminating torsion is
then found to be
\begin{eqnarray}
G_{\mu\nu} &=& 8\pi G T_{\mu\nu}
 +  (8\pi G)^2  \bigg\{
 2 \tau_{\beta\alpha(\mu} \tau^{\alpha\phantom{\nu)}\beta}_{\phantom{\alpha}\nu)} 
-2\tau_{\alpha} \tau^{\alpha}_{\;\;(\mu\nu)} 
 -  \tau_{\alpha\mu\beta} \tau^{\alpha\;\;\beta}_{\;\;\nu} 
\nonumber 
\\
&&
+ \frac{1}{2} \left[2 \tau^{\mu\alpha\beta} \tau_{\alpha\mu\beta} 
+ \tau^{\alpha\mu\beta} \tau_{\alpha\mu\beta} 
-2 \tau_\alpha \tau^\alpha
\right] g_{\mu\nu}
\bigg\},
\label{ECSK_elim}
\end{eqnarray}
where $G_{\mu\nu}$ is the Einstein tensor of $g_{\mu\nu}$, obtained in
the usual way from the Levi-Civita connection, and $\tau_\mu =
\tau^\beta_{\;\;\beta\mu}$.
\newline
\newline
\noindent
{\it Consequences of the  ECSK theory}
\newline

We introduced the ECSK theory as a minimal modification of GR through
the introduction of torsion and the application of the Palatini procedure. As was shown by Kibble~\cite{Kibble},
and then later by Hehl \etal~\cite{ECSK}, the ECSK theory emerges as
the minimal description of space-time when one constructs a local gauge
theory of the Poincar{\'e} group. 
There are, however, a number of unsatisfactory features of this
interpretation. Quantisation of the theory fares no better than in
GR. It is still non-renormalizable (the spin contact interactions are
reminiscent of 4-fermion interaction terms in quantum field theory),
and the torsion has vanishing canonical momentum, which makes it hard
to construct a quantum description. 
The reader is referred to~\cite{ECSK}  for more details.

One may then ask {\it when is the ECSK theory different from GR}? By
inspection, Eq. (\ref{ECSK_elim}) tells us that the spin-potential
modifications enter with an additional power of the Planck mass. 
Thus, we only expect the ECSK theory to deviate from GR at very high
spin-densities of matter. For electrons the critical mass density such
that spin effects are important is 
$\sim 10^{38} $kg~m$^{-3}$ while for neutrons it is $\sim 10^{45}
$kg~m$^{-3}$~\cite{HehlVonDerHeyde1973}. Conditions for such high
densities can exist in the early universe.

Cosmology with spin and torsion has been studied in the hope that the
cosmological singularity may be avoided. However, it appears that only
under very unrealistic circumstances can one avoid the cosmological
singularity in the ECSK theory
(see~\cite{ECSK,Kerlick1976,Tafel1975,Kuchowicz1975}).

\subsubsection{Scalar-Tensor-Vector Theory}

As a matter of completeness, we will briefly present the
Scalar-Tensor-Vector theory of gravity (STVG)  proposed in
\cite{Moffat2006}. STVG is a theory that contains a vector field,
$\phi_\mu$, three scalar fields, $G$, $\omega$ and $\mu$, and a
metric, $g_{\alpha\beta}$. A key characteristic of this theory is that
the modified acceleration law for weak gravitational fields has a
repulsive Yukawa force, as well as the normal Newtonian force. 
The action for STVG takes the form
\begin{eqnarray}
S=S_{Grav}+S_{\phi}+S_S+S_M,
\end{eqnarray}
where
\begin{eqnarray}
S_{Grav}&=&\frac{1}{16\pi}\int
d^4x\sqrt{-g}\left[\frac{1}{G}(R+2\Lambda)\right], \\ 
S_\phi &=&\int d^4
x\sqrt{-g}[\omega(\frac{1}{4}B^{\mu\nu}B_{\mu\nu}+V(\phi))], \\ 
S_S=&=&\int d^4
x\sqrt{-g}\left[\frac{1}{G^3}\left(\frac{1}{2}g^{\mu\nu}\grad_\mu
G\grad_\nu G-V(G)\right)\right] \nonumber\\ 
& & \int d^4
x\sqrt{-g}\left[\frac{1}{G}\left(\frac{1}{2}g^{\mu\nu}\grad_\mu
\omega\grad_\nu \omega-V(\omega)\right)\right] \nonumber \\ 
& &\int d^4 x\sqrt{-g}\left[\frac{1}{\mu^2
G}\left(\frac{1}{2}g^{\mu\nu}\grad_\mu \mu\grad_\nu
\mu-V(\mu)\right)\right], 
\end{eqnarray}
and where $B_{\mu\nu}=\partial_\mu\phi_\nu-\partial_\nu\phi_\mu$. 

This action has been studied in \cite{Moffat2006} where the full field
equations, the weak field limit, and cosmological solutions are
presented. As would be expected from the complexity of the action, it
is difficult to completely solve this system in the detail required to
make accurate predictions. Hence, the author of \cite{Moffat2006} has
posited a certain number of scaling relations that translate into
spatial dependences for $G$, $\omega$ and $\mu$. It is argued in
\cite{Moffat2006} that the classical action for this theory could be
considered as an effective field theory for a
renormalisation-group-flow-quantum-gravity scenario. The reader is
referred to \cite{Moffat2006} for further details.

\newpage 

\section{Higher Derivative and Non-Local Theories of Gravity}
\label{HD}

Recall from Section \ref{sec:lovelock-thm} that General Relativity
represents the most general theory describing a single metric that
in four dimensions has field equations with at most second-order
derivatives \cite{love1, love2}.  One way to extend GR is therefore to
permit the field equations to be higher than second order. Indeed,
such a generalisation might be considered desirable as it will cause
the graviton propagator to fall off more quickly in the UV, thereby
improving the renormalisability properties.  Modifying gravity in this
way, however, also has a number of drawbacks.  In particular, it can
introduce instabilities into the theory, such as ghost-like degrees of
freedom (see Sections \ref{sec:intro-ghosts}, \ref{frstability}, and
\ref{fgenother}, and \cite{woodard} for an overview).

In this section we consider those gravity theories that are
higher than second-order in derivatives.  Such theories can have
interesting phenomenology, and in many cases can be shown (or, at
least, argued to be) less susceptible to instabilities than one may
have initially suspected.  For example, if the higher derivatives act
only on what would otherwise be non-dynamical modes, then they may simply
render them dynamical, rather than automatically generating a
ghost. This is what happens in $f(R)$ gravity, where the higher order
derivatives act on the conformal mode that does not propagate in
GR. In Ho\v rava-Lifschitz gravity, as another example, one allows for
higher-order spatial derivatives, as opposed to higher time
derivatives, in order to guard against ghost-like instabilities.  In
both of these examples the theory can deviate considerably from
GR, while still maintaining some basic stability properties.

This section also includes galileon and ghost condensate
theories. Strictly speaking these are not higher-derivative theories
since their field equations are at most second order in
derivatives. In fact, the galileon theory in particular is constructed
with this in mind. Nevertheless we include them in this section as
both theories contain non-trivial derivative interaction terms. We will
not discuss theories with infinite derivatives, as occur in
string field theory, or p-adic string theory (see
\cite{Moeller-Zwiebach} for discussion of such theories).  

\subsection{$f(R)$ Theories} \label{sec:fR}

Fourth-order theories of gravity have a long history, dating back
to as early as 1918 \cite{weyl}, only a few years after the first
papers on General Relativity by Einstein.  These theories generalise
the Einstein-Hilbert action by adding additional scalar curvature
invariants to the action, or by making the action a more general function
of the Ricci scalar then the simple linear one that leads to
Einstein's equations.  Here we consider the latter of these options, a
choice that leads by Lovelock's theorem to fourth-order field
equations for anything except
the addition of a constant term to the gravitational Lagrangian.
Such theories, generically referred to as $f(R)$ theories of gravity,
have been intensively studied, and have a number of reviews dedicated
to them \cite{frrev0,frrev1,frrev2,frrev3,frrev4}.  This interest
was stimulated in the 1960s, 70s and 80s by the revelations that the
quantisation of matter fields in an unquantised space-time can lead to
such theories \cite{dewitt}, that $f(R)$ theories of gravity can have improved
renormalisation properties \cite{stelle}, and that they can lead to
a period of accelerating expansion early in the Universe's history
\cite{frstar}.  More recently they have been of considerable interest
as a possible explanation for the observed late-time accelerating
expansion of the Universe.

\subsubsection{Action, field equations and transformations}
\label{freqssec}

The $f(R)$ generalisations of Einstein's equations are derived from a
Lagrangian density of the form
\be
\mathcal{L}=  \sqrt{-g} f(R),  \label{density}
\ee
where the factor of $\sqrt{-g}$ is included, as usual, in order to
have the proper weight.  This is clearly about as simple a
generalisation of the Einstein-Hilbert density as one could possibly
conceive of.  The field equations derived from such an
action are automatically generally covariant and Lorentz invariant for
exactly the same reasons that Einstein's equations are.
Unlike the Einstein-Hilbert term, however, the field equations that
one obtains from the least action principle associated with
Eq. (\ref{density}) depend on the variational principle that one
adopts.  Different possibilities are the `metric variation' where the
connection is assumed to be the Levi-Civita one, the `Palatini
approach' in which Eq. (\ref{density}) is varied with respect to the
metric and connection independently, and the `metric-affine' approach in
which the same process occurs but the matter action is now taken to be a
functional of the connection as well as the metric.  In this section
we will mostly be concerned with the metric variational approach,
although we will also outline how the other procedures work below.
\newline
\newline
\noindent
{\it Metric variational approach}
\newline

Let us now derive the field equations in the metric variational
approach.  Integrating Eq. (\ref{density}) over a 4-volume, including
a matter term and varying with respect to $g_{\mu\nu}$ yields
\be
\label{frex}
\delta S = \int d \Omega \sqrt{-g} \Bigl[ \frac{1}{2} f g^{\mu\nu}
  \delta g_{\mu\nu} + f_R \delta R +\frac{\chi}{2} T^{\mu\nu} \delta g_{\mu\nu} \Bigr],
\ee
where $f_R$ means the functional derivative of $f$ with respect to
$R$, $\chi$ is a constant, and $T_{\mu\nu}$ is the energy-momentum tensor
defined by a variation of the matter action with respect to $g_{\mu\nu}$ in the usual way.
Assuming the connection is the Levi-Civita one we can then write
\be
f_R \delta R \simeq  - [ f_R R^{\mu\nu} +f_{R ; \rho\sigma} (g^{\mu\nu} g^{\rho\sigma} -
g^{\mu \rho} g^{\nu \sigma})] \delta g_{\mu\nu}, 
\ee
where $\simeq$ is used here to
mean equal up to surface terms \cite{surfaceterms}. Looking for a stationary point of the
action, by setting the first variation to zero, then gives
\begin{equation}
\label{R16}
f_R R_{\mu\nu} - \frac{1}{2} f g_{\mu\nu} - f_{R;\mu\nu}+g_{\mu\nu} \square f_R=
\frac{\chi}{2} T_{\mu\nu} .
\end{equation}
These are the $f(R)$ field equations with which we will be primarily concerned
in this section.  It can be seen that for the special case $f=R$ the LHS of
Eq. (\ref{R16}) reduces to the Einstein tensor, and the field
equations are second-order in derivatives of the metric.  For all
other cases, except an additional constant, the equations in
(\ref{R16}) are fourth-order in derivatives.
\newline
\newline
\noindent
{\it Conformal transformation in the metric variational approach}
\newline

As with scalar-tensor theories, the $f(R)$ theories of gravity derived
from the metric variational approach can be
conformally transformed into a frame in which the field equations
become those of General Relativity, with a minimally coupled scalar
field.  This is sometimes referred to as `Bicknell's theorem' in the
case of $f(R)$ theories, particularly when the minimally coupled
scalar field is in a quadratic potential \cite{bick}.  In the general
case we consider conformal transformations of the form \cite{cot, maeda}
\begin{equation}
\bar{g}_{\mu\nu} = f_R g_{\mu\nu},
\end{equation}
together with the definition
\begin{equation}
\phi \equiv \sqrt{\frac{3}{\chi}} \ln f_R,
\end{equation}
which allows the field equations from the metric variational principle,
Eqs. (\ref{R16}), to be transformed into
\begin{equation}
\bar{R}_{\mu\nu}- \frac{1}{2} \bar{g}_{\mu\nu} \bar{R} = \frac{\chi}{2}
\left( \phi_{,\mu} \phi_{,\nu}-\frac{1}{2} \bar{g}_{\mu\nu}
\bar{g}^{\rho \sigma}
\phi_{,\rho} \phi_{,\sigma} - \bar{g}_{\mu\nu} V \right) +\frac{\chi}{2} \bar{T}_{\mu\nu}.
\end{equation}
Here $\bar{T}_{\mu\nu}$ is a non-conserved energy-momentum tensor, and
we have defined
\begin{equation}
V =V(\phi ) \equiv \frac{(R f_R-f)}{\chi f_R^2}.
\end{equation}
Theories derived from an action of the form (\ref{density}) can therefore
always be conformally transformed into General Relativity with a
massless scalar field (as long as $f_{R}\neq 0$), and a non-metric
coupling to the matter fields.
\newline
\newline
\noindent
{\it Legendre transformation in the metric variational approach}
\newline

As well as conformal transformations, one can also perform Legendre
transformations on $f(R)$ theories in the metric variational approach.  Such transformations allow
the field equations of $f(R)$ to take the form of a scalar-tensor
theory (albeit it a slightly strange one).  These transformed
theories maintain the universal metric coupling of the matter
fields, unlike the case of conformal transformations.

The first step here is to notice that the Eq. (\ref{density}) can be
written in the equivalent form
\be
\mathcal{L}=  \sqrt{-g} \left[ f(\chi) + f^{\prime}(\chi) (R-\chi)
  \right],
\label{density2}
\ee
where $\chi$ is a new field, and the prime denotes differentiation.
Variation with respect to $\chi$ then gives 
\be
f^{\prime \prime}(\chi) (R-\chi)=0,
\ee
so that $\chi=R$ for all $f^{\prime \prime}(\chi) \neq 0$.
Substitution of this result back into Eq. (\ref{density2}) then
immediately recovers Eq. (\ref{density}), showing that the two Lagrangian
densities are indeed equivalent.  What is more, the special case
$f^{\prime \prime}(\chi)=0$ can be seen to correspond to the
Einstein-Hilbert action.

Now, if we make the definition
\be
\phi \equiv f^{\prime}(\chi),
\ee
and assume that $\phi(\chi)$ is an invertible function, then we can
define a potential
\be
\Lambda (\phi)\equiv \frac{1}{2} \left[ \chi(\phi) \phi - f(\chi(\phi)) \right].
\ee
In terms of this new scalar field we can then write the density
(\ref{density2}) as
\be
\mathcal{L}=  \sqrt{-g} \left[ \phi R - 2 \Lambda(\phi) \right],
\ee
which is clearly just a scalar-tensor theory, as specified in
Eq. (\ref{ST2}), with vanishing coupling constant, $\omega =0$.  As we
have not transformed the metric, the coupling of this field to
any matter fields that are present remains unchanged.
\newline
\newline
\noindent
{\it The Palatini procedure}
\newline

Starting again from the density (\ref{density}) we can also proceed in a
entirely different way to the metric variational approach just described.
Instead of assuming the connection from which the Ricci scalar is
constructed is the Levi-Civita one, we could instead treat the metric and
connection as independent fields.
For the case of General Relativity a variation with respect to the
connection then simply results in the
connection being shown to be the Levi-Civita one, so that the
difference between the metric variational approach and the Palatini
approach is a semantic one.  For the case of $f(R)$
theories, however, the Palatini approach leads to an entirely
different set of field equations.

Starting with an integral of Eq. (\ref{density}) over some 4-volume,
and extremising with respect to $g_{\mu\nu}$ now gives
\be
\label{pal1}
f_R R_{\mu\nu} - \frac{1}{2} g_{\mu\nu} f = \frac{\chi}{2} T_{\mu\nu},
\ee
where $T_{\mu\nu}$ is once again the usual energy-momentum tensor.  In
this expression $R_{\mu\nu}$ is now defined independently from the metric,
and $R$ should be taken to be given by $g^{\mu\nu} R_{\mu\nu}$.  The next
step is the variation of Eq. (\ref{density}) with respect to
$\Gamma^{\mu}_{\phantom{\mu}\nu \sigma}$, which results in
\be
\label{pal2}
\left( \sqrt{-g} g^{\mu\nu} f_R \right)_{;\sigma} =0,
\ee
where the covariant derivative here should be understood to be taken with
respect to $\Gamma^{\mu}_{\phantom{\mu}\nu \sigma}$, which is not the Levi-Civita
connection unless $f_R=$constant (as is the case in GR).  It is
remarkable that the field equations (\ref{pal1}) do not involve any
derivatives of the metric, and only first derivatives of the
connection.  These are a different set of field equations to
Eq. (\ref{R16}), and should be considered a different set of theories to the
$f(R)$ in which $R$ is {\it a priori} taken to be constructed
from the Levi-Civita connection.

It can be noted from Eq. (\ref{pal2}) that even if the connection is
not compatible with the metric $g_{\mu\nu}$, we can still define a new
metric, $\bar{g}_{\mu\nu} = f_R g_{\mu\nu}$, with which it is compatible.
Rewriting the full field equations under this conformal
transformation we see that we recover General Relativity with a minimally
coupled scalar field in a potential, but no kinetic term:
\be
\mathcal{L} = \sqrt{-\bar{g}} \left[ \bar{R} - 2 V(\phi) \right],
\ee
where $\phi\equiv f_R$ and
$
V(\phi) = (R(\phi) \phi -f(\phi))/2\phi^2.
$
Here $R(\phi)$ and $f(\phi)$ are given by inverting the
definition of $\phi$, and $\bar{R}$ should be understood to be
constructed from the
metric connection compatible with $\bar{g}_{\mu\nu}$.  Transforming back
to the original conformal frame this theory then can be shown to be
equivalent to a scalar-tensor theory with $\omega=-3/2$ and $\Lambda =
(R f_R-f)/2$ \cite{Olmo05}.

While being an interesting variant on the metric variational
incarnation of the $f(R)$ theories, there are a number of very severe
problems in proceeding with the Palatini procedure in this
way.  Not least of these is the apparent ill-posed nature of the
Cauchy problem in the presence of most matter fields, which is
discussed in \cite{frcauchy}.  Without a well-posed
initial value problem $f(R)$ gravity in the Palatini formalism lacks
predictive power, and hence is not a good candidate for a viable
theory of gravity.  Furthermore, the Palatini approach to $f(R)$
gravity also appears to introduce problematic strong couplings between
gravity and matter fields at low energies \cite{frpal1,frpal2},
and singularities in systems that are usually well described by weak
fields \cite{frpal3,frpal4,frpal5}.  For these reasons
we will focus on $f(R)$ theories in the metric variational approach in the
sections that follow.  

For further details of the Palatini approach to $f(R)$
gravity, and the results that follow from it, the reader is referred
to \cite{frrev2}.  For studies of weak field gravity in the Palatini
formalism the reader is
referred to
\cite{frp1,frp2,frp3,Olmo05,frp5,frp6,frp7,olmo07,frp9,frp10,frp11,frp12,frp13},
and for cosmology to
\cite{frpc1,frpal1,frpc3,frpc4,frpc5,frpc6,frpc7,frpc8,frpc9,frpc10,frpc11,frpc12,nocar3,frpc14,frpc15,frpc16,frpc17,frpc18,frpc19}.  An interesting class of
theories that interpolate between the Palatini approach to $f(R)$
theories and the metric variational approach to $f(R)$ theories is
investigated in \cite{Cmodels1,Cmodels2}.
\newline
\newline
\noindent
{\it The metric-affine approach}
\newline

One further approach to the $f(R)$ theories of gravity is the
`metric-affine' formulation.  Here one again considers the metric and
connection to be independent, as in the Palatini procedure, but now
allows the matter action to be a function of both metric and
connection (rather than metric alone, as is the case in Palatini and
metric variational formalisms).  The relevant action for the theory
then becomes \cite{Sot07}
\be
\label{ma}
S= \int d\Omega \sqrt{-g} f(R)
+S_m(g_{\mu\nu},\Gamma^{\mu}_{\phantom{\mu}\nu \sigma},\Psi),
\ee
where $R=g^{\mu\nu} R_{\mu\nu}$, and $R_{\mu\nu}$ is taken to be a function of the
connection only, as in the Palatini procedure.  One can therefore think of the
action (\ref{ma}) as a generalisation of the Palatini action, which is
recovered when the dependence of the matter action, $S_m$, on
$\Gamma^{\mu}_{\phantom{\mu}\nu \sigma}$ vanishes.

As is the case in General Relativity, the invariance of the
Ricci scalar under the projective transformation
$\Gamma^{\mu}_{\phantom{\mu}\nu \sigma} \rightarrow
\Gamma^{\mu}_{\phantom{\mu} \nu \sigma}
+\lambda_{\nu} \delta^{\mu}_{\phantom{\mu} \sigma}$ can lead to inconsistency of the
field equations, as matter fields do not generically exhibit an
invariance of this type.  This invariance can be cured by adding to
the action an additional Lagrange multiplier term of the form
$S = \int d\Omega \sqrt{-g} B^{\mu} \Lambda^{\nu}_{\phantom{\mu} \left[\nu
    \sigma \right]}$, 
and results in the field equations
\be
\label{ma1}
f_R R_{\mu \nu} - \frac{1}{2} g_{\mu\nu} f = \frac{\chi}{2} T_{\mu\nu},
\ee
together with $\Gamma^{\mu}_{\phantom{\mu} [\mu \nu]} = 0$, and
\ba
\frac{1}{\sqrt{-g}} \left[ \left(\sqrt{-g} f_R g^{\mu \sigma} \right)_{;\sigma}
    \delta^{\nu}_{\phantom{\nu}\rho} -\left( \sqrt{-g} f_R g^{\mu\nu}
      \right)_{;\rho} \right] + 2 f_R g^{\mu \sigma} \Gamma^{\nu}_{\phantom{\nu}
      \left[\sigma \rho\right]} \nonumber \\
= \frac{\chi}{2} \left[ \Delta_{\rho}^{\phantom{\rho}
	  \mu\nu} - \frac{2}{3} \Delta_{\sigma}^{\phantom{\sigma} \sigma [ \nu}
	  \delta^{\mu ] }_{\phantom{\mu]}\rho} \right],
\label{ma2}
\ea
where $\Delta_{\mu}^{\phantom{\mu} \nu \rho} \equiv -(2/\sqrt{-g}) \delta
S_m/\delta \Gamma^{\mu}_{\phantom{\mu} \nu \rho}$.  It can then be shown that
$\Delta_{\mu}^{\phantom{\mu} [\nu \rho]}=0$ corresponds to a
vanishing of the torsion, and $\Delta_{\mu}^{\phantom{\mu} (\nu \rho)} \neq 0$
introduces non-zero non-metricity \cite{Sot07}.  The metric-affine
approach has not been studied as intensively as the other approaches
to $f(R)$ gravity that we have already discussed, and will not feature heavily
in the sections that follow.

\subsubsection{Weak-field limit}
\label{frweak}

The weak field limit of $f(R)$ theories of gravity has been
extensively studied in the literature.  Here we will first consider
perturbations about Minkowski space, and then perturbations around
de Sitter space, time dependent backgrounds, and inhomogeneous backgrounds.
\newline
\newline
\noindent
{\it Perturbations about Minkowski space}
\newline

The first attempt at finding the
Newtonian limit of an $f(R)$ theory appears to have been
performed by Pechlaner and Sexl for the case of $f=R+\alpha R^2$
\cite{ps}.  The first step 
here is to write down the perturbed line-element as
\be
ds^2=-(1+2 \psi) dt^2 +(1-2 \phi)(dx^2+dy^2+dz^2),
\ee
neglecting time derivatives and second order terms in $\phi$ and
$\psi$ one then finds that the Ricci scalar, $R=-2 \Delta
\psi+4\Delta \phi$, obeys
\be
\label{r2trace}
6 \alpha \delta R- R = -\frac{\chi}{2} \rho,
\ee
and that the potential $\psi$ obeys
\be
\label{r2field}
(1+6 \alpha \Delta) \Delta \psi = \frac{\chi}{4} (1+8 \alpha \Delta )\rho,
\ee
where $\Delta$ is the Laplacian on Euclidean 3-space.
The derivatives of $\rho$ in this last equation occur due to replacing
terms containing $\phi$ with those obtained from taking derivatives of
Eq. (\ref{r2trace}), and do not occur in the actual field equations themselves.

Inserting a delta function source, $\rho=m \delta$, and integrating
Eq. (\ref{r2field}), using the solution to Eq. (\ref{r2trace}) to find
$\phi$, one then gets the solutions \cite{ps}
\ba
\label{r2phi}
16 \pi \psi &=& -\frac{\chi m}{r} \left(1+\frac{e^{-m_0 r}}{3}\right),\\
16 \pi \phi &=& -\frac{\chi m}{r} \left(1-\frac{e^{-m_0 r}}{3}\right),
\label{r2psi}
\ea
where boundary conditions at infinity have been imposed to eliminate
exponentially increasing modes, and where we have defined the mass
\be
m_0^2 \equiv \frac{1}{6\alpha}.
\ee
Mass terms similar to this continue to exist for more general
theories, and more general situations, as we will outline shortly.
One can see straight away that for large masses, with $m_0 \gg r^{-1}$, the
Yukawa potentials in Eqs. (\ref{r2phi}) and (\ref{r2psi}) are
exponentially suppressed, and we recover $\psi=\phi \propto - m/r$.  In the
limit of small masses, however, when $m_0 \ll r^{-1}$, we instead find
that $\psi = 2 \phi \propto -m/r$.  For the case of $f=R+\alpha R^2$
gravity we therefore already expect the PPN parameter $\gamma$ to be
$1$ when the mass of the scalar degree of freedom is large, and $1/2$
when it is small.

One is, of course, also interested in other functions of $f(R)$, and can consider
the case of analytic $f(R)$ theories that can be expanded as
\be
\label{franal}
f(R) = \sum_{i=1}^{\infty} c_i R^i,
\ee
where the $c_i$ are constants. To full post-Newtonian order the weak field solution in the
presence of a perfect fluid is then given in full generality as \cite{cliftonppn}
\begin{align}
\nonumber
g_{00} = &-1 + \frac{2}{c_1} \left(U+ c_2 R \right)
-\frac{2}{c_1^2} U^2 + 2 \frac{c_2^2}{c_1^2} R^2 - \frac{16 c_2}{3
  c_1^2} U R - \frac{7}{18 \pi c_1} \mathcal{V}(UR) \nonumber \\ &+ \frac{3 c_2}{4 \pi c_1} \mathcal{V}(R^2) +
\frac{64}{9 c_1^2} \mathcal{V}(\rho U) - \frac{44 c_2}{3 c_1^2} \mathcal{V}( \rho R)-
\frac{40 c_2}{3 c_1^3} \mathcal{V}(\grad \rho \cdot \grad U) \nonumber \\ & + \frac{40
  c_2^2}{c_1^3} \mathcal{V}(\grad \rho \cdot \grad R) + \frac{2}{c_1} \mathcal{V}(\rho
\Pi) + \frac{4}{c_1} \mathcal{V}(\rho v^2) + \frac{6}{c_1} \mathcal{V}(P) + \frac{1}{6 \pi
  c_1} X(UR)  \nonumber \\ &- \frac{1}{4\pi} \left( \frac{c_2}{c_1}
- \frac{c_3}{2 c_2} \right)X(R^2)
- \frac{4}{3 c_1^2} X(\rho U)+
\frac{8 c_2}{3 c_1^2}X( \rho R) \nonumber \\ & + \frac{8 c_2}{3
  c_1^3}X(\grad \rho \cdot \grad U)  
- \frac{8 c_2^2}{c_1^3}X(\grad \rho
\cdot \grad R) -\frac{2}{c_1} X(P)+ \frac{2}{3c_1}X(\rho \Pi)\\ \label{g0i}
g_{0i} = &-\frac{7 V_i}{2 c_1} - \frac{W_i}{2 c_1} + \frac{X(\rho v_i)}{6 c_1}
 - \frac{Y_i}{6 c_1} -\frac{Z_i}{6\sqrt{6 c_1 c_2}}
\\ \label{gij}
g_{ij} = &\left(1+\frac{2}{c_1} \left(U -c_2 R \right) \right) \delta_{ij}
\end{align}
where $U$ is the usual Newtonian potential, and the other potentials
are defined as
\begin{table*}[!h]
\begin{tabular}{l l}
$V_i \equiv \int \frac{\rho(\mathbf{x}^\prime) v_i(\bf{x}^\prime)}{\vert
  \bf{x}-\bf{x}^\prime \vert} d^3x^\prime$ & $W_i \equiv \int \frac{\rho(\mathbf{x}^\prime) (\mathbf{v}(\mathbf{x}^\prime) \cdot
  (\mathbf{x}-\mathbf{x}^\prime)) (x-x^\prime)_i}{\vert \mathbf{x}-\mathbf{x}^\prime
  \vert^3} d^3x^\prime$ \\
$X(Q) \equiv \int \frac{Q^\prime e^{-\sqrt{\frac{c_1}{6 c_2}} \vert
  \bf{x}-\bf{x}^\prime \vert}}{\vert \mathbf{x}-\mathbf{x}^\prime
  \vert} d^3x^\prime$ & $Y_i \equiv \int \frac{\rho^\prime \mathbf{v}^\prime \cdot
  (\mathbf{x}-\mathbf{x}^\prime) (x-x^\prime)_i}{\vert
  \mathbf{x}-\mathbf{x}^\prime \vert^3 } e^{-\sqrt{\frac{c_1}{6 c_2}} \vert
  \bf{x}-\bf{x}^\prime \vert} d^3x^\prime$ \\
$\mathcal{V}(Q) \equiv \int \frac{Q^\prime}{\vert \mathbf{x}-\mathbf{x^\prime}
  \vert} dx^{\prime 3}$ & $Z_i \equiv \int \frac{\rho^\prime \mathbf{v}^\prime \cdot
  (\mathbf{x}-\mathbf{x}^\prime) (x-x^\prime)_i}{\vert
  \mathbf{x}-\mathbf{x}^\prime \vert^2 } e^{-\sqrt{\frac{c_1}{6 c_2}} \vert
  \bf{x}-\bf{x}^\prime \vert} d^3x^\prime$ \\
$\hat{\chi} \equiv \int \rho^\prime e^{-\sqrt{\frac{c_1}{6 c_2}} \vert
  \bf{x}-\bf{x}^\prime \vert} d^3x^\prime$ & $R = \frac{1}{3 c_2} \int \frac{\rho(\bf{x}^\prime)}{\vert
  \bf{x}-\bf{x}^\prime \vert} e^{-\sqrt{\frac{c_1}{6 c_2}} \vert
  \bf{x}-\bf{x}^\prime \vert} d^3x^\prime$,\\
\end{tabular}
\end{table*}

\noindent where $R$ is the Ricci scalar, $P$ is pressure, $\rho$ is
the rest-mass energy density, and $\Pi$ is the specific energy density
(as defined in \cite{tegp}).
The terms in Eqs. (\ref{g0i})-(\ref{gij}) that are functionals of
derivatives of $\rho$, $U$ and $R$ can be recast into a form where
such derivatives do not appear by further manipulations \cite{cliftonppn}.

From the above it can be seen that the results of General Relativity
are recovered, to the appropriate order, when $c_2\rightarrow c_3
\rightarrow 0$.  For non-vanishing values of the these constants a
large number of extra Yukawa potentials are present, and for large
enough values of $c_2$ we can again see that $\gamma\rightarrow 1/2$,
as the scalar degree of freedom
becomes massless.  

The study of gravitational waves about a Minkowski space background in
$f(R)$ theories has been 
undertaken by Berry and Gair in \cite{frwaves}.  Here the authors consider analytic
functions of the type prescribed by Eq. (\ref{franal}), and find that
an extra mode of oscillation is possible.  The gravitational waves
emitted by extreme-mass-ratio inspirals are then calculated, and the authors
conclude that current laboratory bounds (that result in $\vert c_2/c_1
\vert < 10^{-9} m^2$) mean that the extra oscillatory mode they find cannot be
excited by astrophysical sources.
\newline
\newline
\noindent
{\it Perturbations about de Sitter backgrounds}
\newline

As well as the usual expansions about Minkowski space, in order to
determine the post-Newtonian behaviour of a theory, one is also
interested in perturbations about other backgrounds.  Here we will
consider de Sitter space as a background.  This
is not meant to be an elucidation of cosmological perturbation theory,
but rather a consideration of weak field expansions as applicable to
systems such as the solar system and binary pulsars.  Minkowski space
is not always a stable solution of $f(R)$ theories of gravity that
attempt to produce self-accelerating cosmologies at late-times, and in
these cases time-dependent backgrounds, and de Sitter space
in particular, become of increasing interest for weak field studies.

Much work has been performed on establishing the weak field limit of
$f(R)$ theories about a de Sitter background, as relevant for theories
that try and account for late-time accelerating expansion without dark
energy, see e.g. \cite{frds1,frds2,frds3,frds4,frds5,frds6,frds7,frds8}.
The majority of these studies conclude that, in the absence of extra
mechanisms to mask such behaviour, one should expect to find
$\gamma=1/2$.  This was shown in an early paper on the subject
in \cite{chiba03}, and is the familiar limit of theories in
which a scalar degree of freedom has low mass.  Here we will briefly
sketch out the method by which such a result is found for more general
$f(R)$, following the approach of \cite{chiba}.

The first step here is to identify a de Sitter solution with constant
Ricci curvature $R=R_0=12 H_0^2$.  The line-element is then perturbed
as
\be
ds^2= - (1+2 \psi -H_0^2 r^2)dt^2+(1-2 \phi + H_0^2 r^2)dr^2 +r^2
d\Omega^2,
\ee
where we have chosen to present the de Sitter background using a
static coordinate patch.  We then proceed by perturbing the Ricci
scalar as
\be
R=R_0 +R_1,
\ee
where $R_1 \ll R_0 $.
The perturbative expansion then linearises all field equations with
respect to $\phi$, $\psi$, $R_1$ and $H_0^2 r^2$, and their derivatives, while
neglecting all time derivatives.  The case of spherical symmetry is
considered for simplicity.

To lowest order the trace of the field equations is now
\be
\label{frtrace}
\Delta R_1-\left( \frac{f_{R}-f_{RR} R_0}{3 f_{RR}} \right) R_1 =
-\frac{\chi \rho}{6 f_{RR}}  ,
\ee
where $f_R$ and $f_{RR}$ should be understood to be the value of these
quantities at $R=R_0$ (implicit here is an assumption that these
quantities are all of the same order of magnitude as $R_0$, and that $R_0$ in the
weak-field systems under consideration takes the same value as in the
cosmological background solution).  From cosmological
considerations the second term on the left-hand side of
Eq. (\ref{frtrace}) is then neglected, as the factor in brackets
corresponds to the mass squared of the scalar degree of freedom, 
which must be small compared to $r^{-1}$ in order to have late-time
accelerating expansion.  The resulting form of $R_1$ is therefore
found to be
\be
\label{r1fr}
R_1 \simeq \frac{\chi m}{24 \pi f_{RR} r},
\ee
where $m$ is the mass of the object at the centre of symmetry.
Applying the same approximations to the $(t,t)$ component of the field
equations results in
\be
\Delta \psi = \frac{\chi \rho}{2 f_R}-
\frac{R_1}{2}+\frac{f_{RR}}{f_R} \nabla^2 R_1,
\ee
which on substitution of the expression for $R_1$ gives to lowest order
\be
\psi \simeq -\frac{\chi m}{12 \pi f_R r}.
\ee
The remaining field equations then give $\phi$, to the same order of
approximation, as
\be
\phi \simeq -\frac{\chi m}{24 \pi f_R r} \simeq \frac{\psi}{2}.
\ee
This calculation has not been performed in the PPN gauge, which uses
an isotropic spatial coordinate system, but nevertheless one can
verify that when interpreted within the standard PPN framework
it does indeed give \cite{olmo07}
\be
\gamma = \frac{1}{2}.
\ee

In this section we have discussed de Sitter space as a
background to perturb around.  However, establishing whether de Sitter
space is, in fact, a  stable asymptotic solution of $f(R)$ theories,
and establishing the genericity of initial conditions that lead to de
Sitter space at late times, has not yet been discussed.  We will consider this subject
in following sections.
\newline
\newline
\noindent
{\it Perturbations about other backgrounds}
\newline

Having considered the maximally symmetric Minkowski space and de
Sitter space backgrounds, we can now also consider less symmetric
spaces to perturb around.  This enterprise is hindered by our ability
to find less symmetric solutions to the field equations (\ref{R16}).
We can, however, make progress with some simple cases.

If we consider $f(R)$ theories in which $f=R^{1+\delta}$ then one
can find exact non-static, homogeneous and isotropic vacuum solutions
\cite{capfrw}.
Such solutions can be shown to be, under certain conditions, stable
asymptotic attractors for the general class of spatially flat, vacuum
FLRW solutions \cite{capfrw2}.  They will be
discussed further in the cosmology section that follows.  For this
same class of $f(R)$ theories exact static, spherically
symmetric vacuum solutions are also known \cite{power},
which can also be seen to be generic asymptotes of the general
solution, with the specified symmetries applied \cite{power}.  We are
now in possession of two exact solutions, for the same theories, which
have less than maximal symmetry, and which can be used as backgrounds
to perturb around.


Spherically symmetric, time independent perturbations around
the homogeneous, time-dependent background found in \cite{capfrw} are
given to linear order by \cite{cliss}
\begin{equation}
\nonumber
ds^2=-(1+2 \psi) dt^2+a^2(t) (1-2\phi) (dr^2+r^2 d\Omega^2),
\end{equation}
where $a(t)=t^{\frac{\delta (1+2 \delta)}{(1-\delta)}}$. The
perturbations $\phi(r)$ and $\psi(r)$ are then found to be 
\begin{eqnarray*}
\psi &=& -\frac{c_1}{r} +\frac{2 c_2 (1-6 \delta+4 \delta^2+4
  \delta^3)}{(5-14 \delta-12 \delta^2)} r^2\\
\phi &=& -\frac{(1-2 \delta) c_1}{r} -\delta c_2 r^2.
\end{eqnarray*}
The corresponding perturbative analysis about the static, spherically
symmetric background found in \cite{power} gives
\ba
\nonumber
ds^{2}&=&-r^{2\delta \frac{(1+2\delta )}{(1-\delta )}}(1+V(r))dt^{2}\\
&&+\frac{
(1-2\delta +4\delta ^{2})(1-2\delta -2\delta ^{2})}{(1-\delta )^{2}}
(1+W(r))dr^{2}+r^{2} d\Omega^2,
\nonumber
\ea
where $V(r)$ and $W(r)$ are given in full generality by
$V(r) = c_3 V_1(r)+c_4 V_2(r)+ c_5 V_3(r)$ and 
$W(r) = -c_3 V_1(r)+c_4 W_2(r)+c_5 W_3(r)$,
where
\be
V_1 =  -r^{-\frac{(1-2 \delta+4 \delta^2)}{(1-\delta)}} 
\ee
and where $V_2$, $V_3$, $W_2$ and $W_3$ are oscillatory modes \cite{power}.
It can immediately be seen that the
form of the linearised perturbations around these two backgrounds
are quite different to each other, even though the field equations they obey
are identical. 
%
%
One can verify that an observer in the homogeneous,
time-dependent background should measure
\be
\nonumber
\beta=1 \qquad \text{and} \qquad \gamma=1-2 \delta,
\ee
which gives $\delta = -1.1 \pm 1.2 \times 10^{-5}$ when the constraint
derived from the Cassini space probe on $\gamma$ is applied \cite{bit}.
On the other hand an observer in the static background should measure an anomalous
extra gravitational force that goes like \cite{power}
\be
\nonumber
F \sim - \frac{\delta}{r}.
\ee
When subjected to constraints from observations of the perihelion
precession of Mercury, the presence of this extra force gives $\delta
= 2.7 \pm 4.5 \times 10^{-19}$ \cite{power}. 
The different forms of the gravitational potentials and forces in these
examples show that the choice of symmetries of the background
space-time can have important consequences for its weak field limit,
and the constraints on the underlying theory that are derived from it.
\newline
\newline
\noindent
{\it Chameleon mechanism}
\newline

As with a variety of other modified theories of gravity, the
`Chameleon Mechanism' has been applied to $f(R)$ theories.  This
mechanism was outlined in Section \ref{scalartensorsection2}, where a
summary of some of the accumulated literature on it was outlined.
Here we will simply reiterate the basic point that this mechanism
potentially allows a means by which theories with a light effective
scalar degree of freedom can evade solar system and binary pulsar
tests of the PPN parameter $\gamma$ by allowing the scalar to acquire
a higher mass in the locale of high mass concentrations,
such as the Sun and Earth.

The chameleon mechanism has been applied to specific $f(R)$ theories,
and its behaviour in this application considered further, in e.g. 
\cite{frcham1,frcham2,frcham3,frcham4,frcham5,cham10,frcham7,frcham8}.  As with other applications of the
Chameleon mechanism, if a `thin shell' is present then the mass of the
scalar degree of freedom in these theories is thought to be able to be
supressed enough to satisfy solar system tests of gravity.

\subsubsection{Exact solutions, and general behaviour}

Having discussed the weak field solutions which are of interest for
inferring constraints on $f(R)$ theories from observations of
gravitational phenomena in the solar system and binary pulsar systems,
let us now consider the behaviour of solutions to the full non-linear
field equations.  Here we will be concerned with exact solutions, which can
be obtained in some simple cases, as well as what can be inferred about the
general behaviour of non-linear solutions by other methods, and
what theorems can tell us about the behaviours that are possible.  The
relatively simple structure of $f(R)$ theories make such
considerations a feasible proposition. The geodesic deviation equation
in general $f(R)$ theories has been studied in \cite{GCT}.
\newline
\newline
\noindent
{\it Isolated masses, and black holes}
\newline

Progress was made into understanding the static spherically symmetric
vacuum solutions of $f(R)$ theories of gravity by Mignemi and Wiltshire in
\cite{frmw}.  These authors consider theories with higher powers of the Ricci scalar
added to the Einstein-Hilbert Lagrangian, and use a dynamical systems analysis to
determine the behaviour of the general solutions with the specified
symmetries. They find the asymptotes of these solutions, for a
variety of different cases, and show that the only static spherically
symmetric solutions of the theories they consider that have regular
horizons are the Schwarzschild solutions.  They further find that by
dropping the requirement of regularity the Schwarzschild solution is
also the only solution to these theories that is asymptotically flat.

The black hole `no-hair' theorems have been considered in the context of
$f(R)=R+ \alpha R^2$ theories by Whitt \cite{frwhitt}.
Collapse to a black hole, however, has not been as extensively studied in $f(R)$
theories of gravity as it has in Brans-Dicke theory, where direct
numerical calculations have been performed \cite{BDhole}.
Nevertheless, the same logic that tells us that the vacuum black hole
solutions of general 
relativity are the only vacuum black hole solution of Brans-Dicke
theory that can result from gravitational collapse, also suggest that
this should be 
true for $f(R)$ gravity.  In particular, most of the results of Hawking on
this subject only rely on inequalities of the form
\be
R_{\mu\nu} l^{\mu} l^{\nu} \geq 0,
\ee
where $l^{\mu}$ is a null vector, and not on the details of the field
equations themselves \cite{BDhole2}.  This null energy
condition is true of the conformally transformed scalar fields
equations in Brans-Dicke theory, and is also true in $f(R)$ gravity.
It therefore seems reasonable to expect that the vacuum black hole
solutions of $f(R)$ gravity should also be the same as the vacuum
black hole solutions of General Relativity.  The subject of black hole
radiation in the context of $f(R)$ gravity has been studied in
\cite{frbh1,frbh2,frbh3,frbh4,frbh10} and
\cite{BDentropy2} where it was shown that black holes in $f(R)$
gravity have a entropy given by
\be
S= \frac{f_R A}{4}.
\ee
The subject of the de Sitter no-hair theorems and isotropisation  in
$f(R)$ gravity has been considered by Barrow and Ottewill \cite{bo}
and Goheer, Leach and Dunsby \cite{gld}, where it was shown 
that flat FLRW isotropic points can exist in the phase plane of
Bianchi solutions, and that
de Sitter space can be a stable asymptote of $f(R)$ theories of
gravity.  One should, however, be aware that such behaviour depends on
the theory in question, and the 
initial conditions.  For example, for theories with negative powers of
$R$ in a series expansion of their Lagrangian one may generically
expect such terms to become important asymptotically.  In this case
accelerating power-law expansion is an attractor solution
\cite{higherp}.  This will be discussed further in the cosmology
section below.

As well as the black hole solutions of General Relativity, it is known
that other vacuum solutions to $f(R)$ theories of gravity that can
describe isolated masses also exist.  Due to the complicated nature of
the field equations in these theories, however, only a few exact solutions that
describe these situations have been found.  For the case of
$f(R)=R^{1+\delta}$ solutions are known that correspond to an
isolated mass in a homogeneous and time dependent background, and an
isolated mass in a static, spherically symmetric background.  The
former of these solutions is given by the line-element \cite{cliss}
\be
\label{frhomo}
ds^2= -A_1(r) dt^2 +a^2(t) B_1(r) (dr^2 +r^2 d \Omega^2),
\ee
where $a(t) = t^{\delta \frac{(1+2 \delta)}{(1-\delta)}}$, and where $A_1(r)$
and $B_1(r)$ are given by
\be
\nonumber
A_1(r) = \left(
\frac{1-\frac{C_1}{r}}{1+\frac{C_1}{r}}\right)^{\frac{2}{q}},
\qquad \text{and} \qquad
B_1(r) = \left(1+\frac{C_1}{r}\right)^4 A(r)^{q+2 \delta-1},
\ee
where $q^2 = 1-2 \delta +4\delta^2$.  The latter solution is given by
\cite{power}
\be
\label{frstat}
ds^2=-A_2(r)dt^{2}+\frac{dr^{2}}{B_2(r)}+r^2 d\Omega^2
\ee
where 
\ba
\nonumber
A_2(r) &=& r^{2\delta \frac{(1+2\delta )}{(1-\delta )}}+\frac{C_2}{r^{\frac{
(1-4\delta )}{(1-\delta )}}}, \\
B_2(r) &=& \frac{(1-\delta )^{2}}{(1-2\delta +4\delta ^{2})(1-2\delta (1+\delta
))}\left( 1+\frac{C_2}{r^{\frac{(1-2\delta +4\delta ^{2})}{(1-\delta )}}}
\right). \nonumber
\ea
The constants $C_1$ and $C_2$ appear in these solutions as mass
parameters, and it can be seen that both Eq. (\ref{frhomo}) and Eq. 
(\ref{frstat}) reduce to the Schwarzschild solution when $\delta
\rightarrow 0$.  The problem of static, spherically symmetric
solutions in general $f(R)$ has been considered in \cite{ncgd}, where
a covariant 
formalism was developed for studying the problem, and the
non-uniqueness of the Schwarzschild solution was demonstrated.  The
$\delta=1/4$ case of Eq. (\ref{frstat}) was rediscovered in
\cite{d14}.  Black holes coupled to Yang-Mills fields have been
studied in \cite{fextra3}, where an exact solution was found for the
case $f=\sqrt{R}$.

These solutions are interesting for a number of
reasons.  Firstly, they show that the generalisation of the solutions
of General Relativity to other theories of gravity is not always
unique; i.e. there can be multiple solutions in modified theories of
gravity that reduce to the same solution in the limit of general
relativity.  It may therefore be the case that one needs to understand
the symmetries of the background space-time to a greater extent than
is necessary in General Relativity, in order to fully understand which
solution should be used to model a given situation.
Secondly, Eq. (\ref{frhomo}) shows explicitly that
Birkhoff's theorem is not valid in general, when one considers
generalisations of Einstein's theory.  Spherically symmetric vacuum
solutions of modified theories of gravity can therefore be
time-dependent, which can lead to new phenomenology. Birkhoff's
theorem, in the context of $f(R)$ gravity, has been considered in
\cite{frbirkfar}.  Thirdly,
Eq. (\ref{frstat}) displays non-trivial asymptotic behaviour as $r
\rightarrow \infty$. Such behaviour is unexpected in general
relativity, and again opens the window to new phenomenology.  The
results of Mignemi and Wiltshire \cite{frmw} even suggest that such
behaviour is generic.  A fourth point is that the solution given in
Eq. (\ref{frhomo}) has been shown in \cite{farsing} to exhibit a naked
singularity.  This has clear implications for the applicability of the
cosmic censorship hypothesis to modified theories of gravity.
The Misner-Sharp energy in spherically symmetric space-times is
considered in \cite{Ohta20}.
\newline
\newline
\noindent
{\it Cosmological solutions}
\newline

A variety of cosmological solutions in $f(R)$ theories of gravity are
known, and have had their stability analysed.  Here we will briefly
review and provide references to studies of these solutions.

The conditions for existence and stability of de Sitter solutions in
$f(R)$ gravity appears to have been first studied in \cite{bo}.  One can show that for any theory for which there 
is a value of $R$ which satisfies
\be
\label{frds}
f_R (R_{dS}) R_{dS}=2 f(R_{dS})
\ee
there exists a de Sitter solution with $R_{dS}=4 \Lambda$.  The
stability of de Sitter solutions in $f(R)$ gravity was studied in
\cite{bo,frex6,frdsfar}.  These solutions are of obvious importance for cosmology at
both early and late times.  One can note that with $f(R)\propto R^2$
Eq. (\ref{frds}) is satisfied with any value of $R$.  Theories with
$R^2$ terms in their Lagrangian's have been studied extensively, due to
the naturalness of adding an $R^2$ term to the Einstein-Hilbert
action, and due to their improved renormalisation properties \cite{stelle}.
They were also introduced and studied by Starobinsky for cosmological
purposes, and in particular their ability to give rise to an early
non-singular period of accelerating expansion in a natural way \cite{frstar}.

Less symmetric cosmological solutions than de Sitter space can also be
found for some $f(R)$ theories.  In particular, theories of the type
$f(R)=R^{1+\delta}$ are again of interest here, as they admit simple exact
solutions.  As mentioned in the preceding section, a power-law exact
solution for a spatially flat vacuum FLRW solution is known to be given
by \cite{capfrw} 
\be
\label{rnfrw}
a(t) = t^{\delta \frac{1+2 \delta}{1-\delta}  },
\ee
and a spatially flat solution in the presence of a perfect barotropic
fluid with equation of state $P=w \rho$ is also known \cite{capfrw2}
\be
a(t) = t^{\frac{2 (1+\delta)}{3 (1+w)}}.
\ee
The stability of these solutions, and their properties as asymptotes
of the general solution, have been investigated in \cite{capfrw2} and
\cite{power}.  In fact, it has been shown that these are the only
power-law perfect fluid FLRW solutions that exist for any $f(R)$
gravity theory \cite{frdun3}. 
Explicit non-power-law general solutions with FLRW symmetries were
found in \cite{cgen}, both with spatial curvature, and in the presence
of a perfect fluid.  These solutions show explicitly that in the early
universe both non-singular and inflationary behaviour are possible.
The energy conditions in FLRW solutions have been considered in
\cite{frenergy}, and braneworld cosmology in these theories have been
considered in \cite{farh4}.

Beyond exact solutions, FLRW cosmological solutions have also been studied
in $f(R)$ theories of gravity using dynamical systems analysis.
This has been done for the case of a number of particular $f(R)$
theories in \cite{capfrw2,power,frmp3,frdyn3,frdyn4} and also in the
general case in \cite{frgenphase,higherp,frgen1,frgen2}. 
The dynamical systems approach has even been applied to perturbed FLRW
solutions in \cite{frdyna}.  We will discuss perturbed FLRW further in
the Cosmology section that follows.  These studies find a variety of
interesting cosmological behaviours at both early and late-times.  In
particular non-singular and accelerating behaviour in the early
universe is again identified, as well as late-time accelerating expansion, and
the non-sequential domination of higher powers in the Ricci curvature,
for analytic $f(R)$, as the initial singularity is approached.  The
conditions required for a non-singular `bounce' are given in
\cite{frbounce}, and oscillating solutions were considered in \cite{frosc}.
There have also been
some concerns expressed as to whether a matter dominated epoch is
generically expected to exist after radiation domination
\cite{frin3,frmd2,frgen1,frmd3,frmd4}.  The inverse
problem of finding 
particular forms of $f(R)$ that result in pre-specified cosmological
evolutions has been considered in \cite{frin1,frin2,frin3,frin4,frin5,frin6,frdun2,frex4}.  Such
inversions do not always specify $f(R)$ uniquely
\cite{frin2}, and it has been shown that to reproduce
exact $\Lambda$CDM evolution with dust only one is forced towards the
Einstein-Hilbert action with a cosmological constant \cite{frdun1}.

Exact Bianchi cosmological solutions were discovered for $f(R)=R^n$
theories in \cite{bcfr}, and been studied further in
\cite{frbianchi2,gld}, where shear dynamics and isotropisation are
discussed.  The special case of $n=2$ was studied in \cite{frkasner}.
Bianchi type $I$ and $V$ solutions have been considered in \cite{frb1}
and \cite{frb2}, and Bianchi $VII_A$ solutions in \cite{frb7a}.
Kantowski-Sachs solutions have been studied in \cite{frks1}.
Other know exact solutions are the Einstein static
universe \cite{existence,frmp3,fres1,fres2}, and the G\"{o}del universe
\cite{existence,frex5}.  These studies explore the stability 
of the Einstein static universe, and the existence of closed time-like
curves in the G\"{o}del solution.

\subsubsection{Cosmology}

Much of the recent motivation for studying $f(R)$ gravity has come
from the need to explain the apparent late-time accelerating expansion
of the Universe.  Previous motivation for studying $f(R)$ gravity has
also come from cosmological considerations, including the
presence of an initial singularity, and early universe inflationary
expansion.  We will therefore now present an overview of what
we consider to be some of the most relevant aspects of $f(R)$ gravity
for physical cosmology.  In terms of the viability of FLRW geometry in
$f(R)$ gravity, the Ehlers-Geren-Sachs theorem of General Relativity
has been extended to cover these theories by Rippl, van Elst, Tavakol,
and Taylor in \cite{fregs1}, and more recently by Faraoni in \cite{fregs}.
\newline
\newline
\noindent
{\it Field Equations}
\newline

To describe the cosmology, up to scalar perturbations, we first define line-element
\be
ds^2 = -(1+2 \Psi) dt^2 + a^2(t) (1-2 \Phi) q_{ij} dx^i dx^j,
\ee
and the energy-momentum tensor,
\ba
T^0_{\phantom{0} 0} &=& -\rho -\delta \rho \\
T^0_{\phantom{0} i} &=& -(\rho+P) \grad_i \theta\\
T^{i}_{\phantom{i} j} &=& P \delta^i_{\phantom{i} j} + \delta P
\delta^i_{\phantom{i} j} +(\rho+P)  D^i_{\phantom{i} j} \Sigma,
\ea
where $\theta$ is the peculiar velocity, $\delta P$ is the
pressure perturbation, and $\Sigma$ is the anisotropic stress.
At zeroth order the Friedmann equations are
\ba
\label{frF1}
H^2 &=& \frac{1}{3 F} \left[ 8 \pi \rho - \frac{1}{2} (f-RF)-3 H \dot{F}
  \right] - \frac{\kappa}{a^2}\\
\label{frF2}
\dot{H} &=& -\frac{1}{2F} ( 8 \pi \rho +8 \pi P +\ddot{F} - H \dot{F})
+ \frac{\kappa}{a^2}
\ea
where the Ricci scalar is given by $R = 6 ( 2 H^2 +\dot{H} + \kappa/a^2 )$,
and energy-momentum conservation gives, as usual,
\be
\dot{\rho} + 3 H (\rho +P) =0,
\ee
where $F=f_R$, over-dots denote derivatives with respect to $t$, and
$\kappa$ is spatial curvature.  

Now let us consider the first-order scalar perturbation equations,
which are given in \cite{frpert}.  Here it is convenient to define a new
quantity
\be
\chi \equiv 3 H \Psi + 3 \dot{\Phi}.
\ee
The perturbation equations are then \cite{frpert}
\ba
\label{frperteq1}
&&\dot{\chi} + \left( 2 H + \frac{\dot{F}}{2F} \right) \chi +
\frac{3}{2} \frac{\dot{F}}{F} \dot{\Psi} +\left[ 3 \dot{H} +
  \frac{3}{2F} (2 \ddot{F} + H \dot{F}) -\frac{k}{a^2} \right] \Psi
\nonumber \\&=&
\frac{1}{2F} \left[ 8 \pi \delta \rho + 24 \pi \delta P + 3 \delta
  \ddot{F} +3 H \delta \dot{F} + \left( \frac{k^2-6 \kappa}{a^2} -6H^2
  \right) \delta F \right]
\ea
and
\ba
&&\delta \ddot{F}+3 H \delta \dot{F} + \left( \frac{k^2}{a^2} -
\frac{R}{3} \right) \delta F
\nonumber \\ &=&
\frac{8 \pi}{3} (\delta \rho-3 \delta P) + \dot{F} (\chi+\dot{\Psi}) +
\left(2 \ddot{F} +3 H \dot{F} \right) \Psi - \frac{F}{3} \delta R
\ea
with fluid evolution equations
\be
\delta \dot{\rho} + 3 H (\delta \rho +\delta P) = (\rho+P)\left( \chi-3 H \Psi-
\frac{k^2 \theta}{a} \right)
\ee
and
\be
\frac{(a^4 (\rho+P) k \theta)\dot{\;}}{a^4 (\rho+P)} = \frac{k}{a} \left[
  \Psi +\frac{1}{(\rho+P)} \left( \delta P-\frac{2}{3}
  (k^2-3\kappa) (\rho+P) \Sigma \right) \right].
\ee
Here the perturbation to the Ricci scalar, $\delta R$, is given by
\be
\delta R = -2 \left[\dot{\chi}+4 H \chi - \left( \frac{k^2}{a^2}-3
  \dot{H} \right) \Psi +2 \frac{(k^2-3 \kappa)}{a^2} \Phi \right],
\ee
and we have the constraint equations
\be
\chi +\frac{3}{2} \frac{\dot{F}}{F} \Psi = \frac{3}{2F} \left[ 8\pi a
  (\rho +P) \theta + \delta \dot{F} -H \delta F \right]
\ee
and
\ba
&&\left( H+\frac{\dot{F}}{2F} \right) \chi + \frac{(k^2-3 \kappa)}{a^2} \Phi
+ \frac{3 H \dot{F}}{2F} \Psi \nonumber \\&=& -\frac{1}{2F} \left[ 8 \pi \delta \rho
  - 3 H \delta \dot{F} + \left( 3 \dot{H} + 3H^2- \frac{k^2}{a^2}
  \right) \delta F\right].
\ea
Furthermore, we again have that $\Psi \neq \Phi$, in general.
Instead it is the case that
\be
\Psi-\Phi = - \frac{8 \pi a^2 (\rho+P) \Sigma }{F} - \frac{\delta F}{F}.
\label{frperteq9}
\ee
The equivalent equations to those given above can also be derived 
in the covariant approach to cosmological perturbation theory
\cite{frcovariant}.  In the rest of this section we will consider the
consequences of these equations for various cosmological phenomena.
\newline
\newline
\noindent
{\it Inflation}
\newline

The existence of inflation has provided considerable motivation for
the study of $f(R)$ theories of gravity.
The pioneering work on this subject was that of Starobinsky in 1980, who found
that theories with $R^2$ corrections to their
gravitational Lagrangian can have an early period of de Sitter expansion
\cite{frstar}.  The spectrum of scalar and tensor
fluctuation generated during this type of inflation have  been
studied in \cite{star3,star4,star5,star6} where they were found to compatible with observations of
the CMB. Quantum initial conditions
(``tunnelling from nothing''), as well as the process of reheating,
were also considered in \cite{star2}. 

\begin{figure}[htbp]
\begin{center}
\epsfig{figure=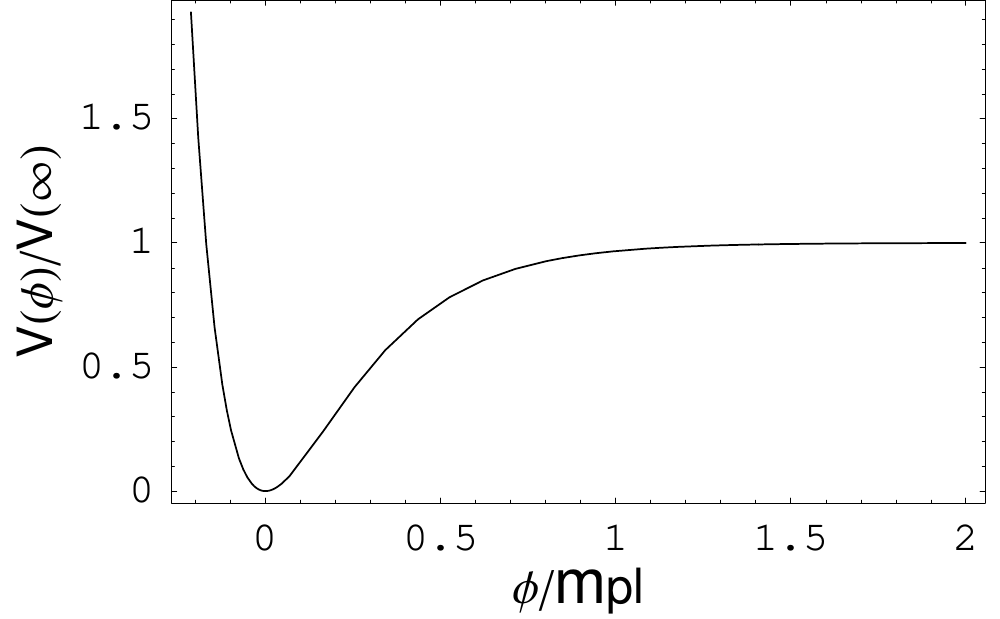,width=9cm}
\end{center}
\caption{The potential given in Eq. (\ref{frpot}), normalised by its
  asymptotic value as $\phi \rightarrow \infty$.}
\label{frfig1}
\end{figure}

Inflation in $f(R)$ gravity is particularly
transparent in the Einstein conformal frame.  Here, for the
Starobinsky model with \cite{frstar}
\be
\label{frstareq}
f(R) = R+ \frac{R^2}{6 M^2},
\ee
the conformally transformed theory in vacuum is one in which the
minimally coupled scalar field exists in a potential
\be
\label{frpot}
V(\phi ) = \frac{3 M^2}{2 \chi} \left( 1- e^{-\sqrt{\frac{\chi}{3}} \phi}
\right).
\ee
This potential is displayed in Figure \ref{frfig1}, where it can be
seen that slow-roll inflation is likely to occur in the region $\phi
\gtrsim m_{pl}$, and reheating is feasible during oscillations around the
minimum at $\phi=0$.  This is, of course, exactly the type of
behaviour that one wants for a viable inflaton field.

In fact, for the theory specified in Eq. (\ref{frstareq}) one can show that
inflation is the transient attractor of the general solution \cite{frinf}, and
that in the region $\phi \gg m_{pl}$ slow-roll inflation occurs with
\be
\epsilon = - \frac{\dot{H}}{H^2} \simeq \frac{M^2}{6 H^2},
\ee
and proceeds for $N \simeq (2 \epsilon)^{-1}$ e-foldings.  We will not
proceed with showing the details of reheating in this model, but only
note that around $\phi \simeq 0$ the potential given in
Eq. (\ref{frpot}) is well approximated by $V \simeq \frac{1}{2} M^2
\phi^2$.  For details of how reheating occurs in this potential the
reader is referred to \cite{star2,frinf2}.  Pre-heating in
$f(R)$ inflationary models has been considered in \cite{frpre}.
Quantum cosmology, instantons, and their implications for inflation have been
studied in \cite{frTav1,frTav2}.
\newline
\newline
\noindent
{\it Dark Energy}
\newline

As well as accelerating expansion in the early universe, $f(R)$
theories of gravity are also capable of producing late-time
accelerating expansion.  There have been a large number of papers
on this subject.  There have also been attempts
to construct quintessence-like $f(R)$ models which produce both early
and late acceleration \cite{qfr1,qfr2,qfr3,qfr4,qfr5}.

An easy way to see the potential for late-time accelerating expansion is to
consider the Friedmann equations (\ref{frF1}) and (\ref{frF2}) in
vacuo.  One can then identify effective density and pressure
parameters by analogy to the Friedmann equations of general
relativity.  These are
\ba
8 \pi \rho_{\textrm{eff}} &=& \frac{RF-f-6 H \dot{F}}{2 F}\\
8 \pi P_{\textrm{eff}} &=& \frac{2 \ddot{F} +4 H \dot{F} +f -RF}{2F}.
\ea
The equation of state of this effective fluid is then given by
\be
\label{freqstate}
w = \frac{2 \ddot{F} +4 H \dot{F} +f -RF}{RF-f-6 H \dot{F}},
\ee
and one can then determine what is required to achieve $w < -1/3$, and
hence accelerating expansion. 

One example of this is the now much
considered theory of Carroll, Duvuri, Trodden and Turner \cite{carfr}
\be
\label{freqs}
f(R)=R-\frac{\mu^{2 (n+1)}}{R^n},
\ee
where $\mu$ is a constant.  For power-law evolution the effective
equation of state, (\ref{freqstate}), is then given at late-times by \cite{carfr}
\be
w = -1+\frac{2 (n+2)}{3 (2 n+1) (n+1)},
\ee
so that if $n=1$, and the extra term in the gravitational Lagrangian
is inversely proportional to $R$, then one achieves an equation of
state with $w=-2/3$, and hence accelerating expansion.  In fact, this
just corresponds to the power-law solution given in Eq. (\ref{rnfrw}),
for a theory with $f(R) \propto R^{-n}$.

The theory specified in Eq. (\ref{freqs}) is now known to have a
number of deficiencies that make it non-viable
\cite{Olmo05,frmd3,frmd4,nocar1,nocar2,nocar3,nocar4,nocar5,frmpc,frmp2}.  Some of
these have to do with the weak-field limit, which we have
already discussed, others come from cosmology, and yet more are due to
stability issues, which we will address in Section \ref{frstability}.  Many of
these problems can be traced back to the value of the effective mass
in the scalar degree of freedom of this theory, which is thought to be
either too small for validity of gravitational physics in the solar
system, or imaginary, leading to some of the instabilities just alluded to.
Models which have been constructed to try and over-come these
problems, while still leading to accelerating expansion at late-times,
are those of Starobinsky \cite{frcham3}:
\be
\label{frstarobinsky}
f(R)=R-\mu R_c \left[ 1- \left(1+\frac{R^2}{R_c^2} \right)^{-n}
  \right],
\ee
Hu and Sawicki \cite{frcham8}:
\be
f(R)=R- \frac{ \mu R_c }{1+(R/R_c)^{-2n}},
\ee
and Battye and Appleby \cite{Battye-App}:
\be
f(R)=R+R_c \log\left[e^{-\mu}+(1-e^{-\mu})e^{-R/R_c} \right]
\ee
where $\mu$, $n$ and $R_c$ are all positive constants.  Attempts
to construct viable models that include an early stage of
inflationary expansion, as well as late-time accelerating expansion,
have been made in \cite{newviab1,newviab2}.  All of these theories
rely on the chameleon mechanism to satisfy solar system constraints on
gravity.
\newline
\newline
\noindent
{\it Observational Probes}
\newline

As with many modifications to gravity, cosmological observables can be
used to constrain $f(R)$ theories of gravity.  Here we will briefly survey
the literature on this subject.

Primordial nucleosynthesis has been studied in $f(R)$ gravity in
\cite{power,brookfield,lambiase,evans}.  Due
to the conformal equivalence between these theories and general
relativity, the behaviour of cosmological solutions during the
radiation dominated epoch are considerably simplified:  They evolve
in a similar way to the radiation dominated solutions of GR, but with a
different value of the effective gravitational constant.  This
situation is familiar from studies of primordial nucleosynthesis in
the scalar-tensor theories of gravity outlined in Section
\ref{scalartensorsection}.  In 
the present case the relevant effective gravitational `constant' is
inversely proportional to $f_R$.  The value of $f_R$ evolves
throughout the matter dominated and accelerating epochs, but is
constant during radiation domination.  Observations of the abundances
of light elements then provide constraints on the allowed values of $f_R$
during the radiation dominated epoch, and hence constrain the rate of
evolution of this quantity that is allowed during the rest of the
Universe's history.

Other probes of the background expansion of an FLRW universe are the
peak positions of the CMB spectrum of temperature fluctuations, and
baryon acoustic oscillations.  Observations of these quantities allow the form of
$a(t)$ to be constrained, but due to the freedom in the choice of
$f(R)$ are not able to falsify the most general form of these
theories directly
\cite{frin1,frin2,frin3,frin4,frin5,frin6,frdun2,frex4}.  To go
further using cosmological observations we must
therefore consider the solutions to the perturbation equations given
above.

The first thing that one may wish to consider is the growth of
density perturbations, $\delta = \delta \rho/\rho$.  In a spatially
flat universe, manipulation of
Eqs. (\ref{frperteq1})-(\ref{frperteq9}) allows one to write
\cite{pzhang,frmp,frmpb}
\be
\label{frdel}
\ddot{\delta} + 2 H \dot{\delta} - \frac{4 \pi \delta \rho}{3 f_R} \frac{(4+3(a/k)^2 M^2)}{(1+ (a/k)^2 M^2)}=0,
\ee
where the mass parameter $M$ is given, just as in the weak field limit
discussed in Section \ref{frweak}, as
\be
M^2 = \frac{f_R - R f_{RR}}{3 f_{RR}}.
\ee
From the third term on the LHS of Eq. (\ref{frdel}) it can be seen
that the evolution of $\delta$ depends on the magnitude of $M$, and,
  in particular, is different in the two regimes $M^2 \gg k^2 /a^{2}$
  and $M^2 \ll k^2 /a^{2}$.  When the former is true, the density
  perturbations evolve as they do in General Relativity, with an
  effective Newton's constant given by $G=1/f_R$.  For a matter
  dominated universe this means
\be
\delta \propto t^{\frac{2}{3}}.
\ee
In the latter regime, in which $M^2 \ll k^2/ a^{2}$, this is no longer
  true.  Here, the third term in Eq. (\ref{frdel}) is modified from
  its form in GR by a multiplicative factor of $4/3$, and the
  evolution of $\delta$ during the matter dominated era is
  consequently modified to
\be
\delta \propto t^{(\sqrt{33}-1)/6}.
\ee
The transition between these two limits is theory dependent.  For
studies on this subject the reader is referred to
\cite{frcham3,frcham8,frmpc,frmpc2,frmp,frmpb,PogosianSilvestri2010,frcmb3,frmpd,frmpe,frmpf,frex2}.
Interesting results are that the change in evolution 
between the two regimes discussed above is scale dependent.  That is,
modes with different wave-numbers can evolve in different ways
depending on whether they are larger or smaller than $a^2 M^2$.  This
length scale is therefore imprinted on the density perturbations.
Furthermore, oscillating modes can also become present when $M^2 \gg
k^2/a^2$, which can lead to undesirable singularities
\cite{froscsing1}.  The inclusion of an $R^2$ in the gravitational
Lagrangian was found to remove these
singularities in \cite{froscsing2}.  One can also see that $f_{RR}>0$
is need for the stability of scalar modes.

The modified growth of structure just discussed has consequences for
large-scale structure, and the cosmic microwave background, which
we will now discuss.  The matter power spectrum 
in $f(R)$ theories of gravity has been considered in \cite{frcham4,nocar3,nocar5,frmpc,frpc10,frmp2,frmp3,frdun5,frdun4,frex3,frex1}, and cluster abundances have been used to constrain these
theories in \cite{frcluster,frhu1,frcluster2}. The formation of non-linear structure has also been
considered in \cite{HuSawicki2007,frnl1,frnl2,frnl3,frnl4}.  Cosmic microwave observations are
considered in \cite{frmp3,frsph,frmp2,frhu1}
where it is 
shown that power on large-scales is sensitive to the modified growth
of structure through the integrated Sachs-Wolfe (ISW) effect.  This can lead
to damped power for small deviations from GR, or amplified power if the
deviations are large enough.  Correlating
ISW effects in the CMB with observations of galaxy number density also
leads to tight constraints \cite{frsph,frhu1}, due
to the sign of the CMB temperature fluctuation changing if the
modification to gravity is large enough.

\subsubsection{Stability issues}
\label{frstability}

There are a variety of stability issues that are of concern for $f(R)$
theories of gravity.  These include ghost degrees of freedom, as
evidenced in the Ostrogradski instability, as well as the 
instabilities found by Frolov, and Dolgov and Kawasaki.  Some of these
issues have been mentioned already.  In this section we will discuss
them further.
\newline
\newline
\noindent
{\it Ghosts, and the Ostrogradski instability}
\newline

In Section \ref{sec:intro-ghosts}, we discussed the problems
associated with ghosts -- pathological fields that admit physical
states of negative energy, or negative norm when quantised.  It is known that
ghosts can occur in general higher derivative 
theories of gravity, see
e.g. \cite{fgenstelle,fgenstring2,fghost1,fghost2,fghost3,fghost4,fghost5,fghosta,fghostb}.
They are not, however,
as problematic in $f(R)$ theories as they are for general fourth-order
theories, as we will now outline.  Let us first consider the
existence of negative energy states in the context of Ostrogradski's
theorem \cite{Ostro}.

The Ostrogradski instability states that Lagrangians that contain
second derivatives, and are non-linear in those
second derivatives, are generically unstable.  At first sight such
a result appears 
to be problematic for $f(R)$ theories of gravity, which
are only linear in second-order derivatives of the metric in the case
of General Relativity with a possible cosmological constant.  One can
show, however, that these instabilities do not occur for $f(R)$
theories \cite{woodard}.
This works in the following way:  Let us consider a Lagrangian
\be
L=L(g,\dot{g},\ddot{g}),
\ee
where dots denote derivatives of $g$ with respect to some parameter $\lambda$.
Now define a set of four canonical variables by $Q_1 \equiv g$,
$Q_2 \equiv \dot{g}$, and
\be
P_1 \equiv \frac{\partial L}{\partial \dot{g}} - 
\frac{d}{d\lambda} \frac{\partial L}{\partial \ddot{g} },
\qquad \text{and} \qquad
P_2 \equiv \frac{\partial L}{\partial \ddot{g}}.
\ee
If it is now possible to write $\ddot{g}=f(Q_1,Q_2,P_2)$
then the Hamiltonian of the system can be written as
\be
\label{osth}
H = P_1 Q_1 +P_2 f - L(Q_1,Q_2,f).
\ee
This Hamiltonian, however, is only linear in the
momentum $P_1$, and cannot therefore be stable.  This is
Ostrogradski's instability.  Now, $f(R)$ gravity avoids this
instability because one cannot write down the equivalent of
$\ddot{g}=f(Q_1,Q_2,P_2)$ for each component of the metric.  Instead,
only a single scalar degree of 
freedom contains the higher-order derivatives, and by an appropriate
field redefinition one can remove this extra field so that the
redefined metric appears in the Lagrangian only linearly in its
second-order derivatives.  This is just the conformal transformation
discussed in Section \ref{freqssec}.  The Ostrogradski instability does
not, therefore, apply to $f(R)$ theories of gravity \cite{woodard}.

Let us now consider ghost-like instabilities from the point of view of linear
fluctuations.  In generic fourth-order theories massive 
spin-2 degrees of freedom appear along with a scalar degree of
freedom, and the familiar massless spin-2 degree of freedom from
General Relativity.  It is the massive spin-2 fields in this
situations that present
the generic problem with ghosts.  Such fields are absent from $f(R)$
theories, however, which contain only the massless spin-2 fields of GR, and a
single scalar field.  Again, this is clear from the existence of the
conformal transformations outlined in Section \ref{freqssec}.  The $f(R)$
theories of gravity therefore do not always suffer from the same
problems with ghosts as more general higher-order theories, which will be
discussed in more detail in Section \ref{fgenother}.
\newline
\newline
\newline
\noindent
{\it Frolov instability}
\newline

A potential problem with $f(R)$ theories that modify the infra-red limit of
General Relativity has been identified by Frolov in \cite{froscsing1}.  This
instability is caused by the fact that for the scalar degree of
freedom in $f(R)$ theories curvature singularities can occur at finite
field value and energy level, a phenomenon previously investigated in
\cite{newinst1,newinst2}.

\begin{figure}[htbp]
\begin{center}
\epsfig{figure=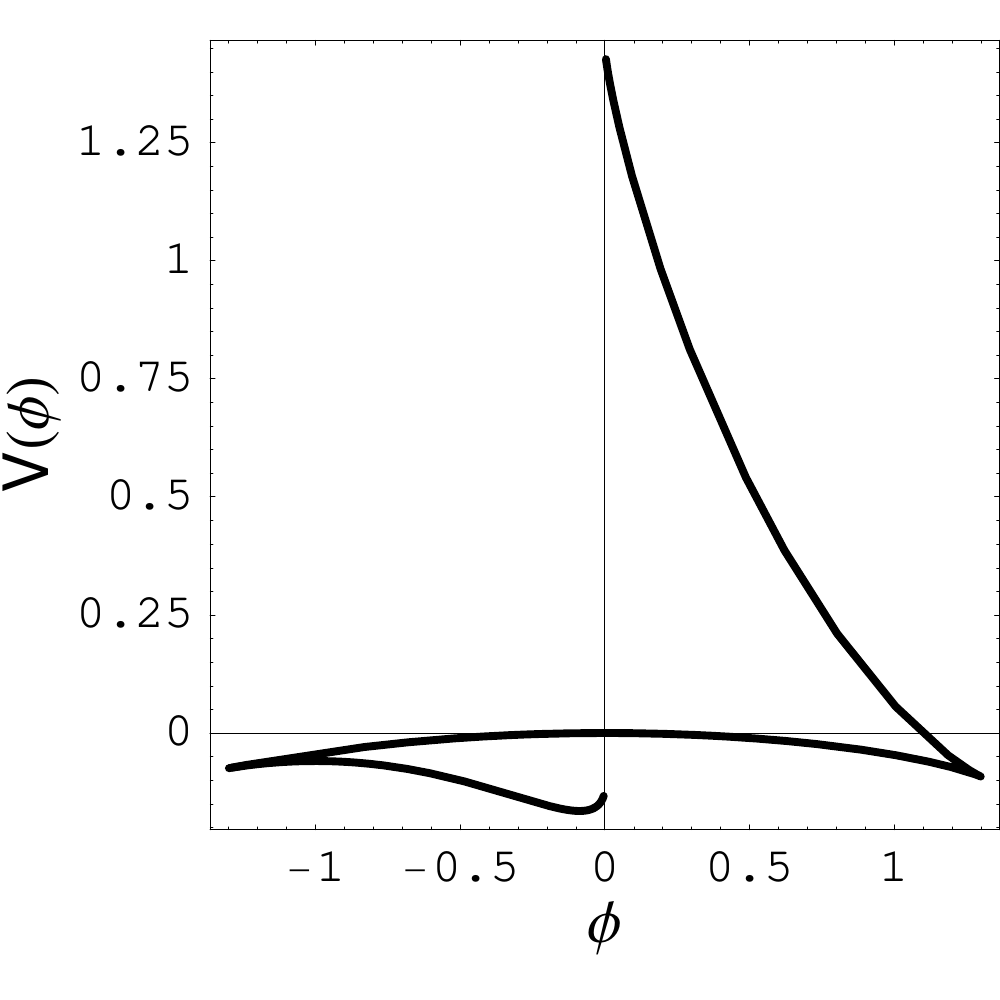,width=7.5cm}
\end{center}
\caption{The potential for the scalar field in Starobinsky's theory,
  Eq. (\ref{frstarobinsky}), with $R_c=1$, $n=1$ and $\mu=2$.}
\label{frfig2}
\end{figure}

To illustrate this problem consider the $f(R)$ proposed by
Starobinsky, Eq. (\ref{frstarobinsky}).  The potential for the
effective scalar field in this theory is shown
in Fig. \ref{frfig2}.  During cosmological expansion, the scalar
fields associated with FLRW cosmologies roll down the slope from
$\phi=0$ to the local minimum at $\phi \simeq -0.1$.  The short
section of curve between this minimum and the singularity at $\phi =0$ is the
only part of the potential the scalar field need experience in the
entire history of a perfect FLRW solution.  Frolov argues, however,
that relatively small perturbation in curvature, caused by collapse of
dust, are enough to push $\phi$ back up the potential to the
singularity.  The existence of such instability is, of course,
undesirable for a physically plausible theory, although it may be
mitigated by the addition of higher power of 
$R$ to the gravitational Lagrangian \cite{newinst2,newinst3,froscsing2}.
\newline
\newline
\newline
\noindent
{\it Dolgov-Kawasaki instability} 
\newline

Finally, let us consider another instability that was initially found
by Dolgov and Kawasaki for the theory given by Eq. (\ref{freqs}) with
$n=1$ \cite{nocar1}.  This was later extended to more general functions
of $f(R)$ that modify gravity in the infra-red limit \cite{nocar2} and formalised better in~\cite{Seifert2007}.

The basic point here is that the trace of the field equations,
(\ref{R16}), acts as the propagation equation for scalar degree of
freedom.  For Eq. (\ref{freqs}) with $n=1$ this equation is
\be
R-\frac{3 \mu^4}{R} - \Box \left( \frac{3 \mu^4}{R^2} \right) = 8 \pi
 \rho,
\ee
where we have taken the matter content to be that of dust.  Now, de
Sitter space is a solution of this equation, with 
\be
R_{dS}=\frac{1}{2} (8 \pi \rho + \sqrt{(8\pi \rho)^2+12 \mu^4} )
\simeq 8 \pi \rho,
\ee
 for cosmologically
relevant $\mu$.  If we now consider perturbations around this solution
with $R=R_{dS}+\delta R$ then we get to lowest order that
\be
\frac{6 \mu^4}{R_{dS}^3} \Box \delta R
+\left(1+\frac{3\mu^4}{R_{dS}^2} \right) \delta R =0.
\ee
Comparing this with the propagation equation of a massive scalar field
gives
\be
m^2 = -\frac{R_{dS}}{2} - \frac{R_{dS}^3}{6 \mu^4} \simeq - \frac{(8
  \pi \rho)^3}{6 \mu^4} \simeq -10^6 \text{GeV},
\ee
where in the last equality the density has been taken to be that of
water, $\rho \simeq 10^3 kg/m^3$, and $\mu$ has been taken to be $\sim
10^{-33} eV$, as required to account for the observed late-time
accelerating expansion of the Universe.  This large negative
mass corresponds to a catastrophic instability that should make itself
apparent on time scales of $\sim 10^{-26}$ seconds.

For more general theories it can be shown that the effective mass
squared in the  relevant scalar field equation takes the same sign as
$f_{RR}$ \cite{nocar2}.  It is 
therefore the negative value of this quantity in the theory of
Eq. (\ref{freqs}) that is responsible for its exhibition of this
instability.  Further, one can show that the addition of higher powers of
$R$ to the gravitational Lagrangian again helps defend it from
instability \cite{frdk3,frds2,frdk5}.  The
Dolgov-Kawasaki instability has been shown not to occur in the
Palatini approach \cite{frdgpal}.  The existence of this
type of instability was rediscovered in
\cite{frmaeda1}, for relativistic stars.  The problem in this context has been
further studied in \cite{frrel1,frrel2,frrel3,frns3,frex8,frex7},
where it was shown that the instability can be avoided by changing the
equation of state of the star, adding a divergence to the scalar field
potential, or including chameleon effects.  Neutron stars in $f(R)$
gravity have been studied in \cite{frns1,frns2}, and
instabilities in systems with time-dependent mass have been studied in
\cite{frtimemass}.

\subsection{General combinations of Ricci and Riemann curvature.}

In the previous section we considered theories that generalise the
Einstein-Hilbert action by replacing the Ricci scalar, $R$, with some
non-linear function, $f(R)$.  Here we go further, and allow the action
to be a function of not only $R$, but of any of 
the three linear and quadratic contractions of the Riemann
curvature tensor\footnote{There
is also a fourth possibility, namely $\epsilon^{\mu\nu\rho\sigma}
R_{\epsilon \tau \mu \nu}
{R^{\epsilon \tau}}_{\rho\sigma}$ \cite{DeW65}.
This contraction, however, is of limited physical interest as
it does not affect the field equations, due to its parity.}: $R$,
$R_{\mu\nu}R^{\mu\nu}$ and $R_{\mu\nu\rho\sigma}R^{\mu\nu\rho\sigma}$.
A systematic approach to studying theories of this type, based on
minimal sets of curvature invariants, is proposed and studied in
\cite{molden1,molden2}.

\subsubsection{Action and field equations}

The most general weight-zero scalar density that one can construct
from $g$, $R$, $R_{\mu\nu}R^{\mu\nu}$ and $R_{\mu\nu\rho\sigma}R^{\mu\nu\rho\sigma}$ alone is
given by
\be
\mathcal{L}=\chi ^{-1}\sqrt{-g}f(R,R_{\mu\nu}R^{\mu\nu},R_{\mu\nu\rho\sigma}R^{\mu\nu\rho\sigma})
\label{densitygen} 
\ee
where $f$ is an arbitrary function of its arguments, and $\chi $ is a
constant which can be determined from the Newtonian limit. The action
is obtained, as usual, by integrating this 
density, together with that of the matter fields, over all space-time. The addition
of supplementary terms to the density (\ref{densitygen}), in order to cancel
total divergences, and which can be transformed to integrals on the boundary, can
be problematic (see e.g. \cite{Mad88}) and so, for simplicity, they
are usually assumed to vanish.

As with $f(R)$ theories one can proceed by using the metric variation
approach (in which the connection is {\it a priori} taken to be given
by the Christoffel symbols), using the Palatini formalism (in which the
metric and connection are taken to be independent fields in the
gravitational action), or the
metric-affine formalism (in which the metric and connection are taken
to be independent fields throughout the entire action).  Here we will
spell out the metric variational approach explicitly, as this is the
most commonly considered form of the theory found in the literature.
For studies involving the Palatini procedure, as applied to the these
theories, the reader is referred to \cite{fgenpa1,fgenpa2,fgenpa3,fgenpa4,fgenpa5,fgenpa6,fgenpa7}.

Varying the action, derived from integrating Eq. (\ref{densitygen})
over all space, with respect to the metric, then gives
\ba
\nonumber
\delta I &=& \chi^{-1} \int d \Omega \sqrt{-g} \Bigl[ \frac{1}{2} f g^{\mu\nu}
  \delta g_{\mu\nu} + f_X \delta X + f_Y \delta Y + f_Z \delta Z \Bigr]\\ 
\label{vargen}
&=& \chi^{-1} \int d \Omega \sqrt{-g} \Bigl[ \frac{1}{2} f g^{\mu\nu}
  \delta g_{\mu\nu} - f_X (R^{\mu\nu} \delta g_{\mu\nu} - g^{\mu\nu} \delta R_{\mu\nu})\\
\nonumber
&& \qquad \qquad \qquad \qquad \qquad \quad - 2 f_Y (R^{\rho (\mu}
  {R^{\nu)}}_{\rho} \delta g_{\mu\nu} - R^{\mu\nu} \delta 
  R_{\mu\nu})\\
\nonumber
&& \qquad \qquad \qquad \qquad \qquad  \quad - 2 f_Z ({R_{\rho \sigma \epsilon}}^{(\nu}
  R^{\mu) \epsilon \sigma \rho} \delta g_{\mu\nu} - 
  {R_{\mu}}^{\nu \rho \sigma} \delta {R^{\mu}}_{\nu \rho \sigma}) \Bigr],
\ea
where we have defined $X \equiv R$, $Y \equiv
R_{\mu\nu}R^{\mu\nu}$ and $Z \equiv R_{\mu\nu\rho\sigma}R^{\mu\nu\rho\sigma}$.  Using
$\delta {R^{mu}}_{\nu \rho \sigma} = \frac{1}{2} g^{\mu \epsilon} (\delta g_{\epsilon \sigma ; \nu \rho} + \delta
g_{\epsilon \nu ; \sigma \rho} - \delta g_{\sigma \nu ; \epsilon \rho} - \delta g_{\epsilon \rho ; \nu \sigma} - \delta
g_{\epsilon \nu ; \rho \sigma} +\delta g_{\rho \nu ; \epsilon \sigma}) $
we can then write
the gravitational part of the action as
\be
\delta I = - \chi^{-1} \int d \Omega \sqrt{-g} P^{\mu\nu}\delta g_{\mu\nu},
\ee
where
\ba  \nonumber
P^{\mu\nu} &\equiv& -\frac{1}{2} f g^{\mu\nu} + f_X R^{\mu\nu}+2 f_Y R^{\rho (\mu} {R^{\nu)}}_{\rho}+2
f_Z R^{\epsilon \sigma \rho (\mu} {R^{\nu)}}_{\rho \sigma \epsilon}
\\ \nonumber&&+f_{X; \rho \sigma}(g^{\mu\nu} g^{\rho \sigma}-g^{\mu
  \rho} g^{\nu \sigma}) 
+\square (f_Y R^{\mu\nu}) + g^{\mu\nu} (f_Y R^{\rho\sigma})_{;\rho\sigma}\\ \label{Pgen}&&-2 (f_Y R^{\rho (\mu})_{;\;
\; \rho}^{\; \nu)}-4 (f_Z R^{\sigma (\mu\nu) \rho})_{;\rho\sigma}.
\ea
The notation $f_N$ here denotes the functional derivative of $f$ with respect
to $N$.  Looking for a stationary point of this action, by setting the first
variation to zero, then gives the field equations 
\be  \label{fequationsgen}
P_{\mu\nu}=\frac{\chi}{2} T_{\mu\nu} - g_{\mu\nu} \Lambda
\ee
where matter fields and a cosmological constant have been included.  Here,
$\Lambda$ is the cosmological constant (defined independent of
$f(X,Y,Z)$) and $T^{\mu\nu}$ is the energy-momentum tensor of the matter fields.
These field equations are generically of fourth-order, with the exception of
cases in which the function $f$ is linear in second derivatives of
the metric \cite{DeW65}, as occurs in GR.

Unlike the case of $f(R)$ theories of gravity, the theories described
by the density given in Eq. (\ref{vargen}) are not, in general,
conformally related to General Relativity with a scalar field.

\subsubsection{Weak-field limit}

Let us consider the weak field limit of theories with additional terms
in their action that are quadratic in curvature invariants:
\be
\label{quadL}
\mathcal{L} = \frac{\sqrt{-g}}{\chi} \left( R + \alpha R^2 
+\beta R_{\mu\nu} R^{\mu\nu}  + \gamma R_{\mu\nu\rho\sigma}R^{\mu\nu\rho\sigma} \right),
\ee
where $\alpha$, $\beta$ and $\gamma$ are constants.  In this case one
can use the well known result that the Gauss-Bonnet combination of
curvature invariants is a total divergence, i.e.
\be
4 R_{\mu\nu}R^{\mu\nu} - R^2 -R_{\mu\nu\rho\sigma} R^{\mu\nu\rho\sigma} = \text{total divergence},
\ee
which in the action integrates to a boundary term that is usually
ignored.  By redefining $\alpha$ and $\beta$ we can therefore write
Eq. (\ref{quadL}) in the equivalent form
\be
\label{quadL2}
\mathcal{L} = \frac{\sqrt{-g}}{\chi} \left( R + \alpha R^2+
\beta R_{\mu\nu} R^{\mu\nu}   \right),
\ee
without any loss of generality in the solutions to the resulting field
equations.

If we now substitute Eq. (\ref{quadL2}) into
Eq. (\ref{fequationsgen}), to get the field equations, we find
that for the perturbed metric
\be
ds^2 = -(1+2 \psi) dt^2 + (1-2 \phi) (dx^2+dy^2+dz^2),
\ee
the lowest order equations in the weak-field and slow-motion
limit are
\ba
\label{fgenfe1}
2 (3 \alpha+\beta) \Delta R-R &=& -\frac{\chi}{2} \rho\\
\label{fgenfe2}
\left(4 \alpha+ \beta\right) \Delta R -R-
2 \Delta (\psi+ \beta \Delta \psi ) &=& -\chi \rho,
\ea
where the Ricci scalar is given as usual by $R=-2 \Delta \psi+4
\Delta \phi$.  
For a delta function source, $\rho = m \delta$, Eqs. (\ref{fgenfe1})
and (\ref{fgenfe2}) can then be seen to have the solutions
\ba
\label{r2phi2}
16 \pi \psi &=& -\frac{\chi m}{r} \left( 1+\frac{e^{-m_1 r}}{3}- \frac{4
  e^{-m_2 r}}{3} \right)\\
\label{r2psi2}
16 \pi \phi &=& -\frac{\chi m}{r} \left( 1-\frac{e^{-m_1 r}}{3}- \frac{2
  e^{-m_2 r}}{3} \right),
\ea
where
\be
\label{m1m2}
m_1^2= \frac{1}{2 (3 \alpha+ \beta)} \qquad \text{and} \qquad m_2^2 =
-\frac{1}{\beta}.
\ee
These are the solutions found by Stelle in 1978
\cite{fgenstelle}.  These solutions can be seen to reduce to
Eqs. (\ref{r2phi}) and (\ref{r2psi}) in the limit $\beta \rightarrow
0_-$.  More generally, however, these theories can be seen to exhibit
massive modes with two different mass parameters.  In order to have
non-oscillatory behaviour in the present case we must therefore
require that both $3 \alpha + \beta \geq 0$ and $\beta \leq 0$ be
simultaneously satisfied.

If the solutions given in Eqs. (\ref{r2phi2}) and (\ref{r2psi2}) are
the correct ones for describing the space-time geometry around
approximately isolated masses, such as the Sun, then one can
immediately see that if $m_1$ and $m_2$ are both large compared to
$1/r$ then one recovers the general relativistic prediction of
$\gamma=1$, just as with $f(R)$ theories.  For small masses, however,
the situation is somewhat different from the $f(R)$ case.  If both
$m_1$ and $m_2$ are small compared to $1/r$ then one has that the
leading order term in Eqs. (\ref{r2phi2}) and (\ref{r2psi2}) is a
constant (which can be absorbed into coordinate redefinitions),
followed by a term proportional to $r$.  This is a considerable
deviation from the behaviour $\gamma \rightarrow 1/2$ that occurs when
$m_1 r$ is small and $m_2 r$ is large, which is the limit of $f(R)$
gravity with a low mass parameter.  It can also be seen that for $m_1
\leq 2 m_2$ gravity is always attractive, while for $m_1 >2 m_2$ it is
attractive over large distances, while being repellent over small distances.

\subsubsection{Exact solutions, and general behaviour}

Let us now discuss what is known about the solutions to these general
fourth-order theories of gravity in the context of both isolated
masses, and cosmological solutions.
\newline
\newline
\noindent
{\it Isolated Masses}
\newline

Motivation for a number of studies in this area has come from
Einstein's particle programme, in which one looks for asymptotically
flat and singularity free vacuum solutions which could be used to model
particles \cite{einpart}.  While it is known that no such solutions
exist in General Relativity (Lichnerowicz's theorem \cite{lich}), it
has been conjectured that they could 
exist in fourth-order theories \cite{bor81}.

By studying the solutions of quadratic theories of the type given in
Eq. (\ref{quadL2}) with $\beta =-3 \alpha$ it has been shown that the
solutions to the linearised vacuum field equations can be both
asymptotically flat as $r \rightarrow \infty$, and smooth as $r
\rightarrow 0$ \cite{fiesch}.  These theories are equivalent to the
sum of an Einstein-Hilbert term and a Weyl term.  Such results would
initially appear to be encouraging for Einstein's programme, but it
was later shown that there are, in fact, no solutions with the
specified properties that exist within a neighbourhood of Minkowski
space \cite{schm85}.  This means that if any non-trivial static
spherically symmetric vacuum solutions to these theories exist, that
are simultaneously asymptotically flat and geodesically complete, then
they must correspond to very large energy densities (exceeding the
energy density of neutron stars by at least 40 orders of magnitude
\cite{schm85}).

The theorems of Lichnerowicz \cite{lich} and Israel \cite{isbh1} have
more recently been considered in the context of fourth-order theories
of the form given in Eq. (\ref{quadL2}) by Nelson \cite{nelsonBH}.
Here it is found that for static space-times with spatial curvature satisfying
\ba
\label{ineqf1}
m_1^2 - \;^{(3)}R &\geq& 0\\
\label{ineqf2}
\bar{R}^{\mu}_{\phantom{\mu} \nu} \bar{R}^{\nu}_{\phantom{\nu} \mu} m_1^2+
 \bar{R}^{\mu}_{\phantom{\mu} \nu} \bar{R}^{\nu}_{\phantom{\nu} \rho}
\bar{R}^{\rho}_{\phantom{\rho} \mu}&\geq& 0,
\ea
the vacuum field equations imply that all asymptotically constant
solutions (or asymptotically flat, if the inequalities are saturated)
obey $R_{\mu\nu}=0$.  The expression for $m_1$ is given in
 Eq. (\ref{m1m2}).  Over-bars here denote quantities projected into
 space-like hyper-surfaces. The spherically symmetric solution to $R_{\mu\nu}=0$
is, of course, the Schwarzschild solution, which is geodesically complete only
for the case of Minkowski space.  Lichnerowicz's theorem can therefore
be extended to all theories that obey the inequalities (\ref{ineqf1})
and (\ref{ineqf2}).  It is then shown in \cite{nelsonBH} that if the
 spatial curvature satisfies
\ba
\label{ineqf3}
\;^{(3)}R &\leq& m_2\\
\label{ineqf4}
\frac{ \bar{R}^{\mu}_{\phantom{\mu} \nu} \bar{R}^{\nu}_{\phantom{\nu} \rho}
\bar{R}^{\rho}_{\phantom{\rho} \mu}}{\bar{R}^{\mu}_{\phantom{\mu} \nu}
 \bar{R}^{\nu}_{\phantom{\nu} \mu}} &\geq& - m_2,
\ea
where $m_2$ is given by Eq. (\ref{m1m2}),
and the space-time is asymptotically constant (or
 asymptotically flat, if the inequalities are saturated), then the
 only solutions with $m_1^2 \geq 0$ that exist in the region exterior
 to a closed 
 spherical null surface also obey $R_{\mu\nu}=0$.  The only asymptotically
 constant vacuum solutions with a horizon, that satisfy the bounds
 (\ref{ineqf3}) 
and (\ref{ineqf4}), are therefore the Schwarzschild solutions.  This
extends the Israel's no-hair theorem for black holes to quadratic
fourth-order theories of gravity.  It is argued in \cite{nelsonBH}
that the inequalities (\ref{ineqf1})-(\ref{ineqf4}) should be
satisfied everywhere where the spatial curvature is smaller than the
scale of the corrections to the Einstein-Hilbert action.  If these
corrections are motivated by quantum considerations, then we should
therefore expect all of the inequalities (\ref{ineqf1})-(\ref{ineqf4}) to
be satisfied for astrophysically interesting systems.  The stability
of Schwarzschild black holes in the quadratic theories (\ref{quadL2})
has been studied in \cite{fgenblack,stelle}.

The initial value problem for quadratic theories, of the type given in
Eq. (\ref{quadL2}), has also been studied in \cite{noakes}, where it was
found to be well-posed.
\newline
\newline
\noindent
{\it Cosmological Solutions}
\newline

There are a number of exact cosmological solutions known to exist for
fourth-order theories containing $R_{\mu\nu}R^{\mu\nu}$ and $R_{\mu\nu\rho\sigma}
R^{\mu\nu\rho\sigma}$.  This simplest of these is, of course, de Sitter space,
which exists for theories with general $f(X,Y,Z)$ in
Eq. (\ref{densitygen}), and a cosmological constant, if \cite{existence}
\be
\frac{1}{2} f - \Lambda = \Lambda f_X + 2 \Lambda^2 f_Y+\frac{4}{3}
\Lambda^2 f_z,
\ee
where $f_N$ denotes differentiation of $f$ with respect to $N$.  The
stability of de Sitter space in quadratic theories, of the type given
in Eq. (\ref{quadL2}), has been studied in \cite{barher1}, and in
the more general case in \cite{fgends1}.  Other known exact
homogeneous and isotropic cosmological solutions are the Einstein
static universe 
and the G\"{o}del universe, the existence of which has been discussed
in \cite{existence} for arbitrary $f(X,Y,Z)$ (together with the
conditions for the existence of closed time-like curves in the case of
latter).  The existence of power-law FLRW solutions, both in vacuum and
in the presence of a perfect fluid, has been discussed by Middleton in
\cite{midd3}.  Power-law scaling FLRW solutions for theories with
$L=R+\alpha \sqrt{R^2-4R_{\mu\nu}R^{\mu\nu}+R_{\mu\nu\rho\sigma}R^{\mu\nu\rho\sigma}}$ have been
investigated in \cite{powergb}.  The extent to which the FLRW solutions
of General Relativity can be reproduced in these theories is discussed
in \cite{fgenfried,fgenfried2}.

As well as isotropic cosmological solutions, a number of studies have
also been performed of anisotropic cosmological solutions in these theories.  The
simplest of these are probably the Bianchi $I$ Kasner-like exact solutions
found in \cite{fgenkasner}, which were used to show that the infinite
sequence of anisotropic oscillations that occurs on approach to the
initial singularity in General Relativity does not occur in
higher-order gravity theories, except in unphysical situations.  This type of
solution was further studied in \cite{midd3}.  Exact Bianchi type $II$
and $VI_h$ solutions were found by Barrow and Hervik in
\cite{barher1}, for quadratic theories of the type (\ref{quadL2}),
and were used to show the lack of validity of the cosmic no-hair
theorems in these theories:  Anisotropic inflation with positive
$\Lambda$ is possible, without de Sitter space as
the late-time asymptote.  These authors also considered the general
behaviour of Bianchi type $I$ and $II$ solutions in quadratic theories,
where the possibility of a stable isotropic singularity was discovered
\cite{barher2}.   Bianchi type $I$, $IV$, $VI_h$ and $VII_h$ universes
have been studied in \cite{barher3},  where it was shown that periods
of anisotropic expansion can occur after a near isotropic expansion,
and before re-isotropisation at late-times.  Bianchi type $VII_A$
solutions have been studied for quadratic theories in \cite{frb7a}, and Bianchi type $IX$
universes have been studied by Cotsakis {\it et al.} in \cite{fgen9},
where the Kasner solution of General Relativity was shown not to be a
stable early asymptote of the quadratic theories given in Eq. (\ref{quadL2}).

Stability of past isotropic attractors has been the subject of study by
Barrow and Middleton \cite{barmid1,barmid2}.  In the first of
these papers the authors demonstrated the stability of past isotropic
solutions to the quadratic theories (\ref{quadL2}) under scalar, vector
and tensor inhomogeneous perturbations \cite{barmid1}.
This supports the hypothesis that small perturbations to the past isotropic
attractor form part of the general cosmological solution to
quadratic theories of fourth-order gravity.  This study is extended
to theories with power-law curvature terms, $(R_{\mu\nu} R^{\mu\nu})^n$, in
their Lagrangian in \cite{barmid2}, where conditions are given for the
stability of early isotropic states.   This study also shows the
instability of the exact solution found in \cite{fgenkasner}, as the
initial singularity is approached. The asymptotic behaviour of theories with
quadratic corrections to the Einstein-Hilbert were studied, in the
context of string cosmology, in \cite{fgenstring1,fgenstring2,fgenstring3}.   Exponential and power-law FLRW solutions in
higher-dimensional string inspired models are found in \cite{Ohta1,Ohta2,Ohta3}.  The evolution of FLRW solutions in generalised
theories has also been studied using a dynamical systems analysis in
\cite{fgenfrw1}. 

\subsubsection{Physical cosmology and dark energy}
\label{fgencossec}

Having discussed various cosmological solutions in these general
fourth-order theories, let us now consider their relevance for
observational cosmology and dark energy.  We will proceed with this by
first discussing studies of more general theories, followed by
theories constructed from the Gauss-Bonnet curvature invariant.  The
Gauss-Bonnet invariant has special properties, which we will discuss
in the Section \ref{fgenother}.
\newline
\newline
\noindent
{\it General Theories}
\newline

In order to construct cosmological models
that can produce late-time accelerating expansion the authors of
\cite{fgende1} considered theories of the type
\be
\label{fgendeL}
L=R + \frac{\mu^{4n+2}}{(a R^2 + b R_{\mu\nu}R^{\mu\nu} + c R_{\mu\nu\rho\sigma}
  R^{\mu\nu\rho\sigma})^n},
\ee
where $\mu$, $n$, $a$, $b$ and $c$ are constants.  It
is found that for these theories there exist power-law attractors for
the general 
spatially flat FLRW solutions, which are given by
\be
\label{fgende1}
a(t) = a_0 \left( \frac{t}{t_0} \right)^{\frac{8n^2+10n+2-3 \alpha \pm
  \sqrt{\Gamma}}{4 (n+1)}},
\ee
where
\ba
\alpha &\equiv& \frac{12 a +4 b+4 c}{12 a+3 b+2 c}\\
\Gamma &\equiv& 9 n^2 \alpha^2- (80 n^3 + 116 n^2 +40 n +4) \alpha
  \nonumber \\&&+64
n^4+160 n^3+132 n^2+40 n+4.
\ea
The smaller of the exponents in Eq. (\ref{fgende1}) can be seen to
  $\rightarrow 0$ as $n \rightarrow \infty$, while the larger tends to
  $4 n$.  For large $n$ it is therefore the case that accelerating
  expansion can occur at late times. This is a generalisation of the
  type of model considered in \cite{carfr}, for $f(R)$ gravity.  In
  \cite{fgende2} it is shown that while the theory given in
  Eq. (\ref{fgendeL}) is capable of explaining the supernova results,
  to do so and still have an acceptable age for the Universe it
  requires the matter content of the Universe to have an equation of
  state $0.07 \leq w \leq 0.21$, to $2 \sigma$.  The FLRW solutions of
  theories with powers of $R$, $R_{\mu\nu}R^{\mu\nu}$ and $R_{\mu\nu\rho\sigma}R^{\mu\nu\rho\sigma}$
  added to the Einstein-Hilbert action were also studied in
  \cite{fgeneas}, where the possibility of late-time accelerating
  expansion was considered.

Primordial nucleosynthesis in theories with powers of $R_{\mu\nu}R^{\mu\nu}$
added to the Einstein-Hilbert action have been considered in
\cite{fhigh12}, where constraints from observed element
abundances are imposed.  As with scalar-tensor theories, the
constraints imposed from big bang nucleosynthesis are largely due to
the different expansion rate during the radiation dominated period due
to a different value of the effective Newton's constant.

The addition of a conformally invariant term to the Einstein-Hilbert
action has been considered in \cite{fgenbe1,fgenbe2,fgenbe3,fgenbe4}.  In this case the gravitational
Lagrangian takes the form given in Eq. (\ref{quadL2}) with $3\alpha +
\beta=0$, and the resulting field equations are sometimes known as the
`Bach-Einstein equations'.  The solutions to these equations have been
studied in the context of inflation \cite{fgenbe1}, the
evolution of background cosmological models \cite{fgenbe2}, the
observational constraints available from pulsars \cite{fgenbe3}, and
weak fields and gravitational waves \cite{fgenbe4}.  Theories of this
type are motivated, in part, from non-commutative
geometry \cite{noncom}. For a review on short scale modifications of
gravity in the context of non-commutative
geometry, see \cite{Nicolini1}.
\newline
\newline
\noindent
{\it Theories with $L=f(R,R^2-4 R_{\mu\nu}R^{\mu\nu} + R_{\mu\nu\rho\sigma} R^{\mu\nu\rho\sigma})$}
\newline

Theories that are functions of the Ricci scalar, $R$, and the Gauss-Bonnet combination,
\be
{\hat G} = R^2-4 R_{\mu\nu}R^{\mu\nu} + R_{\mu\nu\rho\sigma} R^{\mu\nu\rho\sigma},
\ee
have been particularly well studied, as they are motivated by string
theory \cite{Metsaev-order-alpha,gbstring2,gbstring3,gbstring4}, and have improved stability
properties (as will be discussed in Section \ref{fgenother}).  The
linear case of $f=R+{\hat G}$ is known to be equivalent to the
Einstein-Hilbert Lagrangian in 4 dimensions, up to surface terms, but
more general functions, of the type $f(R,{\hat G})$, produce field
equations that differ from those of General Relativity.  The
mathematical properties of the Gauss-Bonnet tensor, that occurs from
varying the action of these theories, as well as the more general
Lovelock tensor, have been studied in \cite{farh1,farh2,farh3,farh10}.

The general behaviour of spatially flat FLRW solution in theories with
$L=R+f({\hat G})$ has been studied by Zhou, Copeland and Saffin in
\cite{fgenzcs} using a phase plane analysis.  In this case the
Friedmann equations become
\ba
3 H^2 &=& {\hat G} f_{\hat G}-f-24 H^3 \dot{f}_{\hat G}+\frac{\chi}{2} \rho\\
2 \dot{H} &=& 8 H^3 \dot{f}_{\hat G}-16 H \dot{H} \dot{f}_{\hat G} - 8 H^2
\ddot{f}_{\hat G}+\frac{\chi}{2} (\rho+P),
\ea
where $R=6 (\dot{H}+2H^2)$ and ${\hat G}=24 H^2 (\dot{H}+H^2)$. The existence of both
stable de Sitter space, and phantom-like accelerating solutions
to the above equations can be demonstrated, as well as trajectories in
the phase 
space that mimic the evolution of the standard $\Lambda$CDM universe
through radiation and matter dominated periods \cite{fgenzcs}.  The stability
of de Sitter space, as well as radiation and matter dominated epochs
has also been studied by de Felice and Tsujikawa in \cite{fgends}, where the conditions $f_{{\hat G}{\hat G}}
>0$ and $f_{{\hat G}{\hat G}} \rightarrow 0_{+}$ as $\vert {\hat G} \vert \rightarrow \infty$
were found to be required for models to be viable.  These authors
suggest the following functional forms for $f({\hat G})$ as examples that satisfy these conditions,
and could produce acceptable expansion histories for the Universe:
\ba
f({\hat G}) &=& \lambda \frac{{\hat G}}{\sqrt{{\hat G}_*}} \tan^{-1} \left( \frac{{\hat G}}{{\hat G}_*}
\right) -\frac{\lambda}{2} \sqrt{{\hat G}_*} \ln \left( 1+\frac{{\hat G}^2}{{\hat G}_*^2}
\right) - \alpha \lambda \sqrt{{\hat G}_*}\\
f({\hat G}) &=& \lambda \frac{{\hat G}}{\sqrt{{\hat G}_*}} \tan^{-1} \left( \frac{{\hat G}}{{\hat G}_*}
\right) - \alpha \lambda \sqrt{{\hat G}_*},
\ea
where $\alpha$, $\lambda$ and ${\hat G}_*$ are constants.  It is further
claimed that these forms of $f({\hat G})$ are compatible with solar system
observations \cite{gbweak1}, producing corrections to the
Schwarzschild metric that are of the form $\sim H^2 r_s^2 (r/r_s)^p$,
where $r_s$ is the Schwarzschild radius of the Sun, $H$ is the Hubble
rate and $p$ is a model dependent quantity.  Much larger correction
to General Relativistic predictions are claimed in \cite{gbweak2} for
theories with polynomial additions of the Gauss-Bonnet term, ${\hat G}^n$, to
the Einstein-Hilbert action.  The cosmologies of these theories, and
theories with inverse powers of $\alpha {\hat G}+\beta R$ added to the
Einstein-Hilbert action (where $\alpha$ and $\beta$ are constants), have
been considered in \cite{gbinv1,gbbg3}, while the FLRW
solutions of other $R+f({\hat G})$ theories have been considered
in \cite{gbbg1,gbbg2}. The `inverse problem',
of finding FLRW  solutions that behave like $\Lambda$CDM has been
considered for $R+f({\hat G})$ theories in \cite{gbinv2}.
The phase space of FLRW solutions to $L=f(R,{\hat G})$ theories, and the
transition from deceleration to acceleration, has also been
studied in \cite{fgenal1,fgenal2,fgenal3}.
Supernova, BAO and CMB observations have been used to constrain
$L=R+f({\hat G})$
theories in \cite{molden3}.

Linear perturbations around spatially flat FLRW universes have been
studied in $L=f(R,{\hat G})$ theories by de Felice, G\'{e}rard
and Suyama in \cite{fgenperts}, and in $L=R+f({\hat G})$ theories
in particular by Li, Barrow and Mota \cite{fgenperts3}.  The former of
these studies uses the velocity potentials and variational principle 
approach of Schutz \cite{schutz}, while the latter uses the
covariant formulation of Ellis and Bruni \cite{ellisbruni}.  In the
case of general $f(R,{\hat G})$ it is found that scalar perturbations
in these theories have, in general, six degrees of freedom, two of
which propagate on small scales with group velocity \cite{fgenperts}
\be
\label{gbpert} 
v_g^2 \simeq - \frac{256}{3} \frac{\dot{H}^2 (f_{RR}f_{{\hat G}{\hat G}}-
  f^2_{R{\hat G}})}{(8 f_{R{\hat G}} H^2 +16 H^4 f_{{\hat G}{\hat G}}+f_{RR})(f_R+4 H
  \dot{f}_{\hat G})} \frac{k^2}{a^2}.
\ee
There can be seen to be a $k$-dependence in Eq. (\ref{gbpert}), which
  does not occur in General Relativity.  Such a relation had
  previously been found for the vacuum case in \cite{fgenpertsvac},
  where it was argued that the space-time is unstable if $v_g^2 <0$,
  or has super-luminal modes in short wavelength modes if $v_g^2
  >0$.  These features are problematic, but can be avoided in theories
  that satisfy $f_{RR} f_{{\hat G}{\hat G}}-f^2_{R{\hat G}} =0$.  Such theories have
  scale independent propagation speeds only, and include the cases of
  $L=f(R)$, $f({\hat G})$, ${\hat G}+f(R)$ and $R+f({\hat G})$.  The latter case, being the
  subject of study in \cite{fgenperts3}, has also been shown to suffer
  from matter instabilities.  This is due to the evolution equation
  for perturbations to $f_{\hat G}$, which we write as $\epsilon$, and which
  obeys \cite{fgenperts3}
\be
\ddot{\epsilon} + \left( \theta + \frac{4 \dot{\theta}}{\theta}
  \right) \dot{\epsilon}+\left[ \left( 1 + \frac{4
  \dot{\theta}}{\theta^2} \right) 
  \frac{k^2}{a^2} - 2
  \left(\dot{\theta}+\frac{\dot{\theta}^2}{\theta^2} +\frac{2\theta^2}{9}
  \right) -  \frac{27 (3-4 \dot{f_{\hat G}} \theta
  )}{48 \theta^4 f_{{\hat G}{\hat G}}} \right] \epsilon =S,
\ee
where $\theta=3H$ is the expansion scalar, and the reader is referred
  to \cite{fgenperts3} for the form of the source term $S$.  For
  stability it is required that the third term in the square
  brackets be positive, and remain dominant over the 
  first, which is expected to be negative during matter domination.
  This requires $f_{{\hat G}{\hat G}} \geq 0$, and for $f_{{\hat G}{\hat G}}H^6$ to
  remain suitably small, in order to avoid instabilities
  \cite{fgenperts3,fgenperts2}.  These are strong constraints
  on the forms of $f({\hat G})$.  It  
  was further shown in  \cite{fgenperts} that vector modes in the
  general case of $f(R,{\hat G})$ decay, and that the propagation of tensor
  modes in these theories is model dependent.

\subsubsection{Other topics}
\label{fgenother}

Let us now consider some remaining topics in fourth-order gravity, that
have yet to be discussed.
\newline
\newline
\noindent
{\it Theories with $L=f(R,\phi,R^2-4 R_{\mu\nu}R^{\mu\nu} + R_{\mu\nu\rho\sigma} R^{\mu\nu\rho\sigma})$}
\newline

There has been some study of fourth-order theories that include a scalar
field, as well as the Ricci scalar and the Gauss-Bonnet scalar, in the
gravitational action.  This is motivated by the dilaton that arises in
string theory compactifications \cite{Metsaev-order-alpha}, and has been studied
in terms of `pre-big bang' cosmology in \cite{gbstring2,gbstring3,gbstring3b}.  In
this scenario there is an early period of very rapid expansion due to
the kinetic term of the scalar field.  Black holes in these
theories, and their extensions, have been studied in \cite{gbbh1,gbbh2,gbbh3,gbbh4,gbbh5,gbbh6,Ohta6,Ohta7,Ohta8,Ohta9,Ohta10,Ohta11,Ohta12,Ohta13,Ohta14,Ohta15,Ohta16}.  Further string motivated
study of FLRW cosmology in the context of these theories has also been
performed in \cite{gbsca0,gbsca0b,gbsca4,gbinv2,Ohta4}.

Late-time acceleration has also been studied in theories where a
scalar field has been included, along with $R$ and $G$, in the
gravitational action  \cite{gbsca1,gbsca1ba,gbsca1bb,gbsca1b,gbsca2,gbsca3,gbsca4,gbde99,gbinv2,gbsca4b,gbsca4c,Ohta5}.  These  
papers have considered the evolution of FLRW space-times, as well as
inflation, structure formation, and the constraints that can 
be imposed upon them from supernovae, CMB, BAO, solar system
observations, and primordial nucleosynthesis.  These observations place
strong constraints on the theories.
\newline
\newline
\noindent
{\it Greater-than-fourth-order Theories}
\newline

Another option that has been considered in the literature is that the
action itself could contain derivatives of curvature invariants, so
that \cite{buchhigh}
\be
L=f(R,\Box R, \Box^2 R, \dots , \Box^n R),
\ee
where $\Box$ is the D'Alembertian.  Extremising the action associated
with this Lagrangian, by varying the metric, gives the field equations
\cite{fhigh0}
\be
\label{fhigheq}
\mathcal{Y} R_{\mu\nu} -\frac{1}{2} f g_{\mu\nu} -
\mathcal{Y}_{;\mu\nu} +g_{\mu\nu} \Box \mathcal{Y}
+\mathcal{X}_{\mu\nu} = \frac{\chi}{2} T_{\mu\nu},
\ee
where
\ba
\nonumber
\mathcal{X}_{\mu\nu} &\equiv& \sum_{i=1}^n  \left[ \frac{1}{2} g_{\mu\nu} (\mathcal{Z}_{i} (\Box^{i-1}
  R)^{;\sigma})_{;\sigma}-\mathcal{Z}_{i;(\mu} \left(\Box^{i-1} R \right)_{;\nu)}\right],\\
\nonumber
\mathcal{Y} &\equiv& \sum_{i=0}^n \Box^{i} \frac{\partial f}{\partial
  (\Box^i R)},\\
\nonumber
\mathcal{Z}_i &\equiv& \sum_{j=i}^n \Box^{j-i} \frac{\partial f}{\partial
  (\Box^j R)}.
\ea
The field equations (\ref{fhigheq}) can be seen to generically contain
derivatives of the metric of order $2n+4$, so that the familiar
fourth-order theories discussed above are recovered when $n=0$.
Theories with infinite $n$ have also been considered in the
literature, and have been claimed to be ghost-free
\cite{fhigh9,Kbounce}.

Greater-than-fourth-order theories can be shown to be equivalent,
under a conformal
transformation, to General Relativity with two scalar fields
\cite{fhigh1}, and their Newtonian limit has been considered in
\cite{fhigh4}, where it was found that the familiar form of Newtonian
potentials and Yukawa potentials can be present.
Their consequences for inflation have been studied in \cite{fhigh2,fhigh5,fhigh6,fhigh7,fhigh8,qfr3,Kbounce}, and the attractor nature of de Sitter
space established. 
Bouncing cosmologies in these theories have been considered in
\cite{fhigh9,fhigh10,Kbounce}, and the form of the CMB has been
investigated in \cite{fhigh10}.   Primordial nucleosynthesis has been
considered in \cite{fhigh12}, and the consequences of this type of
theory for dark energy have been considered in \cite{qfr3}.  For
a more detailed overview of theories with greater than four
derivatives of the metric in their field equations the reader is
referred to \cite{frrev1}. 
\newline
\newline
\noindent
{\it Conformal Gravity}
\newline

One more possibility is to completely abandon the Einstein-Hilbert
action, even as a limiting case of the fundamental action. Such a
proposal has been advocated by Mannheim \cite{Mannheim2011b} in the
case of conformal Gravity. Here the Einstein-Hilbert action is replaced by
\begin{eqnarray}
S_C=-\alpha_G\int d^4
x\sqrt{-g}C_{\lambda\mu\nu\kappa}C^{\lambda\mu\nu\kappa},
\end{eqnarray}
where $C_{\lambda\mu\nu\kappa}$ is the Weyl tensor, given by
\begin{eqnarray}
C_{\lambda\mu\nu\kappa}&=&R_{\lambda\mu\nu\kappa}+
\frac{1}{6}R^\alpha_{\phantom{\alpha}\alpha}  
[g_{\lambda\nu}g_{\mu\kappa}-g_{\lambda\kappa}g_{\mu\nu}] \nonumber
\\&&-\frac{1}{2}[g_{\lambda\nu}R_{\mu\kappa} 
-g_{\lambda\kappa}R_{\mu\nu}-g_{\mu\nu}R_{\lambda\kappa}+g_{\mu\kappa}R_{\lambda\nu}],
\nonumber 
\end{eqnarray}
and $\alpha_G$ is a dimensionless gravitational self-coupling
constant. It has been shown that such an action can be obtained from
the path integral of fermionic degrees of freedom for the conformal
and gauge invariant action of a fermionic field.  

The case has been made that such a theory has a number of desirable properties eluding other
higher derivative theories of gravity. Even though the equations are
fourth order, signalling the presence of negative energy states, it
has been shown that such states are completely decoupled from what 
the authors dub the {\it physical} sector
\cite{Mannheim2011a}. Conformal gravity can then be held up as a
viable theory of quantum gravity. Furthermore, the peculiar
ultraviolet properties of conformal gravity have been argued to lead
to a solution to the cosmological constant problem.  

At a classical level, conformal gravity has been shown to have
intriguing properties. For a start, the non-relativistic limit of the
field equations leads to a fourth order differential equation for the
gravitational potential, $\Phi$, in which the usual Newtonian potentials that drop
off as $1/r$ are but one possibility. The general weak gravity potential is of the form
\begin{eqnarray}
\Phi(r)=-\frac{A}{r}+B r \nonumber.
\end{eqnarray}
Such a form, it has been argued, can be tuned to fit a range of galaxy
rotation curves.  As in some other theories of modified
gravity, this is achieved by fixing universal parameters. This may be
contrasted with the usual dark matter prescription in which, for each
galaxy, one can choose the properties of the dark matter halo.  

The situation becomes more complicated once one adds couplings to
matter fields \cite{Mannheim2001}. Conformal symmetry is spontaneously
broken through a new scalar degree of freedom, $S$, such that the
general (conformal) matter action is 
\begin{eqnarray}
I_M=-\hbar\int d^4x\sqrt{-g}[\frac{1}{2}S^\mu S_\mu- \frac{1}{12} S^2R^\mu_{\phantom{\mu}\mu}+\lambda S^4+i{\bar \psi}
\gamma^\mu\nabla_\mu\psi-gS{\bar \psi}{\psi}], \nonumber
\end{eqnarray}
where $\psi$ is a fermionic field. One can extend this to
more general actions containing scalar and fermionic fields.
The vacuum expectation value of the scalar field is what then sets the
Gravitational constant and the coupling to matter. It can also be used
to renormalise the cosmological constant.

Although some of the quantum properties of conformal gravity have been
worked out, a fully consistent and complete analysis of their
cosmology is still lacking. In particular, and in its current 
incarnation, in which no dark matter is invoked, it is unclear how the
correct angular diameter distance for the CMB can be obtained.
\newline
\newline
\noindent
{\it Theories with $L=f(T)$}
\newline

An interesting variant on generalisations of the Einstein-Hilbert
action are the $L=f(T)$ theories, where 
$T$ is a contraction of the torsion tensor (defined below).  These
theories generalise 
the `teleparallel' approach to General Relativity, which corresponds
to a Lagrangian $L=T$, from which Einstein's equations can be derived
\cite{fT1}.  Here $T$ is defined by
\be
T = \frac{1}{4} T^{\mu\nu\rho}T_{\mu\nu\rho} + \frac{1}{2} T^{\mu\nu\rho}T_{\rho\nu\mu}- T_{\mu
  \nu}^{\phantom{\mu \nu} \mu} T^{\nu\rho}_{\phantom{\nu\rho} \rho},
\ee
where $T^{\mu}_{\phantom{\mu}\nu\rho}$ is the torsion tensor, defined in terms
of the vierbein from $g_{\mu\nu}=\eta_{\alpha\beta} h^{\alpha}_{\mu} h^{\beta}_{\nu}$, as
\be
T^{\mu}_{\phantom{\mu}\nu\rho} = h^{\mu}_{\alpha} \left( \partial_{\nu} h^{\alpha}_{\rho} -
\partial_{\rho} h^{\alpha}_{\nu} \right).
\ee
This definition is equivalent to setting $T^{\mu}_{\phantom{\mu}\nu\rho}$ equal
to the antisymmetric part of the Weitzenbock connection.  
Now, varying the action
\be
S= \int \frac{\sqrt{-g}}{16 \pi G} f(T) d \Omega
\ee
with respect to the vierbein fields gives the field equations
\ba
\label{fTfield}
&&f_T \left( R_{\mu\nu}-\frac{1}{2} g_{\mu\nu} R \right) + \frac{1}{2} g_{\mu\nu}
\left(f-f_tT \right) 
\\ \nonumber &+&
\left( \frac{1}{2} \left( T_{\mu\nu\rho}+T_{\rho\nu\mu}+T_{\nu\mu\rho} \right) - g_{\mu\rho}
T^{\sigma}_{\phantom{\sigma}\nu\sigma} + g_{\mu\nu} T^{\sigma}_{\phantom{\sigma}\rho\sigma}
\right) f_{TT} \nabla^{\sigma} T 
= 8\pi G \Theta_{\mu\nu},
\ea
where we have called the energy-momentum tensor $\Theta_{\mu\nu}$, to
distinguish it from the torsion tensor.  It can be seen that in the
case $f(T)=T$ the field equations (\ref{fTfield}) reduce to Einstein's
equations, so that the theory $L=T$ is equivalent to the
Einstein-Hilbert action, as stated above.  For $f \neq T$, however,
the teleparallel approach outlined here gives different field
equations to the fourth-order theories we have so far considered.

It was within the framework of these generalised equations that
Bengochea and Ferraro suggested that the late-time accelerating
expansion of the Universe could be accounted for without dark energy
\cite{fT2}.  These authors considered the particular case
\be
f=T - \frac{\alpha}{(-T)^n},
\ee
where $\alpha$ and $n$ are constants, and constrained the resulting
FLRW cosmology they found with supernovae, BAOs and the CMB.  They
found the best fitting model has $n=-0.10$, $\Omega_m=0.27$, and has
the required radiation, matter and accelerating epochs.  A large number of papers
have followed \cite{fT2} in a short space of time, exploring the
transition from deceleration to acceleration, observational
constraints, conformal transformations, and structure formation
\cite{fT3,fT4,fT5,fT6,fT7,fT8,fT9,fT10,fT11,fT12,fT13,fT14,fT15,fT16}.  

It has been shown, however, that these
theories do not respect local Lorentz invariance, and have
a number of extra degrees of freedom that are not present in General
Relativity as a result \cite{fT17}.  This can seen by noticing that one can
write the Ricci scalar in terms of $T^{\mu}_{\phantom{\mu}\nu\sigma}$ as
\be
\label{fTR}
R= -T -2 \nabla^{\mu} \left( T^{\nu}_{\phantom{\nu}\mu\nu} \right).
\ee
Now, while $R$ is of course a Lorentz scalar, one can show that
$\nabla^{\mu} \left( T^{\nu}_{\phantom{\nu}\mu\nu} \right)$ is not. It therefore
follows that $T$ is not a Lorentz scalar either, and so the $f(T)$ theories do
not exhibit local Lorentz symmetry.  The exceptional case is $f=T$, in
which case the non-Lorentz invariant part of the action can be seen from
Eq. (\ref{fTR}) to be a total divergence, which does not affect the
field equations.
\newline
\newline
\noindent
{\it Stability Issues}
\newline

There are serious concerns with the stability of general theories of
the type $L=f(R,R_{\mu\nu}R^{\mu\nu}, R_{\mu\nu\rho\sigma}R^{\mu\nu\rho\sigma})$.  Not least of these
is the presence of `ghosts', or perturbative modes with negative
norm, as well as tachyonic instabilities in massive modes
\cite{fgenstelle,fgenstring2,fghost1,fghost2,fghost3,fghost4,fghost5,fghosta,fghostb}.

Let us now outline how ghost terms arise in these theories,
following the discussion of \cite{fghost1,fghost2,fghost3}.  This starts by considering the quadratic theory
\be
\label{fghosteq1}
\mathcal{L}=\sqrt{-g} \left[R+\frac{1}{6m_0^2} R^2 -\frac{1}{2 m_2^2}
C^2 \right],
\ee
where $m_0$ and $m_2$ are constants, and
$C^2=C_{\mu\nu\rho\sigma}C^{\mu\nu\rho\sigma}=R_{\mu\nu\rho\sigma}R^{\mu\nu\rho\sigma}-2R_{\mu\nu}R^{\mu\nu} +R^2$ is the
square of the Weyl tensor.  Any quadratic theory can be written in
this form, up to boundary terms, because of the Gauss-Bonnet
identity.  We now want to identify the scalar and spin-two degrees of
freedom in Eq. (\ref{fghosteq1}), for which it is convenient to
introduce auxiliary fields that play these roles, and to transform
so that they have canonical form.  We will do this now, following
\cite{fghost1}.
By introducing an auxiliary scalar field, $\varphi$, and
performing a conformal transformation $\tilde{g}_{\mu\nu} = e^{\varphi}
g_{\mu\nu}$, Eq. (\ref{fghosteq1}) can be rewritten as \cite{fghost1}
\be
\label{fghosteq2}
\mathcal{L} = \sqrt{-\tilde{g}} \left[ \tilde{R} - \frac{3}{2} \left(
  \tilde{\nabla} \varphi \right)^2- \frac{3m_0^2}{2}
  \left(1-e^{-\varphi}\right)^2 - \frac{\tilde{C}^2}{2m_2^2} \right].
\ee
By extremising this equation with respect to $\varphi$, and substituting the
resulting value of $\phi$ back into the Lagrangian density, one
recovers Eq. (\ref{fghosteq1}).  It can be seen from
  Eq. (\ref{fghosteq2}) that $\varphi$ now has the kinetic term of a
  canonical scalar field.
By introducing a second auxiliary field, $\pi_{\mu\nu}$, 
and transforming the metric so that 
$\bar{g}^{\mu\nu} = \tilde{g}^{\mu\rho} A_{\rho}^{\phantom{\rho} \nu}/\sqrt{\vert A \vert}$,
where $A_{\rho}^{\phantom{\rho} \nu}=(1+\frac{1}{2}\phi) \delta_{\rho}^{\phantom{\rho}\nu}
-\phi_{\rho}^{\phantom{\rho} \nu}$, we can then rewrite Eq. (\ref{fghosteq2})
as \cite{fghost1}
\ba
\label{fghosteq3}
\mathcal{L} &=& \sqrt{-\bar{g}} \Bigg[ \bar{R} - \frac{3}{2}
  \left(A^{-1}\right)_{\mu}^{\phantom{\mu} \nu}
  \bar{\nabla}^{\mu} \varphi \bar{\nabla}_{\nu} \varphi -
  \frac{3m_0^2}{2\sqrt{\vert A \vert}} \left(1-e^{-\varphi}\right)^2\\
 \nonumber && \qquad \qquad - \bar{g}^{\mu\nu} \left[C^{\rho}_{\phantom{\rho} \mu
  \sigma} C^{\sigma}_{\phantom{\sigma} \nu \rho} 
  - C^{\rho}_{\phantom{\rho} \mu \nu} C^{\sigma}_{\phantom{\sigma} \sigma \rho} \right]
  +\frac{m_2^2}{4\sqrt{\vert A \vert}} \left(\phi_{\mu\nu} \phi^{\mu\nu}
  -\phi^2\right) \Bigg]
 \\ \label{fghosteq4} &\simeq& \sqrt{-\bar{g}} \Bigg[ \bar{R} - \frac{3}{2}
  \bar{\nabla}^{\mu} \varphi \bar{\nabla}_{\mu} \varphi -
  \frac{3m_0^2}{2} \varphi^2 +\frac{m_2^2}{4} \left(\phi_{\mu\nu}
  \phi^{\mu\nu} -\phi^2\right) \\
 \nonumber && \qquad \;\; - \frac{1}{4} \Big( \bar{\nabla}_{\mu} \phi
  \bar{\nabla}^{\mu} \phi - \bar{\nabla}_{\mu} \phi_{\nu\rho} \bar{\nabla}^{\mu}
  \phi^{\nu\rho} 
+2 \bar{\nabla}^{\mu}
  \phi_{\mu\nu} \bar{\nabla}^{\nu} \phi -2 
  \bar{\nabla}_{\mu} \phi_{\nu\rho} \bar{\nabla}^{\rho} \phi^{\nu\mu} \Big)
\Bigg],
\ea
where in the second equality we have expanded out to quadratic order
  in $\varphi$ and $\phi_{\mu}^{\phantom{\mu}\nu}$ around zero, so that we are
  considering theories that are close to GR.
The field $\phi_{\mu}^{\phantom{\mu}\nu}=\pi_{\mu}^{\phantom{\mu}\nu}$ has been
  introduced here to make clear 
  with respect to which metric the indices are being raised or
  lowered:  those of $\pi_{\mu}^{\phantom{\mu}\nu}$ are raised and lowered
  with $\tilde{g}_{\mu\nu}$, while those of $\phi_{\mu}^{\phantom{\mu}\nu}$ are
  raised and lowered with $\bar{g}_{\mu\nu}$.  The quantities
  $C^{\mu}_{\phantom{\mu} \nu \rho}$ are defined as
\be
C^{\mu}_{\phantom{\mu} \nu \rho} = \frac{1}{2} \left( \tilde{g}^{-1} \right)^{\mu
  \sigma} \left( \bar{\nabla}_{\nu} \tilde{g}_{\rho \sigma} +\bar{\nabla}_{\rho}
  \tilde{g}_{\nu \sigma}-\bar{\nabla}_{\sigma} \tilde{g}_{\nu \rho} \right).
\ee
The Lagrangian given in Eq. (\ref{fghosteq4}) now has scalar, $\varphi$,
  and spin-2 modes, $\phi_{\mu}^{\phantom{\mu}\nu}$, both in canonical form.
  It can be seen from Eq. (\ref{fghosteq4}) that for real
  $\phi_{\mu}^{\phantom{\mu}\nu}$ the spin-2 field does
  indeed have the wrong sign before its kinetic term, and is therefore
  generically a ghost, while for real $\varphi$ the scalar mode is not
  \cite{frwhitt}.  What is more, if $m_0^2<0$ or $m_2^2<0$ then the
  scalar or spin-2 modes exhibit tachyonic instabilities, respectively.

Having outlined the proof for the generic existence of spin-2 ghosts
in quadratic fourth-order theories of gravity, (\ref{fghosteq1}), let
us now extend this to more general theories of the form
\be
\label{fghosteq5}
\mathcal{L}=\sqrt{-g} f(X,Y,Z),
\ee
where $X \equiv R$, $Y\equiv R_{\mu\nu}R^{\mu\nu}$ and $Z \equiv R_{\mu\nu\rho\sigma}
R^{\mu\nu\rho\sigma}$.  This demonstration proceeds by showing that the particle
content of theories of the type (\ref{fghosteq5}) are the same as the
quadratic theories (\ref{fghosteq1}) (at least, when considering
fluctuations around de Sitter space), and then using the result
derived above, that these theories generically contain spin-2 ghosts.
The first step here is to introduce auxiliary fields $\phi_1$,
$\phi_2$ and $\phi_3$ so that the Lagrangian density (\ref{fghosteq5})
becomes \cite{fghost2,fghost3}
\be
\mathcal{L} = \sqrt{-g} \left[ f + f_1
  (X-\phi_1) +f_2 (Y-\phi_2) +f_3 (Z-\phi_3) \right],
\ee
where $f_i \equiv \partial f/\partial \phi_i$, and $i=\{1,2,3\}$.  As
long as $f_{i}$ is non-degenerate, extremising with respect to the
these three new fields then gives $\phi_1=X$, $\phi_2=Y$ and
$\phi_3=Z$, so that Eq. (\ref{fghosteq5}) is recovered.  Using the
Gauss-Bonnet combination, and discarding boundary terms, this equation
can then be written as \cite{fghost3}
\ba
\mathcal{L} &=& \sqrt{-g} \Big[ (f-\phi_1 f_1-\phi_2 f_2-\phi_3 f_3)
  +f_1 R + \frac{1}{3} (f_1+f_3) R^2 \\&& \qquad \;\;\nonumber  +
  \frac{1}{2} (f_2+4f_3) C_{\mu\nu\rho\sigma} 
  C^{\mu\nu\rho\sigma} \\&& \qquad \;\;\nonumber-\frac{1}{2} (f_2+2 f_3) \left(
  C_{\mu\nu\rho\sigma} C^{\mu\nu\rho\sigma} - 2 
  R_{\mu\nu}R^{\mu\nu} +\frac{2}{3} R^2 \right) \Big],
\ea
where $C_{\mu\nu\rho\sigma}$ is the Weyl tensor, and the last quantity in brackets
is the Gauss-Bonnet combination.  At this point the theory has been
shown to be equal to a scalar-scalar-scalar-tensor theory, with the
three scalars non-minimally coupled to quadratic curvature invariants.
Finally, we can choose to expand the
density above up to second order in fluctuations around a de
Sitter background with constant Ricci curvature, $R_0$, so that
\cite{fghost4} 
\be
\label{fghosteq6}
\mathcal{L} = \sqrt{-g} \left[ -\Lambda + \alpha R + \frac{1}{2 m_0^2}
  R^2 - \frac{1}{2 m_2^2} C^2 \right],
\ee
where boundary terms have been ignored, and we have defined
\begin{eqnarray*}
\Lambda &\equiv& \langle f-X f_X +X^2 (\frac{1}{2}f_{XX}- \frac{1}{4} f_Y- \frac{1}{6} f_Z)\\&&
\;\;\nonumber +X^3 
(\frac{1}{2} f_{XY}+ \frac{1}{3} f_{XZ})  +X^4 (\frac{1}{8} f_{YY}+\frac{1}{18} f_{ZZ}+ \frac{1}{6} f_{YZ}) \rangle_0\\ 
\alpha &\equiv& \langle f_X-X f_{XX} -X^2 (f_{XY}+\frac{2}{3} f_{XZ})\\&&
\;\;\nonumber -  X^3 ( \frac{1}{4}f_{YY} +\frac{1}{9}f_{ZZ} +\frac{1}{3}f_{YZ})\rangle_0\\
m_0^{-2} &\equiv& \langle (3 f_{XX}+2 f_{Y} +2 f_{Z} ) +X(3 f_{XY} +2
f_{XZ}) \\&& \;\;\nonumber+X^2 (\frac{3}{4} f_{YY} +\frac{1}{3}f_{ZZ} +f_{YZ}) \rangle_0\\
m_2^{-2} &\equiv& - \langle f_{Y} +4 f_{Z} \rangle_0,
\end{eqnarray*}
where $\langle \dots \rangle_0$ denotes the value of the quantity
inside the brackets on the de Sitter background.  It should be clear
that Eq. (\ref{fghosteq6}) is identical to Eq. (\ref{fghosteq1}), up
to the values of $\alpha$ and $\Lambda$.  The particle content
of general theories of the type (\ref{fghosteq5}), on a de
Sitter background, therefore also has a scalar mode with mass $m_0$, and a
ghost-like spin-2 mode with mass $m_2$.

As mentioned in Section \ref{fgencossec}, it has been suggested that
theories that are only functions of the Ricci scalar, $R$, and the
Gauss-Bonnet combination, $G=R^2-4 R_{\mu\nu}R^{\mu\nu} + R_{\mu\nu\rho\sigma}R^{\mu\nu\rho\sigma}$, can
evade the ghost problem outlined above \cite{fghost4}.  The reason
given for this is that the mass term $m_2^{-2} \rightarrow 0$ as $f_Y
\rightarrow 4 f_Z$, a condition that is satisfied for theories of the
type $L=f(X,Z-4Y)$, or, equivalently, $L=f(R,G)$.  When $m_2^{-2}$
vanishes it can be seen from
Eq. (\ref{fghosteq6}) that the term responsible for the ghost spin-2
fluctuations will also disappear.  Further requirements for the
non-existence of ghosts in $f(R,G)$ theories are discussed in
\cite{fghost5}, with 
particular reference to the model of \cite{fgende1}.  Such theories
may still be subjected to constraints on their parameters by the
possible existence of tachyonic instabilities, if $m_0^2<0$.  For
theories with $L=f(R,G)$ it can be seen that $m_0^{-2} = \frac{1}{3} R^2 f_{GG}$,
so that the condition $m_0^2>0$ is equivalent to the stability
condition $f_{GG}>0$ found in \cite{fgends,fgenperts2} and
\cite{fgenperts3} in the context of cosmology.

\subsection{Ho\v rava-Lifschitz Gravity}

Ho\v rava-Lifschitz (HL) gravity was proposed as a toy model of
quantum gravity \cite{Horava1, Horava2, Horava3}.  The model is
non-relativistic and relies on anisotropic scaling between space and
time in the UV to help render the theory asymptotically safe.
Furthermore, it was claimed that General Relativity could be recovered
in the infra-red by including additional relevant operators. HL
gravity in its various guises has been reviewed in a number of
articles \cite{Padilla-good, Sotiriou-HL-status, Weinfurtner-nutshell,
  Mukohyama-HL-review, Blas-good}. 

To understand the idea behind HL gravity we must first understand why
perturbative General Relativity is not UV complete.  The
non-renormalisability arises because the coupling constant has
negative mass dimension, $[G]=-2$, and the graviton propagator scales
as $1/p^2$.  Consider the following scalar field theory,
\be \label{phiact}
S=\int d^4 x\left[-\half (\del \varphi)^2+\lambda \varphi^6\right].
\ee
Again, the propagator scales like  $1/p^2$, and the coupling constant
has mass dimension $[\lambda]=-2$, so schematically at least, one
might expect this theory to be non-renormalisable too. To render the
theory asymptotically safe,  we need to improve the UV behaviour of
the propagator. One might do this by adding relativistic higher
derivative terms to the action, but this is known to introduce an
additional ghost-like degree of freedom. The reason for the existence
of this ghost can be traced back to higher order {\it time}
derivatives as opposed to {\it 
  space} derivatives. This observation suggests an alternative approach.
Let us abandon Lorentz invariance and introduce higher order spatial
derivatives without introducing any higher order time derivatives. The
former should improve the UV behaviour of the propagator, whereas the
latter guarantees the absence of ghosts. We therefore  modify the
kinetic term 
\be
-\frac{1}{2} (\partial \varphi)^2 \to \frac{1}{2} \dot
\varphi^2-\frac{1}{2} \varphi (-\Delta)^z \varphi, \label{zphi} 
\ee
where $\Delta$ is the spatial Laplacian. We now have a
non-relativistic dispersion relation $w^2 \propto k^{2z}$,  which
means that time and space scale differently,  
\be
x \to l x, \qquad {\rm and} \qquad t \to l^z t,
\ee
For large enough $z$, it follows that the coupling constant has a
non-negative scaling dimension, $[\lambda]=4z-6$  so we   expect the
theory to be power counting renormalisable, and ghost-free. On the
flip side, we have broken Lorentz invariance, which is well tested at
low energies.  However, we can cope with this by adding a relevant
operator of the form ${\cal L}_{rel}=\frac12 c^2 \varphi \Delta
\varphi$. This leaves the good UV physics unaffected, but allows
Lorentz invariance to be restored as an emergent symmetry in the IR,
with an emergent speed of light $c$. 

In HL gravity, one applies similar logic to the relevant perturbative
degrees of freedom,  schematically replacing $\phi$ with the graviton,
$h_{ij}$.  Since we will require time and space to scale differently
in this model, we must first choose a preferred time, $t$, which in
the language of General Relativity means making an ADM split
\cite{MTW} 
\be
ds^2=-N^2 c^2 dt^2+q_{ij} (dx^i +N^i dt) (dx^j +N^j dt),
\ee
where $q_{ij} (x, t)$ is the spatial metric and $N^i(x, t)$ is the
shift vector.  For the lapse function we consider two separate
scenarios: (i) the {\it projectable} case where the lapse $N=N(t)$ is
homogeneous and (ii) the {\it non-projectable} case where the lapse
$N=N(x, t)$ can depend on space.  Having chosen a preferred time, we
no longer have the full diffeomorphism group, Diff$({\cal M})$, but a
subset known as foliation preserving diffeomorphisms, Diff$({\cal M},
{\cal F})$,  generated by 
\be
\delta t=f(t), \qquad {\rm and} \qquad \delta x^i=\xi^i(x, t).
\ee
Diff$({\cal M}, {\cal F})$ is defined by the following set of
infinitesimal transformations 
\ba
\delta N &=&\del_t (Nf)+\xi^i \del_j N \label{dN} \\
\delta N^i &=& \del_t (N^if+\xi^i) +{\cal L}_\xi N^i \label{dNi} \\
\delta q_{ij} &=& f\del_t q_{ij} +{\cal L}_\xi  q_{ij}. \label{dgij}
\ea
Note that this hard breaking of diffeomorphism invariance is at the
root of many of the problems facing HL gravity as it allows additional
degrees of freedom to propagate \cite{Cai-dynamical,
  Charmousis-strong}.  To see the extra degree of freedom emerge it is
convenient to perform a Stuckelberg trick \cite{Stuck}, and
artificially restore full diffeomorphism invariance at the expense of
introducing a new field -- the Stuckelberg field.  This field becomes
strongly coupled as the parameters of the  low energy theory run
towards their diff-invariant values \cite{Charmousis-strong} (see also
\cite{Blas1, Koyama-Arroja, Sot-Pap}).  

We can think of the  lapse and shift as playing the role of gauge
fields in Diff$({\cal M}, {\cal F})$. It follows that  the projectable
case is the more natural since then the gauge fields have the same
space-time dependence as the corresponding generators. Having said that
one might expect it to be easier to match the non-projectable case to
General Relativity in the infra-red.  

In any event, the action from these theories is built from objects
that are covariant with 
respect to Diff$({\cal M}, {\cal F})$. These are the spatial metric,
$q_{ij}$, and the extrinsic curvature, $K_{ij}=\frac{1}{2N} \left[ \dot
  q_{ij} -2 \grad_{(i} N_{j)} \right]$, where $\grad_i$ is the
spatial covariant derivative. In the non-projectable case one should
also consider terms built from $a_i=\grad_i \log N$ \cite{Blas1}. To
build the gravitational analogue of the action (\ref{phiact}), we
replace the kinetic term  such that
\be
\frac12 \dot \varphi^2 \to \frac{1}{\kappa_G} \sqrt{q}N
(K_{ij}K^{ij}-\lambda K^2),
\ee
where $\kappa_G$ is the gravitational coupling with scaling dimension
$[\kappa_G]=z-3$, and $\lambda$ is a dimensionless parameter that also
runs with scale. Clearly, for the $z=3$ theory the gravitational
coupling constant is dimensionless, which may lead one to suspect the theory to
be power counting renormalisable.  For $z=3$ the  leading order term
in the  UV part of the action becomes
\be \label{pot}
-\frac12 \varphi (-\Delta)^3 \varphi \to  - \kappa_G \sqrt{q}N V_6,
\ee
where the dimension six contribution to the potential is
\be
V_6=\beta \grad_k R_{ij} \grad^k R^{ij} + \ldots .
\ee
Here $\beta$ is a dimensionless parameter, $R_{ij}$ is the spatial
Ricci tensor and  ``$\ldots$" denotes any  of the other possible
dimension $6$ operators that    one might wish to include,  e.g. $R^3,
R \Delta R, (a^ia_i)^3$ etc.

Now let us consider the type of relevant operators one might add. If
we demand our action to be invariant under spatial parity $x^i\to
-x^i$ and time reversal $t \to -t$, then we only need consider even
dimensional operators, 
\be
{\cal L}_{rel}=-\frac{1}{\kappa_G} \sqrt{q} N (V_2+\kappa_G V_4).
\ee
At dimension four these are \cite{Kimpton}
\begin{multline}
V_4=A_1 R^2 +A_2 R_{ij}R^{ij} ++B_1R a^i a_i+B_2R_{ij}a^ia^j+B_3R
a^i{}_i \\+C_1 (a^i a_i)^2+C_2 (a^i a_i)a^j{}_j+C_3 (a^i{}_i)^2+C_4
a^{ij} a_{ij},  
\end{multline}
whereas at dimension two,
\be 
V_2=-\alpha (a^i a_i)-c^2 R.
\ee
Here we have introduced the notation $a^{i_1 \ldots i_n}=\grad^{i_1}
\cdots \grad^{i_n} \log N$. Of course, terms of this form are only
relevant in the non-projectable case.  

The full $z=3$ gravitational theory is now given by
\begin{multline} \label{z=3act}
S=\int dt d^3 x   ~\frac{1}{\kappa_G} \sqrt{q}N \left[
  K_{ij}K^{ij}-\lambda K^2+c^2 R+\alpha (a^i a_i)^2-\kappa_G V_4
  -\kappa_G^2 V_6\right]+S_m,
\end{multline}
where $S_m[N, N_i, q_{ij}; \Psi]$ is the  matter part of the action,
and $V_6$ is the relevant dimension 6 operator.  Note that in the
absence of full diffeomorphism invariance we do not require matter
to satisfy energy-momentum conservation
\cite{Charmousis-strong}. Indeed, in general we expect to see
violation of energy conservation since Diff$({\cal M})$ breaking
operators in the gravity Lagrangian  will induce Diff$({\cal M})$
breaking quantum corrections to the matter Lagrangian \cite{Kimpton}. 

Let us compare this to the Einstein-Hilbert action, written in terms
of the ADM variables as
\be  \label{ehact} 
S_{GR}=\frac{c^4}{ 16 \pi G } \int dt d^3 x   ~ \sqrt{q}N
\left[\frac{K_{ij}K^{ij}-K^2}{c^2}+R\right] .
\ee
The claim is that $\lambda=1$ and $\alpha=0$ are the infra-red fixed
points of the renormalisation group flow. Of course, the parameter
$\alpha$ plays no role  in the projectable theory, since any terms
containing $a_i=\grad_i \log N$ will vanish for $N=N(t)$.  For both
the projectable and non-projectable theories, the free parameters run
to their infra-red fixed points at low energies, so that the HL action
(\ref{z=3act}) tends towards the Einstein-Hilbert action (\ref{ehact})
with  an emergent speed of light, $c$, and an emergent Newton's
constant, $G =\kappa_G c^2/16\pi$. 

Before delving further into the different manifestations of HL
gravity, let us pause to make a few general comments. The first
of these is with regard to the large number of terms appearing in the
potential.  We have
not bothered to present the contributions from dimension six operators
since they are two numerous\footnote{Note that the full set of
  inequivalent terms up to dimension six has been presented  for the
  projectable case in \cite{Sotiriou-phen, Sotiriou-qg}.}. To reduce
the number of terms, Ho\v rava originally borrowed the notion of {\it
  detailed balance} from condensed matter theory \cite{Horava2}, but
this has since been shown to lead to phenomenological
problems \cite{Charmousis-strong, Sotiriou-phen, Sotiriou-qg}. Of
course, the large number of terms appearing in the potential is
really only an aesthetic concern. 

 A second, more serious, concern, involves the fine tuning of light
 cones for each field. Since Lorentz invariance is not exact there is
 no symmetry guaranteeing that all fields see the same emergent light
 cone. We would like there to be  some mechanism suppressing Lorentz
 violating operators at low energies but preliminary investigations
 suggest that fine tuning is required \cite{Pospelov-on-lorentz}. It
 is possible that supersymmetry may be able help with this to some extent
 \cite{GrootNibbelink}. 
 
 Another issue that has yet to be fully explored concerns possible
 equivalence principle violations in HL gravity \cite{Kimpton}. To see
 how these might arise it is convenient to go to the Stuckelberg
 picture. The Stuckelberg trick was developed in the context of
 massive gauge theories \cite{Stuck}, but it has proven very useful in
 elucidating some of the key physics of HL gravity as well (see, for
 example, \cite{Charmousis-strong, Blas1, Kimpton}).  Recall that the
 anisotropic scaling of space and time requires a hard breaking of
 Diff$({\cal M})$ down to Diff$({\cal M, F})$.  We can {\it
 artificially} restore full diffeomorphism invariance by redefining
 the ADM slicing in terms of the Stuckelberg field $\phi(x, t)$. That
 is, the slicing goes from
\be
t= \textrm{constant} \qquad \to \qquad  \phi(x, t)=\textrm{constant}.
\ee
The unit normal to the spatial surfaces is covariant under Diff$({\cal
 M})$ and given by  the space-time gradient $n_\mu=\nabla_\mu
 \phi /\sqrt{-(\nabla \phi)^2}$. We can now express HL gravity as a
 relativistic theory involving the space-time metric, $g_\mn$, {\it
 and the Stuckelberg field} \cite{Blas1, Kimpton, Germani-HL}. We  do
 this by defining  the space-time analogue of the spatial metric and
 the extrinsic curvature in terms of the projection tensor
 $q_\mn=g_\mn+n_\mu n_\nu$ and its Lie derivative $\frac{1}{2}
 {\cal L}_n q_\mn$ \cite{Germani-HL}. Violations of the EP can occur
 because the Stuckelberg field can mediate a force between matter
 fields carrying Stuckelberg ``charge". As shown in \cite{Kimpton},
 Stuckelberg ``charge'' is a measure of violation of energy-momentum
 conservation,  schematically given by
\be
\Gamma \sim \sqrt{\left|\frac{\nabla_\mu T^\mn \nabla_\alpha
 T^\alpha_\nu }{T_\mn T^\mn} \right|} \neq 0. 
\ee
 This is allowed by foliation
 preserving diffeomorphisms which  simply require \cite{Kimpton}
\be
q_{\alpha \nu} \nabla_\mu T^{\mu\nu}=0, \qquad {\rm and} \qquad
 \frac{1}{\sqrt{-g}}\frac{\delta S_m}{\delta \phi}=-\frac{n_\nu
 \nabla_\mu T^{\mu\nu}}{\sqrt{-(\nabla \phi)^2}}. 
\ee 
Even if the Stuckelberg charges, $\Gamma_1, \Gamma_2, \ldots$ are
 always small, violation of the EP can still be large since the relevant E\"{o}tv\"{o}s
 parameter $\eta \sim \frac{\Gamma_1-\Gamma_2}{\Gamma_1+\Gamma_2}$
 really only cares about the charge ratios.

\subsubsection{The projectable theory} \label{sec:proj}

We now focus on the projectable version of HL gravity, for which the
lapse function is homogeneous, $N=N(t)$.  The action is then given by
\be \label{projact} 
S=\int dt d^3 x   ~\frac{1}{\kappa_G} \sqrt{q}N \left[
  K_{ij}K^{ij}-\lambda K^2-V\right]+S_m,
\ee
where the potential $V=-c^2R+$ higher derivative operators. Since the
condition of projectability is imposed at the level of the theory
itself, it follows that the Hamiltonian constraint is non-local:
\be
\int d^3x \sqrt{q}\left[ K_{ij}K^{ij}-\lambda
  K^2+V\right]=\frac{\delta S_m}{\delta N}.
\ee
In comparison with GR where the Hamiltonian constraint is local, this
admits a much larger class of solutions. Indeed, it has been suggested
that the resulting integration constant can account for dark matter
\cite{Mukohyama-dark}, although this may lead to the formation of
caustics and the break down of the theory \cite{Blas1}.
\newline
\newline
\noindent
{\it Dark matter as an integration constant}
\newline

To see how this might emerge, we rewrite the action (\ref{projact}) in
the following  form 
\begin{multline}
S_=\frac{c^4}{16\pi G} \int dt d^3 x   \sqrt{q}N \left[
  \frac{K_{ij}K^{ij}-K^2}{c^2}
  +R\right]+S_m\\+(1-\lambda)\frac{c^2}{16 \pi G} \int dt d^3 x
\sqrt{q}N K^2 +\textrm{UV corrections}.
\end{multline}
Focusing on the low energy theory, the resulting field equations are
\cite{Mukohyama-dark} 
\ba
\int d^3 x \sqrt{-g}\left( G^{(4)}_\mn-\frac{8\pi G}{c^2}
T_\mn\right) n^\mu n^\nu &=& {\cal O}(1-\lambda) \label{nneq}\\ 
\left(G^{(4)}_{i \mu}-\frac{8\pi G}{c^4} T_{i \mu}\right) n^\mu
&=&{\cal O}(1-\lambda) \label{ineq}\\ 
G^{(4)}_{ij}-\frac{8\pi G}{c^4} T_{ij} &=& {\cal O}(1-\lambda), \label{ijeq}
\ea
where $g_\mn$ is the full space-time metric, $G^{(4)}_\mn$ is the
corresponding Einstein tensor, and $n^\mu=\frac{1}{N}(1, -N^i)$ is the
unit normal to hyper-surfaces of constant $t$. The stress energy tensor,
$T_\mn$, is not necessarily conserved, as previously stated.  Note that
the non-local Hamiltonian constraint, (\ref{nneq}),  and the local
momentum constraint, (\ref{ineq}), are preserved by the dynamical
equations (\ref{ijeq}). 

Now, these equations can be rewritten as follows \cite{Mukohyama-dark}:
\be
G^{(4)}_\mn=\frac{8\pi G}{c^4} (T_\mn+T^{HL}_\mn)+{\cal O}(1-\lambda),
\ee
where $T^{HL}_\mn= \rho^{HL} n_\mu n_\nu$ and $\int d^3 x \sqrt{q}
\rho^{HL}=0$. Note that this latter condition does not require
$\rho^{HL}$ to vanish at all points in space, and one might wish to
identify $T^{HL}_\mn$ with a pressureless fluid moving with
$4$-velocity $n^\mu$. Taking $\rho^{HL}>0$ in our Hubble patch, we may
associate this integration constant with dark matter
\cite{Mukohyama-dark}. 

This scenario has been criticised in \cite{Blas1}, where it is argued
that the cosmological fluid $T^{HL}_\mn$ will inevitably lead to the
formation of caustics and the break down of the theory. To see why
this might be the case it is convenient to go to the Stuckelberg
picture, where we identify the unit normal with the space-time gradient
$n_\mu=\nabla_\mu \phi / \sqrt{-(\nabla \phi)^2}$.    Now
$T^{HL}_\mn$ behaves like a pressureless fluid, and in General
Relativity it is well known that this will lead to the formation of
caustics, due to the attractive nature of gravity. This is not a
problem for real dust, as virialisation can occur.  However, in the
scenario of \cite{Mukohyama-dark}, the fluid is characterised by the
gradient $\nabla_\mu \phi$ which  is problematic as the Stuckelberg
field is not differentiable at the caustic. These conclusions have
been disputed in \cite{Mukohyama-caustic} where it is argued that as
the putative caustic begins to form we enter a UV regime and   the
parameter $\lambda$ runs away from its IR fixed point  at
$\lambda=1$. At $\lambda \neq 1$ it is claimed that an extra repulsive
force could ultimately prevent caustics from forming. 
\newline
\newline
\noindent
{\it Perturbation theory -- ghosts, tachyons, and strong coupling}
\newline

Let us now consider linearised perturbations about a Minkowski
background. In what follows we will work in units where the emergent
speed of light is given by $c=1$. Given the residual diffeomorphism,
in Eqs. (\ref{dN}) to (\ref{dgij}), we can choose a gauge defined by
\be
N=1, \qquad N_i=\del_i B+n_i, 
\quad {\rm and} \quad q_{ij}=(1+2\zeta) \delta_{ij} +h_{ij},
\ee
where $n_i$ is a divergence-free vector, and $h_{ij}$ is a
transverse-tracefree tensor. To study the propagating degrees of
freedom we neglect the matter contribution,  and  integrate out the
constraints. As it is non-local, the Hamiltonian constraint does not
affect the local propagating degrees of freedom. Meanwhile, the
momentum constraint yields 
\be \label{momcon}
B=-\frac{1}{c_s^{ 2} \Delta} \dot \zeta, \qquad {\rm and} \qquad n_i=0,
\ee
where  
\be \label{cs}
c_s^2=\frac{1-\lambda}{3\lambda-1 }.
\ee
This will shortly be identified with the scalar speed of sound at low energies.
Plugging this into the action and expanding to quadratic order one
finds \cite{Sotiriou-qg, Mukohyama-HL-review}
\be\label{projact2}
S=\frac{1}{8\pi G}\int dt d^3 x\left[ -\frac{1}{c_s^{2} }\left( \dot
  \zeta^2+\zeta {\cal O}_s \zeta \right)+\frac{1}{8} \left( \dot
  h_{ij}^2+h^{ij} {\cal O}_t h_{ij} \right)\right]+S_\textrm{int},
\ee
where $S_\textrm{int}$ denotes the interactions and
\be
{\cal
  O}_s=c_s^2\left(\Delta+\frac{\Delta^2}{k_{UV}^2}+\frac{\Delta^3}{k_{UV}^4}\right), \qquad {\cal O}_t=\Delta+\mu \frac{\Delta^2}{k_{UV}^2} +\nu\frac{\Delta^3}{k_{UV}^4},
\ee
where $\mu$ and $\nu$ are dimensionless parameters of order one. Here
we assume that all Lorentz symmetry breaking terms in $V_4$ and $V_6$
depend on roughly the same scale, $k_{UV} \lesssim 1/\sqrt{8\pi
  G}$. The dispersion relation for the scalar is given by 
\be
w^2=c_s^2 \left(k^2+\frac{k^4}{k_{UV}^2}+\frac{k^6}{k_{UV}^4}\right).
\ee
Now we see a problem: At low energies, $k \ll k_{UV}$, we require
$c_s^2>0$ to avoid a tachyonic instability, where we identify $c_s$
with the speed of propagation of the scalar waves.
However, as we see from the action (\ref{projact2}), $c_s^2>0$
yields a ghost, which is
far more troubling. We therefore take $c_s^2<0$, but with $|c_s|$
being small so as to render the tachyonic instability mild. But how
small does $|c_s|$ need to be? The timescale of this instability is
$t_s \sim 1/|c_s |k>1/|c_s |k_{UV}$, and as this would need to exceed
the age of the 
universe, we infer $|c_s|<H_0/k_{UV}$,  where $H_0$ is the current
Hubble scale \cite{Blas1}. Furthermore, since modifications to
Newton's law have been tested down to the meV scale, we may impose
$k_{UV}>~$meV. This gives  $|c_s|<10^{-30}$, or equivalently
$|1-\lambda| <10^{-60}$, on scales $H_0\ll k\ll k_{UV}$.

What about the interaction terms? At low energies, $k \ll k_{UV}$, and
working up to  cubic order we find that \cite{Koyama-Arroja,
  Sotiriou-HL-status} 
\begin{multline}
S_\textrm{int} =\frac{1}{8\pi G}\int dt d^3 x \left\{ \zeta (\del_i
 \zeta)^2-\frac{2}{c_s^4} \del_t \zeta \del_i \zeta \frac{\del^i
 }{\Delta}\del_t  \zeta 
 \right.\\\left.
+\frac{3}{2}\left[\frac{1}{c_s^4} \zeta \left(\frac{\del_i
 \del_j}{\Delta} \del_t  \zeta\right)^2 
-\left(\frac{2}{c_s^2}+\frac{1}{c_s^4}\right) \zeta( \del_t
 \zeta)^2\right]\right\}+\ldots .
\end{multline}
To estimate the strength of these interactions we canonically
normalise the quadratic term (\ref{projact2}) in the infra-red by
redefining 
\be
t=\frac{\hat t}{|c_s|}, \qquad {\rm and} \qquad \zeta=\sqrt{8\pi G
  |c_S|} \hat \zeta,
\ee
so that
\begin{multline}
S_\textrm{int} =\sqrt{\frac{8\pi G}{ |c_s|^3}}\int d\hat t d^3 x
 \left\{ c_s^2 {\hat \zeta} (\del_i {\hat \zeta})^2-2 \del_{\hat t}
 {\hat \zeta} \del_i {\hat \zeta} \frac{\del^i }{\Delta}\del_{\hat t}
 {\hat \zeta} 
 \right.\\\left.
+\frac{3}{2}\left[{\hat \zeta} \left(\frac{\del_i \del_j}{\Delta}
 \del_{\hat t}  {\hat \zeta}\right)^2 
-\left(2c_s^2+1\right) {\hat \zeta}( \del_{\hat t} {\hat
 \zeta})^2\right]\right\}+\ldots .
\end{multline}
For small $|c_s|$, we see that the largest cubic interactions become
strongly coupled at a scale $\Lambda_{sc} \sim \sqrt{|c_s|^3/8\pi
  G}$. Imposing the constraint,  $|c_s|<10^{-30}$ and taking
$1/\sqrt{8\pi G} \sim 10^{18}$ GeV, we find $\Lambda_{sc} \lesssim
10^{-18} \textrm{eV}$.  This  lies well below scale of the UV
corrections given by $k_{UV} >$ meV so we can certainly trust the
effective low energy description we have used to derive this
scale. The implications for the theory are profound. The scale
$\Lambda_{sc}$ represents the scale at which perturbative quantum
field theory breaks down in Minkowski space. For scattering processes
above  $\Lambda_{sc} \lesssim 10^{-18} \textrm{eV}$ we must sum up the
contribution from all multi-loop diagrams. Since the claims of
renormalisability are based on the validity  of the perturbative
description at all energies, we see that much of the motivation for
studying this theory is lost. We also note that any notion of
Minkowski space is meaningless below distances $1/\Lambda_{sc} \gtrsim
10^8$km since one would require a scattering process at energies above
$\Lambda_{sc}$ to probe its structure. In analogy with DGP gravity
(see section \ref{sec:dgp}) one might hope to raise the scale of
strong coupling by considering fluctuations on curved backgrounds, for
example, on the background gravitational field generated by the Sun
\cite{Mukohyama-HL-review}. However, this seems optimistic since
Minkowski space is an excellent approximation\footnote{At distances
  $r\gtrsim 10^8$km from the Sun, the Newtonian potential  $V(r)
  \lesssim 10^{-8}$.} to the background geometry at distances of order
$ 1/\Lambda_{sc} \gtrsim 10^8$km, so the derived scale, $\Lambda_{sc}
\sim \sqrt{ |c_s|^3/8\pi G}$ should still be reliable.

\subsubsection{The non-projectable theory} \label{sec:nonproj}
We now consider the non-projectable theory for which the lapse
function can depend on space, $N=N(x, t)$, just as in General
Relativity. This means the Hamiltonian constraint is now local and
that  terms depending on $a_i=\grad_i \log N$  could play an
interesting role in the dynamics \cite{Blas2, Blas3}. Just as in the
projectable case, the absence of full diffeomorphism invariance allows
an extra scalar mode to propagate \cite{Charmousis-strong}. The action
is given by 
\begin{multline}
S_=\frac{c^4}{16\pi G} \int dt d^3 x   \sqrt{q}N \left[
  \frac{K_{ij}K^{ij}-K^2}{c^2} +R\right]+\alpha \frac{c^2}{16 \pi G}
  \int dt d^3 x \sqrt{q}N  a_i a^i\\ 
+(1-\lambda)\frac{c^2}{16 \pi G} \int dt d^3 x \sqrt{q}N K^2 +S_m
  +\textrm{UV corrections}.
\end{multline}
Again we consider small vacuum perturbations on a Minkowski
background, working in units where $c=1$. Given the reduced set of
diffeomorphisms, Eqs. (\ref{dN}) to (\ref{dgij}), we cannot  gauge
away the fluctuations in the lapse function that depend on
space. Instead, we choose a gauge 
\be
N=1+\chi, \qquad N_i=\del_i B+n_i, 
\quad {\rm and} \quad q_{ij}=(1+2\zeta) \delta_{ij} +h_{ij},
\ee
where, as before, $n_i$ is a divergence free vector, and $h_{ij}$ is a
transverse-tracefree tensor. Now integrating out the momentum
constraint, one finds 
$$
B=-\frac{1}{c_s^{ 2} \Delta} \dot \zeta, \qquad n_i=0,
$$
where $c_s^2$ is given by equation (\ref{cs}), but, as we will see, we
do not identify it with the speed of sound. The Hamiltonian constraint
yields \cite{Blas2, Sot-Pap}
$$
\chi=-\frac{2}{\alpha} \zeta.
$$
Expanding to quadratic order gives \cite{Blas2, Sot-Pap}
\be\label{nonprojact2}
S=\frac{1}{8\pi G}\int dt d^3 x\left[ -\frac{1}{c_s^{2} }\left( \dot
  \zeta^2+\zeta\tilde  {\cal O}_s \zeta \right)+\frac{1}{8} \left(
  \dot h_{ij}^2+h^{ij} \tilde{\cal O}_t h_{ij}
  \right)\right]+S_\textrm{int},
\ee
where $S_\textrm{int}$ denotes the interactions and
\be
\tilde {\cal
  O}_s=c_s^2\left(\frac{\alpha-2}{\alpha}\Delta+\frac{\Delta^2}{k_{UV}^2}+\frac{\Delta^3}{k_{UV}^4}\right), \quad {\rm and} \quad \tilde {\cal O}_t=\Delta+\tilde \mu \frac{\Delta^2}{k_{UV}^2}+\tilde \nu\frac{\Delta^3}{k_{UV}^4},
\ee
where $\tilde \mu$ and $\tilde \nu$ are dimensionless parameters of
order one. Again we assume that all Lorentz symmetry breaking terms in
$V_4$ and $V_6$ depend on roughly the same scale, $k_{UV} \lesssim
1/\sqrt{8\pi G}$. We now write down the dispersion relation for the
scalar
\be
w^2=c_s^2 \left(\frac{\alpha-2}{\alpha} k^2+\frac{k^4}{k_{UV}^2}+\frac{k^6}{k_{UV}^4}\right).
\ee
It follows that the speed of propagation of the scalar waves in the
non-projectable theory are given by
\be
\tilde c_s^2=c_s^2 \left(\frac{\alpha-2}{\alpha}\right).
\ee
The ghost and the tachyon problems can now be avoided by
simultaneously taking \cite{Blas2, Sot-Pap}
\be
c_s^2<0, \qquad {\rm and} \qquad 0<\alpha<2.
\ee
Recall that this was not possible in the projectable case, where we
had to accept the tachyonic instability and use the fact that this
should be slow relative to a Hubble time to place strong bounds on
$|1-\lambda|$. Since in the non-projectable case we no longer have
such concerns regarding tachyonic instabilities, the strongest bounds
on $|1-\lambda|$ and $|\alpha|$ come from preferred frame effects in
the Solar System, requiring \cite{Will2006}
\be \label{bounds}
|1-\lambda|, |\alpha| \lesssim 10^{-7}.
\ee
For $\lambda$ and $\alpha$ satisfying this bound, the speed of
propagation of the scalar is 
$$
\tilde c_s^2 \approx \frac{\lambda-1}{\alpha},
$$
which should not be too slow. If the scalar graviton fluctuations
propagated significantly slower than other fields to which they
couple, then those fields would decay into scalar gravitons via the
Cerenkov process. This would be particularly worrying for light since
it would result in photon decay and the absence of cosmic rays. To
avoid this problem we take $\tilde c_s^2 \sim 1$, and so $|1-\lambda|
\sim  |\alpha|$.  Solar system constraints on these theories are
considered further in \cite{HLsolar2, HLsolar3, HLsolar4}, based on
the solutions presented in \cite{HLsolar1}, and black holes are
studied in \cite{Ohta17,Ohta18,Ohta19}.

What about the interactions? Again, focusing on the low energy theory
we find that the quantum fluctuations on Minkowski space become
strongly coupled at the scale $\tilde \Lambda_{sc} \sim  \sqrt{
  |\lambda-1| /8\pi G} \sim  \sqrt{ |\alpha| /8\pi G}$ \cite{Sot-Pap,
  Kimpton}.  Note that this result can be derived using a direct
method similar to the one presented for the projectable case in the
previous section \cite{Sot-Pap}, or using the Stuckelberg method in
the decoupling limit \cite{Kimpton}. Now given the bounds in Eq.
(\ref{bounds}), it follows that the strong coupling scale is $\tilde
\Lambda_{sc} \lesssim 10^{15}$ GeV.  If $k_{UV} > \Lambda_{sc}$ we can
trust our low energy description, and the derivation of this scale.  As
explained in the projectable case, strong coupling casts serious
doubts on the claims of renormalisability as these rely on the validity
of perturbative quantum field theory.  However, if $k_{UV} <
\Lambda_{sc}$, we cannot trust our derivation of the strong coupling
scale, since the low energy description would not be valid there
\cite{Blas3}. In this scenario, new physics that softens the
interactions kicks in at $k \sim k_{UV}$. This situation is reminiscent of
the case in string theory where we introduce the string scale just below the
Planck scale, where strong coupling would otherwise occur. 

Whilst this seems promising there are some issues facing the
non-projectable theory. One of these relates to the formal structure
of the theory, and in particular  the constraint algebra, which  is
dynamically inconsistent. This manifests itself through the lapse
function vanishing asymptotically for  generic solutions to the
constraint equations \cite{hen}. The asymptotically flat solutions we
have just discussed represent a non-generic subset of measure zero in
the space of all solutions.

In the quantum version of the theory, its has been  claimed that one
must take $\lambda < 1/3$ in order to have a stable vacuum
\cite{shu}. This is  incompatible with phenomenological requirements
for the following reason:  We expect there to be 3 fixed points in the
renormalisation  group flow for $\lambda$. These are $\lambda=1$ (diff
invariance), $\lambda=1/3$ (conformal invariance) and $\lambda=\infty$
\cite{Horava2}. Now, at low energies we require  $|\lambda -1| <
10^{-7}$ and $c_s^2=\frac{1-\lambda}{3\lambda-1 }<0$. This suggests
that $\lambda$ flows from infinity in the UV to $\lambda=1$  in the
IR.

\subsubsection{Aspects of Ho\v rava-Lifschitz cosmology}

HL gravity was first applied to cosmology in
\cite{Kiritsis-HL-cosmology, Calcagni-cosmology}. As we have seen, in
the UV, the relevant degrees of freedom have an anisotropic dispersion
relation $w^2  \approx k^6/k_{UV}^4$. This  is often at the root of much of
the interesting cosmology  that has subsequently arisen, including (i)
a scale invariant spectrum of cosmological perturbations, without
early time acceleration \cite{Mukohyama-scale-invariant,
  Chen-scale-invariant},  (ii) cosmological bounces \cite{bounce1,
  bounce2, bounce3, bounce4, bounce5}, (iii) dark matter as an
integration constant \cite{Mukohyama-dark}, (iv) chirality of primordial gravity waves \cite{soda1} and  (v) enhancement of
baryon asymmetry, abundance of gravity waves, dark  matter, and so
on~\cite{Mukohyama-phen}. These latter effects occur because the
modified dispersion relation results in radiation scaling like $1/a^6$
as opposed to $1/a^4$ in the UV regime. We refer the reader to the
following review articles on this subject \cite{Mukohyama-HL-review,
  Saridakis-aspects}. 

For an FLRW universe, the background cosmology in both the projectable
and non-projectable theories is qualitatively very similar. Choosing
units where the emergent speed of light is $c=1$,  the Friedmann
equation takes the following form:
\be \label{HLfrw}
\frac{3 \lambda-1}{2} H^2=\frac{8 \pi G}{3}
\left(\rho+\frac{C(t)}{a^3} \right)+V(a),
\ee
where 
\be
V(a)=\frac{\kappa}{a^2}\left[1+m\left(\frac{\kappa}{k_{UV}^2
    a^2}\right)+n\left(\frac{\kappa}{k_{UV}^2a^2}\right)^2\right],
\ee
and where $m$ and $n$ are dimensionless parameters of order one.  Here
$\kappa=0,\pm 1$ is the spatial curvature. We assume that the matter
component with energy density, $\rho$, and  pressure, $P$, satisfies the
usual energy-conservation law, $\dot \rho+3H(\rho +P)=0$, although
this is not necessarily required in HL gravity, as we have already
discussed. 

The contribution from $C(t)/a^3$ depends on the theory in
question. For the projectable theory it corresponds to the ``dark
matter integration constant" \cite{Mukohyama-dark}, with $C(t) \to $
constant at low energies. For the non-projectable theory there is no
such contribution and $C(t) \equiv 0$.  

We immediately notice that the effective Newton's constant seen by
cosmology differs from the one derived by comparing the low energy
effective action to the Einstein-Hilbert action:
\be
G_{cosmo}=\frac{2}{3\lambda-1} G.
\ee
Although as $\lambda \to 1$, at low energies, we see that $G_{cosmo}
\to G$. 

To see how HL cosmology can admit a bounce, consider the limiting
behaviour of the right hand side of Eq. (\ref{HLfrw}). Neglecting
$C(t)/a^3$ and assuming $\rho$ scales like $1/a^3$ or $1/a^4$ we see
that this goes like $\sim n k_{UV}^{-4} \left({\kappa}/{a^2}\right)^3$, which is negative if $n \kappa<0$. By 
continuity, this suggests that there exists $a_{*}$ for which the
right-hand side of Eq. (\ref{HLfrw}) is zero at $a=a_{*}$. This
corresponds to the position of the bounce, since at this point $H=0$. 

Observational constraints on $|\lambda-1|$ coming from  the background
cosmology have been studied in \cite{Dutta1, Dutta2} using
BAO+CMB+SN1a, but they are not particularly strong.  At $1\sigma$
confidence level they find that $|\lambda-1| \lesssim 0.02$, which  is
far weaker than the bounds presented in previous sections. Recall that
in the projectable theory, stability considerations require
$|\lambda-1| \lesssim 10^{-60}$, whereas in the non-projectable theory
preferred frame effects require  $|\lambda-1| \lesssim 10^{-7}$. 

Cosmological perturbations in HL gravity have also been considered
(see, for example,  \cite{Mukohyama-scale-invariant,
  Chen-scale-invariant, HLperts1,
  HLperts2,HLperts3,HLperts4,HLperts5,HLperts6, HLpertsdS, Izumi-nonlin}).  Indeed,
for the projectable theory, it has been claimed that scalar
fluctuations on cosmological backgrounds are stable. This is in
contrast to the corresponding fluctuations on Minkowski space
\cite{HLpertsdS} which are known to suffer from either a ghost or
tachyonic instability, as we saw in section \ref{sec:proj}. Whilst this may be relevant to long wavelength modes, it is of no consequence on sub-horizon scales where we can trust the perturbative analysis about Minkowski
space, to a good approximation. Also, gravity waves produced during inflation have been found to be chiral in HL gravity, thereby representing a robust prediction of the theory \cite{soda1}.

\subsubsection{The $\Theta$CDM model}
HL gravity represents the UV completion of an interesting cosmological model, dubbed $\Theta$CDM \cite{ThCDM}. In this model, it is assumed that the old cosmological constant problem is solved in some way, such that  the net contribution to the cosmological constant is vanishing. The model then seeks to explain the tiny, but non-zero, amount of cosmic acceleration that is currently observed, without any fine tuning. Indeed, it is shown that  the model  allows for a technically natural small contribution to cosmic acceleration, without any corrections from other scales in the theory.

A key assumption corresponds to the fact that Lorentz invariance is broken in the gravitational sector.  Thus the theory contains a unit time-like vector field which may be  generic (as 
in Einstein-Aether theory) or expressed in terms of the gradient of a scalar field defining a 
global time (sometimes called the {\it khronon}\footnote{The khronon field is naturally identified with the Stuckelberg mode in HL gravity at low energies.}). The proposed 
acceleration mechanism appears generically when we  assume the existence of another field, $\Theta$, which is taken to be invariant 
under  shift transformations. The model is a valid effective field  theory up to a high cut-off just a few orders of magnitude below the Planck scale, with a UV completion offered by HL gravity in the khronon case. 

 In the absence of any matter sources (including the cosmological constant) 
the model possesses two solutions corresponding to Minkowski and de Sitter space-times. 
The former solution is unstable and the presence of an arbitrarily small amount of matter 
destroys it. The cosmological evolution of a matter-filled universe is driven to the de Sitter 
attractor, with effective equation of state $w = 
-1$. The value of the effective cosmological 
constant on the de Sitter branch is determined by the lowest dimension coupling between the 
Goldstone field and the khronon. Remarkably, it is technically natural to assume this coupling to be small as it is protected from radiative corrections by a discrete symmetry. Thus, 
in the absence of a contribution from the cosmological constant, the current value of cosmic 
acceleration would not present any fine-tuning problem.

 Interestingly, the evolution 
of cosmological perturbations is different in the $\Theta$CDM and $\Lambda$CDM models. In particular, the 
growth of linear perturbations is enhanced in $\Theta$CDM as compared to the standard $\Lambda$CDM 
case. The enhancement is most prominent at very large scales of order a few gigaparsecs, but 
extends also to shorter scales. Another difference is the appearance of an effective anisotropic 
stress, resulting in a non-trivial gravitational slip at very large scales. In principle, these effects  may allow one to discriminate between  $\Theta$CDM and $\Lambda$CDM in the near future. 

\subsubsection{HMT-da Silva theory}

We have discussed in Sections \ref{sec:proj} and \ref{sec:nonproj} how
the original versions of HL gravity are plagued with problems at the
level of both theory and  phenomenology. The root of this is the
breaking of diffeomorphism invariance and the additional scalar degree
of freedom that propagates as a result. With this in mind, Ho\v rava
and Melby-Thompson (HMT) proposed a modified version of the
projectable theory possessing an additional $U(1)$ symmetry
\cite{HMT}.  It is claimed that this extra symmetry  removes the
troublesome scalar degree of freedom, so that one is left with a
spin-2 graviton as the only propagating mode.  The HMT action is given
by 
\begin{multline} \label{hmt} 
S_{HMT}=\frac{1}{\kappa_G} \int dt d^3 x   ~ \sqrt{q}\left\{N \left[
  K_{ij}K^{ij}- K^2-V(q_{ij},  R_{ij}, \grad_k R_{ij})  
\right.\right.\\\left.\left.
+\nu \Theta^{ij} (2K_{ij}+\grad_i \grad_j \nu)\right]
-A(R-2\Omega)\right\},
\end{multline}
where $\Theta^{ij}=R^{ij}-\half R q^{ij} +\Omega q^{ij}$, and $\Omega$
is a dimensionful coupling constant that can run with scale. This
constant controls the scalar curvature of spatial slices, and can be
thought of as a second cosmological constant. As with the original
projectable theory, we assume $N=N(t)$, with a potential
$V=-c^2R+\ldots$ containing the usual terms up to dimension six in
order to guarantee the $z=3$ scaling in the UV.  In addition, however,
HMT  theory contains two new fields given by $A=A(x, t)$ and
$\nu=\nu(x, t)$. These are important in extending the symmetry group
to $U(1) \times \textrm{Diff}({\cal M}, {\cal F})$. 

The action is invariant under Diff$({\cal M}, {\cal F})$, with
\ba
\delta N &=&\del_t (Nf) \label{dN1} \\
\delta N^i &=& \del_t (N^if+\xi^i) +{\cal L}_\xi N^i \label{dNi1} \\
\delta q_{ij} &=& f\del_t q_{ij} +{\cal L}_\xi  q_{ij} \label{dgij1} \\
\delta A &=&\del_t (Af)+\xi^i \del_j A \\
\delta \nu&=& f\del_t \nu +\xi^i \del_j \nu.			
\ea
Note that $\nu$ transforms as a scalar, whereas $A$ transforms like a
spatial scalar and a temporal vector. Indeed, $A$ transforms exactly
as the lapse function would in a non-projectable theory. This is not a
coincidence. One can think of $A$ as being the next to leading order
term in the non-relativistic expansion of the lapse. Of course, one
ought to ask why we have not included the parameter $\lambda$ in front
of the $K^2$ term in the action, as in previous versions of HL
gravity. According to \cite{HMT}, the parameter $\lambda$ is fixed to
be equal to one by requiring the action to be invariant under a local
$U(1)$ symmetry:
\ba
\delta A &=& \dot \psi-N^i \grad_i \psi \\
\delta \nu &=&-\psi \\
\delta N^i&=&N \grad^i \psi.
\ea
It is this  symmetry that removes the scalar graviton. Furthermore,
fixing $\lambda=1$ ensures no conflict with  observational tests of
Lorentz violation at low energies. The HMT model has been applied to
cosmology in \cite{Wang-hmt}.

Recently, da Silva  has argued that in contrast to the claims of
\cite{HMT},  one can accommodate $\lambda \neq 1$ and still retain the
$U(1) \times \textrm{Diff}({\cal M}, {\cal F})$ invariance
\cite{daSilva}. Indeed, he  proposed the following action:  
 \begin{multline} 
S_\textrm{daSilva}=\frac{1}{\kappa_G} \int dt d^3 x   ~ \sqrt{q}\left\{N
\left[ K_{ij}K^{ij}- \lambda K^2-V(q_{ij},  R_{ij}, \grad_k R_{ij})  
\right.\right.\\\left.\left.
+\nu \Theta^{ij} (2K_{ij}+\grad_i \grad_j \nu)+
(1-\lambda)[(\Delta \nu)^2+2K \Delta \nu]\right] 
-A(R-2\Omega) \right\},
\end{multline}
which is invariant under the same symmetries as (\ref{hmt}).  Now we
must subject $\lambda$ to the same constraints  as before, in
particular those coming from preferred frame effects.  There are
claims that the extra symmetry will eliminate the scalar graviton even
when $\lambda \neq 1$ \cite{daSilva}, although more detailed study is
required to be sure. Preliminary investigations on this subject have
been carried out in \cite{Wang-sc}. In any event,  strong coupling problems have recently
 been shown to infect the matter sector of this theory \cite{Wang-sc1}, unless one introduces a low scale of Lorentz violation, in a way that is reminiscent of  \cite{Blas2, Blas3}.

\subsection{Galileons}
\label{galileons}

Galileon theory \cite{NicolisRattazziTrincherini2008} was originally developed
by Nicolis {\it et al.} to facilitate a model independent analysis of
a large class of modified gravity models.  In each case,
General Relativity on perturbed Minkowski space is modified by an additional
single scalar field, the galileon, with derivative
self-interactions. Although the galileon and the graviton both couple
to matter, any direct coupling between them is neglected to leading
order.  The resulting vacuum Lagrangian  is invariant under the
following shift in the galileon field
\be \label{galsym}
\pi \to \pi+b_\mu x^\mu +c.
\ee
This symmetry corresponds to a generalisation of  {\it Galilean}
invariance, hence the name. The inspiration for the model comes from
DGP gravity \cite{dgp}. In Section \ref{sec:dgp}, we will see how the
{\it boundary effective theory} on the DGP brane is well described by
the following action, 
\be
S^\textrm{DGP}_\textrm{eff} = \int d^4 x \left[{\cal L}_{GR}+{\cal
    L}^\textrm{DGP}_\pi\right], \label{seffgal}
\ee
 where
\ba
{\cal L}_{GR}  &=&\frac{1}{16\pi G} \left\{\frac{1}{4} \tilde h^\mn \left[\partial^2
  \left(\tilde h_\mn -\half \tilde h \eta_\mn\right)
  +\ldots\right]\right\}+\half \tilde h_\mn T^\mn, \label{LGR}  \\
{\cal L}^\textrm{DGP}_\pi &=&\frac{1}{16\pi G}  \left\{\frac{1}{2}\left[ 3  \pi \partial^2
  \pi-r_c^2 (\del  \pi)^2 \partial^2 \pi\right]\right\}+\frac{1}{2}\pi T. \label{pilag1}
 \ea
 The Lagrangian (\ref{seffgal}) has two components: a linearised GR
 piece, ${\cal L}_{GR}$, and a modification due to the brane bending
 mode, ${\cal L}^\textrm{DGP}_\pi$. It is  valid in the so-called
 decoupling limit in which all interactions go to zero except the
 scalar self-interactions.  Focusing on the $\pi$-Lagrangian, ${\cal
   L}^\textrm{DGP}_\pi$, Nicolis {\it et al.} observed that the vacuum
 field equations are built exclusively out of second derivatives,
 $\del_\mu \del_\nu \pi$. In particular, this means that there are no
 terms higher than second order, ensuring a well defined Cauchy
 problem and avoiding any of the potential problems arising from
 ghosts in higher derivative theories.  In addition there are  no
 first or zero derivative terms which means that the $\pi$ Lagrangian
 possesses the Galilean symmetry. This  is inherited from Poincar\'e
 invariance in the bulk \cite{bigal1}.  

One might expect that almost {\it any} co-dimension one braneworld
model with large distance deviations from GR will be described, in
part, and in some appropriate limit, by a generalised $\pi$ Lagrangian
possessing the Galilean symmetry. This essentially follows from the
fact that the extrinsic curvature of the brane is $K_\mn \approx
\del_\mu \del_\nu \pi$, on scales where we can neglect background curvature.

We  should also note that even if there is no direct coupling to matter and therefore no modification of gravity, galileons are of interest in their own right as a source of energy-momentum. In particular, one can potentially obtain violations of the null energy condition without introducing any instability \cite{Nicolis-NEC, Creminelli-galilean-genesis}.  Generically, however, a single galileon will result in superluminality, although the situation may be improved by going to multi-galileon theory (see section \ref{sec:multigal}).

\subsubsection{Galileon modification of gravity}
To see how to generalise the decoupling limit of DGP to a larger class
of modified gravity theories, let us consider the amplitude, ${\cal
  A}$, for  the exchange of one graviton between two conserved
sources, $T_\mn$ and $T'_\mn$.  In General Relativity, this amplitude
is given by
\be
 {\cal A}_{GR}=   \frac{16\pi G}{p^2} \left[T_\mn
   T'^\mn-\frac{1}{2} TT' \right],
\ee
where $T=T^\mu_\mu$. We are interested in the case where gravity is modified by
an additional scalar, so that locally we have 
\be
\delta {\cal A}={\cal A}- {\cal A}_{GR}=-\frac{1}{\alpha p^2} TT'.
\ee
Such a theory can be described by the following action
\be
S=\int d^4 x \frac{1}{16\pi G} \sqrt{-\tilde g} R(\tilde g)-\frac{\alpha}{2}
\pi \partial^2 \pi +\half \tilde h_\mn T^\mn +\pi T+\textrm{interactions},
\ee
where $\tilde g_\mn =\eta_\mn +\tilde h_\mn$. The fluctuation $\tilde
h_\mn$ is identified with the GR graviton, and as such, for a given
source and boundary conditions, it coincides with the linearised
solutions of GR. This statement is true to all orders in the ``decoupling" limit
\be \label{dec}
M_{pl}, \alpha, T^\mn \to \infty, \qquad
\frac{\alpha}{M_{pl}^2}=\textrm{const}, \quad {\rm and} \quad
\frac{T^\mn}{M_{pl}}=\textrm{const},
\ee
where $M_{pl}=\sqrt{1/8\pi G}$. Note that matter is
minimally coupled to the metric $g_\mn=\eta_\mn +h_\mn$, where the
{\it physical} graviton is $h_\mn=\tilde h_\mn +2\pi \eta_\mn$. 
\newline
\newline
\noindent
{\it Galileon Action and  Equations of Motion}
\newline

Now suppose we consider the decoupling limit (\ref{dec}) with the
additional assumption that the strength of some of the scalar
self-interactions can be held fixed. This amounts to neglecting the
back-reaction of the scalar onto the geometry so that we can consider
it as a field on Minkowski space. We retain some of the scalar
self-interactions for the following reason: we are interested in an
${\cal O}(1)$ modification of GR on cosmological scales, but we would
like this to be screened down to $\lesssim {\cal O}(10^{-5})$ on solar
system scales. As we will see, the derivative self-interactions can
help shut down the scalar at short distances through Vainshtein
screening\footnote{See section \ref{sec:dgp-sc} for a detailed
  discussion of the Vainshtein mechanism in DGP gravity.}. In the
decoupling limit, the action is given by  
\be
S\left[\tilde h_\mn , \pi\right] =\int d^4x {\cal L}_{GR}+{\cal
  L}_{\pi}, \label{galact}
\ee
where  ${\cal L}_\pi={\cal L}_{gal}(\pi, \del \pi, \del \del \pi) +\pi
T$ represents the  generalisation of the $\pi$-Lagrangian in DGP
gravity.

The vacuum part of the generalised  $\pi$-Lagrangian, ${\cal
  L}_{gal}(\pi, \del \pi, \del \del \pi)$ gives  second order field
equations, and  is assumed to be Galilean invariant in the sense that
  ${\cal L}_{gal} \to {\cal L}_{gal}+$ total derivative, when $\pi \to
  \pi +b_\mu x^\mu +c$. What is the most general Lagrangian  with this
  property? The answer is remarkably simple, and in four dimensions is
  given by \cite{NicolisRattazziTrincherini2008, Deffayet-covariant} 
\be
{\cal L}_{gal}(\pi, \del \pi, \del \del \pi)=\sum_{i=1}^5 c_i {\cal
  L}_i (\pi, \del \pi, \del \del \pi),
\ee
where the $c_i$ are constants, and\footnote{We define $(\del \del \pi
  )^n =(\del_{\alpha_1} \del^{\alpha_2} \pi )(\del_{\alpha_2}
  \del^{\alpha_3} \pi ) \ldots (\del_{\alpha_{n}} \del^{\alpha_1}
  \pi)$. Note that we have presented the simpler expressions as
  suggested by \cite{Deffayet-covariant}.} 
\ba
{\cal L}_1 &=& \pi \\
{\cal L}_2 &=& -\frac{1}{2} (\del \pi)^2 \\
{\cal L}_3 &=& -\frac{1}{2} \partial^2 \pi  (\del \pi)^2  \\
{\cal L}_4 &=& -\frac{1}{2}\left[ (\partial^2 \pi)^2  -(\del \del
  \pi)^2\right](\del \pi)^2  \qquad \\ 
{\cal L}_5 &=& -\frac{1}{2} \left[ (\partial^2 \pi)^3 -3 (\partial^2 \pi)(\del
  \del \pi)^2 +2 (\del \del \pi)^3 \right](\del \pi)^2. 
\ea
By construction, the variation of each component is built exclusively
  out of second derivatives, 
$$\frac{\delta}{\delta \pi} \int d^4x {\cal L}_i(\pi, \del \pi, \del
\del \pi)={\cal E}_i ( \del \del \pi),$$ 
where 
$${\cal E}_i( \del \del \pi)=(i-1)! \delta^{\mu_1}_{[\nu_1} \ldots
  \delta^{\mu_{i-1}}_{\nu_{i-1}]} (\del_{\mu_1} \del^{\nu_1} \pi)
\dots (\del_{\mu_{i-1}} \del^{\nu_{i-1}} \pi).  $$
Specifically, 
\ba
{\cal E}_1 &=& 1 \\
{\cal E}_2 &=& \partial^2 \pi \\
{\cal E}_3 &=& (\partial^2 \pi)^2 -(\del \del \pi)^2 \\
{\cal E}_4 &=& (\partial^2 \pi)^3 -3 \partial^2 \pi (\del \del \pi)^2 +2 (\del
  \del \pi)^3 \\ 
{\cal E}_5 &=&(\partial^2 \pi)^4 -6 (\partial^2 \pi)^2 (\del\del \pi)^2 +8 \partial^2
  \pi (\del \del \pi)^3+3 [ (\del \del \pi)^2]^2-6 (\del \del
  \pi)^4. \qquad 
 \ea
It follows that the field equations for the galileon model are
  therefore given by the following:
\ba
-\half \partial^2 \left(\tilde h_\mn -\half \tilde h \eta_\mn\right) +\ldots
  &=& 8\pi G T_\mn, \label{grlin} \\
\sum_{i=1}^5 c_i {\cal E}_i (\pi, \del \pi, \del \del \pi)&=& -T. \label{scalareom}
\ea
Equation (\ref{grlin}) corresponds to the linearised Einstein
  equations , and so their solution, $\tilde h_\mn$, corresponds to
  the standard GR solution for a given source and boundary
  conditions. The modification of GR is encoded entirely in the
  solution of the scalar  equation of motion, (\ref{scalareom}).
\newline
\newline
\noindent
{\it Galileon cosmology as a weak field}
\newline

The galileon theory has been constructed in terms of a tensor and a
scalar propagating on a Minkowski background. Whilst it is
straightforward to understand the weak gravitational field in the
solar system using this description, it is not clear how one should
describe cosmology. Fortunately, at distances below the curvature
scale any metric is well approximated by a local perturbation about
Minkowski space.    In what follows, {\it local} will mean local in
both space and time, which for cosmological solutions will correspond
to sub-Hubble distances and sub-Hubble times.  

Let us consider a spatially flat FLRW space-time. If we take our
position to be given by $\vec x=\vec y=0$ and $t=\tau=0$, then for
$|\vec x| \ll H^{-1}$ and $|t| \ll H$ we have \cite{NicolisRattazziTrincherini2008}
\be 
ds^2=-d\tau^2+a(\tau)^2 d\vec y^2 \approx \left[1-\half H^2 |\vec
  x|^2+\half (2\dot H+H^2) t^2\right] (-dt^2+d\vec x^2),
\ee
where the Hubble scale $H$ and its time derivative $\dot H$ are
evaluated  {\it now}.  We recognise this as a perturbation on
Minkowski space in Newtonian gauge, 
\be
ds^2 \approx -(1+2\Psi)dt^2+(1-2\Phi)d\vec x^2\,,
\ee
where the Newtonian potentials are~\footnote{It may look like that since $\Phi - \Psi = 2\Phi\ne0$ that this construction introduces anisotropic stress.
This is not the case, however, as strictly speaking the condition for the absence of anisotropic stress is $D^i_{\;\;j} (\Phi-\Psi)=0$. It is
easy to show that $D^i_{\;\;j} |\vec{x}|^2=0$, hence, no anisotropic stress is present.}
\be
\Psi=-\frac{1}{4}H^2 |\vec x|^2+\frac{1}{4}(2\dot H +H^2)t^2\,, \quad
	   {\rm and } \quad \Phi=-\Psi\,.
\ee
For a given cosmological fluid, the corresponding GR solutions have
Hubble parameter $H_{GR}$. Since $\tilde h_{\mu\nu}$ agrees with the
linearised GR solution, we have $\tilde h_{tt}=-2\Psi_{GR}$, and
$\tilde h_{ij}=2\Psi_{GR}\delta_{ij}$, where 
 \be
\Psi_{GR}=-\frac{1}{4}H_{GR}^2 |\vec x|^2+\frac{1}{4}(2\dot H_{GR}
+H_{GR}^2)t^2\,. 
\ee
Now in our modified theory  the physical Hubble parameter (associated
with $h_{\mu\nu}$) differs from the corresponding GR value, $H \neq
H_{GR}$. Since $h_\mn=\tilde h_\mn +2\pi \eta_\mn$, we have a
non-trivial scalar
\be
\pi= \Psi-\Psi_{GR}\,.
\ee
Note that a Galilean transformation $\pi \to \pi+b_\mu x^\mu+c$ merely
corresponds to a coordinate transformation $x^\mu\to
x^\mu-cx^\mu+\frac{1}{2}(x_\nu x^\nu b^\mu-2b_\nu x^\nu x^\mu)$ in the
physical metric. 
\newline
\newline
\noindent
{\it Self-accelerating solutions}
\newline

Of particular interest are self-accelerating solutions. A
self-accelerating vacuum is one that accelerates even in the absence
of any sources for the fields $\tilde h_{\mu\nu}$ and $\pi$. These are
familiar to us from DGP gravity (see Section \ref{sec:dgp}), where the
self-accelerating solution is haunted by ghosts.  Is the same true in
a general galileon scenario? Or can we identify a scenario that admits
a {\it consistent} self-accelerating solution? 

There is some ambiguity as to what is actually meant by
`self-acceleration'  if the tadpole term $\int d^4 x c_1 \pi$ is
present in the galileon Lagrangian. The point is that at the level of
the graviton equations of motion, the source corresponds to the vacuum
energy, $\lambda$. However, at the level of the scalar equations of
motion the tadpole term, $\int d^4 x ~c_1 \pi$, has the effect of
renormalising the vacuum energy seen by the $\pi$ field, $\lambda \to
\lambda +c_1$. Indeed, in a braneworld context, one might associate
the tadpole with the vacuum energy in the bulk. To avoid considering a
simple cosmological constant, we set the bare vacuum energy
$\lambda=0$, and require the $\pi$-tadpole term to vanish,
$c_1=0$. This  guarantees that Minkowski space is a solution for the
physical metric since the field equations are solved by $\tilde
h_{\mu\nu}=0$, and $\pi =0$. Note that this Minkowski solution need
not be stable, as in the ghost condensate scenario. On the contrary,
our interest is in stable de Sitter solutions. Given the constraints
$\lambda=0$, and $c_1=0$, any de Sitter solution is necessarily
self-accelerating.

We now consider maximally symmetric vacua  in the absence of vacuum
energy, $\lambda=0$, and the tadpole, $c_1=0$. The corresponding GR
solution is {\it always} Minkowski space, with $\tilde
h_{\mu\nu}=0$. Non-trivial solutions for the scalar $\pi=\bar \pi(x)$,
however, could give rise to self-acceleration. A self-accelerating
vacuum with de Sitter curvature $H^2$ would have $\bar
\pi=-\frac{1}{4} H^2  x_\mu x^\mu$.  Plugging this into the field
equations (\ref{scalareom}), with $c_1=0$, and $T=0$, gives
\cite{NicolisRattazziTrincherini2008}
\be
-2c_2H^2+3c_3 H^4 -3c_4 H^6 +\frac{3}{2} c_5 H^8=0.
\ee
Clearly non-trivial solutions exist for suitable choices of the
parameters $c_i$, so self-accelerating solutions also exist.

Are these vacua consistent? To investigate this we need to consider
fluctuations  $\tilde h_\mn$, and $\delta \pi=\pi-\bar \pi$ about the
self-accelerating vacuum. Because of the Galilean symmetry, the
galileon structure is preserved in  the  effective theory describing
fluctuations,  
$$
S\left[\tilde h_\mn , \delta \pi\right] =\int d^4x {\cal L}_{GR}+{\cal
  L}_{\delta \pi},
$$
where $ {\cal L}_{\delta \pi}=\sum_{i=1}^5 d_i {\cal L}_i(\delta \pi,
\del \delta\pi, \del \del \delta\pi)+\delta \pi T$.  The coefficients
can be obtained from the coefficients in the underlying theory via a
linear map $d_i =\sum_i M_{ij} c_j$, where the matrix $M_{ij}$ depends
on the background curvature, $H^2$ \cite{NicolisRattazziTrincherini2008}. 

There are two immediate things to consider: (i) does the spectrum of
fluctuations contain a ghost, and (ii) does the scalar get screened on
solar system scales?   For a general galileon theory,  to  avoid the
ghost we must choose parameters such that $d_2>0$. In DGP gravity,
where the  ghost is known to be present on the self-accelerating
branch, one would find $d^\textrm{DGP}_2 <0$ (see Section
\ref{sec:dgpsa-lin} for further details).  

 In order to screen the scalar at solar system scales  one must appeal
 to the Vainshtein mechanism. Again, we discuss the Vainshtein
 mechanism in detail in the context of DGP gravity in Section
 \ref{sec:dgp-sc}. The mechanism works in exactly the same way in a
 general galileon theory.  For simplicity we assume spherical symmetry
 for fluctuations on the self-accelerating background and consider the
 profile outside of  a heavy non-relativistic source,
 $T_\mn=diag(\rho(r), 0, 0, 0)$. Now, it is well known that the GR
 solution is given by the standard Newtonian potential 
 \be
 |\tilde h_\mn (r)| \sim \frac{GM}{ r},
 \ee
 where the mass of the source $M=\int \rho(r) dV$. The galileon
 solution $\delta \pi(r)$ is given by \cite{NicolisRattazziTrincherini2008, Burrage-revisiting} 
 \be
 d_2 \left(\frac{\delta \pi'}{r} \right)+2d_3\left(\frac{\delta
 \pi'}{r} \right)^2 +2d_4\left(\frac{\delta \pi'}{r}
 \right)^3=\frac{M}{4\pi r^3}.  \label{galeqn}
 \ee
 Note that ${\cal E}_5$ is identically zero when evaluated on a
 spherically symmetric field. At large distances one can neglect the
 higher order terms in Eq. (\ref{galeqn}) and derive the linearised solution
 \be
 \delta \pi^{lin}(r) =-\frac{M}{4\pi d_2 r}.
 \ee
 Now  $|\tilde h_\mn (r)|  \sim |\delta \pi^{lin}(r)|$ so we have an
 ${\cal O}(1)$ modification of GR.  At shorter distances, the
 non-linear terms in Equation (\ref{galeqn}) become important and
 start to dominate. This happens at the so-called `Vainshtein radius',
 given  by \cite{Burrage-revisiting}
 $$r_V \sim \textrm{max} \left\{\left(\frac{d_3
 M}{d_2^2}\right)^{1/3}, \left(\frac{d_4 M^2}{d_2^3}\right)^{1/6}
 \right\}.$$ 
 Depending on which of the non-linear terms dominates, the profile of
 the galileon field changes to  
  $$\delta \pi^{nonlin}(r) \sim
 \begin{cases}\left(\frac{M}{d_3}\right)^{1/2}\sqrt{r} & \textrm{if
 the term  with $d_3$ dominates} \\
 \left(\frac{M}{d_4}\right)^{1/3} r  & \textrm{if the term with $d_4$
 dominates} \end{cases} 
 $$
 For a suitable choice of parameters one can have $|\tilde h_\mn (r)|
 \gg |\delta \pi^{nonlin}(r)|$ on solar system scales, and might claim
 that  the modification of gravity does indeed get screened.  However,
 it is important to note that the Vainshtein mechanism itself has yet
 to be properly understood in a well defined and fully covariant
 theory. We discuss some aspects of this at the end of Section
 \ref{sec:dgp-sc}.
  
Nonetheless, our galileon analysis indicates that self-accelerating
solutions that are ghost-free and exhibit some form of Vainshtein
screening on solar system scales could exist. However, there are also
other concerns. Firstly, we should consider the question of
back-reaction. Our galileon description holds provided we can neglect
the effect of the scalar field back onto the geometry. This turns out
not to be problematic provided we take $|d_i| \lesssim
M_{pl}^2/H^{2i-4}$ \cite{NicolisRattazziTrincherini2008}.  

More serious concerns appear when we study fluctuations about the
spherically symmetric solutions we have just described. These can
cause problems at both the classical and the quantum levels.  At the
quantum level, one must identify the scale at which the quantum
fluctuations become strongly coupled, and the  radius at which one can
no longer trust the classical background.  As the background solution
changes with scale, so does the strong coupling scale. This means that
there exists a critical radius at which the quantum effects start to
dominate and one can no longer trust the classical solution. Aspects
of strong coupling in DGP gravity are discussed in Section
\ref{sec:dgp-sc}. Here we note that for a general galileon model the
critical  radius at which the theory enters a quantum fog can
sometimes be unacceptably large\footnote{Larger than the Schwarzschild
  radius of the Sun}. At the classical level, we find that
fluctuations at short distances can sometimes propagate  extremely
slowly, so much so that  a huge amount of Cerenkov radiation would be
emitted as the earth moves through the solar galileon field. Indeed,
to simultaneously avoid problems with Cerenkov emission and a low
scale of strong coupling in a ghost-free theory with self-accelerating
solutions, one must introduce a tadpole. As we have already explained,
this could be considered undesirable as  a tadpole will renormalise
the vacuum energy seen by the galileon.

Yet another problem concerns radial fluctuations at large
distances. These can propagate at superluminal speeds, indicating
problems for causality\footnote{Note, however, that it has been suggested that  causal paradoxes associated with superluminality do not always  manifest themselves in theories with non-linear scalar interactions \cite{vikman-superlum}.}. This is known to be a problem in DGP gravity
\cite{Adams-causality, Hinterbichler-super} and can only be avoided in
the general case by eliminating {\it all} of the interaction
terms. This is unacceptable since  the interaction terms are crucial
to the successful implementation of the Vainshtein mechanism. 

In summary then, while it is possible to obtain self-acceleration in a
general galileon model that avoids some of the problems facing DGP
gravity, {\it  one cannot find a completely consistent
  scenario}. However,  the situation {\it can}  be improved by the
introduction of a second galileon \cite{Padilla-bigal2}, as we will
discuss in Section \ref{sec:multigal}.
\newline
\newline
\noindent  
{\it Conformal galileon}
\newline

 The conformal galileon is constructed in much the same way as the
 pure galileon we have just described, except now we demand that the
 relevant vacuum Lagrangian, ${\cal L}_{gal}^\textrm{conformal}(\pi,
 \del \pi, \del \del \pi)$, is invariant  under the conformal group:
 \ba
\textrm{ dilations}&:& \pi(x^\mu) \to \pi( b x^\mu)+\log b \\
\textrm{  translations}&:& \pi(x^\mu) \to \pi (x^\mu+a^\mu) \\
\textrm{  boosts} &:& \pi(x^\mu)\to \pi(\Lambda^\mu{}_\nu x^\nu) \\
\begin{array}{c} \textrm{  special conformal}\\  \textrm{
 transformations} \end{array}&:& \pi(x^\mu) \to
 \pi(x^\mu+c^\mu|x|^2-2(c\cdot x) x^\mu)-2c\cdot x . \qquad  
 \ea
 It turns out that ${\cal L}_{gal}^\textrm{conformal}=\sum_i c_i {\cal
 L}_i^\textrm{conformal}$, where \cite{NicolisRattazziTrincherini2008, deRham-DBI}\footnote{Our sign conventions agree with
 \cite{NicolisRattazziTrincherini2008}, rather than \cite{deRham-DBI}.} 
 \ba
{\cal L}^\textrm{conformal}_1 &=& e^{4\pi} \\
{\cal L}^\textrm{conformal}_2 &=& -\frac{1}{2}e^{2\pi} (\del \pi)^2 \\
{\cal L}^\textrm{conformal}_3 &=& -\frac{1}{2} \left[\partial^2 \pi
 +\frac{1}{2} (\del \pi)^2\right] (\del \pi)^2 \\ 
{\cal L}^\textrm{conformal}_4 &=&e^{-2\pi}{\cal L}_4  -\frac{1}{20}
 e^{-2\pi}(\del \pi)^2 \left\{4(\del \pi)\cdot (\del\del \pi) \cdot
 (\del\pi)\right.\nonumber\\ 
&&\qquad \left. -4(\del \pi)^2\partial^2 \pi+3[(\del \pi)^2]^2\right\} \qquad \\
{\cal L}^\textrm{conformal}_5 &=&e^{-4\pi}{\cal L}_5+3
 e^{-4\pi}(\del\pi)^2 \Bigg\{{\cal L}_4+\frac{1}{56} [(\del \pi)^2]^3
 \nonumber\\ 
&& +\frac{5}{7}(\del \pi)^2[ (\del \pi)^2 \partial^2\pi-(\del \pi)\cdot
 (\del\del \pi) \cdot (\del\pi)]\Bigg\}. 
\ea
Aspects of the conformal galileon model are studied in
 \cite{Creminelli-galilean-genesis,Nicolis-NEC}. Violations of the null energy
 condition here can drive inflationary expansion without introducing
 instabilities. There are, however, some issues with superluminality.
 
 A supersymmetric version of the conformal galileon has been obtained in \cite{susygal} as a consistent completion of the supersymmetric ghost condensate.

\subsubsection{Covariant galileon}

The galileon action (\ref{galact}) describes fields propagating on a
Minkowski background, and does not represent a fully covariant
theory. Although  galileon theory was originally motivated by
co-dimension one braneworld models, it is interesting to consider the
four dimensional covariant completion of the theory in its own
right. This has been worked out in \cite{Deffayet-covariant,
  Deffayet-generalized-galileons}, and is given by 
\be
S[g_\mn, \pi]= \int d^4 x \sqrt{-g}\left[ \frac{1}{16\pi G} R+{\cal
    L}_{gal}^{cov} \right]+S_{matter}[\hat g_\mn, \psi_i].
\ee
Naturally we recognise the first term in brackets as the standard
Einstein-Hilbert action, $ \frac{1}{16\pi G} \int d^4 x \sqrt{-g} R$. The last
term corresponds to the matter action. Note that the matter fields are
minimally coupled to the metric $\hat g_\mn=f(\pi) g_\mn$,  where the
conformal factor depends on $\pi$. An obvious example would be $\hat
g_\mn= e^{2\pi} g_\mn$ although this is by no means a unique
choice. Neglecting the tadpole, the covariant completion of $ {\cal
  L}_{gal}=\sum_i c_i   {\cal L}_i$ is given by ${\cal
  L}_{gal}^{cov}=\sum_i  c_i {\cal L}^{cov}_i$, where\footnote{We
  define $(\nabla \nabla \pi )^n =(\nabla_{\alpha_1} \nabla^{\alpha_2}
  \pi )(\nabla_{\alpha_2} \nabla^{\alpha_3} \pi ) \ldots
  (\nabla_{a\alpha_{n}} \nabla^{\alpha_1} \pi)$, $\Box\pi=g^\mn
  \nabla_\mu \nabla_\nu \pi$, and $(\nabla \pi)^2=g^\mn \nabla_\mu \pi
  \nabla_\nu \pi$.} 
\ba
{\cal L}^{cov}_2 &=& -\frac{1}{2} (\nabla \pi)^2 \\
{\cal L}^{cov}_3 &=& -\frac{1}{2} \Box \pi  (\nabla \pi)^2  \\
{\cal L}^{cov}_4 &=& -\frac{1}{2}\left[ (\Box \pi)^2  -(\nabla
  \nabla \pi)^2-\frac{1}{4} R(\nabla \pi)^2 \right](\nabla \pi)^2
\qquad \\ 
{\cal L}^{cov}_5 &=& -\frac{1}{2} \Bigg[ (\Box \pi)^3 -3 (\Box
  \pi)(\nabla \nabla \pi)^2 +2 (\nabla \nabla \pi)^3 \nonumber\\ 
&& \qquad \qquad -6G_\mn(\nabla^\mu \pi)(\nabla^{\nu}\nabla_\alpha
  \pi)\nabla^\alpha \pi) \Bigg](\nabla \pi)^2. 
\ea
Note that for the 4th and 5th order terms one must introduce some
non-minimal gravitational coupling to $\pi$. This is necessary since
the naive covariant completion of ${\cal L}_4$ and ${\cal L}_5$, with
minimal couplings, results in equations of motion containing higher
derivatives.  The non-minimal coupling helps to eliminate those higher
derivatives. Now although the field equations in our covariant theory
remain at most second order in derivatives, Galilean invariance {\it
  is} broken. We will not present the field equations here since they
are long and complicated, especially for the higher-order terms. The
interested reader can find them in \cite{Deffayet-covariant}, but
should be mindful of the fact that the formulae for ${\cal L}^{cov}_4$
and ${\cal L}^{cov}_5$ presented here differ from those in
\cite{Deffayet-covariant}  by an overall factor of $4$ and $5$,
respectively.

In a very recent paper, covariant galileon terms are seen to arise in
Kaluza-Klein compactifications of Lovelock actions
\cite{vanacoleyen-galileons}. This might have been expected since the
underlying theory has at most second-order fields, and this is
inherited by the dimensionally reduced theory.   We discuss aspects of
Lovelock gravity, and in particular, Gauss-Bonnet gravity, in Section
\ref{sec:egb}.

\subsubsection{DBI galileon} \label{sec:dbigal}

The galileon Lagrangian ${\cal L}_{gal}$ can also be obtained from the
non-relativistic limit of a probe brane in five dimensional Minkowski
space \cite{deRham-DBI}. The probe brane action corresponds to a
generalisation of the DBI action, as we will now explain. We take our
bulk coordinates to be $(x^\mu, y)$, and place the probe brane  at
$y=\pi(x)$.  The induced metric on the brane is then given by
$g_\mn=\eta_\mn+\del_\mu \pi \del_\nu \pi$, from which we deduce that
the DBI action is
\be
S_\textrm{DBI}= -\lambda \int d^4x\sqrt{-g}=\int d^4 x
-\lambda\sqrt{1+(\del \pi)^2}. 
\ee
For a slowly moving brane $(\del \pi)^2 \ll 1$, the leading order
dynamical piece goes like $-\frac{\lambda}{2} (\del \pi)^2$. To
generalise this, we first consider objects that transform covariant on
the brane, and then build a Lagrangian from them that gives  rises to
field equations that are at most second-order. The relevant covariant
objects are the extrinsic curvature, $K_\mn$, the induced curvature,
$R_{\mu\nu \alpha \beta}$, and the covariant derivatives of these
quantities. The generalised DBI action required to guarantee
second-order field equations is
\be
S_\textrm{gen-DBI}=\sum_i c_i S_i,
\ee
where\footnote{We write $(\del^{\mu_1} \pi)(\del_{\mu_1} \del^{\mu_2}
  \pi) \ldots   (\del_{\mu_n} \del^{\mu_{n+1}} \pi)(\del_{\mu_{n+1}}
  \pi)=(\del \pi)\cdot (\del \del \pi)^n \cdot (\del \pi)$.} 
\ba
S_2&=&-\int d^4 x \sqrt{-g} \\&\to& -\int d^4 x \sqrt{1+(\del \pi)^2} \\
S_3&=&\int d^4 x \sqrt{-g} K \\&\to& -\int d^4 x \gamma \left[\partial^2
  \pi-\gamma^2 (\del \pi)\cdot (\del\del \pi) \cdot (\del\pi)\right] 
\ea
and
\ba
S_4 &=&-\int d^4 x \sqrt{-g} R \\
 &\to& -\int d^4 x \gamma \Big[ (\partial^2 \pi)^2-(\del \del \pi)^2
\nonumber \\  &&+2 \gamma^2 \left( (\del \pi)\cdot (\del\del
\pi)^2  \cdot (\del \pi) -\partial^2 \pi  (\del \pi)\cdot (\del\del \pi)
\cdot (\del \pi)\right)\Big]\\
S_5&=&\frac{3}{2} \int d^4 x \sqrt{-g} (J-2G^\mn K_\mn) \\
&\to&  -\int d^4 x \gamma^2 \Big[ (\partial^2\pi)^3+2(\del \del \pi)^3 
-3(\partial^2 \pi)(\del \del \pi)^2 \nonumber\\ \nonumber 
&&+6\gamma^2((\partial^2\pi)(\del \pi)\cdot (\del\del \pi)^2 \cdot (\del
\pi) -(\del \pi)\cdot (\del\del \pi)^3 \cdot (\del \pi) ) 
\\ 
&&-3\gamma^2((\partial^2\pi)^2-(\del \del \pi)^2)(\del \pi)\cdot (\del\del
\pi) \cdot (\del \pi)\Big],
\ea
with the Lorentz factor $\gamma=1/\sqrt{1+(\del \pi)^2}$. The
expressions for $S_3$ and $S_5$ can be identified with the boundary
terms in General Relativity and in Gauss-Bonnet gravity,
respectively. Of course, the former is the Gibbons-Hawking term, and
the latter is the Myers boundary term \cite{Myers-higher}, discussed
in more detail in Section \ref{egb-generic}. Now, for a slowly moving
probe in Minkowski space, it can be shown that $S_i \approx \int d^4 x
{\cal L}_i$, which means that, neglecting the tadpole,
$S_\textrm{gen-DBI} \approx \int d^4 x{\cal L}_{gal}$
\cite{deRham-DBI}. One can then recover the conformal galileon by
considering a probe brane in AdS, and the covariant galileon by
considering a general bulk geometry \cite{deRham-DBI}.

This procedure has recently been extended to probe branes that are curved, giving rise to a more general class of effective theories on curved space \cite{kurt1, kurt2} (see, also \cite{clare}). These represent the analogues of galileons and DBI theories living on $dS_4$ and $AdS_4$, retaining the same number of symmetries as their flat space counterparts. There is a rich structure and in some cases the symmetries can even admit non-trivial potentials beyond the usual tadpoles.

\subsubsection{Galileon cosmology}

Galileon cosmology encompasses much more than the original model and
its covariant completion.  The cosmological behaviour of a number of
models that are {\it inspired} by the galileon have also been
investigated  (see, for example, \cite{Silva-sa, Kobayashi-evolution,
  Kobayashi-cosmic, De-Felice-generalized, Kobayashi-g-inflation,
  Deffayet-imperfect}). These include the {\it braiding} model
\cite{Kobayashi-g-inflation, Deffayet-imperfect, Pujolas1},  which is described
by  the following action 
\be
S=\int d^4 x \sqrt{-g}\left[\frac{1}{16\pi G} R+ K(\phi, X)+G(\phi, X)
  \Box \phi\right],
\ee
where $X=-(\nabla \phi)^2$. Note that for $K=\frac{c_2}{2} X $ and
$G=\frac{c_2}{2} X$ we recover the covariant galileon action up to
cubic order. This model still gives rise to second-order field
equations and admits some rich phenomenology. It is claimed that the
scalar equation of state can cross the phantom divide without
introducing any instabilities, and results in a blue tilt for the
spectrum tensor perturbations. Constraints on the model coming from
large scale structure and  non-Gaussianity have been obtained in
\cite{Kimura-LSS,Mizuno-primordial}, respectively. Non-Gaussianity in DBI galileon inflation has been studied in \cite{Renaux1}. An even more general class of scalar tensor theories yielding second order field equations has recently been presented in \cite{cedric}, and is now known to be equivalent to Horndeski's general theory \cite{horndeski} in four dimensions \cite{Kobayashi1}. 

It has been argued that some of these generalised models are
perhaps {\it too} general \cite{Burrage-galileon-inflation}.  The
point is that there is no symmetry protecting the theory from
large  radiative corrections. This can spoil the functional form of
the Lagrangian so much so that we require more input parameters  than
we can measure, and we lose all predictivity. In
contrast, the pure galileon, conformal galileon and DBI galileon
theories are safe against radiative corrections since they possess
additional symmetries that control the form of the derivative
interactions.  For this reason, in the remainder of this section we
will restrict attention to those  models for which the Galilean
invariance is only weakly broken, so that any radiative corrections
that break the galileon symmetry are suppressed. 

Let us begin with early universe cosmology and  the covariant
galileon. This can give rise to inflation even in the absence of a
potential \cite{Burrage-galileon-inflation}. The theory is radiatively
safe because the terms that break Galilean symmetry are suppressed by
powers of $\Lambda/M_{pl}$, where $\Lambda$ is the naive
cut-off\footnote{This corresponds to the scale of the  galileon self-interactions.}.  As with DBI inflation, fluctuations about the quasi
de Sitter background  will result in large non-Gaussianities  at low
sound speeds  \cite{Mizuno-primordial, Burrage-galileon-inflation}. It
has been argued that  what sets this model apart is the fact that the
non-Gaussianity is not constrained to obey $f_{NL} \sim 1/c_s^2$,
making it distinguishable from DBI inflation
\cite{Burrage-galileon-inflation}, although this claim has been
disputed in \cite{Creminelli-galilean}. Indeed, it is interesting to
note that the authors of \cite{Creminelli-galilean} adopt the
effective field theory approach to inflation, imposing Galilean
symmetry on the small perturbations around the inflationary
background. This permits additional interactions compared with
\cite{Burrage-galileon-inflation}, but maintains stability against
radiative corrections. They find  that one can have large (observable)
four point functions even when the three point function is small.

We now turn to the cosmology of the late universe.  The late time
 cosmology of the covariant galileon has been studied for up to cubic
 \cite{Chow-galileon, Mota-cosmology}, quartic
 \cite{Gannouji-galileon, Ali-modified}, and quintic scalar
 interactions \cite{DeFelice-cosmology-of,Nesseris-observational,
 DeFelice-matter}. In the latter model, we focus on the role of the
 galileon as a dark energy field --  it deviates  slightly from the
 original galileon scenario \cite{NicolisRattazziTrincherini2008} because the
 galileon coupling to matter disappears as $M_{pl} \to \infty$. In any
 event, the model is described by the following action 
 \be
 S=\int d^4 x \sqrt{-g} \left[\frac{1}{16\pi G} R+{\cal L}_\phi +{\cal
 L}_m(g_\mn, \psi)\right],
 \ee
 where 
 \begin{multline}
 {\cal L}_\phi=\frac{c_2}{2} (\nabla \phi)^2+\frac{c_3}{2 \Lambda^3}
 (\nabla \phi)^2 \Box \phi + 
 \frac{c_4}{2 \Lambda^6}(\nabla \phi)^2 \left[ 2 (\Box \phi)^2-2
 (\nabla \nabla \phi)^2-\half R (\nabla \phi)^2 \right] \\ 
+ \frac{c_5}{2 \Lambda^9} (\nabla \phi)^2 \left[ (\Box \phi)^3 -3
 (\Box\phi) (\nabla \nabla \phi)^2+2 (\nabla \nabla
 \phi)^3-6G_\mn(\nabla^\mu \phi)(\nabla^{\nu}\nabla_\alpha
 \phi)\nabla^\alpha \phi \right]. 
 \end{multline}
As there is no potential, late time acceleration must be driven by the
kinetic terms. There is a late time de Sitter solution characterised
by $H=H_{dS}=$constant and $\dot \phi=\dot \phi_{dS}=$constant. The
existence of this fixes a relationship between $c_2, c_3, c_4$ and
$c_5$ such that there are only two free parameters, given by
$$\alpha=c_4 x_{dS}^4, \qquad {\rm and } \qquad \beta=c_5 x_{dS}^5,$$
where $x_{dS}=\sqrt{8 \pi G}\left(\dot \phi_{dS}/H_{dS}\right)$. In
\cite{DeFelice-cosmology-of} conditions are derived that guarantee the
absence of ghosts and imaginary sounds speeds in both the tensor and
the scalar sector. The  viable region of parameter space, $(\alpha,
\beta)$, where these conditions are met is presented. As regards the
cosmological evolution, we see that there exists a tracker solution
that approaches the late time de Sitter attractor. 
\begin{figure}
\begin{center}
\epsfig{file=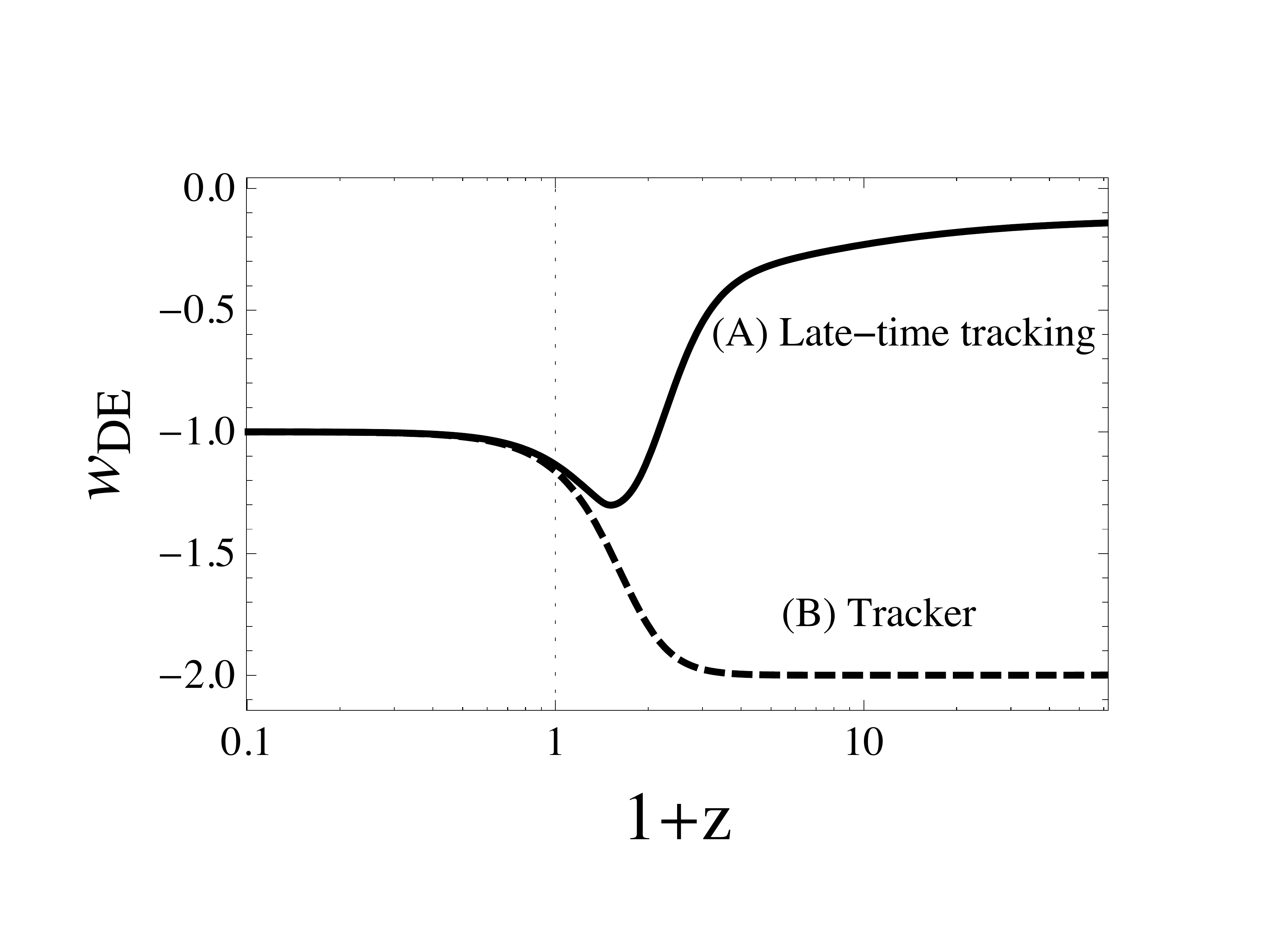,width=10cm,}
\caption{Taken from Figure 1 in \cite{DeFelice-matter}. The equation
  of state of the galileon field, $w_{DE}$, versus redshift, $z$, for
  $\alpha=1.37$ and $\beta=0.44$. The evolution is given for generic
  initial conditions (A) and for the tracker (B).} 
\label{fig:wDE} 
\end{center}
\end{figure}
In Figure \ref{fig:wDE}, we plot the evolution of the galileon
equation of state for the tracker solution, and for generic initial
conditions. Note that the tracker has a phantom equation of
state. Indeed, even for generic initial conditions, the galileon field
is drawn into a phantom phase by the tracker. It turns out that the
tracker solution is disfavoured by a combined data analysis (SNe, BAO,
CMB). The generic case fares rather better, especially if we have
non-zero curvature, $\Omega_k$, although it is still disfavoured with
respect to $\Lambda$CDM \cite{Nesseris-observational}.

Matter density perturbations have been studied in detail within the
context of this model in \cite{DeFelice-matter}, where it is shown
that the growth rate of matter perturbations is larger than in
$\Lambda$CDM. In the generic case, for suitable choices of $\alpha$
and $\beta$, we typically find that the growth index today is
$\gamma_0 <0.4$, with large variations at earlier times. This makes
the model easily distinguishable from $\Lambda$CDM. Another
distinguishing feature is the effective gravitational potential
changing with time, even during matter domination. 

\subsubsection{Multi-galileons} \label{sec:multigal}

The extension of the galileon scenario to include multiple  scalar
 fields  \cite{bigal1, Padilla-bigal2,  Padilla-multi,
 Deffayet-generalized-galileons} and even arbitrary p-forms
 \cite{Deffayet-arbitrary}  has recently been developed (see
 \cite{Fairlie-universal, Fairlie-euler} for earlier work). A general
 multi-galileon theory, in four dimensions, with $N$ real scalar
 degrees of freedom is given by the Lagrangian \cite{Padilla-multi,
 Zhou-goldstones} 
\be \label{lagsonf}
\mathcal{L}_\textrm{N-gal} = \sum_{m=1}^5 \alpha^{i_1...i_m}\,
 \delta^{\phantom{[}\mu_2...\mu_{m}\phantom{]}}_{[\nu_2...\nu_{m}]}
\pi_{i_1}\del_{\mu_2}\!\del^{\nu_2}
 \pi_{i_2}...\del_{\mu_{m}}\!\del^{\nu_{m}} \pi_{i_m},
\ee
where $\{\alpha^{i_1 ... i_m}\}$ are free parameters of the theory and
 $\delta^{\phantom{[}\mu_1...\mu_{m}}_{[\nu_1...\nu_{m}]}=m!\delta^{\mu_1}_{\phantom{\mu_2}\![\nu_1}...\delta^{\mu_m}_{\phantom{\mu_m}\!\nu_{m}]}$.  
 As usual, summation over repeated Lorentz (Greek) and galileon
 indices (Latin) should be understood to be implicit. Note  further
 that we {\it define} the $m=1$ term of expression (\ref{lagsonf}) to
 be $\alpha^{i_1}\pi_{i_1}$.  The Lagrangian (\ref{lagsonf}) is
 constructed so that it is invariant under 
\be \label{mgalsym}
 \pi_i \to \pi_i +(b_i)_\mu x^\mu +c_i, \qquad i=1, \ldots, N.
\ee
 One might expect this to appear in the decoupling limit of some
 co-dimension $N$ braneworld scenarios, with $\pi_1, \ldots, \pi_N$
 corresponding to the position of the brane in the $N$ transverse
 directions \cite{bigal1}. Indeed, one can generalise the formalism
 discussed in Section \ref{sec:dbigal} to probe a brane of
 co-dimension $N$ and recover the multi-galileon theory in the
 non-relativistic limit \cite{Hinterbichler-multi}. 
 
How many free parameters are there in this theory? We can  always
choose $\alpha^{i_1 ... i_m}$ to be  symmetric so  the total number
free parameters is given by 
\be
\sum_{ m=1}^{5} \left(\begin{array}{c} N+m-1 \\ m \end{array}
\right)=\sum_{m=1}^{5} \frac{(N+m-1)!}{m!(N-1)!}. 
\ee 
Even for $N=3$, this corresponds to 55 free parameters. To reduce the
number of parameters one can consider imposing internal symmetries on
the galileon fields \cite{Padilla-multi}, although this will break the
galileon symmetry (\ref{mgalsym}).  The phenomenology of spherically
symmetric solutions with an internal SO(N) has been studied and found
to suffer from problems with instabilities and superluminality, at
least for standard non-derivative matter coupling
\cite{Andrews-instabilities}. One can also prove a generalised form a
Goldstone's theorem when internal symmetries are present
\cite{Zhou-goldstones}.
 
Of course, the large number of free parameters is less of an issue in
the simple case of $N=2$, dubbed {\it bigalileon theory}. The
phenomenology of this theory was developed in detail in
\cite{Padilla-bigal2}. Let us summarise the main results.  In direct
analogy with the single galileon case, the bigalileon  theory is
formulated on sub-horizon scales as fields propagating on Minkowski
space. This time we have our  GR graviton, $\tilde h_\mn$, and {\it
  two} scalar galileons, $\pi$ and $\xi$. Only one of the scalars,
$\pi$, say, couples directly to the trace of the energy-momentum
tensor. The other scalar, $\xi$, couples indirectly through its mixing
with $\pi$, so it still has an important role to play. The
governing action is given by \cite{bigal1, Padilla-bigal2} 
\be
S\left[ \tilde h_\mn, \pi, \xi\right]=\int d^4 x {\cal L}_{GR}+{\cal
  L}_{\pi, \xi}, 
\ee
where ${\cal L}_{GR}$ is given by Eq. (\ref{LGR}), and
\be
{\cal L}_{\pi, \xi}=\sum_{0 \leqslant m+n \leqslant 4} (\alpha_{m, n}
\pi+\beta_{m, n} \xi) {\cal E}_{m, n}(\del \del \pi, \del \del
\xi)+\pi T,
\ee
with
\begin{multline} \label{eommn}
\mathcal{E}_{m,n}= (m+n)! \delta^{\mu_1}_ {[\nu_1} \ldots
             \delta^{\mu_m} _{\nu_m} 
             ~\delta^{\rho_1} _{\sigma_1} \ldots \delta^{\rho_n}_{
             \sigma_n]} 
           (\del_{\mu_1}\del^{\nu_1}\pi) \ldots\\
           \ldots  (\del_{\mu_m}\del^{\nu_m}\pi)
             ~(\del_{\rho_1}\del^{\sigma_1}\xi) \ldots
             (\del_{\rho_n}\del^{\sigma_n}\xi)\,. 
\end{multline}
 The {\it physical} metric  is given by
 $g_{\mu\nu}=\eta_{\mu\nu}+h_{\mu\nu}$, where  $h_{\mu\nu}=\tilde
 h_{\mu\nu}+2\pi \eta_{\mu\nu}$. Given a source $T_{\mu\nu}$, $\tilde
 h_{\mu\nu}$ gives the usual perturbative GR solution, and so $2\pi
 \eta_{\mu\nu}$ gives the modified gravity correction. The field
 equations for the scalars are 
\be
T+\sum_{0 \leqslant m+n\leqslant 4} a_{m,n}\mathcal{E}_{m,n}=0,\quad
 {\rm and} \quad 
\sum_{0 \leqslant m+n\leqslant 4} b_{m,n}\mathcal{E}_{m,n}=0, 
\ee
where\footnote{We define $\alpha_{-1,n}=\beta_{m, -1}=0$.} $a_{m,
 n}=(m+1)(\alpha_{m,n}+\beta_{m+1,n-1})$ and
 $b_{m,n}=(n+1)(\beta_{m,n}+\alpha_{m-1,n+1})$. 

In contrast to the single galileon case, self-accelerating solutions
can be consistent in some bigalileon theories. Indeed, one can choose
parameters such that  we simultaneously satisfy each of the following:
(i) there is no tadpole, (ii) there is a self-accelerating vacuum,
(iii) fluctuations about the self-accelerating vacuum do not contain a
ghost, (iv) spherically symmetric excitations about the
self-accelerating vacuum undergo Vainshtein screening in the solar
system, (v) fluctuations about the spherically symmetric solutions are
never superluminal, (vi) fluctuations about the spherically symmetric
solutions never lead to trouble with excessive emission of Cerenkov
radiation, (vii) there is not an unacceptably low momentum scale for
strong coupling on the spherically symmetric solution, and (viii)
there are no problems with back-reaction on the spherically symmetric
solution (or the vacuum). This supports the case for considering
bigalileon theories as a viable alternative to dark energy.

One can also develop models of self-tuning in bigalileon theory, where
the vacuum energy does not affect the four dimensional
curvature. These models get around Weinberg's no-go theorem by
breaking Poincar\'e invariance. Unfortunately, in order for them to
remain compatible with solar system tests one must limit the amount of
vacuum energy to be $\lesssim$ meV.  

\subsection{Other Theories}

Let us now consider some further theories that have yet to be
discussed.  These are ghost condensate theories, non-metric theories,
and the dark energy from curvature corrections approach of Piazza.

\subsubsection{Ghost condensates}

Ghost condensate theories involve introducing into the gravitational
sector an extra scalar field, $\phi$, with shift symmetry
\be
\phi\rightarrow \phi + {\rm constant}.
\ee
The only terms in the action that can obey this symmetry are derivative
ones, and so the building block for this theory is taken to be
\be
X=\partial_\mu\phi\partial^\mu\phi.  
\ee
In \cite{Hamedetal2004} it was shown that if the leading order term
in the action has the wrong sign, so that $\phi$ is a ghost field,
it is still possible to construct a theory that is stable to small
fluctuations by including terms that push $X$ to a fixed value, so
that
\be
\langle X\rangle=C.
\ee
Theories of this type have a number of interesting properties. For a
start, the non-zero vacuum expectation value of the ghost field signals a
spontaneous breaking of Lorentz invariance. What is more, fluctuations
in the ghost field about the vacuum expectation value appear
linearly in the energy-momentum tensor, meaning that anti-gravity is
possible.  A further interesting point is that in the weak field
limit large ghost condensate clumps move more slowly than small
clumps, with potentially interesting phenomenological consequences.

It has been argued in \cite{Hamedetal2004} that ghost condensate
fields act like the gravitational counterpart to the Higgs field of the
standard model of particle physics.  This is because gravitational
fields propagating 
through the condensate acquire a massive mode, much like particles
acquire mass while propagating through the Higgs field.  Ghost condensates
also introduce oscillatory correction to the gravitational potential,
with a Jeans instability that grows with time.  For mass parameters of
the order $10^{-3}$eV, these corrections occur on spatial and
temporal scales greater than $H_0^{-1}$. If the massive modes are of
order $10$MeV, however, then corrections can occur on length
scales as small as $1000$km, but, again, only on time scales greater
than $H_0^{-1}$.

The action for ghost condensate theories can be written
\be
\label{ghostaction}
S=\int d^4x\sqrt{-g}\left(\frac{R}{16\pi G}+M^4P(X)\right),
\ee
where $M$ is a mass scale (confined to be $1{\rm meV}\le M\le 10 {\rm
  MeV}$), and $P(X)$ is a function that must have a non-zero minimum at
$X=C$ in order to be a ghost. For stability we then require 
\begin{eqnarray}
P'(C) &\ge& 0\\
P'(C)+2CP''(C) &\ge& 0.
\end{eqnarray}
Extremisation of the action (\ref{ghostaction}), with respect to the
metric, yields field equations of the form
\begin{eqnarray}
G_{\mu\nu}=8\pi G T_{\phi,\mu\nu},
\end{eqnarray}
where
\begin{eqnarray}
T_{\phi,\mu\nu}=M^4\left(P(X)g_{\mu\nu}+2P'(X)\partial_\mu\phi\partial_\nu\phi
\right),
\end{eqnarray} 
and for simplicity we have not included a term in the action for
normal matter fields.

To see why this theory is considered a modified theory of gravity we
look at the perturbed field equations.  Writing the metric as
$g_{\mu\nu}=\eta_{\mu\nu}+h_{\mu\nu}$, and the ghost field as
$\phi=\sqrt{C}t+\pi$, the Lagrangian density of the theory becomes 
\begin{eqnarray}
\label{lghost}
{\cal L}=\sqrt{-\gamma}\left[F_0(X)+F_1(X)K^2+F_2(X)K^{ij}K_{ij}+
  \cdots\right],
\end{eqnarray}
where we have chosen a unitary gauge, and where $\pi$ has been set to
zero.  The $F_n$ here are functions of $X$ that are derived from
$P(X)$, and $K$ is the extrinsic curvature of the 3 dimensional
hyper-surfaces of constant $\phi$.  Diffeomorphism invariance in
Eq. (\ref{lghost}) can be seen to have been explicitly broken.

Let us now consider cosmology.  For a homogeneous and isotropic
Universe, the equation of motion for $\phi$ is 
\begin{eqnarray}
\label{ghostprop}
\frac{d}{dt}\left(a^3{\dot \phi}P'(X)\right)=0.
\end{eqnarray}
If we assume that $X\rightarrow 0$ and $P'(C)\rightarrow 0$, as
$t\rightarrow \infty$, then Eq. (\ref{ghostprop}) tells us that
$\phi\rightarrow\pm\sqrt{C}t$.  The Friedmann equation for this theory is
\begin{eqnarray}
H^2=\frac{m^2}{3}\left(2XP'(X)-P(X)\right),
\end{eqnarray}
where the new mass parameter is $m\equiv\sqrt{8\pi
  G}M^2=M^2/M_{Pl}$. The Raychaudhuri equation is
\begin{eqnarray}
\frac{\ddot a}{a}=-\frac{m^2}{3}\left(XP'(X)+P(X)\right).
\end{eqnarray}
Now, it can be shown that the effect of the ghost field on the expansion of the Universe is
such it can mimic radiation domination, matter domination and vacuum domination. Indeed,
the simple choice of $P(X)=\frac{1}{2}(X-C)^2$ leads to
$a(t)\propto (Cmt)^{1/2}$ at early times, and $a(t)\propto (Cmt)^{2/3}$
at late times. Adding a constant, such that
$P(X)=\frac{1}{2}(X-C)^2+\Lambda/m^2$, leads to a period of vacuum
domination.  

A general class of solutions, with matter sources included, has been
studied in \cite{Krauseetal2006}.  Some of these solutions combine
dark matter and dark energy-like behaviour, at the background
level. The behaviour of scalar perturbations in the ghost condensate
theory was worked out in detail in \cite{Mukohyama2006}.  Modified
Newtonian potentials were discovered with 
\begin{eqnarray}
\Phi&=&\Phi_{GR}+\Phi_{mod}\\
\Psi&=&\Psi_{GR}+\Psi_{mod},
\end{eqnarray}
where $\Phi_{GR}$ and $\Psi_{GR}$ take their standard form from
General Relativity, while  $\Phi_{mod}$ and $\Psi_{mod}$
are corrections due to the ghost condensate that occur in the limit where the
wavelength of the fluctuation is larger than the symmetry breaking scale. If we
consider the case of de Sitter space, where  
$\Phi_{mod}=\Psi_{mod}$, then the evolution equations are
\begin{eqnarray}
\partial^2_t\Phi_{mod}+3H_0\partial_t\Phi_{mod}+\left(\frac{\alpha}{M^2}
  \frac{k^4}{a^4}-\frac{\alpha 
  M^2}{2 M_{Pl}^2}\frac{k^2}{a^2}+2H^2_0\right)\Phi_{mod}
=\frac{\alpha M^2}{2M^2_{Pl}}\frac{k^2}{a^2}\Phi_{GR},
\end{eqnarray}
where $\alpha$ is a combination of dimensionless coefficients of
${\cal O}(1)$ from the action.  These equations shows that  
the Newtonian part of the potential seeds the modified part.

\subsubsection{Non-metric gravity}

We will now describe the non-metric gravity theory that deforms GR
while keeping only two dynamical degrees of
freedom~\cite{Bengtsson1990,Bengtsson1993,Bengtsson1995,Krasnov2007,Krasnov2008a}.
 In this theory the fundamental gravitational object is no longer the
 metric but a triple of $2$-forms $\twoform{B}^i = B^i_{\mu\nu} dx^\mu
 \wedge dx^\nu$, where lower-case Latin indices denote internal
 $SU(2)$ indices and take values from $1$ to $3$. The space-time metric
is an emergent variable and is given in terms of $B^{i}_{\mu\nu}$ as
\begin{equation}
\sqrt{-g} g_{\mu\nu} \propto
\tilde{\epsilon}^{\alpha\beta\gamma\delta} B^i_{\mu\alpha}
B^j_{\nu\beta} B^k_{\gamma\delta} \epsilon_{ijk} 
\label{Urbantke}
\end{equation}
where $\tilde{\epsilon}^{\alpha\beta\gamma\delta}$ is the completely
antisymmetric tensor density having components $\pm 1$ in any
coordinate system. The proportionality symbol is used above, rather
than equality, because the metric is defined only up to conformal
rescalings.   The reason for this is that $B^{i}_{\mu\nu}$ is self-dual, i.e. 
$\frac{1}{2} \epsilon_{\mu\nu}^{\phantom{\mu\nu}\rho\sigma}
B^i_{\rho\sigma} = i B^i_{\mu\nu}$ is a conformally invariant relation.

The class of theories we will now describe contains only two
propagating degrees of freedom~\cite{Bengtsson2007,Freidel2008}, just
like GR.  Spherically symmetric solutions, as well as black holes,
have been
studied~\cite{KrasnovShtanov2007,KrasnovShtanov2008,IshibashiSpeziale2009},
and extensions of these ideas to bimetric theories have also been
considered~\cite{Speziale2010}. Let us now describe the kinematical setup
of the theory before proceeding to discuss its dynamics.
\newline
\newline
\noindent
{\it Kinematics}
\newline

Consider the set of $1$-forms $\{\oneform{e}^0,\oneform{e}^I\}$, the
tetrad, where the capital Latin indices denote internal $SO(3)$
indices (note that this is a different space than $SU(2)$ considered
above) such that
\begin{equation}
ds^2 = g_{\mu\nu} dx^\mu dx^\nu = -\oneform{e}^0 \oneform{e}^0 +
\delta_{IJ} \oneform{e}^I \oneform{e}^J. 
\end{equation}
From the tetrad we can also define the self-dual $2$-forms
$\twoform{\Sigma}^I_{(+)}$, and similarly the anti self-dual $2$-forms
$\twoform{\Sigma}^I_{(-)}$), by
\begin{equation}
\twoform{\Sigma}^I_{\pm} = i \oneform{e}^0 \wedge \oneform{e}^I \mp
\frac{1}{2} \epsilon^I_{\phantom{I}JK} \oneform{e}^J \wedge
\oneform{e}^K .
\end{equation}
Any other self-dual $2$-form can then be decomposed in terms of
$\twoform{\Sigma}^I_{(+)}$.  In particular, we may write 
\begin{equation}
 \twoform{B}^i = B^i_{\phantom{i}I}  \twoform{\Sigma}^I_{(+)}.
\end{equation}
From $\twoform{B}^i$ we can then define the connection one-forms
$\oneform{A}^i$ as\footnote{Strictly speaking this defines a
  three-form which is then dualised to a one-form.} 
\begin{equation}
d\twoform{B}^i + \epsilon^{ijk} \oneform{A}^j \wedge \twoform{B}^k = 0.
\label{dB}
\end{equation}
The above equation can be solved to get
\begin{equation}
A^i_\mu = \frac{1}{2\det B} B^{i\alpha\beta} \delta_{jk}
B^j_{\mu\alpha} \nabla^\nu B^k_{\nu\beta} ,
\end{equation}
where $\det B = -\frac{1}{24} \epsilon^{ijk} B^{i\;\nu}_{\mu}
B^{j\;\rho}_{\nu}  B^{k\;\mu}_{\rho}$.  Now, since $\twoform{B}^i$ is
conformally invariant, we know that $\oneform{A}^i$ is too.
We can now proceed and define the curvature two-forms $\twoform{F}^i$
of  $\oneform{A}^i$ as
\begin{equation}
\twoform{F}^i = d\oneform{A}^i + \frac{1}{2}
\epsilon^{i}_{\phantom{i}jk} \oneform{A}^j \wedge \oneform{A}^k.
\label{dA}
\end{equation}
\newline
\newline
\noindent
{\it Dynamics}
\newline

The action for this theory takes on the form of the well known BF-theory:
\begin{equation}
\label{nonloc1}
S[B,A] = \frac{i}{8\pi G} \int \left[\delta_{ij} \twoform{B}^i\wedge
  \twoform{F}^j[A] - \frac{1}{2} V(\twoform{B}^i\wedge
  \twoform{B}^j)\right] + S_m,
\end{equation}
where $V({\mathbf M})$ is a holomorphic function of a complex
symmetric $3\times 3$ matrix, ${\mathbf M}$, that is required to be
homogeneous of degree one (i.e. $V(\lambda {\mathbf M}) = \lambda V(
{\mathbf M})$) so that when it is applied to a four-form such as
$\twoform{B}^i\wedge \twoform{B}^j$ the result is also a four-form. We
introduce the internal metric $h^{ij} = B^i_{\phantom{i}I}
B^j_{\phantom{j}J} \delta^{IJ}$ and  further decompose it into trace
and traceless parts as $h^{ij} = \frac{1}{3} h\left(\delta^{ij} +
H^{ij}\right)$,  where $h = \delta_{ij} h^{ij}$ and
$\delta_{ij}H^{ij}=0$. We can then write $V$ as $V(h^{ij}) =
\frac{1}{3} h U(H^{ij})$, and expand $U$ as
\begin{equation}
U({\mathbf H}) = \Lambda_0 - \frac{1}{8\ell^2} \tr{\mathbf H}^2 +
{\cal O}({\mathbf H}^3),
\end{equation}
where the constant $\Lambda_0$ plays the role of the cosmological
constant, while the constant $\ell$ is a new scale that describes
deviations from GR.  The minus sign in the 2nd term above is required
to avoid instabilities.  In particular, as $\ell\rightarrow 0$ the
theory reduces to the Plebanski formulation of GR with a
cosmological constant (see Section \ref{sec:otherGR}).

As discussed above, the metric here is defined only up to conformal
transformations.  In order to couple the theory to matter
fields we have to fix this ambiguity, which can be achieved by the
introduction of a new function $R(h^{ij})$ that is also homogeneous of
degree one. The conformal freedom is then fixed by
requiring\footnote{This method of fixing the conformal ambiguity can
be shown to arise naturally by 
considering the motion of a test body~\cite{Krasnov2008b}.}
$R({\mathbf h}) = 1$.  In a similar fashion to $V$, we can then
decompose $R$ as $R = \frac{1}{3} h U_m({\mathbf H})$,  and expand $U_m$ as
\begin{equation}
U_m = 1 - \frac{g}{2} \tr{\mathbf H}^2 + {\cal O}({\mathbf H}^3),
\end{equation}
where $g$ is a dimensionless constant that can be of any sign.  This
new parameter measures the departure from the Urbantke metric, given
by an equality in Eq. (\ref{Urbantke}). Rather than the two new
parameters, $g$ and $\ell$, it is sometimes convenient
to use the two dimensionless parameters $\beta$ and $\gamma$, defined by
\begin{equation}
\beta = g - \frac{1}{3}, \qquad \qquad \textrm{and} \qquad \qquad
\gamma = \frac{1}{\ell^2\Lambda_0} - \frac{4}{3}.
\end{equation}
Cosmological consideration then tell us that $0< g < 1$, and hence $
-\frac{1}{3} < \beta < \frac{2}{3}$ and
$\gamma>0$~\cite{KrasnovShtanov2010}.  General Relativity is recovered
in the limit $\gamma \rightarrow \infty$.

Variation of Eq. (\ref{nonloc1}) with respect to $\oneform{A}$ gives
Eq. (\ref{dB}), while variation with respect to $\twoform{B}$ gives
\begin{equation}
\delta_{ij} B^i_{\phantom{i}I} \twoform{F}^j = 
 B^{i}_{\phantom{i}I}\left[ \frac{\partial U}{\partial H^{ij}}   +
 \frac{1}{3} \Lambda \delta^{ij} - 2 \pi G T  \left(  \frac{\partial
 U_m}{\partial H^{ij}} + \frac{1}{3} \Lambda_m \delta^{ij} \right) 
 \right] B^{j}_{\phantom{j}J} \twoform{\Sigma}^J_{(+)} 
 - 8 \pi G T_{IJ} \twoform{\Sigma}^J_{(-)},
\end{equation}
where $\Lambda$ and $\Lambda_m$ are the Legendre transforms of $U$ and
$U_m$ respectively, i.e. 
\begin{equation}
\Lambda = U - \frac{\partial U}{\partial H^{ij}} H^{ij} \qquad \qquad
\textrm{and} \qquad \qquad \Lambda_m = U_m - \frac{\partial
  U_m}{\partial H^{ij}} H^{ij}, 
\end{equation}
and where $T^{IJ} =  \tilde{T}^\mu_{\;\;\nu}
\Sigma^{I}_{\mu\lambda}\Sigma^{J\nu\lambda}$, where  $
\tilde{T}^\mu_{\;\;\nu}$ is the traceless part of the 
energy-momentum tensor, $T_{\mu\nu}$~\cite{Krasnov2009}.
\newline
\newline
\noindent
{\it Cosmology}
\newline

The cosmology of this theory has been analysed by Krasnov and
Shtanov at the level of perturbed FLRW solutions~\cite{KrasnovShtanov2010}.
Let us first consider the FLRW solutions of this theory.
For homogeneous and isotropic spaces we have $B^{iI} = \delta^{iI}$. 
Hence, we can drop the distinction between $i$ and $I$ and let
$\twoform{B}^i = \twoform{\Sigma}^i$.  We also have $H^{ij} = 0$, so
$U=\Lambda_0$ and $U_m = 1$, resulting in $R = \frac{1}{3} h$ and $V =
\frac{1}{3} h \Lambda_0$.  The fixing condition $R=1$ then gives
$h=3$ and $V= \Lambda_0$.  Under these conditions the field equations
can be written
\be
\twoform{F}^i = \frac{1}{3} \Lambda_0  \twoform{\Sigma}^i 
- 2 \pi G (P - \frac{\rho}{3}) \twoform{\Sigma}^i 
-  2 \pi G (\rho+P)  \twoform{\tilde{\Sigma}}^i .
\ee

For the homogeneous and isotropic space-time we also have
$\oneform{e}^0 = a d\tau$ and $\oneform{e}^i = a dx^i$.  After some algebra
this gives $\oneform{A}^j = i {\cal H} dx^j$ and
 $\twoform{F}^i =  i {\cal H}'  d\tau \wedge dx^j - \frac{1}{2}
\epsilon^{ijk} {\cal H}^2 dx^j \wedge dx^k$.  The field equations can
then be written
\begin{eqnarray}
3 {\cal H}^2 = 8\pi G a^2  (\rho + \rho_\Lambda  ),
\end{eqnarray}
where $P_\Lambda = - \rho_\Lambda$ with $\Lambda_0 = 8\pi G \rho_\Lambda $, and
\begin{eqnarray}
 -2 {\cal H}'  -  {\cal H}^2 = 8\pi G a^2 ( P + P_\Lambda),
\end{eqnarray}
where $\rho$ and $P$ do not include the cosmological constant. Thus,
for metric backgrounds the FLRW solutions of this theory are the same
as those of General Relativity.  The situation changes, however, when
we consider linear fluctuations.  In this case one gets departures
from $\Lambda$CDM that depend on $g$ and $\ell$.

We now consider the perturbed space-time metric in the conformal
Newtonian gauge.  The perturbation for $B^i_{\phantom{i}j}$ is then
given (after some convenient gauge-fixing) in terms of a new scalar mode $\chi$ as
\begin{equation}
B^i_{\phantom{i}j} = \delta^i_{\phantom{i}j} + \frac{1}{2a^2} D^i_{\phantom{i}j} \chi.
\end{equation}
The perturbed field equations are then given by~\cite{KrasnovShtanov2010}
\begin{eqnarray}
&&
-k^2 \Phi  = 4 \pi G a^2 \rho \left[\delta + 3 {\cal H} (1+w)
  \theta\right] + \frac{1}{3a^2} k^2 \left[ k^2 \chi - 3 {\cal H}
  \chi' \right], 
\\
&&
\Phi' + {\cal H} \Psi = 4 \pi G a^2 (\rho + P) \theta - \frac{1}{3a^2}
k^2 \chi' ,
\\
&&
\Phi'' + 2 {\cal H} \Phi' + {\cal H} \Psi'+ \left(2 {\cal H}' + {\cal
  H}^2\right)\Psi + \frac{k^2}{3}(\Phi - \Psi) = 4 \pi G a^2 \delta P
+ \frac{k^2}{9a^2} \chi ,
\\
&&
\Phi - \Psi =  8\pi G a^2 (\rho +P)\Sigma + \frac{1}{a^2}\left( \chi''
+ \frac{k^2}{3} \chi\right) ,
\end{eqnarray}
and one can show that the Bianchi identities are satisfied
independently of the $\chi$ terms. In this sense the $\chi$ field is
non-dynamical. The remaining equations determine $\chi$ in terms of
$\Phi$ and $\Psi$ as
\begin{equation}
\chi'' - 2 {\cal H}\chi' - \left[\lap+ 4 {\cal H}^2  +\Lambda_0 \gamma
  + 8\pi G \beta(\rho - 3P) \right]\chi + a^2 (\Phi + \Psi) = 0 ,
\end{equation}
where $\rho$ and $P$ does not include the cosmological constant.
This equation can be solved to get $\chi$ in terms of $\Phi+\Psi$,
hence the $\chi$ terms in the field equations can be thought of as
non-local modifications of the Einstein equations.

Krasnov and Shtanov also find the vector and tensor mode
equations~\cite{KrasnovShtanov2010}.  Furthermore, they study the
evolution of perturbations during inflation, a well as radiation,
matter and $\Lambda$ dominated epochs, and estimate the effects of the
modifications on the matter power spectrum.

\subsubsection{Dark energy from curvature corrections}

A proposal for IR modifications of gravity has been put forward by
Piazza~\cite{Piazza2009a,Piazza2009b}. The starting point for this is
the usual semi-classical gravity, where matter fields are quantised on
a curved background manifold.  The operators of the matter field
theory are then modified in the IR in a way we will now describe.
Schematically, in a cosmological setup, operators corresponding to
Fourier modes of physical momentum $k$ are corrected by terms of order
$H^2/k^2$, where $H$ is the Hubble parameter.  These modifications
lead to the apparent existence of Dark Energy, but without introducing
a new scale in the problem.

To illustrate this idea consider the vacuum expectation value of the
 local energy density of a massless field\footnote{see
 e.g.\cite{ZeldovichStarobinsky1971,FullingParker1974a}
 for the explicit expression of a massive scalar on flat FLRW
 background.}:
\begin{eqnarray}
 \langle T^0_{\;\;0}(t,\vec{x})\rangle_{\mbox{bare}} &=& \int
 d^3k\left[ k + \frac{f_{\mbox{quad}}(t)}{k} +
 \frac{f_{\mbox{log}}(t)}{k^3}+ \ldots \right] 
\label{T_bare}
\\
 &=&  \mbox{local terms} + \mbox{non-local terms}, \nonumber
\end{eqnarray}
where spatial homogeneity has been assumed for simplicity. The local
terms can be removed by local gravitational counter-terms, while the
non-local pieces represent the genuine particle/energy content of the
chosen ``vacuum'' state. The first term contributes to the
cosmological constant, and in flat space-time can be removed by the
usual procedure of normal-ordering (the $f$'s vanish in flat
space-time). In curved space-time, however, the presence of the
time-dependent $f$'s makes the normal-ordering procedure meaningless.
The conjecture of~\cite{Piazza2009a,Piazza2009b} is that there exists
a theory that resembles semi-classical GR on small scales, but that
has an IR-completion that prohibits the time dependent pieces in
(\ref{T_bare}). If that is the case then one can still deal with the
cosmological constant term by the usual procedure of normal-ordering,
as in flat space-time.

To try to construct such a theory~\cite{Piazza2009a,Piazza2009b}
propose what they call the \emph{Ultra Strong Equivalence Principle}:
For each matter field or sector sufficiently decoupled from all other
matter fields, there exists a state (the ``vacuum'') for which 
the expectation value of the (bare) energy-momentum tensor is the same
as in flat space, regardless of the configuration of the gravitational
field.

What this principle aims to achieve is to remove the time-dependent
terms in Eq. (\ref{T_bare}) by appropriate modifications of
semi-classical gravity that manifest themselves when
the Fourier modes have wavelengths comparable to the inverse extrinsic
curvature (i.e. the inverse Hubble radius $H^{-1}$). At the present, a
complete theory that implements this idea is lacking, but a toy-model
with massive scalar fields has been considered in~\cite{Piazza2009a,Piazza2009b}.
Letting $\vec{n}$ be the comoving momentum that labels operators in
Fourier space (related to physical momentum as $\vec{n}/a$), the
modification to $O(H^2a^2/n^2)$ is given by the modified commutation relation
\begin{equation}
\left[A^{(1)}_{\vec{n}}, A^{(1)\dagger}_{\vec{n}'} \right] =
\delta^{(3)}(\vec{n} - \vec{n}') \left( 1 - \frac{H^2 a^2}{2n^2} +
\ldots \right)  ,
\end{equation}
where $A^{(1)}_{\vec{n}}$ is the annihilation operator. This
prescription is equivalent to using the standard commutation relation
$\left[A^{(0)}_{\vec{n}}, A^{(0)\dagger}_{\vec{n}'} \right]=
\delta^{(3)}(\vec{n} - \vec{n}')$ for the standard operator
$A^{(0)}_{\vec{n}}$, but with a modified comoving momentum given by
$\vec{k} = \vec{n}\left( 1 - \frac{H^2 a^2}{2n^2}\right)$ that locally
defines the infinitesimal translations. In a local neighbourhood
(smaller than a Hubble patch)  the above prescription can be shown to
cancel the quadratically divergent piece 
$f_{\mbox{quad}}(t)$ in Eq. (\ref{T_bare}). Note that the momentum
$\vec{k}$ is not conserved, but $\vec{n}$ is. 

To extend the above to the global picture one can use the translation
operator $e^{-i \lambda P^{(1)}}$, where $\vec{P}^{(1)}= \int d^3n
\;\vec{n}\; A^{(1)\dagger}_{\vec{n}}A^{(1)}_{\vec{n}} = \vec{P}^{(1)}=
\int d^3n \; \vec{k} \; A^{(0)\dagger}_{\vec{n}}A^{(0)}_{\vec{n}}$ is
the momentum operator constructed with the modified Fourier modes, and
$\lambda$ is the comoving proper distance to a point far away from the origin.
In GR the comoving distance $\lambda = d(t)/a(t)$ is a constant given
by the ratio of the physical distance, $d(t)$, to the scale factor,
$a(t)$. However, in the present theory one finds instead that
\begin{equation}
\dot{\lambda} =  \frac{1}{4} \lambda^3 \frac{d}{dt}(a^2H^2)  \  \ + \mbox{higher orders}.
\label{USEP_lambda}
\end{equation}
Comoving distances obeying Eq. (\ref{USEP_lambda}) are, in fact,
already strongly disfavoured by
observations~\cite{NesserisPiazzaTsujikawa2009}. One may, however, try
to explore further whether the dynamical Hubble scale $H(t)$ itself
could provide the scale required by cosmic acceleration by considering
the more general expansion
\begin{eqnarray}
\dot{\lambda} &=&   
A_1 \lambda H + A_2 (\lambda H)^2 +  \ldots
\nonumber 
\\
&& \ \ \ \ + B_1 \lambda^2 \frac{d}{dt}(aH) +  B_2 \lambda^3
\frac{d}{dt}(a^2H^2)  + \ldots ,
\end{eqnarray}
where $A_i$ and $B_i$ are a set of constants. The authors find that
certain regions of the resulting parameter space can fit the data as well as
$\Lambda$CDM. 

\newpage 

\section{Higher Dimensional Theories of Gravity}
\label{HD2}

The first systematic studies of higher dimensional geometry date back
to the likes of Riemann, Cayley and Grassmann in the mid nineteenth
century. It lies at the heart of General Relativity, where space and
time form part of a curved $3+1$ dimensional manifold, as described in
Section \ref{GR}.  Of course, Riemannian geometry is not restricted to
3+1 dimensions, so we have the tools to study gravitational theories
in higher dimensions. Indeed, this is more than just a theoretical
curiosity.  Superstring theory, arguably our best candidate for a
quantum theory of gravity,  can only be formulated consistently in 10
dimensions.

The problem now is a phenomenological one: Gravity does not behave
like a 10 dimensional force in our experiments and
observations. Perhaps the simplest observation along these lines is
the stability of earth's orbit. In $D$ dimensions of space-time, the
Newtonian potential due to a point source will typically go like
$1/r^{D-3}$. For $D \neq 4$, it follows that we cannot have stable
planetary orbits, and so it is clear that gravity should not {\it
  appear} 10 dimensional on solar system scales. We use the word {\it
  appear}, because there exist gravitational models  where  the extra
dimensions are hidden from experiment, but which open up at shorter
and/or larger distances.

In this section we will review various models of higher dimensional
gravity that have been proposed. We will only discuss the case of
extra spatial dimensions, although extra temporal dimensions have been studied (see eg \cite{Shtanov1}). One might worry that extra temporal dimensions lead to problems
with causality, as they permit closed time-like curves in the form of
circles in the plane of the two temporal directions.

\subsection{Kaluza-Klein Theories of Gravity}

Kaluza-Klein (KK) theory grew out of an attempt to unify
gravity and electrodynamics \cite{Nordstrom, Kaluza, Klein1,
  Klein2}. The basic idea was to consider General Relativity on a
$4+1$ dimensional manifold where one of the spatial dimensions was
taken to be small and compact. One can perform a harmonic expansion of
all fields along the extra dimension, and compute an effective $3+1$
dimensional theory by integrating out the heavy modes. This idea has
been embraced by string theorists who compactify $10$ dimensional
string theories and $11$ dimensional supergravity/M-theory on  compact
manifolds of $6$ or $7$ dimensions respectively, often switching on
fluxes and wrapping branes on the compact space (see
\cite{compact-review} for a review). Each different compactification
gives a different effective $4$-dimensional theory, so much so that we
now talk about an entire landscape of effective theories
\cite{landscape}. 


Assuming that the extra dimensions have been stabilised, the late-time
dynamics of KK theories is most easily understood at the level of the
$4D$ effective theory.  As we will show, this  will generically
correspond to a $4D$ gravity theory with extra fields, examples of
which are studied in detail in Section \ref{extrafields}. At early
times, when the $3$  dimensional space is comparable in size to the
extra dimensions, the effective description clearly breaks down. This
forms the basis of KK cosmology where one can ask the deeply profound
question of why  and how the $3$ extended dimensions of space were
able to grow large, while the extra dimensions remained
microscopically small. It seems fair to say that a fully satisfactory
answer to this question has yet to emerge.

We will now discuss some aspects of KK theory, starting with an
overview of dimensional reduction and effective theory before moving
on to a discussion of KK cosmology at early times. For a more detailed
review of KK theory see \cite{KKlove, KKwess}.

\subsubsection{Kaluza-Klein compactifications}
\label{sec:KKcompactification}

To understand the generic features of KK compactifications, it is
sufficient to describe the dimensional reduction  of General
Relativity on a circle, $S^1$. We first define General Relativity in
$D=d+1$ dimensions, via the generalised Einstein-Hilbert action
\be \label{genEH}
S[\gamma]=\frac{1}{16\pi G_D} \int d^{D}X \sqrt{-\g} {\cal R},
\ee
where $G_D$ is Newton's constant in $D$ dimensions, $\gamma_\AB$ is
the $D$ dimensional metric with corresponding Ricci tensor,
${\R}_{AB}$, and Ricci scalar, $\R=\g^{AB} \R_{AB}$.  Note that we are
neglecting the matter Lagrangian for brevity. We are assuming that one
of  the spatial dimensions is compactified on a circle of radius
$L/2\pi$. To this end we can define coordinates $X^A=(x^\mu, z)$,
where the coordinate $z$ lies along the compact direction, such that
$0\leq z< L$.

We can expand the metric as  a Fourier series of the form
\be \label{KKaction}
 \gamma_{AB}(x,z)=\sum_n \g_{AB}^{(n)}(x)e^{inz/ L}.
\ee
We find that this gives an infinite number of extra fields in $d$
dimensions. Modes with $n \neq 0$ correspond to massive fields with
mass $|n|/L$, whereas the zero mode corresponds to a massless field. As we take
$L$ to be smaller and smaller we see that the mass of the first
massive field becomes very large. This means that if we compactify on
a small enough circle we can truncate to massless modes in the
$4$-dimensional theory. Massive modes will only get excited by
scattering processes whose energy lies at or above the
compactification scale $1/L$. This also applies to matter fields
arising in particle physics. Indeed, particle physics imposes by far
the strongest constraints on the size of the extra dimension. Standard
Model processes have been well tested with great precision  down to
distances of the order $\sim$(TeV)${}^{-1}$, with no evidence of extra
dimensions yet emerging \cite{PPbounds}.  Assuming that the extra
dimensions are universal, that is the  Standard Model fields can
extend all the way into them, we infer that  $L \lesssim 10^{-19}$
m. The natural scale of the compact dimensions is usually taken be
Planckian, $L \sim l_{pl}$.

Let us now focus on the zero modes, $\g^{(0)}_{AB}(x)$.  We could
define $\g^{(0)}_{\mu\nu}$, $\g^{(0)}_{\mu z}$ and  $\g^{(0)}_{zz}$ to
be the $d$-dimensional fields $g_{\mu\nu}(x)$, $A_{\mu}(x)$ and
$\phi(x)$.  In effective field theory language, these will correspond
to the metric, gauge field, and dilaton, respectively. In
order that our results are more transparent we will actually define
the components of the metric in the following way:
\be
\g^{(0)}_{\mu\nu}=e^{2 \alpha \phi}g_{\mu\nu}+e^{2 \beta
\phi}A_{\mu}A_{\nu}, \qquad
\g^{(0)}_{\mu z}= e^{2 \beta
\phi}A_{\mu}, \qquad \g^{(0)}_{zz}=e^{2 \beta
\phi},
\ee
where $\alpha=1/\sqrt{2(d-1)(d-2)}$, and $\beta=-(d-2)\alpha$.  Since we have
truncated to the massless fields, we can integrate out the $z$ part of the
action given in Eq. (\ref{KKaction}). We find that the $d$-dimensional effective
action is then given by
\be
S_{\textrm{eff}}[g, A, \phi]=\frac{L}{16 \pi G_D}\int d^{d}x \sqrt{-g} \left(R-\half(\nabla
\phi)^2-\quarter e^{-2(d-1)\alpha\phi } F^2 \right),
\ee
where  $F^2=  F_\mn F^\mn$ and $F_{\mu\nu}=\nabla_\mu A_\nu-\nabla_\nu
A_\mu$ is the electromagnetic field strength. The curvature associated
with the  $d$ dimensional metric, $g_{\mu\nu}$, is  described by the
Ricci tensor, ${R}_{\mu\nu}$, and Ricci scalar, $R=g^{\mu\nu}
R_{\mu\nu}$. What we now have is an Einstein-Maxwell-Dilaton system in
$d$ dimensions.  Of course, Kaluza and Klein were particularly
interested in the case of $d=4$. They were frustrated by the presence
of the dilaton, $\phi$, in the resulting $4$-dimensional effective
theory. The point is that one cannot simply set the dilaton to zero
and retain a non-trivial  Maxwell field, since this would be in
conflict with the field equations  arising from Eq. (\ref{KKaction}),
\ba
&&G_\mn= \half\left[\nabla_\mu \phi \nabla_\nu\phi-\half (\nabla
  \phi)^2 g_\mn+ e^{-2(d-1)\alpha
    \phi}\left(F_{\mu\alpha}F_\nu{}^\alpha-\quarter F^2 g_\mn
  \right)\right],\\
&&\nabla^\mu \left(e^{-2(d-1) \alpha \phi} F_\mn \right)=0, \\
&&\Box \phi = -\half (d-1) \alpha e^{-2(d-1)\alpha \phi} F^2, 
\ea
where $G_\mn=R_\mn-\half R g_\mn$ is the Einstein tensor in $d$
dimensions.  In the usual jargon, switching off the dilaton does not
represent a {\it consistent truncation} of the higher dimensional
theory \cite{truncations}.  We should also note that the {\it
  physical} size of the compact dimension is not necessarily given by
$L$, but by $L e^{\beta \phi(x)}$.  If $L$ is to represent an accurate
measure of the compactification scale, we are therefore implicitly
assuming that   $\phi$ is stabilised close to zero.   For this to
happen we need to generate a potential for $\phi$ that admits a stable
solution-- this is known as the problem of moduli stabilisation. In
more general compactifications, moduli potentials can be generated by
Casimir effects of fields in the compact space \cite{chodos1, chodos2,
  Candelas, Ed1, Ed2}, but the moduli remain unstable \cite{Ed2}.  In
fact, the problem of moduli stabilisation has only recently been
solved by switching on fluxes to stabilise the volume of the compact
space \cite{KST, KKLT}.

There are, of course, many different compactifications that have been
studied in the literature, a detailed analysis of which is clearly
beyond the scope of this review (see \cite{compact-review}). However,
aside from details such as the inclusion of fluxes and branes on the
compact space, the general scheme of each compactification is the
same as the one we have just described.  Typically, a
compactification of, say, $11$ dimensional super-gravity down to four
dimensions will give rise to a gravity theory with a plethora of
extra fields. These extra fields include scalars, pseudo-scalars,
vectors, pseudo-vectors, and arbitrary p-forms.   Modifications of
gravity due to extra fields are studied in detail in Section
\ref{extrafields}.

\subsubsection{Kaluza-Klein cosmology}

As we have just discussed,  for a phenomenologically viable theory
with compact (and stabilised) extra dimensions, the  characteristic
size, $L$,  of the compact manifold  should not exceed the scale
probed by modern collider experiments, which is currently around
$10^{-19}$ m \cite{PPbounds}.  It is amusing to compare this to the
characteristic size of the $3$ large spatial dimensions, which is at
least a Hubble length $c/H_0 \sim  10^{26}$m, or in other words at
least $45$ orders of magnitude greater. Of course, it was not always
like that. In the very early universe, at times $t \lesssim L/c$, one
might expect that all spatial dimensions were of the same scale, each
playing an equally important role in the dynamical evolution. This
begs the question: Why did the Universe evolve into a state where just
$3$ spatial dimensions grew to macroscopic scales? Put another way,
how does one achieve a dynamical compactification mechanism in the
early Universe such that $3$ spatial dimensions expand exponentially to
an extremely large size, in contrast to the remaining spatial
dimensions? Were those extra dimensions somehow prevented from growing
beyond a certain size, or did they grow initially and later contract
towards their current state?

These questions have led many authors (see, e.g. \cite{Freund, Alvarez,
  Kolb, Sahdev1, Sahdev2, Sahdev3, Abbott1, Abbott2,  Abbott3, Okada,
  Greene}) to consider the dynamics of anisotropic cosmologies in
$D=d+1$ dimensions, where $d=n+\tilde n$. Indeed, consider the
Bianchi-type metric
\be
ds^2=\g_{ab} dX^a dX^b=-dt^2+a^2(t)q_{ij}(x) dx^i dx^j+\tilde a^2(t)
  \tilde q_{mn}(\tilde x) d \tilde x^m d \tilde x^n, 
\ee
where the coordinates $x^i$ run over the  $n$ spatial dimensions  and
the coordinates $\tilde x^m$ run over the $\tilde n$  spatial
dimensions.  The $n$-dimensional metric, $q_{ij}(x)$, is taken to
have constant curvature $\kappa$, whereas the $\tilde n$-dimensional
metric, $\tilde q_{mn}(\tilde x)$, is taken to have   constant
curvature $\tilde \kappa$.  The growth of these two spaces is
controlled by the relevant scale factors $a(t)$ and $\tilde
a(t)$. Naturally, we will be interested in the case of $n=3$, but for
the moment let us keep things general. 

We now apply Einstein's equations in $D=d+1$ dimensions, 
\be
{\cal G}_{ab}\equiv {\cal R}_{ab}-\frac{1}{2} {\cal R}
\gamma_{ab}=8\pi G_D T_{ab},
\label{G=T}
\ee
where the energy-momentum tensor is given by an anisotropic fluid, 
$$
T^a_b=\textrm{diag} \left(-\rho, \overbrace{P, \ldots, P}^n, \overbrace{\tilde
  P, \ldots, \tilde P}^{\tilde n}\right).
$$
As usual, $\rho(t)$ is the energy density, whereas $p(t)$ is the
pressure along the $n$ dimensions and $\tilde P(t)$ is the pressure
along the $\tilde n$ dimensions.  Einstein's Equations (\ref{G=T})
then yield the following \cite{Freund}
\ba
&\frac{n}{2}(n-1) \left(H^2+\frac{\kappa}{a^2}\right)+\frac{\tilde
  n}{2}(\tilde n-1) \left(\tilde H^2+\frac{\tilde \kappa}{\tilde
  a^2}\right)+n \tilde nH \tilde H = 8 \pi G_D \rho,& \label{cos1}\\
&\frac{\ddot a}{a}  +(n-1)\left(H^2+\frac{\kappa}{a^2}\right)+\tilde
nH \tilde H =  \frac{8 \pi G_D}{d-1}\left[\rho+(\tilde n-1)P-\tilde
  n\tilde P\right], &  \label{cos2}\\
&\frac{\ddot {\tilde a}}{\tilde a} +(\tilde n-1)\left(\tilde
H^2+\frac{\tilde \kappa}{\tilde a^2}\right)+nH \tilde  H =  \frac{8
  \pi G_D}{d-1}\left[\rho-nP+(n-1)\tilde P\right],&  \label{cos3}
\ea
where $H=\dot a/a$ and $\tilde H=\dot{\tilde  a}/\tilde a$ are the
Hubble parameters of the two expanding/contracting spaces. Of course,
we also have energy conservation, which gives
\be
\dot \rho +nH(\rho+P)+\tilde n\tilde H (\rho +\tilde P)=0. \label{cons}
\ee
Note that we do not necessarily have to assume that the cosmological
dynamics is governed by $D$-dimensional General Relativity. We can
also consider modifications of  GR where additional fields are
present. For example, in string gas cosmology \cite{Bran0, Tseytlin},
we consider the action
\be
S=\frac{1}{16\pi G_D}  \int d^Dx \sqrt{-\gamma} e^{-2\phi} \left({\cal
  R}-4 (\nabla \phi)^2-c\right) +S_m[\gamma, \Psi_n],
\ee
where $S_m$ is the matter part of the action, containing the string
gas, and the constant $c$ vanishes in the critical case\footnote{ For
  the bosonic string the critical dimension is  $D=26$, whereas  for
  the superstring the critical dimension is $D=10$.},  but not
otherwise. The resulting field equations can be written in the form
$G_{ab}=8\pi G_D T_{ab}$, where
\be
T_{ab}=\frac{1}{8\pi G_D}\left[-\frac{c}{2}\gamma_{ab}+8 \nabla_a \phi
  \nabla_b \phi -6 \g_{ab} (\nabla
  \phi)^2-2(\nabla_a\nabla_b-\g_{ab}\Box)\phi\right]+e^{2\phi}
T^{(m)}_{ab},
\ee
and $T^{(m)}_{ab}=-\frac{2}{\sqrt{-\g}}\frac{\delta S_m}{\delta
  \g^{ab}}$. The scalar equation of motion just follows from energy
conservation, $\nabla^aT_{ab}=0$.

\begin{figure}
\begin{center}
\epsfig{file=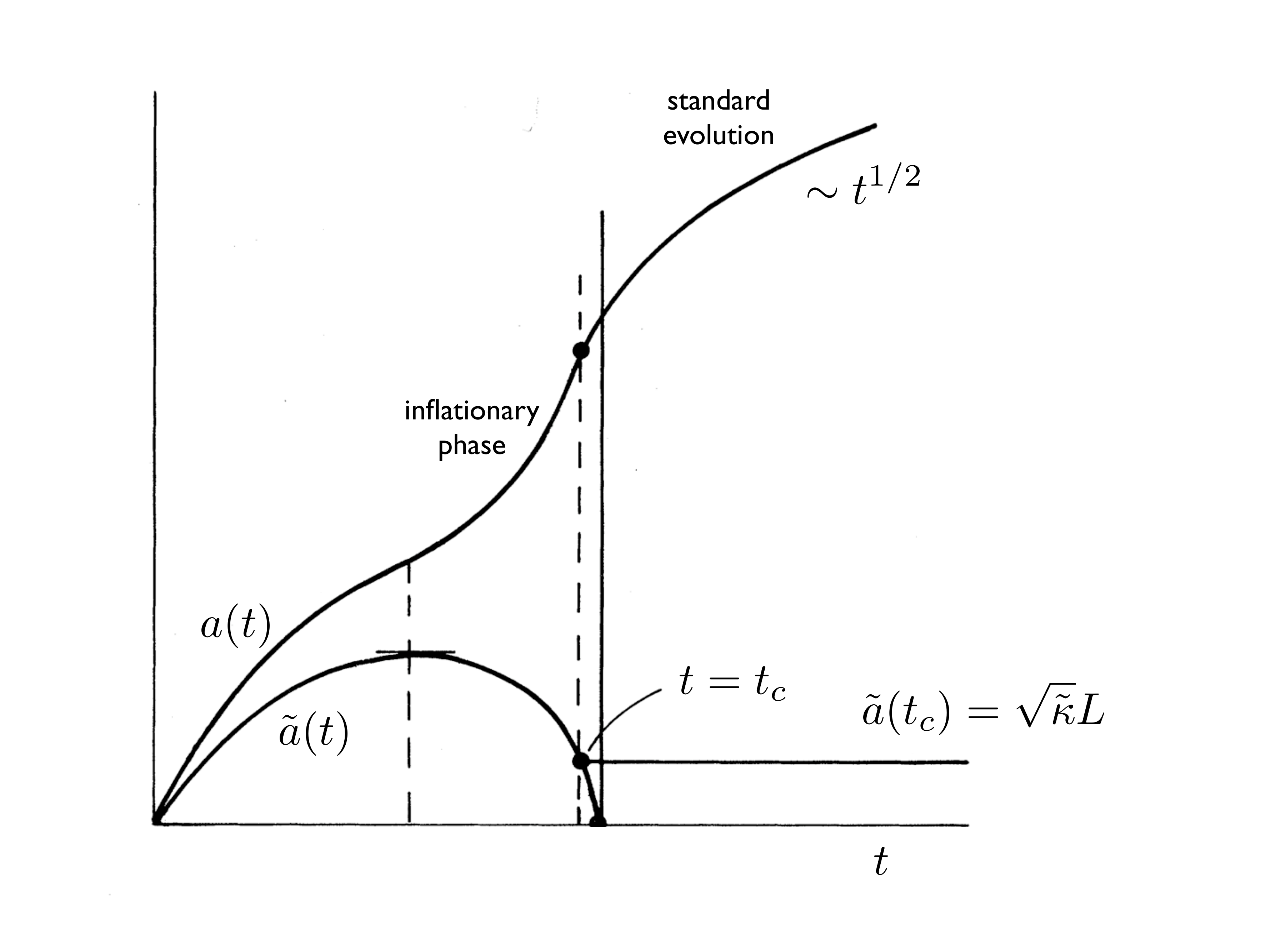,width=10cm,}
\caption{Adapted from Figure 1 in \cite{Abbott1}. The typical
  evolution of the scale factors in the two different sub-spaces. The
  scale factor $a(t)$ along the $n$ dimensions grows large, entering
  an inflationary phase as the remaining $\tilde n$ dimensions begin
  to recollapse. The scale factor $\tilde a(t)$ along the $\tilde n$
  dimension is assumed to be stabilised by quantum gravity effects at
  some time $t_c$.}
\label{fig:abbott1}
\end{center}
\end{figure}

Let us return to Equations (\ref{cos1})-(\ref{cos3}) with a view
toward dynamical compactification. Many of the earlier works
\cite{Alvarez,   Kolb, Sahdev1, Sahdev2, Sahdev3, Abbott1, Abbott2}
focus on isotropic perfect fluids, for which $P=\tilde P=w\rho$.  For
simplicity and definiteness, let us follow the analysis of Abbott,
Barr and Ellis \cite{Abbott1}. We consider an epoch in which we have
radiation domination, $w=1/{(n+\tilde n)}$, so that  the evolution
equations read
\ba
&\frac{n}{2}(n-1) \left(H^2+\frac{\kappa}{a^2}\right)+\frac{\tilde
  n}{2}(\tilde n-1) \left(\tilde H^2+\frac{\tilde \kappa}{\tilde
  a^2}\right)+n \tilde nH \tilde H = 8 \pi G_D \rho,& \label{cos1a}\\
&\frac{\ddot a}{a}  +(n-1)\left(H^2+\frac{\kappa}{a^2}\right)+\tilde
nH \tilde H =  \frac{8 \pi G_D}{d}\rho,&  \label{cos2a}\\
&\frac{\ddot {\tilde a}}{\tilde a} +(\tilde n-1)\left(\tilde
H^2+\frac{\tilde \kappa}{\tilde a^2}\right)+nH \tilde  H =  \frac{8
  \pi G_D}{d}\rho.&  \label{cos3a}
\ea
Now, if the $\tilde n$ dimensions are  taken to be an $\tilde
n$-sphere $(\tilde \kappa>0)$, it is clear from Equation (\ref{cos3a})
that they will  reach a  {\it maximum} size when $\tilde H=0$, and
will subsequently start to recollapse. In contrast, we can take the
$n$ dimensions to be flat or hyperbolic $(\kappa \leq 0)$, so that
these dimensions will never turn around. In fact, one can show that as
we start to approach the singularity of the collapsing sub-space
$(\tilde a(t) \to 0)$, the $n$ dimensions  enter a phase of
accelerated expansion.  To see this note that $\tilde H$ starts to
become large  and negative, and so it is clear from Equation
(\ref{cos2a}) that  we will enter a phase with $\ddot a>0$.  The
typical evolution of the two scale factors is shown in Figure
\ref{fig:abbott1}.
Note that we can even allow for a sufficiently small $\kappa>0$ and
still retain this qualitative behaviour.  The upper bound on $\kappa$
follows from demanding that the turnaround in $a(t)$ occurs after the
turnaround in $\tilde a(t)$.  The bound is not strong enough to be
interesting:  It merely implies  that today's Universe is larger than
the horizon \cite{Abbott1}.

Of course, it is clear that the classical equations will start to
break down in the neighbourhood of the singular point. The physical
radius of the $\tilde n$-sphere is $\tilde a/\sqrt{\tilde \kappa}$, so
we certainly would not expect to trust our field equations when
$\tilde a(t) \lesssim {\sqrt {\tilde \kappa}} L_{D}$, where $L_{D}
\propto G_D^{1/(D-2)}$ is the fundamental Planck length in $D$
dimensions.  In \cite{Abbott1}, it is assumed that quantum gravity
effects will ultimately stabilise the size of the internal space,
ending the inflationary phase  at some time $t_c$, where $\tilde
a(t_c)= {\sqrt {\tilde \kappa}} L$, for some compactification scale $L
\gtrsim L_D$ .  Albeit without much justification,  let us accept this
assumption for the moment, and consider the physically interesting
case of $n=3$. One might hope that the inflationary  phase is
sufficiently long  to offer a solution to the flatness, entropy and
horizon problems of the standard cosmology.  Consider the entropy
problem in particular.  Entropy is indeed released from the extra
dimensions into the usual $3$ dimensions of space \cite{Alvarez,
  Abbott1}, but only as much as $\log S  \sim |{\cal O}(1)|
\log(L_D/L)$ \cite{Kolb}.  Since we demand that $L \gtrsim L_D$, this
is clearly way short of the  total required to solve the entropy
problem,   $\log S \sim 88$.  In short, KK inflation does not last
long enough to provide an alternative to scalar field driven inflation.
 
To get the required expansion of $3$-dimensional space, we must
 therefore include some additional scalar fields. Adapting
 \cite{Holman} slightly, we can mimic a period of slow-roll inflation
 by plugging a cosmological constant, $\Lambda$, into the Field
 Equations (\ref{cos1})-(\ref{cos3}). Setting $P=\tilde P=-\rho$,
 where $\rho=\Lambda/8\pi G_D$, we find 
 \ba
&\frac{n}{2}(n-1) \left(H^2+\frac{\kappa}{a^2}\right)+\frac{\tilde
 n}{2}(\tilde n-1) \left(\tilde H^2+\frac{\tilde \kappa}{\tilde
 a^2}\right)+n \tilde nH \tilde H = \Lambda,& \label{cos1b}\\
&\frac{\ddot a}{a}  +(n-1)\left(H^2+\frac{\kappa}{a^2}\right)+\tilde
 nH \tilde H =  \frac{8 \Lambda}{d-1},&  \label{cos2b}\\
&\frac{\ddot {\tilde a}}{\tilde a} +(\tilde n-1)\left(\tilde
 H^2+\frac{\tilde \kappa}{\tilde a^2}\right)+nH \tilde  H =
 \frac{2\Lambda}{d-1}.&  \label{cos3b}
\ea
Again, by taking the $n$ dimensions to be flat ($\kappa=0$), and the
$\tilde n$ dimensions to be positively curved ($\tilde \kappa>0$), we
find a solution for which the flat directions grow exponentially, and
the spherical dimensions remain fixed with $H=H_*$ at $\tilde
a=\tilde a_*$, where $H_*^2=\frac{\tilde n-1}{n} \frac{\tilde
\kappa}{\tilde a_*^2}=\frac{2 \Lambda}{n(d-1)}$.  The radius of the
extra dimensions lies at the {\it maximum} of its potential, so this
solution is unstable. Indeed, fluctuations reveal that the spherical
dimensions collapse to zero size over a time scale $ \Delta t \sim
(1+\sqrt{1+8/n})/4H_*$, after which we cannot count on exponential
growth in the flat directions.
 
We now consider the phenomenologically interesting case of $n=3$. To
get the required number of $\sim 65$ e-folds of inflation along the
$3$ flat directions we need $H_* \Delta t \sim  65$.  For $n=3$, $H_*
\Delta t \sim  0.729$, so once again inflation is cut short far too
early. We could imagine getting around this problem if we could alter
the  potential for the radius $\tilde a$, such that it develops a
minimum as well as a maximum, by switching on fluxes \cite{KST,
  KKLT}. Generically, it is still very difficult to get enough
exponential growth along the familiar $3$ dimensions without causing
the extra dimensions to grow alongside them \cite{Holman}.   For
further details on the latest attempts to  embed inflation in  higher
dimensional theories,  we refer the reader to \cite{Lidsey, Quevedo,
  Linde, Kallosh, Burgess, McAl, Buck}. Bounds on the variation of fundamental constants for dynamical compactifications have been studied in \cite{Gunther1, Gunther2}, while PPN parameters for KK models in the solar system were computed in \cite{Zhuk3}.  Note that Kaluza-Klein cosmologies have also recently been applied to the dark energy problem \cite{Zhuk1, Zhuk2}.   
 
We end our discussion of Kaluza-Klein cosmology by asking the
question: Why are there $3$ large spatial dimensions?  We have already
alluded to  an anthropic explanation demanding the existence of stable
planetary orbits\footnote{On the subject of planetary orbits,  it is
  amusing to note that Kepler himself reasoned that the 3-fold nature
  of the Holy Trinity was responsible for the perceived dimensionality
  of space.  Ptolemy is reputed to have offered some alternative ideas
  in his work {\it On Dimensionality}, but they have since been
  lost.}.  To this we could add the existence of stable atoms
and chemistry, both key to the development of intelligent life, and
both requiring no more than $3$ (large) spatial dimensions.  

Modern attempts at a {\it dynamical} understanding of the
dimensionality of space include String Gas Cosmology \cite{Bran0,
  Tseytlin} (for reviews see, e.g., \cite{Bran1, watson, Bran3}). Here
the  spatial dimensions are taken to be compact and precisely $3$
dimensions are allowed to grow large due to the annihilation of
strings wrapping around those dimensions. The point is that strings
winding around compact dimensions oppose their expansion since the
energy of the string winding modes increases with radius. To allow the
compact dimensions to grow large the winding modes must therefore
collide and annihilate with the anti-winding modes.  Generically, we
would only expect collisions of  $1+1$ dimensional strings in at most
$3+1$ dimensions. Thus, the dimensionality of the string controls the
dimensionality of space by allowing at most $3$ spatial dimensions to
grow to macroscopic scales. Note that this result is not spoilt by the
inclusion of branes wrapping compact directions, as these happen to
fall out of equilibrium before the strings \cite{Alexander}.

Whilst this idea has some appeal at first glance, it has not stood up
to intense scrutiny. More detailed quantitative analyses suggest that
the desired outcome is not at all generic, and requires highly fine
tuned initial conditions \cite{Easther1, Easther2, Easther3}. Whilst
one can engineer an anisotropic  set-up  allowing $3$ dimensions to
grow large as desired, typically the internal dimensions also grow to
large sizes, just at a slower rate \cite{Easther1, Easther2}. In fact,
it turns out that either all dimensions grow large since the string
gas eventually annihilates completely, or all dimensions stay small
since the string gas gets frozen out \cite{Easther3}. There are also
problems at the level of cosmology.  For example, when properly
calculated, the scalar perturbations have a blue power spectrum with
$n=5$, which is strongly ruled out by observations \cite{Kaloper-sg}.
It has been argued that a near scale invariant spectrum can be
obtained if the dilaton gets frozen during the strong coupled Hagedorn
phase in the very early Universe \cite{Bran2}. However,  such claims
still rely on a semi-classical treatment of cosmological perturbations
that cannot be trusted during the Hagedorn phase, as the strings are
strongly interacting.

In the context of $10$ dimensional string theory, other attempts to
explain the dimensionality of our Universe consider that for integer
values of $n$,  the inequality $2n <10 \implies n\leq 4$ \cite{durrer,
  randall}.  This is interesting because it means that the world
volume of $3+1$ dimensional branes (known as $3$-branes) are less
likely to intersect than those of larger branes. In particular, Karch
and Randall \cite{randall} have shown that an FLRW universe initially
filled with equal numbers of  branes and anti-branes will ultimately
come to be dominated by $3$-branes and $7$-branes. This analysis
accounts for the fact that larger branes dilute more slowly, as well
as the likelihood of intersections and annihilations (hence the
importance of $7$-branes). In a braneworld scenario, this could
explain why we might be more likely to find ourselves living on a
$3$-brane, as opposed to a larger brane. The consequences of living on
a $3$-brane are discussed in detail in the Section \ref{sec:braneworlds}.

Finally, we note that for toroidal compactifications, 3 large spatial dimensions  can be linked to  the stability of the small extra dimensions, at least in the presence of solitonic strings/branes that correspond to point masses in the   large  dimensions \cite{Zhuk4}.

\subsection{The Braneworld Paradigm}
\label{sec:braneworlds}

The braneworld paradigm \cite{Akama, Rub-Shap, ADD1, AADD, ADD2}
represents a radical alternative to the standard Kaluza-Klein
scenario, discussed in the previous section. In the KK scenario, the
extra dimensions must be small and compact, the size of the internal
space constrained by collider experiments to be below the inverse TeV
scale. In the braneworld scenario the extra dimensions can be much
larger, perhaps even infinite in extent.  This is made possible by
relaxing the assumption of universal extra dimensions.

In the braneworld picture the Standard Model fields are {\it not}
universal, rather they are confined to lie on a $3+1$ dimensional
hyper-surface, known as the {\it brane}, embedded in some higher
dimensional space-time, known as the {\it bulk}.  Tests of Standard
Model processes can only constrain how far the brane may extend into
the bulk, or, in other words, the brane thickness. They do not
constrain the size of the bulk itself.  Such constraints can only come
from gravitational experiments, as gravity is the only force that
extends into the bulk space-time.  As is well known, on small scales
gravity is much weaker than the other three fundamental forces, making
it difficult to test at short distances. In fact, the gravitational
interaction has only been probed down to  $\sim 0.1$ mm,  with
torsion-balance tests of the inverse square law \cite{Adelberger}. It
is too simplistic, however, to suggest that this translates into an
upper bound on the radius of the bulk. Gravity is intimately related
to geometry, and, as we shall see, one can warp the bulk geometry such
that an infinitely large extra dimension is still allowed by
experiment. For an excellent introduction to large extra dimensions
see \cite{rub-intro}.

\begin{figure}
\begin{center}
\epsfig{file=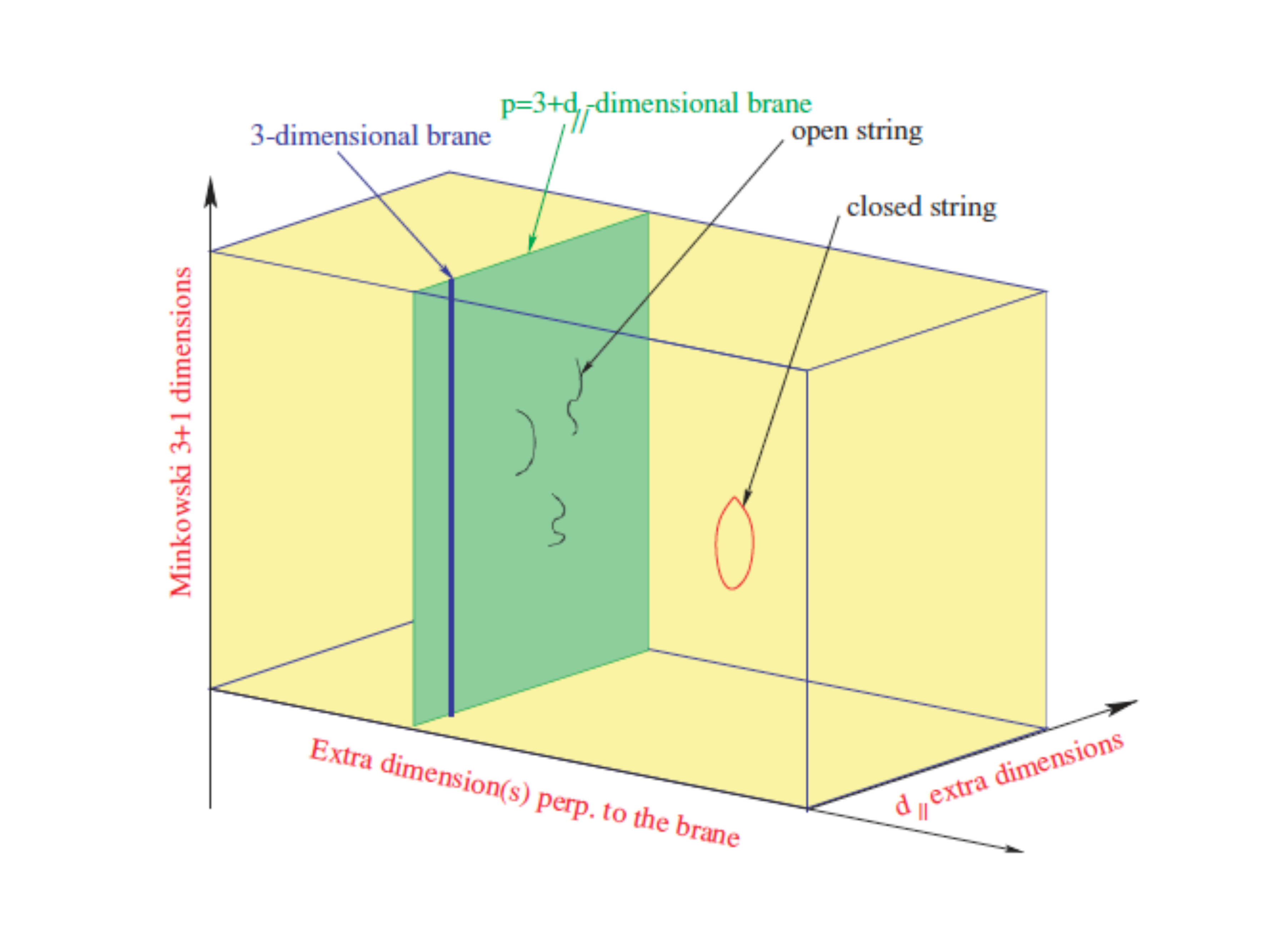,width=11cm,}
\caption{Taken from \cite{branepic}. Here the brane has $3$
  macroscopic dimensions, and $d_\parallel$ compact dimensions. The
  open strings end on the D branes, whereas the closed strings
  propagate through the bulk.}
\label{fig:dbranes}
\end{center}
\end{figure}

Before delving into a detailed discussion of the various models, we
note in passing that the braneworld paradigm is well motivated by
string theory \cite{Horava-Witten, LOSW, AADD}. As well as fundamental
strings, string theory contains fundamental objects known as D-branes
\cite{dbranes}. These are extended objects upon which open strings can
end. The braneworld set-up therefore has a natural interpretation  in
terms of a stack of D-branes embedded in a higher dimensional target
space (see Figure \ref{fig:dbranes}). Open strings, with their ends
attached to the D-branes, can be identified with the Standard Model
fields bound to the brane. Only closed strings can propagate through
the bulk, and these are identified with the gravitational
interactions.

\subsubsection{The ADD model}
\label{sec:ADD}

The braneworld paradigm really began to gather momentum with the
seminal work of Arkani-Hamed, Dimopoulos and Dvali \cite{ADD1}, in
which the large extra dimension is exploited in order to explain the
vast hierarchy between the electro-weak scale, $M_{EW} \sim$ TeV, and
the Planck scale, $M_{pl} \sim 10^{16}$ TeV.   In this scenario the
hierarchy does not go away completely, rather it is reformulated as a
hierarchy between the scale of the extra dimensions, $\mu \sim 1/L$,
and the electro-weak scale. The set-up is as follows:  Standard Model
fermions and gauge bosons are localised on a $3+1$ dimensional domain
wall,  in an (effective) $D=4+n$ dimensional space-time. We should
clarify that $D$ counts the number of {\it macroscopic} dimensions in
this scenario. Any {\it microscopic} dimensions, with characteristic
size given by the fundamental Planck length, behave as in the standard
KK scenario described previously.

Now, the width of the wall cannot exceed the inverse TeV scale, as
explained above.  The bulk space transverse to the wall is compact but
much larger than the width of the wall ($L\gg$ TeV${}^{-1}$), so much
so that we can treat the wall as an infinitely thin $3$-brane. In the
simplest construction, the bulk action is then described by the
generalised Einstein-Hilbert action
\be 
S_{\textrm{bulk}}=\frac{M_D^{D-2}}{2}
 \int d^{D}X \sqrt{-\g} {\cal R},
\ee
where $M_D$ is the fundamental Planck scale in $D$ dimensions,
$\gamma_\AB$ is the $D$ dimensional metric with corresponding Ricci
tensor, ${\R}_{AB}$, and Ricci scalar, $\R=\g^{AB} \R_{AB}$. The
Planck mass is related to the fundamental Newton's constant in $D$
dimensions by $8 \pi G_D=M_D^{2-D}$. At large distances, gravitational
interactions along the brane are mediated by the graviton zero mode,
which has a homogeneous profile over the extra dimensions.  Truncating
to the zero mode, we can compute the four-dimensional effective action
describing long distance gravity along the brane by integrating out
the macroscopic extra dimensions. This result in
\be 
S_{\textrm{eff}}=\frac{M_D^{D-2} V_n}{2}  \int d^{4}x \sqrt{-g} { R},
\ee
where $g_{\mu\nu}$ is the four-dimensional metric on the brane, with
Ricci tensor $R_{\mu\nu}$, and Ricci scalar $R=g^{\mu\nu} R_{\mu\nu}$,
and where the volume of the extra dimensions is given by $V_n \propto
L^n$. The effective four-dimensional Planck scale, as seen by an
observer on the brane, is then given by
\be
M_{pl}^2 \sim   M_D^{2+n}  L^{n}.
\ee
By taking the macroscopic extra dimensions to be sufficiently large,
we can eliminate the standard hierarchy in $D$ dimensions,  $M_D \sim
M_{EW} \sim 1$ TeV, and replace it with a new hierarchy involving the
scale of the extra dimensions, $\mu \sim 1/L \ll M_{EW}$. This is not
in violation of short distance gravity tests, at  least in $D \geq 6$
dimensions. Indeed, in six dimensions one can eliminate the hierarchy
even for millimetre size extra dimensions.

In actual fact, the strongest constraints on the ADD model do not come
from short distance gravity tests, but from astrophysics and cosmology
\cite{ADD2}. The problem arises because the Kaluza-Klein modes can be
extremely light, $m_{KK} \gtrsim 1/L \gtrsim 10^{-4}$ eV, and
extremely numerous, $N_{KK} \sim M_{pl}^2/M_D^2 \lesssim
10^{32}$. This means that even though each mode is only very weakly
coupled, with strength $1/M_{pl}$, scattering processes along the
brane,  at energies $E \gtrsim m_{KK}$, can produce a copious number
of KK gravitons.

The strongest astrophysical constraint comes from  the possible
emission of KK modes during the collapse of SN1987a. Requiring this to
not be the dominant cooling processes imposes a lower bound  on the
fundamental Planck scale.  For example, with $n=2$ we have $M_D \geq
50$ TeV, whereas for $n=3$ we have $M_D \geq 3$ TeV \cite{sn1987a}.

In cosmology, one has to worry about over-production of KK modes at
high  temperatures, since this may destroy the standard Big Bang
picture.  In order to be consistent with Big Bang Nucleosynthesis, and
the current composition of the Universe, one must identify a maximum
temperature for the early Universe for a given fundamental
scale. Taking the fundamental scale  to be $M_D \sim 1$ TeV imposes a
temperature bound $T \leq 10$ MeV for $n=2$, rising to $T \leq 10$ GeV
for $n=6$ \cite{hall}.  A higher fundamental scale will raise the
maximum temperature, but then one loses much of  the appeal  of the
original model.  While a low maximum temperature is not in
contradiction with cosmological data, it does present a challenge to
models of baryogenesis and inflation. The temperature bounds can be
weakened considerably if one does not require the bulk to be flat. For
example, when the bulk manifold is a compact hyperbolic space, the
maximum temperature  can be pushed beyond the GeV scale even for $n=2$
\cite{nemanja}.

\subsection{Randall-Sundrum Gravity}
As we have already suggested, in a generic braneworld set-up there is
no obvious reason why one should demand that the bulk space should be
flat, as in the ADD model. In perhaps the most celebrated braneworld
model,  developed soon after ADD by Randall and Sundrum \cite{rs1,
  rs2}, the bulk is an anti-de Sitter space.  There are  two versions
of the Randall-Sundrum model, generally referred to as Randall-Sundrum
I (RS1) \cite{rs1} and Randall-Sundrum II (RS2) \cite{rs2}. Somewhat
confusingly, the RS1 model contains two branes, whereas the RS2 model
only contains a single brane.

The RS1 model \cite{rs1} was also proposed as a resolution to the
hierarchy problem. It improves on the ADD model as the compact extra
dimension need not be so large as to introduce a new hierarchy.  This
is achieved by exploiting the exponential warp factor to generate a
large bulk  volume from a small compactification radius.  In contrast,
the RS2 model \cite{rs2} contains a single brane and a non-compact
extra dimension -- it is infinite in extent. This time the bulk warp
factor ensures that gravity is localised close to the brane, so that a
brane observer only sees the gravitational effects of the extra
dimension above scales set by the bulk curvature.

Although the phrase {\it Randall-Sundrum  gravity} really refers to
these two original models, here we extend the definition to include
any model with similar features. In this section we are particularly
interested in five-dimensional models containing $3$-branes, with some
non-trivial geometry , or ``warping", present in the bulk.    We begin
with an overview of some of the models. For further details  see, for
example,  \cite{thesis, livingrev}.

\subsubsection{The RS1 model}
\label{sec:rs1}

\begin{figure}
\begin{center}
\epsfig{file=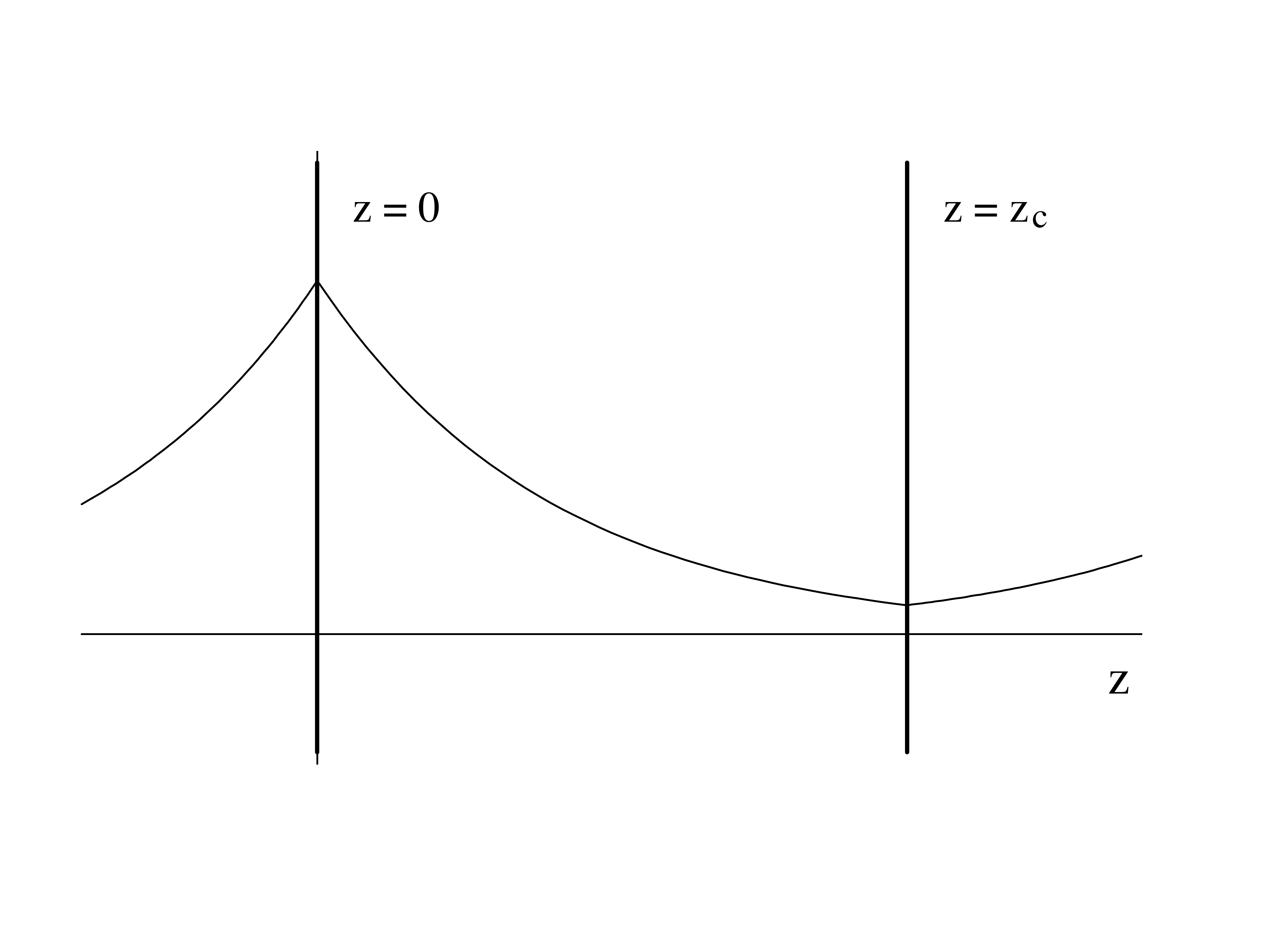,width=8cm,}
\caption{The behaviour of the warp factor in the RS1 model.} \label{fig:RS1}
\end{center}
\end{figure}

In RS1, we have two 3-branes separated by a region of five-dimensional
anti-de Sitter space \cite{rs1}. The branes are located at $z=0$, and
$z=z_c$, and we impose $\mathbb{Z}_2$ symmetry across each
brane. Neglecting Gibbons-Hawking boundary terms \cite{GH}, the action
describing this model is given by
\begin{multline} \label{eqn:RS1action}
S = \frac{M_5^3}{2}\int d^4x \int_{-z_c}^{z_c} dz \sqrt{-\gamma}\left(\R - 2\Lambda
\right)~\\
 -\sigma_+ \int_{z=0}  d^4x \sqrt{-g^{(+)}} ~ -\sigma_- \int_{z=z_c}
d^4x \sqrt{-g^{(-)}}~,
\end{multline}
where $\g_{ab}$ is the bulk metric, and  $g^{(+)}_{\mu\nu}$ and
$g^{(-)}_{\mu\nu}$ are the metrics on the branes at $z=0$ and $z=z_c$,
respectively.  $M_5$ is  the five-dimensional Planck scale and is
related to the five-dimensional Newton's constant via the standard
relation $G_5=1/8\pi M_5^3$.   We also include a  {\it negative} bulk
cosmological constant, $\Lambda=-6/l^2$.  If we fine-tune the brane
tensions against $\Lambda$, such that
$$
\sigma_+=-\sigma_-=\frac{6M_5^3}{l}=\frac{3 }{4 \pi G_5l},
$$
then we admit a background solution in which the branes exhibit
four-dimensional   Poincar\'e  invariance:
\begin{equation} \label{eqn:RS1metric}
ds^2=e^{-2|z|/l}\eta_{\mu\nu}dx^{\mu}dx^{\nu}+dz^2,
\end{equation}
for $-z_c \leq z \leq z_c$.  The $\mathbb{Z}_2$ symmetry about $z=0$
is explicit, whereas the other boundary condition should be understood
to be implicit. The line-element given in Eq. (\ref{eqn:RS1metric})
contains an exponential warp factor that is displayed graphically in
Figure~\ref{fig:RS1}.
In between the branes we recognise the geometry to be anti-de Sitter
space, written in Poincar\'e coordinates. Notice the peak in the warp
factor at the positive tension brane, and the trough at the negative
tension brane. Although only a toy model, the RS1 set-up is well
motivated by a number of  string theory/super-gravity
constructions~\cite{Horava-Witten, LOSW, Duff:susyRS}.

By integrating out the $4D$ zero-mode we are able to derive  the $4D$
effective Planck scale  on a given brane \cite{rs1, thesis}:
\be \label{Mpm}
M_{\pm}^2=\pm M_5^3 l \left(1-e^{\mp 2z_c/l}\right),
\ee
where $\pm$ labels the sign of the corresponding brane tension. In
terms of the effective Newton's constants we have
\be
G_\pm=\frac{G_5}{l} \left(\frac{ \pm 1}{1-e^{\mp 2z_c/l}}\right).
\ee
Now suppose we live on the {\it negative} tension brane. If we take
the fundamental Planck scale $M_5 \sim$ TeV, the  bulk curvature scale
to be just below $1/l \sim 0.01 M_5$, and the distance between the
branes to be such that  $z_c/l \sim 35$, we recover the desired
effective Planck scale, $M_- \sim m_{pl} \sim 10^{16}$ TeV. Thus, the
hierarchy problem has been eliminated altogether, and not just shifted
around, as in the ADD model. In contrast, the hierarchy is not
eliminated if we live on the positive tension brane   since then
the effective Planck mass  is given by $M_+ \sim e^{-z_c/l} M_-$
\cite{rs1,thesis}.

As it stands, the RS1 model is incomplete. The problem is that on
either brane the low energy $4D$ effective theory is not GR, but
Brans-Dicke gravity. The extra scalar comes from fluctuations in the
brane separation, and is sometimes referred to as the {\it radion}
\cite{GW1,  garriga, radion}. The value of the Brans-Dicke parameter
depends on the brane, and  is given by  \cite{garriga}
\be \label{wpm}
w_{BD}^{(\pm)} =\frac{3}{2}  \left(e^{\pm 2z_c/l}-1\right).
\ee
Observations require this parameter to be large ($w_{BD}>40000$, see
Section \ref{bdgravitysection}). Note that for the positive tension
brane $w_{BD}^{(+)}$ can be made arbitrarily  large with increasing
brane separation. The same cannot be said for $w_{BD}^{(-)}$ on the
negative tension brane. If we want to live on the negative tension
brane we must   generate a mass for the radion to suppress its
fluctuations. The Goldberger-Wise mechanism does exactly that, and
thus stabilises the distance between the branes \cite{GW2}.

\subsubsection{The RS2 model} 
\label{sec:rs2}

\begin{figure}
\begin{center}
\epsfig{file=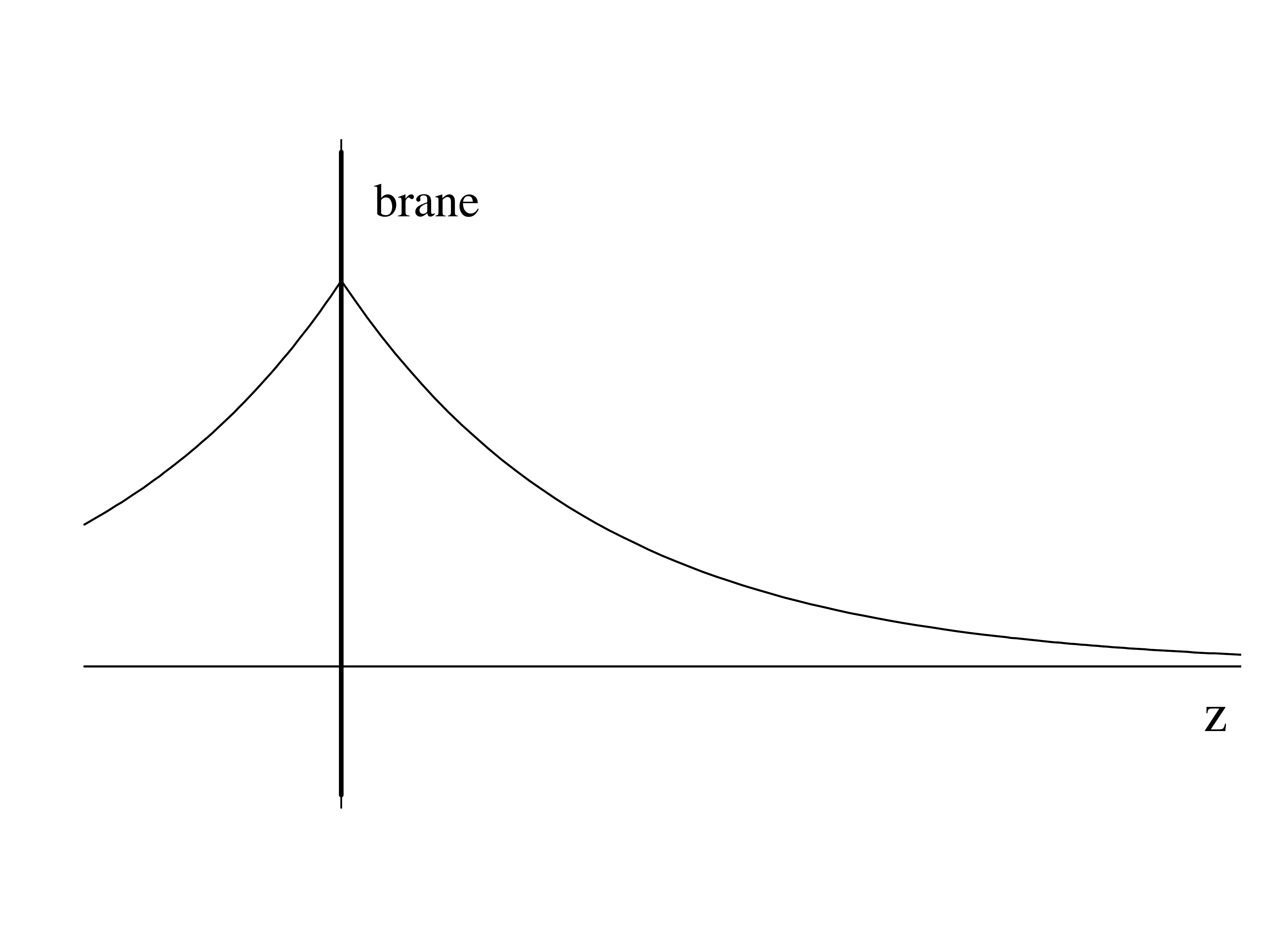,width=8cm,}
\caption{The behaviour of the warp factor in the RS2 model.} \label{fig:RS2}
\end{center}
\end{figure}

The RS2 model is obtained from RS1 by taking the negative tension
brane off to infinity \cite{rs2}. The geometry is then described by
the metric in Eq. (\ref{eqn:RS1metric}) with $z_c \to \infty$.  The
corresponding warp factor is shown in Figure \ref{fig:RS2}, the single
peak at $z=0$ indicating that we have a single brane with positive
tension. In this limit $M_+ \to M_5^3l$, so we cannot eliminate the
hierarchy problem as in the RS1 model. Rather, the situation is more
akin to the  five-dimensional ADD model, with the curvature scale,
$1/l$, playing the role of the compactification scale, $\mu \sim 1/L$. 

What makes the RS2 model interesting is the way in which $4D$ GR is
recovered on the brane. As we have just seen, in RS1 the observer on
the positive tension brane sees a low energy gravity theory
corresponding to Brans-Dicke gravity, with a BD parameter
$w_{BD}^{(+)}=\frac{3}{2}  \left(e^{2z_c/l}-1\right)$. In the RS2
limit of infinite brane separation the BD scalar decouples and one is
left with $4D$ GR.  As we will see in detail in Section
\ref{sec:RSperts}, even though the bulk is infinite in extent, gravity
is localised on the brane at energies below the bulk curvature scale,
$1/l$.  As a method to screen the extra dimensions from the low energy
observer, this represents a radical alternative to the standard method
of Kaluza-Klein compactifications.

The key to gravity localisation is that the bulk volume is finite even
though it has infinite extent. This ensures that there is a
normalisable graviton zero mode, which, in the absence of any other
massless modes, guarantees $4D$ GR at low enough energies.  The finite
bulk volume arises because  the warp factor falls off exponentially as
we move away from the brane. Intuitively, gravity localisation occurs
because the warping makes it difficult for the graviton to propagate
too far away from the brane, so much so that the region of the bulk
with $z \gtrsim l$ has no influence on low energy brane interactions.

Since $1/l$ sets the scale at which the brane observer starts to
become sensitive to the bulk,  table top experiments of the inverse
square law  impose the limit $1/l\gtrsim 10^{-4}$ eV. This translates
into a lower bound on the fundamental Planck scale, $M_5 \gtrsim 10^5$
TeV, which is well above the electro-weak scale.  Since TeV scale
gravity is not  phenomenologically viable in this case,  we  abandon
any discussion of the hierarchy problem for the single brane scenario.

Up until now our discussion has centred around weak gravity on the
brane. What about strong gravity in the presence of localised sources?
Whilst there have been some interesting numerical studies (see, e.g.,
\cite{rsstars, smallbhs}) exact strong gravity solutions are rare in
RS2. One exception are the solutions for a domain wall localised on
the brane \cite{bwdw, bwinst}. In contrast, an exact solution
describing a braneworld black hole remains elusive\footnote{There have
  been many attempts to find such a solution, most of which remain
  unpublished.}. The difficulty arises because the brane is an
accelerated surface, so any black hole residing on the brane must
follow an accelerated trajectory. Such an accelerated black hole
requires knowledge of the AdS C-metric, but this  solution is unknown
in five dimensions. For a nice review of the search for braneworld
black hole solutions see \cite{ruthbwbh}, and more recently
\cite{Nori-review}.
\newline
\newline
\noindent
{\it RS2 and AdS/CFT}
\newline

Emparan, Fabbri and Kaloper (EFK) have suggested that a static
braneworld black hole does not exist \cite{efk}. To understand their
argument we must first  recall the {\it AdS/CFT correspondence} in
which type IIB string theory on $AdS_5 \times S^5$ is conjectured to
be dual to ${\cal N}=4$ $SU(N)$ super-Yang-Mills, in the large $N$
limit \cite{maldacena}. This suggests an alternative description of
RS2 gravity \cite{gubser}: `{\it Gravity on an RS2 brane is dual to a
  strongly coupled conformal field theory (CFT)  cut off in the UV,
  minimally coupled to $4D$ gravity}'. There is plenty of evidence for
this  holographic description of  RS2 (see, for example, \cite{gubser,
  savonije, duff-liu, non-crit, exactbwh, covbwh, Hawking-brane-new,
  ads/cft-nohair, Kiritsis2}, and for a review \cite{thesis}). Another way of
describing the correspondence  is to say \cite{efk}: `{\it A
  classical source on the brane is dual to a quantum corrected source
  in four dimensions, with the quantum corrections coming from the
  strongly coupled CFT}'.  The quantum corrections are large because
of the large number of degrees of freedom in the large-$N$ limit. When
applied to the problem of finding a black hole  on the brane, this
suggests that the solution should not be static since it should
include the back-reaction of the  Hawking radiation \cite{efk} (see
also \cite{tanaka-bhevap}).  Fitzpatrick, Randall and Wiseman have
disputed this interpretation, pointing out that the CFT is strongly
coupled and may therefore carry fewer degrees of freedom
\cite{frw}. At this point it is fair to say that as yet there has been
no consensus, and the subject remains an active area of debate (see
also \cite{dwpujolas, robinbwbh, bwiso, toby1, toby2}).

Note that if we accept the EFK conjecture,  we can improve the bound
on the bulk curvature by an order of magnitude,  $1/l \gtrsim 10^{-3}$
eV. This is  based on the existence of long-lived  black hole X-ray
binaries  \cite{egk}. For smaller values of $1/l$ these binaries would
have already decayed.

\subsubsection{Other RS-like models}
\label{sec:otherRS}

One of the characteristic features of the Randall-Sundrum models is
the structure of the bulk geometry, described by a non-factorisable,
or warped, metric. One can embrace this structure and consider a whole
slew of interesting generalisations. Here we consider a class of
models described by  warped geometries of the form
\be
ds^2=a^2(z)\bar g_{\mu\nu}(x) dx^{\mu}dx^{\nu}+dz^2.
\ee
Many of the most interesting RS-like models exhibit  {\it
  quasi-localisation} and give rise to large-distance modifications of
gravity. Perhaps the most celebrated of these is the DGP model
\cite{dgp}, which will be discussed in detail in Section
\ref{sec:dgp}. Other interesting examples include the GRS model
\cite{GRS}, the asymmetric brane model \cite{asymm1, asymm2} and the
CGP model \cite{stealth}. We begin with the simplest generalisation,
however, proposed by Karch and Randall \cite{kr} (see also
\cite{Kaloper-bent}).
\newline
\newline
\noindent
{\it The Karch-Randall model}
\newline

Here we take  the RS2 model and de-tune the brane tension, $\sigma_+
\neq 6M_5^3 /l$, so that one no longer has Poincar\'e invariance along
the brane. For an excess tension, $\sigma_+>6M_5^3 /l$, the metric
$\bar g_{\mu\nu}$ is de Sitter, whereas for a tension deficit,
$\sigma_+<6M_5^3 /l$, $\bar g_{\mu\nu}$ is anti-de Sitter.  The
effective cosmological constant on the brane is given by
\be
\Lambda_4=3 \left[\left(\frac{\sigma_+}{6
    M_5^3}\right)^2-\frac{1}{l^2}\right].
\ee
The behaviour of the warp factor also changes. We find that 
\begin{align}
&a(z)=A\cosh (c-|z|/l), & c=\cosh^{-1}(1/A), && \textrm{for anti-de Sitter}, \\
&a(z)=A\sinh (c-|z|/l), & c=\sinh^{-1}(1/A), & &\textrm{for de Sitter},
\end{align}
where $A=l\sqrt{|\Lambda_4/3|}$. The decay of the warp factor away
from the brane is greatest for the dS brane, cutting off the
space-time at $z=lc$. This means that   gravity is more strongly
localised than in the standard RS2 scenario.   In contrast, for the
AdS brane the warp factor turns around at some finite value of
$z$. This means the bulk volume is infinite, and gravity is not
localised at all on the AdS brane as there is no normalisable zero
mode. Actually, when $|\Lambda_4|$ is small (compared with $M_5$), the
AdS brane exhibits quasi-localisation. This is because there is a
normalisable mode that is {\it ultra-light}, with mass
$m^2_{ultra-light} \sim \Lambda_4^2/M_5  \ll | \Lambda_4|$
\cite{schwartz}. At intermediate energies, $k \gg E \gg
m_{ultra-light}$, the light mode behaves as if it were effectively
massless and one recovers $4D$ gravity. Note that there is no issue
with the vDVZ discontinuity in AdS when
$m_{ultra-light}^2/||\Lambda_4| \lesssim 0.1 $\cite{novdvz}.  We say
that gravity is only {\it quasi}-localised because the extra-dimension
opens up in the far infra-red, at energies $E \lesssim
m_{ultra-light}$.

Now consider what happens when we introduce a second AdS brane, along
the lines of an AdS generalisation of the RS1 model
\cite{bigravity}. The first thing to note is that both branes can have
positive tension since the warp factor turns around. The second thing
to note is that the bulk volume is rendered finite, and so we have a
zero mode as well as the ultra-light mode.  At  energies above the
mass of both modes gravity is mediated by the exchange of two spin-2
fields, one massless and one massive. This corresponds to a braneworld
realisation of the bigravity scenarios discussed in Section
\ref{bigravitysection} (see also \cite{2000, multi-branes,
  bigravity-me}).
\newline
\newline
\noindent
{\it The GRS model}
\newline

The GRS model \cite{GRS} was developed by Gregory, Rubakov and
Sibiryakov. The set-up contains  three Minkowski branes, one with
positive tension and two with negative tension. The positive tension
brane is $\mathbb{Z}_2$ symmetric and is flanked on either side by a
section of anti-de Sitter space as far as a negative tension
brane. Beyond the negative tension brane lies an infinite region of
Minkowski space. The warp factor goes like \cite{GRS}
\be
a(z)=\begin{cases} e^{-|z|/l} \qquad & |z|<z_c \\
e^{-z_c/l} \qquad & |z|>z_c.
\end{cases}
\ee
As the bulk volume is infinite, gravity is not localised, although the
decaying warp factor around the positive tension brane gives some
degree of quasi-localisation. Again, although there is no zero mode,
there is an ultra-light mode, and the extra dimension opens up at very
large distances. Unfortunately, the GRS model is known to be unstable
due to the presence of a ghost in the spectrum of linearised
fluctuations \cite{pilo}.
\newline
\newline
\noindent
{\it The asymmetric brane model}
\newline

The asymmetric brane model \cite{asymm1, asymm2} (see also
\cite{Stoica-cosmo}) is a single brane mode,l like RS2, only without
$\mathbb{Z}_2$ symmetry imposed across the brane. Indeed, the
fundamental parameters in the bulk are allowed to differ on either
side of the brane, including the bulk cosmological constant and the
bulk Planck scales. Allowing the bulk Planck scales to differ might
seem strange, but not if we imagine a string compactification down to
five dimensions in which the dilaton is stabilised at different values
on either side of a domain wall (the brane).  If the bulk cosmological
constant and Planck scales are given by 
\be
\Lambda=\begin{cases} -6/l_1^2 \qquad & z>0 \\ -6/l_2^2 \qquad & z<0
\end{cases}, \qquad \qquad M=\begin{cases} M_1 \qquad & z>0 \\ M_2
\qquad & z<0. \end{cases}
\ee
The asymmetric model then admits Minkowski branes for a suitably tuned
brane tension, $\sigma=3(\epsilon_1 M_1^3 /l_1+\epsilon_2 M_2^3
/l_2)$, where $\epsilon_{1}=\pm1$ and $\epsilon _2=\pm 1$. The
corresponding solutions have a warp factor of the form \cite{asymm2}
\be
a(z)=\begin{cases} e^{-\epsilon_1 z/l_1} \qquad & z>0 \\
e^{\epsilon_2  z/l_2} \qquad & z<0.
\end{cases}
\ee
The parameters $\epsilon_1$ and $\epsilon_2$ control whether the warp
factor grows ($\epsilon=-1$) or decays  ($\epsilon=-1$) away from the
brane in a given direction.  The model includes RS2 as a special case.

It is, however, more interesting to consider the case where one of the
warp factors grows away from the brane while the other decays
(e.g. $\epsilon_2=-\epsilon_1=1$). The bulk volume is then infinite so
that there is no zero mode, but by choosing the scales appropriately
one can engineer a degree of quasi-localisation.  The point is that on
the growing side the graviton localises at the AdS boundary, where its
sees an effective $4D$ Planck scale $M_1^3 l_1$. By taking this scale
to be very large, the effect of localisation close to the AdS boundary
is almost decoupled from the gravitational dynamics near the brane. We
should  note that the asymmetric  model also admits self-accelerating
solutions, just as in the DGP model. In fact, the model shares a
number of features with the DGP model. This is no coincidence, since
the DGP model can be obtained as a limiting case of the asymmetric
model \cite{koyama-koyama}. It is known that the self-accelerating
solutions of DGP contain ghost-like instabilities \cite{dgpspec,
  dgpspecrev, dgppersp, koyama-more}, so the same is expected to be
true for the asymmetric model.
\newline
\newline
\noindent
{\it The CGP model}
\newline

The CGP model \cite{stealth}, developed by Charmousis, Gregory and
Padilla, also exhibits quasi-localisation. It combines the main
features of both the asymmetric model and the DGP model in that there
is an induced gravity term, and asymmetry across the
brane. Interestingly, the model contains a new type of cosmological
solution that tends to Minkowski space at very late times, but
undergoes an intermediate period of cosmic acceleration  in the
presence of ordinary matter. In fact, it corresponds to a braneworld
realisation of the Cardassian cosmology, with $H^2 \approx \frac{8 \pi
  G}{3}( \rho+c\sqrt{\rho})$ \cite{cardassian}. Although a ghost is
present when we introduce a small positive vacuum energy, this
decouples in the Minkowski limit \cite{stealth-ghosts}.

\subsubsection{Action and equations of motion}
\label{sec:genaction}

In each of the models described in our overview, we have a five
dimensional bulk space split into a series of domains separated from
each another  by 3-branes.   Here we consider the action and field
equations for generic models of this type. The $3$-branes may be
thought of as the boundaries of the various domains so that the action
is given by 
\be\label{gen-action}
S =  \int_{\textrm{bulk}} d^5 x \sqrt{-\gamma}\left(\frac{M_5^3}{2} \R +{\cal
  L}_{\textrm{bulk}}\right)+\sum_{\textrm{branes}} \int_{\textrm{brane}} d^4 x \sqrt{-g}\left[-
  \Delta  \left( M_5^3 K\right)+{\cal L}_{\textrm{brane}}\right],
\ee
where $\g_{ab}$ is the bulk metric with corresponding Ricci scalar,
$\R$, $M_5$ is the bulk Planck scale, and ${\cal L}_{\textrm{bulk}}$ is the
Lagrangian density describing the bulk field content. In principle
both $M_5$ and ${\cal L}_{\textrm{bulk}}$  can vary from domain to domain. For
each  brane $g_{\mu\nu}$ is the induced metric and ${\cal L}_{\textrm{brane}}$
is the Lagrangian density describing the field content on that
particular brane. $K=g^{\mu\nu} K_{\mu\nu}$ is the trace of extrinsic
curvature, $K_{\mu\nu}$. This  should be evaluated on either side of
the brane as it can differ from side to side. Labelling the two sides
of a given brane, using $L$ and $R$, we define $K_{\mu\nu}|_{L,
  R}=\frac{1}{2} {\cal L}_{n|_{ L, R}} g_{\mu\nu} $, i.e. extrinsic
curvature is given by the Lie derivative of the induced metric, with
respect to the unit normal $n^a|_{L, R}$. The unit normal on  both
sides points from $L$ to $R$. Note that what appears in the action is
the jump\footnote{Henceforth we define the jump of any quantity $Q$
  across a brane as $\Delta Q=Q|_R-Q|_L$.}
$$
\Delta \left(M_5^3 K\right)=  M_5^3 K|_R- M_5^3 K|_L.
$$
This corresponds to the Gibbons-Hawking boundary term \cite{GH} for
the bulk domains on each side of the brane.  The $\Delta$ here is not
to be confused with the 3-dimensional Laplacian.

Now there are two (completely equivalent) ways to treat the brane
contributions at the level of the field equations. One approach is to
treat them as delta-function sources in the Einstein
equations. However, our preferred approach is to explicitly separate
the field equations in the bulk from the boundary conditions at the
brane. The  bulk equations of motion are then given by the bulk Einstein equations
\be
{\cal G}_{ab}=\R_{ab}-\half \R\g_{ab}=\frac{1}{M_5^3} T^{\textrm{bulk}}_{ab},
\ee
where $T^{\textrm{bulk}}_{ab}=-\frac{2}{\sqrt{-\g}} \frac{\delta }{\delta
  \g^{ab}} \int_{\textrm{bulk}} d^5 x \sqrt{-\g} {\cal L}_{\textrm{bulk}}$ is the bulk
energy-momentum tensor. The boundary conditions at $\Sigma_i$ are
given by the Israel junction conditions \cite{Israel}
\be \label{israel}
\Delta\left[M_5^2(K_{\mu\nu}-K g_{\mu\nu})\right]=-T^{\textrm{brane}}_{\mu\nu},
\ee
where  $T^{(i)}_{\mu\nu}=-\frac{2}{\sqrt{-g}} \frac{\delta }{\delta
  g^{\mu\nu}} \int_{\textrm{brane}} d^4 x \sqrt{-g} {\cal L}_{\textrm{brane}}$ is the
brane energy-momentum tensor.  Note that in each of these examples,
the bulk geometry is only sourced by a cosmological constant,
$T^{\textrm{bulk}}_{ab}=-M_5^3 \Lambda \g_{ab}$.

\subsubsection{Linear perturbations in RS1 and RS2}
\label{sec:RSperts}

We now consider the theory of linear perturbations. For brevity, we
will  restrict attention to RS1 and RS2, although the methods we use
are fairly standard, and should apply to all RS-like models (for
further details, see \cite{garriga, radion, pilo, bigravity-me}). 
\newline
\newline
\noindent
{\it Weak gravity on a RS1 brane}
\newline

It is enough to consider RS1, as RS2 can be readily obtained by taking
the negative tension brane to infinity. Recall that the background
metric, $\bar \g_{ab}$, is given by Equation (\ref{eqn:RS1metric}),
with the positive tension  brane (the ``$+$" brane) fixed at $z=0$ and
the negative tension  brane  (the ``$-$" brane) fixed at  $z=z_c$.  We
see that the induced metric on the ``$+$" brane is given by $\bar
g_{\mu\nu}^{(+)}=\eta_{\mu\nu}$, and on the   ``$-$" brane by $\bar
g_{\mu\nu}^{(-)}=e^{-2z_c/l} \eta_{\mu\nu}$. Note that we have $
\mathbb{Z}_2$ symmetry across the branes, so we can restrict attention
to  $0 \leq z \leq z_c$.

We now consider  small perturbations about the background, so that the
metric is given by $\g_{ab}=\bar \g_{ab}+\delta \g_{ab}$. It is
convenient to  choose  Gaussian Normal (GN) gauge, defined by
\be \label{eqn:GN}
\delta \g_{\mu z}=\delta \g_{zz}=0.
\ee
Actually, this is only a partial gauge fixing. Since we have no
additional bulk matter, we can also take the metric to be transverse
and trace-free {\it in the bulk}. In other words, $\delta
\g_{\mu\nu}=\chi_{\mu\nu}(x, z)$, where
\be
 \del_\nu \chi^{\nu}_{\mu}=\chi^{\mu}_{\mu}=0.
\ee
This is known as Randall-Sundrum (RS) gauge \cite{rs2}. In RS gauge,
the linearised bulk equations of motion, $\delta {\cal
  G}_{ab}=\frac{6~}{l^2} \delta \g_{ab}$, yield
\be \label{eqn:linbulk}
\left[ e^{2z/l}\del^2   + \frac{\del^2}{\del
z^2}-\frac{4}{l^2}\right] \chi_{\mu\nu}=0,
\ee
where $\del^2=\del_\mu \del^\mu$.

Unfortunately, we can no longer assume that the branes
are fixed at $z=0$  and $z= z_c$. The presence of matter on the branes
will cause them to bend~\cite{garriga} so that they  will now be
positioned at $z=f_+(x)$ and $z=z_c+f_-(x)$, for some functions $f_\pm$ that depend
only on the coordinates $x^{\mu}$. This makes it difficult to apply the Israel
junction conditions at the branes. To get round this we can apply a gauge
transformation that fixes the position of the  ``$+$"  brane, and
another that fixes  the position of the ``$-$"  brane~\cite{radion},
without spoiling the Gaussian Normal condition~(\ref{eqn:GN}). This gives rise to two
coordinate patches that are related by a gauge transformation in the
region of overlap. We will call them the ``$+$" patch and the ``$-$" patch accordingly.

We first consider the ``+" brane. To fix its position we make the
following coordinate transformation 
\be
z \to z-f_+(x), \qquad x_\mu \to x_\mu +\frac{l}{2} (1-e^{-2z/l})\del_\mu f_+.
\ee
The  ``$+$"  brane is now fixed at $z=0$, although the other brane is
now  at $z=z_c+f_--f_+$. It follows that the metric perturbation in
the ``$+$'' patch is given by
\be \label{g+}
\delta \g_{\mu\nu}=\chi^{(+)}_{\mu\nu}(x, z)=\chi_{\mu\nu}(x, z)-l(1-e^{-2z/l}) 
\del_\mu \del_\nu f_+
-\frac{2}{l}f_+\bar \g_{\mu\nu}.
\ee
Similarly, to fix the position of  the ``$-$" brane we let
\be
z \to z-f_-(x), \qquad  x_\mu \to x_\mu +\frac{l}{2}
\left(1-e^{-2(z-z_c)/l}\right)\del_\mu f_-.
\ee
Now we have  the ``$-$"  brane at $z=z_c$, but with the  ``$+$"  brane
at $z=f_+-f_-$.  The metric perturbation in the ``$-$'' patch is given
by
\be \label{g-}
\delta \g_{\mu\nu}=\chi^{(-)}_{\mu\nu}(x, z)=\chi_{\mu\nu}(x,
z)-l\left(1-e^{-2(z-z_c)/l}\right) \del_\mu \del_\nu f_-
-\frac{2}{l}f_-\bar \g_{\mu\nu}.
\ee
Now the induced metric on the ``$+$" brane is given by
$g_{\mu\nu}^{(+)}=\bar g_{\mu\nu}^{(+)}+\delta g_{\mu\nu}^{(+)}$,
where
\be
\delta g_{\mu\nu}^{(+)}=\chi^{(+)}_{\mu\nu} (x, 0)=\chi_{\mu\nu}(x, 0)
-\frac{2}{l} f_+\bar g_{\mu\nu}^{(+)},
\ee
whereas on the ``$-$" brane it is given by $g_{\mu\nu}^{(-)}=\bar
g_{\mu\nu}^{(-)}+\delta g_{\mu\nu}^{(-)}$, where
\be
\delta g_{\mu\nu}^{(-)}=\chi^{(-)}_{\mu\nu}(x, z_c)=\chi_{\mu\nu}(x,
z_c)-\frac{2}{l}f_-\bar g_{\mu\nu}^{(-)}.
\ee
We are now ready to make use of the linearised Israel junction
conditions given by Eq. (\ref{israel}) at each brane to find
\be
\Delta\left[M_5^2\delta (K_{\mu\nu}-K
  g_{\mu\nu})\right]^{(\pm)}=\sigma_\pm  \delta
g^{(\pm)}_{\mu\nu}-{\cal T}^{(\pm)}_{\mu\nu},
\ee
where $\sigma_\pm=\pm 6M_5^3/l $ is the tension on the ``$\pm$'
brane, and ${\cal T}^{(\pm)}_{\mu\nu}$ is the energy-momentum tensor
for matter excitations. Now, owing to the $\mathbb{Z}_2$ symmetry,
the extrinsic curvature simply changes by a sign when evaluated on
either side of a given brane. It follows that the linearised boundary
conditions at each brane are given by
\ba 
\left( \frac{\del}{\del
z}+\frac{2}{l} \right)\chi_{\mu\nu}\Big |_{z=0} &=& -S_{\mu\nu}^{(+)},  \label{eqn:linbc1} \\
\left( \frac{\del}{\del
z}+\frac{2}{l} \right)\chi_{\mu\nu}\Big |_{z=z_c} &=& -S_{\mu\nu}^{(-)}, \label{eqn:linbc2} 
\ea
where 
\be
S^{\pm}_{\mu\nu}(x)=\pm \frac{1}{M_5^3}\left[
  \mathcal{T}^{(\pm)}_{\mu\nu}-\frac{1}{3}\mathcal{T}^{(\pm)} \bar
  g_{\mu\nu}^{(\pm)}\right]  -2\del_\mu \del_\nu f_\pm,
\ee
and where ${\cal T}^{(\pm)}=g_{(\pm)}^{\mu\nu}
\mathcal{T}^{(\pm)}_{\mu\nu}$ is the trace of the appropriate
energy-momentum tensor.  Indeed, taking the trace of these equations,
and using the fact that $\chi_{\mu\nu}$ is traceless, we clearly see
that matter on a brane causes it to be bend, such that
\be \label{feqn}
\frac{\del^2 f_\pm}{a_\pm^2}=\mp \frac{{\cal T}^{(\pm)}}{6M_5^3},
\ee
where $a_\pm$ gives the warp factor at the ``$\pm$" brane
(i.e. $a_+=1$ and $a_-=e^{-z_c/l}$). Equations (\ref{eqn:linbulk}),
(\ref{eqn:linbc1}) and (\ref{eqn:linbc2}) give the governing
differential equations, and a complete set of boundary conditions for
the graviton mode $\chi_{\mu\nu}$.  We now  take Fourier transforms
along the brane directions, $Q(x, \ldots) \to \tilde Q(p,
\ldots)=\frac{1}{(2\pi)^2}\int d^4 x~  e^{-i p_\mu x^\mu} Q(x,
\ldots)$, to find that
\be
\chi_{\mu\nu}(x, z)=\frac{1}{(2\pi)^2} \int d^4 p~ e^{i p_\mu x^\mu}
\tilde \chi_{\mu\nu}(p, z),
\ee
where
\be
\left[ -p^2 e^{2z/l}   + \frac{\del^2}{\del
z^2}-\frac{4}{l^2}\right] \tilde \chi_{\mu\nu}=0,
\ee
and
\be
\left(\frac{\del}{\del
z}+\frac{2}{l} \right)\tilde \chi_{\mu\nu}\Big |_{z=0} = -\tilde
S_{\mu\nu}^{(+)}(p), \qquad \left(\frac{\del}{\del 
z}+\frac{2}{l} \right)\tilde \chi_{\mu\nu}\Big |_{z=z_c} =-\tilde S_{\mu\nu}^{(-)}(p).
\ee
This system is easily solved to give
\be
\tilde  \chi_{\mu\nu}(p, z)=C^{(+)}(p, z) \tilde
S_{\mu\nu}^{(+)}-C^{(-)}(p, z) \tilde S_{\mu\nu}^{(-)},
\ee
where 
\be
C^{(+)}(p, z) =-\frac{1}{p \det
   A(p)}\left[{I_1\left(ple^{z_c/l}\right)  K_2
     \left(ple^{z/l}\right)   +  K_1\left(ple^{z_c/l}\right)  I_2
     \left(ple^{z/l}\right)  }\right],
\ee
and
\be
 C^{(-)}(p, z) = -\frac{1}{p \det
   A(p)}e^{-z_c/l}\left[{I_1\left(pl\right)  K_2
     \left(ple^{z/l}\right)   +  K_1\left(pl\right)  I_2
     \left(ple^{z/l}\right) }\right].
 \ee
Note that $I_n,$ and $K_n$ are modified Bessel's functions of integer
order $n$ \cite{Abram-Stegun}, and
\be 
\det
A(p)=I_1\left(pl\right)K_1\left(ple^{z_c/l}\right)-K_1\left(pl\right)I_1\left(ple^{z_c/l}\right).
\ee
From Eq. (\ref{feqn}) we also have 
\be
f_\pm (x)=\pm \frac{1}{(2\pi)^2} \frac{a_\pm^2}{M_5^3} \int d^4 p~
e^{i p_\mu x^\mu} \frac{\tilde {\cal T}^{(\pm)}}{6p^2}.
\ee
To compute the metric perturbation on each of the branes we simply use
Eqs.(\ref{g+}) and (\ref{g-}), given our knowledge of
$\chi_{\mu\nu}(x, z)$ and $f_\pm(x)$. At the positive tension brane we
then have 
\begin{multline}
\chi^{(+)}_{\mu\nu}=\frac{1}{(2\pi)^2} \frac{1}{M_5^3} \int d^4 p~
 e^{ip_\mu x^\mu} \left\{C^{(+)}(p, 0) \left[\tilde {\cal
 T}^{(+)}_{\mu\nu}-\frac{\alpha_{+}}{2} \tilde {\cal
 T}^{(+)} \bar g_{\mu\nu}^{(+)}\right] \right.\\\left.
-\frac{e^{-z_c/l}}{lp^2 \det A(p)}\left[\tilde {\cal T}^{(-)}_{\mu\nu}-\frac{1}{3} \tilde {\cal
 T}^{(-)} \bar g_{\mu\nu}^{(-)}\right]\right\}+\textrm{pure gauge terms},
\end{multline}
whereas at the negative tension brane we have
\begin{multline}
\chi^{(-)}_{\mu\nu}=\frac{1}{(2\pi)^2} \frac{1}{M_5^3} \int d^4 p~
 e^{ip_\mu x^\mu} \left\{C^{(-)}(p, z_c) \left[\tilde {\cal
 T}^{(-)}_{\mu\nu}-\frac{\alpha_{-}}{2} \tilde {\cal
 T}^{(-)} \bar g_{\mu\nu}^{(-)}\right] \right.\\\left.
 -\frac{e^{-z_c/l}}{lp^2  \det A(p)} \left[\tilde {\cal T}^{(+)}_{\mu\nu}-\frac{1}{3} \tilde {\cal
 T}^{(+)} \bar g_{\mu\nu}^{(+)}\right]\right\}+\textrm{pure gauge terms}.
\end{multline}
The parameters $\alpha_\pm$  are crucial as they control the tensor
structure of the propagator on the ``$\pm$" branes. They are given by
\be
\alpha_+=\frac{2}{3}\left(1+\frac{1}{lp^2 C^{(+)}(p, 0)}\right),
\qquad \alpha_-=\frac{2}{3}\left(1-\frac{e^{-2z_c/l}}{lp^2 C^{(-)}(p,
  z_c)}\right).
\ee
Using the properties of modified Bessel functions \cite{Abram-Stegun},
we can show that at low energies, $p \ll ke^{-z_c/l}$, we have
\cite{garriga}
\begin{multline}
\chi^{(+)}_{\mu\nu}\approx \frac{1}{(2\pi)^2} \int d^4 p~ e^{ip_\mu
 x^\mu} \left\{  \frac{2}{M_+^2 p^2}\left[\tilde {\cal
 T}^{(+)}_{\mu\nu}-\left(\frac{w_{BD}^++1}{2w_{BD}^++3}\right) \tilde
 {\cal
 T}^{(+)} \bar g_{\mu\nu}^{(+)}\right] \right.\\\left.
+\frac{2}{M_-^2 p^2}\ \left[\tilde {\cal T}^{(-)}_{\mu\nu}-\frac{1}{3} \tilde {\cal
 T}^{(-)} \bar g_{\mu\nu}^{(-)}\right]\right\}+\textrm{pure gauge terms},
\end{multline}
and 
\begin{multline}
\chi^{(-)}_{\mu\nu}\approx \frac{e^{-2z_c/l}}{(2\pi)^2} \int d^4 p~
 e^{ip_\mu x^\mu} \left\{  \frac{2}{M_-^2 p^2}\left[\tilde {\cal
 T}^{(-)}_{\mu\nu}-\left(\frac{w_{BD}^-+1}{2w_{BD}^-+3}\right) \tilde
 {\cal T}^{(-)} \bar g_{\mu\nu}^{(-)}\right] \right.\\\left.
+\frac{2}{M_+^2 p^2}\ \left[\tilde {\cal T}^{(+)}_{\mu\nu}-\frac{1}{3} \tilde {\cal
 T}^{(+)} \bar g_{\mu\nu}^{(+)}\right]\right\}+\textrm{pure gauge terms},
\end{multline}
where $M_\pm$ and $w_{BD}^\pm$  are the $4D$ effective Planck scale
and Brans-Dicke parameter on the ``$\pm$" brane, respectively. They
are given by Equations (\ref{Mpm}) and (\ref{wpm}).  We can now see
explicitly how BD gravity emerges as the low energy effective theory
in RS1, on both branes, as claimed in Section \ref{sec:rs1}.
\newline
\newline
\noindent
{\it Graviton spectrum} 
\newline

Let us now pause to comment on the mass spectrum for the graviton.
The spectrum can be obtained by identifying the poles in the
propagator. These are given by the solutions to  $p^2=-m^2$, where
$\det A(p)=0$.  This gives 
\begin{multline}
I_1\left(iml\right)K_1\left(imle^{z_c/l}\right)-K_1\left(iml\right)I_1\left(imle^{z_c/l}\right)=0 \\
\implies
J_1\left(ml\right)Y_1\left(mle^{z_c/l}\right)-Y_1\left(ml\right)J_1\left(mle^{z_c/l}\right)=0,
\end{multline}
where $J_1$ and $Y_1$ are Bessel's functions of order one. By solving
this equation we see that there is a zero mode, and a tower of heavy
Kaluza-Klein modes with mass splitting $\Delta m \sim
1/l(e^{z_c/l}-1)$. 
\newline
\newline
\noindent
{\it Radion effective action}
\newline

The finite Brans-Dicke parameter indicates the presence of a massless
scalar in addition to the massless graviton. This is due to the
radion, or brane bending mode, $\delta f=f_+-f_-$. In some RS-like
models (e.g. GRS), the radion can exhibit pathological behaviour that
can only be revealed by computing the effective action \cite{pilo}.
We will now briefly outline the procedure for doing this.  

Let us consider vacuum fluctuations (${\cal T}^{(\pm)}_{\mu\nu}=0$) in
the scalar sector. The first thing to note is that the Field Equations
(\ref{eqn:linbulk}), (\ref{eqn:linbc1}) and (\ref{eqn:linbc2})  admit
a solution of the form \cite{radion}
\be
\chi_{\mu\nu}^{(\textrm{rad})}=-\frac{l^2}{4} e^{2z/l} \del_\mu \del_\nu \psi,
\ee
where $\del^2 \psi=0$ and the vacuum boundary conditions require
\be \label{fvac}
f^{\textrm{vac}}_\pm=-\frac{l}{2 a_\pm^2} \psi,
\ee
where $f_\pm^{\textrm{vac}}$ is the vacuum fluctuation in $f_\pm$. Note that
Equation (\ref{feqn}) implies $\del^2 f^{\textrm{vac}}_\pm=0$, and so Equation
(\ref{fvac}) is consistent with $\del^2 \psi=0$. Equation (\ref{fvac})
also imposes a relation between $f_+^{\textrm{vac}}$ and $f_-^{\textrm{vac}}$,
resulting in a single free scalar degree of freedom, which we take to
be the physical radion mode, $\psi$.  Note that the radion profile in
the bulk is localised close to the ``$-$'' brane, in contrast to the
graviton zero mode which is localised close to the ``$+$'' brane. We
now work in  the ``$+$"  patch, which has the branes positioned at
$z=0$ and $z=z_c-\delta f^{\textrm{vac}}$. Focusing solely on the scalar
sector, the metric perturbation is given by
\be \label{grad}
\delta \g_{\mu\nu}=-\frac{l^2}{4} e^{2z/l} \del_\mu \del_\nu \psi
-l (1-e^{-2z/l}) 
\del_\mu \del_\nu f_+^{\textrm{vac}}
-\frac{2}{l} f_+^{\textrm{vac}}\bar \g_{\mu\nu}, \qquad \delta  \g_{\mu z}= \delta  \g_{zz}=0.
\ee
In order to integrate out the extra
dimension, it is convenient to have {\it both} branes fixed. We can do
this with the following coordinate transformation
\be
z \to z+ B(z)\delta f^{\textrm{vac}} ,\qquad  x_{\mu} \to
x_{\mu}-e^{-2z/l}\del_{\mu} (\delta f^{\textrm{vac}}) \int^z_0e^{2y/l} B(y) dy,
\ee
where $B(z)$ is some differentiable function for $0 \leq z \leq z_c$,
satisfying $B(0)=0$ and $B(z_c)=1$. While this transformation ensures
that $\delta \gamma_{\mu z}$ is still zero, the price we pay for fixed branes
is that we now have non-vanishing $\delta \gamma_{zz}$. More precisely,
\begin{eqnarray}
&& \delta \g_{\mu\nu}=h_{\mu\nu}=-\frac{l^2}{4} e^{2z/l} \del_\mu \del_\nu \psi-l (1-e^{-2z/l}) 
\del_\mu \del_\nu f_+^{\textrm{vac}}-\frac{2}{l}\left[f_+^{\textrm{vac}}-(\delta
  f^{\textrm{vac}}) B(z)\right]\bar \g_{\mu\nu} \nonumber \\  &&\hspace{5cm}+2e^{-2z/l} 
\del_\mu\del_\nu(\delta f^{\textrm{vac}}) \int_0^z e^{2y/l} B(y) dy, \\
&& \delta \g_{zz}=h_{\mu z}=-2\delta f^{\textrm{vac}} B^\prime(z). \label{fixed}
\end{eqnarray} 
To quadratic order, the effective action is given by 
\begin{multline} \label{eqn:effaction}
S_{\textrm{eff}}=-\frac{M_5^3}{2} \int_0^{z_c} dz \int d^4x \sqrt{-\bar \g} ~
h^{ab}\delta \left[{\cal G}_{ab}-\frac{6~}{l^2} \g_{ab}\right] \\
+\half \int_{z=0} d^4x \sqrt{-\bar g^{(+)}} ~ 
h^{\mu\nu} \left[\Delta\left(M_5^2\delta (K_{\mu\nu}-K
  g_{\mu\nu})\right)^{(+)}-\sigma_+ h_{\mu\nu}\right] \\ 
+\half\int_{z=z_c} d^4x \sqrt{-\bar g^{(-)}} ~
h^{\mu\nu} \left[\Delta\left(M_5^2\delta (K_{\mu\nu}-K
  g_{\mu\nu})\right)^{(-)}-\sigma_-  h_{\mu\nu}\right].
\end{multline}
It turns out that $\delta \left[{\cal G}_{\mu\nu}-\frac{6~}{l^2}
  \g_{\mu\nu}\right]$ and $\Delta\left(M_5^2\delta (K_{\mu\nu}-K
g_{\mu\nu})\right)^{(\pm)}-\sigma_\pm h_{\mu\nu} $
are identically zero~\cite{pilo}. After integrating out a
total derivative in $z$, we arrive at the following effective action
for the radion:
\be
S_{\textrm{radion}}=-\frac{3}{2}M_-^2\int d^4x ~(\del_\mu \psi)^2,
\ee
where $M_-$ is given by Equation (\ref{Mpm}). Thus, as expected, the
radion, $\psi$, behaves as a massless scalar, and, being localised
close to the ``$-$" brane, its coupling strength is controlled by
the scale $M_-$.  As we move the branes further and further away
from one another, $M_-$ increases and  the radion starts to decouple
(decoupling completely in the RS2 limit).  Note that in some models,
such as GRS \cite{GRS}, the radion effective action comes in with
the wrong overall sign, signalling the presence of a physical ghost
\cite{pilo}. There is no such pathology in RS1 or RS2.
\newline
\newline
\noindent
{\it Weak gravity on a RS2 brane}
\newline

Let us now focus on RS2 by taking the negative tension brane to
infinity.  As $z_c \to \infty$, the metric on the remaining positive
tension brane is given by
\begin{multline}
\chi^{(+)}_{\mu\nu}=\frac{1}{(2\pi)^2}  \int d^4 p~ e^{ip_\mu x^\mu}
 \frac{K_2\left(pl\right)}{M_5^3pK_1\left(pl\right)} \left[\tilde
 {\cal T}^{(+)}_{\mu\nu}-\frac{1+\frac{K_1\left(pl\right)}{pl
 K_2\left(pl\right)} }{3} \tilde {\cal
 T}^{(+)} \bar g_{\mu\nu}^{(+)}\right] \\+\textrm{pure gauge terms}.
\end{multline}
At low energies, $p \ll k$, we again use the properties of modified
Bessel functions \cite{Abram-Stegun} to show that
\be
\chi^{(+)}_{\mu\nu} \approx \frac{1}{(2\pi)^2}  \int d^4 p~ e^{ip_\mu
  x^\mu} \frac{2}{M_4^2p^2}\left[\tilde {\cal
    T}^{(+)}_{\mu\nu}-\frac{1 }{2} \tilde {\cal
 T}^{(+)} \bar g_{\mu\nu}^{(+)}\right]+\textrm{pure gauge terms}.
\ee
It follows that gravity is indeed localised on the brane at low
energies, with a $4D$ effective Planck scale $M_4^2=M_5^3l$, or,
equivalently, with an effective Newton's constant $G_4=G_5 /l$. Note
that the tensor structure of the propagator matches that of General
Relativity, with a factor of $-\half$ in front of the trace term. This
is where the brane bending mode $f_+(x)$ plays a crucial role.  It
cancels   part of the graviton zero mode in just the right way to
guarantee good agreement with solar system gravity tests. It is worth
noting that this neat cancellation of terms does not always happen,
even in single brane scenarios. We will see this in DGP gravity, for
example, where one has to argue for some sort of Vainshtein effect to
pass observational tests.

Let us finally consider massive modes.  In RS2 there is actually a
continuum of massive modes, consistent with the fact that the extra
dimension is no longer compact.  However, as we have seen, we still
recover $4D$ GR to leading order. The next to leading order
corrections are obtained by integrating over the continuum, or,
equivalently, by considering the next to leading order expansion in
$pl$ above. The result is that the Newtonian potential
reads\footnote{Note that this result differs slightly from the original one
  quoted in \cite{rs2}.}
$$
V(r) \propto \frac{1}{r} \left(1+\frac{2l^2}{3  r^2}+{\cal O}(l^3/ r^3) \right).
$$
We refer the reader to \cite{garriga, thesis, duff-liu}  for details
of the derivation. Note that the $1/r^3$ correction can be obtained in
a dual picture as the one loop CFT correction to the graviton
propagator \cite{duff-liu}.

\subsection{Brane Cosmology}
\label{sec:BWcos}

There are two obvious reasons why cosmology offers an interesting
arena in which to develop the braneworld paradigm. The first is that
cosmological branes possess a high degree of symmetry, and this makes
it possible to solve the field equations.  The second is that
cosmological physics can be tested by a number of observations,
ranging from supernova data to the abundance of light elements. In
this section we will study the cosmology of co-dimension one branes,
focusing on the  RS2 scenario with a single $\mathbb{Z}_2$ symmetric
brane. We will review the background dynamics \cite{Binetruy-noncon,
  Christos-gencosmo} before moving on to cosmological perturbation
theory \cite{Kodama-gauges, Lukas-perts}. Further details, including
generalisations to multi-brane scenarios with bulk scalar fields, can
be found in the following review articles \cite{Quevedo, livingrev,
  Ida-review, Langlois-review, Brax-review, Anne-review} (see also
\cite{thesis, Csaki-cosmo, Csaki-radion, Shiromizu-cov2brane}).  Other generalisations include: the
cosmology of branes without $\mathbb{Z}_2$ symmetry \cite{asymm1, Stoica-cosmo, Shtanov3}; anisotropic braneworlds \cite{Maartens1, bwiso}; and branes for which energy is  explicitly transferred between bulk and brane \cite{Kiritsis1,Kiritsis2,Umezu}. Note that any anisotropy is seen to dissipate on the brane in RS gravity, a feature that can be identified with CFT particle production in the holographic picture \cite{bwiso}.  Bulk-brane energy transfer has been used to account for dark energy \cite{Kiritsis1,Kiritsis2,Umezu}. 

Braneworld cosmology can be studied using two different formalisms:
The {\it brane based} formalism, and the {\it bulk based}
formalism. These two approaches are completely equivalent and yield a
background cosmology governed by the following Friedmann equations
\cite{Binetruy-noncon, Binetruy-brane-cosmological, Christos-gencosmo}
\ba
H^2+\frac{\kappa}{a^2}&=&\frac{\Lambda_4}{3}+\frac{8 \pi
  G_4}{3}\rho\left( 1+\frac{\rho}{2\sigma} \right)+\frac{\mu}{a^4}, \label{friedmann}\\
\dot H-  \frac{\kappa}{a^2} &=& -4 \pi G_4 (\rho+P)\left(
1+\frac{\rho}{\sigma} \right)-\frac{2\mu}{a^4}, \label{raych}
\ea
where $H=\dot a/a$ is the Hubble parameter along the brane,  $a(t)$ is
the scale factor, and $\kappa=0, \pm 1$ describes the spatial
curvature.  The brane is sourced by a tension, $\sigma$, and  a
cosmological fluid with  energy density, $\rho(t)$,  and  pressure,
$P(t)$.  The parameters $\Lambda_4$ and $G_4$ denote the effective
cosmological constant and Newton's constant on the brane,
respectively. As in the standard scenario, the Raychaudhuri Equation
(\ref{raych}) follows from  the Friedmann Equation (\ref{friedmann})
and energy conservation,
\be \label{Econs}
\dot \rho+3H(\rho+P)=0.
\ee

We will now review the derivation of this cosmology using the two
equivalent formalisms.  For the moment, however, let us comment on a
few of its important features.  From Equations (\ref{friedmann}) and
(\ref{raych}) we see that the corrections to the standard cosmology
manifest themselves in a term $\propto \rho^2$, and a {\it dark
  radiation} term $\frac{\mu}{a^4} =\frac{8 \pi G_4}{3}\rho_{\textrm{weyl}}$.
The latter corresponds to a non-local  ``Weyl" contribution and can
only be fixed by specifying the bulk geometry. In the holographic
description of RS2,  the $\rho^2$ corrections contribute to the
conformal anomaly \cite{ads/cft-nohair}, while the dark  radiation is
identified with thermal excitations of  the CFT \cite{gubser,
  savonije, non-crit, exactbwh, covbwh, Hawking-brane-new}.

Both corrections will strongly affect Big Bang nucleosynthesis (BBN),
so their magnitude  can be constrained by the  abundance of light
elements. These require that the dark radiation can be  at most 10\%
of the photon energy density, $\rho_{\textrm{weyl}}/\rho_\g \lesssim 0.01$, in
the period following BBN \cite{braneBBN}. In addition, the $\rho^2$
corrections to the cosmological evolution should be negligible after
BBN, which imposes a constraint\footnote{Assuming a RS2 scenario with
  fine tuned tension, $\sigma=12M_5^3 /l$,  the BBN constraint is
  significantly weaker than the constraint arising from table top
  experiments quoted in Section \ref{sec:rs2}, i.e. $1/l \gtrsim
  10^{-4}$ eV, giving $\sigma \gtrsim (100
  \textrm{GeV})^4$\cite{Brax-review}.  Here we have used the fact that
  $M_4^2=M_5^3l \sim 10^{16}$ TeV.} on the tension $\sigma \gtrsim
(\textrm{MeV})^4 $ \cite{Binetruy-noncon}.

The $\rho^2$ corrections play the dominant role in the very early
universe and will have a big impact on  the inflationary
dynamics. Assuming the inflaton is confined to the brane along with
all the  Standard Model fields, one finds that the slow roll
parameters are given by \cite{Maartens-chaotic}
\ba
\epsilon&=&-\frac{\dot H}{H^2}=\frac{1}{16\pi G_4}
\left(\frac{V'}{V}\right)^2\left[\frac{1+\frac{V}{\sigma}}{
    \left(1+\frac{V}{2\sigma}\right)^2}\right],\\
\eta &=& -\frac{\ddot \phi}{H\dot \phi}=\frac{1}{8\pi
  G_4}\left(\frac{V''}{V}\right)\left[\frac{1}{1+\frac{V}{2\sigma}}\right],
\ea
where $V(\phi)$ is the inflaton potential. At low energies, $V \ll
\sigma$, these match the standard formulae of $4D$ General Relativity,
$$
\epsilon \sim \epsilon_{GR} =\frac{1}{16\pi G_4}
\left(\frac{V'}{V}\right)^2,\qquad \eta\sim \eta_{GR}=\frac{1}{8\pi
  G_4}\left(\frac{V''}{V}\right).
$$
However, at high energies we have 
\be
\epsilon \sim \epsilon_{GR} \frac{4\sigma}{V} \ll \epsilon_{GR},
\qquad \eta \sim \eta_{GR} \frac{2\sigma}{V} \ll \eta_{GR}.
\ee
This means we can get away with steeper potentials in the braneworld
case \cite{Maartens-chaotic}. This is essentially because Hubble
friction gets enhanced by the $\rho^2$ terms.  However, such
potentials are incompatible with observational  constraints as they
lead to a large tensor-to-scalar ratio \cite{Liddle-obs}.


\subsubsection{Brane based formalism -- covariant formulation} 
\label{sec:branebased}

We shall now derive the background cosmology, Eqs. (\ref{friedmann})
and (\ref{raych}), using  the brane based formalism. This makes use of
the Gauss-Codazzi equations \cite{Gauss, Codazzi}, and the Israel
junction conditions to derive the Einstein tensor on the brane
\cite{ShiromizuMaedaSasaki1999}.  We will assume $\mathbb{Z}_2$ symmetry,
although we refer the reader to \cite{Battye-Einstein} for a
non-$\mathbb{Z}_2$ symmetric generalisation.

The Gauss-Codazzi equations are the fundamental equations of embedded
hyper-surfaces.  The  brane can be thought of as an embedding,
\be
x^a=X^a(\xi^\mu),
\ee
in the bulk geometry, $\g_{ab}(x)$. By $\mathbb{Z}_2$ symmetry this is
the same on both sides of the brane.  We can now define tangent
vectors $V^a_\mu=\del X^a/\del \xi^\mu$, and the outward pointing unit
normal, $n^a$, satisfying 
$$
\g_{ab} n^a n^b=1, \qquad \g_{ab}n^a V^b_\mu=0.
$$
It follows that the induced metric on the brane is given by 
\be
g_{\mu\nu}=\g_{ab}(X) V^a_\mu V^b_\nu.
\ee
For such an embedding,   the Gauss-Codazzi equations \cite{Gauss, Codazzi} give
 \ba
 R_{\mu\nu\alpha\beta} &=&{\cal R}_{abcd} V^a_\mu V^b_\nu V^c_{\alpha}
 V^d_{\beta} +K_{\mu\alpha} K_{\nu\beta}-K_{\mu\beta} K_{\nu\alpha},
 \label{gauss}\\ 
 \nabla^\mu ( K_{\mu\nu} -K g_{\mu\nu}) &=& \R_{ab} n^a V^b_\nu, \label{codazzi} \\
 R-K^2+K_{\mu\nu}K^{\mu\nu} &=&  -2 {\cal G}_{ab} n^a n^b,
\ea
where $\R_{abcd}$ is  the Riemann tensor in the bulk,
$R_{\mu\nu\alpha \beta}$ is  the Riemann tensor on the brane, and we
recall that the extrinsic curvature of the brane is given by
$K_{\mu\nu}=\half {\cal L}_n g_{\mu\nu}$, the Lie derivative of the
induced metric, with respect to the normal.

Now, in  the RS2 scenario the dynamics of the brane are governed by
the bulk equations of motion  and the $\mathbb{Z}_2$ symmetric
junction conditions
\be
{\cal G}_{ab}=\frac{6~}{l^2} \g_{ab}, \qquad K_\mn -K \g_\mn=3\sigma_*
g_\mn -4 \pi G_5 {\cal T}_\mn, \ \label{rseqns}
\ee
with $\sigma_*=\frac{4 \pi G_5\sigma}{3}$, where $\sigma$ is the brane
tension and ${\cal T}_\mn$ is the energy-momentum tensor of additional
matter excitations. Of course, in the RS2 scenario we have a
fine-tuned tension such that $\sigma_*=1/l$, but we will keep things
general for the moment.

Consider the Codazzi Equation (\ref{codazzi}). Because the bulk is
only sourced by a cosmological constant, the right hand side of this
equation  is identically zero. This is important, because the junction
conditions in Eq. (\ref{rseqns}) now imply the usual conservation law
along the brane, $\nabla^\mu \T_\mn=0$. To extract information about
the Einstein tensor on the brane, we contract the Gauss Equation
(\ref{gauss}), and plug in Eq. (\ref{rseqns}), to give \cite{ShiromizuMaedaSasaki1999}
\be \label{Gmn}
G_\mn=-\Lambda_4 g_\mn+8 \pi G_4{\cal T}_\mn+ (4\pi G_5)^2
\Pi_\mn-E_\mn ,
\ee
where $\Lambda_4=3\left(\sigma_*^2-\frac{1~}{l^2}\right)$ is the
effective cosmological cosmological constant on the brane, and the
effective Newton's constant is given by
\be
G_4=G_5\sigma_* .
\ee
The corrections to standard $4D$ gravity are encoded in a local
contribution, $\Pi_\mn$, and a non-local ``Weyl" contribution,
$E_\mn$. The local piece is a  quadratic combination of the
energy-momentum tensor
\be
\Pi_\mn=-\T_\mu^\alpha T_{\nu\alpha}+\frac{1}{3} \T \T_\mn+\half
\T^{\alpha\beta}\T_{\alpha\beta} g_\mn-\frac{1}{6} \T^2 g_\mn ,
\ee
whereas the non-local ``Weyl" piece is
\be
E_\mn=C_{abcd} n^a V^b_\mu n^c V^d_\nu ,
\ee
where $C_{abcd}$ is the Weyl tensor of the bulk.  It is often referred
to as the {\it electric} part of the Weyl tensor, and in general one
must solve the bulk equations of motion first in order to evaluate it
in full. We should also note that $E_\mn$ has vanishing trace, and its
divergence is sourced by the local quadratic contribution
\be \label{Econstraints}
g ^\mn E_\mn=0, \qquad \nabla_\mu E^\mu_\nu=(4\pi G_5)^2   \nabla_\mu
\Pi^\mu_\nu .
\ee
The latter equation follows from the divergence of Equation
(\ref{Gmn}), making use of the Bianchi identity and the local
conservation of energy-momentum, $\nabla^\mu \T_\mn=0$.

Our interest here lies in the cosmology, so let us now assume spatial
homogeneity and isotropy on the brane. The induced metric is given by
the usual  Friedmann-Lema\^{i}tre-Robertson-Walker (FLRW) metric
\be \label{frw-metric}
ds^2=g_\mn d\xi^\mu d\xi^\nu=-dt^2+a(t)^2 q_{ij} dx^i dx^j,
\ee
where $q_{ij}(x)$ is the metric of a hyper-surface of constant
curvature, $\kappa=0, \pm 1$.  The matter excitations contribute a
cosmological fluid with  energy density, $\rho(t)$,  and  pressure,
$P(t)$, such that its energy-momentum tensor is given by
\be \label{Tmn}
\T^\mu_\nu=\textrm{diag}(-\rho(t), P(t), P(t), P(t)).
\ee
In the absence of non-trivial sources in the bulk $T^{\mu}_{\nu}$ is
conserved, which means we have the standard relation given by Equation
(\ref{Econs}).

It remains to compute the non-local piece, $E_\mn$. Although in
general we must solve the bulk equations of motion to evaluate this,
we can exploit the large amount of symmetry to avoid doing so in the
current instance. Making use of the Constraints Equations
(\ref{Econstraints}), we can show that 
$$
E^\mu_\nu=\frac{\mu}{a^4} \textrm{diag}\left(-1, \frac{1}{3}, \frac{1}{3},
\frac{1}{3}\right),
$$
where $\mu$ is an integration constant that should be fixed by the
bulk geometry (we will see in the next section that it can be
identified with the mass of a bulk black hole).  Note that $E^\mu_\nu$
is conserved on a cosmological background, by virtue of the fact that
$\nabla_\mu \Pi^\mu_\nu =0$.

The modified Friedmann Equations (\ref{friedmann}) and (\ref{raych})
now follow automatically from Equation (\ref{Gmn}).

\subsubsection{Bulk based formalism -- moving branes in a static bulk}
\label{sec:bulk-based}

The principle limitation of the brane based formalism we have just
described is that it suppresses physics deep inside the bulk. This can
be dangerous since it is not at all obvious if a particular choice of
the non-local ``Weyl" term, $E_\mn$, at the brane will evolve into a
pathological bulk geometry.  Such problems can be avoided by solving
for the bulk geometry first. For example, in the brane based
formalism, the parameter $\mu$ giving rise to dark radiation, is just
an integration constant and can take any sign. Using the bulk based
formalism, we are able to identify $\mu$ with the mass of a bulk black
hole, and use this to constrain it to be positive.

We will now develop the bulk based formalism for cosmological branes.
Again, we will assume $\mathbb{Z}_2$ symmetry across the branes, for
brevity.  The reader can refer to \cite{thesis} for the
non-$\mathbb{Z}_2$ symmetric generalisation. The bulk based formalism
requires us to solve for the bulk geometry. Since we are interested in
cosmological branes (with constant curvature Euclidean 3-spaces), we
study the Einstein equations, $\G_\ab=\frac{6~}{l^2}\g_\ab$, with the
following metric ansatz \cite{Christos-gencosmo}
\begin{equation} \label{eqn:Birkhoffmetric}
ds^2 = \g_\ab dx^a dx^b=e^{2\nu} A^{-2/3} (-dt^2 + dz^2)+ A^{2/3}
q_{ij} dx^i dx^j,
\end{equation}
where $A$ and $\nu$ are undetermined functions of $t$ and $z$, and as
before $q_{ij}(x)$ is the metric of a hyper-surface of constant
curvature, $\kappa=0, \pm 1$.  Now, in an extremely elegant
calculation, Bowcock {\it et al.} \cite{Christos-gencosmo} were able
to prove a generalised form of Birkhoff's theorem, showing that the
bulk geometry is necessarily given by
\begin{equation} \label{eqn:schmetric}
ds^2=-V(r)d\tau^2+\frac{dr^2}{V(r)}+r^2q_{ij} dx^i dx^j,
\end{equation}
where
\begin{equation}
V(r)=\frac{ r^2}{l^2}+\kappa-\frac{\mu}{r^2}.
\end{equation}
For $\mu>0$, the metric in Eq. (\ref{eqn:schmetric}) takes the form of
a (topological) Schwarzschild black hole in anti-de Sitter space.
Here we have written the solution in an explicitly time independent
coordinate system, meaning that we can no longer say that we have a
static brane sitting  at a fixed coordinate position. On the contrary, we now have a
dynamic brane, whose trajectory in the these coordinates is more
complicated.  Braneworld cosmology from this perspective was first
studied by Ida~\cite{Ida-review},  although moving branes in a static
anti-de Sitter bulk were considered earlier by Kraus~\cite{Kraus-dynamics}.  

To construct the brane solution, we treat it as an embedding,  
\begin{equation}
\tau=\tau(t), \qquad r=a(t),
\end{equation}
of the bulk geometry given in Eq. (\ref{eqn:schmetric}). The induced
metric on the brane is then
\be
ds^2=g_\mn d\xi^\mu d\xi^\nu =\left(-V(a) \dot \tau^2+\frac{\dot
  a^2}{V(a) }\right) dt^2+a^2(t)q_{ij} dx^i dx^j 
\ee
where over-dots denote $\del/ \del t$. We are free to choose $t$ to correspond
to the proper time with respect to an observer comoving with the
brane. This imposes the condition
\begin{equation} \label{cond}
-V(a) \dot \tau^2+\frac{\dot a^2}{V(a) }=-1,
\end{equation}
ensuring that the brane takes the standard FLRW form, as in
Eq. (\ref{frw-metric}). The function $a(t)$ is then immediately
identified with the scale factor along the brane.

The boundary condition at the brane are given in
Eq. (\ref{rseqns}). We must compute the extrinsic curvature,
$K_\mn=\half {\cal L}_n g_{\mu\nu}$, defined as the Lie derivative of
the normal pointing {\it into} the bulk.  Assuming we cut away the AdS
boundary and retain the region $r<a(t)$, we find that the inward
pointing unit normal is given by
\be
n_a=(-\dot a, \dot \tau, 0, 0, 0),
\ee
where we  are free to specify that $\dot \tau>0$.  The components of
extrinsic curvature are then given by
\be
K^i_{j}=\frac{V \dot \tau}{a} \delta^i_j, \qquad
K^t_t=-\left(\frac{\ddot a+V'/2}{V\dot \tau}\right).
\ee
In the presence of a cosmological fluid, as given by Equation
(\ref{Tmn}), the junction conditions in Eq. (\ref{rseqns}) yield the
following:
\begin{equation} \label{eqn:ext1}
\frac{V \dot \tau}{a}=\sigma_*(1+\frac{\rho}{\sigma}),
\end{equation}
\begin{equation} \label{eqn:ext2}
\frac{\ddot a+\frac{1}{2}V'}{V \dot
\tau}=\sigma_*\left[1-2\frac{\rho}{\sigma}-3\frac{P}{\sigma} \right].
\end{equation}
Making use of  Equation (\ref{cond}) we then arrive at the modified
Friedmann Equations (\ref{friedmann}) and (\ref{raych}).

\subsubsection{Cosmological perturbations}  
\label{sec:branecosperts}

While the theory describing  cosmological perturbations in braneworld
gravity has been well developed in recent years (see, for example,
\cite{Maartens-cosmological-dynamics, Langlois-large-scale,
  Maartens-chaotic, Copeland-steep, Sahni-relic, Nunes-tracking,
  Liddle-curvaton-reheating, Bridgman-cosmological, Bridgman-cosmic,
  Gordon-density-perts, Langlois-gravitational-waves,
  Gorbunov-gravity-waves, Mukohyama-gauge-inv,
  Mukohyama-perturbation-junction, Mukohyama-doubly,
  Mukohyama-integro, Hawking-brane-new, Kodama-gauges,
  Langlois-brane-cosmological, Lukas-perts,
  Koyama-evolution,vandebruck-cosmological-consequences,
  Kobayashi-quantum-fluc, Kobayashi-primordial-gravitational,
  Kodama-behavior, Langlois-evolution, Hawking-trace-anomaly,
  Deruelle-perturbations,   Brax-bulk-scalars, Dorca-cosmological,
  Neronov-metric, Chen-rotational, Chung-lensed,
  Deffayet2002,Riazuelo-gauge-inv, Leong-1+3, Cardoso-scalar}),
even approximate solutions to the resulting field equations have been
notoriously hard to come by. The problem stems from the fact that one
has to solve the fully coupled system of brane and bulk, which is, in
general, a far from trivial task.   Indeed, this will generically
render the brane based formalism somewhat  incomplete without making
{\it ad hoc} assumptions about the perturbations of the non-local
``Weyl" contribution, $E_\mn$. The bulk based formalism is better in
this respect, but the resulting system is virtually intractable, and
only in some special cases, where the bulk and brane equations become
separable, has progress been made \cite{garriga, Hawking-brane-new,
  Bridgman-cosmological, Bridgman-cosmic,
  Langlois-gravitational-waves}. Note that, unlike in other sections,
we will on occasion include vector and tensor perturbations, as well
as scalars. This is because the bulk can source vector and tensor
modes on the brane, giving qualitatively different behaviour to that
seen in standard $4D$ cosmology.
\newline
\newline
\noindent
{\it Cosmological perturbations in the brane based formalism}
\newline

We shall now review some aspects of cosmological  perturbation theory
using the brane based formalism introduced in Section
\ref{sec:branebased}. Further details can be found in \cite{livingrev,
  Maartens-cosmological-dynamics}. The dynamics on the brane are
governed by Equation (\ref{Gmn}), and so perturbations about the
background cosmology on the brane satisfy
\be \label{dGmn}
\delta G^\mu_\nu=8 \pi G_4{\delta \cal T}^\mu_\nu+ (4\pi G_5)^2 \delta
\Pi^\mu_\nu-\delta E^\mu_\nu .
\ee
To study this we use the standard four-dimensional formalism
\cite{Bardeen1980,KodamaSasaki1984,ellisbruni,MukhanovFeldmanBrandenberger1992},
decomposing the system into scalar, vector and tensor perturbations
with respect to the spatial diffeomorphism group in the background
cosmology.   Working with conformal time as opposed to proper time,
the perturbed metric is given by
\begin{multline}\label{dgmn}
ds^2=a^2\left[ -(1+2\Psi) d\tau^2-2(\beta_i+\grad_i \beta) dx^i d \tau \right.\\
\left.+\left[(1-2\Phi)q_{ij}+D_{ij} \nu+2\grad_{(i} A_{j)}+h_{ij} \right] dx^i dx^j \right],
\end{multline}
%
where $A_i$ and $\beta_i$ are transverse vectors on $q_{ij}$, and
$h_{ij}$ is a transverse and trace-free tensor.  Recall that the
operator $D_{ij}=\grad_i \grad_j -\frac{1}{3} q_{ij} \Delta$. The
fluctuations energy-momentum on the brane are written in terms of the
fluctuations in density, $\delta$, pressure, $\delta P$, fluid
3-velocity, $v_i$, and anisotropic stress, $\Sigma^i_j$, in the usual
way,
\be \label{Tdecomp}
\delta \T^0_0=-\rho \delta, \qquad \delta \T^0_i=-(\rho+P)v_i, \qquad
\delta \T^i_j=\delta P \delta^i_j+(\rho+P)\Sigma^i_j .
\ee
The fluid 3-velocity and anisotropic stress can then be decomposed with
respect to their scalar, vector and tensor components,
\ba 
v_i &=&\theta_i^{(\textrm{vector})}+\grad_i \theta , \label{vi}\\
\Sigma_{ij} &=& \Sigma_{ij}^{(\textrm{tensor})}+\grad_{(i}
\Sigma^{(\textrm{vector})}_{j)}+D_{ij} \Sigma . \label{Sigmaij}
\ea
The fluctuation in the quadratic piece is also given in terms of $\rho
\delta, ~\delta P \ldots$ etc., according to
\ba
&& \delta \Pi^0_0=\frac{2}{3} \rho \delta \T^0_0, \qquad \delta
\Pi^0_i=\frac{2}{3} \rho \delta \T^0_i , \nonumber\\
&&  \delta \Pi^i_j =-\frac{1}{3} (\rho+3P) \delta
\T^i_j-\frac{2}{3}(\rho+P) \delta^i_j \left(\delta \T^0_0-\half \delta
\T^k_k \right) ,
\ea
where $\delta T^\mu_\nu$ are given by Equations (\ref{Tdecomp}).

Now we consider the contribution from the non-local ``Weyl"
perturbation, and identify it with the fluctuation in some dark energy
energy-momentum tensor
$$ 
\delta E_\mn=-8 \pi G_4{\delta \cal T}^{\textrm{weyl}}_\mn .
$$
In direct analogy with Equations (\ref{Tdecomp}), we can read off
corresponding fluctuations in {Weyl} energy density,  $\delta^{\textrm{weyl}}$,
{ Weyl} pressure, $\delta P^{\textrm{weyl}}$, {Weyl} fluid 3-velocity,
$v^{\textrm{weyl}}_i$, and {Weyl} anisotropic stress,
$\Sigma_{ij}^{\textrm{weyl}}$. We
might hope to determine each of  these  in terms of the local matter
fluctuations, $ \delta, ~\delta P \ldots$ etc., by making use of the
Constraints Eqs. (\ref{Econstraints})  to linear order. We find that
for scalar perturbations $\delta E^\mu_\mu=0$ gives\footnote{The remaining formulae in this
  section are taken from \cite{Langlois-large-scale}, where one should
  identify $\delta\rho=\rho\delta, ~\delta q=-( \rho+P) \theta,
  ~\delta \pi=(\rho+P)\Sigma, ~{\cal R}=\Psi, ~A=-\Phi, ~E=\half \nu,
  ~B=-\beta$.}
\be
\delta P^{\textrm{weyl}}=\frac{1}{3} \rho^{\textrm{weyl}} \delta^{\textrm{weyl}},
\ee
and $\nabla_\mu \left[\delta E^\mu_\nu-(4\pi G_5)^2
  \Pi^\mu_\nu\right]=0$ gives
\ba
&& ({\delta}^{\textrm{weyl}} )'+4\Psi'+\frac{4}{3}  \Delta \left[\beta+\half
  \nu'-\theta^{\textrm{weyl}}\right]=0, \label{gradE1}\\ 
&& -\frac{4}{3} \rho^{\textrm{weyl}} \left[ (\theta^{\textrm{weyl}})'+\Phi-\frac{2}{3}
  (\Delta+3\kappa) \Sigma^{\textrm{weyl}}-\frac{\delta^{\textrm{weyl}}}{4}\right]=
\nonumber \\ 
&& \hspace{3cm}\frac{\rho+P}{\sigma}\left\{ \rho
\delta+(\rho+P)[3{\cal H} \theta-(\Delta+3\kappa) \Sigma] \right\} , \label{gradE2}
\ea
%
where $'$ denotes differentiation with respect to conformal time, and
${\cal H}=a'/a=aH$. Clearly we have too many Weyl unknowns and not
enough equations. The bottom line is that we need to know the Weyl
anisotropic stress $\delta \pi_{ij}^{\textrm{weyl}}$ explicitly, and for that
we need to abandon the brane based formalism and solve the bulk
equations of motion \cite{Maartens-cosmological-dynamics}.

Some progress can be made at super-horizon scales, since then we can
neglect the spatial gradients in Eqs. (\ref{gradE1}) and
(\ref{gradE2}), and solve for the Weyl energy density and Weyl
momentum in terms of $\rho \delta, ~\delta P \ldots$ etc., thereby
closing the system \cite{Maartens-cosmological-dynamics,
  Gordon-density-perts, Langlois-large-scale}. This simplification has
been applied  to the study of  both density perturbations and vector
perturbations at large scales, revealing qualitatively different
behaviour to that in General Relativity \cite{Gordon-density-perts,
  Maartens-geometry}. Regarding density perturbations, it can be shown
that the quantity $ \rho a^4 \Delta  \delta$ will grow during
slow-roll inflation on super-horizon scales (it stays constant in GR)
\cite{Gordon-density-perts}.  It can be shown that vector
perturbations can be non-vanishing, even in the absence of vorticity
\cite{Maartens-geometry}.

We can also solve for the (total) curvature perturbation on large
scales \cite{Langlois-large-scale}. Unfortunately, this does not mean
we can compute large-scale CMB anisotropies. The problem is that to
evaluate the (non-integrated) Sachs-Wolfe equation we need knowledge
of the metric perturbations. These are sourced by the dark anisotropic
stress, according to
\be
\hat \Phi-\hat \Psi=8\pi G_4 a^2 \left(\frac{4}{3} \rho^{\textrm{weyl}}\right) \Sigma^{\textrm{weyl}},
\ee
where $\hat \Psi$ and $\hat \Phi$ are the Bardeen gauge invariants for
the metric perturbations, and we have neglected the local anisotropic
stress. The braneworld corrections to the Sachs-Wolfe effect are given
by \cite{Langlois-large-scale}
\begin{multline}
\frac{\delta T}{T}=\frac{\delta T}{T}\Big |_{GR}-\frac{8}{3}
\left(\frac{\rho_{\g}}{\rho_{cdm}}\right) {\cal S}_{\textrm{weyl}} \\ 
-8\pi G_4 a^2 \left(\frac{4}{3}
\rho^{\textrm{weyl}}\right)\Sigma^{\textrm{weyl}}+\frac{16 \pi G_4}{a^{5/2}}\int da
~a^{7/2} \left(\frac{4}{3} \rho^{\textrm{weyl}}\right) \Sigma^{\textrm{weyl}} ,
\end{multline}
where ${\cal S}_{\textrm{weyl}}$ is the Weyl entropy perturbation, determined
by $\rho \delta^{\textrm{weyl}}$.
\newline
\newline
\noindent
{\it Cosmological perturbations in the bulk based formalism}
\newline

As we saw in the previous section, a proper treatment of cosmological
perturbation theory in brane cosmology  requires us to solve the
coupled system of brane and bulk. We will now present the details of
this in the bulk based formalism, essentially following
\cite{Kodama-gauges}  (see also \cite{Lukas-perts,
  Mukohyama-gauge-inv, Mukohyama-perturbation-junction,
  Mukohyama-doubly, Mukohyama-integro}). Note that in this section we
deviate from our usual convention of treating cosmological
perturbations with respect to conformal time on the brane, preferring
instead to use proper time of comoving observers, in keeping with the
majority of the relevant literature. For consistency, however, we do
define the brane quantities as in the previous section.

We begin with some notation on the background. The bulk metric is given by
\be \label{lambdaq}
ds^2=\bar \g_{ab} dx^a dx^b=\lambda_{\alpha \beta} dx^\alpha
dx^\beta+r^2 q_{ij}dx^i dx^j ,
\ee
where $\lambda_{\alpha \beta}$ is some two-dimensional metric, and
$\lambda_{\alpha\beta}$ and $r$ depend only on the first two
coordinates, $x^\alpha$.  This corresponds to a section of
(topological) AdS-Schwarzschild, and one can choose a gauge such that
\be
\lambda_{\alpha \beta} dx^\alpha
dx^\beta=-V(r)d\tau^2+\frac{dr^2}{V(r)}, \qquad
V(r)=\frac{r^2}{l^2}+\kappa-\frac{\mu}{r^2} .
\ee
However, here we will leave the choice of gauge unspecified so that
our analysis can be applied in other gauges (e.g. Gaussian-Normal
gauge). The background embedding equation is now given by $x^a=\bar
X^a(\xi^\mu)$, where $\xi^\mu=(t, \xi^i)$ are the coordinates along the brane. We take
$$
\bar X^\alpha=\bar X^\alpha (t), \qquad \bar X^i=\xi^i ,
$$
and $\lambda_{\alpha \beta}(\bar X) \dot {\bar X}^\alpha \dot {\bar
  X}^\beta=-1$ to ensure that the induced metric on the brane has the
standard FLRW form, as in Eq. (\ref{frw-metric}), with scale factor
$a(t)=r(\bar X)$. Over-dots here denote differentiation with respect
to proper time on the brane, $t$.

We shall now specify the perturbations in the bulk. Decomposing the
bulk metric in terms of scalar, vectors and tensor components (with
respect to $q_{ij}$), we write $\delta
\g_\ab=h^{\textrm{scalar}}_{ab}+h^{\textrm{vector}}_{ab}+h^{\textrm{tensor}}_{ab}$, where,
\begin{align}
&h^{\textrm{scalar}}_{\alpha\beta}=\chi_{\alpha \beta},  &&h^{\textrm{scalar}}_{\alpha
    i}=r\grad_i \chi_\alpha, &&h^{\textrm{scalar}}_{ij}=2r^2\left[A
    q_{ij}+D_{ij} E \right],  \\
&h^{\textrm{vector}}_{\alpha\beta}=0, && h^{\textrm{vector}}_{\alpha i}=r B_{\alpha i},
  &&h^{\textrm{vector}}_{ij}=2r^2\grad_{(i} H_{j)}, \\
&h^{\textrm{tensor}}_{\alpha\beta}=0, & &h^{\textrm{tensor}}_{\alpha i}=0,  && h^{\textrm{tensor}}_{ij}=r^2 H_{ij},
\end{align}
where $D_{ij}=\grad_i \grad_j -\frac{1}{3} q_{ij} \Delta$.  Here
$B_{\alpha i}$ and $H_i$ are transverse, and $H_{ij}$ is transverse
and trace-free. One can identify the following gauge invariants in the
bulk \cite{Kodama-gauges}:
\ba
\textrm{scalars:} &&   Y_{\alpha\beta}=\chi_{\alpha\beta}-2D_{(\alpha}
Q_{\beta)}, \qquad Z=A-\frac{1}{3} \Delta E-Q^\alpha \del_{\alpha} \ln
r, \qquad \\
\textrm{vectors:} && F_{\alpha i}=B_{\alpha i}-r D_{\alpha} H_i ,\\
\textrm{tensors:} &&H_{ij},
\ea
where $D_\alpha$ is the covariant derivative on $\lambda_{\alpha
  \beta}$, and $Q_{\alpha}=r(\chi_\alpha-r\del_{\alpha} E)$. For
simplicity we will assume that there are no exceptional modes
\cite{Kodama-gauges} in any sector. Now, the bulk equations of motion,
$\delta G^a_b=0$, are extremely complex in general (see
\cite{Kodama-gauges}).
From a practical perspective, the only way to proceed is to assume
that the background bulk is maximally symmetric anti de Sitter space,
so there is no bulk black hole (i.e. $\mu=0$)
\cite{Mukohyama-gauge-inv}. The vector and scalar perturbations in the
bulk can then be expressed in terms of a corresponding `master
variable'. For example, the scalar gauge invariants $Y_{\alpha \beta}$
and $Z$ can be written as
\be
Y_{\alpha\beta}=\frac{1}{r} \left[ D_\alpha D_\beta-\frac{2}{3}
  \lambda_{\alpha \beta}\left(D^2-\frac{1}{2l^2} \right) \right]
\Omega, \qquad Z=-\half
Y=\frac{1}{6r}\left(D^2-\frac{2}{l^2}\right)\Omega ,
\ee 
where the scalar master variable $\Omega$ satisfies
\be \label{mastersc}
\left[ D^2-3 \del_\alpha \ln r D^\alpha+\frac{\Delta+3\kappa}{r^2}
  +\frac{1}{l^2}\right] \Omega=0 .
\ee
For the vectors, we introduce the vector master variable, writing the gauge invariant as
\be
F_{\alpha i}=\frac{1}{r^2} \epsilon_{\alpha \beta} D^\beta \Omega_i ,
\ee
where $\epsilon_{\alpha\beta}$ is the total antisymmetric Levi-Civita
tensor on $\lambda_{\alpha \beta}$, and  $\Omega_i$ satisfies
\be \label{mastervec}
D^\alpha \left[ r^5 D_\beta \left(\frac{D^\beta
    \Omega_i}{r^3}\right)+(\Delta+2\kappa) \Omega_i\right]=0 .
\ee
The tensor gauge invariant, $H_{ij}$, is essentially its own master
variable, and satisfies 
\be \label{tensorbulk}
\left[D^2 +3 D^\alpha \ln r D_\alpha+\frac{1}{r^2}(\Delta-2\kappa) \right] H_{ij}=0 .\\
\ee
To solve for these master variables we need to specify boundary
conditions at the brane, as well as the asymptotic boundary
conditions. The latter correspond to the condition of normalisability.
Let us now consider the boundary conditions at the brane. The first
thing to note here is that the brane position can  fluctuate, so that
it now corresponds to an embedding $x^a=\bar X^a+f^a(\xi^\mu)$. We
decompose this fluctuation in terms of scalars $f^\alpha, \epsilon$,
and a transverse vector  $\epsilon^i$, such that $f^a=(f^\alpha,
\epsilon^i+\grad^i \epsilon)$. Secondly, in applying the boundary
conditions, it is important to express all fields in terms of
covariant objects along the (background) brane. This suggests that all
fields of the form $W^\alpha$ and $W^{\alpha \beta}$  should be
decomposed in terms of their components parallel to the tangent vector
$ \dot{\bar X}^\alpha$ and along the normal $\bar n^\alpha$, as
follows:
\ba
W^\alpha &=&-W_\parallel \dot{\bar X}^\alpha+W_\perp \bar n^\alpha ,\\
W^{\alpha \beta}&=&W_{\parallel \parallel} \dot{\bar
  X}^\alpha\dot{\bar X}^\beta-W_{\parallel \perp} \dot{\bar
  X}^{\alpha} \bar n^{\beta}-W_{\perp \parallel}\dot{\bar X}^{\beta}
\bar n^{\alpha}+W_{\perp\perp}\bar n^\alpha \bar n^\beta .
\ea
The induced metric on the brane then takes the form
\begin{multline}\label{dgmn1}
ds^2= -(1+2\Psi) dt^2-2a(\beta_i+\grad_i \beta) dx^i d t \\
+a^2\left[(1-2\Phi)q_{ij}+D_{ij} \nu+2\grad_{(i} A_{j)}+h_{ij} \right] dx^i dx^j ,
\end{multline}
with
\begin{align}
&\Psi = -\half Y_{\parallel \parallel}-\dot{U}_\parallel+\bar K^t_t
  U_\perp , && \beta =- \frac{U_\parallel}{a}-\frac{a \dot \nu}{2} , \\ 
&\Phi = -Z-\frac{1}{3} \Delta \nu+H U_\parallel -(D_\perp \ln r)
  U_\perp , && \nu = 2(E+\epsilon) , \\
& \beta_i = -F_{\parallel i}-a\dot A_i , && A_i=H_i+\epsilon_i , \\
& h_{ij} =H_{ij} ,
 \end{align}
where $U^\alpha=Q^\alpha+f^\alpha$. Note that $U^\alpha, \nu$ and
$A_i$ are invariant under gauge transformations in the bulk, so these
expressions have been written entirely in terms of bulk gauge
invariants.
 
The gauge invariants on the brane are the ones familiar to us from
standard cosmological perturbation theory in four dimensions. Given
the fluctuation in the brane energy-momentum tensor
\be \label{Tdecomp1}
\delta \T^t_t=-\rho \delta, \qquad \delta \T^t_i=-a(\rho+P)v_i, \qquad
\delta \T^i_j=\delta P \delta^i_j+(\rho+P)\Sigma^i_j ,
\ee
and Equations (\ref{vi}) and (\ref{Sigmaij}), we note that the
components of the anisotropic stress $\Sigma, \Sigma^{(\textrm{vector})}$ and
$\Sigma_{ij}^{(\textrm{tensor})}$ are all gauge invariant on the brane, along
with the following scalars,
\ba
\hat \Phi&=&\Phi+\frac{1}{6} \Delta \nu+\dot a \sigma_g =-Z-(D_\perp
\ln r)U_\perp ,\\
\hat \Psi&=&\Psi-(a \sigma_g)\dot{} =-\half Y_{\parallel \parallel}+\bar K^t_t U_\perp ,\\
\hat \theta&=& \theta-\sigma_g ,\\
\hat \delta &=& \delta+3Ha \left(1+\frac{P}{\rho}\right)\theta ,\\
\Gamma &=& \rho \delta- c_s^2 \delta P ,
\ea
the vectors $\theta_i$ and $\hat \beta_i= \beta_i +a\dot
A_i=-F_{\parallel i}$, and the tensor, $h_{ij}$.   Note that
$$
\sigma_g=\frac{a\dot \nu}{2}+\beta=-\frac{U_\parallel}{a}.
$$
The goal is now to solve for the master variables, subject to boundary
conditions set by normalisability and the values of the non-dynamical
gauge invariants on the brane, namely, $\Gamma, \Sigma,
\Sigma_i^{(\textrm{vector})}$ and $\Sigma_{ij}^{(\textrm{tensor})}$. We can then use
this knowledge to derive  the dynamical gauge invariants, $\hat
\Phi,\hat  \Psi, \hat \beta_i, h_{ij}, \hat \theta, \hat \delta$, and
$\theta_i$.  We are already able to express the metric invariants in
terms of the master variables (and $U_\perp$) as follows:
\ba
\hat \Phi &=& \frac{1}{2a}\left[H \dot \Omega-(D_\perp \ln r) D_\perp
  \Omega +\left(\frac{1}{l^2} +\frac{(\Delta+3 \kappa)}{3a^2}
  \right)\Omega \right]\\ 
&&\hspace{2cm}-(D_\perp \ln r)U_\perp , \nonumber \\
\hat \Psi &=& -\frac{1}{2a} \left[ \ddot \Omega-2H\dot
  \Omega+\left(2D_\perp \ln r-\bar K^t_t \right) D_\perp
  \Omega-\left(\frac{1}{l^2} +\frac{2(\Delta+3 \kappa)}{3a^2}
  \right)\Omega\right]\qquad\qquad\\
&&\hspace{2cm}+\bar K^t_t U_\perp , \nonumber \\
\hat \beta_i &=& -\frac{1}{a^2} D_\perp \Omega_i . \label{Sigmai}
\ea
To fix $U_\perp$, and the remaining dynamical gauge invariants, we
need to impose boundary conditions at the brane, given by the
linearised Israel junction conditions $\delta (K^\mu_\nu-K
\delta^\mu_\nu)=-4\pi G_5 \delta \T^\mu_\nu$. For the scalars this gives
\ba
(\Delta+3\kappa)D_\perp \left(\frac{\Omega}{r}\right) &=&8\pi
G_5a^2\left[\rho \hat \delta-(\Delta+3\kappa)[(\rho+P)\Sigma]\right],  \label{sc1}\\
(D_\perp \Omega)\dot{}-\bar K^t_t \dot \Omega&=&  8\pi G a^2
\left[(\rho+P)\hat \theta-\del_t [a(\rho+P) \Sigma]~\right],  \label{sc2} \\
U_\perp &=& 4\pi G_5 a^2 (\rho+P)\Sigma ,  \label{sc3}\\
-\Gamma &=& c_s^2 \rho \hat \delta+(\rho+P)\left(\hat
\Psi-\del_t\left(a\hat \theta\right) + \frac{2}{3} (\Delta+3\kappa)
\Sigma\right) . \qquad  \label{sc4}
\ea
We should now view these equations as follows  \cite{Kodama-gauges}:
Equations (\ref{sc1}) and (\ref{sc2}) fix $\hat \delta$ and $\hat
\theta$ given knowledge of the anisotropic stress, $\Sigma$ and the
master variable, $\Omega$, while  Equation (\ref{sc3}) fixes
$U_\perp$.  One can then substitute the expression for $\hat \Psi,
~\hat \delta$ and $\hat \theta$ (in terms of $\Omega$ and $\Sigma$)
into Equation (\ref{sc4}) to derive a Neumann type boundary condition
on $\Omega$. In principle, one should now be able to solve the Master
Equation (\ref{mastersc}), with a suitable boundary condition derived
from Eq. (\ref{sc4}), and feed the solution back into the expressions
for $\hat\Phi, \hat \Psi, \hat \theta$ and $\hat \delta$.

For vectors, the Israel junction conditions yield
\ba
(\Delta+2\kappa) \Omega_i &=& 8\pi G_5 a^4(\rho+P) \theta_i , \label{vec1}\\
\dot \Omega_i &=& 4 \pi G_5 a^3 (\rho+P)\Sigma_i^{(\textrm{vector})} . \label{vec2}
\ea
This time we view Equation (\ref{vec2}) as a Dirichlet type boundary
condition on the master variable. We then solve for $\Omega_i$ using
Equation (\ref{mastervec}), and subsequently fix $\hat \beta_i$ and
$\theta_i$ using Equations (\ref{Sigmai}) and (\ref{vec1}).

Finally, we consider the tensor. Here we solve Equation
(\ref{tensorbulk}), subject to the following junction condition at the
brane:
\be
D_\perp H_{ij}=-4\pi G_5 (\rho+P)\Sigma_{ij}^{(\textrm{tensor})}.
\ee

We have now presented the formalism in full, but the task of solving
the system explicitly is another matter altogether. To do so one must
choose coordinates for the background bulk. Whilst the static
coordinate system might seem the simplest from the bulk perspective,
it is rarely used owing to the fact that a Gaussian Normal (GN)
coordinate system makes it much easier to specify the boundary
conditions. For example, when $\mu=0$ and $\kappa=0$, the metric in GN
coordinates is given by \cite{Binetruy-brane-cosmological}
\ba
&&\lambda_{\alpha \beta} dx^\alpha dx^\beta=dz^2-N^2(t, z) dt^2, \\
&& r(t, z)=a(t)\left[\cosh \left( |z|/l
  \right)-\left(1+\frac{\rho(t)}{\sigma}\right)\sinh \left(|z|/l
  \right) \right] ,
\ea
where $N=\frac{\dot r(t, z)}{\dot a(t)}$. Even in this case analytic
solutions can only be obtained in the case of a background de Sitter
brane, when the Master Equation (\ref{mastersc}) becomes separable
\cite{Bridgman-cosmological, Langlois-gravitational-waves}. Numerical
solutions, however, have been obtained  for scalar perturbations on  a
radiation dominated brane \cite{Cardoso-scalar}. It was found that
short wavelength density perturbations are amplified relative to their
value GR during horizon reentry, but not so much that they cause an
observable effect in the CMB or in large-scale structure. 

\subsection{Dvali-Gabadadze-Porrati Gravity}
\label{sec:dgp}

The most celebrated braneworld model exhibiting {\it
  quasi}-localisation and an {\it infra-red} modification of gravity is
without doubt the DGP model, developed by Dvali, Gabadadze and Porrati
\cite{dgp}. The model admits two distinct sectors: The {\it normal}
branch and the {\it self-accelerating} branch. The latter has
generated plenty of interest since it gives rise to cosmic
acceleration without the need for dark energy
\cite{Deffayet2000, Deffayet-accelerated}. However, as we will
see, fluctuations about the self-accelerating vacuum  suffer from
ghost-like instabilities \cite{dgpspec, dgpspecrev,
  dgppersp,koyama-more}. Although the normal branch is less
interesting phenomenologically, it is fundamentally more healthy and
is the closest thing we have to a consistent non-linear completion of
massive gravity. Here the graviton is a resonance of finite width,
$1/r_c$, as opposed to a massive field. At short distances, $r \ll
r_c$, the brane dynamics do not feel the width of the resonance, and
the theory resembles 4D GR. At large distances, $r \gg r_c$, however,
the theory becomes five dimensional as the resonance effectively
decays into continuum Kaluza-Klein modes. It is claimed that the
Vainshtein mechanism works well on the normal branch of DGP, screening
the longitudinal graviton without introducing any new pathological
modes, in contrast to massive gravity \cite{Deffayet-ghosts,
  Deffayet-nonperturbative}.  The breakdown of classical perturbation
theory at the Vainshtein scale can be linked to quantum fluctuations
on the vacuum becoming strongly coupled at around $1000$ km
\cite{LutyPorratiRattazzi2003, Rubakov-strong, Dvali-predictive}.

\subsubsection{Action, equations of motion, and vacua}
\label{sec:dgp-actionetc}

The DGP model contains a single 3-brane embedded in an otherwise empty
five dimensional bulk space-time.  Generalisations that include a bulk cosmological constant and/or bulk branes have been studied \cite{Kiritsis4,bigravity-me}.  However, the original DGP action is
given by
\be\label{dgp-action}
S = M_5^3 \int_{\cal M} d^5 x \sqrt{-\gamma}  \R + \int_{\del {\cal
    M}} d^4 x \sqrt{-g}\left[-2 M_5^3 K+\frac{M_4^2}{2} R-\sigma
  +{\cal L}_{\textrm{matter}}\right] ,
\ee
where, by $\mathbb{Z}_2$ symmetry across the brane,   we identify the
entire bulk space-time with two identical copies of $\cal M$, and the
brane with the common boundary, $\del \cal M$.  The bulk metric is
given by $\g_{ab}$, with corresponding Ricci scalar, $\R$, and $M_5$
is the Planck scale in the bulk.   The induced metric on the brane is
given by $g_{\mu\nu}$, and $K=g^{\mu\nu} K_{\mu\nu}$ is the trace of
extrinsic curvature, $K_{\mu\nu}=\half {\cal L}_n g_\mn$. Here we
define  the unit normal  to point {\it into}  ${\cal M}$.

The  brane has a bare vacuum energy, or tension, $\sigma$, and
additional matter contributions contained in ${\cal L}_{matter}$.
However, the key feature in the brane Lagrangian is the induced
curvature term, given by $M_4^2 R$. Such a term can be generated by
matter loop corrections \cite{Birrell-quantum}, or finite width
effects \cite{Carter-curvature}. The mass scale, $M_4$, is taken to be
Planckian, $\sim 10^{18}$ GeV. There is a hierarchy between this scale
and the bulk Planck scale, $M_4 \gg M_5$, which has proven difficult
to derive from fundamental theory. Nevertheless, the hierarchy enables
us to identify a crossover scale, $r_c \sim M_4^2/2M_5^3$, below which
the theory looks four dimensional, and above which it looks five
dimensional.

The  bulk equations of motion  are given by the vacuum Einstein equations, 
\be \label{dgo-einstein}
{\cal G}_{ab}=\R_{ab}-\half \R \g_{ab}=0,
\ee
and the boundary conditions at the brane are given by the Israel junction conditions,
\be \label{dgp-israel}
2M_5^2(K_{\mu\nu}-K g_{\mu\nu})=M_4^2 G_\mn+\sigma g_\mn-\T_\mn ,
\ee
where  $\T_{\mu\nu}=-\frac{2}{\sqrt{-g}} \frac{\delta }{\delta
  g^{\mu\nu}} \int_{\del \cal M} d^4 x \sqrt{-g} {\cal L}_{\textrm{matter}}$ is
the energy-momentum tensor for the additional matter.  

Let us now consider the vacua of this theory, corresponding to
maximally symmetric brane solutions with $\T_\mn=0$. For a given
tension,  there exist two distinct  vacua \cite{Deffayet2000, Deffayet-accelerated}. Assuming $\sigma>0$ for definiteness, one
finds that these correspond to de Sitter branes with intrinsic
curvature,
\be \label{dgpH}
H=\frac{M_5^3}{M_4^2} \left[\epsilon+\sqrt{1+\frac{M_4^2 \sigma}{6
      M_5^3}} \right] ,
\ee
where $\epsilon=\pm 1$. In conformal coordinates, the full solution can be written as
\be \label{dgpds2}
ds^2=\bar \g_{ab} dx^a dx^b=e^{2\epsilon Hy} (dy^2+\bar g_\mn dx^\mu
dx^\nu) ,
\ee
where $\bar g_\mn dx^\mu dx^\nu=-dt^2+e^{2Ht} d{\bf x}^2$ is the $4D$
de Sitter line-element written in Poincar\' e coordinates. The domain
${\cal M}$ corresponds to $0<y<\infty$, while the brane is located on
the boundary at $y=0$.

The two branches of this solution are often referred to as the {\it
  normal} branch ($\epsilon=-1$), and the {\it self-accelerating}
branch ($\epsilon=1$). The latter is so-called because as we take the
limit of vanishing vacuum energy, $\sigma \to 0$, the metric on the
brane is still asymptotically de Sitter. The limiting de Sitter
curvature is given by the cross-over scale,  $H=2M_5^3/M_4^2 \sim
1/r_c$. If this were to account for dark energy today, the crossover
from four to five dimensions would have to occur at the horizon scale,
placing the fundamental  Planck scale at $M_5 \sim 10$ MeV or
so. However, as we will discuss in Section \ref{sec:dgpsa-lin}, this
branch of solution is unstable as it contains physical ghost
excitations. The normal branch is asymptotically Minkowski in the
limit $\sigma \to 0$, and does not suffer from a ghostly
pathology. On this branch it is phenomenologically more interesting
to have the cross-over scale at distances below the horizon scale.

\begin{figure}
\begin{center}
\epsfig{file=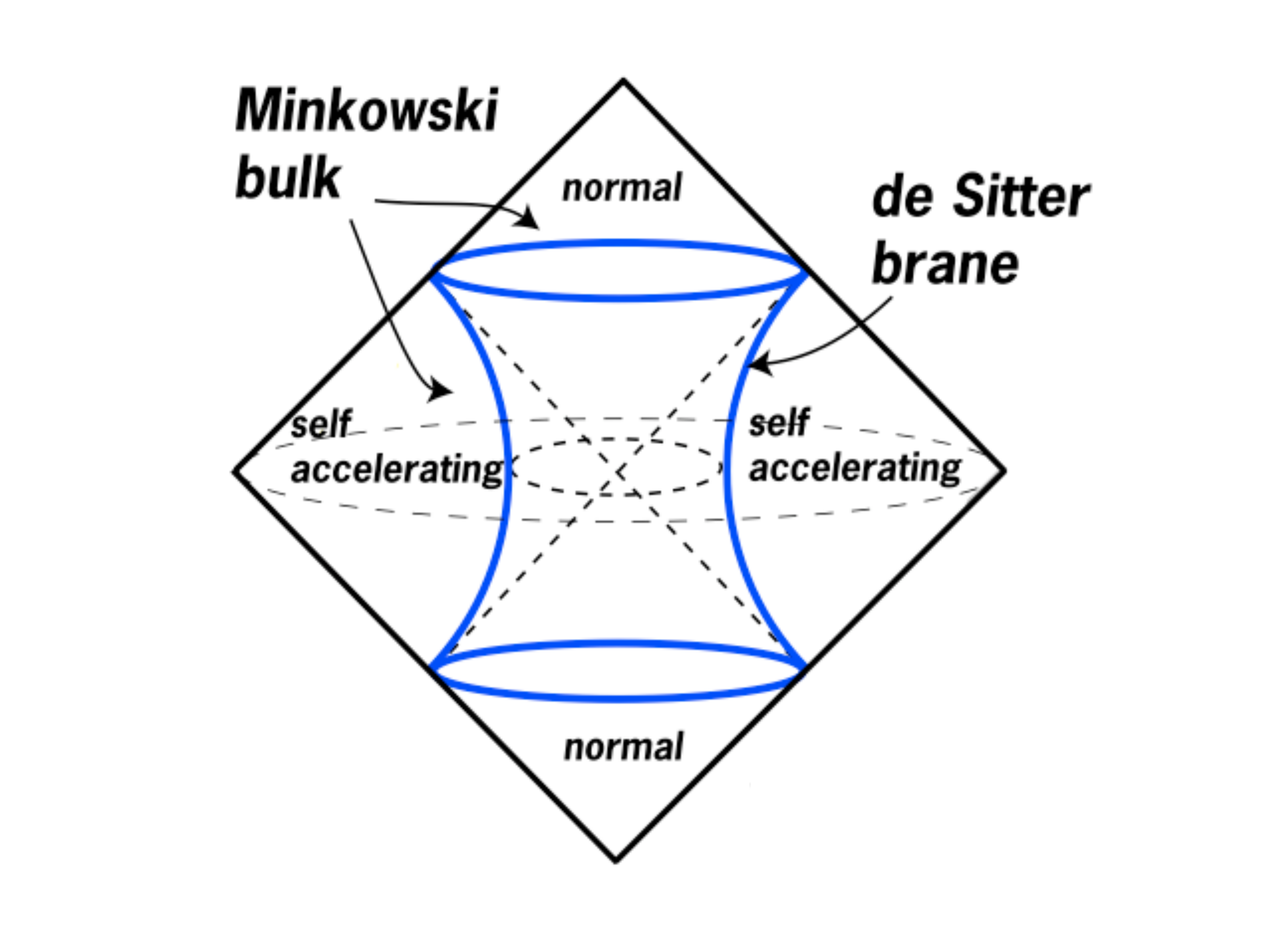,width=9cm,}
\caption{Embedding of a de Sitter brane in a flat 5D bulk, taken from
  \cite{dgpspec, dgppersp}. The braneworld volume is the hyperboloid
  in the Minkowski bulk. The normal branch corresponds to keeping the
  interior of the hyperboloid, while the self-accelerating branch
  corresponds to the exterior. } \label{fig:dgp}
\end{center}
\end{figure}

The distinction between branches is best understood by considering
their embedding in the $5D$ Minkowski bulk. In each case, the brane
can be viewed as a $4D$ hyperboloid of radius $1/H$ (see Figure
\ref{fig:dgp}).
The choice of sign, $\epsilon=\pm 1$, corresponds to the choice in
whether one retains the interior ($\epsilon=-1$) or the exterior
($\epsilon=1$) of the hyperboloid. Note that retaining the exterior
ensures that the brane behaves like a  domain wall with  {\it
  negative} effective tension, even when $\sigma \geq 0$
\cite{dgppersp}. This is an early warning sign that the
self-accelerating branch could be pathological.

\subsubsection{Linear perturbations on the normal branch} 
\label{sec:dgp-lin-nb}

We now consider linear perturbations on the normal branch. For
simplicity and brevity, let us consider the limiting case of vanishing
vacuum energy, $\sigma=0$, corresponding to fluctuations about a
Minkowski brane in a $5D$ Minkowski bulk, $\g_{ab}=\eta_{ab}+h_{ab}$.
It is convenient to choose harmonic gauge, 
\be \label{dedonder}
\del_a h^a_b=\half \del_b h^a_a ,
\ee
so that the linearised bulk equations of motion take the simple form
\be
[\del_y^2+\del^2 ]h_{ab}=0 ,
\ee
where $\del^2=\del_\mu \del^\mu$. Taking Fourier transforms with
respect to the brane coordinates, so that 
$$
Q(x, y) \to \tilde Q(p, y)=\frac{1}{(2\pi)^2} \int d^4x e^{i p_\mu
  x^\mu} Q(x, y) ,
$$
it can be seen that the normalisable solution is given by
\be
\tilde h_{ab}(p, y)=e^{-py}\tilde h_{ab}(p, 0) .
\ee
We have not yet fully fixed the gauge as there is still a residual
symmetry corresponding to gauge transformations of the form $x^a \to
x^a +\xi^a$, where $(\del_y^2 +\del^2 )  \xi^a=0$, or, equivalently,
$\tilde \xi^a(p, y)=e^{-py} \tilde \xi^a (p, 0)$.  We now  use $\xi^y$
to fix the brane position to lie at $y=0$, and $\xi^\mu$ to set
$h_{\mu y}=0$.  It follows from the $y$ component of Equation
(\ref{dedonder}) that $h_{yy}=h$, where $h=\eta^\mn h_\mn$. The $\nu$
component of Eq. (\ref{dedonder}) then gives $\del^\mu h_\mn=\del_\nu h$.

The junction conditions at the brane, 
$$
2M_5^3\delta (K_{\mu\nu}-K g_\mn)=M_4^2 \delta G_\mn- \T_\mn ,
$$
now yield the following solution for the metric on the brane:
\be
\tilde h_\mn (p, 0)= \frac{2}{M_4^2 p^2+2M_5^3 p} \left[\tilde
  \T_\mn-\frac{1}{3}  \tilde \T \eta_\mn\right] +\textrm{pure gauge
  pieces}.
\ee
At low energies, $p \ll 1/r_c =2M_5^3/M_4^2$, this becomes
\be
\tilde h_\mn (p, 0)\sim \frac{1}{M_5^3 p} \left[ \tilde
  \T_\mn-\frac{1}{3}  \tilde \T \eta_\mn\right] ,
\ee
which suggests that the potential for a point mass will scale like
$V(r) \propto  1/r^2$ at large distances, $r \gg r_c$. This is
consistent with $5D$ gravity, and is to be expected for an infinitely
large $5$th dimension. In contrast, at high energies, $p \gg 1/r_c$,
we have 
\be
\tilde h_\mn (p, 0) \sim  \frac{2}{M_4^2 p^2} \left[ \tilde
  \T_\mn-\frac{1}{3}  \tilde \T \eta_\mn\right] .
\ee
This can be compared with the propagator in four-dimensional General
Relativity,  $\tilde h_\mn^{GR} = \frac{2}{M_{pl}^2 p^2} \left[ \tilde
  \T_\mn-\frac{1}{2}  \tilde \T \eta_\mn\right]$. Thus we {\it almost}
have agreement. Indeed,  since the scaling with momenta is the same,
so the potential for a point mass will go like $V(r) \propto 1/r$ for
$r \ll r_c$.  We can understand this as follows: The graviton
mediating the interaction between two point particles along the brane
is bound close to the brane by the  induced curvature there. It is
only at very large distances, when the induced curvature is
insignificant, that the graviton is able to probe the extra
dimension.

However, at this stage it would be premature to say that we have
localisation of gravity at short distances.  The problem lies with the
tensor structure of the propagator.  As we can see, the GR result
differs from DGP in that it has a factor of $\half$ as opposed to
$\frac{1}{3}$. This is due to the longitudinal mode of the graviton
propagating in DGP but not in GR, and will inevitably lead to
disagreement with observation. To see this, consider the amplitude for
the interaction between two sources, $ \T_\mn$ and $T_\mn'$, which is
given by ${\cal A}=h_\mn \T'^\mn$. In GR, we have
\be
\tilde {\cal A}_{GR}=   \frac{2}{M_{pl}^2 p^2} \left[ \tilde T_\mn(p)
  \tilde T'^\mn(-p)-\frac{1}{2}  \tilde \T(p)  \tilde \T'(-p)\right] ,
\ee
whereas in DGP, at high energies, we have
\be
\tilde {\cal A}_{DGP} \sim    \frac{2}{M_{4}^2 p^2} \left[ \tilde
  T_\mn(p)  \tilde T'^\mn(-p)-\frac{1}{3}  \tilde \T(p)  \tilde
  \T'(-p)\right] .
\ee
To get the same potential for the interaction of two non-relativistic
sources, we require that\footnote{To verify this result, simply insert
  the appropriate energy-momentum tensors for non-relativistic
  sources, $ \T_\mn=diag(\rho, 0, 0, 0)$, and $\T'_\mn=diag(\rho', 0, 0,
  0)$, in each expression for the amplitude, and match the two results.}
$M_4^2=\frac{4}{3} M_{pl}^2$. Now consider what happens when one of
the sources is relativistic, say $\T'_\mn$. We then have $ T'=0$, and
so the amplitudes differ by a factor of $\frac{3}{4}$. So, although we
can recover the standard Newtonian potential, the prediction for light
bending around the Sun differs from GR by a factor of $\frac{3}{4}$.

This is  reminiscent of the vDVZ discontinuity in massive gravity
\cite{vdvz1, vdvz2}. In fact, it is sometimes said that DGP gravity
(on the normal branch)  is the closest thing we have to a non-linear
completion of massive gravity. Strictly speaking,  the behaviour is
more like  a massive resonance of width $1/r_c$, with decay into a
continuum of massive modes occurring at $r > r_c$. However, the
similarities with massive gravity do suggest a possible resolution to
this issue of tensor structure, and  the resulting phenomenological
problems. In massive gravity, it has been argued that the linearised
analysis around a heavy source breaks down at the so-called  {\it
  Vainshtein} radius, which can be much larger than the Schwarzschild
radius \cite{vainshtein}.  Below the Vainshtein radius, it is claimed
that non-linear effects help to screen the longitudinal scalar and
match the theory to GR without a vDVZ discontinuity. Similar claims
have been made in DGP gravity \cite{Deffayet-nonperturbative}, as we
will discuss in Section \ref{sec:dgp-sc}.

\subsubsection{Linear perturbations (and ghosts) on the
  self-accelerating branch} 
\label{sec:dgpsa-lin}

Let us consider linearised perturbations and ghosts about the self-accelerating
vacuum described in Section \ref{sec:dgp-actionetc} (see also
\cite{dgpspec, dgppersp, dgpspecrev, koyama-more, Koyama-ghosts}). To
reveal the ghost in the cleanest way possible, it is convenient to
retain non-zero tension on the background.  It is also enough to
consider vacuum fluctuations, so we will set $ \T_\mn=0$ for brevity
(a more complete discussion including matter fluctuations  can be
found in~\cite{dgpspec}).

Recall that the self-accelerating background solution is given by the
geometry in Eq. (\ref{dgpds2}), with $\epsilon=+1$ and the brane
positioned at $y=0$. A generic perturbation can be described by
$\g_{ab}=\bar \g_{ab}+\delta \g_{ab}$, with the brane position
shifted to $y=F(x)$. We will work in Gaussian Normal (GN)
coordinates, so that
\be
\delta \g_{yy}=\delta \g_{\mu y}=0, \qquad {\rm and} \qquad \delta
\g_\mn=e^{Hy/2} h_\mn(x, y) .
\ee 
The tensor $h_\mn$ can be decomposed in terms of the irreducible
representations of the $4D$ de Sitter diffeomorphism group as
\be
h_{\mu\nu} = h^{  TT}_{\mu\nu} + \nabla_\mu A_\nu + \nabla_\nu A_\mu +
\nabla_\mu \nabla_\nu \phi - \frac{1}{4} \bar g_{\mu\nu} \Box \phi +
\frac{h}{4} \bar g_{\mu\nu},  \label{decomps} 
\ee
where $\nabla$ is the covariant derivative associated with the $4D$ de
Sitter metric, $\bar g_\mn$.  The tensor $h^{  TT}_{\mu\nu}$ is
transverse and trace-free, $\nabla^\mu h^{{  TT} }{}_\mn = \bar
g^\mn h^{  TT}_\mn = 0$, and  $A_\mu$ is transverse, $\nabla^\mu
A_\mu = 0$. In addition, we have two scalars,  $\phi$ and $h = \bar
g^\mn h_\mn$.

Following a similar approach to the one outlined for Randall-Sundrum
gravity in Section \ref{sec:RSperts}, we can now fix the position of
the brane to be at $y=0$ while remaining in GN gauge by making the
following gauge transformation:
\be
y \to y-Fe^{-Hy}, \qquad {\rm and} \qquad x^\mu \to
x^\mu-\frac{e^{-Hy}}{H} \nabla^\mu F .
\ee
Although the brane position is now fixed at $y=0$,  the original brane
position $F(x)$ still enters the dynamics through a book-keeping term,
$h_\mn^{(F)}$, that  modifies the metric perturbation as
\be
h_\mn \to  h^{  TT}_{\mu\nu} + \nabla_\mu A_\nu + \nabla_\nu A_\mu +
\nabla_\mu \nabla_\nu \phi - \frac14 \bar g_{\mu\nu} \Box \phi +
\frac{h}{4} \bar g_{\mu\nu}+h_\mn^{(F)} .
\ee
The book-keeping term is pure gauge in the bulk, and is given by
\be
h_\mn^{(F)}=\frac{2}{H}e^{Hy/2}\left(\nabla_\mu \nabla_\nu+H^2\bar
g_\mn\right)F .
\ee
We can now substitute our modified expression for $h_\mn$ into the
linearised fields equations in the bulk, $\delta \G_{ab}=0$, and on
the brane, $2M_5^3\delta (K_{\mu\nu}-K g_\mn)=M_4^2 \delta G_\mn$. It
turns out that the transverse vector, $A_\mu$, is a free field in the
linearised theory and can be set to zero. In addition, the $yy$ and
$y\mu$ equations in the bulk imply that one can consistently choose a
gauge for which
\be
h=0, \qquad {\rm and} \qquad (\Box+4H^2)\phi=0 .
\ee
Note that we now have $h_\mn= h^{ 
  TT}_{\mu\nu}+h_\mn^{(\phi)}+h_\mn^{(F)}$, where the contribution
from $\phi(x, y)$ has been  rewritten as
\be
h_\mn^{(\phi)}=\left(\nabla_\mu \nabla_\nu+H^2\bar g_\mn\right)\phi(x, y).
\ee
This mode is now entirely transverse and trace-free in its own
right. In the absence of any additional matter on the brane,
$\T_\mn=0$, the same is true of the book-keeping mode,
$h_\mn^{(F)}$. This is because the trace of the Israel equation now
implies that
\be
(\Box+4H^2)F=0.
\ee
The entire perturbation $h_\mn(x, y)$ is therefore now completely
transverse and trace-free. This greatly simplifies the bulk and brane
equations of motion, giving
\begin{eqnarray}
\left[ \Box-2H^2+\del_y^2-\frac{9H^2}{4}\right]h_\mn(x, y)=0 &&\qquad \textrm{ for $y>0$,} \\
\left[M_4^2(\Box-2H^2)+2M_5^3\left(\del_y-\frac{3H}{2}\right)\right]h_\mn=0
&& \qquad\textrm{ at $y=0^+$.}
\end{eqnarray}
Variables in the tensor and scalar fields can now be separated as follows:
\be
h^{  TT}_{\mu\nu}(x, y)=\sum_{m} u_m(y) \chi^{(m)}_\mn(x), \qquad
\phi(x, y)=W(y)\hat \phi(x) ,
\ee
where $\chi_\mn^{(m)}$ is $4D$ tensor field of mass $m$ satisfying
$(\Box-2H^2)\chi_\mn^{(m)}=m^2\chi_\mn^{(m)}$. Note that $\hat \phi$
is a $4D$ tachyonic field satisfying $(\Box+4H^2)\hat \phi=0$. This is
a mild instability, related to the repulsive nature of inflating
domain walls.

Assuming that the tensor and scalar equations of motion can be treated
independently, we find that there is a continuum of normalisable
tensor modes with mass $m^2 \geq 9H^2/4$. In addition, there is also a
discrete tensor mode with mass
\be
m_d^2=\frac{1}{r_c}\left(3H-\frac{1}{r_c}\right),
\ee
and normalisable wave-function
$u_{m_d}(y)=\alpha_{m_d}e^{-y\sqrt{\frac{9H^2}{4}-m_d^2}}$. Now, for
{\it positive} brane tension, $\sigma >0$, one can easily check that
$0<m_d^2<2H^2$. For massive gravitons propagating in $4D$ de Sitter,
it is well known that masses lying in this range result in the
graviton containing a helicity-$0$ ghost~\cite{Higuchi-forbidden}.
The lightest tensor mode in this case is therefore perturbatively
unstable. For {\it negative} brane tension we find $m_d^2>2H^2$, and
there is no helicity-$0$ ghost in the lightest tensor.

Now consider the scalar equations of motion. The first thing to note
is that $h_\mn^{(\phi)}$ behaves like a transverse trace-free mode
with mass $m^2_\phi=2H^2$, because $(\Box-2H^2)h_\mn^{(\phi)}=2H^2
h_\mn^{(\phi)}$. Since none of the tensor modes have this mass, they
are all orthogonal to $h_\mn^{(\phi)}$. This means it was consistent
to assume that the scalar and tensor equations of motion could indeed
be treated independently. It turns out that the scalar has a {\it
  normalisable} wave-function $W(y)=e^{-Hy/2}$, and  the $4D$ scalar
$\hat \phi$ is sourced by $F$ by the relation
\be
\hat \phi(x)=\alpha F(x), \qquad
\alpha=-\left[\frac{2H-\frac{1}{r_c}}{H\left(H-\frac{1}{r_c}\right)}\right] .
\ee 
This is well defined for $\sigma \neq 0$, as in this case $H \neq
1/r_c$. $h_\mn^{(\phi)}(x, y)$ may now be thought of as a genuine
radion mode, measuring the physical motion of the brane with respect
to infinity. It does {\it not} decouple even though we only have a
single brane. This property is related to the fact that the background
warp factor, $e^{2Hy}$, grows as we move deeper into the bulk.

We have now identified the  helicity-$0$ mode of the lightest tensor
as a ghost when $\sigma >0$. When $\sigma<0$, a calculation of the
$4D$ effective action reveals the ghost to be the radion (see
\cite{dgpspec, dgpspecrev}). Given that there is always a ghost for
non-zero tension, by continuity one may expect this to remain the case
when $\sigma=0$.
\newline
\newline
\noindent
{\it The limit of vanishing tension}
\newline

To study this limit more closely, let us first ask whether we can
trust the above solutions in the limit where $\sigma \to 0$.  In this
limit $H\to 1/r_c$, and the quantity $\alpha$ becomes ill-defined.  To
understand what has gone wrong, note that the mass of the lightest
tensor has the limit $m_d^2 \to 2H^2$. This means that it is no longer
orthogonal to the radion, $h_\mn^{(\phi)}$, and so we cannot treat the
tensor and scalar equations of motion independently. This behaviour
can be traced back to  an additional symmetry that appears in the
linearised theory in the limit of vanishing brane tension. It is
analogous to the ``partially massless limit" in the theory of a
massive graviton propagating in de Sitter
space~\cite{Higuchi-forbidden}. In {\it that} theory, the equations of
motion are invariant under the following redefinition of the graviton field:
\be
\chi_\mn^{(\sqrt{2} H)}(x)\to \chi_\mn^{(\sqrt{2} H)}(x)+(\nabla_\mu
\nabla_\nu+H^2\bar g_\mn)\psi(x) .
\ee
This field redefinition has the effect of extracting out part of the
helicity-$0$ mode from $\chi_\mn^{(\sqrt{2} H)}$, and as a result of
the symmetry this mode disappears from the spectrum. In DGP gravity
this shift must be accompanied by a shift in the scalar field, $\phi$,  
\be
\phi(x, y)\to \phi(x, y)-\alpha_{\sqrt{2}H} e^{-Hy/2}\psi(x)=\phi(x,
y)-\lim_{\sigma \to 0}\alpha_{m_d}
e^{-y\sqrt{\frac{9H^2}{4}-m_d^2}}\psi(x) ,
\ee
in order to render the overall perturbation, $h_\mn(x, y)$, invariant.
These $\psi$ shifts can be understood as extracting part of the
helicity-$0$ mode from $\chi_\mn^{(\sqrt{2} H)}$, and absorbing it
into a renormalisation of $\phi$. The symmetry will have the effect of
combining the helicity-$0$ mode and the radion into a single degree of
freedom. It is only after fixing this $\psi$ symmetry that we can
treat the scalar and tensor equations of motion independently of each
another. We could also consider extracting the entire helicity-$0$
mode and absorbing it into $ \phi$, or {\it vice versa}.

Actually, there exists a clever choice of gauge that enables us  to
take a smooth limit as $\sigma \to 0$~\cite{koyama-more}.   One can
then readily  calculate the $4D$ effective action \cite{dgpspec}, and
derive the corresponding Hamiltonian \cite{koyama-more}.  It turns out
that the Hamiltonian is unbounded from below, signalling a ghost-like
instability \cite{koyama-more}. This ghost is a combination of the
radion and helicity-$0$ mode, and represents the residual scalar
degree of freedom left over after fixing the aforementioned $\psi$
symmetry. 
\newline
\newline
\noindent
{\it Ghosts}
\newline

We have shown that for any value of the brane tension, perturbations
about the self-accelerating branch of DGP contain a ghost. As
explained in Section \ref{sec:intro-ghosts}, the ghost will generate
catastrophic instabilities as it couples  universally with
gravitational strength to the Standard Model fields and the remaining
gravity modes. The existence of the ghost can be trusted  as long as
we can trust our effective description.  It had been argued that this
breaks down at the Hubble energy scale, casting doubt on the ghost's
existence at sub-horizon distances
\cite{Deffayet-perturbations}. However the analysis of
\cite{KoyamaSilva2007} suggests that the cut-off for the effective
theory is actually at a much higher energy,  being  the same as that
on the normal branch. As we will see in the next section, for
fluctuations on the vacuum this corresponds to $\Lambda_{\textrm{cut-off}}
\sim 10^{-13}$ eV. As this is well above the characteristic scale of
the self-accelerating vacuum, $H_0 \sim 10^{-33}$ eV, the ghost  will
cause this vacuum to be infinitesimally short-lived due to a divergent
rate of particle creation\footnote{Note that it has been argued that
  this choice of vacuum state explicitly breaks de Sitter invariance
  \cite{Izumi-no-de-sitter}, and so one is free to impose a Lorentz
  non-invariant  cut-off in the $3$-momentum such that the creation
  rate is no longer divergent.}.

It is also worth pointing out that the ghost can manifest itself
beyond perturbation theory.  In the presence of a self-accelerating
brane one can accommodate a Schwarzschild bulk with {\it negative}
mass, without introducing any naked singularities. This demonstrates
that the five-dimensional energy is unbounded from below, as suggested
by the perturbative analysis. Furthermore,  standard Euclidean
techniques indicate that the spontaneous nucleation of
self-accelerating branes is unsuppressed. This is problematic even on
the normal branch, and suggests that self-accelerating branes should
be projected out the  from the theory altogether \cite{dgppersp}.

\subsubsection{From strong coupling to  the Vainshtein mechanism} 
\label{sec:dgp-sc}

We now return to the normal branch of DGP gravity. In Section
\ref{sec:dgp-lin-nb}, we saw some similarities between this theory and
massive gravity.  Indeed, it happens that the brane to brane graviton
propagator is given by 
\be
D^{  DGP}_{\mn \alpha \beta}(p)=D^{  \textrm{massive}}_{\mn \alpha \beta}(p, p/r_c),
\ee
where $D_{\mn \alpha \beta}^{  \textrm{massive}}(p, m^2)$ is the propagator for
$4D$ massive gravity. Thus, just as massive gravity suffers from the
vDVZ discontinuity \cite{vdvz1, vdvz2} as $m \to 0$, so does DGP
gravity (normal branch) in the limit $p \ll 1/r_c$. This means the
linearised theory is not reduced to GR at short distances, but to a
four-dimensional scalar-tensor theory. If this description can be
trusted at solar system scales, then it leads to wildly inaccurate
predictions for the bending of light around the Sun.

In massive gravity, it turns out that the linearised theory cannot
necessarily be trusted  at solar system scales. This is because it
breaks down at the Vainshtein radius, $r_V^{  \textrm{massive}} \sim
(r_s/m^4)^{1/5}$, where $r_s \sim  3$ km is the Schwarzschild radius
of the Sun. For quantum fluctuations in vacuo, the breakdown of
classical perturbation theory at $r_V^{  \textrm{massive}}$ translates into a
strong coupling scale $\Lambda_{  \textrm{cut-off}} \sim (M_{pl} m^4)^{1/5}$
\cite{ArkaniHamed-effective}. Note that the equivalent scales in
four-dimensional General Relativity are  simply the Schwarzschild
radius and the Planck scale.
\newline
\newline
\noindent
{\it Strong coupling in DGP gravity}
\newline

The situation in DGP gravity (normal branch) is similar to the case of
massive gravity, as we will now show. While aspects of the strong coupling problem were first identified in \cite{Kiritsis3, Rubakov-strong}, the derivation of  the strong
coupling scale for quantum fluctuations on the Minkowski brane is most
elegantly presented in \cite{LutyPorratiRattazzi2003}.  Here we review the results
of this study\footnote{Correcting a few typos along the way.} (see
also \cite{bigal1}). Their analysis involves a computation of the
boundary effective action\footnote{The boundary effective action,
  $\Gamma[\phi]$,  is obtained by integrating  the full action,
  $S_{\cal M}[\Phi]+S_{\del \cal M}[\phi]$, over bulk fields, $\Phi$,
  satisfying the boundary condition $\Phi |_{\del \cal M}=\phi$, i.e. 
\be
e^{i \Gamma[\phi]}=\int_{\Phi|_{\del \cal M}=\phi} d[\Phi]
e^{i\left(S_{\cal M}[\Phi]+S_{\del \cal M}[\phi]\right)} .
\ee
}, which is found to be
\begin{multline}
\Gamma[ h_\mn, N_\mu, h_{yy}]=\int d^4
x~\frac{M_4^2}{4}\left[\frac{1}{2}
  h^\mn\left(\del^2-\frac{\sqrt{-\del^2}}{r_c}\right)
  h_\mn-\frac{1}{4}
  h\left(\del^2-\frac{\sqrt{-\del^2}}{r_c}\right)h\right.  \\
\left. -\frac{1}{r_c} N^\mu \left(\sqrt{-\del^2}+\frac{1}{r_c} \right)
N_\mu+\frac{1}{2r_c} h \sqrt{-\del^2} h_{yy}-\frac{1}{r_c} N^\mu
\del_\mu h_{yy}-\frac{1}{4r_c} h_{yy} \sqrt{-\del^2} h_{yy}\right] \\ 
+\half \int d^4x ~h_\mn  \T^\mn+\Gamma_{  int}[ h_\mn, N_\mu, h_{yy}],
\end{multline}
where $N_\mu=h_{\mu y}$ and $\del^2=\del_\mu \del^\mu$. The function
$\Gamma_{  int}$ contains all the higher-order interaction terms to be
discussed shortly. For now let us focus on the quadratic term. This
can be diagonalised by means of the following field redefinition
\be
h_\mn=\tilde h_\mn+\pi \eta_\mn, \qquad N_\mu=\tilde N_\mu+r_c
\del_\mu \pi, \qquad h_{yy} =-2r_c\left(
\sqrt{-\del^2}+\frac{1}{r_c}\right)\pi .
\ee
The effective action can then be written as
\begin{multline}
\Gamma[\tilde h_\mn, \tilde N_\mu, \pi]=\int d^4
x~\frac{M_4^2}{4}\left[\frac{1}{2} \tilde
  h^\mn\left(\del^2-\frac{\sqrt{-\del^2}}{r_c}\right) \tilde
  h_\mn-\frac{1}{4} \tilde
  h\left(\del^2-\frac{\sqrt{-\del^2}}{r_c}\right)\tilde h\right. \\ 
\left. -\frac{1}{r_c} \tilde N^\mu \left(\sqrt{-\del^2}+\frac{1}{r_c}
\right) \tilde N_\mu+3 \pi
\left(\del^2-\frac{\sqrt{-\del^2}}{r_c}\right) \pi\right]\\ 
+\half \int d^4x ~(\tilde h_\mn  \T^\mn+\pi \T)+\Gamma_{  int}[ \tilde
  h_\mn, \tilde N_\mu, \pi] .
\end{multline}
The next step is to write everything in terms of the canonically normalised fields
\be
\hat h_\mn=\frac{M_4}{{2}} \tilde h_\mn, \qquad \hat
N_\mu=\frac{M_4}{\sqrt{2r_c}} \tilde N_\mu, \qquad \hat \pi
=\sqrt{\frac{3}{2}} M_4 \pi .
\ee
This gives
\begin{multline}
\Gamma[\hat h_\mn, \hat N_\mu, \hat \pi]=\int d^4 x~\frac{1}{2} \hat
h^\mn\left(\del^2-\frac{\sqrt{-\del^2}}{r_c}\right) \hat
h_\mn-\frac{1}{4} \hat
h\left(\del^2-\frac{\sqrt{-\del^2}}{r_c}\right)\hat h \\ 
 -\frac{1}{2} \hat N^\mu \left(\sqrt{-\del^2}+\frac{1}{r_c} \right)
 \hat N_\mu+\frac{1}{2}\hat \pi
 \left(\del^2-\frac{\sqrt{-\del^2}}{r_c}\right) \hat \pi\\ 
+ \int d^4x ~\left(\frac{1}{M_4}\hat  h_\mn  \T^\mn+\frac{1}{\sqrt{6}
  M_4}\hat \pi \T \right)+\Gamma_{  \textrm{int}}[ \hat h_\mn, \hat N_\mu, \hat
  \pi] .
\end{multline}
Turning our  attention to the interaction piece, we note that it
contains terms that schematically take  the form  
\be
\int d^4x M_5^3 \sqrt{-\del^2} ( h_\mn)^a (N_\mu)^b (h_{yy})^c, \qquad
\int d^4x M_4^2 \sqrt{-\del^2}^2 (h_\mn)^d ,
\ee
where $a+b+c \geq 3$, and $d \geq 3$. By writing these in terms of the
canonically normalised fields  one can easily check that the largest
interaction comes from the term with $a=0$ and $b+c=3$. Indeed, for
$\sqrt{-\del^2} \gg 1/r_c$, the interaction term goes like
\cite{LutyPorratiRattazzi2003, bigal1} 
\be \label{Gamma}
\Gamma_{ \textrm{int}}=-\frac{1}{3\sqrt{6} \Lambda^3} \int d^4 x (\del \hat
\pi)^2 \del^2 \hat \pi+\textrm{sub-leading interactions} ,
\ee
where 
\be
\Lambda=(M_4/r_c^2)^{1/3}.
\ee
This corresponds to the scale at which quantum fluctuations in vacuum
become strongly coupled. For $M_4 \sim M_{pl}$, and $r_c \sim 1/H_0$,
we have  $\Lambda \sim 10^{-13}$ eV $\sim 1/(1000 {~km})$. In other
words, for scattering processes above $10^{-13}$ eV perturbative
quantum field theory in a vacuum is no longer well defined. Of course,
it is important to realise that we do not actually live in a vacuum,
and it has been demonstrated that the strong coupling scale can be
raised on curved backgrounds \cite{Nicolis-classical}. At the surface
of the earth, the new scale corresponds to  about $10^{-5}$ eV,
indicating the presence of  large quantum corrections to the $\pi$
field at distances below a centimetre. This will not impact much on
table top experiments, however, as the $\pi$ field is expected to be
overwhelmed by the classical graviton at this scale, due to
Vainshtein screening.
\newline
\newline
\noindent
{\it The $\pi$-Lagrangian}
\newline

To see the emergence of the new strong coupling scale most succinctly,
it is convenient to take the limit in which the troublesome $\pi$
field decouples from the graviton. This will also help us in our
discussion of the breakdown of classical perturbation theory around
heavy sources and the  Vainshtein effect. In any event, we take the
so-called decoupling limit \cite{Nicolis-classical}
\be
M_4, r_c, {\cal T}_\mn \to \infty, \qquad {\rm and} \qquad \Lambda,
~\frac{\cal T_\mn}{M_4}=\textrm{fixed},
\ee
so that all the sub-leading interactions in Eq. (\ref{Gamma}) go to
zero. The limiting theory should be valid at intermediate scales,
$$ 
\textrm{max}\left(L_4, r_s\right) \ll r\ll  r_c ,
$$ 
where $L_4 \sim 1/M_4$ is the Planck length, and $r_s$ is the
Schwarzschild radius of the source (if present).  The action goes like 
$$
\Gamma \sim  \int d^4 x \left[{\cal L}_{GR}+{\cal L}_\pi\right],
$$
where
\ba
{\cal L}_{GR}&=& \frac{M_4^2}{4}\left[ \frac{1}{2}  \tilde h^\mn\del^2
  \tilde  h_\mn-\frac{1}{4}  \tilde h\del^2 \tilde h\right]
+\frac{1}{2} \tilde h_\mn \T^\mn , \label{GRlag} \\
{\cal L}_\pi &=& \frac{M_4^2}{4}\left[ 3  \pi \del^2  \pi-r_c^2 (\del
  \pi)^2 \del^2 \pi\right]+\frac{1}{2}\pi \T , \label{pilag}
 \ea
and we have set the free field $N_\mu=0$. Note that ${\cal L}_{GR}$ is
just the standard Einstein-Hilbert Lagrangian, expanded to quadratic
order about Minkowski space. The modification to GR is encoded in
${\cal L}_\pi$. This is often referred to as the $\pi$-Lagrangian and
much of the interesting phenomenology of DGP gravity, from strong
coupling and Vainshtein effects on the normal branch to ghosts on the
self-accelerating branch, can be studied using this Lagrangian.

In vacuum ($\T_\mn=0$) the $\pi$-Lagrangian possesses a symmetry $\pi
\to \pi+a_\mu x^\mu +b$, where $a_\mu$ and $b$ are constants. This is
sometimes referred to as Galilean invariance and is the inspiration
for {\it galileon} models \cite{NicolisRattazziTrincherini2008}. The most general
Galilean invariant Lagrangians and will be studied in detail in
Section \ref{galileons}.  

We now return to the question of the strong coupling scale in the
theory described by Eq. (\ref{pilag}) in the presence of a classical
source.  The first thing to note is that our classical background is
no longer the vacuum, but the solution $\pi=\pi_{cl}(x)$  of the
classical field equations  
\be
3\del^2 \pi -r_c^2\left[ (\del_\mu \del_\nu \pi) ( \del^\mu \del^\nu
  \pi)-(\del^2 \pi)^2\right]=-\frac{\T}{M_4^2} . \label{pieqns}
 \ee
Our interest now lies in the quantum fluctuations,
$\pi=\pi_{cl}(x)+\varphi(x)$, about this solution. Plugging this into
the Lagrangian given in Eq. (\ref{pilag}) we find
\be \label{varphilag}
{\cal L}_\pi = \frac{M_4^2}{4} \left[-Z^\mn(x)\del_\mu \varphi
  \del_\nu \varphi-r_c^2 (\del \varphi)^2 \del^2 \varphi\right]+\half
\varphi \delta \T ,
\ee 
where 
\be
Z^\mn(x)=3 \eta^\mn-2r_c^2\left(\del^\mu \del^\nu -\eta^\mn\del^2
\right)\pi_{cl}  .
\ee
For non-trivial solutions $\pi_{cl}(x)$ is not constant, and so
neither is $Z^\mn(x)$. However, assuming the background varies slowly
(relative to the fluctuations) we can treat $Z^\mn$ as approximately
constant in a neighbourhood of a point in space-time.  Further
assuming that the eigen-values of  $Z^\mu_\nu(x)$ are all of similar
magnitude, $\sim Z(x)$, we can identify a localised strong coupling
scale,
\be
\Lambda_* (x) \sim \Lambda \sqrt{Z(x)} \gg \Lambda .
\ee
As we will see shortly, for spherically symmetric solutions inside the
Vainshtein radius, $r_V \sim (r_s r_c^2)^{1/3}$, we have $\pi_{cl} \sim
\sqrt{r_s r}/r_c$, and so $Z(r) \sim (r_V/ r)^{3/2}$. On the sphere of
radius $r$ the local strong-coupling scale is 
$$
\Lambda_*(r) \sim \Lambda \left(\frac{ r_V}{r} \right)^{3/4}.
$$
It follows that the classical background, $\pi_{cl}$, ceases to make
sense below a critical length, $r_\textrm{min}$, where $r_\textrm{min}
\Lambda_*(r_\textrm{min}) \sim 1$. This is because scattering
processes would need to exceed the local strong coupling scale to
probe the structure of the background at $r< r_\textrm{min}$. This
short distance cut-off will typically lie well within the
Schwarzschild radius of the source, and therefore outside of the
regime of validity of the $\pi$-Lagrangian.

We should also consider the implication of the local strong coupling
scale for energetic processes taking place on the surface of the
Earth. Here the Earth's gravitational field has the dominant effect,
so we have  $\Lambda_* \sim (\sqrt{r_c/r_H}) 10^{-5}$ eV, where $r_H
\sim 1/H_0 \sim 10^{26}$ m is the current Hubble radius. For $r_c \sim
r_H$ this means the fluctuations in the $\pi$ field becomes strongly
coupled at around $1$ cm.  However it is important to realise that it
is only the $\varphi$ self-interactions that grow large at this scale.
The direct coupling to the graviton and  the coupling to matter are
negligible, with the latter going like $G_4/Z(r) \sim 10^{-15}
G_4$. Although the scalar fluctuations  enter a quantum fog at the
centimetre scale they do  not spoil our classical description as this
is dominated by the graviton, which is fifteen orders of magnitude
larger \cite{Nicolis-classical}.

Finally, let us note another important feature of the scalar dynamics:
As Poincar\'e invariance is broken by the classical background,
$\pi_{cl}$, the fluctuations no longer have to propagate on the light
cone. Indeed, one can explicitly show that angular fluctuations about
spherically solutions will be super-luminal, causing problems for
causality \cite{Adams-causality, Hinterbichler-super}.
\newline
\newline
\noindent
{\it The Vainshtein effect in DGP gravity}
\newline

Let us now consider the Vainshtein effect directly.  As we will see,
this is a mechanism in which an additional  scalar mode is screened at
short distances by non-linear interactions, thereby eliminating the
troublesome vDVZ discontinuity. To see how the scalar is screened in
DGP gravity, we consider the classical solution to the field equations
around a heavy non-relativistic source, $T_\mn=\textrm{diag}(\rho(r), 0,0,0)$,
assuming spherical symmetry for simplicity. If we are to screen the
scalar and recover GR at a given scale, the classical graviton
solution, $\tilde h_\mn^{(cl)}$,  should dominate over the
corresponding scalar solution, $\pi_{cl}$.   In other words, we should
schematically have 
$$
|\tilde h_\mn^{(cl)}| \gg |\pi_{cl}|.
$$
Since $\tilde h_\mn^{(cl)}$ is derived from Eq. (\ref{GRlag}), it is
just the standard linearised GR solution given by the Newtonian
potential, 
\be
|h_\mn^{(cl)}| \sim \frac{r_s}{r} , \label{hsoln}
\ee
where $r_s=2G_4M$ is the Schwarzschild radius of the source,  $M=\int
\rho(r) dV$ is its mass, and $G_4=1/8\pi M_4^2$ is the four
dimensional Newton's constant.

In general, the dynamics of the scalar are governed by the Field
Equations (\ref{pieqns}). It follows that the spherically symmetric
solution, $\pi_{cl}(r)$, around our heavy source satisfies 
\be \label{pieqns1}
\pi_{cl}'+\frac{2r_c^2}{3r} \pi_{cl}'^2=\frac{r_s}{3r^2},
\ee 
where we have integrated once over a sphere centred at the origin and
enclosing the entire source. Now at large distances we expect the
linear term to dominate, giving a solution
\be
\pi^{(\textrm{lin})}_{cl}(r)=-\frac{r_s}{3r} . \label{pilin}
\ee
This scales in the same way as the corresponding graviton solution in
Eq. (\ref{hsoln}), and so we have an ${\cal O}(1)$ modification of
General Relativity. However, the linearised description breaks down
once the non-linear piece becomes comparable, or, equivalently, when 
\be
\left|   (\pi^{(\textrm{lin})}_{cl})' \right| \sim \left| \frac{r_c^2}{
  r}(\pi^{(\textrm{lin})}_{cl})'^2 \right| .
\ee
Substituting our solution from Eq. (\ref{pilin}), we see that the
linearised theory breaks down at the Vainshtein radius,
\be
r_V \sim  (r_s r_c^2)^{1/3}.
\ee
For the Sun we have $r_s^{\odot} \sim 3$ km, and so
\be
r_V^\odot \sim \left(\frac{r_c}{r_H}\right)^{2/3} 10^{18} \textrm{m}.
\ee
As long as the cross-over scale is not too far inside the cosmological
horizon, the Vainshtein effect extends beyond the edges of the solar
system, given by the Oort cloud at an average of $10^{16}$ m or so.
For the earth we have $r_s^{\oplus} \sim 9$ mm, and so 
\be
r_V^\oplus \sim \left(\frac{r_c}{r_H}\right)^{2/3}10^{17} \textrm{m} .
\ee
In each case, the breakdown of classical perturbation theory has the
same origin as the strong coupling of quantum fluctuations in vacuum,
namely the self-interaction of the scalar, $\pi$, with scale-dependent
coupling. Note that the Vainshtein scale can always be obtained from
the strong coupling scale by trading the Planck length for the
Schwarzschild radius of the source.
  
Now let us ask whether or not the scalar gets screened below the
Vainshtein radius. For $r \ll r_V$, the linearised solution in
Eq. (\ref{pilin}) is no longer a good approximation. On the contrary,
the non-linear part of Equation (\ref{pieqns1}) will dominate, so
that at short distances we have the solution
\be
\pi^{(\textrm{nonlin})}_{cl}(r)=\frac{\sqrt{2r_s r} }{~r_c}.
\ee
This implies that  at short distances the scalar correction to the
Schwarzschild solution goes like \cite{Gruzinov-on-the}
\be
\frac{\delta V}{V} \sim\frac{|\pi^{(\textrm{nonlin})}_{cl}|}{|h_\mn^{(cl)}|}
\sim \left(\frac{r}{r_V} \right)^{3/2} .
\ee
Given that $r_V \to \infty$ as  $r_c \to \infty$, we see that the
scalar does appear to  get screened in this limit, and that one is
able to recover GR without any vDVZ discontinuity.

We can now use the correction to the Newtonian force to place bounds
on the cross-over scale.  At $r \sim 5$ AU fractional corrections to
the Sun's gravitational field should be $\lesssim 10^{-8}$
\cite{planet}, which implies $r_c \gtrsim 10^{-4} r_H$.  Similarly,
the corrections to the Earth's gravitational field may have an
observable effect on the precession of the moon, where corrections to
the potential go like $ \delta V/V |_\textrm{moon} \sim 10^{-13}
r_H/r_c$ \cite{Dvali-moon}.

While we have demonstrated aspects of the Vainshtein mechanism at the
level of the $\pi$-Lagrangian, for simplicity, it is worth noting that
it was originally discussed  using  the full theory
\cite{Deffayet-nonperturbative, Gruzinov-on-the}. In
\cite{Deffayet-nonperturbative} it was emphasised how one must choose
the correct expansion parameter at a given scale.  Standard
perturbation theory corresponds to performing an expansion in $r_s/r$,
but this is not a good expansion parameter in the limit $r_c \to
\infty$, as then the next-to-leading order terms become singular. At
short distances, the claim is that one should expand in powers of
$r/r_V$ around the Schwarzschild solution \cite{Gruzinov-on-the}, so
the result can be smoothly patched onto the  standard perturbative
solution in some neighbourhood of the Vainshtein radius
\cite{Deffayet-nonperturbative}.


It is fair to say that the Vainshtein mechanism has not yet been fully
explored as a viable concept. While it is clear that there is a
breakdown of linear perturbation theory at the Vainshtein radius in
both massive gravity and DGP, the arguments for a smooth transition to
GR are at best promising, but hardly conclusive.  N-body simulations
of large-scale structure in  DGP gravity, however, support the case
for its successful implementation \cite{Scoccimarro2009,ChanScoccimarro2009}. We will discuss this further in the next section.

One should also worry about the {\it elephant problem}\footnote{We
  thank Nemanja Kaloper for this colourful observation.}: An elephant
is an extended object made up of many point particles each with their
own Vainshtein radius. Is the Vainshtein radius of the elephant the
same as the Vainshtein radius of a point particle with the same mass
  located at the centre of mass of the elephant?  While this may not
  be a problem when the size of the source particle distribution is
  much smaller than the Vainshtein radius, it clearly becomes an issue
  when the particles are well distributed over the whole Vainshtein
  sphere. How does one account for such many particle systems? The
  answer is not known and given the role of non-linearities in the
  Vainshtein mechanism the problem may well be very complicated,
  especially in the full covariant theory. See \cite{Kaloper-hair} for a recent discussion.

\subsubsection{DGP cosmology}

In this section we describe DGP cosmology, from the  FLRW background,
to linear perturbations and non-linear studies. A thorough review of
DGP cosmology up to 2006 can be found in~\cite{Lue2005}.

As discussed in Section \ref{sec:branebased},  Shiromizu, Maeda and
Sasaki~\cite{ShiromizuMaedaSasaki1999} found the $4D$ Einstein
equations of a $3$-brane world embedded in a $5D$ bulk with $Z_2$
symmetry.  Applying this formalism to DGP we get
\begin{equation}
G_{\mu\nu} = (16 \pi G r_c)^2 \Pi_{\mu\nu}  - E_{\mu\nu},
\label{G_DGP}
\end{equation}
where
\begin{equation}
\Pi_{\mu\nu} = -
\frac{1}{4}\TT_{\mu\lambda}\TT^\lambda_{\phantom{\lambda}\nu} +
\frac{1}{12} \TT \TT_{\mu\nu} + \frac{1}{24}\left[3  \TT^{\alpha\beta}
  \TT_{\alpha\beta}  - \TT^2 \right] g_{\mu\nu} ,
\end{equation}
and
\begin{equation}
\TT_{\mu\nu} =  \T_{\mu\nu} - \frac{1}{8\pi G} G_{\mu\nu}.
\end{equation}
The Bianchi identities give
\begin{equation}
\nabla_\nu E^\nu_{\phantom{\nu}\mu}=  (16 \pi G r_c)^2\nabla_\nu
\Pi^\nu_{\phantom{\nu}\mu},
\end{equation}
while the matter stress-energy tensor satisfies local energy-momentum
conservation: $\nabla_\nu T^\nu_{\phantom{\nu}\mu}=0$.

The cosmological solutions for an FLRW space-time were found by
Deffayet~\cite{Deffayet2000} by considering the embedding of the DGP
brane into a Minkowski bulk. This corresponds to setting
$E_{\mu\nu}=0$ and  plugging an FLRW metric into our field equations
(\ref{G_DGP}). Note that the global structure of the cosmological
solutions has been investigated by Lue~\cite{Lue2002}, while some
exact solutions in special cases have been found by
Dick~\cite{Dick2001}. If one were to generalise the bulk to include $5D$ Schwarzschild, one can run into an unusual class of singularity on account of the branch cut in Friedmann equation \cite{Shtanov2, dgppersp}. Cosmological branes in generalised DGP gravity can also mimic a phantom equation of state \cite{Sahni1,Lue1}. 

Staying with the simplest case of a Minkowski bulk and a tensionless brane, the $00$ component of the $4D$ field equations (with $E_{\mu\nu}=0$)
gives the Friedmann equation as
\begin{equation}
H^2 + \frac{\kappa}{a^2} - \frac{\epsilon}{r_c}\sqrt{H^2 +
  \frac{\kappa}{a^2}} = \frac{8\pi G}{3}\rho ,
\label{DGP_friedman}
\end{equation}
where $\epsilon=-1$ for the normal branch, and $\epsilon =1$ for the
self-accelerating  branch.  Clearly, if $\rho=0$ then the normal
branch implies Minkowski space, while the self-accelerating branch
gives de Sitter space with Hubble constant $H_\infty=
\frac{1}{r_c}$. This self-accelerating model has been proposed as an
alternative to dark energy~\cite{Deffayet-accelerated}, but as we
have already seen, the theory contains a ghost in its perturbative
spectrum. A phenomenological extension of Eq. (\ref{DGP_friedman}) has
been considered by Dvali and Turner~\cite{DvaliTurner2003}, where for
$\kappa=0$ and $\epsilon =1$ the term $\frac{H^{2\alpha}}{r_c^{2-\alpha}}$
is added to Eq. (\ref{DGP_friedman}).

After using Eq. (\ref{DGP_friedman}) the $ij$ component can be
re-worked into a modified Raychaudhuri equation, as
\begin{equation}
2\frac{dH}{dt} + 3 H^2 + \frac{\kappa}{a^2}  = -\frac{ 3 H^2 +
  \frac{3\kappa}{a^2} - 2 \epsilon r_c \sqrt{H^2 + \frac{\kappa}{a^2}}
  \;  8\pi G P}{1 - 2 \epsilon r_c \sqrt{H^2 + \frac{\kappa}{a^2}} } .
\label{DGP_ray}
\end{equation}
It is instructive to cast the background equations into a form
resembling an effective dark-energy fluid with density $\rho_E$ and
pressure $P_E$.  This gives
\begin{equation}
X =  8\pi G \rho_E =  \frac{3\epsilon}{r_c} \sqrt{H^2 + \frac{\kappa}{a^2}} ,
\end{equation}
and
\begin{equation}
Y =  8 \pi G P_E =  -\epsilon\frac{ \frac{dH}{dt} + 3 H^2 +
  \frac{2\kappa}{a^2} }{ r_c \sqrt{H^2 + \frac{\kappa}{a^2} }},
\end{equation}
so that we can define the equation of state, $w_E$, as
\begin{equation}
w_E = \frac{P_E}{\rho_E} = -\frac{ \frac{dH}{dt}  + 3 H^2 +
  \frac{2\kappa}{a^2}}{ 3H^2 + \frac{3\kappa}{a^2}  } .
\end{equation}
The Bianchi identities, and energy-momentum conservation, ensure that 
$$
\frac{d\rho_E}{dt} +3H (\rho_E + P_E) \equiv 0.
$$
As a consequence, during radiation domination we get $w_E \approx
-\frac{1}{3}$, during matter domination we get $w_E \approx
-\frac{1}{2}$, during a possible spatial curvature dominated era we
get $w_E \approx -\frac{2}{3}$, and eventually during the
self-accelerating era we get $w_E \rightarrow -1$. 

Let us now look at the perturbed FLRW universe in DGP. Perturbation
theory for DGP was originally worked out in detail by
Deffayet~\cite{Deffayet2002}, and has since been developed by a number
of authors. Although braneworld cosmological perturbation theory is
discussed in detail in Section \ref{sec:branecosperts}, we will now
present the formalism for DGP gravity explicitly, in the interests of
self-containment.

Let us consider scalar perturbations in the conformal Newtonian gauge. 
The components of $E_{\mu\nu}$ are then\footnote{Note that this is
  consistent with our general 
  analysis for brane world cosmological perturbations, provided one
  identifies $\mu_E=-8\pi G_4 \rho^{weyl}\delta^{weyl},
  ~\Theta_E=-8\pi G_4\left(\frac{4}{3} \rho^{weyl}\right)
  \theta^{weyl}, \Sigma_E=-8\pi  G_4\left(\frac{4}{3}
  \rho^{weyl}\right) \Sigma^{weyl}$.}
\begin{eqnarray}
E^0_{\;\; 0} &=& -\mu_E,
\\
E^0_{\;\; i} &=& - \grad_i \Theta_E,
\\
E^i_{\;\; j} &=&  \frac{1}{3}\mu_E \delta^i_{\;\;j}  +  D^i_{\;\;j} \Sigma_E.
\end{eqnarray}
We find the perturbed Einstein equations are then given by
\begin{eqnarray}
2 \left(\Delta + 3 \kappa \right) - 6 {\cal H} \left(\Phi' + {\cal
  H}\Psi\right) &=& A_D 8\pi G a^2 \rho \delta  + B_D a^2 \mu_E,
\qquad
\\
\left(\Phi' + {\cal H} \Psi\right) &=& A_D 8\pi G a^2 \left(\rho +
  P\right)\theta  + B_D a^2 \Theta_E,
\qquad
\end{eqnarray}
\begin{eqnarray}
&&
{\Phi}''+ \adotoa{\Psi}' +2\adotoa {\Phi}' + \left(2\adotoa'+  \adotoa^2 + \frac{1}{3}\Delta\right)\Psi -\left( \frac{1}{3}\Delta + \kappa \right) \Phi
\nonumber
\\
&=&
A_D 4\pi G a^2 \delta P
- \frac{B_D}{2} \left[
 (1 + w_E)   A_D \left( 8\pi G a^2 \rho \delta  + a^2 \mu_E\right)  - \frac{a^2}{2}\mu_E
\right],
\end{eqnarray}
\begin{eqnarray}
\Phi - \Psi &=& a^2 \frac{  r_c^2 (X+3Y) 8\pi G (\rho + P)\Sigma  - 3
  \Sigma_E  }{3 + r_c^2 (X+3Y)}, 
\end{eqnarray}
where we recall that primes denote differentiation with respect to
conformal time, and ${\cal H}=\frac{a'}{a}=aH$. Note that we have set
$A_D = \frac{2X r_c^2}{2Xr_c^2 - 3}$ and $B_D =   \frac{3}{2Xr_c^2 -
  3} =  A_D - 1$, for simplicity. In the limit $r_c \rightarrow
\infty$ one recovers the familiar perturbation equations of General
Relativity from the above.

We can now use the Bianchi identities to find the field equations for
$\mu_E$ and $\Theta_E$.  They are
\begin{equation}
{\mu}_E' + 4 {\cal H} \mu_E  - \Delta \Theta_E =0,
\label{DGP_delta_dot}
\end{equation}
and
\begin{eqnarray}
&&
{\Theta}_E' + 4 {\cal H} \Theta_E  - \frac{1}{3} \mu_E + \left(1 +
w_E\right)\left(\mu_E + 3 \frac{\cal H}{a} \Theta_E \right)
\nonumber
\\
&&
+ \frac{\Delta + 3\kappa}{1+3w_E} \left\{ \frac{4}{3} \Sigma_E  +
2\frac{1+w_E}{a^2} \left[ \left(2 + 3 w_E\right)\Phi - \Psi \right]
\right\}
=0.
\label{DGP_theta_dot}
\end{eqnarray}
It is clear that the above equations are not closed due to the
presence of the bulk anisotropic stress, $\Sigma_E$.  Several authors
have applied a variety of approximations to solve the linear
perturbations in the DGP model. Sawicki and
Carroll~\cite{SawickiCarroll2005} assumed that the Weyl perturbations
are zero.  It is clear, however, from Eq. (\ref{DGP_theta_dot}) that
this is not a consistent approximation.

To fully determine the DGP perturbations one must use the
five-dimensional equations. For maximally symmetric $5D$ space-times,
the bulk scalar mode perturbations can be deduced using a single
master variable, $\Omega$, as we discussed in Section
\ref{sec:BWcos}. Assuming that the bulk cosmological constant is zero,
the bulk metric is given by Equation (\ref{lambdaq}), with
\ba
&\lambda_{\alpha\beta}=dz^2-n^2(t, z)dt^2,  \qquad n(t, z) = 1 +
\frac{\epsilon\left( H^2 + \frac{dH}{dt}\right)}{\sqrt{H^2 +
    \frac{\kappa}{a^2}}} |z|, &\\ 
& r(t, z)= a\left(1 + \epsilon \sqrt{H^2 + \frac{\kappa}{a^2}} |z|\right).&
\ea
Using the Master Equation (\ref{mastersc}), we find
\begin{equation}
\frac{\partial}{\partial t} \left[\frac{1}{n b^3} \frac{\partial
    \Omega}{\partial t} \right] 
- \frac{\partial}{\partial z} \left[\frac{n}{b^3}
    \frac{\partial\Omega}{\partial z}\right] 
 - \frac{\Delta + 3\kappa}{b^2}\frac{n}{b^3}\Omega = 0.
\label{DGP_master}
\end{equation}
The energy-momentum tensor of the Weyl fluid on the brane can then be
related to $\Omega$. For the case of a spatially flat universe ($\kappa=0$) one finds
\begin{eqnarray}
\mu_E &=& - \frac{1}{3} \frac{k^4}{a^5} \Omega\bigg|_{z=0} ,
\\
\Theta_E &=&  \frac{1}{3} \frac{k^2}{a^4}
\left(\frac{\partial\Omega}{\partial t}- H \Omega \right)\bigg|_{z=0},
\\
\Sigma_E &=&  -\frac{1}{6a^3} \left(
3\frac{\partial^2\Omega}{\partial t^2} - 3H
\frac{\partial\Omega}{\partial t} 
+ \frac{k^2}{a^2}  \Omega - \frac{3}{H} \frac{dH}{dt}
\frac{\partial\Omega}{\partial z} \right)\bigg|_{z=0}  ,
\label{DGP_weyl}
\end{eqnarray}
where $k$ is the 3-momentum on the homogeneous background. Thus, in
general one has to solve Eq. (\ref{DGP_master}) with appropriate
initial and boundary conditions, and then use Eq. (\ref{DGP_weyl}) in
the perturbed Einstein equations.  In practise, one can however apply
various approximations. 

Koyama and Maartens~\cite{KoyamaMaartens2006} assume the small-scale
approximation $k/a\gg r_c, \cal H$.  Under this assumption
Eq. (\ref{DGP_delta_dot}) implies that $\Theta_E = 0$.
In quasi-static situations we also have $\del_t\Omega\approx 0$ and
$\frac{1}{H}\frac{dH}{dt}\ll \frac{1}{\cal H}$, so that the master equation becomes
$\del^2_z\Omega - \frac{2\epsilon H}{n} \del_z \Omega -
\frac{k^2}{a^2n^2} \Omega=0$.  Assuming that the solution of this
last equation is regular as $z\rightarrow \infty$, it can then be
shown that $\Omega = \Omega_{br} (1 + \epsilon H
 z)^{-\frac{k}{aH}}$ when $aH/k\ll 1$. Therefore, using Eq. (\ref{DGP_weyl}) 
one finds that $\mu_E = 2 k^2 \Sigma_E$ on the brane, in the quasi-static limit. 
Inserting this into Eq. (\ref{DGP_theta_dot}) gives $\Sigma_E$, and
therefore $\mu_E$, in terms of the potentials.  This
in turn allows us to eliminate all the Weyl perturbations, to get
\begin{equation}
 - k^2 \Phi = 4 \pi G \left( 1 - \frac{1}{3\beta}\right)  \rho a^2 \delta_M,
\label{DGP_QS_Phi}
\end{equation}
and
\begin{equation}
 - k^2 \Psi = 4 \pi G \left( 1 + \frac{1}{3\beta}\right)  \rho a^2 \delta_M,
\label{DGP_QS_Psi}
\end{equation}
where $\beta = 1 + 2 \epsilon H r_c w_E$.

Sawicki, Song and Hu~\cite{SawickiSongHu2007} propose a scaling ansatz
to close the Bianchi identities, and solve the master equation on the
self-accelerating branch near the cosmological horizon. This allows
them to move away from the quasi-static regime. The ansatz they use is
$\Omega = A(p) a^s G(z/z_{hor})$, where $z_{\textrm{hor}} = aH\int_0^a
\frac{d\tilde{a}}{\tilde{a}^2H(\tilde{a})^2}$, and $s$ is an exponent
that is approximately constant during times when a particular fluid
dominates the expansion.  The master equation then becomes an ordinary
differential equation with independent variable $z/z_{\textrm{hor}}$, and can
be solved iteratively by assuming the boundary conditions $G(0)=1$ and
$G(1) = 0$.  They find that for super-horizon modes $s=6$ during
radiation domination, decreasing to $s=4$ during matter domination,
and finally approaching $s=1$ during $\Lambda$ domination. For all
sub-horizon modes $s=3$, which reproduces the quasi-static
approximation of Koyama and Maartens~\cite{KoyamaMaartens2006}. By iteration, one can
correct for the time-dependence of $s$, as is necessary during transitions
between cosmological eras.  Song has performed a similar analysis on
the normal branch~\cite{Song2008}.

Cardoso \etal~\cite{CardosoEtAl2008} solve the master equation
numerically by employing null coordinates in the $\{t,z\}$ plane,
thus obtaining the linearised DGP solutions without resorting to any
approximations. They find that the quasi-static approximation is
valid to within $5\%$ for $k\ge0.01hMpc^{-1}$.  Seahra and
Hu~\cite{SeahraHu2010} develop analytic solutions for both branches
of DGP based on the scaling ansatz, and compare these with numerical
solutions in~\cite{CardosoEtAl2008}. They find that the
analytic/scaling solutions are accurate to within a few percent all
the way to the horizon, and therefore that the use of the scaling ansatz
in the observational constraints imposed
in~\cite{FangEtAl2008,LombriserEtAl2009} is justifiable.

The FLRW background evolution on the self-accelerating branch has been
extensively tested, using a variety of different
cosmological observations. Alcaniz~\cite{Alcaniz2002} used measurements of the
angular size of high redshift compact radio
sources~\cite{GurvitsKellermannFrey1998} to place constraints on
$r_c$. He finds that $\frac{1}{4r_c^2H_0^2} \ge 0.29$ at $1\sigma$,
with a best-fit value of $r_c\sim 0.94 H_0^{-1}$.  Jain, Dev and
Alcaniz~\cite{JainDevAlcaniz2002} find that gravitationally lensed
QSOs require $r_c\ge 1.14 H_0^{-1}$ at $1\sigma$.  Deffayet
\etal~\cite{DeffayetEtAl2002b} use SN-Ia
from~\cite{PerlmutterEtAl1998}, and the angular diameter distance to
recombination of pre-WMAP CMB data, while assuming a flat
universe. Their SN analysis gives 
$\frac{1}{4r_c^2H_0^2}	= 0.17^{+0.03}_{-0.02}$ at $1\sigma$, which
translates to $r_c = 1.21^{+0.09}_{-0.09} H_0^{-1}$. Including
CMB data increases the preferred value of $\Omega_M$, and leads to a
best fit model with $r \sim 1.4H_0^{-1}$. Fairbairn and
Goobar~\cite{FairbairnGoobar2005} use the SNLS SN-Ia
data~\cite{AstierEtAl2005} and BAO data~\cite{EisensteinEtAl2005} to
show that the self-accelerating model is not compatible with a flat
universe at the $99\%$ level. For the generalised model
of~\cite{DvaliTurner2003} they find that $-0.8 < \alpha < 0.3$ at
$1\sigma$ (the self-accelerating model corresponds to $\alpha = 1$).
Maartens and Majerotto~\cite{MaartensMajerotto2006} used
SN-Ia~\cite{RiessEtAl2004,AstierEtAl2005}, a CMB shift
parameter~\cite{WangMukherjee2006}),
and BAO data~\cite{EisensteinEtAl2005} to place constraints on the
self-accelerating FLRW background.  They find that the
self-accelerating model is consistent with these data sets to within
$2\sigma$, but is worse fit than $\Lambda$CDM. This
puts tension on the self-accelerating model, but as they point out, it
is not necessarily reliable to use the BAO data for models other than
$\Lambda$CDM, as $\Lambda$CDM is used throughout the analysis in
\cite{EisensteinEtAl2005}.  Rydbeck, Fairnbairm and
Goobar~\cite{RydbeckFairbairnGoobar2007} repeated the analysis
of~\cite{EisensteinEtAl2005} by including SN-Ia data from
ESSENCE~\cite{MiknaitisEtAl2007,WoodVaseyEtAl2007}, and constraints on
the CMB shift parameter from WMAP-3~\cite{SpergelEtAl2006} to disfavour the
self-accelerating model at the $1-2\sigma$ level, depending on whether
a peculiar velocity error on the SN-Ia data is included or not. The
inclusion of BAO data from \cite{EisensteinEtAl2005} further supports this result.

Moving away from using distance measurements alone, a number of 
studies have been performed on the growth of linear structure using
the small-scale quasi-static approximation of Koyama and
Maartens~\cite{KoyamaMaartens2006}. Song, Sawicki and
Hu~\cite{SongSawickiHu2007} used the angular diameter distance to
recombination from WMAP-3~\cite{SpergelEtAl2006}, SN-Ia data
from~\cite{RiessEtAl2004,AstierEtAl2005}, and constraints on the
Hubble constant to exclude the spatially flat self-accelerating model
at $3\sigma$.  By allowing non-zero spatial curvature, however, they
show that consistency with the data can be improved, but is still
marginally worse than spatially flat $\Lambda$CDM.  They then use
data coming from BAOs, the linear growth of structure, and ISW and
galaxy-ISW correlations to show that any self-accelerating models that
shares the same expansion history as a quintessence-CDM models are
strongly disfavoured.  As these authors argue, one must properly take
into account the perturbations on the self-accelerating branch in
order to make firm conclusion.  Xia considered the generalised 
model of Dvali and Turner~\cite{DvaliTurner2003} in conjunction with
Union SN-Ia data set~\cite{KowalskiEtAl2008}, CMB distance measurements, 
GRB data~\cite{Schaefer2006}, and a collection of data on the growth
of linear structure~\cite{HawkinsEtAl2002, VerdeEtAl2001,
  TegmarkEtAl2006, RossEtAl2006, GuzzoEtAl2008, daAngelaEtAl2006,
  McDonaldEtAl2004}.  Using the quasi-static approximation of Koyama-
Maartens~\cite{KoyamaMaartens2006} to calculate the growth function,
Xia finds $\alpha < 0.686$ at $2\sigma$.  This excludes the
self-accelerating value of $\alpha = 1$.

Fang \etal.~\cite{FangEtAl2008} perform a thorough test of the
self-accelerating model using the PPF approach
of~\cite{Hu2008,HuSawicki2007}.  Using CMB and large-scale structure
data these authors find that the spatially flat model is a $5.3\sigma$
poorer fit than $\Lambda$CDM, while that open model is
$4.8\sigma$ worse.  In the latter case non-zero spatial curvature
improves the fits for distance measurements, but worsens those
involving the growth of linear structure.  One may speculate that
changes to the initial power spectrum may be able to improve the this
situation, but the required reduction in large-scale power also
produces unacceptable reductions in power in the large-scale CMB
polarisation spectrum.  Changes of this type cannot therefore save the
model.  In a follow-up study Lombriser \etal~\cite{LombriserEtAl2009}
constrained both branches of the DGP model by employing the PPF
description of the CMB and large-scale structure.  These authors find
that either brane tension or $\Lambda$ is required for these models to
fit the data well, but that both cases the best fitting
models are practically indistinguishable from $\Lambda$CDM. They
further find that the cross-over scale is $H_0r_c > 3$ if spatial curvature is
included, and $H_0r_c > 3.5$ in the spatially flat case.

The growth index on small scales, $\gamma$, has been calculated by
Wei using a Taylor series expansion in the fractional density
parameters $\Omega_{DGP}$ and $\Omega_{\kappa}$~\cite{Wei2008}.
Ferreira and Skordis have also calculated this quantity, within a more
general frame-work of modified gravity models, to find $\gamma =
\frac{11}{16} - \frac{7}{5632} \Omega_{DGP} +
\frac{93}{4096}\Omega_{DGP}^2 +
O(\Omega_{DGP}^3)$~\cite{FerreiraSkordis2010}.  This analytic result
is in excellent agreement with numerical studies\footnote{The terms
  $\frac{3}{16}\Omega_{DGP}$ and $\frac{1}{16}\Omega_{DGP}^2$ in the
  Wei~\cite{Wei2008} result are incorrect, and appear to have come
  from erroneously dropping terms proportional to $\Omega_{DGP}$ when
  going from Eq. (21) to Eq. (22) of that paper.}.

Although the DGP model is strongly constrained by the observations we
have already discussed, it is still instructive to consider the
substantial work that has been performed on constructing and studying
non-linear structure formation scenarios in these models
~\cite{LueScoccimarroStarkman2004,KoyamaTaruyaHiramatsu2009,Scoccimarro2009,ChanScoccimarro2009,Schmidt2009a,Schmidt2009b,KhouryWyman2009,WymanKhoury2010,SchmidtHuLima2009,SeahraHu2010}.
Lue, Scoccimarro and Starkman ~\cite{LueScoccimarroStarkman2004} have
considered spherical perturbations on sub-horizon scales, and derived
the gravitational force law in a collapsing top-hat model embedded in
an expanding background.  They find that the non-linear CDM density
contrast evolves as
\begin{equation}
\frac{d^2\delta_M}{dt^2} + \left[2H - \frac{4}{3} \frac{1}{1+\delta_M}
  \frac{d\delta_M}{dt} \right]\frac{d\delta_M}{dt} = 4 \pi G_{\textrm{eff}}
\bar{\rho}_M \delta_M (1 + \delta_M) , 
\label{NL_delta_DGP}
\end{equation}
where
\begin{equation}
G_{\textrm{eff}} =  G \left[  1 + \frac{2}{3\beta} \frac{1}{\epsilon_D} \left(
  \sqrt{ 1 + \epsilon_D  } - 1 \right)  \right] ,
\label{Geff_DGP}
\end{equation}
and $\epsilon_D = \frac{8}{9\beta^2}
\frac{\Omega_M}{\Omega^2_{DGP}}\delta_M = \frac{8}{9}
\frac{(1+\Omega_M)^2}{(1+\Omega_M^2)^2}\Omega_M \delta_M$. Their model
uncovers a transition point at $\epsilon_D\sim 1$, below which gravity
behaves as in GR 
(in accordance with the Vainshtein mechanism). This gives a Vainshtein
radius of $r_\star = \left(\frac{16 G M r_c^2}{9
  \beta^2}\right)^{1/3}$, where $M$ is the mass of the spherically
symmetric object.  This result was subsequently re-derived by Koyama
and Silva~\cite{KoyamaSilva2007} without the restrictive assumptions
of~\cite{LueScoccimarroStarkman2004}.  It can be shown that the
solutions to Eq. (\ref{NL_delta_DGP}) are a factor of two larger than
the corresponding solutions in theories that obey Birkhoff's law, and
that have similar expansion histories.

The form of $G_{\textrm{eff}}$ in Eq. (\ref{Geff_DGP}) is due to an additional
degree of freedom in DGP: The brane-bending mode, as uncovered
in~\cite{LutyPorratiRattazzi2003,NicolisRattazziTrincherini2008}. To
expand upon this we may rewrite the RHS of Eq. (\ref{NL_delta_DGP})
using $4\pi G_{\textrm{eff}} \delta \rho_M = \Delta \Psi$.  The potential
$\Psi$ can then be written in terms of a Poisson potential, $\Psi_P$,
and a brane-bending mode, $\varphi$, as  $\Psi= \Psi_P +
\frac{1}{2}\varphi$.  These two new potentials, $\Psi_P$ and $\varphi$,
satisfy $\Delta\Psi_P = 8\pi  G\delta \rho_M$, and $\Delta\varphi =
8\pi (G_{\textrm{eff}}  - G) \delta \rho_M$. Ignoring non-local contributions,
the brane-bending mode can then be shown to obey the following
equation in the decoupling limit, and in the sub-horizon and quasi-static regime:
\begin{equation}
\Delta\varphi + \frac{r_c^2}{3\beta}\left[ \left(\Delta\varphi
  \right)^2 - \left(\grad_i\grad_j \varphi\right) \left(
  \grad^i\grad^j \varphi \right) \right] = \frac{8\pi G}{3\beta}
\delta \rho_M .
\label{DGP_BBM}
\end{equation}
This  equation is closely related to the Field Equation
(\ref{pieqns}), following from the $\pi$-Lagrangian discussed in
Section \ref{sec:dgp-sc}.

Schmidt, Hu and Lima~\cite{SchmidtHuLima2009} used the spherical
collapse model in DGP to study the halo mass function, bias and the
non-linear matter power spectrum. They find that top-hat spherical
collapse in DGP requires a new, more general method for defining the
virial radius that does not rely on energy conservation. To obtain the
comoving number density of halos per logarithmic interval in the
virial mass, and the linear bias, they use the Sheth-Tormen
method~\cite{ShethTormen1999}, while they use the
Navarro-Frenk-White~\cite{NavarroFrenkWhite1995} form for halo
profiles.  In this way they find that the spherical collapse model agrees well with the
halo mass function and bias obtained from N-body simulations, for
both the normal and the self-accelerating branch. For the non-linear power
spectrum, the spherical collapse model in the self-accelerating branch
also matches the simulation results very well.  This is not true,
however, for the normal branch, although even in this case the spherical
collapse model predictions are better than those obtained from
HALOFIT~\cite{SmithEtAl2002}.

N-body simulations of DGP have been conducted by three independent
groups: Schmidt~\cite{Schmidt2009a,Schmidt2009b}, Chan and
Scoccimarro~\cite{ChanScoccimarro2009}, and Khoury and Wyman
\cite{KhouryWyman2009}. The general result, common to
all of these studies, is that the brane-bending mode on the
self-accelerating branch provides a repulsive force that greatly
suppresses the growth of structure, while the opposite effect occurs
on the normal branch.  All three simulations also display the
Vainshtein effect.

The N-body simulations conducted by Schmidt were on both the
self-accelerating~\cite{Schmidt2009a}, and the normal
branch~\cite{Schmidt2009b}. Rather than assuming Eq. (\ref{Geff_DGP}),
Schmidt use a relaxation solver on Eq. (\ref{DGP_BBM}) to show that
the Vainshtein effect is recovered without making any assumptions
about symmetry.  Specifically, he finds that the Vainshtein effect is
weakened for non-spherically symmetric situations, and in general sets
in at smaller scales than is found in~\cite{KhouryWyman2009}. Like
~\cite{KhouryWyman2009}, however, he also finds that the HALOFIT model
does not correctly describe the non-linear DGP matter power spectrum. 
Schmidt then proceeds to calculate the halo mass function, and shows
that the abundance of massive halos in self-accelerating DGP is much
smaller than in CDM models. This last result puts strong constraints
on the self-accelerating model from cluster abundance measurements,
independent from of the CMB and large-scale structure constraints
discussed above~\cite{FangEtAl2008,LombriserEtAl2009}). For the normal branch,
structure is enhanced, and the abundance of massive halos is larger
than a CDM model~\cite{Schmidt2009b}. In this case, the halo profiles
were also obtained, and departures from the predictions of GR were 
seen outside of the halo virial radius.  Finally, Schmidt calculates
the bispectrum in both the normal and self-accelerating branches. The
self-accelerating (normal) branch  bispectrum is found to be enhanced
(suppressed) for equilateral configurations, but not for squeezed
configurations.  This is in agreement
with~\cite{ChanScoccimarro2009}, and illustrates the diminishing strength of
the Vainshtein effect for squashed matter configurations.

Scoccimarro~\cite{Scoccimarro2009} derives the linear and non-linear
equations for the growth of structure in DGP without using the
Mukohyama formalism.  This results in a set of equations that includes
non-local terms. For example, in the quasi-static limit it is found that
\begin{equation}
 \left[\Delta -
 \frac{a}{r_c}\sqrt{-\Delta}\right]\left(\Phi+\Psi\right) = 8 \pi G
 a^2 \rho \delta ,
\end{equation}
rather than $\Delta(\Phi+\Psi) = 8 \pi G a^2\rho\delta$, as implied by
adding the quasi-static expressions given in Eqs. (\ref{DGP_QS_Phi})
and (\ref{DGP_QS_Psi}). Scoccimarro finds that in the linearised
quasi-static limit the bulk behaviour decouples from the brane
behaviour, and thus that the non-local operators can be safely
ignored.  This ensures the validity of the Koyama-Maartens
result~\cite{KoyamaMaartens2006} on small scales. On larger scales,
however, the non-local terms become more important.  On very scales
the linear approximation for the brane-bending mode breaks down, and
Scoccimarro finds a non-local and non-linear equation for the
potential, $\Psi$, and the density perturbation, $\delta$.
Chan and Scoccimarro~\cite{ChanScoccimarro2009} perform N-body
simulations by accounting for the non-local operators $\sqrt{-\Delta}$
and $1/\sqrt{-\Delta}$.  These operators contribute to the equation
for the brane bending mode, $\varphi$, and the potential,
$\Psi$~\cite{Scoccimarro2009}.  They uncover the Vainshtein mechanism
through a broad transition around $k\sim 2h$Mpc$^{-1}$ for $z=1$, and
$k\sim 1h$Mpc$^{-1}$ for $z=0$. They also compute the non-linear matter
power spectrum and bi-spectrum, the CDM mass function, and the halo
bias.  The results of all this are in broad agreement with those of
Schmidt~\cite{Schmidt2009a,Schmidt2009b}. 

The simulations of Khoury and Wyman~\cite{KhouryWyman2009} were
improved upon in~\cite{WymanKhoury2010}, where Eq. (\ref{DGP_BBM}) was
solved.  It was found that in DGP, and higher dimensional cascading gravity
models, peculiar velocities are enhanced by $24-34\%$
compared to CDM~\cite{LeeKomatsu2010}.  This corresponds to an
enhancement by four orders of magnitude in the probability of
the occurrence of high velocity merging system such as the ``bullet cluster". 

Scoccimarro~\cite{Scoccimarro2009} and Koyama, Taruya and
Hiramatsu~\cite{KoyamaTaruyaHiramatsu2009} have also developed two
independent techniques based on higher-order perturbation theory in
order to find the non-linear power spectrum.
Scoccimarro~\cite{Scoccimarro2009} shows that the non-linearities
coming from the brane-bending mode can be described by a time and
space dependent gravitational constant.  He then goes on to develop a
re-summation scheme to calculate the non-linear power spectrum and the
bi-spectrum.  Koyama, Taruya and Hiramatsu
~\cite{KoyamaTaruyaHiramatsu2009} develop a general perturbation
theory method that can be applied in the quasi non-linear regime of
any theory that has an additional scalar degree of freedom (such as
Brans-Dicke, $f(R)$ and DGP). For the case of DGP, they find that
their perturbative method recovers the extreme non-linearity of the
Vainshtein mechanism. Their technique has compared with the HALOFIT
mapping~\cite{SmithEtAl2002}, and the non-linear PPF fit of Hu and
Sawicki~\cite{HuSawicki2007}.  It is found that HALOFIT over-predicts
power on small-scales, in agreement with the findings
of~\cite{KhouryWyman2009,Schmidt2009a,Schmidt2009b}, while the
non-linear PPF fit works well within the regime of validity of the
perturbation theory.

\subsection{Higher Co-Dimension Braneworlds} 
\label{sec:highercod}

An important characteristic of any braneworld model is the brane's
{\it co-dimension}. This is given by the difference between the
dimension of the bulk and the dimension of the brane.  Up to this
point we have been concentrating on co-dimension one  models.  These
are by far the most well developed, mainly because they are much more
tractable.  In this section we will discuss braneworld models with
co-dimension $n \geq 2$.  In this case the gravitational description
typically becomes much more complicated.

We have already seen higher co-dimension branes  in Section
\ref{sec:ADD},  when discussing the ADD model \cite{ADD1} in six or
more dimensions. This is a particularly simple example of a higher
co-dimension braneworld set-up, corresponding to a tension-less brane
in a Minkowski bulk.  In general, of course, one must consider more
complicated scenarios, including the full  gravitational back-reaction
of brane sources, both in the bulk and on the brane.  Indeed, attempts
to solve for a higher co-dimension delta function source in Einstein
gravity will generically result in a singular bulk geometry unless the
energy-momentum of the source is given by pure tension
\cite{Cline-cosmology-of-codim}. This is problematic if we insist on
using that infinitely thin defect as a model of our $4$-dimensional
Universe.  However, in a realistic set-up the defect will not really
be a delta function source (it will have some finite thickness, and
the would-be singularity will be automatically
resolved). Alternatively, one could consider a genuine delta function
source and avoid issues with the singularity by including higher-order
operators in the bulk gravity theory \cite{Dvali-seesaw,
  Bostock-Einstein, Charmousis-matching,Charmousis-Einstein}.

Despite being extremely difficult to study, higher co-dimension
braneworld models have been  discussed in the literature (see, for
example, \cite{Carroll-sidestepping, Cline-cosmology-of-codim,
  Vinet-can-codim, Nilles-selftuning, Aghababaie-towards,
  Dvali-gravity, Dvali-seesaw,Dvali-diluting,
  Gabadadze-softly,Dubovsky-BIG, Kaloper-charting,Kaloper-BIG,
  deRham-casc1, deRham-casc2, deRham-casc3, deRham-intro-casc,
  deRham-flat3, Minamitsuji-self,
  Agarwal-casc,Corradini-induced1,Corradini-induced2,Bostock-Einstein,
  Charmousis-self-properties, Charmousis-properties,
  Charmousis-consistency, Papantonopoulos-induced,Papantonopoulos-reg,
  CuadrosMelgar-black,CuadrosMelgar-perturbations}). Much of the
interest in this field lies in the fact that these models offer new
ways to think about the cosmological constant problem
\cite{Carroll-sidestepping, Cline-cosmology-of-codim,
  Vinet-can-codim,Aghababaie-towards, Dvali-diluting,
  Gabadadze-softly}.  For example, in Section \ref{sec:degrav} we will
briefly review the importance of higher co-dimension branes in the
degravitation scenario \cite{degrav1}.



Co-dimension two branes are particularly interesting as there is
reason to believe that within them one may be able to realise
self-tuning of the vacuum energy.  This is because  a maximally
symmetric co-dimension two brane {\it generically} behaves like a
cosmic string in the bulk by forming a conical
singularity\footnote{There are exceptions that give rise to curvature
  singularities as opposed to conical, even for maximal symmetry on
  the brane \cite{Tolley-bulk-singularities}.}. Changes in tension
then alter the deficit angle, as opposed to the geometry, of the
defect \cite{Cline-cosmology-of-codim}. Braneworld models with two
extra dimensions shaped like a rugby ball have been developed with
this in mind \cite{Carroll-sidestepping}. The curvature of the brane
is now completely determined by the bulk cosmological constant and the
magnetic flux, independent of the brane tension. One might then infer
that phase transitions on the brane that  alter the brane tension do
not alter the curvature. This is not the case, however, as such a
transition can affect the brane curvature via the backdoor, by
altering magnetic flux, so there is no self-tuning
\cite{Vinet-can-codim, Nilles-selftuning}. The situation can sometimes
(but not always \cite{Burgess-bulk-axions, Burgess-large}) be improved
by including super-symmetry and dilaton dynamics in the bulk to
protect the relationship between the bulk cosmological constant and
the flux \cite{Aghababaie-towards,Burgess-sled1,Burgess-sled2,
  Burgess-bulk-axions, Burgess-large}. A low scale of bulk
super-symmetry breaking, set by the size of the rugby ball, can then
account for a small amount of dark energy \cite{Burgess-sled1}. Other
notable contributions to the literature on rugby ball
compactifications with fluxes and/or scalars  include the following: A
detailed discussion of the brane-bulk matching
conditions\cite{Burgess-effective, Bayntun-cod}, an analysis of the
low energy effective theory \cite{Kobayashi-curvature}, and a study of
cosmological evolution on the brane
\cite{Papantonopoulos-cosmological}.

The rugby ball described above represents a compact extra dimension,
with potentially troublesome/interesting light moduli fields
\cite{Albrecht-exponentially,Albrecht-natural}. In any event, it is
worth pursuing alternative braneworld models with non-compact extra
dimensions if they are able to reproduce $4D$ gravity at some scale.
Co-dimension two branes in Gauss-Bonnet gravity will be discussed in
this context in Section \ref{sec:GB-cod2}.  On the other hand, one
might expect that an infinite bulk should be compatible with some form
of {\it quasi}-localisation of gravity by considering generalisations
of DGP gravity \cite{dgp} in any number of dimensions.  Such
generalisations are sometimes referred to as Brane Induced Gravity
(BIG) models \cite{Dvali-gravity, Dvali-seesaw, Dvali-diluting,
  Gabadadze-softly, Dubovsky-BIG,deRham-casc1, deRham-casc2,
  deRham-intro-casc, Corradini-induced1, Corradini-induced2,
  Kaloper-charting,Kaloper-BIG}. In the simplest scenario  we can
consider BIG models on a single  3-brane of co-dimension $n$,
described by the following action \cite{Dvali-gravity}:
\be
S=\frac{M_*^{2+n}}{2} \int_{\textrm{bulk}} d^{4+n} x \sqrt{-\gamma}{\cal
  R}+\frac{M_{pl}^{2}}{2} \int_{\textrm{bulk}} d^{4} x \sqrt{-g}{R} .
\ee
Using higher derivative operators to resolve the singularity in the
bulk propagator\footnote{Note that this theory becomes
  higher-dimensional  at large distances, so the finer details of how
  the theory behaves there will be sensitive to  how we resolve the
  singularity in the bulk \cite{Dvali-gravity}.}, we find that  for $n
=2, 3$ the Green's function in momentum space along the brane takes
the form \cite{Dvali-seesaw}
\be \label{seesawG}
\tilde G(p, 0) =\frac{2}{M_{pl}^2 p^2 +\frac{M_*^{2+n}}{D(p,0)}}, 
\ee
where
\be
D(p, 0)=\int \frac{d^n q}{(2\pi)^n} \frac{1}{p^2+q^2+(p^2+q^2)^2/M_{uv}^{2}},
\ee
and where $M_{uv} \lesssim M_* $ represents the regularisation
scale. Taking $M_{uv} \sim 10^{-3}$ eV, this theory reproduces normal
$4D$ gravity at intermediate scales $M_{uv}^{-1} \ll r \ll r_c$,  but
is modified at sub-millimetre scales and  at very large distances. It
has therefore been dubbed {\it seesaw} gravity in
\cite{Dvali-seesaw}. The large distance cross-over scale can be
computed from Equation (\ref{seesawG}), 
\be
r_c \sim \begin{cases} \frac{M_{pl}}{M_*^2} \sqrt{\ln(M_{uv} r_c)} &
  \textrm{for $n=2$} \\
\frac{M_{pl}}{M_*^2} \sqrt{\frac{M_{uv}}{M_* }} & \textrm{for $n=3$.} \end{cases}
\ee
Unfortunately, these constructions suffer from the presence of ghosts
in the spectrum of fluctuations \cite{Dubovsky-BIG}.

\subsubsection{Cascading gravity}

More sophisticated BIG proposals make the transition from $4+n$
dimensional gravity to $4$ dimensional gravity via a series of
intermediate steps, $(4+n)D\to (4+n-1)D \to \ldots \to 5D \to 4D$
\cite{deRham-casc1, deRham-casc2, deRham-intro-casc,
  Corradini-induced1, Corradini-induced2,
  Kaloper-charting,Kaloper-BIG}. This could potentially help with the
ghost problem.  These are sometimes referred to as {\it cascading}
gravity models as the effective description tumbles down a cascade of
extra dimensions as we come in from large distances, and increase the
resolution of our description.  Let us now turn our attention to this
class of  gravity models.
\newline
\newline
\noindent
{\it The Kaloper-Kiley model}
\newline

The first braneworld model to explore a cascade from $6D \to 5D \to
4D$ gravity was developed by Kaloper and Kiley \cite{
  Kaloper-charting,Kaloper-BIG}. They considered a landscape of models
describing  BIG  on a $3$-brane in six dimensions. The  singularities
are resolved  by modelling the $3$-brane as a cylindrical $4$-brane
with compact radius $r_0$.  The brane has tension $\sigma_5$,  and
some induced curvature weighted by a $5D$ Planck scale, $M_5$. The
effective $4D$ tension and  Planck scale at distances $r \gg r_0$ are
then given by $\sigma_4^\textrm{eff}=2\pi r_0\sigma_5$ and
$M_4^\textrm{eff}=2\pi r_0 M_5^3 $, respectively. Note that an axion
flux is used to cancel the vacuum pressure in the compact direction
$q^2=2\sigma_5r_0^2$.

\begin{figure}
\begin{center}
\epsfig{file=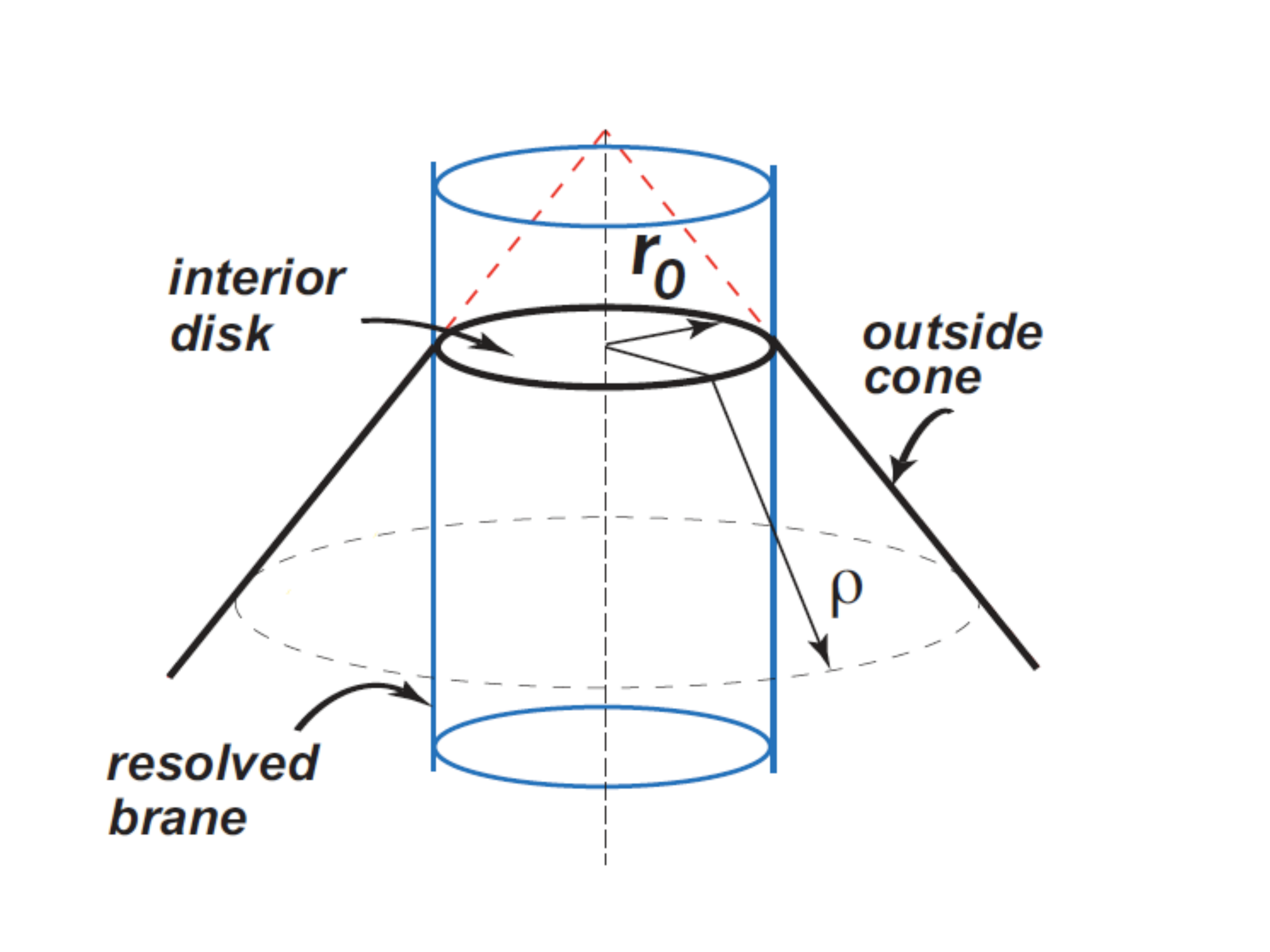,width=7.3cm,}
\caption{Taken from \cite{Kaloper-charting}. An illustration of the
  $2D$ bulk geometry formed around the resolved brane in vacuum. }
\label{fig:Kaloper-charting}
\end{center}
\end{figure}

The resulting vacua resemble a cone in the bulk, with the conical
singularity cut away (see Figure \ref{fig:Kaloper-charting}). The
metric is given by
\be
ds^2_6=\eta_\mn dx^\mu dx^\nu +d\rho^2+f(\rho)^2 d\phi^2,
\ee
where 
$$
f(\rho)=\begin{cases} \rho & 0 \leq \rho < r_0 \\
(1-b)\left(\rho+\frac{b}{1-b} r_0\right ) & \rho >r_0. \end{cases}
$$

The brane tension does not affect the geometry along the brane's
non-compact directions. Instead, it controls the deficit angle
measured at infinity, according to $b=2\sigma_5 r_0/M_6^2$, where
$M_6$ is the fundamental scale of gravity in the bulk. When the
tension lies below a critical value, $\sigma_{\textrm{crit}}=M_6^2/2r_0$, the
deficit angle is less than $2\pi$, and the bulk space is infinite. For
sub-critical branes, $\sigma_5 <\sigma_{crit}$,  the theory generically
resembles the seesaw gravity theory.  We have $4D$ (Brans-Dicke)
gravity at intermediate scales $r_0 <r<r_c$, and a crossover to $6D$
gravity at a scale $r_c \sim  \frac{M_{4}^\textrm{eff}}{(1-b) M_6^2}
\sqrt{\ln(2(1-b) r_c/r_0)} $. Note that the scalar fluctuations
naively indicate the possible presence of a ghost, but this conclusion
cannot be trusted since the perturbative theory breaks down due to
strong coupling \cite{Kaloper-charting}. This is consistent with the
conclusions drawn in \cite{Dubovsky-on-models} for thick brane
regularisations of seesaw gravity.

\begin{figure}
\begin{center}
\epsfig{file=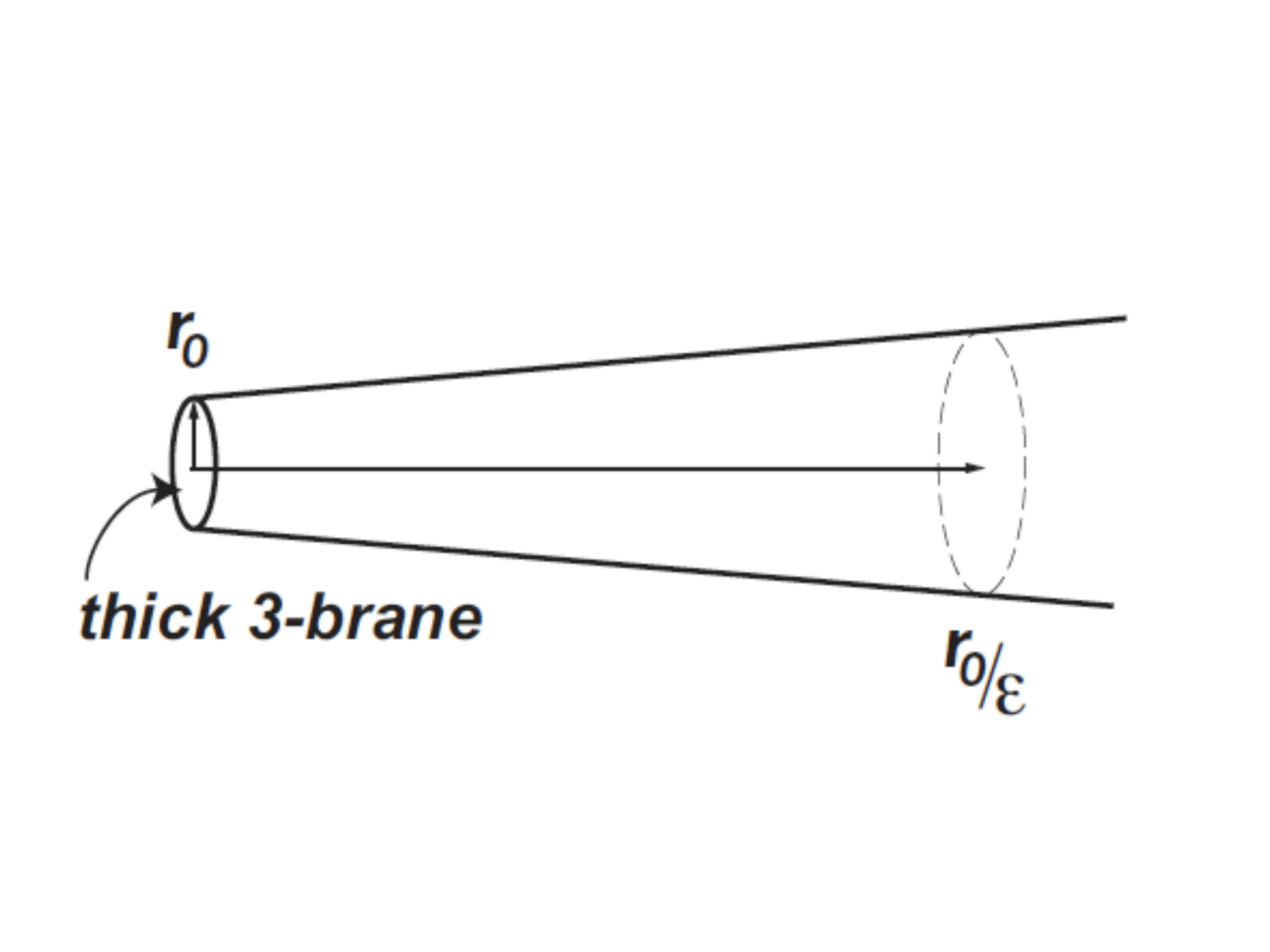,width=7.3cm,}
\caption{Taken from \cite{Kaloper-charting}. An illustration of the
  $2D$ bulk geometry formed around the resolved brane in vacuum, in
  the near critical limit.}
\label{fig:throat}
\end{center}
\end{figure}

Things are much more interesting in the near critical limit, for which
$b=1-\epsilon$, where $0<\epsilon \ll 1$.  In this case the brane lies
inside a very deep throat, such that the angular dimension is
effectively compactified up to distances of the order $r_0/\epsilon$
(see Figure \ref{fig:throat}).
It turns out that perturbation theory is under much better control in
this limit, and that there are no ghosts. The improved behaviour is
due to the fact that we make a series of transitions from $4D \to 5D
\to 6D$ gravity. Indeed, the theory looks like four dimensional
Brans-Dicke gravity at scales $r_0 <r<r_c$, becoming {\it five}
dimensional beyond $r_c \sim M_4^\textrm{eff}/M_6^4 r_0$. The
transition to six dimensions only occurs at very large distances,
$r>r_0/\epsilon$,  at which point the cylinder decompactifies.
 
Note that this model does {\it not} represent a solution to the
cosmological constant problem. Instead it recasts it in terms of a
fine-tuning of the hierarchy between cross-over scales at each
transition.
\newline
\newline
\noindent 
{\it The cascading DGP model}
\newline

A more recent BIG model exploring these ideas was developed by de Rham
{\it et al.}, who coined the phrase {\it cascading gravity}
\cite{deRham-casc1, deRham-casc2, deRham-intro-casc}. In the simplest
model  \cite{deRham-casc1}, one has a DGP $3$-brane within a DGP
$4$-brane within a six dimensional bulk.  This situation is described
by the following action:
\be
S=\frac{M_6^4}{2} \int_{\textrm{bulk}} d^6 x\sqrt{-g_6} R_6+\frac{M_5^3}{2}
\int_{\textrm{4-brane}} d^5 x\sqrt{-g_5} R_5+\frac{M_4^2}{2} \int_{\textrm{3-brane}} d^4
x\sqrt{-g_4} R_4 ,
\ee
where $(g_6)_{AB}, (g_5)_{ab}$ and $(g_4)_\mn$ are the metrics in the
bulk, on the $4$-brane and on the $3$-brane, respectively. The
corresponding Ricci scalars are given by $R_6, R_5$ and $R_4$. The
model contains two important mass scales corresponding to the
following ratios:
\be \label{ratios}
m_5=\frac{M_5^3}{M_4^2}, \qquad {\rm and} \qquad m_6=\frac{M_6^4}{M_5^3}.
\ee
The claim is that the intermediate DGP $4$-brane can  help to resolve
the singularity in the bulk propagator at the location of the
$3$-brane \cite{deRham-casc1}.  Let us now consider fluctuations due
to a conserved source, $T_\mn$, about a bulk and branes that are all
Minkowski. The field  can then be decomposed in terms of a scalar,
$\pi$, and a transverse and trace-free tensor, $h^\textrm{TT}_\mn$,
such that 
\be
h_\mn=h^\textrm{TT}_\mn+\pi \eta_\mn+\textrm{gauge terms}.
\ee
Working in momentum space one finds that the Fourier transformed
fields are given by \cite{deRham-casc1, deRham-casc2}
\ba
\tilde h^\textrm{TT}_\mn(p)
&=&\frac{2}{M_4^2}\left(\frac{1}{p^2+g(p^2)}\right)\left[\tilde
  T_\mn-\frac{1}{3} \tilde T \eta_\mn+\frac{p_\mu p_\nu}{p^2} \tilde
  T\right], \\
\tilde \pi(p) &=&-\frac{1}{3M_4^2}\left(\frac{1}{p^2-2g(p^2)}\right) \tilde T,
\ea
where\footnote{Note that the various formulae for propagators and
  amplitudes in \cite{deRham-casc1,deRham-casc2}
  differ. One must take $m_{5,6} \to \frac{1}{2} m_{5,6}$ in going
  from \cite{deRham-casc1} to \cite{deRham-casc2}. The difference is
  because only one half of the 6D bulk, and one half of the 4-brane
  are considered in \cite{deRham-casc2}. This is a perfectly
  legitimate truncation by $\mathbb{Z}_2$ symmetry across each of the
  branes. Here, however, we will adopt the conventions of \cite{deRham-casc1} when
  expressing our formulae.}
\be
g(p^2)=\begin{cases}\frac{\pi
  m_5}{2}\frac{\sqrt{|p^2-4m_6^2|}}{\tanh^{-1}\left(\sqrt{\frac{|p-2m_6|}{p+2m_6}}\right)} &p<2m_6 \\
\frac{\pi
  m_5}{2}\frac{\sqrt{|p^2-4m_6^2|}}{\tan^{-1}\left(\sqrt{\frac{|p-2m_6|}{p+2m_6}}\right)} &p>2m_6. \end{cases}
\ee
The amplitude between two conserved sources on the brane, $T_\mn$ and
$T'_\mn$, now takes the surprisingly simple form 
 \be
 {\cal A}=\frac{2}{M_4^2}\left(\frac{1}{p^2+g(p^2)}\right)\left[\tilde
   T_\mn \tilde T'^\mn-\frac{1}{3} \tilde T \tilde T'-
   \frac{1}{6}\frac{p^2+g(p^2)}{p^2-2g(p^2)} \tilde T\tilde T' \right] .
\ee
The last term represents the contribution from the scalar amplitude,
whereas the remainder is the  contribution from the massive spin-2
field. The coefficient  of scalar amplitude changes sign as we flow
from the IR to the UV,
\be
\frac{p^2+g(p^2)}{p^2-2g(p^2)} \to \begin{cases} -\frac{1}{2} &
  \textrm{as $p \to 0$} \\ 1 &\textrm{as $p \to \infty$,}
\end{cases}
\ee
indicating the presence of a ghost in the UV theory, consistent with
the results of \cite{Dubovsky-BIG}. To eliminate this ghost we must
add additional operators to the $3$-brane action that modify the
scalar propagator in the UV\footnote{It has also been argued that
  thickening of the brane can help to eliminate the ghost in this
  model \cite{deRham-casc2}}. It turns out that this can be achieved
simply by adding  tension, $\lambda$, to the $3$-brane
\cite{deRham-casc1, deRham-casc3}.  As the $3$-brane is co-dimension
2,  its vacuum solution is unaffected by the change in tension (it is
still Minkowski). In contrast, the $3$-brane tension {\it does } alter
the profile of the fields in the bulk by creating a deficit angle.
This means that kinetic terms describing fluctuations on the new
vacuum now receive corrections from the non-linear bulk interactions.
The result is that the scalar propagator on the $3$-brane is
modified. One might expect that such effects will necessarily be
suppressed by the five and six dimensional Planck scales, but this is
not necessarily so because of strong coupling of the scalar in the
five dimensional boundary effective field theory \cite{deRham-casc1,
  deRham-casc3}.

In any event, at higher energies, $p \gg m_6$,  the modified amplitude goes like
\be
{\cal A}_\lambda \sim
\frac{2}{M_4^2}\left(\frac{1}{p^2+2m_5p}\right)\left[\tilde T_\mn
  \tilde T'^\mn-\frac{1}{3} \tilde T \tilde T'-
  \frac{1}{6}\frac{p^2+2m_5 p}{p^2\left(1-\frac{3\lambda}{2 m_6^2
      M_4^2}\right)-4m_5 p} \tilde T\tilde T' \right] .
\ee
For large enough tension, $\lambda>\frac{2m_6^2 M_4^2}{3}$, the scalar
amplitude always has the correct sign, so there is no ghost. However,
the tension cannot be arbitrarily large since we require that the bulk
deficit angle is less than $2\pi$.  This places an upper bound
$\lambda <2\pi M_6^4=2\pi m_5 m_6 M_4^2$. Provided $m_6 < m_5$ we
therefore have a window,
\be \label{casc-noghosts}
\frac{2m_6^2 M_4^2}{3}<\lambda <2\pi m_5 m_6 M_4^2,
\ee
for which the theory is ghost-free in the UV \cite{deRham-casc1,
  deRham-casc3}. The condition $m_6 < m_5$ has added significance in
that it permits a transition from $4D$ to $5D$ at energies $p \sim
m_5$. In particular, we have $4D$ scalar-tensor gravity in the far UV,
for $p > m_5>m_6$,
\be
{\cal A}^{(4D)}_\lambda \sim  \frac{2}{M_4^2 p^2}\left[\tilde T_\mn
  \tilde T'^\mn- \frac{1}{2}\left(\frac{1-\frac{\lambda}{ m_6^2
      M_4^2}}{1-\frac{3\lambda}{2 m_6^2 M_4^2}}\right) \tilde T\tilde
  T' \right] ,
\ee
and $5D$ scalar-tensor gravity for $m_5 >p>m_6$,
\be
{\cal A}^{(5D)}_\lambda \sim  \frac{1}{M_5^3p}\left[\tilde T_\mn
  \tilde T'^\mn-\frac{1}{4} \tilde  T\tilde T' \right] .
\ee
 In the far infra-red, $p<m_6$, we recover  $6D$ gravity,
\be
{\cal A}^{(6D)} \sim\frac{\ln(m_6/p)}{\pi M_6^4} \left[\tilde T_\mn
   \tilde T'^\mn-\frac{1}{4} \tilde T \tilde T'\right] .
\ee
Thus we have a cascade from $6D \to 5D \to 4D$ gravity as we move
from large to short distances. It is interesting to note that the
existence of this cascade seems to be closely related to the absence
of ghosts.  By taking $M_4 \sim M_{pl}$, $M_5 \sim 10$ MeV and $M_6
\sim$ meV, the cascade occurs at the current horizon scale, as $m_6
\sim m_5 \sim H_0$.  Note that a cascade from seven dimensions has
also been studied in some detail in \cite{deRham-flat3}.
 
For tensions in the range allowed by Eq. (\ref{casc-noghosts}) the
$4D$ theory at short distances is a Brans-Dicke theory that is
incompatible with solar system constraints. The authors of
\cite{deRham-casc1} suggest that some sort of Vainshtein mechanism
could account for the recovery of GR at short scales.  Indeed, the
role of non-linearities in cascading DGP has not yet been properly
investigated, and one may be concerned that there are issues
surrounding the validity of the linearised results (as was found in the
sub-critical regime of the Kaloper-Kiley model
\cite{Kaloper-charting}).

Another aspect of cascading DGP that requires further study is
cosmology (see \cite{Minamitsuji-self, Agarwal-casc} for some
preliminary work). In \cite{Agarwal-casc}  a five dimensional proxy
theory is used that is expected to capture many of the salient
features of the original model.  This proxy model corresponds to a
$5D$ scalar-tensor theory that, in the limit where the scalar and
tensor degrees of freedom decouple, agrees with the decoupling limit of the $5D$ boundary
effective field theory of cascading DGP presented in
\cite{deRham-casc1}.  Of course, it is by no means unique in this
respect.
\newline
\newline
\noindent 
{\it Induced gravity on intersecting braneworlds}
\newline

A third BIG model that is very closely related to cascading DGP was
developed by Corradini, Koyama and Tasinato \cite{Corradini-induced1,
Corradini-induced2}. Here a six dimensional bulk contains {\it two}
intersecting DGP $4$-branes, each with induced curvature. At the
intersection we place a DGP $3$-brane, also with some induced
curvature. The cosmological evolution  is derived using a formalism
based on mirage cosmology \cite{Kehagias-mirage}. The resulting
Friedmann equation on the $3$-brane is then given by \cite{ Corradini-induced2}
\be
\rho=2M_4^2 H^2+6M_5^3 k_2\sqrt{1+\frac{H^2}{k_2^2}}+4 M_6^4 \tan^{-1}
\left[ \tan \alpha \sqrt{1+\frac{H^2}{k_2^2}}\right] ,
\ee
where $\alpha$ is the angle between the two $4$-branes, and $k_2$ is a
constant that encodes information about the warped geometry in the
bulk (and can be derived in a non-trivial way from the parameters of
the theory).  The model claims to admit self-accelerating and
self-tuning vacua. For fluctuations on a Minkowski $3$-brane one may
also expect a cascade from $6D \to 5D \to 4D$ gravity as we move from
large to short distances, for suitably chosen scales. A thorough
perturbative analysis has yet to be done that confirms this
expectation, or the possible existence of ghosts and strong coupling.
 
\subsubsection{Degravitation} 
\label{sec:degrav}

Higher co-dimension braneworld models and cascading gravity are
expected to play an important role in realising the degravitation
scenario \cite{degrav1}. The idea of degravitation is best understood
by rephrasing the cosmological constant problem. Instead of asking
``{\it why is the vacuum energy so small?}", we ask ``{\it why does
  the vacuum energy hardly gravitate at all?}".  In other words, we
accept what our particle physics models are saying and take the vacuum
energy to be up at the $(\textrm{TeV})^4$ scale or beyond. We then try
to develop a gravity theory that prevents this large vacuum energy
from generating a large amount of curvature. 
 
A phenomenological description of degravitation is given by \cite{degrav1}
\be \label{eqn:degrav}
{\cal G}^{-1} (L^2 \Box) G_{\mu\nu}=8 \pi T_{\mu\nu},
\ee
where Newton's constant, $G$, has been promoted to a differential
operator, ${\cal G}(L^2 \Box)$, depending on a length scale, $L$, and
the covariant d'Alembertian operator, $\Box$. The idea is that this
operator behaves like a high pass filter characterised by the scale
$L$. Sources with short characteristic wavelengths, $l \ll L$,  pass
through the filter and gravitate normally. Sources with  long
characteristic  wavelengths, $l \gg L$,  such as the cosmological
constant, are filtered out and hardly gravitate at all. These
considerations amount to imposing the following limits: ${\cal G}\to
G$ for $L^2(-\Box) \to \infty$ and  ${\cal G}\to 0$ for $L^2(-\Box)
\to 0$.

Since the covariant derivative does not commute with the
d'Alembertian, Equation (\ref{eqn:degrav}) suggests that the
energy-momentum tensor is not conserved. However, it is important to
realise that  this equation is only expected to describe the
linearised dynamics of the helicity-2 mode of the graviton, given by
$g_{\mu\nu}=\eta_{\mu\nu}+h_{\mu\nu}$. The source {\it is} conserved
with respect to the {\it full} metric, $\hat
g_{\mu\nu}=\eta_{\mu\nu}+h_{\mu\nu}+\ldots$, where the ellipsis
denotes the  additional (Stuckelberg) modes that necessarily appear in
a fully covariant theory \cite{degrav2}. It is clear that the extra
modes must play an important role in the filtering process at long
wavelengths. The transition to normal gravity for short wavelength
modes occurs precisely because these modes get screened by the
Vainshtein mechanism \cite{vainshtein} (see our discussion of the
Vainshtein mechanism in the context of DGP gravity in Section
\ref{sec:dgp-sc}).

The filter operator is often parametrised as follows:
\be
{\cal G}^{-1} (L^2 \Box)=\left(1-\frac{m^2(\Box)}{\Box}\right)G^{-1},
\ee
where $m^2 (\Box) \propto L^{2(\alpha-1)} \Box ^\alpha$, and $0\leq
\alpha<1$. The upper limit is a necessary, but not  sufficient,
condition for  degravitation,  whereas the lower limit is required by
unitarity \cite{degrav2}. Massive gravity (see Section
\ref{sec:massive}) corresponds to $\alpha=0$, whereas DGP
gravity\footnote{The full DGP theory does not exhibit degravitation,
  but  this is not necessarily the unique theory with $\alpha=1/2$.}
(see Section \ref{sec:dgp}) corresponds to $\alpha=1/2$. It turns out
that models with co-dimension $n>2$ are expected to correspond to
$\alpha=0$ \cite{deRham-casc2}. 

Using Equation (\ref{eqn:degrav}) as its starting point,
degravitation can be demonstrated in a cosmological context by the
ratio of scalar curvature to $\Lambda$ scaling as $t^{-1/2}$ for a
broad class of filters, where $t$ is the proper time of comoving
observers \cite{degrav2}. However, the corresponding cosmological
solution arising from a higher co-dimension brane scenario has yet to
be found explicitly.

Working at the level of the phenomenological equations, N-body
simulations  for degravitation scenarios have been studied by Khoury
and Wyman~\cite{KhouryWyman2009}. Their simulation is based on a
modified Poisson equation, given by
\begin{equation}
-\left[\frac{k^2}{a^2} + \frac{1}{L^2} \left(\frac{k L}{a}
 \right)^{2\alpha}\right] \Psi = 4 \pi G\rho .
\end{equation}
Thus, the emergence of the Vainshtein mechanism is encoded in the simulations 
via Eq. (\ref{Geff_DGP}), rather than being recovered from the
simulation itself. They find that the matter power spectrum determined
from the simulation agrees well 
with the HALOFIT~\cite{SmithEtAl2002} formula for $k <\sim
0.2h$Mpc$^{-1}$. However, for smaller scales the HALOFIT formula
over-predicts power because HALOFIT does not capture the Vainshtein
effect. Nevertheless, they find a way to recalibrate the HALOFIT
parameters such that it agrees with their simulation on all scales. 

\subsection{Einstein Gauss-Bonnet Gravity} 
\label{sec:egb}

In Section \ref{sec:lovelock-thm} we outlined how Lovelock's theorem
\cite{love1, love2} constrains the class of theories that can be
constructed from the metric tensor alone:  In four dimensions, the
most general rank-2 tensor that can be derived from a variational
principle and is (i) symmetric, (ii) divergence free, and (iii) built
out of the metric and its first two derivatives only is given by a
linear combination of the metric and the Einstein tensor, i.e.
\be
{\cal E}_{\mn}^{(4D)}=-\half \alpha_0 g_{\mn} +\alpha_1 G_{\mn}.
\ee
Such a term arises from the variation of the Einstein-Hilbert action
in the presence of a cosmological constant, 
\be
{\cal E}_{\mn}^{(4D)}=\frac{1}{\sqrt{-g}}\frac{\delta}{\delta g^{\mn}}
\int d^4 x\sqrt{-g} \left( \alpha_0+\alpha_1 R\right).
\ee
In four dimensions the Einstein-Hilbert action therefore gives the most general field
equations with the desired properties.  In more than four dimensions,
however, this result no longer holds. For $D=5$ or $6$ dimensions the
most general rank-2 tensor satisfying the same three conditions is
given by \cite{love1} (see also \cite{lanczos, zumino})
\be
{\cal E}_{ab}^{(5D/6D)}=-\half \alpha_0 \g_{ab} +\alpha_1 {\cal
  G}_{ab}+\alpha_2 {\cal H}_{ab},
\ee
where (in $D$ dimensions) $\g_{ab}$ is the metric, ${\cal
  G}_{ab}=\R_{ab}-\half \R \g_{ab}$ is the Einstein tensor, and we
have introduced the {\it Lovelock tensor} \cite{love1}
\be
{\cal H}_{ab}=2\R\R_{ab}-4 \R_{a\alpha} \R_b^c-4\R_{ac bd}\R^{cd}+2
\R_{acd e}\R_b{}^{cde}- \frac{1}{2} \hat G \g_{ab},
\ee
where $\hat G=\R_{abc d}\R^{ab c d}-4\R_{ab} \R^{ab}+\R^2$, and
$\R_{abcd}$ is the Riemann tensor constructed from $\g_{ab}$. The
Lovelock tensor is obtained by variation of the {\it Gauss-Bonnet}
action,
\be
{\cal H}_{ab}=\frac{1}{\sqrt{-\g}}\frac{\delta}{\delta \g^{ab}} \int
d^D x\sqrt{-g}\hat G.
\ee
These results can generalised still further as the dimensionality of
space-time is increased. To see how, note that $d^D x \sqrt{-\g}\R$
and  $d^D x \sqrt{-\g} \hat G$ correspond to the Euler classes of
order one and order two \cite{Nakahara}, and are topological in $D
\leq 2$ and $D \leq 4$ respectively\footnote{One can check explicitly
  that the Einstein tensor is identically zero in two dimensions, and
  that the Lovelock tensor is identically zero in four dimensions.}.
To generalise Lovelock's theorem to even higher dimensions we must
then add the higher order Euler classes to the corresponding action,
and compute the metric variation. The Euler class at order $k$ depends
on $k$th powers of curvature. In $D$ dimensions we must therefore
include Euler classes up to order $[(D-1)/2]$, where the square
brackets denote the integer part.  The reader is referred to
\cite{Charmousis-higher-order}, and references therein, for further
details.

\subsubsection{Action, equations of motion, and vacua}

In five or six dimensions General Relativity is just a special case of
a broader class of theories  commonly referred to as
Einstein-Gauss-Bonnet (EGB) gravity. These theories are described by
the action
\be \label{EGBaction}
S=\frac{1}{16 \pi G_D}\int d^Dx \sqrt{-\g} \left(\R-2\Lambda+\alpha
\hat G\right)+\int d^Dx {\cal L}_m (\g_{ab}, \psi) ,
\ee
where $ {\cal L}_m$ is the Lagrangian density of the higher
dimensional matter fields that are minimally coupled to the
metric. The corresponding field equations  are given by 
\be
\G_{ab}+\Lambda \g_{ab}+\alpha {\cal H}_{ab}=8\pi G_D T_{ab},
\ee
where $T_{ab}=-\frac{2}{\sqrt{-\g}} \frac{\delta}{\delta \g^{ab}}\int
d^Dx {\cal L}_m (\g_{ab}, \psi)$ is the energy-momentum tensor of the
matter fields.  As we will see shortly, although $\Lambda$ acts like a
bare cosmological constant, it differs from the {\it effective}
cosmological constant, $\Lambda_\textrm{eff}$, seen by the geometry.

The Gauss-Bonnet corrections are weighted by the parameter
$\alpha$. This  has dimensions of $[length]^2$ and is often associated
with the slope parameter, $\alpha'$, in heterotic string theory. To
see why, consider the effective theory describing the dynamics of the
heterotic string on the 10 dimensional target space. Working at tree
level in the string coupling and performing a perturbative  expansion
in the inverse string tension, $\mu_F^{-1} \sim 2\pi \alpha'$, we see
that the leading order term gives the standard Einstein-Hilbert
action, and the next to leading order term, after some field
redefinitions, gives the Gauss-Bonnet action \cite{Gross-the-quartic,
  Metsaev-order-alpha}. In fact, one may have expected such a result
based on the fact that string theory is known to be ghost free. As
noted by Zwiebach \cite{Zwiebach}, second-order corrections to the
Einstein-Hilbert action will necessarily give rise to ghosts unless
they appear in the Gauss-Bonnet combination. Note that if we do
identify EGB with a stringy generalisation of General Relativity, we
should restrict attention to $\alpha \geq 0$ for consistency. 
\newline
\newline
\noindent
{\it Maximally symmetric vacua}
\newline

We now consider the maximally symmetric vacuum solutions, $\bar
\g_{ab}$, satisfying $\bar
\R_{abcd}=\frac{2\Lambda_\textrm{eff}}{(D-1)(D-2)}(\bar \g_{ac} \bar
\g_{bd}-\bar \g_{ad}\bar \g_{bc})$, where $\Lambda_\textrm{eff}$ is
the effective cosmological constant seen by the curvature.  There are
two possible values for $\Lambda_\textrm{eff}$, given by
\cite{Charmousis-the-instability}
\be
\Lambda_\textrm{eff}^\pm=\Lambda_{CS} \left(1\pm
\sqrt{1-\frac{2\Lambda}{\Lambda_{CS}}}\right) ,
\ee
where
\be
\Lambda_{CS}=-\frac{1}{4\alpha} \frac{(D-1)(D-2)}{(D-3)(D-4)}.
\ee
For these vacua to be well defined the bare cosmological constant must
satisfy the bound $\Lambda/\Lambda_{CS}\leq 1/2$. It is easy to check
that $\Lambda_{  \textrm{eff}}^+/\Lambda_{CS}\geq 1 \geq \Lambda_{
  \textrm{eff}}^-/\Lambda_{CS}$, with equality when
$\Lambda=\Lambda_{CS}/2$. This is known as the  Chern-Simons limit (at
least in odd dimensions) \cite{Zanelli-lecture-notes,
  Crisostomo-black-hole-scan}, and corresponds to the case where the
two roots coincide. Note that only the lower root has a smooth limit,
$\Lambda_{  \textrm{eff}}^- \to \Lambda$, as $\alpha \to 0$, and as such is
often referred to as the ``Einstein'' branch. In contrast, the upper
root, $\Lambda_{  \textrm{eff}}^+$, is not smooth as $\alpha \to 0$, and
represents a distinct new feature of EGB  gravity that is completely
absent in higher dimensional General Relativity. For this reason, this
branch is often referred to as the ``stringy'', or ``Gauss-Bonnet'', branch. 

We now consider metric perturbations about these vacua,  $\g_{ab}=\bar
\g_{ab}+\delta \g_{ab}$. The linearised field equations then take the
remarkably simply form
\be
\delta {\cal G}_{ab}+\Lambda_\textrm{eff} \delta \g_{ab}=8 \pi
G_\textrm{eff} T_{ab} ,
\ee
where $\delta {\cal G}_{ab}$ is the linearised Einstein tensor. Thus,
we have perturbative Einstein gravity with an effective Newton's
constant given by 
\be \label{Geff}
G_\textrm{eff}= \frac{G_D}{1-\frac{\Lambda_\textrm{eff}}{\Lambda_{CS}}}.
\ee
Assuming that the bare Newton's constant is positive, $G_D >0$, it
follows that perturbative  gravity on the Einstein branch
($\Lambda_\textrm{eff}/\Lambda_{CS}<1$) is essentially well behaved,
as $G_\textrm{eff} >0$.   In contrast, on the Gauss-Bonnet branch
($\Lambda_\textrm{eff}/\Lambda_{CS}>1$), we have  $G_\textrm{eff} <0$
indicating the presence of a perturbative ghost
\cite{Charmousis-the-instability}. We refer the reader to the closing
paragraphs of Section  \ref{sec:dgpsa-lin} for a discussion of the
pathologies associated with ghosts.

The above conclusions  regarding stability are robust provided we can
trust our effective perturbative description. Here we expect this
description to be valid at energies up to a cut-off,
$E_\textrm{cut-off} \sim 1/(G_\textrm{eff})^{1/(D-2}$. As we approach
the Chern-Simons limit  ($\Lambda \to \Lambda_{CS}/2$), where the two
branches coincide, it follows that the cut-off for the effective
description should have the limit $E_\textrm{cut-off} \to 0$. This
indicates strong coupling, and a breakdown of perturbation theory. To
analyse the stability of either branch close to this limit one must
study  non-perturbative phenomena such as instanton transitions. This
reveals that transitions between branches are unsuppressed in the near
Chern-Simons regime, and that there is very strong mixing between the
two (almost degenerate) vacua. We conclude that  {\it neither} of them
can accurately describe the true quantum vacuum state in this regime,
as both will quickly become littered with bubbles of the other vacuum
\cite{Charmousis-the-instability}.
\newline
\newline
\noindent
{\it Spherically symmetric solutions}
\newline

Static spherically symmetric solutions to the vacuum field equations
were first discovered by Boulware and Deser \cite{Boulware-string},
\be \label{schw}
ds^2=-V(r) dt^2+\frac{dr^2}{V(r)} +r^2 d\Omega_{D-2},
\ee
where $d\Omega_{D-2}$ is the metric on a unit $(D-2)$-sphere.  We have
two branches for the  potential, $V(r)$, given by
\cite{Boulware-string, Charmousis-the-instability} 
\be \label{egbbh}
V_\pm (r)=1+\frac{2 \Lambda_{CS}r^2}{(D-1)(D-2)}\left(1 \pm
\sqrt{1-\frac{2\Lambda}{\Lambda_{CS}} -\frac{M(D-1)}{\Lambda_{CS}
    \Omega_{D-2}r^{D-1}}}\right) ,
\ee
where $\Omega_{D-2}$ is  the volume of the unit $(D-2)$-sphere. $M$ is
an integration constant that one can identify with the mass of a
spherically symmetric source (possibly point-like). As above, the
lower root corresponds to the Einstein branch, and has a smooth limit
as $\alpha \to 0$.  The upper root corresponds to the Gauss-Bonnet
branch, and does not have a smooth limit. Of course, the two roots are
degenerate in the Chern-Simons case ($\Lambda=\Lambda_{CS}/2$).

We shall stay away from the Chern-Simons limit in the remainder of
this section. Note that a generalised form of Birkhoff's theorem now
holds that guarantees that Equation (\ref{schw}) represents the most
general spherically symmetric solution, even without the assumption of
staticity \cite{Charmousis=general-GB}.  Let us consider the
properties of this solution for a source of positive mass, $M>0$. On
the Einstein branch the singularity at $r=0$ is shielded by an event
horizon, and there are no obvious pathologies. This is not the case on
the Gauss-Bonnet branch, where we have a naked time-like singularity.

At asymptotically large radii the metric functions take the form
\be
V_\pm  \approx -\frac{2\Lambda^\pm_\textrm{eff} r^2}{(D-1)(D-2)}+1 \pm
\frac{M}{\Lambda_{CS} \Omega_{D-2}r^{D-3}} .
\ee
The asymptotic solution here can be seen to resemble the generalised
Schwarzschild-Tangherlini solution in $D$ dimensional GR
\cite{tangherlini-schwarzschild}, with cosmological constant
$\Lambda_\textrm{eff}^{\pm}$, and  mass $\pm M$.  In particular, on
the Gauss-Bonnet branch, it appears as if our solution has negative
gravitational energy even for a positive energy source.  This
conclusion, however, is incorrect. A proper computation of the
gravitational energy taking into account all of the  Gauss-Bonnet
corrections reveals  the mass of the solution in Eq. (\ref{schw}) is
$+M$ on {\it both} branches \cite{Deser-Tekin1, Deser-Tekin2,
  Padilla-surface}.  For an in-depth discussion of this, and related
stability issues, the reader is referred to
\cite{Charmousis-the-instability}.  For an excellent review of
Lovelock gravities, and their black hole solutions, see
\cite{Charmousis-higher-order}.

\subsubsection{Kaluza-Klein reduction of EGB gravity}
In Section \ref{sec:KKcompactification} we discussed compactification
of higher dimensional GR on a circle.  The same operation can now be
performed with the EGB action, given in Eq. (\ref{EGBaction}). We
start in five dimensions with coordinates $(x^\mu, z)$, where the $z$
direction is compact.  We can then dimensionally reduce down to four
dimensions using the following ansatz 
\be
\g_{\mu\nu}=g_{\mu\nu}+e^{2\phi} A_{\mu}A_{\nu}, \qquad
\g_{\mu z}= e^{2\phi} A_{\mu}, \qquad {\rm and} \qquad \g_{zz}=e^{2\phi},
\ee
where the metric, $g_{\mu\nu}$, gauge field, $A_\mu$ and the dilaton,
$\phi$, are all independent of $z$. Integrating out the compact
dimension we arrive at the following effective theory
\cite{Mueller-Hoissen-dimensionally}:
\be
S_\textrm{eff}=\frac{1}{16\pi G_5} \int d^4 x\sqrt{-g} {\cal L}_{\textrm{eff}},
\ee
where 
\begin{multline} \label{Leff}
 {\cal L}_{\textrm{eff}}=
 e^\phi \left[ R-\frac{1}{4} e^{2\phi} F^2-2\Lambda   +\alpha
 (R_{\mu\nu\alpha\beta}R^{\mu\nu\alpha\beta}-4R_\mn R^\mn+R^2)\right] 
\\ 
+\alpha e^{3\phi}\left[-\frac{3}{8} e^{2\phi}\left(F^\mu{}_\nu
 F^\nu{}_\alpha F^\alpha{}_\beta F^\beta{}_\mu -\frac{1}{2} (F^2)^2
 \right)\right.\\ 
-\left(F^\mu{}_\alpha F^\nu{}_\beta+F^\mn
 F_{\alpha\beta}\right)\left(R^{\alpha\beta}{}_\mn-4R^\alpha_{[\mu}\delta^\beta_{\nu]}
 +R\delta^\alpha_{[\mu}\delta^\beta_{\nu]}\right)\\ 
 -2( \nabla_\mu F_{\alpha\beta})(\nabla^\alpha F^{\beta \mu})-2
 (\nabla^\mu F_{\alpha\mu} )(\nabla_\nu F^{\alpha \nu})
 -12(F_{\mu\alpha}\nabla^\alpha \phi)( F^{\mu\beta}\nabla_\beta
 \phi)+6F^2 (\nabla\phi)^2 \\ 
\left. +4 (F_{ \alpha\beta} \nabla_\mu \phi )\nabla^\mu F^{\alpha
 \beta}-4 (F_{ \alpha\beta} \nabla_\mu \phi )\nabla^\alpha F^{\beta
 \mu}-12(F_{ \alpha\beta} \nabla^\beta \phi )\nabla_\mu F^{\alpha
 \mu}\right.\Bigg]\\ 
+\alpha e^\phi\left[ 8 e^{-\phi} \nabla_{\mu}\nabla_\nu
 e^\phi-2e^{2\phi}F_{\mu\alpha}F_\nu{}^\alpha\right]G^\mn  +3\alpha
 e^{2\phi}\left[2 F^{\alpha\beta}F_{\beta}{}^\mu
 \nabla_{\mu}\nabla_{\alpha}e^\phi+F^2\Box e^\phi \right] ,
\end{multline}
and $F^2=F_\mn F^\mn$, with $F_\mn=\nabla_\mu A_\nu -\nabla_\nu
A_\mu$. Note that for $\Lambda=0$ and $\alpha=0$ we recover the
results of Section \ref{sec:KKcompactification}, provided we perform a
conformal transformation to the Einstein frame and canonically
normalise the dilaton.

One property of the Lagrangian in Eq. (\ref{Leff}) is that it gives rise to
field equations that are at most second order in derivatives.  This
is, of course, inherited from the underlying theory, and enables us
to consider an interesting class of  scalar-tensor theories by
freezing the gauge field such that $A_\mu=0$. We refer the reader to
Section \ref{scalartensorsection}, and \cite{Amendola-solar}, for
further discussions on four-dimensional theories of this type.

\subsubsection{Co-dimension one branes in EGB gravity}

Interest in EGB gravity has really taken off in recent years, largely
due to its application to braneworlds. We  discussed the braneworld
paradigm in some detail in Section
\ref{sec:braneworlds}. Schematically, the physics of $3$-branes in
five dimensional EGB gravity is more or less the same as in General
Relativity, as we will now discuss.
\newline
\newline
\noindent
{\it Generic action and equations of motion} \label{egb-generic}
\newline

As in the GR case discussed in Section \ref{sec:genaction}, we split
our five dimensional bulk into a series of domains separated by a
series of $3$-branes, so that the action is given by 
\begin{multline}\label{EGBgen-action}
S =  \int_{\textrm{bulk}} d^5 x \sqrt{-\gamma}\left[\frac{M_5^3}{2} (\R+\alpha
  \hat G ) +{\cal L}_{\textrm{bulk}}\right] \\ 
+\sum_{\textrm{branes}} \int_{\textrm{brane}} d^4 x \sqrt{-g}\left\{- \Delta  \left[
  M_5^3 \left(K+2\alpha(J-2G^\mn K_\mn)\right) \right]+{\cal
  L}_{\textrm{brane}}\right\} ,
\end{multline}
where $M_5=(1/8\pi G_5)^{1/3}$ is the bulk Planck scale. In comparison
with Equation (\ref{gen-action}) we have a Gauss-Bonnet correction in
the bulk and the corresponding Myers boundary term \cite{Myers-higher}
on each brane. The latter depends on the extrinsic curvature, $K_\mn$,
the induced Einstein tensor $G_\mn=R_\mn-\half R g_\mn$, and the trace
$J=g^\mn J_\mn$, where \cite{Davis-generalized}
\be
J_\mn=\frac{1}{3}\left( 2K K_{\mu\alpha}
K^\alpha_\nu+K_{\alpha\beta}K^{\alpha\beta} K_\mn-2 K_{\mu \alpha}
K^{\alpha\beta} K_{\beta \nu}-K^2 K_\mn\right) .
\ee
The bulk field equations are then given by
\be \label{EGB-bulkeom}
\R_{ab}-\half R \g_{ab}+\alpha {\cal H}_{ab}=\frac{1}{M_5^3} T_{ab}^{\textrm{bulk}},
\ee
with the boundary conditions at the brane given by the ``DGW" junction
conditions \cite{Davis-generalized, Gravanis-israel} (see also
\cite{Deruelle-smooth}),
\be \label{dgw}
\Delta\left\{M_5^2\left[K_{\mu\nu}-K g_{\mu\nu}+2\alpha (3J_\mn -J
  g_\mn-2P_{\mu\alpha \nu \beta} K^{\alpha
    \beta})\right]\right\}=-T^{\textrm{brane}}_{\mu\nu} .
\ee
Here we introduce the {\it  double dual} of the Riemann tensor, defined as \cite{MTW}
\be
P^{\mu\nu}_{\phantom{\mu\nu} \alpha \beta} = -
R^{\mu\nu}_{\phantom{\mu\nu}\alpha \beta} 
+2 R^{\mu}_{\phantom{\mu} [\alpha} \delta^{\nu}_{\phantom{\nu} \beta]}
-2 R^{\nu}_{\phantom{\nu} [\alpha} \delta^{\mu}_{\phantom{\mu} \beta]}
-R \delta^{\mu}_{\phantom{\mu} [ \alpha} \delta^{\nu}_{\phantom{\nu} \beta]}.
\ee
In what follows we will assume that   the bulk geometry is
only\footnote{For generalisations with bulk scalar
  fields and bulk Maxwell fields see \cite{Charmousis-scalar-brane}
  and \cite{Lidsey-brane-word-cosmology}, respectively.} sourced by a  cosmological
constant $T^{\textrm{bulk}}_{ab}=M_5^3 \Lambda \g_{ab}$, which we take to be
negative, as in the Randall-Sundrum model, $\Lambda=-6k^2$.  We will
consider a single brane with tension $\sigma$, and some induced
curvature, $M_{\textrm{ind}}^2 R$.  Its  energy momentum tensor  is then given by
\be
T^{\textrm{brane}}_{\mu\nu}=-\sigma g_\mn -M_{\textrm{ind}}^2 G_\mn +\T_\mn ,
\ee
where $\T_\mn$ denotes the contribution from any additional matter
excitations on the brane. For simplicity we will  impose
$\mathbb{Z}_2$ symmetry across the brane (see \cite{asymm1, asymm2}
for a discussion of asymmetric configurations). 
\newline
\newline
\noindent
{\it Weak gravity on a Minkowski brane}
\newline

Let us seek Randall-Sundrum-like vacua, corresponding to a Minkowski
brane in an AdS bulk,
\be
ds^2=\bar \g_{ab} dx^a dx^b =dz^2+e^{-2 |z|/l_\textrm{eff}} \eta_\mn
dx^\mu dx^\nu .
\ee
The effective AdS curvature in the bulk is then determined by the bulk
equations of motion (\ref{EGB-bulkeom}). We find
\be
\frac{1}{l^2_\textrm{eff}}=\frac{1}{4\alpha} \left(1 \pm
\sqrt{1-\frac{8\alpha}{l^2} } \right) ,
\ee
where the lower root corresponds to the Einstein branch and the upper
root to the GB branch. Meanwhile, the vacuum DGW junction conditions
at $z=0$ impose the following constraint on the brane tension
\be
\sigma=6\frac{M_5^3}{
  l_\textrm{eff}}\left(1-\frac{4\alpha}{3l_\textrm{eff}^2} \right). 
\ee
Since we are interested in weak gravity on the brane, we consider
small fluctuations in the metric and the brane position. As we saw
previously, the linearised equations of motion are identical to those
found in perturbative General Relativity, 
\be
\delta {\cal G}_{ab}-\frac{6~}{l^2_\textrm{eff}} \delta \g_{ab}=0.
\ee
The linearised junction conditions also take a remarkably simple form,
\be
2(M_5^\textrm{eff})^3 \delta (K^\mu_\nu
-K\delta^\mu_\nu)=M_\textrm{ren}^2 \delta G^\mu_\nu-\T^\mu_\nu ,
\ee
where we identify the {\it effective} five dimensional Planck scale,
\be
(M_5^\textrm{eff})^3=M_5^3\left(1-\frac{4\alpha}{ l_\textrm{eff}^2}\right)
\ee
and the renormalised induced curvature scale,
\be
M^2_\textrm{ren}=M_{\textrm{ind}}^2+\frac{8 \alpha M_5^3}{ l_\textrm{eff}} .
\ee
The effective Planck scale is consistent with the effective Newton's
constant introduced in Equation (\ref{Geff}),
$(M_5^\textrm{eff})=(1/8\pi G_\textrm{eff})^{1/3}$. It follows that
our weak gravity description is identical to the corresponding
description in General Relativity, provided we make use of the
effective cosmological constant and Planck scale in the bulk, and
renormalise the induced curvature on the brane. For further details
the reader is referred to \cite{Charmousis-higher-order}, and
references therein (see also  \cite{Deruelle-newtons-law}).

The characteristic behaviour of weak gravity on the brane is as
follows \cite{Charmousis-higher-order}: At large distances we  recover
four dimensional General Relativity, with an effective four
dimensional Planck scale
\be
M_{pl}^\textrm{IR}=\sqrt{M_{\textrm{ind}}^2+M_5^3 l_\textrm{eff}(1+4\alpha/
  l_\textrm{eff}^2)}=\sqrt{M_{\textrm{ren}}^2+(M_5^\textrm{eff})^3
  l_\textrm{eff}} .
\ee
This holds provided $1/l_\textrm{eff} \neq 0$.  For
$1/l_\textrm{eff}=0$ the large distance behaviour is five dimensional
owing to the absence of a normalisable zero mode in the Minkowski
bulk.

At short distances we recover four dimensional Brans-Dicke gravity,
with a different effective Planck scale\footnote{There is a
  typo in the corresponding expression for $M^\textrm{UV}_{pl}$ in
  \cite{Charmousis-higher-order}.}, $M^\textrm{UV}_{pl}=\sqrt{M_{\textrm{ind}}^2+8 \alpha M_5^3
  /l_\textrm{eff}}=M_\textrm{ren}$, and a Brans-Dicke parameter given
by
\be
2w+3=\frac{3}{4} \left(\frac{M_{\textrm{ind}}^2/l_\textrm{eff}+M_5^3(1+4\alpha/
  l_\textrm{eff}^2)}{M_5^3(1-4\alpha/
  l_\textrm{eff}^2)}\right)=\frac{3}{4}\left(\frac{(M_{pl}^\textrm{IR})^2}{(M_5^\textrm{eff})^3l_\textrm{eff}}\right) .
\ee
This holds provided $M_{\textrm{ren}} \neq 0$. If $M_{\textrm{ren}} =0$ then the short
distance behaviour is five dimensional since there is no effective
induced curvature giving rise to quasi-localisation near the brane.

These results are consistent with the behaviour that would be expected
in linearised theory on a brane in five dimensional General
Relativity, with bulk cosmological constant,
$\Lambda_\textrm{eff}=6/l^2_\textrm{eff}$, bulk Planck scale,
$M_5^\textrm{eff}$, and brane induced curvature scale, $M_{\textrm{ren}}$. It
is interesting to note that the Brans-Dicke parameter gets large close
to the Chern-Simons limit as $M_5^\textrm{eff} \to 0$. However, as we
saw earlier, quantum fluctuations in the bulk become strongly coupled
at $\sim M_5^\textrm{eff}$, so this prediction may be
unreliable.
\newline
\newline
\noindent
{\it Brane cosmology in EGB gravity}
\newline

The methods used in Section \ref{sec:BWcos} to derive the cosmological
evolution of co-dimension one branes apply equally well to branes in
EGB gravity, provided we use the DGW junction conditions give in
Eq. (\ref{dgw}). Recall that brane cosmology can be studied using
either the {\it brane-based formalism} or the {\it bulk-based
  formalism}. The brane-based formalism relies on the covariant
formulation of the effective Einstein equation on the brane. This has
been worked out for EGB gravity and applied to cosmological branes
\cite{Maeda-covariant} (see also \cite{Kobayashi-low-energy}). In the
bulk based formalism, the generalised form of Birkhoff's theorem
\cite{Charmousis=general-GB} ensures that the bulk geometry around a
FLRW brane is given by
\begin{equation} \label{eqn:schmetricGB}
ds^2=-V(r)d\tau^2+\frac{dr^2}{V(r)}+r^2q_{ij} dx^i dx^j,
\end{equation}
where $V(r)$ takes the form
\be
V_\pm (r)=\kappa+\frac{r^2}{4\alpha} \left(1 \pm
\sqrt{1-\frac{8\alpha}{ l^2}-\frac{8\alpha \mu}{r^{4}}}\right) ,
\ee
and $q_{ij}(x)$ is the metric of a 3-space of constant curvature,
$\kappa=0, \pm 1$.  Note that this generalises the black hole solution
given in Eq. (\ref{egbbh}) to different ``horizon" topologies
\cite{Cai-topological}. Let us once again assume $\mathbb{Z}_2$
symmetry across the brane for simplicity. As in Section
\ref{sec:bulk-based}, we  treat the brane as an embedding
$\tau=\tau(t)$, and $r=a(t)$ in the bulk geometry, identifying $t$
with the proper time of comoving observers on the brane, and $a(t)$
with the scale factor. After imposing the DGW boundary conditions
given in Eq. (\ref{dgw}), we find that the Hubble parameter, $H=\dot
a/a$, obeys  \cite{Charmousis-higher-order} (see also
\cite{Davis-generalized,
  Charmousis=general-GB,Gregory-brane-world-holography,
  Maeda-braneworld-dynamics, Kofinas-brane-cosmology})
\ba
\left[1+\frac{4
    \alpha}{3}\left(2\left(H^2+\frac{\kappa}{a^2}\right)+\frac{\kappa-V(a)}{a^2}\right)\right]\sqrt{H^2+\frac{V(a)}{a^2}}&=&\frac{\rho_{\textrm{brane}}}{6M_5^3}\nonumber\\&=&\frac{\sigma+\rho-3M_{\textrm{ind}}^2 \left(H^2+\frac{\kappa}{a^2}\right)}{6M_5^3}.\qquad
\ea
This can be recast as a cubic equation in $H^2+\frac{\kappa}{a^2}$,
and solved analytically to give a modified Friedmann equation. The
most commonly studied scenario has $M_{\textrm{ind}}=0$, with the  bulk taken
to lie on the Einstein branch.  We then find \cite{Davis-generalized}
\be
H^2+\frac{\kappa}{a^2}=\frac{c_++c_--2}{8\alpha},
\ee
where
\be
c_\pm= \left[\sqrt{\left(1-\frac{8\alpha}{ l^2}+\frac{8\alpha
      \mu}{a^4}\right)^{3/2}+\frac{\alpha(\sigma+\rho)^2}{2M_5^6}}\pm
  \frac{(\sigma+\rho)}{M_5^3}\sqrt{\frac{\alpha}{2}} \right]^{2/3} .
\ee
This cosmology can give rise to rapid inflation, just as in
Randall-Sundrum cosmology \cite{Lidsey-inflation-in-GB}. However, the
Gauss-Bonnet corrections do introduce some new features at  the level
of  cosmological perturbations. For example, in Randall-Sundrum
cosmology the consistency relation between the tensor-to-scalar ratio,
$r$, and the tensor spectral index, $n_T$, agrees with the standard GR
result, $r=-8n_T$. This  relationship is broken by the GB corrections
as the amplitude for the tensor perturbations are no longer
monotonically increasing with scale \cite{Dufaux-cosmo-perts}.  We
also find that observational constraints of GB brane inflation are
typically softened relative to Randall-Sundrum cosmology, such that
certain ``steep" potentials are no longer ruled out
\cite{Tsujikawa-observational}.

\subsubsection{Co-dimension two branes in EGB gravity} 
\label{sec:GB-cod2}

In Section \ref{sec:highercod} we mentioned that higher-order
operators could be added to the bulk gravity to  regularise the
singularities that appear when we have an infinitely thin braneworld
of co-dimension $\geq 2$. One of the ways to do this is to add
Gauss-Bonnet corrections in the bulk \cite{Bostock-Einstein,
  Charmousis-matching,Charmousis-Einstein}.  The simplest and most
well studied scenario is to consider co-dimension-2 branes in  six
dimensional EGB gravity without any additional sources in the bulk
(see, for example, \cite{Bostock-Einstein, Charmousis-consistency,
  Kanno-quasi, Kanno-on-the,  Charmousis-properties,
  Charmousis-self-properties, CuadrosMelgar-black,
  CuadrosMelgar-perturbations}). This gives
\be\label{EGB6d}
S =  \int_{\textrm{bulk}} d^6 x \sqrt{-\gamma}\left[\frac{1}{16\pi G_6}
  (\R+\alpha \hat G ) \right]+ \int_{\textrm{brane}} d^4 x {\cal L}_{\textrm{brane}} .
\ee
If we assume axial symmetry in the bulk we get
\be
ds^2=\gamma_{AB} dx^A dx^B=dr^2+L^2(x, r)d\theta^2+g_\mn(x, r) dx^\mu
dx^\nu, 
\ee
and it turns out that the Einstein tensor on the brane is given by
\cite{Bostock-Einstein, Charmousis-consistency} 
\be
G_\mn=-\frac{1}{4\alpha}g_\mn+\frac{G_6}{\alpha(1-\beta)}T^{\textrm{brane}}_\mn+f(\beta)
W_\mn ,
\ee
where
\be
W_\mn=K_\mu^\lambda  K_{\nu \lambda}-K K_\mn+\frac{1}{2} g_\mn
(K^2-K_{\lambda \sigma}K^{\lambda \sigma}) ,
\ee
and $K_\mn=\frac{1}{2} \del_r g_\mn$.  The parameter $\beta$ is the
deficit angle on the brane, and is often assumed to be constant,
although this need not be case \cite{Charmousis-consistency}. A field
dependent deficit angle, $\beta=\beta(x)$,  will lead to two important
effects: Transfer of  energy between bulk and brane, and an effective
four dimensional Newton's ``constant" that can vary as
$G_\textrm{eff}=\frac{G_6}{8\pi \alpha (1-\beta(x))}$
\cite{Charmousis-consistency}. These can clearly be constrained by
observation, but it would be very interesting to study the role
varying $\beta$ could play in attempts to self-tune the vacuum
curvature in co-dimension 2 models.

The function $f(\beta)$ depends on the mathematical technique used to
derive the boundary conditions at the distributional source
\cite{Charmousis-consistency, Kanno-quasi, Kanno-on-the}.  The
boundary conditions derived in \cite{Bostock-Einstein} lead to the
condition $W_\mn |_{\textrm{brane}}= 0$, suggesting that Einstein gravity
should be recovered on the brane at {\it all scales}, even for an
infinitely large bulk.  However, it is now understood that these
conclusions  rely on the assumption $\del_r K_\mn =0$ at the brane,
which is too constraining. For constant $\beta$  the Friedmann
equation on the brane is given by  \cite{Charmousis-consistency} 
\be
H^2+\frac{\kappa}{a^2}=\frac{8 \pi G_\textrm{eff}}{3}
\rho_{\textrm{brane}}-\frac{1}{12\alpha} +\frac{c^2}{\rho_{\textrm{brane}}^2 a^8} ,
\ee 
where $c^2$ is an integration constant.  However, as we already
emphasised,  in general  $\beta$ can vary, and in this case one cannot
find a closed system of equations on the brane
\cite{Charmousis-consistency}.

\newpage 

\section{Parameterised Post-Friedmannian Approaches and Observational
Constraints}
\label{PPF}

As expounded in this review, there now exists a vast range of
candidate theories of gravity that modify Einstein's theory of General
Relativity in one way or another.  It is also seems clear that many
more such theories are likely to be proposed in the future. If these theories
are to be of any value in understanding and resolving the problems
associated with the Dark Universe then they must be confronted with
cosmological data. This is, in principle, straightforward but
time-consuming. It involves working out the perturbation equations for
each and every theory, incorporating them into the Einstein-Boltzmann
solvers and N-body codes, and calculating a list of observables.

This situation is analogous to the experimental study of General
Relativity in the early 1970s, during what has become known as the
`Golden Age' of General Relativity. There, one had a plethora of
alternative theories of gravity that needed to be confronted with
constraints from Solar System measurements. The Parameterised
Post-Newtonian (PPN) method was invented in this case as an
intermediate step between theory and experiment.  It involves a set of
generic parameters that can be easily constrained by experiments
\cite{ThorneWill1971}.  Using the PPN method one can then take any
given theory, calculate the PPN parameters it predicts, and compare
them with observational constraints.  This process is outlined in
Section \ref{sec:PPN}.

Over the last few years, the idea of creating such an intermediate
step when considering cosmological constraints has starting taking
hold. It has been dubbed by some the `Parameterised
Post-Friedmannian approach', and it attempts to encompass, at the linear
level, the behaviour of a wide array of alternative theories of gravity.
We will now outline the basic idea behind this approach.

The bulk of the cosmological data that can be used to constrain
modifications of gravity can be interpreted in terms of perturbations
about a Friedmann-Lema\^{i}tre-Robertson-Walker universe. Throughout this
review we have presented how the evolution of cosmological
perturbations is modified in these theories, relative to their
behaviour in General Relativity. One can now ask oneself if there is a
{\it general} way of modifying the equations of cosmological
perturbations such that it will encompass all the theories we have
previously discussed.

The simplest approach, that has been in vogue for the past few years,
is to modify two of the four Einstein field equations as follows:
\begin{eqnarray}
-2k^2\Phi  &=&  8 \pi \mu G a^2 \sum_X \rho_X \left[\delta_X+3(1+w_X)
 \adotoa \theta_X \right]
\label{G} 
\\  
\Phi-\Psi &=& \zeta\Phi, \label{slip}
\end{eqnarray}
where two new functions have been introduced: The effective Newton
constant, $G_{\rm eff}=\mu G$, and the gravitational slip, $\zeta$.
One can interpret $G_{\rm eff}$ as the inclusion of a form of
gravitational screening, reducing (or enhancing) the local
gravitational force on cosmological scales. The gravitational slip
phenomenologically parametrises the shear that seems to arise
frequently in scenarios of modified gravity. This parametrisation is
incredibly useful for quantifying deviations from General Relativity,
and a number of authors have used it in their analyses of cosmological
data \cite{Bertschinger2006, Caldwell2007, HuSawicki2007, Hu2008,
  Amendola2008, Daniel2008, Daniel2010, Beanetal2010, Zhaoetal2010}.

It is instructive to see in what circumstances such a parametrisation
might arise, and to do this we will develop a consistent formalism in what
follows.

\subsection{The Formalism}
\label{the_formalism}

In order to generalise the perturbed Einstein equations we follow the
approach and notation used in \cite{Skordis2009, SkordisProcMod,FerreiraSkordis2010,
  Baker2011}.  Here we split the field equations of the theory in
question into a set of evolution equations for the metric, evolution equations for the
additional gravitational fields (if any are present), and a set of
constraint equations.  The evolution equations for the metric, and the
constraint equations, can then be written schematically as
\begin{equation}
{\delta G}^{mod}_{\mu\nu} \equiv {\delta G}_{\mu\nu} - \delta U_{\mu\nu}=8\pi
G \left[ \delta T_{\mu\nu}+ \delta T^{E}_{\mu\nu} \right], 
\end{equation}
where ${\delta G}_{\mu \nu}$ is the perturbed Einstein tensor, $\delta
U_{\mu\nu}$ is the contribution of any other terms that involve
perturbations of the metric, $\delta T_{\mu\nu}$ is the perturbed
energy-momentum tensor of matter fields in the space-time, and $\delta
T^{E}_{\mu\nu}$ is the contribution of any terms that involve
perturbations to the additional gravitational fields.

Let us now be more specific. We can define the following new
variables: 
$U_\Delta \equiv -a^2 \delta U^0_{\phantom{0}0}$, 
$\grad_iU_\Theta \equiv -a^2 \delta U^0_{\phantom{0}i}$,  
$U_P \equiv \delta U^i_{\phantom{0}i}$
and $D^i_{\phantom{i}j}U_\Sigma \equiv a^2(  \delta U^i_{\phantom{0}j}
-   \frac{1}{3} \delta U^k_{\phantom{0}k} \delta^i_{\phantom{i}j}) $,
as well as the new gauge invariant
\begin{equation}
\GammaGI \equiv \frac{1}{k}\left(\PhiGI' + \adotoa \PsiGI\right),
\label{GammaGI}
\end{equation}
where the Bardeen potentials $\PhiGI$ and $\PsiGI$ are defined in
Eqs. (\ref{BarGR1}) and (\ref{BarGR2}) of Section
\ref{sec:cosmology}.  Dropping the hats (i.e. working in the conformal
Newtonian gauge), we can write the two
constraint equations coming from $G_{00}$ and $G_{0i}$ as
\begin{eqnarray}
 -2 k^2 (\Phi + 3 \adotoa_k \Gamma) &=& 8\pi G a^2  \sum_X \rho_X  \delta_X +U_\Delta
\label{mod_Delta}
\\
 2k \Gamma  &=&  8\pi G a^2 \sum_X (\rho_X + P_X)\theta_X + U_\Theta,
\label{momentum}
\end{eqnarray}
and the two evolution equations coming from the trace and traceless
parts of $G_{ij}$ as
\begin{eqnarray}
k \Gamma'  + 2k\adotoa \Gamma + \left(\adotoa'  - \adotoa^2 -
 \frac{k^2}{3}\right) \Psi + \frac{k^2}{3}  \Phi &=&
 4\pi G a^2 \sum_X \delta P_X  +  \frac{1}{6} U_P 
\label{modP} \\
 \Phi - \Psi &=&  8\pi G a^2 \sum_X (\rho_X + P_X)\Sigma_X + U_\Sigma.
\label{modshear}
\end{eqnarray}
We can then combine Eqs. (\ref{mod_Delta}) and (\ref{momentum}) to
find a modified Poisson equation:
\be
 -2 k^2 \Phi  = 8\pi G a^2  \sum_X \rho_X \left[ \delta_X + 3 \adotoa(1+w_X)\theta_X\right]
+U_\Delta + 3 \adotoa U_\Theta .
\label{mod_Poisson}
\ee

Assuming that the theory in question has at most $N$-time
derivatives in its field equations, and bearing in mind from
Eqs. (\ref{BarGR1}), (\ref{BarGR2}) and (\ref{GammaGI}) that $\PhiGI$
and $\GammaGI$ have one time derivative when expressed in an arbitrary
gauge, the components of the tensor $U_{\alpha\beta}$ can be written as
\begin{eqnarray}
U_{\Delta} &=& \sum^{N-2}_{n=0} k^{2-n} \left[A_n \PhiGI^{(n)} +  E_n  \GammaGI^{(n)} \right],
\label{Udelta}
\\
U_{\Theta} &=& \sum^{N-2}_{n=0} k^{1-n} \left[B_n \PhiGI^{(n)} +  F_n  \GammaGI^{(n)} \right],
\label{Utheta} 
\\
U_{P} &=& \sum^{N-1}_{n=0} k^{2-n} \left[C_n \PhiGI^{(n)} +  I_n  \GammaGI^{(n)} \right],
\label{Up}
\\
U_{\Sigma} &=& \sum^{N-1}_{n=0} k^{-n} \left[D_n \PhiGI^{(n)} +  J_n  \GammaGI^{(n)} \right],
\label{Usigma}
\end{eqnarray}
where $\PhiGI^{(n)} \equiv \frac{d^n}{d\tau^n}\PhiGI$, and similarly for $\GammaGI$.
The coefficients $A_n$-$J_n$ depend on time and scale through the
scale factor, $a$, and wavenumber, $k$.  For the sake of brevity we
will refrain from explicitly stating these dependences.

Now, although we have defined $U$ above in the Newtonian gauge, the
individual terms appearing in the expressions above are all gauge
invariant.  This fact, however, imposes further constraints because
the gauge-invariant variable $\GammaGI$ contains second derivatives of
the scale factor, when expressed in an arbitrary gauge.  Hence, to
avoid higher derivatives of the background appearing in the field
equations, we have to set
\begin{eqnarray}
 E_{N-2} =  F_{N-2} = I_{N-1}  = J_{N-1} = 0.
\end{eqnarray}
We assume that the evolution equations for the matter fields remain
unchanged, and that these equations are supplemented by additional
evolution equations for the extra gravitational fields.  Finally, the
field equations are closed by imposing the Bianchi identities.  This
imposes one of two possible options: (i) $\nabla^\alpha
T^E_{\alpha\beta} = \nabla^\alpha U_{\alpha\beta}=0$, which imposes a
series of constraints such that the theory remains consistent, or (ii)
$\nabla^\alpha (T^E_{\alpha\beta} + U_{\alpha\beta})=0$, which is the
more general situation.  For a detailed discussion of these issues see
\cite{Skordis2009,SkordisProcMod,Baker2011}.

\subsubsection{Evolution of perturbations on super-horizon scales}
\label{Bertschinger_construction}

In principle it may seem that one should explore the long wavelength
behaviour of cosmological perturbations on a case by case basis.  It
turns out, however, that the infinite wavelength mode can be studied
simply by considering the evolution of the background equations.  This
observation is due to Bertschinger \cite{Bertschinger2006}, and
proceeds as follows.  Consider a background with scale factor
$a(\tau,\kappa)$ where $\kappa$ is the spatial curvature of a
hyper-surface of constant $\tau$:
\begin{eqnarray}
ds^2 = a^2\left[-d\tau^2 + d\chi^2  +   \frac{1}{\kappa}
  \sinh_\kappa^2(\sqrt{\kappa}\chi) d\Omega \right]
\end{eqnarray}
where $\sinh_\kappa(x)$ equals $\sin(x)$ for $\kappa>0$,  equals $x$ for
$\kappa=0$, or equals $\sinh(x)$ for $\kappa<0$. We can now perturb
this space-time as $\kappa\rightarrow \kappa(1+\delta_\kappa)$ , and
compensate by a change the coordinates $\tau\rightarrow\tau+\alpha$
and $\chi\rightarrow\chi(1- \frac{1}{2}\delta_\kappa)$. Here
$\delta_\kappa$ is a constant, while $\alpha=\alpha(\tau)$. Note that
in this case the scale factor $a(\tau,\kappa)$ is perturbed as $a(\tau
+ \alpha, \kappa(1 + \delta_\kappa))$.  In words, the scale factor
will come out as the solution of some generalised set of Friedmann
equations, and will depend on the spatial curvature, $\kappa$.

We can now write this new geometry in the form of a perturbed FLRW metric,
with background curvature $\kappa$: 
\begin{eqnarray}
ds^2=a^2\left\{-(1+2\Psi)d\tau^2+(1-2\Phi)\left[d\chi^2+
  \frac{1}{\kappa} \sinh_\kappa^2(\sqrt{\kappa}\chi)
  d\Omega\right]\right\}, 
\end{eqnarray}
where 
\begin{eqnarray}
\Psi(\tau) &=&  \frac{\partial \ln a}{\partial \ln \kappa}
\delta_\kappa + \alpha' + \adotoa \alpha, \\
\Phi(\tau) &=& \left( \frac{1}{2} -    \frac{\partial \ln a}{\partial
  \ln \kappa} \right) \delta_\kappa - \adotoa \alpha .
\end{eqnarray}
One can now eliminate $\alpha$ to find a generic evolution equation that
relates $\delta_\kappa$ with $\Phi$ and $\Psi$ {\it without}
specifying any particular theory of gravity:
\begin{eqnarray}
\frac{1}{a^2}\frac{d}{d\tau}\left(\frac{a^2\Phi}{\adotoa}\right) &=&
   \Phi - \Psi  +
   \left[\frac{1}{2a}\frac{d}{d\tau}\left(\frac{a}{\adotoa}\right) -
   \frac{\partial\ln \adotoa}{\partial \ln \kappa} \right]
   \delta_\kappa ,
\label{bert_large}
\end{eqnarray}
where entropy perturbations have been neglected\footnote{Including
  entropy perturbations is straightforward, and for this the reader is
  referred to~\cite{Bertschinger2006}.}.  The constant
  $\delta_\kappa$ that remains in the equation above has a direct
  physical interpretation: It is twice the comoving curvature
  perturbation.  Note that we have also assumed that shear
  perturbations are negligible on large scales, that local
  energy-momentum us conserved, and that spatial gradients can be discarded. 
By choosing $\adotoa(a,\kappa)$, and a relation between $\Phi$ and
  $\Psi$, one now has the evolution equation completely defined.
This is a powerful statement, as it means it is possible to determine
the evolution of large-scale perturbations without delving into the
details of the theory.  

However, in order to complete the system here one still needs to specify a
relation between $\Phi$ and $\Psi$.  As we will describe below, it
has become standard practise to assume the simplest form of the PPF
parametrisation, that we described above, on super-horizon scales.  Within this
approach comparisons have been made between choices of $\mu$ and $\zeta$,
and the outcomes of numerical solutions for specific theories.  These
comparisons show reasonable agreement, but, as yet, there is no
compelling argument for applying the simplest PPF parametrisation on
large scales.  In other words, there is no guarantee that a simple
relation of the form $\Psi = \zeta\Phi$ can encompass all possible
theories of gravity.  In fact, it would appear that
Eq. (\ref{bert_large}) does not allow such an interpretation.


\subsubsection{The simplified PPF approach, and its extensions}

Let us now revisit the simplest, and by far the most popular, version
of the Parameterised Post Friedmannian approach
\cite{Bertschinger2006,Caldwell2007,HuSawicki2007,Hu2008,Amendola2008}.
Here one considers only Eqs. (\ref{modshear}) and
(\ref{mod_Poisson}). The further assumption that perturbations have no
anisotropic stress (i.e. $\Sigma_X=0$) then allows one to reorganise
them in the form of Eqs. (\ref{G}) and (\ref{slip}). These equations
can then be rewritten in terms of the $U_{\alpha\beta}$ tensor by choosing 
\begin{eqnarray}
U_\Delta+3{\cal H}U_\theta&=&2 k^2\left[\frac{1-\mu}{\mu}\right]\Phi , \nonumber \\
U_\Sigma&=&\zeta\Phi. \nonumber
\end{eqnarray}
The coefficients $A_i$, $B_i$, $C_i$ and $D_i$ must now be chosen to
satisfy this condition.

It is understandable why this approach is popular, as it has a number
of benefits.  Firstly, it is applicable in the quasi-static regime the
arises when $k\rightarrow \infty$ (i.e. on small scales relative
to the cosmological horizon). Furthermore, the perturbation equations
now form a closed system.  This means that if, for example, we
restrict ourselves to a dust-filled universe then Eqs. (\ref{G}) and
(\ref{slip}), together with
\begin{eqnarray}
\delta_M'&=& -k^2 \theta_M+ 3\Phi'
\\
\theta_M'&=& - \adotoa \theta_M +\Psi  ,
\end{eqnarray}
form a complete system of differential equations that can be
straightforwardly solved.

It is instructive at this point to specify the various versions of the
simplified PPF formalism that are currently in use:
\begin{itemize}
\item Caldwell, Cooray and Melchiorri~\cite{Caldwell2007, Daniel2008} introduced the system\footnote{The initial convention was that 
 $\varpi={\varpi}_0 \frac{\Omega_{0DE}}{\Omega_{0M}} a^3$, but the
 authors changed their convention in subsequent publications.}
\begin{eqnarray}
 \mu &=& 1 \nonumber \\
\zeta &\equiv& \varpi = {\varpi}_0  a^3, \label{Caldwell1}
\end{eqnarray}
which was later extended to~\cite{Daniel2010}
\begin{eqnarray}
\mu&=&1+\mu_0 a^3 \label{Caldwell2}.
\end{eqnarray}
\item Bertschinger and
  Zukin~\cite{Bertschinger2006,BertschingerZukin2008} proposed a
  reduced parametrisation that takes into account the conservation of
  long wavelength curvature perturbations. Only one parameter is
  considered\footnote{The symbol used was actually $\gamma$, to which we have
  added a subscript "$BZ$" to distinguish it from other parameters
  with the same name. A similar approach will be used
  with the other frame-works presented here.}: 
\begin{eqnarray}
\gamma_{BZ} &=& 1- \zeta,
\label{BertschingerZukin}
\end{eqnarray}
that is further refined as $\gamma_{BZ} = 1 + \beta a^s$, where
$\beta$ and  $s$ are constants.  These authors make the additional
assumptions that $\gamma_{BZ}$ depends only on time, even on
sub-horizon scales, and that the Bertschinger long-wavelength
construction (see Section \ref{Bertschinger_construction}) can also be
extended to sub-horizon scales. The latter of these assumptions allows
them to solve for $\Phi$, from which $\Psi$ and $\delta$ are then
determined. The assumption of scale-independence is later relaxed
further, such that the following parameterisation can be made 
\begin{eqnarray}
\gamma_{BZ} &=& \frac{1 + \beta_1 k^2 a^s}{1 + \beta_2 k^2 a^s}
\nonumber \\
G_\Phi &=& \mu(1-\zeta)  = G  \frac{1 + \alpha_1 k^2 a^s}{1 + \alpha_2
  k^2 a^s}, \nonumber
\end{eqnarray}
where $\alpha_i$ and $\beta_i$ are constants~\footnote{We kept the original symbol $G_\Phi$ here, although
  under our conventions it would be more accurately written as
  $G_\Psi$, as it plays the role of an effective gravitational
  constant for a modified Poisson equation for $\Psi$.}. 
\item Amendola, Kunz and Sapone~\cite{Amendola2008} modify the Poisson
  equation and the slip by considering the two functions $Q_A$ and $\eta_A$:
\begin{eqnarray}
 Q_A &=&   \mu , \nonumber \\
 \eta_A &=& - \zeta , 
\label{amendola}
\end{eqnarray}
while a third variable, $\Sigma_A = Q_A(1 + \eta_A/2)$, is also
introduced in order to simplify the relevant expression for
gravitational lensing: $\Phi + \Psi = 2\Sigma_A\Phi$. A comparison
of this approach is made with DGP and scalar-tensor theories, both on
sub-horizon scales.
\item Zhang \etal~\cite{Zhang2007} make the parameterisation
\begin{eqnarray}
\eta_{ZL} &=&\frac{1}{1-\zeta}, \nonumber \\
\tilde{G}_{eff} &=& \mu(1 - \zeta/2).
 \label{ZLBD}
\end{eqnarray}
These authors also introduce the $E_G$ statistic, which we describe further below.
\item Zhao \etal~\cite{Zhaoetal2010} modify the Poisson equation using
  $\Psi$.  They introduce the two parameters
\begin{eqnarray}
\eta_{ZG} &=&\frac{1}{1-\zeta} , \nonumber \\
\mu_{ZG} &=&\mu(1-\zeta) ,
\label{Zhaoetal}
\end{eqnarray}
as well as a third derived parameter $\Sigma_{ZG} = \mu_{ZG}(1 +
\eta_{ZG})/2 = \Sigma_A$. Further refinements are made by considering
two specific models:  A case where $\mu_{ZG}=\mu_{ZG}(\tau)$ and
$\eta_{ZG}=\eta_{ZG}(\tau)$ transit from their GR values in the early
universe to modified constant values today, and a case where the two
functions are pixelised in the $\{\tau,k\}$ plane by $2\times2$ pixels
per function.
\item A parametrisation specifically designed for small scales was
  proposed by Amin, Blandford and Wagoner~\cite{Amin2008}, where
\begin{eqnarray}
 B_{A}(\adotoa_k,a) &=& (1 - \zeta) \mu  = \beta_0(a) + \beta_1(a)
 \adotoa_k + \beta_2(a) \adotoa_k^2 + \ldots 
\nonumber 
 \\
  \Gamma_A(\adotoa_k,a) &=& \mu  = \gamma_0(a) + \gamma_1(a) \adotoa_k
 + \gamma_2(a) \adotoa_k^2 + \ldots ,
\label{Amin}
\end{eqnarray}
while the matter density fluctuation is likewise parametrised by
\begin{equation}
 \delta_m(a,k)  = \delta(k)_i \left[ \delta_0(a) + \delta_1(a)
 \adotoa_k + \delta_2(a) \adotoa_k^2\right] ,
\end{equation}
where $\delta(k)_i$ is specified by initial conditions.  In this
approach different theories correspond to different sets of the
functions $\{\beta_i,\gamma_i,\delta_i\}$.  The authors find the
appropriate functions for $\Lambda$CDM, scalar-tensor theories,
quintessence models, $f(R)$ theories, and DGP.  They also note,
however, that not every theory can be adequately matched to this
expansion (e.g. k-essence).
\end{itemize}

The simplified PPF approach, though useful as a phenomenological tool
that can be used to constrain $\mu$ and $\zeta$, is not without its
problems.  For a start, it is not clear what theories a specific
choice of parameters actually encompasses.  In
\cite{Skordis2009,FerreiraSkordis2010,Baker2011} this problem has been
addressed in detail, and it is shown that if one wishes to consider
theories with second order field equations only then the PPF equations become:
\begin{eqnarray}
-2k^2\Phi  &=&  8 \pi \mu G a^2 \sum_X \rho_X \left[\delta_X+3(1+w_X) \adotoa \theta_X\right]
\label{Gext} 
\\  
\Phi-\Psi &=& \zeta\Phi+\frac{g}{k} \Phi' , \label{slipext}
\end{eqnarray}
with the constraint $\mu=1-\frac{g}{k}{\cal H}$.  In other words, if
the field equations of the theory are second order then the simplified
PPF approach is not applicable.

Let us now consider what the simplified PPF approach {\it does}
correspond to in the field equations. Taking the expansion for
$U_{\alpha\beta}$, and considering terms up to the lowest acceptable order, we have:
\begin{eqnarray}
U_{\Delta} &=& A_0 k^2 \Phi+ A_1 k \Phi' + E_0 k^2\Gamma , \nonumber \\
U_{\Theta} &=& B_0 k \Phi+ B_1  \Phi' + F_0 k\Gamma , \nonumber \\
U_{P} &=& C_0 k^2 \Phi+ C_1 k \Phi' + C_2 \Phi'' +  I_0 k^2\Gamma  +
I_1 k \Gamma' , \nonumber \\
U_{\Sigma} &=& D_0 \Phi .  \nonumber 
\end{eqnarray}
Applying the Bianchi identities one can then determine the above
functions in terms of $\zeta = D_0$ and $\tilde{g} = \frac{1}{\mu} -
1$.  This gives $F_0 = E_0 = I_0 = I_1 = 0$ (i.e. no  $\Gamma$ terms appearing), and
\begin{eqnarray}
 k B_0 = 2 k^2 \frac{\tilde{g}' + \adotoa(\tilde{g} + \zeta)}{
   3\adotoa' - 3 \adotoa^2 -  k^2} \; , 
   & A_0 = 2\tilde{g} - 3\adotoa_k B_0 \; ,
   & C_0 = 2\zeta + 3\left( \frac{1}{k} B_0' + 2 \adotoa_k B_0\right) ,
\nonumber \\
 B_1 =   2k^2\frac{\tilde{g}}{ 3\adotoa' - 3 \adotoa^2 -k^2} \;,
   & A_1 =  - 3\adotoa_k B_1 \;,  
   & kC_1 =  3\left( B_1' + 2 \adotoa B_1 + k B_0\right) . \nonumber
\end{eqnarray}
Since  $A_1 + 3\adotoa_k B_1=E_0=F_0=0$, the form of the Poisson
Equation (\ref{mod_Poisson}) is retained.  However, since $A_1$, $B_1$
and $C_2$ do not vanish, the field equations contain third time
derivatives in arbitrary gauge.  This is due to the presence of
$\Phi''$, and corresponds to a higher-order gravitational 
theory\footnote{Since the form of the Poisson equation is retained,
  however, both $\Phi$ and $\Psi$ remain non-dynamical.  Hence, the
  higher derivatives do not introduce additional propagating degrees
  of freedom.}.  In other words, one to simplify the equations for the
gravitational slip, but only at the cost of introducing higher-order
terms in the other evolution equations (to enforce consistency).

The above way of reconstructing the $U$'s from the simplified approach
is not unique. One can also relax the condition $\nabla^\alpha
T^E_{\alpha\beta} = \nabla^\alpha U_{\alpha\beta}=0$, but this moves
us further away from the simplified approach.

A further problem of the simplified approach is that it is impossible
to determine $\mu$ and $\zeta$ for a specific theory without solving
the field equations for a specific choice of initial conditions.  What
we would like to have is a one-to-one correspondence between the
functions that appear in the parametrised frame-work and those that
appear in the theories themselves, without having to solve the field
equations every time (just as in the PPN formalism, discussed in
Section \ref{GR}).  This can be achieved by constructing the field
equations as in Section \ref{the_formalism}.

\subsubsection{The Hu-Sawicki frame-work}

Hu and Sawicki introduced a frame-work~\cite{HuSawicki2007}, later
generalised by Hu~\cite{Hu2008}, that depends on a function of time
and space, $g_{HS}(\tau,k)$, two functions of time, $f_\zeta(\tau)$
and $f_G(\tau)$, and a constant, $c_\Gamma$.  Their frame-work goes
beyond the simplified PPF frame-works described above, and tends to
them in either the super-horizon or sub-horizon limits of the late
Universe\footnote{At late times the matter content of the Universe is
  effectively described by dust and dark energy, for which the dust
  has no anisotropic stress.}.  The field equations in this case are
written as
\begin{eqnarray}
\Phi+\Psi &=& -\frac{8\pi G a^2}{c_\kappa k^2}\sum_X \rho_X
\left[\delta_X+ 3\adotoa (1+w_X)\theta_X  \right] -8\pi G a^2\sum_X
(\rho_X  + P_X)\Sigma_X + 2 \Gamma_{HS}  ,
\nonumber \\
\Phi - \Psi &=&    8\pi G a^2 \sum_X (\rho_X + P_X)\Sigma_X   +
g_{HS}  \left(\Phi+\Psi\right)  ,
 \label{HuSawicki}
\end{eqnarray} 
where $c_\kappa = 1 - \frac{3\kappa}{k^2}$. The variable $\Gamma_{HS}$
is obtained by solving the differential equation 
\begin{eqnarray}
&&
2(1 + g_{HS})\left(1 + \frac{c_\Gamma}{\adotoa_k^2}  \right)\left[
  \Gamma_{HS}' + \adotoa \Gamma_{HS} + \frac{c_\Gamma k
  }{\adotoa_k}\left(\Gamma_{HS} + \frac{1}{2} f_G (\Phi + \Psi)
  \right)\right] 
\nonumber 
\\
&=& (2\adotoa g_{HS} - g_{HS}') (\Phi+\Psi)
+ 8\pi G a^2 g_{HS} \sum_X \left[\left(  (\rho_X + P_X) \Sigma_X
  \right)' + \adotoa (\rho_X + P_X) \Sigma_X \right] 
\nonumber 
\\
&&
- \frac{8\pi G a^2}{k^2}  \left[g_{HS}(1+f_\zeta) + f_G \right]\sum_X(\rho_X + P_X) \theta_X
+ \frac{8\pi G a^2}{k^2} (\rho_E + P_E)\theta_T ,
\label{GammaHS}
\end{eqnarray}
where $\theta_T = \frac{\sum_X (\rho_X + P_X)\theta_X}{\sum_X \rho_X +
  P_X}$ is calculated in conformal Newtonian gauge\footnote{In the
  original formulation the comoving gauge was used, but we translated
  the equations here to make it more consistent with the other
  approaches we described.}. In both Eq. (\ref{HuSawicki}) and
  Eq. (\ref{GammaHS}) $\rho_X$ and $P_X$ do not include contributions
  from the dark energy.  Taking the sub-horizon limit, for which
  $\adotoa_k\rightarrow 0$, we get that $\Gamma_{HS} \rightarrow
  -\frac{1}{2}f_G(\Phi + \Psi)$, and the system then reduces to the
  simplified frame-work with
\begin{eqnarray}
\mu  &=& \frac{1 + g_{HS}}{1 + f_G} ,
\nonumber \\
\zeta &=& \frac{2g_{HS}}{1 + g_{HS}} .
\end{eqnarray}
In the super-horizon limit the system obeys the Bertschinger
construction (see Section \ref{Bertschinger_construction}), 
with $g_{HS} = g_{HS}(\tau)$ while $f_\zeta(\tau)$ provides the next
order correction (beyond the Bertschinger construction).  We see that the sub-horizon limit depends only on
$g_{HS}$ and $f_G$, while the super-horizon limit depends only on
$g_{HS}$ and $f_\zeta$.  The constant $c_\Gamma$ controls the
transition scale between these two limits.

As in the simplified approaches, it is not clear from the outset what
kind of theories this frame-work encompasses. One can find fitting
functions $g_{HS}$, $f_G$ and $f_\zeta$ that reproduce the solutions
for specific theories, but only after some experimentation.

\subsection{Models for $\mu$ and $\zeta$ on Sub-Horizon Scales}

On sub-horizon scales the situation is greatly simplified. In this
case one can make a quasi-static approximation, and discard all time variations in
the perturbed fields.  Let us now consider this regime in a few
different models.

To start with we can consider $f(R)$ theories. Following
\cite{PogosianSilvestri2010} we then have
\begin{eqnarray}
-k^2(\Phi+\Psi)&=&4\pi \frac{G a^2\rho_M}{f_R}\delta_M , \nonumber \\
\Psi-\Phi&=&-\frac{2f_{RR}k^2}{3f_{RR}k^2+a^2f_R}f_R(\Phi+\Psi) , \nonumber
\end{eqnarray}
which can be rewritten in the $\mu$-$\zeta$ form given above. Note
that there is a scale in these equations, the ``Compton wavelength'' given by
\begin{eqnarray}
Q\equiv3 \frac{k^2}{a^2}\frac{f_{RR}}{f_R} . \nonumber
\end{eqnarray}
For large wavelengths $Q\rightarrow 0$, and $\Psi-\Phi\rightarrow 0$. For small
wavelengths $Q\gg 1$, and $\Psi-\Phi\simeq -(2/3) f_R (\Phi+\Psi)$. 

Let us now consider DGP models.  In this case the expression for
gravitational slip can be found to be \cite{KoyamaMaartens2006}
\begin{eqnarray}
\mu &=& 1 - \frac{1}{3\beta} , \nonumber \\
\zeta &=& \frac{3\beta -2 }{3\beta - 1} ,
\end{eqnarray}
where $\beta = 1 + 2\epsilon H r_c w_E$ (for more details see Section \ref{sec:dgp},
and in particular Eqs. (\ref{DGP_QS_Phi}) and (\ref{DGP_QS_Psi})).

Finally, let us consider scalar-tensor theories.  For a generalised
coupling parameter, $F(\phi)$, it can be shown that \cite{Amendola2008}
\begin{eqnarray}
\mu&=&\frac{G^*}{FG_{\rm cav}}\frac{2(F+F^{'2})}{2F+3F^{'2}} \nonumber \\
\zeta&=&\frac{F^{'2}}{F+F^{'2}} \nonumber
\end{eqnarray}

The expression above can be very useful in studying the evolution of
structure on observable scales.  They can, for example, be used to
correctly reproduce the growth rate of structure, and the effects of lensing, on
scales of up to hundreds of Mpc.

\subsubsection{The importance of shear}

An important issue arises if one wishes to distinguish between the
effects of dark energy and the effects of modified gravity. In
\cite{KunzSapone2007} the authors constructed an explicit example of
how anisotropic stress could mimic the effect of gravitational slip. 
To see this let us consider Eq. (\ref{modshear})
with $U_\Sigma=D_0\Phi$. There is clearly a degeneracy between the two
terms on the left hand side. In particular, if we focus on DGP then we
have that $\Sigma=0$, but $U_\Sigma = \frac{3\beta -2 }{3\beta - 1}
\Phi$\footnote{Actually, as discussed in Section \ref{sec:dgp}, DGP has shear but 
the quasi-static limit imposes the condition $\mu_E = 2 k^2 \Sigma_E$
between energy density associated with the perturbation of the Weyl
tensor and the shear.  This in turn generates the term $U_\Sigma$
above.}.   However, we can of course consider an alternative model
where gravity is unmodified, but the dark energy has anisotropic
stress given by $(1 + w_E)\Sigma_{E} = -\frac{1}{2k^2}\frac{3\beta -2
}{3\beta - 1} \delta_M $.   This would be an odd form of dark energy,
but it could also explain observations that may otherwise have been
attributed to gravitational slip. Distinguishing between such models 
is, of course, made easier when considering a wide variety of
different scales.

\subsubsection{The growth function}

The primary effect of modified gravity will be on the growth of
structure. The time evolution of the density field can be a sensitive
probe of not only the expansion rate of the Universe but also its
matter content. In a flat, matter dominated universe we have that
$\delta_M$, the density contrast of matter, evolves as
$\delta_M\propto a$. We can parametrise deviations from this behaviour
in terms of the growth function, $f$, given by 
\begin{eqnarray}
f\equiv\frac{\delta_M'}{ \adotoa \delta_M} , 
\label{f_growth}
\end{eqnarray}
or by the introduction of a parameter $\gamma$ through
\cite{Peebles1980,Linder2005a,Lee09a,Lee09b}
\begin{eqnarray}
f = \Omega_M^\gamma .
 \label{gamma}
\end{eqnarray} 
Note that $\gamma$ and $f$ are not parameters in the usual sense, but
are derived quantities (indirect observables, to some extent).  It is,
however, sometimes convenient to parameterise some results or
processes in terms of these quantities.

For standard growth in the presence of a cosmological constant one has
$\gamma\approx6/11$ to a very good approximation,  although this can
change with $\Omega_M$.  For the case of General Relativity with a
dark energy component with equation of state $P=w\rho$ (where $w$ is
constant) we have
\begin{eqnarray}
\gamma=\gamma_0+\gamma_1\Omega_E + O(\Omega_E^2)
\label{gamma_growth}
\end{eqnarray}
where $\Omega_E = 1 - \Omega_M$, and where $\gamma_0$ and $\gamma_1$ are given by 
\begin{eqnarray}
\gamma_0  &=&    \frac{3(1 - w) }{5-6w} , \nonumber\\
\gamma_1 &=& \frac{\gamma_0}{2w}  \left[  \frac{6w^2-7w+2}{5-12w}  -
  \frac{2-3w}{3}\gamma_0\right] . \label{highernohk}
\end{eqnarray}
In the case of a cosmological constant (i.e. with $w=-1$) these
expressions reduce to
\begin{eqnarray}
\gamma=  \frac{6}{11} + \frac{15}{2057} \Omega_E + \ldots , \nonumber
\end{eqnarray}
which gives the first order correction to the expression $\gamma
\approx 6/11$, given above.

A natural question to ask is how the growth parameter, $\gamma$,
depends on modifications of gravity.  There have been a number of
attempts at finding analytic expressions that relate $\gamma$ to
parameters in the underlying theory. This has focused on specific
theories, as well as the extended PPF approach. We now discuss some of
the results.

An expression for the growth parameter in the quasi-static limit of
$f(R)$ theories was found in \cite{ApplebyWeller2010}.  Here one can
define a time dependent mass scale:
\begin{eqnarray}
M^2(a)=\frac{1}{3{\bar f}_{RR}} , \nonumber
\end{eqnarray}
where ${\bar f}_{RR}$ is the second functional derivative of $f$ with
respect to the Ricci scalar, $R$, evaluated at the General Relativistic value of $R$.
A non-local expression for $\gamma$ can then be found of the form
\begin{eqnarray}
\gamma=\frac{6}{11}-\frac{\Omega_\Lambda k^2}{2\Omega_M}a^{-11/2}\int_0^a
\frac{{\tilde a}^{3/2}d{\tilde a}}{k^2+{\tilde a}^2M^2({\tilde a})} . \nonumber
\end{eqnarray}
One can see that for $k\ll aM(a)$ this expression for $\gamma$ remains the
same as in General Relativity, but that once $k$ crosses the mass
threshold modifications start to kick-in.

In DGP models the correct expression for $\gamma$ is given
by\footnote{Incorrect values for $\gamma$ in DGP models have been presented
  in~\cite{LinderCahn2007,Amendola2008,Wei2008}.}~\cite{FerreiraSkordis2010}
\begin{equation}
\gamma= \frac{11}{16} + \frac{7}{5632} \Omega_E - \frac{93}{4096}\Omega_E^2 + O(\Omega_E^3).
\end{equation}
This result is in excellent agreement with numerical studies (to
within $2\%$ or better for $\Omega_E<0.8$, and to $5\%$ for $\Omega_E
<0.9$).

\begin{figure}[htbp]
\begin{flushleft}
\vspace{-15pt}
\epsfig{figure=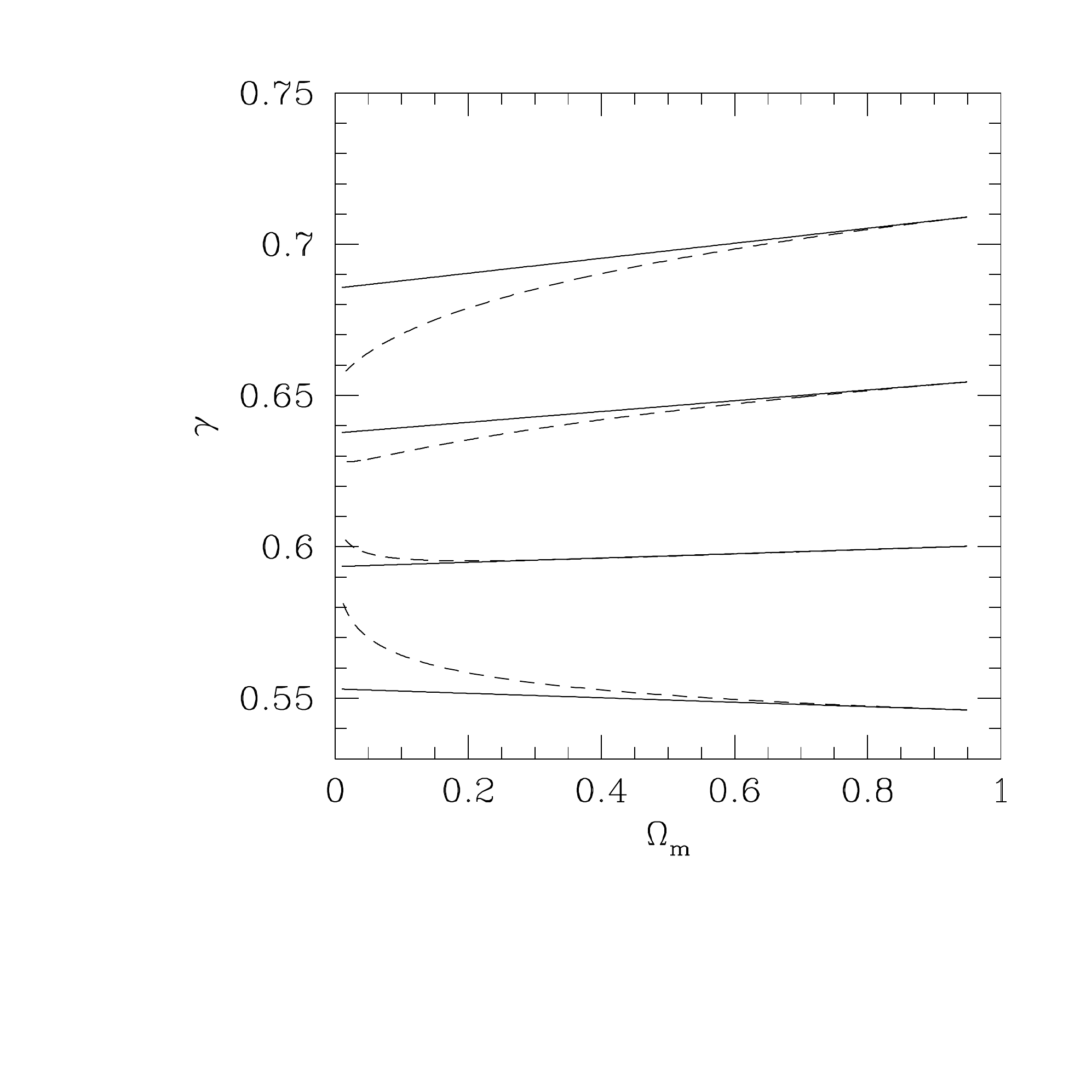,width=9cm}
\end{flushleft}
\vspace{-60pt}
\caption{The growth parameter, $\gamma$, for a selection of gravitational
slip parameters of the form $\zeta=\zeta_1\Omega_E$, as
a function of $\Omega_M$. The dashed curves are the numerical results for
 $\zeta_1=0$, $0.2$, $0.4$ and $0.6$ in ascending order, and the
corresponding analytic approximations are plotted as solid lines.}
\label{PPFFig1}
\vspace{-1pt} 
\end{figure}

An attempt at finding approximate analytic expressions for $\gamma$
was presented in \cite{FerreiraSkordis2010}. In this report we will
restrict ourselves to limiting cases where
$\zeta\simeq\zeta_1\Omega_E$.  If we then assume that we can Taylor
expand $\gamma$ as in Eq. (\ref{gamma_growth}) then we find the coefficients
\begin{eqnarray}
\gamma_0  &=&    \frac{3(1 - w + \zeta_{1}) }{5-6w} , \label{0gamma} \\
\gamma_1 &=& \frac{3w}{2}  \left[  - 3Y_1   -(2 - 3w) Y_1^2 + 4 Y_2
  \right] , 
\end{eqnarray}
where we have defined
\begin{eqnarray}
Y_1 &\equiv& \frac{1-w+\zeta_1}{w(5-6w) } ,
\nonumber 
\\
Y_2 &\equiv& \frac{ ( 1 - w  )  ( 15 w^2   - 4 w -  1 ) + \zeta_{1}  (
  9 w^2   + 2 w -  2 )}{2w^2(12w-5)(5-6w)} - \frac{\zeta_{1}^2
}{2w^2(12w-5)(5-6w)} .
\nonumber
\end{eqnarray}
A comparison between the approximation described above and the exact numerical
value of $\gamma$ is shown in Figure \ref{PPFFig1}.
More general expression for $\gamma$ that include the impact of $\mu$,
varying $w$,  higher-order terms in $\Omega_E$, as well as scale
dependent correction in $\adotoa_k$ are presented in
\cite{FerreiraSkordis2010}.

\subsubsection{Current constraints on the PPF parameters}

At the time of writing this report, there have only been a few
attempts at constraining the PPF parameters using existing
cosmological data sets~\cite{Beanetal2010,Daniel2010,Zhaoetal2010}.
Let us now consider each of these in turn.

In \cite{Beanetal2010} the authors find constraints on $Q=\mu$ and
$R=1-\zeta$ using the Union set of Supernovae from the
Supernovae Cosmology Project (with an additional sample), the joint
2dF and DR7 SDSS survey estimate of the Baryon Acoustic Oscillations,
the WMAP 7-year CMB Temperature and Polarisation data, the matter
power spectrum from the SDSS DR7 release, auto and cross-correlation
functions from 2MASS and the SDSS LRG catalogue with ISW and the
COSMOS weak lensing survey.

A scale and time dependent form of the PPF parameters is proposed:
\begin{eqnarray}
Q(k,a)=1+[Q_0e^{k/k_c}+Q_\infty(1-e^{k/k_c})-1]a^s , \nonumber \\
R(k,a)=1+[R_0e^{k/k_c}+R_\infty(1-e^{k/k_c})-1]a^s . \nonumber 
\end{eqnarray}
Setting $k_c\rightarrow\infty$ and $s=0$ the authors then find time
independent constraints on the parameters: $0.97<Q<1.01$ and
$0.99<R<1.02$ (both at the 95$\%$ confidence level). Their analysis
shows that primary constraints come from the WMAP7 data, and that
current weak lensing and cross correlations between galaxies and ISW
have a minimal effect.  Allowing time dependence greatly relaxes the
constraints on the parameters: $1.04<Q<2.66$ and $-0.22<R<1.44$ (both
at the 95$\%$ confidence level). If a transition scale of $k_c=0.01$
Mpc$^{-1}$ is considered, parameters are then constrained to be $
0.47<Q_0< 3.49$, $-0.80 <R_0<2.52 $, $ 0.97<Q_\infty<2.65 $, and $
-0.43<R_\infty<1.76 $.

The analysis of Daniel {\it et al} \cite{Daniel2010} extends the above analysis to include
an incarnation of the CFHTLS lensing survey, which has substantially
larger coverage than that of the COSMOS survey. An earlier release of
WMAP data is used, and cross correlations between large scale
structure and ISW are not included. The authors focus on
scale-independent parameters, $\zeta={\tilde \omega}={\tilde
  \omega}_0a^3$ and $\mu=\mu_0a^3$, and find $-1.4<{\tilde
  \omega}_0<2.8$ and $-0.67<\mu_0<2.0$. The authors also consider a
more elaborate time evolution of  both ${\tilde \omega}$
and $\mu$ by dividing it up into bins in the following redshift
ranges: (a) $[0,1]$ , (b) $[1,2]$ and (c)  $[2,9]$.  Setting ${\tilde
	  \omega}=0$ they then find that $\mu$ is constrained to be:
	$-0.074<\mu_{0a}<0.08$,  $-0.058<\mu_{0b}<0.14$,
	$-0.023<\mu_{0c}<0.22$.  Setting $\mu=1$ they find that
	$\tilde \omega$ is constrained to be: $-0.074<{\tilde
	  \omega}_{0a}<0.08$,  $-0.058<{\tilde \omega}_{0b}<0.14$,
	$-0.023<{\tilde \omega}_{0c}<0.22$.

In \cite{Zhaoetal2010} the authors consider two different
parameterisations. In the first they consider a scale independent but
time evolving parametrisation for $\eta_{ZG}$ and $\mu_{ZG}$, given by
Eq. (\ref{Zhaoetal}) as
\begin{eqnarray}
\mu_{ZG}(z)=\frac{1-\mu_0}{2}\left(1+{\rm tanh}\frac{z-z_s}{\Delta
  z}\right)+\mu_0 , \nonumber \\
\eta_{ZG}(z)=\frac{1-\eta_0}{2}\left(1+{\rm tanh}\frac{z-z_s}{\Delta
  z}\right)+\eta_0 . \nonumber
\end{eqnarray}
These authors use essentially the same data sets as in
\cite{Beanetal2010}, albeit with an earlier release of the WMAP data
(5 year) and without BAO constraints. They find that if $z_s=1$ then
the constraints are $0.65<\mu_0<1.9$ and $-0.41<\eta_0<2.18$, while if
$z_s=1$ the constraints are $0.68<\mu_0<1.11$ and $0.7<\eta_0<1.9$.

In a second parametrisation the authors bin the parameters in both
time and scale. They choose not to bin in $\eta$, but rather in
$\Sigma_{ZG}=\frac{\mu_{ZG}(1+\eta_{ZG})}{2}$.  This is the
combination that enters into the calculation of both the ISW effect
and weak lensing through their dependence on $\Phi+\Psi$. 
They consider the following bins: (a) $k\in [0,0.1]$, $z\in[1,2]$, (b)
$k\in [0,0.1]$, $z\in[0,1]$, (c) $k\in [0.1,0.2]$, $z\in[1,1]$, and
(d) $k\in [0.1,0.2]$, $z\in[1,2]$. The resulting constraints are then
as follows: In bin (a) $  0.96<\Sigma_1<1.04 $, $ 0.64<\mu_1<1.42 $,
in bin (b) $  0.93<\Sigma_2< 1.07$, $0.65 <\mu_2<1.34 $, in bin (c) $
0.58<\Sigma_3<1.02 $, $ 0.24<\mu_3< 2.24$, in bin (d) $
0<\Sigma_4<2.23 $, $ 0.05<\mu_4<2.46 $.

\subsubsection{Constraining the growth rate}

Over the past few years there have been attempts to target the growth
rate directly using redshift space distortions. This method involves
measuring either the redshift space correlation function,
$\xi^s(r_{||},r_\perp)$ (where the $r_{||}$ and $r_\perp$ correspond
to parallel or perpendicular directions to the line of sight), or the
redshift space power spectrum, $P^s_g(k)$.  The power spectrum can be
related to the real space power spectrum through an extension of what
is known as the `Kaiser formula':
\begin{eqnarray}
P^s({\bf k})=P({\bf
  k})[1+2\mu^2\beta^2+\mu^4\beta^2]G\left[\frac{k^2\mu^2\sigma_v^2}{H^2(z)}\right]
  , \label{PKRSD}
\end{eqnarray}
where $\beta=\frac{f}{b}$, $b$ is the bias factor, $G(x)$ encodes the non-linear
effect due to velocity dispersion, and $\mu$ is the cosine of the ${\bf k}$ vector
with the line of sight.  Note that $P^s(k)$ becomes anisotropic (as
does the correlation function) and it is through this anisotropy that
one can measure $\beta$ and hence $f$.

\begin{figure}[htbp]
\begin{center}
\epsfig{figure=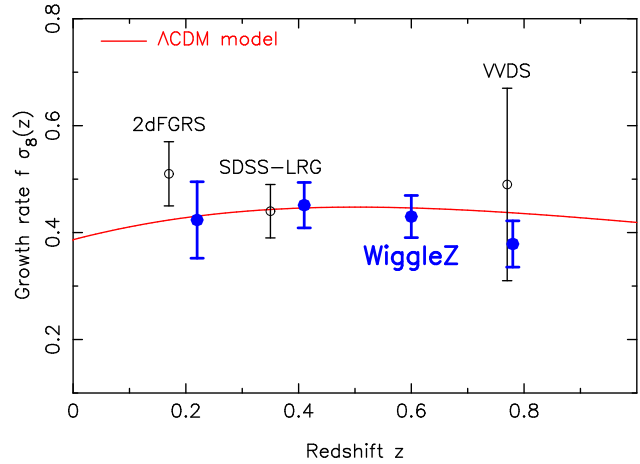,width=10cm}
\end{center}
\vspace{-10pt}
\caption{The growth rate, $f$, as a function of redshift, $z$, for a number
of different models. The symbols and error bars
correspond to constraints from VVDS, 2dFGRS and WiggleZ.  Taken
from~\cite{wigglez}.}
\label{PPFFig2}
\end{figure}

Until recently, measurements of $\beta$ were seen as constraints on
$\Omega_M$.  The reason for this is that in $\Lambda CDM$ we have
$\beta\simeq \Omega_M^{6/11}(z)/b$, and hence measurements of $\beta$
at different redshifts can be used to reconstruct the history of
$\Omega_M$.  In \cite{GuzzoEtAl2008} it is found that the wide part of
the VIMOS-VLT Deep Survey (VVDS) can be used to obtain $\beta
=0.70\pm0.26$ at $z\simeq0.8$. This result is then combined with the
constraint from the 2dFGRS, $\beta=0.49\pm0.09$ at $z\simeq 0.15$, and
the 2dF-SDSS LRG and QSO (2SLAQ) constraint (shown in Figure
\ref{PPFFig2}).  The emphasis in \cite{GuzzoEtAl2008} was on finding
deviations from growth rate in $\Lambda CDM$, and although not done
systematically, their  analysis showed that a number of specific
models could be ruled out.

The analysis of \cite{GuzzoEtAl2008} have led to a number of upcoming
redshift surveys focusing, in part, on redshift space
distortions. In particular, WiggleZ, FMOS, VIPERS, GAMA and BOSS,
many of whom were primarily targeting large scale structure and the
Baryon Acoustic Oscillations, will now all deliver constraints on $\beta$
in the near to medium future.

In \cite{Songetal2010} it was argued that measurements of the growth
of structure through constraints on the large scale peculiar velocity
dispersion, $\sigma^2_v$ (the large-scale redshift space distortion),
can be combined with weak lensing measurements to break the
degeneracies, and target $\mu$ more accurately. With measurements at
different redshifts it should be possible to reconstruct the time
evolution of $\mu$.

\subsubsection{The $E_G$ diagnostic}

The effect of modified gravity on the gravitational potentials, $\Phi$
and $\Psi$, can in principle be teased out with a careful combination
of measurements. In \cite{Zhang2007} it was argued that this should be
possible using a cross-correlation between an estimate of the local
velocity field from redshift space distortions, and the gravitational
potentials inferred from a lensing map from background galaxies.  More
specifically, from a more general form of Eq. (\ref{PKRSD}),
\begin{eqnarray}
P^s({\bf k})=P^s({\bf k})[P_g({\bf k})+2\mu^2P_{g\theta}({\bf
    k})+\mu^4P_\theta({\bf
    k})]G\left[\frac{k^2\mu^2\sigma_v^2}{H^2(z)}\right] , \label{PKRSD2}
\end{eqnarray}
it is possible to estimate the galaxy-velocity power spectrum,
$P_{g\theta}$, in bandpowers $P_\alpha$, and the galaxy-galaxy power
spectrum, $P_g$, in bandpowers $P_i$. From a weak lensing measurement
of the convergence
\begin{eqnarray}
\kappa=\frac{1}{2}\int_0^{\chi_s}
\nabla^2(\Psi-\Phi)W(\chi,\chi_s)d\chi , \nonumber
\end{eqnarray}
where $W$ is the lensing kernel, it is possible to construct a cross
power spectrum with galaxies, $C_{\kappa g}(\ell)\simeq \sum_\alpha
f_\alpha(\ell)P^{(2)}_\alpha$, where $P^{(2)}_\alpha$ is the band
power estimate of $P_{\nabla^2(\Psi-\Phi)g}$. These band powers can
all be collected into one estimator,
\begin{eqnarray}
{\hat E}_G\equiv
\frac{C_{\kappa g}(\ell,\Delta \ell)}{3H_0^2a^{-1}
\sum_\alpha f_\alpha(\ell,\Delta_\ell )P_\alpha} , \nonumber
\end{eqnarray}
where a band averaging over bins of width $\Delta\ell$ has been assumed. It can be shown that
\begin{eqnarray}
\langle {\hat E}_G\rangle=\left
	[\frac{\nabla^2(\Psi-\Phi)}{3H_0^2a^{-1}\beta\delta}\right]_{k=\ell/{\bar \chi}} .
\end{eqnarray}

The diagnostic $E_G$ will take different values depending on the
theory of gravity: $E_G=\Omega_0/\beta$ in $\Lambda$CDM and DGP (with different $\Omega_0$ in 
either case), and $E_G=\Omega_0/(1+f_R)\beta$ in $F(R)$.  In TeVeS
$E_G$ is significantly different from the $\Lambda CDM$ value, and is
scale dependent.  Hence, $E_G$ is in principle a good diagnostic of
the underlying gravitational theory.

In \cite{Reyesetal2010}, the authors attempted to extract an estimate of $E_G$ from
a sample of 70,205 luminous red galaxies (LRGs) from the SDSS. Their estimate
was concentrated at a mean redshift of $z\simeq0.32$ and they found 8 estimates of
$E_G$ across a range of between 2$h^{-1}$ to 50$h^{-1}$ Mpc, with a mean of
$E_G=0.40\pm0.07$. This should be compared with the $\Lambda$CDM prediction
of $E_G=0.408\pm0.029$, the $F(R)$ prediction of $E_G=0.328-0.365$, and an
approximate TeVeS prediction of $E_G\simeq 0.22$. The authors of
\cite{Reyesetal2010} have argued that this is evidence for the
validity of General Relativity on cosmological scales.
The results from \cite{Reyesetal2010} are somewhat tentative and
preliminary, but nonetheless promising.  A judicious choice of
cosmological parameters may indeed be able to tease out the
particularities of how $\Phi$ and $\Psi$ evolve in different theories
of gravity.  With planned mega surveys such LAMOST, LSST and SKA it is
likely that very tight constraints on $E_G$ will be achievable in the future.

\subsection{Forecasting Constraints from Future Surveys}

There is still quite a way to go in terms of constraining deviations
from General Relativity. As mentioned in the previous section, with
the planned mega surveys it should be possible to reconstruct the time
evolution of $\mu$ and $\zeta$ with ever increasing precision \cite{Dossett2010,Dossett2011}. With
the projected quality of the data, it also makes sense to
consider model independent parameterisations of $\mu$ and $\zeta$ using some form
of Principal Component Analysis (PCA). The authors in \cite{Zhaoetal2009} have done
just this, and attempted to quantify how many eigen-modes in an expansion of
$\mu$ and $\zeta$ can be accurately constrained with surveys such as
Planck, DES, LSST, SNAP and JDEM. They find that while with DES it should be possible to
constrain up to 20 independent eigen-modes, LSST will improve this by
a factor of 5, allowing detailed reconstruction of the time evolution
in modified gravity. They find that in the best case scenario LSST
will accurately constrain the time and scale dependence of $\mu$ and
$\zeta$ in the range $0.5<z<2$ and $0.04<k0.16$ h$^{-1}$ Mpc.

\newpage 

\section{Discussion}
\label{discussion}


In this review we have discussed recent advances in gravitational
physics, many of which have been driven by the discovery of
the dark side of the Universe in the late 1990s.  In this final
section let us now briefly consider the outlook for future gravity
research, and, in particular, what we consider to be some of the
upcoming highlights.

From a theoretical perspective, model building is an important part of
understanding and explaining existing data, as well as making
predictions for the future.  Tight constraints already available
on solar system and astrophysical scales, however, mean that model
builders are presented with a choice:  They can either study
minimal deviations away from General Relativity, or must otherwise look for
mechanisms that hide modifications to gravity on the scales probed by
experiment.  The former of these has value for understanding the
special nature of General Relativity, and the consequences of moving
away from it, while the latter provides an exciting opportunity to
try and solve some of the cosmological puzzles that have arisen with
the discovery of dark matter and energy.

Modified gravity necessarily involves additional fields, extra dimensions, or broken
symmetries, since we know that GR is the unique diffeomorphism
invariant theory of a single rank-2 tensor that can be constructed
from the metric variation of an action in four dimensions. An
important consideration is then whether or not these deviations manifest
themselves at the level of the background cosmology, or merely at the
level of perturbations. Of course, if we wish to account for dark
energy, or solve the cosmological constant problem using modified
gravity, these deviations must be manifest in the solutions of the
Friedmann equations.  We must also require, however, that they do not spoil the successful
predictions of the standard cosmology, such as the abundance of light
elements, the peak positions of the CMB acoustic spectrum, or the
predictions for baryon acoustic oscillations. This requires the
background FLRW cosmology of the modified theories to closely mimic the
standard evolution of $\Lambda$CDM from nucleosynthesis through to
matter domination.

Let us now consider the assumptions that go into the standard cosmology.
These include \cite{Bert-talk}
 \begin{itemize}
 \item Einstein's field equations.
 \item The universality of free fall.
 \item Local Lorentz invariance.
 \item Three spatial dimensions (below the electro-weak scale, at least).
 \item Conservation of energy-momentum.
 \item Homogeneity and isotropy of space on large scales.
 \item Matter fields being well modelled by fluids of dust and radiation.
 \end{itemize}
These assumptions, when confronted with observations, require the
presence of dark matter and dark energy.  From a particle physics
perspective this is very discomforting, as we have discussed in
Section \ref{sec:pitfalls}, as it requires the vacuum energy to be at
least $60$ orders of magnitude smaller than expected from particle theory.
This problem is exacerbated by the fact that contributions from the
zero-point energy of massive particles changes as the universe cools
and goes through phase transitions.  To circumvent the difficulties
that this entails, many model builders therefore assume some
unknown symmetry that sets the vacuum energy, plus any contribution
from a bare cosmological constant, to zero.  To account for cosmic
acceleration they then require modifications to gravity on large
scales\footnote{Some approaches, such as degravitation, (see Section
  \ref{sec:degrav}) even use modifications to gravity in order to
  screen the effects of the large vacuum energy, thereby removing the
  need for the unknown symmetry.}.

%

In some respects addressing the puzzles in cosmology is the easy part
of the model builders job.  More difficult is to do this in
a way that is consistent with the well established observations of
relativistic gravitational phenomena.  This often involves identifying a screening
mechanism that can work on solar system scales. The point here is that in
order to modify the cosmological evolution we have to deviate
considerably from GR in the IR, but IR modifications of gravity are
known to cause strong deviations from the predictions of General
Relativity for experiments that involve, for example, the bending or
time delay of light. Deviations on the scales that these observations
are made must therefore be suppressed.  Currently the most popular
examples of this type of mechanism are the Vainshtein mechanism and
the chameleon mechanism, both of which we have discussed in this
review, and neither of these is without its problems.   It is
therefore important for theorists to identify new ways to screen
the deviations from GR at short distances, and we expect this to be an
important avenue of future research.

It is also crucial to ensure that our models obey some basic requirements
of stability.  There is little motivation for carrying out complicated
simulations, and devising expensive experiments, to probe theories that
are fundamentally sick.  For example, we may choose to ask if the theory
contains ghosts, which can lead to catastrophic instabilities unless the
mass of the ghost lies above the effective theory cut-off, as
described in Section \ref{sec:intro-ghosts}.  Ghosts are rife in
modified theories of gravity, and dealing with them is one of the
principle challenges faced by model builders.  Another important thing
to establish is the cut-off for our effective classical theory.
Modified theories of gravity can become strongly coupled at lower
than expected energies, and enter a quantum fog.  If we want an
effective theory description of gravitational fields at the surface of the earth
to be classical, we need to impose that the cut-off is at least of the
order of a few $meV$, since classical gravity has been tested in the lab at
this scale.  Conservatively, one may also wish to impose that the
classical description of gravitational fields around the Sun should be
trustworthy, at least down to its Schwarzschild radius.

Let us now consider the prospect of future input into this field from
experiments and observations. Tests of Lorentz invariance and the weak
equivalence principle, that 
are already well established on Earth, are now being proposed and planned
as space missions.  Space offers a number of benefits over Earth-based
experiments, including a lack of seismic noise, and the fact that cooled
atoms stay in interferometers longer.  There are also theoretical
reasons for wanting to test these foundational issues of gravity in
space, such as the proposal by Khoury and Weltman that extra
gravitational degrees of freedom could have an environmental dependence.  Space
based tests of Lorentz invariance offer the possibility of improving
constraints on violations of this symmetry by orders of magnitudes
(see, e.g. \cite{Lorspace1,Lorspace2}).  Space based tests of
the equivalence are being planned by the French space agency CNES,
under the name MICROSCOPE (MICRO-Satellite \`{a} train\'{e}e
Compens\'{e}e pour l'Observation du Principe d'\'{E}quivalence), and
by ESA and NASA, under the name STEP (Satellite Test of the
Equivalence Principle).  These two missions promise to increase bounds
on violations of the weak equivalence principle to the level of $1$ part
in $10^{15}$, and $1$ part in $10^{18}$, respectively.

With regards to space based tests of metric theories of gravity, there
is also hope for further improvement on current bounds.  The
Bepi-Columbo Mercury orbiter being planned by ESA will, after a
two year mission, be capable of placing constraints on PPN
parameters of the order $\gamma-1 \sim 3 \times 10^{-5}$, $\beta -1
\sim 3\times 10^{-4}$, and $\alpha_1 \sim 10^{-5}$ \cite{bcmercury}.
The bounds on $\gamma$ could be improved further by Gaia, a high
precision space telescope that could constrain $\gamma-1$ to around
$1$ part in $10^6$.  Still greater constraints may be possible with
LATOR (Laser Astronomic Test of Relativity) \cite{latort}.  This
mission consists of two satellites that orbit the Sun at
$1$AU, and will have the potential of being able to constrain
$\gamma-1$ to around $1$ part in $10^8$, and solar frame-dragging to
the level of $\sim 1\%$.  Such constraints are orders of magnitude
greater than those that are currently available.

Lunar laser ranging has played an important part in testing gravity
over the past 4 decades, since retroreflectors were placed on the moon
by Apollo astronauts, and the Soviet Lunokhod rovers \cite{llrt}.
Improvements in ground based technology during this time have improved
the bounds on PPN parameters that reflected lasers have been able to
impose, but we are now reaching the stage where further gains
will be limited by the retroreflectors themselves.  Tightening the
current bounds using this technology will therefore require new space
missions, and, in particular, the possibility of planting
retroreflectors on other planets has great potential.  Laser ranging
of Mars over a $10$ year period would allow $\gamma-1$ to be
constrained to around $1$ part in $10^6$ \cite{llrt2}, and the
Nordtvedt parameter $\eta = 4 \beta-\gamma-3$ to the level of $\sim
6$-$2 \times 10^{-6}$ \cite{llrt3}.  Again, these are order of
magnitude improvements on current bounds.

Moving beyond the solar system, binary pulsars are an excellent test
of relativistic gravity, and offer the possibility of becoming more
constraining than solar system test in the near future \cite{lrpulsar}.
Pulsars also offer the opportunity to test gravity through the
emission of gravitational waves.  In particular, pulsar-white dwarf
systems have great potential to constrain the emission of dipolar
gravitational, which is a generic prediction of a large number of
modified theories of gravity.  Binary systems PSR J1141-6545,
J0751+1807 and J1757-5322 are all recently discovered pulsar-white
dwarf systems.  What is more, continued observation of existing
pulsars also offer the possibility of new tests of gravity, as, for
example, the perihelion of PSR B1913+16 precesses it may
soon allow for tests of the Shapiro time-delay effect.  Future
prospects for testing gravitational physics using pulsars are also
bright due to the large numbers of these objects that are expected to
be found by the Parkes, Arecibo, and Green Bank telescopes, as well,
of course, as the SKA (Square Kilometre Array).  The chance of
detecting a pulsar-black hole  systems increases dramatically with
large-scale observations of this kind. Such a system would be
potentially of great importance for testing strong field gravity.
Finally, the double pulsar PSR J0737-3039A,B also offers a unique test
of gravitational physics, with excellent prospects for improving constraints
on gravity in the future as observations of it continue.

In all of these areas it is likely that the bounds on deviations from
General Relativity will continue to be tightened, with lab tests too
promising continued improvement.  The E\"{o}t-Wash group at the University
of Washington, and others, continue to increase bounds at ever smaller
scales, and even particle experiments using the LHC at CERN are
looking for the signs of the extra-dimensions that are crucial for so
many modern theories of gravity.  The future prospects for
constraining all of these aspects of gravity means that the extra
degrees of freedom in modified theories of gravity will have ever
smaller regions of parameter space in which to hide.  This also,
of course, means that the possibility of making a detection of a
deviation from General Relativity is improved, if any such
deviations really do exist in nature.

It is, of course, the case that there have been tremendous developments in observational cosmology over the past couple of decades, and these observations have put gravity in the spotlight once again. With measurements of
the CMB, weak lensing, and galaxy surveys, as well as probes of the expansion rate
with distant supernovae $I_A$s, a strange Universe has been uncovered in which
more than $95\%$ of the energy budget is in some exotic dark form. The quality of these
observations are such that it is difficult to avoid such a conclusion {\it if} the
gravitational force arises from Einstein's theory of gravity. An alternative point of view is that
these observations are pointing at a flaw in our understanding of the behaviour of the Universe on the largest scales, and of behaviour of gravity at these distances in particular. Indeed, these observations may be a sign that we must think beyond Einstein's General Theory of Relativity, and his field equations.

In the same way that cosmological observations may be hinting at new physics in the gravitational
sector, they can also be used to constrain and even rule out alternatives. With a vast range of experiments
being planned and constructed throughout the world, it seems that we are a critical juncture
in the path to understanding gravity. At ESA, the satellite mission Euclid is
under assessment.  This mission could map out vast regions of space, probing the growth rate and morphology of large-scale structure, both key observables for constraining theories of gravity through their
effects on gravitational collapse. The SKA is in a planning phase with path finders being constructed on two continents. SKA will be a vast radio telescope that will generate a survey of up to a billion radio galaxies, mapping out the evolution of structure back to extremely high redshifts. These observatories will also be competing against the Large Synoptic Survey Telescope (LSST) that will image the sky over a period of ten years, building up a survey of galaxies that will primarily be used for weak lensing.  

These larger experiments will be complemented by smaller, more rapid surveys such FastSound, Weave, Boss, Viper, KIDS, DES and a host of other collaborations that will produce galaxy and weak lensing surveys on smaller scales, targeting different redshifts. Interestingly enough, while the original scientific driver for many of these surveys was to constrain dark energy, they have all taken on board the need to
test gravity. Indeed measuring the growth rate of gravitational collapse, as a test of modified gravity, has
become the core business in all of these surveys. 

Einstein developed his General Theory of Relativity almost a century
ago, and, although it remains a cornerstone of modern physics, one
could argue that of all the fundamental forces of nature it is gravity
that remains the least well understood.  This is almost certainly due 
to the weakness of the gravitational interaction, which makes it
incredibly difficult to test in the lab experimentally.
Inevitably, experiments on the scale of planets, stars, galaxies, and beyond
cannot be performed with the same level of precision and control as
those conducted for the other forces on Earth. Never the less, technology is now starting
to catch up with gravity.   The latter half of the twentieth century may have
belonged to the Standard Model of particle physics, but there is every reason to
suspect that the twenty first century will belong to gravity.


\newpage
\noindent
{\bf Acknowledgements}
\newline

TC and PGF acknowledge the support of the STFC, CERN, the BIPAC,
the Oxford Martin School and Jesus College, Oxford.  AP and CS are
supported by Royal Society University Research Fellowships.  For comments,
discussion and support we wish to thank Tessa Baker, John Barrow,
Cliff Burgess, Christos 
Charmousis, Ed Copeland, Kenny Dalglish, Gregory Gabadadze, Nemanja
Kaloper, Ian Kimpton, Kazuya Koyama, Ed Macaulay, Jo\~{a}o Magueijo, Gustavo Niz, Claudia de Rham,
Paul Saffin, Thomas Sotiriou, Glenn Starkman, Reza Tavakol, Anzhong Wang, Shuang Yong
Zhou, Tom Zlosnik, Joe Zuntz, and Jessica Padilla and her Mum.

\newpage
\noindent
{\bf References}
\newline



%


\end{document}